# THE INTERNATIONAL LINEAR COLLIDER

Technical Design Report | Volume 3.1: Accelerator R&D

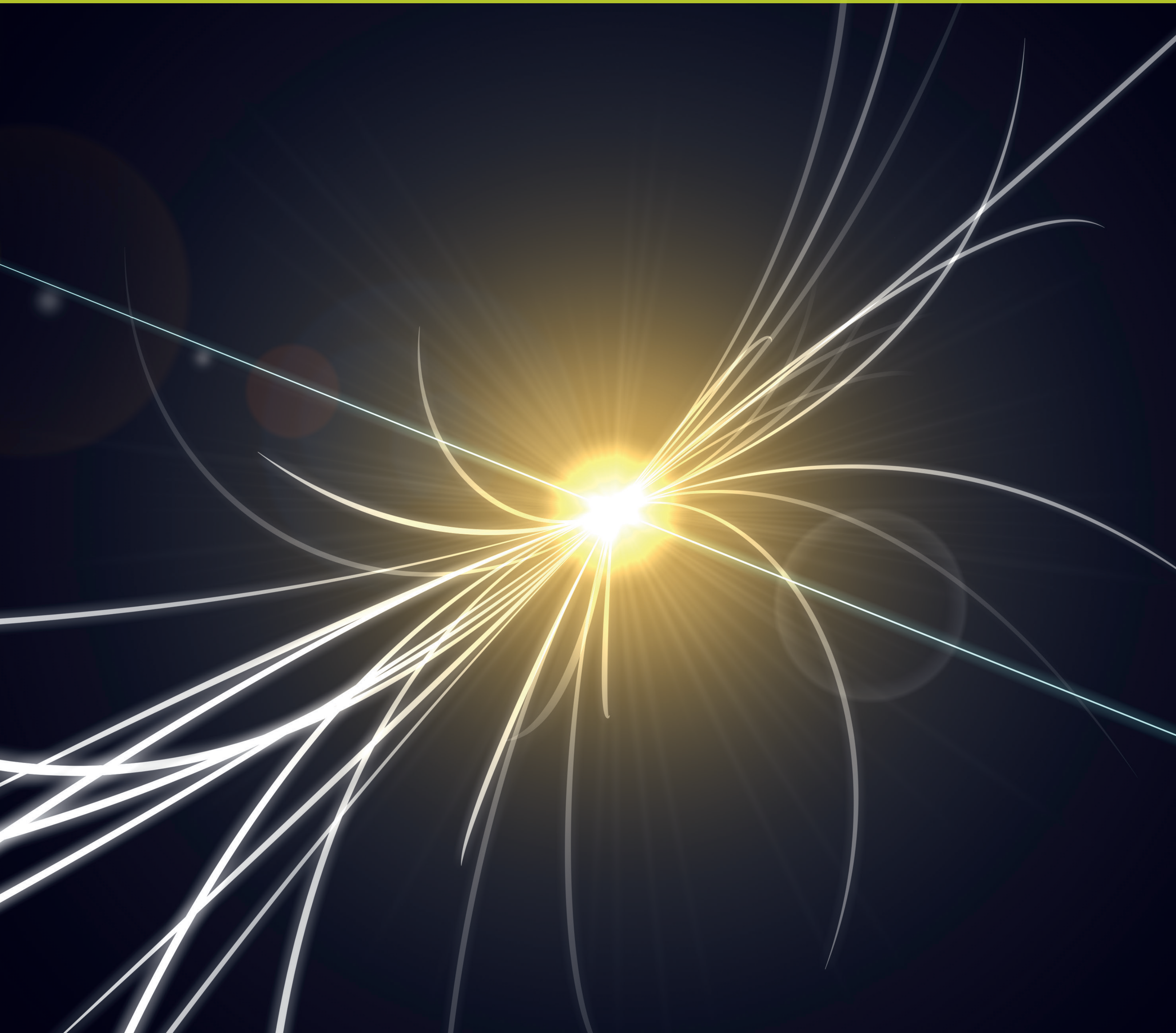

# The International Linear Collider

# Technical Design Report

2013

international linear collider

Volume 1: Executive Summary

Volume 2: Physics

Volume 3: Accelerator

Part I: R&D in the Technical Design Phase

Volume 3: Accelerator

Part II: Baseline Design

Volume 4: Detectors

# Volume 3

# Accelerator

## Part I

## R&D in the Technical Design Phase

Editors

Chris Adolphsen, Maura Barone, Barry Barish, Karsten Buesser, Phil Burrows, John Carwardine (Editorial Board Chair), Jeffrey Clark, Hélène Mainaud Durand, Gerry Dugan, Eckhard Elsen, Atsuto Enomoto, Brian Foster, Shigeki Fukuda, Wei Gai, Martin Gastal, Rongli Geng, Camille Ginsburg, Susanna Guiducci, Mike Harrison, Hitoshi Hayano, Keith Kershaw, Kiyoshi Kubo, Vic Kuchler, Benno List, Wanming Liu, Shinichiro Michizono, Christopher Nantista, John Osborne, Mark Palmer, James McEwan Paterson, Thomas Peterson, Nan Phinney, Paolo Pierini, Marc Ross, David Rubin, Andrei Seryi, John Sheppard, Nikolay Solyak, Steinar Stapnes, Toshiaki Tauchi, Nobu Toge, Nicholas Walker, Akira Yamamoto, Kaoru Yokoya

# Acknowledgements

We acknowledge the support of BMWF, Austria; MinObr, Belarus; FNRS and FWO, Belgium; NSERC, Canada; NSFC, China; MPO CR and VSC CR, Czech Republic; Commission of the European Communities; HIP, Finland; IN2P3/CNRS, CEA-DSM/IRFU, France; BMBF, DFG, Helmholtz Association, MPG and AvH Foundation, Germany; DAE and DST, India; ISF, Israel; INFN, Italy; MEXT and JSPS, Japan; CRI(MST) and MOST/KOSEF, Korea; FOM and NWO, The Netherlands; NFR, Norway; MNSW, Poland; ANCS, Romania; MES of Russia and ROSATOM, Russian Federation; MON, Serbia and Montenegro; MSSR, Slovakia; MICINN-MINECO and CPAN, Spain; SRC, Sweden; STFC, United Kingdom; DOE and NSF, United States of America.

# Contents

# Common preamble to Parts I and II

The International Linear Collider (ILC) is a linear electron-positron collider based on 1.3 GHz superconducting radio-frequency (SCRF) accelerating technology. It is designed to reach 200-500 GeV (extendable to 1 TeV) centre-of-mass energy with high luminosity. The design is the result of over twenty years of linear collider R&D, beginning in earnest with the construction and operation of the SLC at SLAC. This was followed by extensive development work on warm X-band solutions (NLC/JLC) and the pioneering work by the TESLA collaboration in the 1990s on superconducting L-band RF. In 2004, the International Technology Review Panel, set up by the International Committee for Future Accelerators, ICFA, selected superconducting technology for ILC construction. The Global Design Effort (GDE) was set up by ICFA in 2005 to coordinate the development of this technology as a worldwide international collaboration. Drawing on the resources of over 300 national laboratories, universities and institutes worldwide, the GDE produced the ILC *Reference Design Report* (RDR) [1–4] in August 2007. The report describes a conceptual design for the ILC and gives an estimated cost and the required personnel from collaborating institutions.

The work done by the GDE during the RDR phase identified many high-risk challenges that required R&D, which have subsequently been the focus of the worldwide activity during the Technical Design Phase. This phase has achieved a significant increase in the achievable gradient of SCRF cavities through a much better understanding of the factors that affect it. This improved understanding has permitted the industrialisation of the superconducting RF technology to more than one company in all three regions, achieving the TDP goal of 90 % of industrially produced cavities reaching an accelerating gradient of 31.5 MV/m. A further consequence is an improved costing and construction schedule than was possible in the RDR. Other important R&D milestones have included the detailed understanding of the effects of, and effective mitigation strategies for, the "electron-cloud" effects that tend to deteriorate the quality of the positron beam, particularly in the ILC damping rings. The achievement of the R&D goals of the TDR has culminated in the publication of this report, which represents the completion of the GDE's mandate; as such, it forms a detailed solution to the technical implementation of the ILC, requiring only engineering design related to a site-specific solution to allow the start of construction.

Volume 3 (Accelerator) of the *Technical Design Report* is divided into two separate parts reflecting the GDE's primary goals during the Technical Design Phase period (2007–2012):

**Part I: R&D in the Technical Design Phase** summarises the programmes and primary results of the risk-mitigating worldwide R&D including industrialisation activities.

**Part II: Baseline Design** provides a comprehensive summary of the reference layout, parameters and technical design of the accelerator, including an updated cost and construction schedule estimate.

The R&D results and studies of cost-effective solutions for the collider presented in Part I directly support the design presented in Part II, which is structured as a technical reference.

# Chapter 1
# Introduction

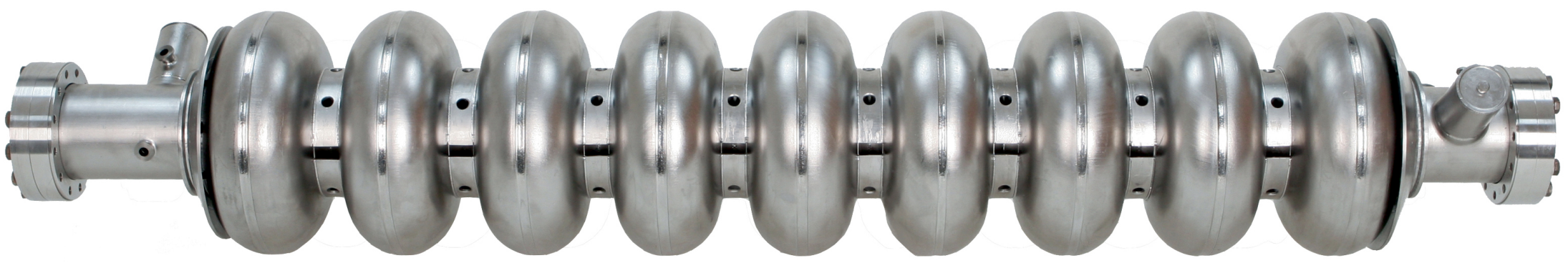

**Figure 1.1.** A superconducting nine-cell 1.3 GHz resonator (cavity).

The ILC *Reference Design Report* [3] published in 2007 documented a baseline design for the ILC that was by intention conservative. Several areas were identified as requiring further R&D both to reduce risk and to contain possible cost increases. This volume of the TDR concentrates on these R&D areas, the four highest priority items being:

1. SRF cavities capable of reproducibly achieving at least 35 MV/m.
2. A cryomodule consisting of eight or more cavities, operating at a gradient of 31.5 MV/m.
3. Linac string test (or integration test) of more than one cryomodule linac with beam.
4. Development of models and mitigation techniques for electron-cloud effects in the positron damping ring.

Other R&D areas (for example in the beam-delivery system and the sources) were also identified.

The first three priority R&D items all relate to the SCRF linear-accelerator technology, the primary cost driver of the machine. Although it was noted by the International Technology Review Panel [5] that TESLA SRF technology was 'mature', the ILC gradient goal had only been achieved in a handful of cavities (one of which had accelerated beam at 35 MV/m in the TESLA Test Facility at DESY – a proof of principle). One of the major technical aims has been the demonstration of large-scale production of reproducible high-gradient SCRF cavities, which required a detailed fundamental understanding of physics involved in the technology. During the five-year R&D programme, more than 200 cavities have been successfully manufactured and processed (Section 2.3). This required the development of the necessary fabrication and test infrastructure in all three regions (Section 2.2). In the USA, the sites were Fermilab, Argonne National Laboratory and Jefferson Lab. The Japanese site was KEK. In Europe, development has been driven by the design and construction of the European X-ray free-electron laser at DESY [6]. This major project uses technology very similar to that used in the ILC and therefore can be considered a large prototype.

An equally important programme has been the development of industrial capacity in each region, which has successfully resulted in multiple vendors in Asia, the Americas and Europe, each capable of producing high-performance ILC cavities. Cryomodule development and integrated systems testing is being pursued at all the primary SCRF sites: The FLASH FEL facility at DESY (Section 3.2) has successfully accelerated an ILC-like electron beam through high-gradient cryomodules, and has demonstrated many of the demanding tuning techniques required by the ILC; further system-test accelerators at Fermilab (NML) (Section 3.4) and at KEK (STF) Section 3.3) have been under development and will see beam operation in 2013. The S1-Global programme at KEK [7] successfully integrated cavities and auxiliary components delivered from DESY, FNAL, INFN and KEK into a single cryomodule, allowing direct comparative tests of different technologies as well as demonstrating the key concept of "plug compatibility", simplifying integration of parallel design efforts worldwide.

Other main-linac components developed by the R&D program include a tuneable high-power RF delivery system and associated low-level RF controls, and a next-generation solid-state modulator (Section 2.8).

During the ILC reference design phase, electron-cloud issues were identified as the major technical risk to the design luminosity. A multi-year study was therefore launched at the CESR accelerator at Cornell [8]. The well understood machine characteristics and highly flexible operating parameters of this facility facilitated an R&D program that has culminated in a definitive report on the physics of high-intensity, positively charged beams. The study also developed and identified acceptable mitigation techniques that are now included in the baseline design (Section 3.5).

The Accelerator Test Facilities (ATF, and ATF-2) at KEK have been successfully constructed and commissioned to demonstrate stable collisions of very small beams (Section 3.6). Other R&D which focuses on specific critical components (for example: polarised electron source; undulator, target and capture device for the positron source; fast kicker systems for the damping rings; high-powered beam dumps for the BDS), have also made significant progress and have demonstrated all key ILC parameters, either directly or via low-risk extrapolation (Chapter 4).

The remainder of this document focuses on describing the R&D programmes whose success is the foundation of the TDR reference design.

# Chapter 2
# Superconducting RF technology

## 2.1 Overview

The R&D for superconducting radiofrequency (SCRF) technology has been a high-priority and prominent global activity for the ILC in the Technical Design Phase (TDP) [9]. This activity builds on and extends the pioneering work accomplished in the decade leading up to the International Technology Review Panel's (ITRP) choice of SCRF technology in 2004 [10]. In that decade, R&D on 1.3 GHz technology carried out by the TESLA Collaboration[1] succeeded in reducing the cost per MeV/m by a large factor over the early 1990s state-of-the-art SCRF [11].

Four critical R&D topics were identified [12,13] by the ILC Program Advisory Committee and were adopted as the technical goals for the TDP. They are summarised in Table 2.1. (The notation *S0-S2* refers to the shorthand for the individual goals set at the beginning of the TDP). The goals include high-gradient operation in individual cavities (S0), assembly of a string of cavities in a cryomodule (S1), the test of a cryomodule with beam acceleration (S2), and to increase the involvement of industries in this development so that eventually major parts of the production can be carried out in industry.

**Table 2.1** The main goals and timeline for SCRF R&D established at the beginning of the Technical Design Phase

| Year | 2007 | 2008 | 2009 | 2010 | 2011 | 2012 |
|---|---|---|---|---|---|---|
| **S0:** Cavity gradient at 35 MV/m in vertical test | → yield 50% | | | | → yield 90% | |
| **S1:** Cavity string at average gradient of 31.5 MV/m in cryomodule | | Global effort for string assembly and test | | | | |
| **S2:** System test with beam acceleration including high- and low-level RF | | FLASH at DESY, ASTA/NML at FNAL, STF2 at KEK | | | | |
| **Industrialisation:** Study and preparation for industrial production of SCRF cavities and cryomodules | | | Production technology R&D | | | |

This chapter describes the major technical progress of the GDE program in pursuit of these goals. Highlights of the R&D are summarised in Table 2.2.

One of the immediate challenges of the SCRF R&D was to identify the technical causes for gradient limitations and to find means of eliminating such effects. The development of such remediation methods thus allows a baseline process for consistent production of 35 MV/m cavities to be defined. This process is needed to achieve both the high-gradient goal and to demonstrate a *production yield* of 90 % worldwide. This goal has been met in the TDP-phase 2 program: a yield of 94 % for cavity production above 28 MV/m and an average gradient of 37.1 MV/m has been achieved. The yield thus corresponds to 94 % for a cavity ensemble with an average gradient above 35 MV/m and complies

[1] now known as TESLA Technology Collaboration (TTC), see http://tesla.desy.de

**Table 2.2**
Main achievements of the SCRF R&D effort.

| Achievements |
|---|
| ▪ Understanding and mitigation of field emission at low gradient. |
| ▪ Deployment of tools to identify and repair quench-causing defects in low gradient cavities. |
| ▪ Establishment of a baseline sequence of cavity fabrication and surface preparation for ILC. |
| ▪ Achievement of a production yield of 94 % for cavities with an average gradient of 37.1 MV/m. The ensemble comprises of cavities with fields satisfying 35 MV/m ± 20 %. The yield for cavities with gradient above 28 MV/m (35 MV/m) is 94 % (75 %). |
| ▪ Achievement of an average field gradient of 32 MV/m in a prototype cryomodule for the European XFEL program. |
| ▪ Demonstration of the technical feasibility of assembling ILC cryomodules with global in-kind contributions. |

with the allowable gradient-spread specification of ±20 %. The yield for cavities with gradient above is 35 MV/m is 75 %. Larger statistics on the yield will soon be obtained from the European XFEL cavity production program, for which at least the first-pass processing is very similar to that of the ILC.

Associated systems such as high- and low-level RF, the cryomodule including quadrupoles and beam-position monitors, and cryogenics have to match these demands. Demonstrating these additional requirements has necessitated extending the SCRF infrastructure which was successfully constructed and commissioned at FNAL, ANL, JLab and KEK in addition to the existing infrastructure at DESY.

There were several noteworthy achievements in the HLRF R&D program. The modulator adopted for the baseline design was developed based on a Marx topology using solid-state switching in a modular design, which has many advantages including high availability, no need for a pulse transformer, vernier control and cost. The local power-distribution system evolved from the RDR to allow for a ± 20 % spread in cavity gradients. For the flat-site single-tunnel design the klystrons are moved to the surface and the RF outputs of clusters of klystrons are combined and transporting to the tunnel via low-loss waveguide for feeding long strings of cryomodules. R&D on this approach was based at SLAC, with primary goals of testing the feasibility of using over-moded waveguides for transporting the very high RF power levels and the ability to incrementally tap-off the power for the cryomodules. The LLRF controls were developed as part of the FLASH program and have exceeded the requirements for beam control.

Given the large number of institutions spread across the world that contribute to ILC SCRF R&D, it was considered important to establish the concept of *plug-compatibility*. This allows the R&D efforts in the collaborating institutes the flexibility to have variants of cavities, flanges etc. as long as they meet agreed interface definitions. A large number of technical interfaces have been defined to enable more efficient R&D. This international effort is exemplified in the construction of a cryomodule, *S1-Global* [14, 15] hosted at KEK, which facilitated the exploration of the plug-compatible design philosophies and comparisons of technologies.

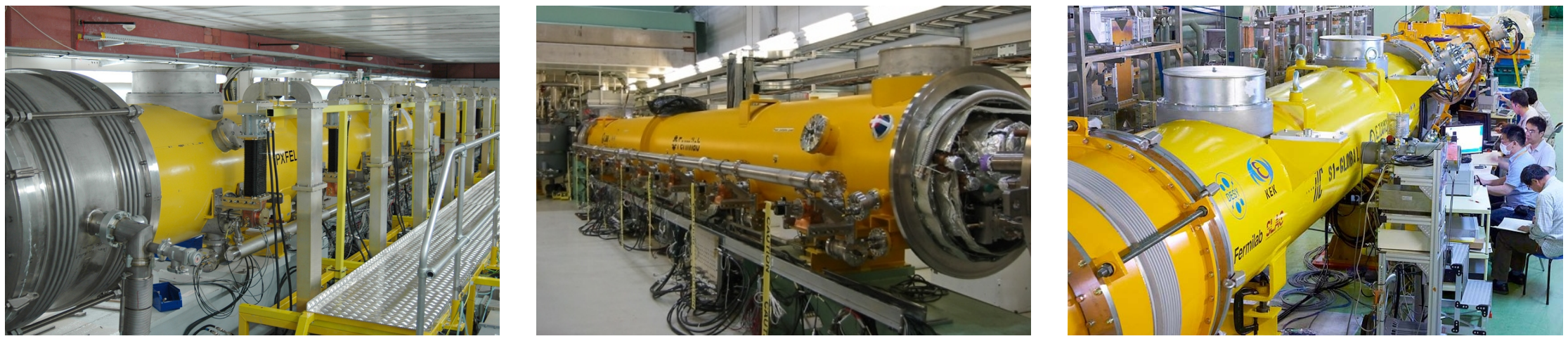

**(a)** PXFEL1 at DESY **(b)** CM2 at FNAL **(c)** S1-Global at KEK

**Figure 2.1.** Cryomodules assembled in all three regions.

The FLASH facility at DESY has been used to demonstrate the S2 goal of controlled beam acceleration in a multi-year series of dedicated beam study periods. More recently beam acceleration has also been achieved in the STF facility in KEK.

An impression of the widespread activities necessary to address the SCRF R&D goals can be obtained from the completion of cryomodules at DESY, FNAL and KEK (see Fig. 2.1). The cryomodule assembly at KEK stands out for its truly global nature. The cavity production facility at KEK is a one-stop facility for cavity production open to users and industry (Fig. 2.2).

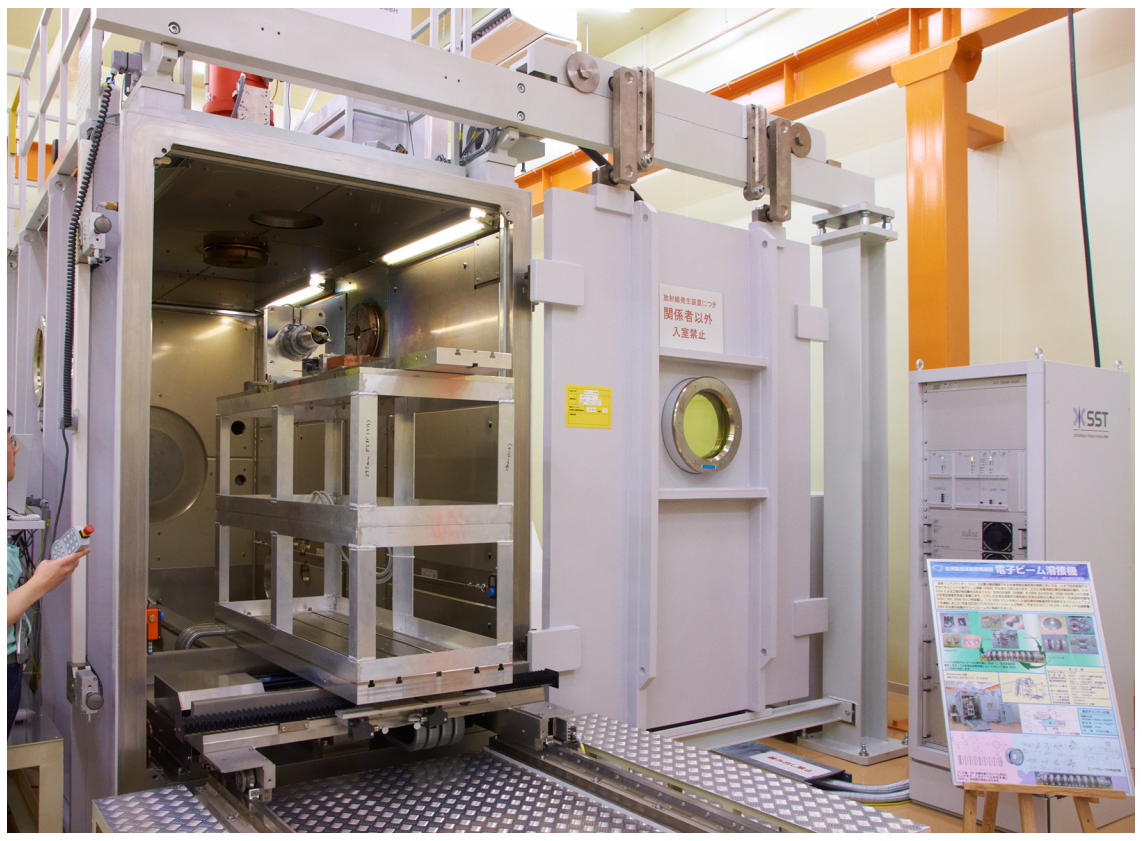

**Figure 2.2**
The Cavity Fabrication Facility at KEK showing the electron-beam-welding setup.

In parallel with the efforts in the laboratories, a technology-transfer program was launched with the goal of increasing the number of qualified vendors in the Americas and Asia regions. Several manufacturers are now engaged in the production of cavities. While originally only two companies could provide cavities meeting the ILC criteria, companies in all three regions are now successfully manufacturing high-gradient cavities. The number of successfully tested cavities achieving the ILC specification has now reached several dozen.

Several studies of SCRF mass-production models were commissioned from industry. These also included R&D for cost-effective industrialisation as an additional development towards a realistic proposal of the machine.

Overall the GDE SCRF R&D program has been very successful. It has achieved its technical goals and moreover demonstrated that tremendous progress can be made with an international effort in which all partners co-ordinate their respective programmes in order to achieve a common goal.

## 2.2 Development of Cavity Infrastructure and R&D

An important component of the worldwide ILC SCRF R&D program has been the design, fabrication, dissemination and utilisation of the best SCRF technologies across the world, such that not only are the best tools available at the originating laboratory or industry, but these best tools are available at all locations. It has only been through the close and cooperative interaction of SCRF researchers worldwide that the outstanding progress in cavity fabrication and performance has been realised in the past five years.

The following sections give an account of the installed infrastructure at the various laboratories participating in SCRF R&D worldwide, demonstrating how new tools or the systematic application of existing tools have produced the progress detailed in the subsequent sections.

### 2.2.1 Production and Test Facilities

Following the development of a standard processing scheme (Section 2.3), cavity-production infrastructure based on the best practices has been created and implemented in all regions. This has allowed for uniformity of treatment practices across the three regions, as well as the exchange of cavities between laboratories for tests and R&D. In addition, as processes became stable and a de facto standard, efforts have been made worldwide to move those manufacturing steps out to industrial partners. In the Americas and Europe this has led to the installation and commissioning of equipment directly at industrial partners; in Asia, this is accomplished through creation of a pilot plant at KEK that is available for industrial partners to use.

In Europe, the mass production of 800 cavities for the European XFEL at DESY has started and begins to provide important information regarding the control of processes in an industrial environment (Section 2.5). This development marks an important milestone since the cavity-treatment process outlined in Section 2.3 was agreed as suitable for mass production. In addition to the European XFEL order, the contracts include 24 cavities that will be available for additional treatment for highest gradients and will be supplemented with couplers and tuners. These cavities are part of the ILC-HiGrade project [16] that addresses the high gradient in the context of cavity mass production.

In the Americas, new chemical-processing facilities have been created at several labs. A joint Argonne/Fermilab facility (see Fig. 2.3) was based on the existing faciliities at JLab. In the Americas, approximately 100 9-cell cavities sourced from industries in both Europe and the United States are being used to verify, improve and simplify the production formula and improve infrastructure at the laboratories and capabilities in industry. The typical process for developing a vendor includes production and test of several single-cell prototypes and, after successful testing of these, progression to production of full 9-cell cavities. To give one example, in 2007 there was one vendor in the Americas qualified to build 9-cell cavities; this vendor has now not only tuned-up their production process to successfully make multiple nine-cell cavities, but a second vendor has recently had its first nine-cell cavity tested, reaching a gradient of 29 MV/m with no field emission. The progress in cavity-production capability in industry worldwide is shown in Table 2.3. In the six years since the launch of the programme, the number of qualified vendors has increased from two to five[2], while the number of laboratories capable of achieving ILC gradients has gone from one to five.

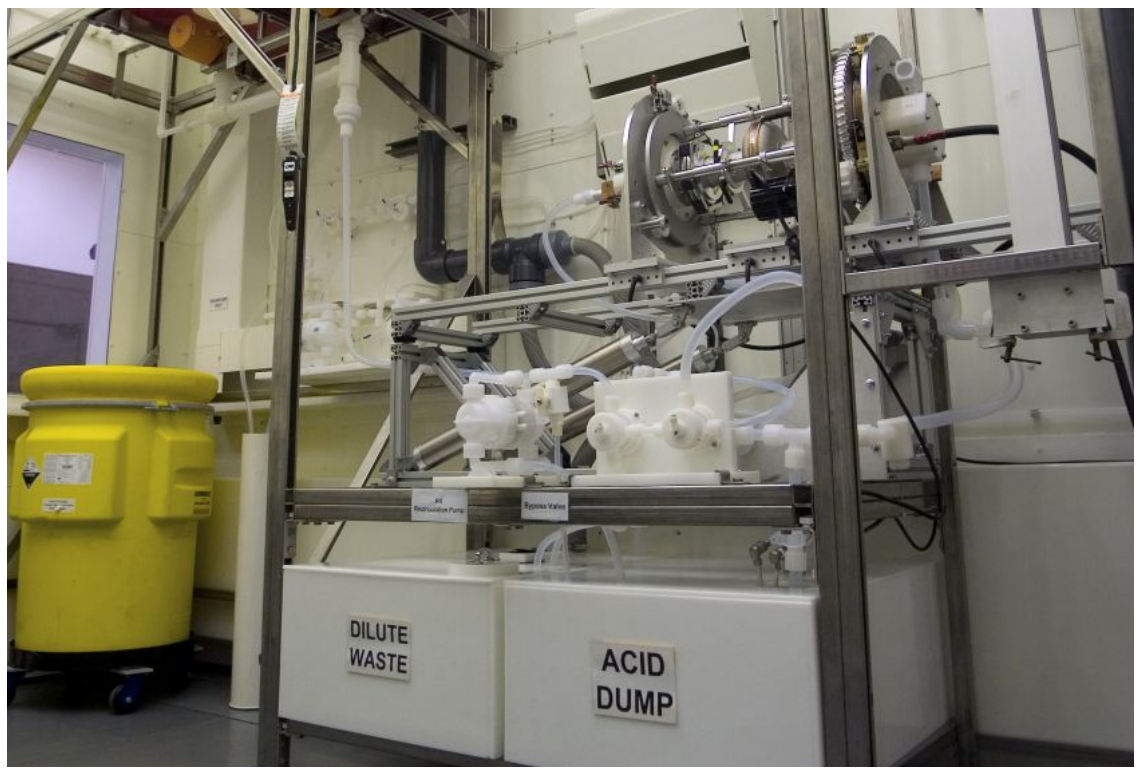

**Figure 2.3**
The joint ANL/FNAL chemical-processing facility.

The creation of more production facilities has included, beyond electro-polishing (EP) facilities, the creation of high-pressure rinsing systems and associated cleanrooms at ANL, and the commissioning of furnaces at Fermilab and Cornell to remove the last large bottleneck in the production of cavities. The engineering of this processing facility, and the standardisation of the chemical-processing steps, has now led to a chemical-polishing facility being assembled in industry, where it is currently being commissioned with the processing of nine-cell cavities.

[2] Advanced Energy Systems (AES), USA; ACCEL – now Research Instruments (RI), Germany; Zanon, Italy; Mitsubishi Heavy Industries (MHI) and Hitachi, Japan.

**Table 2.3**
The progress in qualifying vendors and their success in achieving the 35 MV/m gradient goal in fabrication.

| year | # 9-cell cavities qualified | Labs achieving 35 MV/m processing | Companies achieving 35 MV/m fabrication |
|---|---|---|---|
| 2006 | 10 | DESY | ACCEL, ZANON |
| 2011 | 41 | DESY, JLAB, FNAL, KEK | RI, ZANON, AES, MHI |
| 2012 | (45) | DESY, JLAB, FNAL, KEK, Cornell | RI, ZANON, AES, MHI, Hitachi |

In Asia a new EP facility was constructed and became operational in 2008 (Fig. 2.4) for the cavity process and treatment at STF. In addition to the EP system, the STF infrastructure includes a buffered chemical-polishing utility for small parts and flange-surface etching, ultrasonic rinsing, a high-pressure rinsing system and a cleanroom for cavity assembly. Various diagnostics are provided, and are extensive enough to carry out systematic studies of the EP parameters.

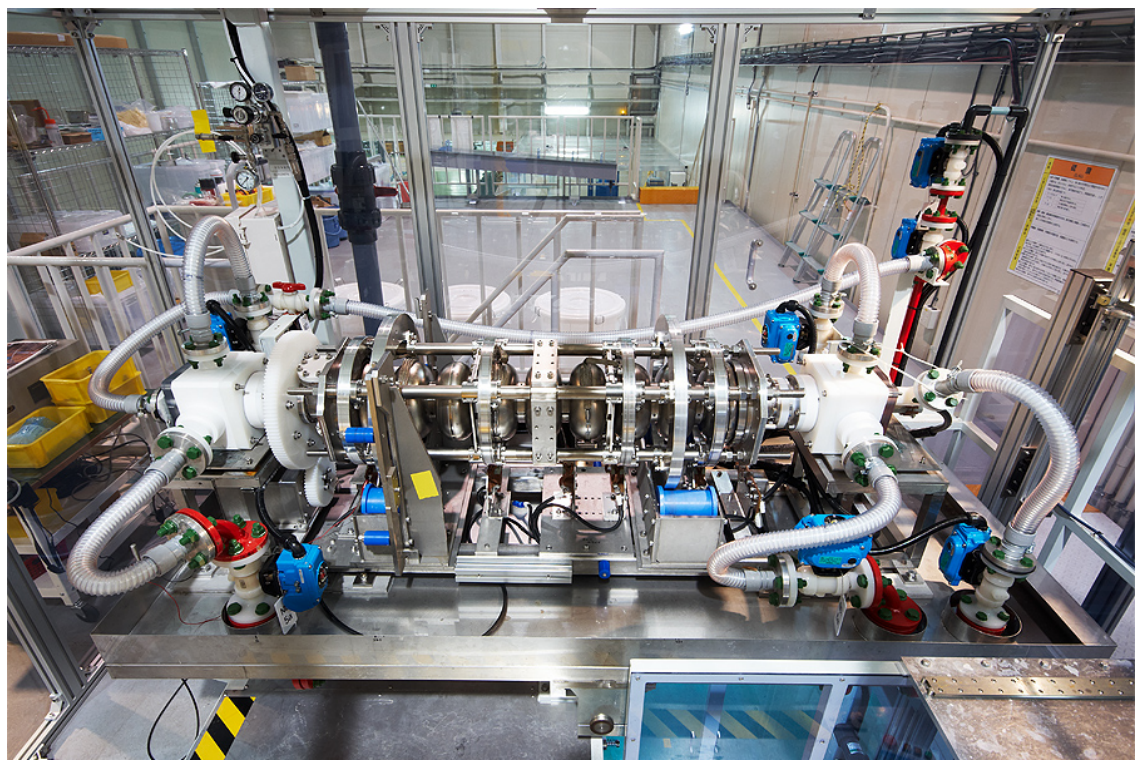

**Figure 2.4**
EP facility at KEK

At KEK, a pilot-plant fabrication facility has been created to study potential improvements to cavity fabrication. Although located at KEK, the facility is open to industries who wish to use its capabilities, including an electron-beam welder, a trimming machine, a press machine and a chemical pre-processing facility. Figure 2.5 shows the general layout. The cavities to be fabricated by using the facility are expected to be installed in cryomodules and tested at STF between 2013 and 2014 (Section 3.3).

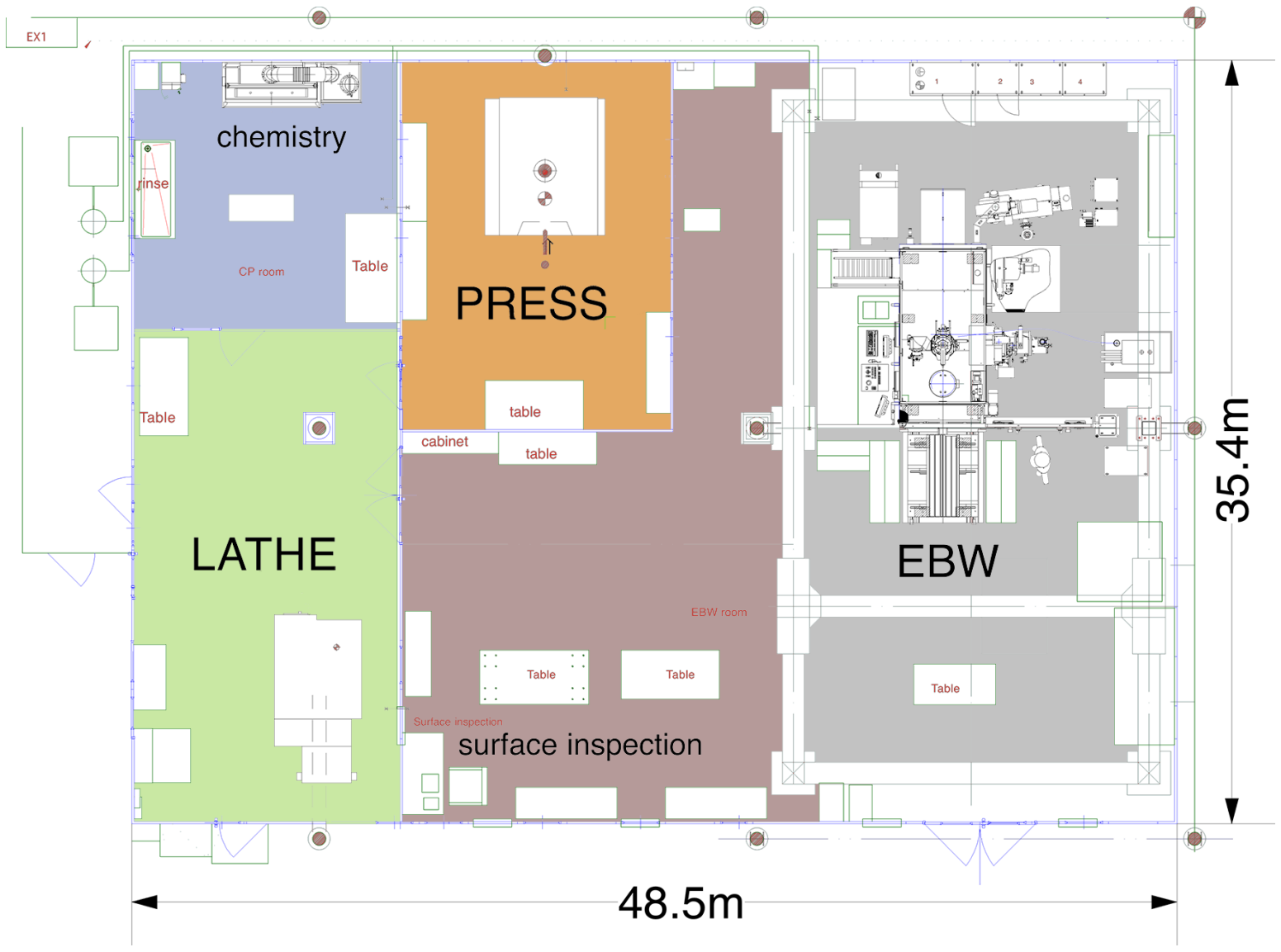

**Figure 2.5**
Layout of the KEK Fabrication Facility for cavities.

At IHEP, Beijing, an R&D program of 1.3 GHz SCRF technology has resulted in the completion of the first large-grain low-loss 9-cell cavity and installation of several key SCRF facilities including a mechanical-polishing machine, a chemical-etching tool, a high-pressure rinsing machine, and a cavity-inspection camera [17]. Beijing University (BKU) has proceeded to develop 1.3 GHz cavities with the primary objective of applications in photon science using facilities in Chinese industry.

At CEA-Saclay a system for vertical EP of cavities is being developed, based on the original Cornell system [18]. The system will simplify the critical surface-treatment process of electropolishing for superconducting niobium cavities. So far, most systems rely on a *horizontal* EP, which necessitates a mechanical rotation of the cavity, which can be avoided in the vertical system making it mechanically simpler. In addition a vertical system allows for easier draining of the processing acids. An initial design has been produced at Saclay/IRFU and is under test (see Fig. 2.6).

**Figure 2.6**
A photograph of the facility for vertical electropolishing at Saclay/IRFU.

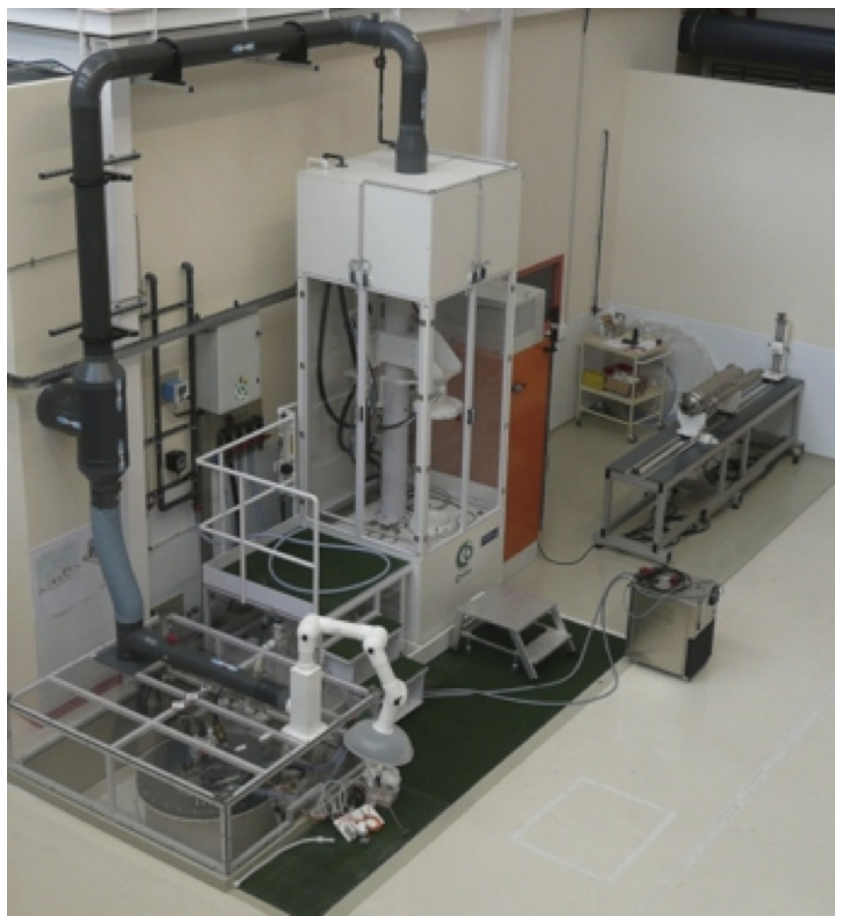

In addition to the standard EP infrastructure that now exists in all three regions, Fermilab has successfully set up a centrifugal mechanical or barrel polishing (CBP) facility, based on the concept originally proposed and tested at KEK in the 1990's [19]. Originally proposed as a remedial technique for low-performance cavities, the recent studies with the new infrastructure have demonstrated excellent performance, pointing the way to the possible use of CBP instead of bulk EP (see Section 2.3.3).

## 2.2.2 Inspection Infrastructure and Capabilities

One of the most important developments has been in the area of optical inspection techniques, both before and after cavity test. A leading cause of quench limits in cavities is the existence of surface defects near the weld equators, developing a high resolution, automated camera system was one of the priorities.

**Figure 2.7**
The setup for use of the Kyoto-KEK camera and an example image.

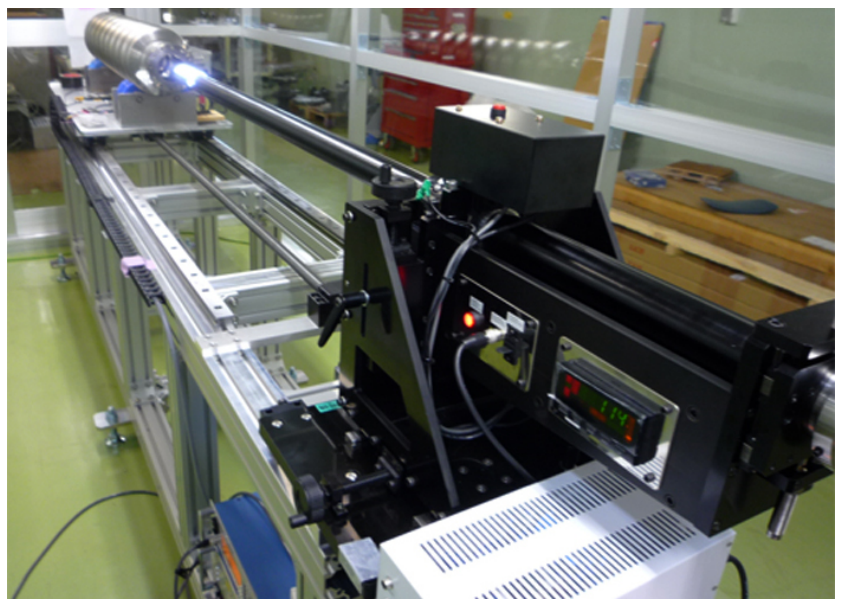

**(a)** Kyoto-KEK optical-inspection camera system.

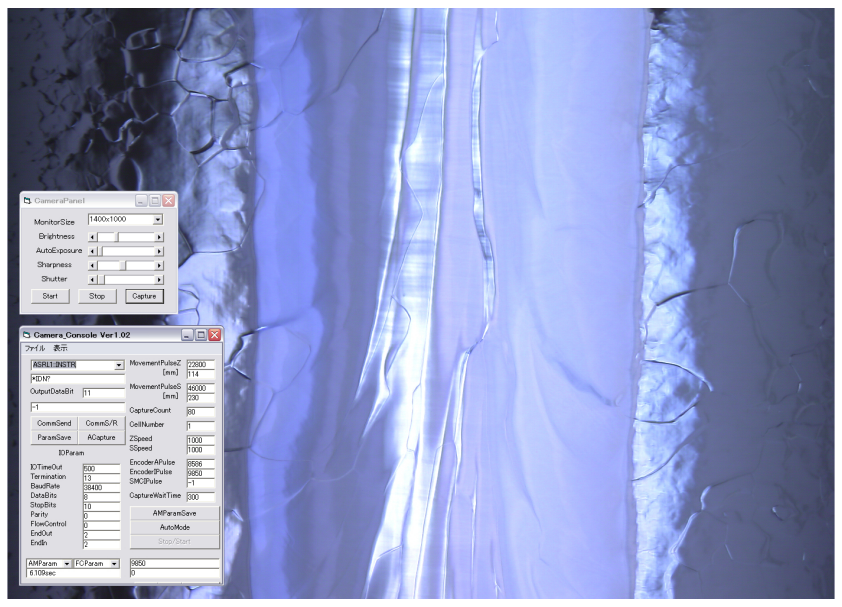

**(b)** Example of automated image-capture software displaying the area of the welding seam.

Two types of high-resolution cavity-inspection tools have been developed in recent years under the globally coordinated ILC gradient R&D effort. The most utilised camera system was developed in a collaboration of KEK and Kyoto University [20]. The camera can resolve surface features down to 10 µm. Some depth profile information can also be obtained. A newly developed high-performance

CMOS camera, when combined with an LED illumination system, resulted in a ten-times increase in brightness and much improved resolution. The other type is based on imaging by a combination of a long-distance microscope and a tiltable mirror inserted into the cavity [21]. Automated image-capturing instruments have been put in place at Cornell, DESY, FNAL, JLab and KEK and defect-finding and surface-characterisation software is being further optimised. Figure 2.7 shows one implementation of the camera system for cavity-surface inspection, together with an example of a high-resolution image of the electron-beam weld.

The Kyoto-KEK camera system is used now for inspection of the interior of most cavities available for research and optimisation. It also plays a vital role in qualifying production lines in industry for the European XFEL. Additionally, in conjunction with measurements taken during tests, the system is used to correlate quench positions with surface irregularities. These correlations allow the use of local remediation techniques to repair the surface and mechanically remove the performance limitation. An ongoing effort [22] is to correlate the irregularities seen before test with the test results, such that the camera can be used in a predictive, rather than reactive, manner.

The camera system is also used to characterise the surface properties in a quantitative manner [22]. Estimators for surface roughness, typical feature size etc. derived from image processing serve to enable the objective comparison of surface-treatment methods.

In addition to the optical inspection camera systems, several labs have developed infrastructure and techniques to examine cavity surface irregularities the surface chemical characteristics. These include (among others) the replica moulding technique [23, 24], and techniques for microscopically analysing samples cut from sacrificed cavities (see for example [25]).

Finally, Fermilab and Kyoto-KEK are investigating in X-ray tomography to identify voids in the cavity material. Tomography has the advantage over the camera system that it finds defects buried below the surface; as some of the limiting surface features are believed to be due to diffusion of sub-surface defects during welding, or voids revealed during processing, tomography is a promising technique for the quality assessment of incoming half-cells before welding in the cavity-manufacturing process.

### 2.2.3 Cavity-Tuning Facilities

**Figure 2.8** The tools in place for the European XFEL for the mechanical adjustment of cavities.

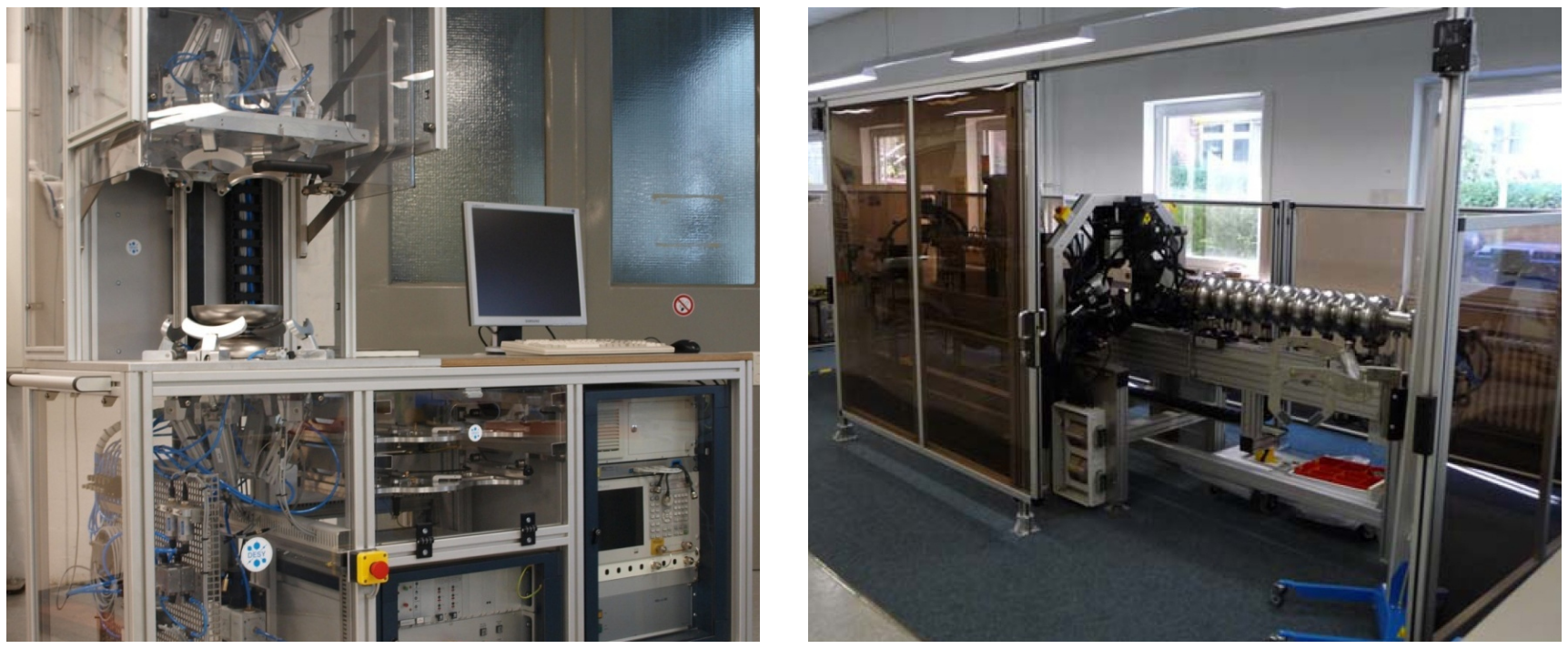

**(a)** HaZeMeMa, the tool for room-temperature RF measurement of half cells

**(b)** The tool for field-flatness tuning and geometric adjustment of 9-cell cavities.

Automatic cavity-tuning machines are also now operational worldwide. Originally designed at DESY [26], the current versions have been created in collaboration by DESY, KEK and Fermilab [27]. The machine has the capability of tuning field flatness and correcting cell-to-cell alignment through linked operation between the automated software, bead-pull, six-jaw cell deforming, and the system to measure cavity eccentricity. In conjunction with the European XFEL production, two European manufacturers have been supplied with tools certified to European standards to perform the tuning.

These automated tools can be operated by non-RF experts and will simultaneously provide a standardised log of the recorded data. Figure 2.8 shows the automated room-temperature RF measurement of half cells ("HaZeMeMa") and the tuning machines in use in the production for the European XFEL.

## 2.2.4 Quench Detection

Two possible approaches have been developed to detect localised quenches in a cavity. A second-sound system, originally conceived and developed at Cornell [28, 29], uses oscillating super leak transducers (OST) and the signal of the phase transition of superfluid helium propagating in the helium to triangulate the quench location at the cavity outer surface during vertical test. The system is relatively simple and locates the quench location to a precision of a couple of centimetres. Presently, Cornell OSTs are implemented in many laboratories such as DESY, FNAL, JLab and KEK.

Alternatively, extensive *temperature mapping* systems [30–35] have been developed and are in widespread use in various implementations. In all cases, carbon resistors are placed on the outer surface of a cavity and register heat fluxes due to localised RF heating at the quench location on the inner surface of the cavity. Mapping is made either by fixing a net of resistors in place on the cavity [30], or by rotating a linear array of thermometers azimuthally around the profile of the cavity [31]. Depending on the origin of the quench and the density of resistors, the system identifies the location of the hot spot to within a few millimetres.

Both the second-sound and temperature-mapping techniques, used in conjunction with the Kyoto-KEK camera, have had great success in identifying surface flaws that have caused the quench (see Section 2.3.3).

## 2.2.5 Test Infrastructure and Measurement Techniques

The first 2 K cold test of a cavity is usually carried out in a cryostat housed vertically in a cylindrical dewar. Such vertical test facilities are now in place at multiple laboratories worldwide. For the European XFEL, the first cold test of the incoming 9-cell cavities will be made at the Accelerator Module Test Facility (AMTF), shown in Fig. 2.9. The cavities (including helium tank) will be lowered into the vertical cryostat in sets of four as shown in Fig. 2.10a to minimise the mounting effort and the number of cooling cycles.

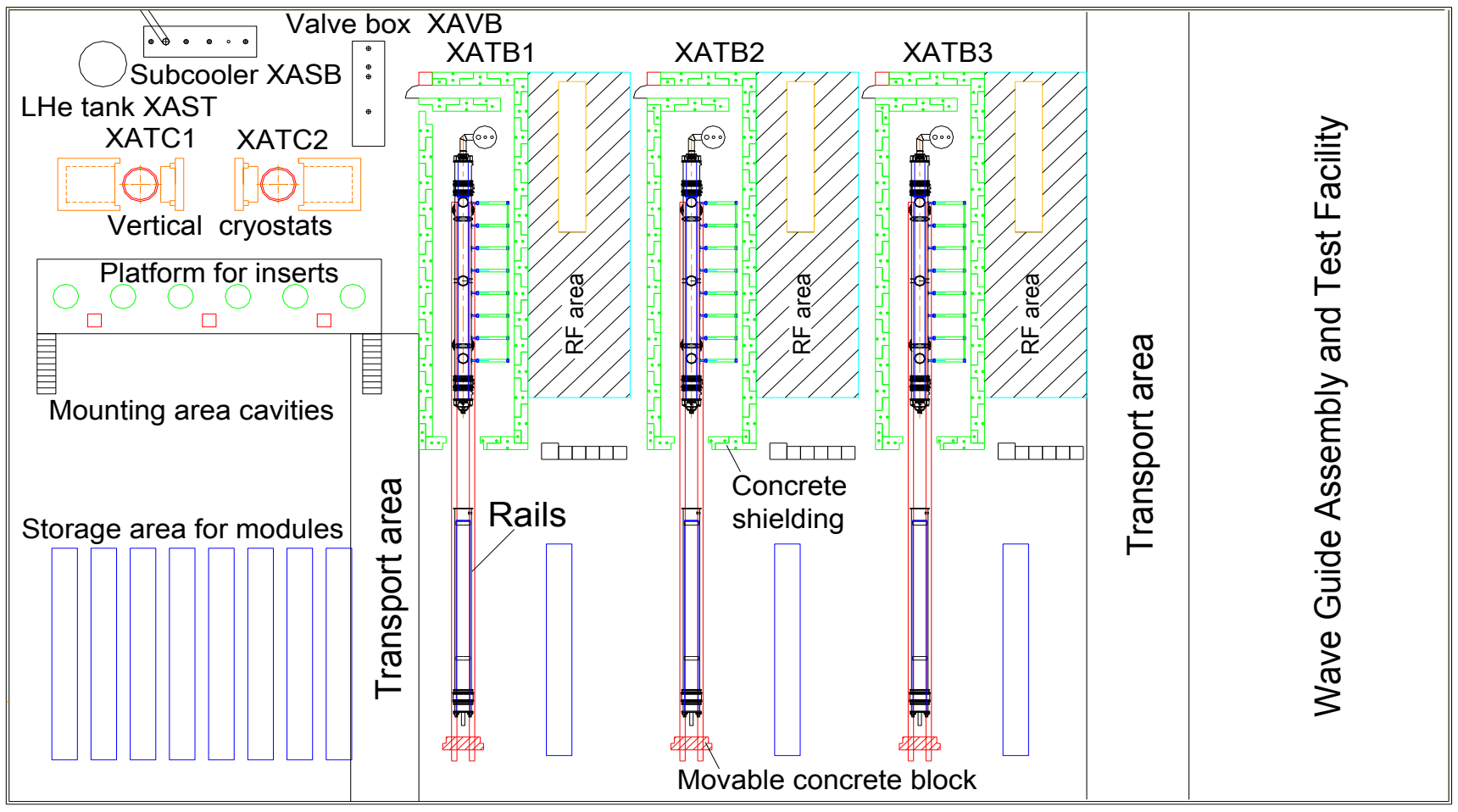

**Figure 2.9** Layout of the Accelerator Module Test Facility hall at DESY.

The facility has been planned for minimal physical handling of the cavities to guarantee high throughput and reproducibility. The test procedure was standardised and aims for rapid characterisation of the cavity performance. Cavities failing the standard test ($\pi$-mode) will be further examined in the vertical test stands where additional RF-modes can be excited for detailed investigations of the cavity.

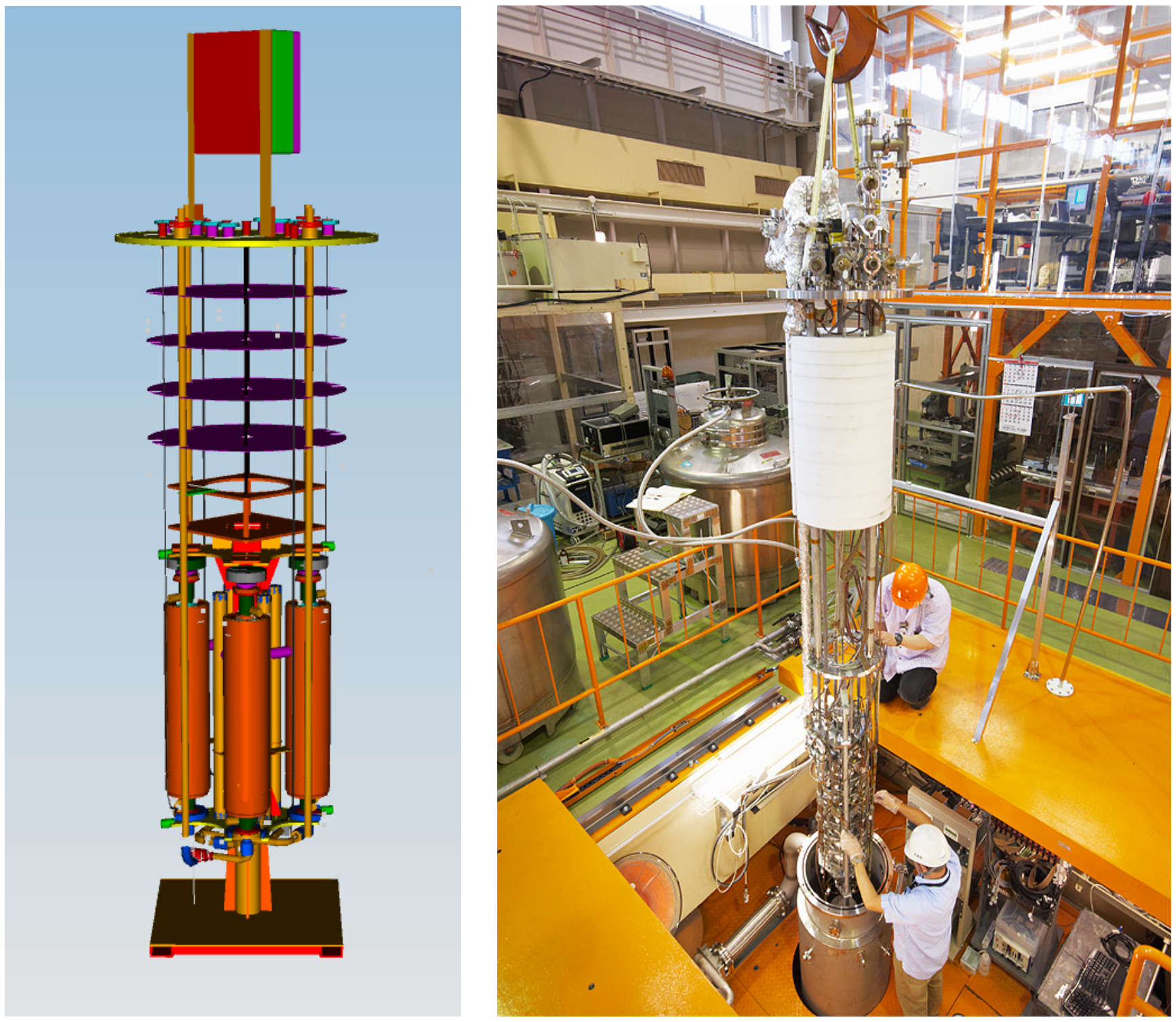

(a) (b)

**Figure 2.10**
(a) Four cavities, with their He vessels, ready to be lowered into the vertical cryostat. (b) The vertical test stand established at KEK-STF

The latter vertical cryostat has been equipped with OST sensors for quench detection.

The first vertical test system at Fermilab was commissioned in 2007. The Fermilab facility is now capable of two cool-downs per week for either 9-cell or single-cell tests [36]. This single dewar has averaged 80 test cycles in the past 2 years, and in 2013 two new, larger dewars will be brought online.

At STF, the vertical test stand for low-power cavity-field testing was commissioned in 2008. For the vertical test, the first 4 m-deep cryostat was assembled together with radiation shielding and a helium-pumping system. The test stand is now routinely operated once per week. Figure 2.10b shows the vertical test facility at KEK-STF.

The Horizontal Test Facility (HTF) at Fermilab was brought online in 2008 and has been used at a rate of approximately one test per month for qualifying the cavity-dressing process before assembly into an ILC-style cryomodule, known as CM-2 [37]. HTF allows for an intermediate system test of the jacketed cavity and coupler system, before string assembly.

### 2.2.6 Diagnostics for Field Emission

It is expected that field emission will continue to be a limiting factor in high gradient cavities. Furthermore, limits on dark current in the linacs will effectively set acceptance criteria for field emission during production of the cavities. Monitoring systems for field-emission remain relatively immature across the various test infrastructures worldwide; developing a reliable measurement system as well as techniques for cross-calibration of infrastructure remains an important on-going R&D item. The most common diagnostic for field emission is the detection of Bremsstrahlung X-rays by radiation sensors placed above the top plate of the cavity-testing dewar. This convenient diagnostic provides information about the onset of field emission and the strength of field emitters. Alternatively, radiation sensors such as silicon diodes are placed adjacent to the cavity (hence immersed in liquid helium). These can be implemented as an X-ray mapping system, either fixed [33] or rotating [35]. The energy spectrum of Bremsstrahlung X-rays is sometimes measured by using scintillators and associated electronics. The end-point energy corresponds to the highest impact energy of field-emitted electrons. Since impacting electrons also deposit heat into the cavity wall, linear heating patterns will appear on the temperature map. The information from X-ray diagnostics, when combined with numerical simulations of electron trajectories, provides insights into the location of field emitters.

### 2.2.7 Remediation Techniques

The ability to mechanical repair the identified cavity surface defects which can limit the performance of a cavity is important in achieving the required production yields at high gradients. Several methods have been developed and applied in various locations worldwide.

Local mechanical grinding has proven very successful in remediating defects which cause cavities to quench (Section 2.3.3) . A miniature grinding mechanism with a very small CCD camera that fits in the cavity has been developed for the purpose as shown (Fig. 2.11).

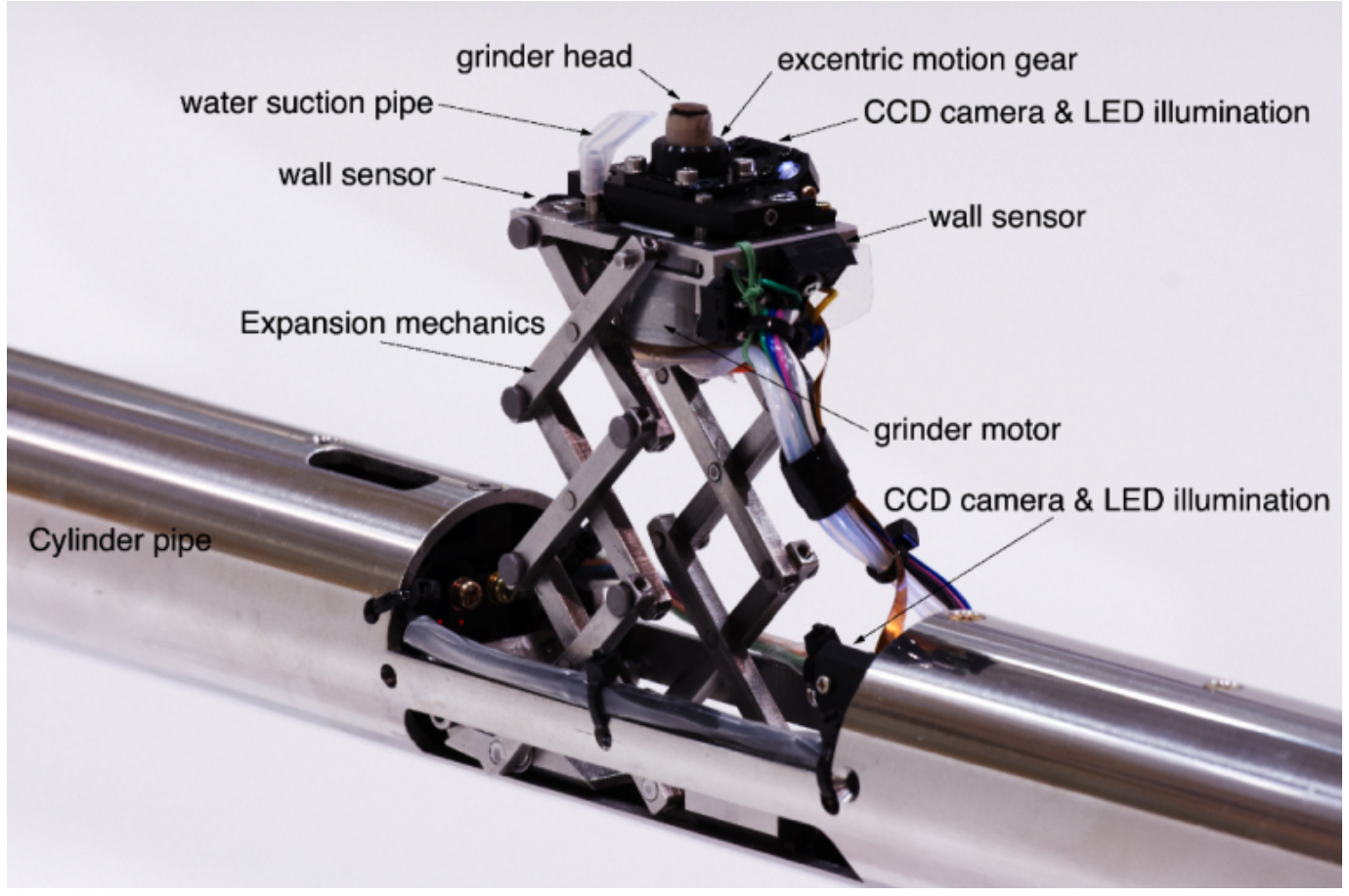

**Figure 2.11**
Local grinding tool with an expandable motor stage installed in a 50 mm-diameter cylindrical housing.

CBP also started as a remediation technique, although it was known that this implies a reset of the whole cavity surface, not just a local area. Consequently new problematic surface features may become uncovered. After several cavities of relatively low performance were treated with the tumbling technique, many of the imperfections, or pits disappeared. The technique has therefore been used for proactive repair before the first processing and test of a cavity at FNAL.

### 2.2.8 Infrastructure for High-power Couplers

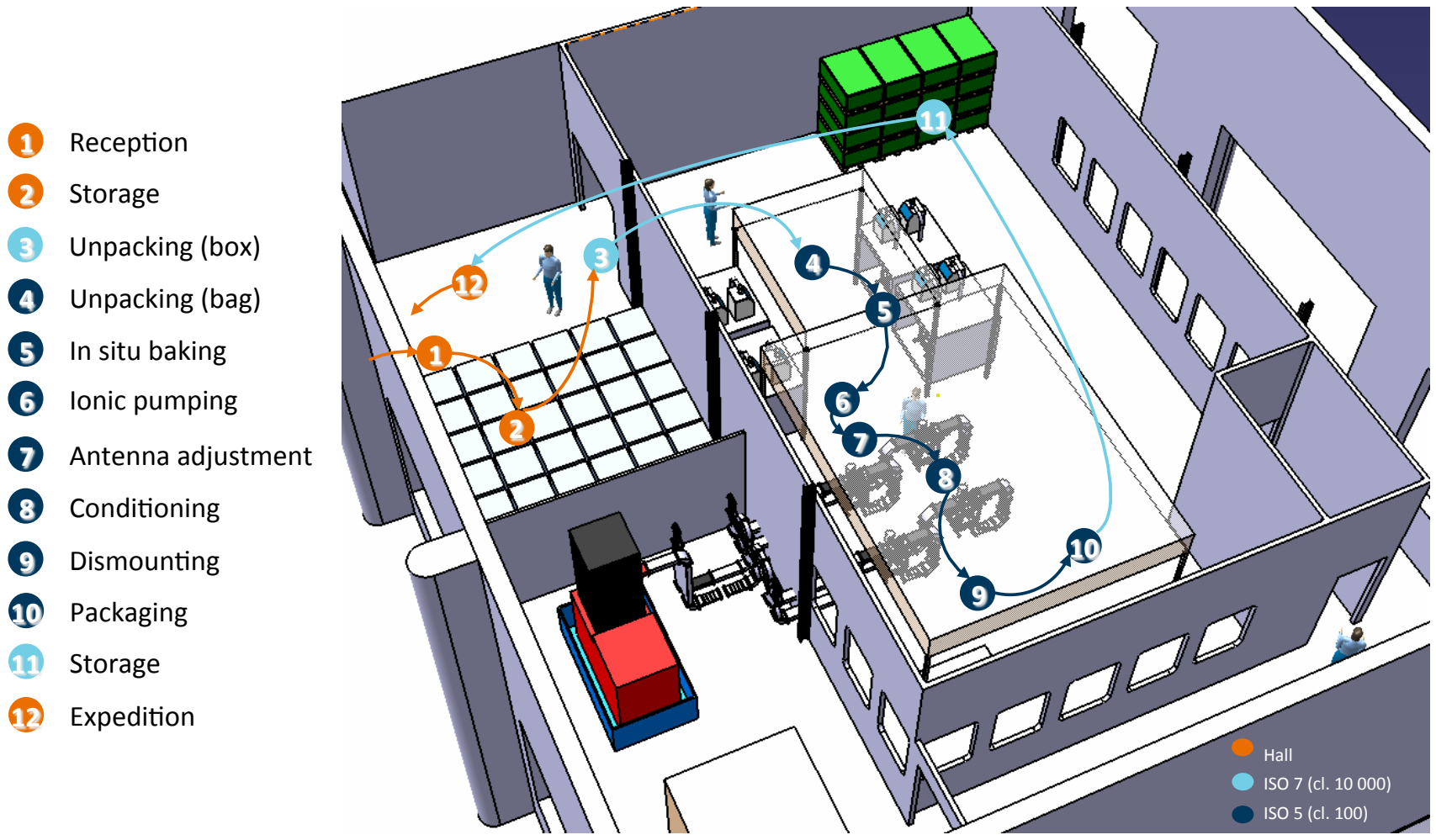

**Figure 2.12**
Layout of the coupler-conditioning facility at CNRS/LAL.

The high-power input couplers for the cavities transfer the RF-power from the waveguide system into the cavity. They provide the power transition between warm exterior and cold interior of the cavity. Since the power loads vary through the couplers of a cryomodule they have to be adjustable and hence constitute a delicate and costly piece of the SCRF equipment. The couplers are now produced in industry.

The assembled couplers have to undergo a careful and time consuming burn-in procedure, during which they are slowly conditioned. Tests at LAL had shown that the procedure could be dramatically shortened in an automatic set-up. The installed infrastructure enabling the test of the mass-produced couplers is shown in Fig. 2.12. An overall speed-up of the conditioning by a factor of four has been achieved by driving several cavities in parallel from the same RF source. Such a setup necessitates an elaborate RF network so that each coupler can be conditioned up to 5 MW. The capacity is expandable: instead of the currently foreseen four coupler pairs that will be tested in parallel, the system could be extended to process eight pairs. The processing time of a single coupler is typically a week.

As a contribution to the US SCRF cryomodule programme centred at FNAL, SLAC has been responsible for inspection, cleaning, assembly and high-power RF processing of industry-produced TTF-III-style fundamental power couplers [38]. Couplers were mounted in pairs side-by-side on a disk-shaped circular $TM_{11}$-mode coupler-processing cavity developed at SLAC [39], as shown in Fig. 2.13.

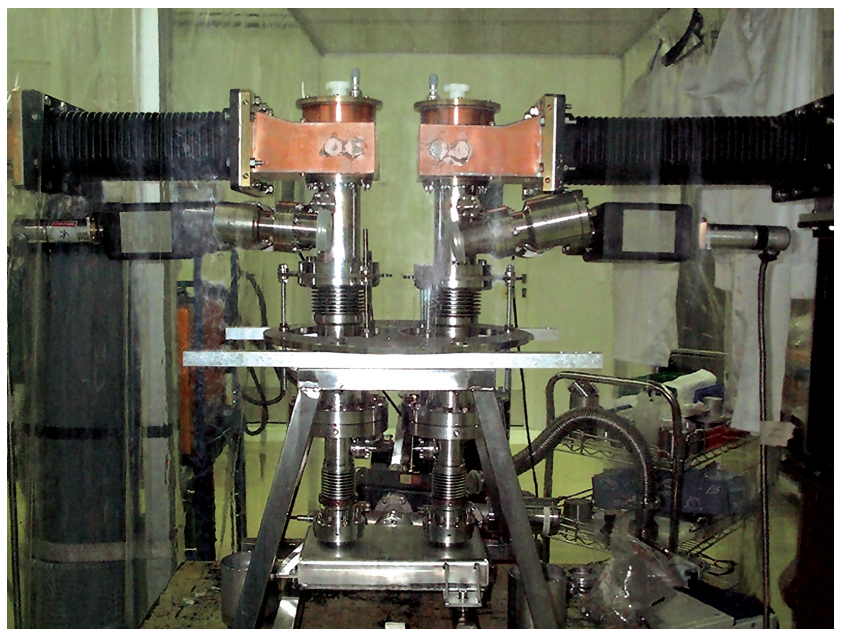

**Figure 2.13**
A pair of TTF-III-type couplers mounted on a coupler-processing cavity and installed for high-power processing.

## 2.3 High-gradient SCRF Cavity Yield

### 2.3.1 Baseline Cavity

The TESLA 9-cell superconducting cavity [40] is the baseline design for the ILC. It has been developed over the past 15 years and achieved the highest gradients to date for multi-cell cavities. The cavities are to be qualified for ILC at an average gradient of 35 MV/m, with a permissible spread of up to ±20%, in a vertical test and operated at an average gradient of 31.5 MV/m in cryomodules.

#### 2.3.1.1 Cavity-Shape Design

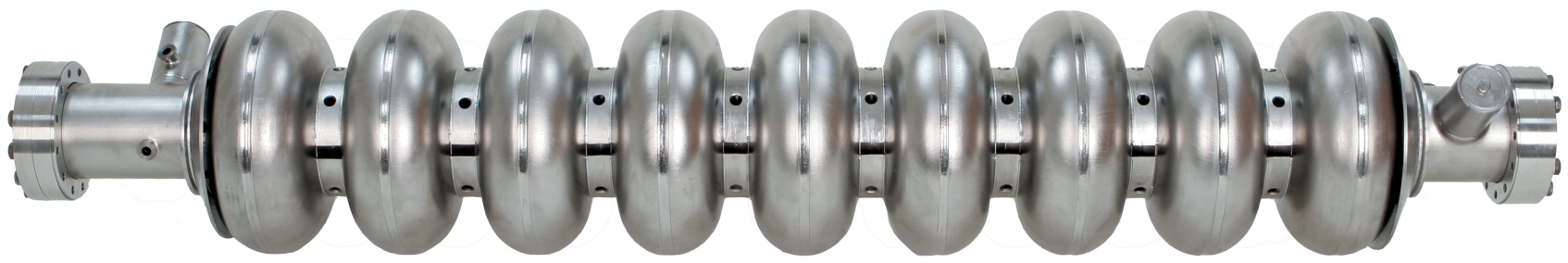

**Figure 2.14**
TESLA-style 9-cell niobium cavity.

Numerous cavities of TESLA-style (or TESLA-like-style) 9-cell cavities (see Fig. 2.14) manufactured by different vendors and processed at various facilities in the three regions (Asia, Americas and Europe) have demonstrated vertical-test results which meet the gradient and $Q_0$ required for ILC [14, 41, 42]. There is significant operational experience with these cavities and it has been demonstrated with beam that accelerating gradients of greater than 35 MV/m are possible after installation in a cryomodule.

The ILC 9-cells cavity works in RF $\pi$-mode at a resonant frequency of 1.3 GHz. Each cavity consists of nine accelerating cells and two end-group sections. One end group has a port for coupling

RF power from the power source into the structure, and the other end has a port for a field-sampling probe used to determine and control the accelerating gradient. Each end group also has a resonant higher-order-mode (HOM) coupler structure with a probe port that is connected to a small electric-field antenna for extracting HOM power and for diagnostics.

Among other factors, the cell shape is optimised for a small ratio of the surface electric field $E_{peak}$ to the acceleration gradient $E_{acc}$ i.e. $E_{peak}/E_{acc}$. At the time of the TESLA cavity shape design, field emission was a principal gradient limitation issue. Hence a logical design optimisation goal is to reduce $E_{peak}/E_{acc}$.

With the recent development of advanced surface processing and cleaning procedures, field emission has been significantly reduced during the vertical test of 9-cell cavities. However, as the ILC gradient goal of 35 MV/m is considerably higher than the original TESLA gradient goal of 25 MV/m, a reduced $E_{peak}/E_{acc}$ remains an attractive feature. This argument becomes more compelling in view of the allowable gradient scatter. To meet the average goal, the gradient of individual cavities may need to go as high as 42 MV/m during the vertical qualification test, which presents a significantly increased risk of field emission.

**Table 2.4**
Parameters of the 9-cell SCRF cavities.

| Parameter | Value |
|---|---|
| Type of accelerating structure | Standing wave |
| Accelerating mode | $TM_{010}$, $\pi$-mode |
| Type of cavity-cell shape | Tesla (or Tesla-like) |
| Fundamental frequency | 1.300 GHz |
| Operation: | |
| – Average gradient (range allowed) | 31.5 MV/m ($\pm$20%) |
| – Quality factor (at 31.5 MV/m) | $\geq 1 \times 10^{10}$ |
| Qualification: | |
| – Average gradient (range allowed) | 35.0 MV/m ($\pm$20%) |
| – Quality factor (at 35 MV/m) | $\geq 0.8 \times 10^{10}$ |
| – Field flatness | $\geq$ 95 % |
| – Acceptable radiation (at 35 MV/m) | $\leq 10^{-2}$ mGy/min |
| Active length | 1038.5 mm |
| Total length (beam flanges, face-to-face) | 1247.4(30) mm |
| Input-coupler pitch distance, including inter-connection | 1326.7 mm |
| Number of cells | 9 |
| Cell-to-cell coupling | 1.87% |
| Iris aperture diameter (inner/end cell) | 70/78 mm |
| Equator inner diameter | ~210 mm |
| $R/Q$ | 1036 Ω |
| $E_{peak}/E_{acc}$ | 2.0 |
| $B_{peak}/E_{acc}$ | 4.26 mT/(MV/m) |
| Tunable range | ±300 kHz |
| $\Delta f/\Delta L$ | 315 kHz/mm |
| Number of HOM couplers | 2 |
| $Q_{ext}$ for high-impedance HOM | $< 1.0 \times 10^5$ |
| Nb material for cavity (incl. HOM coupler and beam pipe): | |
| – RRR | $\geq$300 |
| – Mechanical yield strength (annealed) | $\geq$39 MPa |
| Material for helium tank | Nb-Ti Alloy |
| Max design pressure (high-pressure safety code) | 0.2 MPa |
| Max hydraulic-test pressure | 0.3 MPa |

There are alternative cavity-shape designs (see Section 2.3.4.2) that have a reduced ratio of the peak surface magnetic field $H_{peak}$ to the accelerating gradient, $H_{peak}/E_{acc}$, at the cost of increased $E_{peak}/E_{acc}$. The reduced $H_{peak}/E_{acc}$ makes the alternative shape cavities a preferred choice for reaching the ultimate gradient, because of the higher fundamental theoretical limit to the surface magnetic field. This expected benefit has been successfully demonstrated with many 1-cell cavities [43, 44]. However, the ILC cavity-gradient R&D results in recent years have shown that the gradient in practical 9-cell cavities is predominantly limited by highly localised defects. The full potential of alternative-shape cavities to approach the ultimate gradient is yet to be demonstrated in 9-cell cavities through extended R&D. Initial experience has shown that the alternate-shape 9-cell cavities do indeed have increased field emission due to the higher $E_{peak}/E_{acc}$.

Considering the global experience with 9-cell niobium cavities, the TESLA shape was confirmed as the baseline cavity shape for ILC. Table 2.4 gives a summary of the major RF parameters of the TESLA-shape cavity.

#### 2.3.1.2 Cavity material, Fabrication, Processing and Cryogenic RF Testing

The basic acceleration cells in a baseline cavity are fabricated by using high-purity sheet niobium with a typical RRR (Resistive Resistance Ratio) of 300 or better. These raw materials are commercially produced and are available from vendors in all three regions. By multiple steps of electron-beam melting under vacuum, the purity of a niobium ingot is improved with the reduction of interstitial impurities such as C, N, O. After a series of forging, rolling, annealing and etching processes, the niobium ingot is reduced to fine-grain niobium sheets suitable for forming into half-cells through deep drawing. Table 2.5 gives typical properties of high-purity niobium for use in ILC cavities.

**Table 2.5**
Typical properties of high-purity niobium for use in ILC cavities

| Element | Impurity content in ppm (wt) | Property | Value |
|---|---|---|---|
| Ta | $\leq$500 | RRR | $\geq$300 |
| W | $\leq$70 | Grain size | $\approx$ 50 µm |
| Ti | $\leq$50 | Yield strength | $>$ 50 MPa |
| Fe | $\leq$30 | Tensile strength | $>$100 MPa |
| Mo | $\leq$50 | Elongation at break | 30% |
| Ni | $\leq$30 | Vickers hardness | |
| H | $\leq$2 | HV 10 | $\leq$50 |
| N | $\leq$10 | | |
| O | $\leq$10 | | |
| C | $\leq$10 | | |

As a quality assurance check prior to use, the cell material is eddy-current scanned to a depth of 0.5 mm into the surface of the sheet material. Over decades, the effort of SCRF cavity development has been marked by close information feedback between laboratories and industries, which allowed steady improvement of the quality of niobium sheets in recent years. The niobium-material vendors have accumulated additional experience due to the global ramp-up of the effort in developing the SCRF technology for the ILC, leading to increased consistency in delivered niobium sheets. Although all niobium sheets for the European XFEL cavity production will be eddy-current scanned, it is likely that for ILC production only random spot-checking of sheet niobium will be required as a QA/QC step.

Niobium sheets are formed into cups through deep drawing. The dies are usually made from a high-yield-strength aluminium alloy. The half-cells are machined at the iris and the equator. At the iris, the half-cells are cut to the specified length (allowing for weld shrinkage) while at the equator an extra length of $\sim$ 1 mm is left to retain the possibility of a precise length trimming of the dumbbell after frequency measurements. The half-cells are thoroughly cleaned by ultrasonic degreasing and chemical etching. Two half-cells are then joined at the iris with electron-beam welding (EBW) to form a *dumbbell*. The EBW at the iris is usually done from the inside to ensure a smooth weld seam at the location of the highest electric field in the resonator. Since niobium is a strong Getter material for oxygen, it is important to carry out the EBW in sufficiently good vacuum. Tests have shown that RRR 300 niobium is not degraded by welding at a pressure below 6.7 mPa.

The next step is the EBW of the stiffening ring between cells. Here the welding shrinkage may lead to a significant distortion of the cell shape, which needs to be corrected. Afterwards, frequency measurements are made on the dumbbells to determine the correct amount of trimming at the equators. Then the equator weld is prepared followed by additional chemical etching. Next, the dumbbell RF surface QA/QC is performed through visual inspection; sometimes overnight soaking in de-ionized water is used to disclose iron inclusion originating from the sheet-rolling process. Defects and foreign-material imprints from raw material or previous fabrication steps are removed by mechanical grinding. After the inspection and cleaning (a light etching followed by de-ionized water rinsing and clean-room drying), eight dumbbells and two end-group sub-assemblies are stacked in a fixture to carry out the equator EBWs, which are done from the outside. The weld parameters are chosen to achieve full penetration. A reliable method for obtaining a smooth weld seam of a few mm width at the inner surface is to raster a slightly defocused beam in an elliptic pattern and to apply 50 % of beam power during the first weld pass and 100 % beam power in the second pass.

Figure 2.15 shows the flow diagram of the fabrication steps for the complete 9-cell cavity.

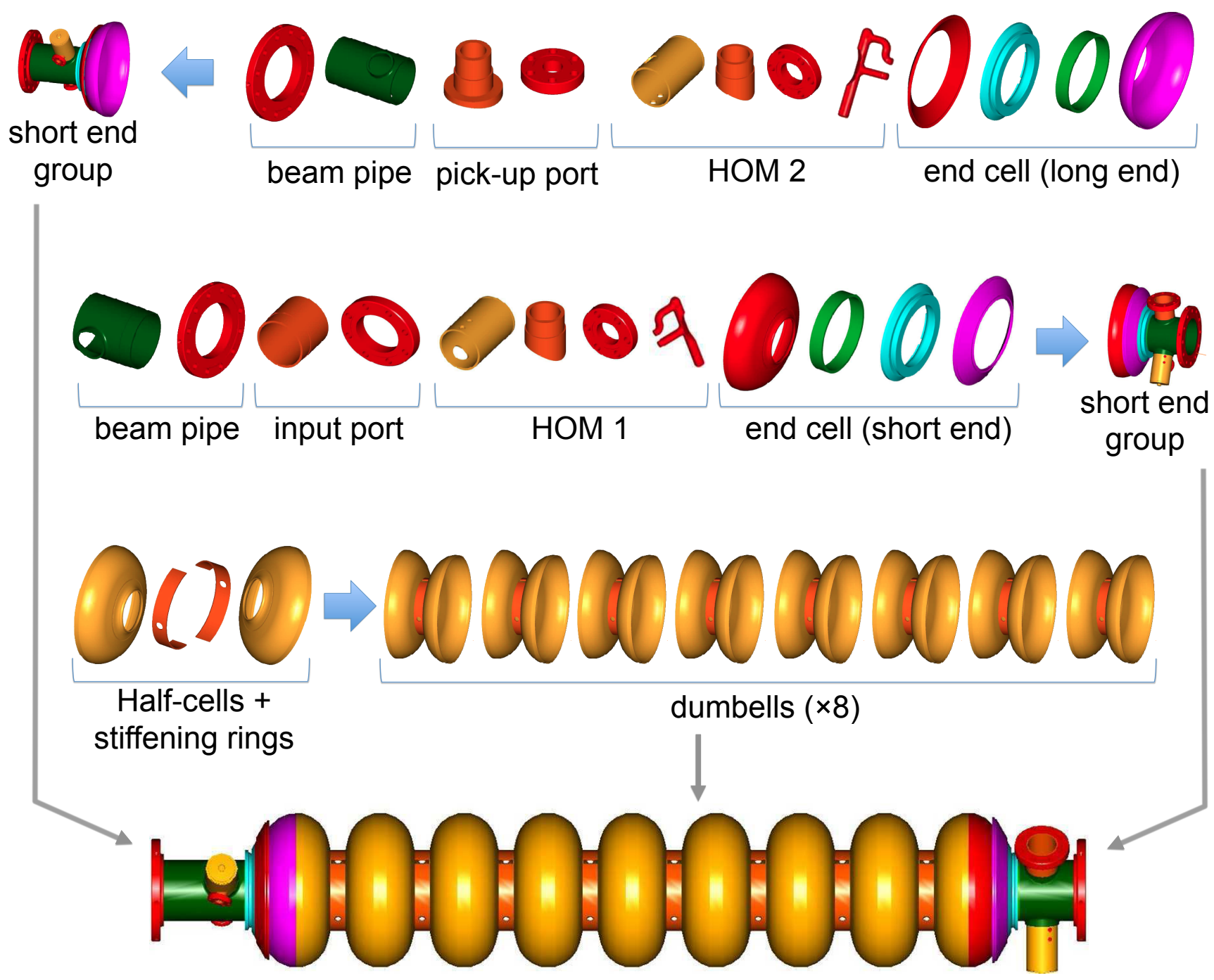

**Figure 2.15** ILC 9-cell-cavity fabrication flow diagram.

As a quality-assurance check prior to cavity processing and testing, the RF surface of a completed 9-cell cavity is optically inspected for irregularities in the EBW regions. It has been shown that defects that cause the most quenches are located in the fusion zone or the heat-affected zone of the equator weld. Many of these defects can be traced back to the fabrication process. Typically, the fusion zone is a few mm in width. The heat-affected zone is a narrow region outside the fusion zone, where grain-growth is apparent due to heat deposited during the welding. The two heat-affected zones (one each on both sides of the weld) plus the fusion zone covers a total zone typically 40–50 mm in width. Due to the high surface magnetic field in this region, quenches can be initiated either through a thermal process (such as strong resistive heating of a normal-conducting inclusion) or a magneto-thermal process (such as strong magnetic field enhancement at sharp edges of pure geometric defects). There is a growing database on the optically identifiable defects, which are correlated to quench during the final cavity-qualification test [45–47]. The optical inspection is very effective in identifying sub-mm sized geometrical defects in the equator region, such as sharp edges, exposed welding pores/holes, unclosed weld prep, rough under-bead and weld spatters etc. This not only provides rapid feedback for production but also allows informed decisions for defect removal by suitable methods, such as local grinding at a distinctive defect area [48] and mechanical polishing over the entire RF surface of a completed cavity [49]. Overall, the QA/QC of the completed cavity by optical inspection is an important step for improved production yield. Ultimately, this leads to cost-effective mass production of niobium cavities.

Prior to the cavity surface processing, a series of mechanical, vacuum and RF inspections are carried out. The cavity is mechanically measured with a coordinate measuring machine to compare dimensional measurements against mechanical tolerances identified on design drawings. All the welding joints are leak checked. The cavity fundamental mode frequency is measured. The frequency of other pass-band modes may be also measured. The field flatness of the fundamental mode may be measured, which is done by pulling a bead through the beam axis of the cavity. This determines the stored energy of each cell and provides an indication of uniformity of the electric field on the beam axis. In case of a severely non-uniform field distribution, a rough tuning is done by mechanically squeezing or stretching the cells in order to reach the fundamental-mode frequency and a total length within the specified tolerances.

Cavity surface processing consists of a series of steps. This has been the focus of the globally coordinated cavity gradient R&D program in the past years [50]. The ILC baseline surface processing is described in detail in Section 2.3.2.

After the completion of the cavity processing steps, the cavity proceeds for qualification RF testing[3]. The cavity is mounted vertically in a cryogenic test stand. RF cables are connected to the cavity probes and the stand is inserted into a cryogenic dewar. The dewar is cooled to 2 K with the entire cavity immersed in superfluid liquid helium. The cavity is tested to determine its gradient, quality factor and limitations. In case of sub-standard performance limited by quench, the pass-band excitation measurement technique may be used during vertical testing for identifying the weakest cell. Additional instrumentation such as Cornell OST's [28] may be used for identification of quench sources with spatial resolution of a few cm for the fundamental mode or other pass-band modes. An important diagnostic associated with cavity testing for field-emission monitoring is X-ray detection. It is customary to place gamma probes outside the dewar. Some radiation sensors such as silicon diodes can be placed on the cavity and immersed in liquid helium. The cryogenic radiation sensors may be permanently attached to a cavity, allowing tracking of the field-emission performance of the cavity from its vertical qualification testing to horizontal testing in the final cryomodule.

Experience has shown that some cavities will not pass the ILC specification during the first qualification test but a second-pass reprocessing can often raise its performance effectively. By allowing the second-pass reprocessing, the overall yield of acceptable cavities is improved in a cost-effective manner. The average gradient also tends to improve. A detailed description of the criterion for the second-pass processing is given in Section 2.3.3.

### 2.3.2 R&D on ILC High-gradient-cavity Processing

The baseline procedure [51] of ILC 9-cell niobium cavity processing and handling is summarised in Table 2.6. A detailed description of the key processing steps is given in the following sections.

**Table 2.6**
Processing and handling of high-purity niobium cavities

| |
|---|
| Light BCP etching (10 μm) |
| Heavy EP (100-120 μm) |
| Post-heavy-EP cleaning |
| Vacuum-furnace outgassing (800 °C for 2 h) |
| RF tuning by no-touch bead-pull |
| Light EP (25 μm) |
| Post-light-EP cleaning |
| First HPR 3 passes ($\sim$ 6 h) |
| First clean room assembly |
| Final HPR 3 passes ($\sim$ 6 h) |
| Final clean-room assembly |
| Leak checking |
| In-situ baking at 120 °C for 48 h |

[3]For the mass production of ILC cavities, the helium vessel will be welded to the cavity and the HOM couplers will be tuned and attached before the cavity qualification test, following the current European XFEL approach.

#### 2.3.2.1 Electrolyte Mixing

Electrolytes for the electropolishing have to be mixed in a suitable molar ratio of HF:$H_2O$:$H_2SO_4$ which is compatible with the original Siemens recipe [52]. As the weight concentration of commercial acids varies across regions, the volume ratio needs to be tailored to the suitable molar ratio. For example, a volume ratio of 1:10 HF(48%):$H_2SO_4$(96%) is used at JLab. Prior to mixing, the electrolyte storage tank and acid piping are flushed with sulphuric acid (96 % concentration) to eliminate water residues. Then 210 l of sulphuric acid (96 %) are transferred into the electrolyte storage tank and chilled while keeping the acid in circulation. (This is done by connecting the cavity acid inlet hose with the cavity acid outlet hose). The first 5 l of hydrofluoric acid (48 %) are added when the electrolyte temperature in the storage tank reaches 15 °C or lower. The acid continues to be circulated until the electrolyte temperature recovers to 15 °C, at which point the second 5 l of hydrofluoric acid are added. The process is repeated till 21 l of hydrofluoric acid have been added.

#### 2.3.2.2 Electropolishing

A horizontally placed EP [53] machine is used for the baseline EP processing in all regions. Three key EP parameters have been established and are monitored during processing, as shown in Table 2.7.

**Table 2.7**
Details of the electropolishing procedure

| Parameter | Explanation |
|---|---|
| Voltage | Nominal at 14.5 V as measured across the body of cavity and the cathode, allowable range 13–17 V |
| Cavity cell temperature | Nominal at 25 °C as measured at the cavity outer surface near the equator region, allowable range 20–30 °C for final light EP and nominal at 30 °C, allowable range up to 35 °C for heavy EP. |
| Acid flow rate | Nominal at 3–4 l/min |

In addition to these key parameters, it is important to keep the purging nitrogen gas flow at the minimum level. The acid supplying holes in the cathode tube face upward. This reduces the disturbance of the viscous layer across the acid-niobium interface.

Several provisions are available for control of the cavity-cell temperature: by steering the chilled water set-point; by duty-cycling the voltage; by irrigating the hot spot on the cavity outer surface with chilled water. Typically, by using one or more of these methods, a standard fine-grain 9-cell cavity EP can be operated within the allowable temperature ranges. A setup to provide water irrigation for the outer surface of the cavity is useful for temperature control in extended ranges. This scheme was first successfully demonstrated at the Cornell vertical EP machine [54]. Figure 2.16 shows an example as used in the JLab horizontal EP machine. An array of nozzles sprays chilled water at the iris and equator regions from the bottom of the cavity. Cavity rotation achieves uniform whole-cavity cooling. This chilled-water irrigation system has been used routinely for electropolishing of 86 7-cell cavities for the CEBAF upgrade, for which a lower cell temperature (20 °C) was desired.

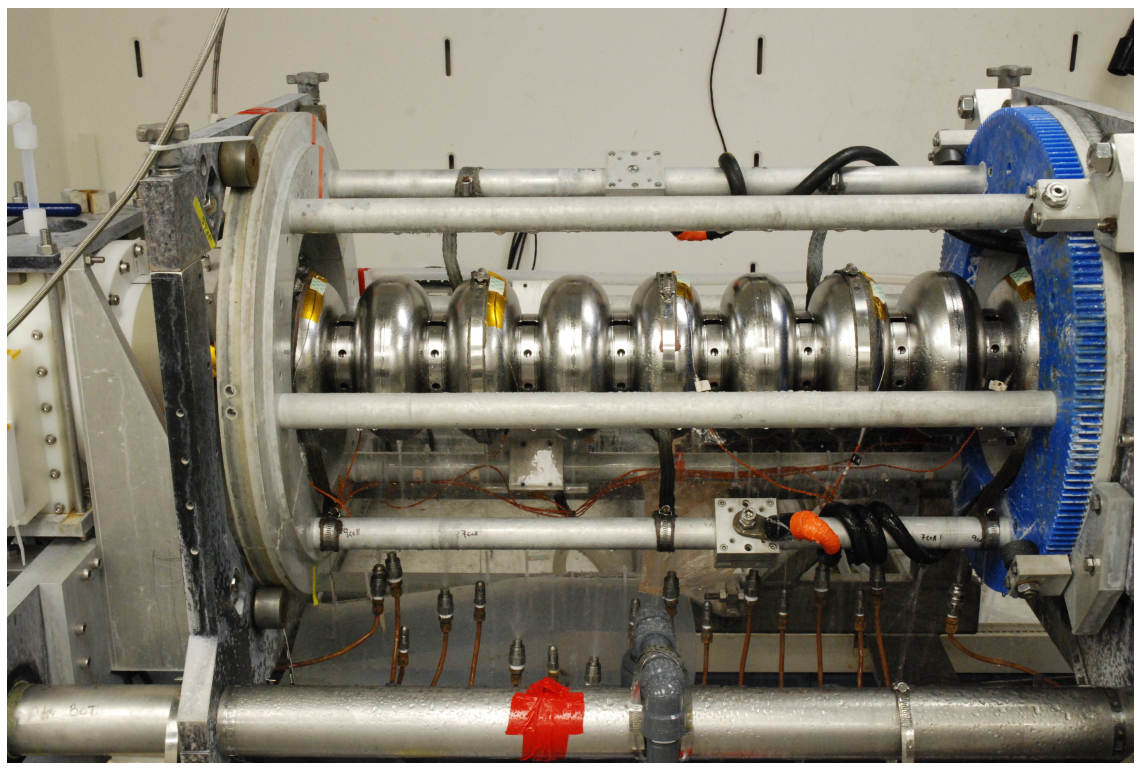

**Figure 2.16**
Cavity outer surface water irrigation setup as implemented in the JLab EP machine. An array of nozzles spray chilled water at selected iris or equator regions.

The best EP result is obtained in the continuous-current-oscillation mode. The appearance of current oscillation also serves as a sensitive in-situ process QA/QC indicator. The disappearance of current oscillation is correlated to excessive loss of HF or excessive $H_2O$ in the electrolyte.

An *HF Auto Rinse* procedure has been found to be beneficial: the acid flow and the cavity rotation continues after the EP voltage is shut off. Under this condition, HF continues to react with the top layer of the cavity inner surface while the anodisation process is stopped. An effect similar to that achieved by HF rinsing has been observed, in which the sulphur-bearing niobium-oxide granules, an inherent contaminant on the finished niobium surface due to the EP process itself [55], are effectively removed from the cavity surface.

#### 2.3.2.3 Post Cleaning after Heavy EP

The cleaning steps applied after heavy EP are summarised in Table 2.8.

**Table 2.8**
Processing and handling after initial heavy EP processing

| | |
|---|---|
| Low-pressure water rinsing | Immediately after the cavity is dismounted from the EP machine, the cavity inner surface is copiously irrigated with de-ionized water. |
| Brushing & wiping | The HOM can and cavity end group are brushed and wiped with soapy water. This procedure cleans the inner surface areas that lack direct line-of-sight access by the high-pressure water jets. |
| Ultrasonic cleaning for 1 h | The entire cavity is immersed in a bath of hot (50 °C) de-ionized water with 2 % Liquinox mixed. |
| Ethanol rinsing | About 5 l of high-purity ethanol is transferred into th sealed cavity. The cavity is tilted and flipped for 10 min. |
| Rinsing | High-pressure water rinsing for two passes, each for about 2 h. |

#### 2.3.2.4 Vacuum-Furnace Heat Treatment

Following post-heavy-EP cleaning, the cavity is heat treated in a vacuum furnace for hydrogen degassing. The baseline furnace vacuum is typically in the mid-$10^{-6}$ Pa range after over-night pumping. The furnace temperature ramps up at a rate of 5 °C/min, remains at 800 °C for 2 h, then ramps down naturally by turning off the heating elements.

#### 2.3.2.5 RF Tuning by No-Touch Bead-Pull

It is necessary to tune the cavity $\pi$-mode field flatness following the heavy EP and vacuum-furnace outgassing. A no-touch bead-pull procedure has been developed for the final tuning prior to the final EP [51].

#### 2.3.2.6 Post Cleaning and Handling after Final EP

The cavity handling steps completing and following the final EP are described in Table 2.9.

#### 2.3.2.7 Results on field emission and quench limitations

The initial S0 effort focused on the issue of field emission, which was identified to be the main cause of gradient variability at the beginning of the ILC Technical Design Phase. The R&D priority was given to improved post-EP cleaning procedures. Three methods are now established for effective field-emission reduction: (1) Ethanol rising was successfully developed and applied at DESY [56]; (2) Ultrasonic cleaning was introduced and optimised at JLab [57]; (3) EP with a fresh acid mixture was found effective for field-emission reduction at KEK [58]. Some of the methods have been successfully transferred across facilities at different labs.

The source of field emission is now better understood. Besides the traditional particulate field emitters, niobium oxide granules are found to be a major field emitter introduced by the electropolishing process itself [59, 60]. Wiping and brushing of end-group components immediately after EP processing reduces niobium-oxide granules (often accompanied by increased sulphur-bearing compounds) in the hidden areas where HPR cleaning is less effective due to lack of direct water-jet bombardment [44]. Streamlined clean-room assembly procedures are now routinely used, resulting in minimum re-contamination by particulates generated by the assembly process itself. *Field-emission-free* performance up to 40 MV/m has been demonstrated in several electropolished 9-cell cavities. Efforts are continuing to develop new and improved cleaning techniques with a view to complete eradication of field emission up to the theoretical quench limit of a niobium cavity.

**Table 2.9**
Post cleaning and handling after Final EP

| | |
|---|---|
| Low-pressure water rinsing | Immediately after the cavity is dismounted from the EP machine, the cavity inner surface is copiously irrigated with de-ionized water. |
| Brushing & wiping | The HOM-can and cavity end group are brushed and wiped with soapy water. This procedure cleans the inner surface areas that lack direct line-of-sight access by the high-pressure water jets. |
| Ultrasonic cleaning for 1 h | The entire cavity is immersed in a bath of hot (50 °C) de-ionized water with 2 % Liquinox. |
| Rinsing | First high-pressure water rinsing. 3 passes, each for about 2 h. |
| Assembly | First clean-room assembly of all (except the bottom flange sub-assembly) auxiliary hardware components including RF probes. |
| Rinsing | Final high-pressure water rising. 3 passes, each for about 2 h. |
| Final Assembly | Final clean-room assembly of the bottom flange sub-assembly. |
| Pump down | Slow pump down. The speed of pump down is controlled to avoid turbulent flow. This prevents transfer of agitated particulates in the vacuum pluming into the cavity. |
| Leak checking | – |
| Baking | Baking at 120 °C for 48 h. The cavity under vacuum is enclosed in an insulated baking box. Hot nitrogen is blown into the box. The temperature of the centre cell is used for process control. |

**Figure 2.17**
Two classes of quench limit in the state-of-the-art 9-cell niobium cavities. Shown are 16 9-cell cavities (10 built by ACCEL/RI ad 6 by AES), processed and tested at JLab since July 2008.

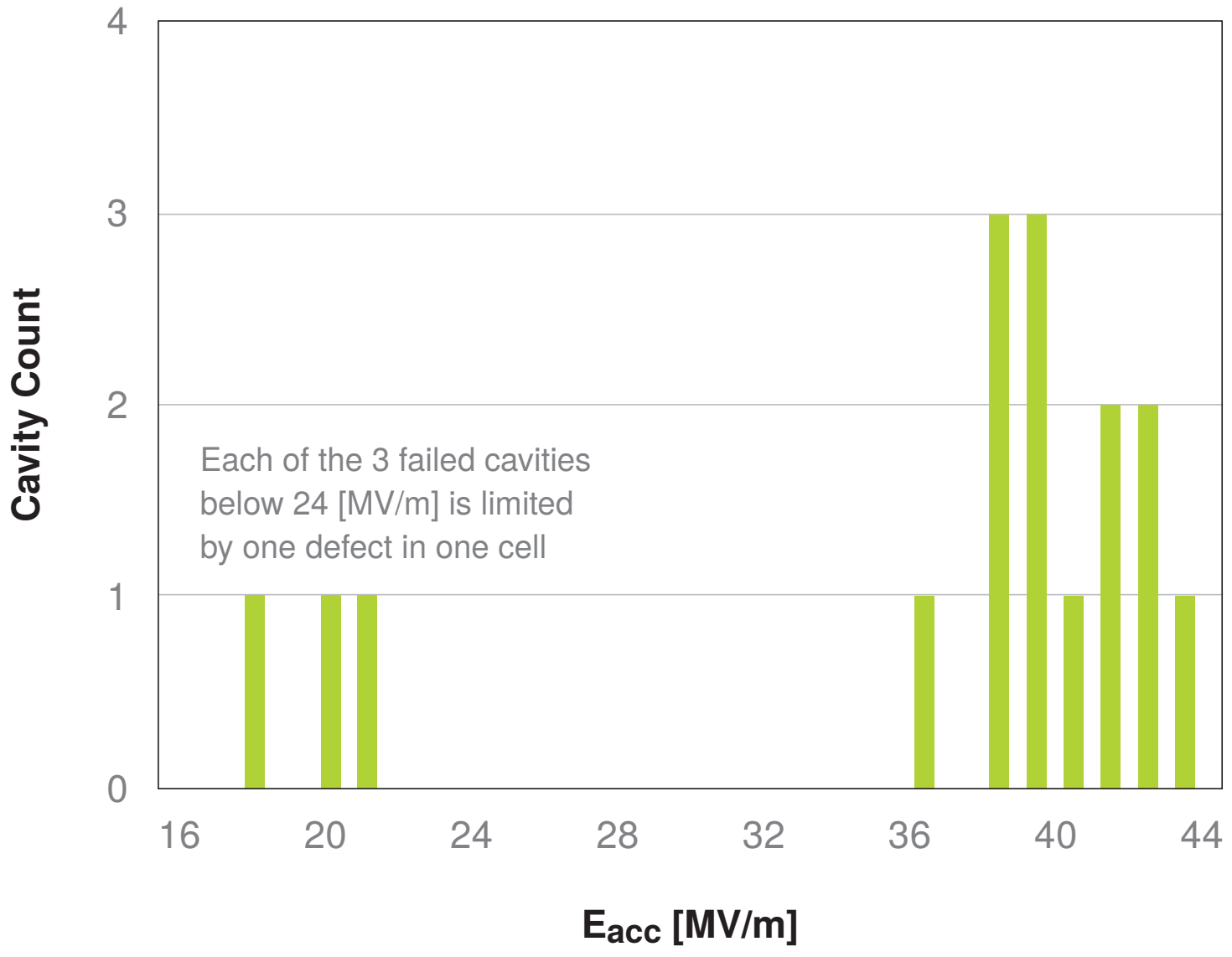

The understanding of the source of quenches is much improved thanks to the routine applications of various diagnostic systems described in Section 2.2.2. Generally speaking, there are two classes of quenches (see Fig. 2.17). Quenches at $< 25$ MV/m are found to be caused predominantly by highly localised geometrical defects near the electron-beam-welded joint at the cavity equator. These defects, referred to as type-I defects, can be roughly categorised as circular pits or bumps with a typical diameter in the range of 200–800 µm. The detailed morphology of type-I defects can be quite complicated, as revealed by replica measurements and by microscopic inspection of small samples cut out from real cavities [25]. There is experimental evidence to show that the effect of local magnetic-field enhancement at sharp edges is responsible for initiation of quenches at geometrical defects [61, 62]. The magnetic-field enhancement factors have been calculated for simple geometrical features [63]. The quench limit at $< 25$ MV/m due to geometrical defects is insensitive to EP even when applied repeatedly. Sometimes, pre-cursor defects are already evident prior to any chemical processing of the cavity surface. Therefore, the most likely cause of these defects lies in the cavity fabrication, in particular the electron-beam welding process where bulk defects such as voids and surface irregularities are possible under certain conditions. It is now recognised that, in order to prevent quenching below 25 MV/m, improved QA/QC in material and fabrication are important. Experience has shown that the performance of cavities limited by type-I defects can be significantly improved by removing the known limiting defects. Several methods have been explored including local grinding [48], targeted mechanical polishing [64], whole-surface mechanical polishing [49], local re-melting by an electron beam [65] and local re-melting by a laser beam [66]. These methods have all shown promising results with 1-cell cavities. However, up to now, only the local grinding and mechanical polishing (targeted or whole-surface) have been successfully applied to 9-cell cavities, as discussed in Section 2.2.7.

Further studies are still needed to understand the source of quenches at higher gradients in the nominal range of $> 30$ MV/m. It is clear that the quenches above 30 MV/m are also caused by highly localised defects near the equator electron-beam weld (like quenches caused by type-I defects). These defects are categorised as type-II defects. Unlike the type-I defects, it appears that there is no observable feature (down to the spatial limit of the present optical-inspection technique, typically a few µm) at the site of the quench-causing defect. This suggests the subtle nature of the source of type-II defects. It has been suggested that the locally suppressed superconductivity due to compositional irregularities or lattice irregularities may be responsible. Additional capabilities of compositional analysis "in-situ" at the predicted quench location are expected to shed light on this issue. Off-line, the microscopic analysis of small samples cut out at the predicted quench location will provide important information for further gradient improvement. It has been also found by experience that the performance of a cavity quench-limited at $> 30$ MV/m is often improved by an additional light EP processing followed by low-temperature baking (120 °C).

### 2.3.2.8 Progress in raising the quality factor

The unloaded quality factor $Q_0$ of the ILC 9-cell niobium cavity is required to exceed $10^{10}$ at the average ILC beam operation gradient of 31.5 MV/m. There are three known mechanisms that may cause $Q_0$ degradation due to surface-field effects, namely Q-disease, field emission and the so-called high-field Q-slope. At 31.5 MV/m accelerating gradient, the peak surface electric and magnetic field for a TESLA-shape cavity are 63 MV/m and 134 mT, respectively. Successful improvement in post-EP cleaning and handling has significantly reduced the field-emission problem during cavity vertical testing (see Section 2.3.2.7). Heat treatment at 800 °C in a vacuum furnace prior to the final surface processing outgases hydrogen and hence removes the Q-disease problem. Surface processing by EP followed by in-situ low-temperature bake at 120 °C for 48 h effectively corrects the high-field Q-slope problem for the peak surface electric-field regime up to 180 mT [51].

With the application of the above cavity treatment and processing procedures, the $Q_0$ of a 9-cell ILC cavity during vertical test is likely to exceed $10^{10}$ at 31.5 MV/m accelerating gradient (Fig. 2.18). However, since additional handling is necessary to assemble qualified cavities into a cavity string for cryomodule inclusion, great care must be exercised to prevent re-contamination of the cavity surface.

Experience has shown that field emitters introduced by re-contamination may cause significant loss of quality factor from vertical qualification tests to the beam operation of cavities.

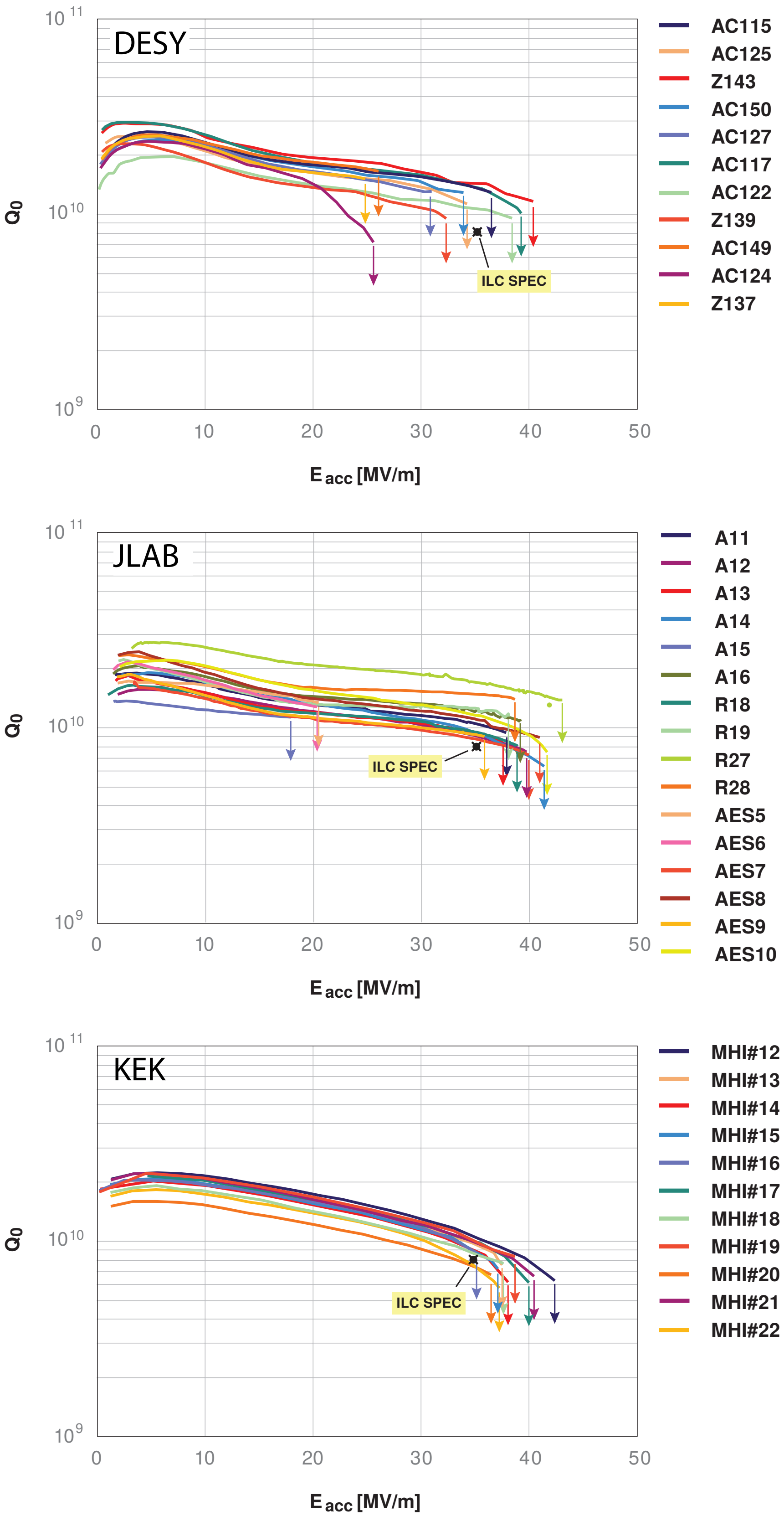

**Figure 2.18** Examples of Q($E_{acc}$) for 9-cell cavities processed following the baseline ILC cavity surface processing in the regions.

### 2.3.3 Yield and cavity selection criterion

A global R&D program was established in 2006 to address the challenge of reaching the ILC gradient specification. The *S0* program, coordinated by the GDE Cavity Group, attracted global participation from ANL, Cornell, DESY, FNAL, JLab, KEK and other labs such as IHEP (Beijing), Peking University, Saclay and TRIUMF. Significant progress in understanding the gradient limit and gradient variation has been made by instrumented cavity testing at cryogenic temperatures and high-resolution optical inspection of the cavity surface. This has been accompanied by steady progress in reproducibility at 35 MV/m and in improved practical gradient limit in 9-cell cavities. At the beginning of the technical design phase, Europe was the only region having demonstrated 35 MV/m cavity fabrication and processing. Today, 35 MV/m cavity fabrication and processing have been demonstrated also in the Asia and Americas regions (Fig. 2.18). A solid SCRF technical base for the ILC on a global scale is now in place.

The global efforts in ILC gradient R&D are rewarded not only by improved gradient yield and reproducibility but also an extended gradient envelope. By mid 2010, a major SCRF gradient R&D milestone of 50 % yield at 35 MV/m has been achieved. The average gradient in state-of-the-art 9-cell cavities has now been raised to $\sim$ 40 MV/m, a steady increase in comparison to the value of 35 MV/m in 2005.

At the start of the GDE cavity R&D programme, goals of a 50 % process yield by 2010 (phase I of the R&D program) and a 90 % production yield were established at 35 MV/m with $Q_0 \geq 8 \times 10^9$ by 2012 (phase II of the program). Since the cavity gradients are given by a distribution rather than a single number, as the Technical Design Phase of the ILC progressed, the need for a clearer definition of the gradient yield was recognised. In order to accomplish this, a globally consistent and up-to-date database for recording test results was established [67]. In 2009, the GDE ILC Cavity Group proposed a clear definition for the gradient yield, which adopted the concept of the first-pass and second-pass yield. The rule for cavity selection was also established. In the meantime, the ILC Global Cavity Database Team was created as a part of the S0 effort. The team included members from Fermilab, DESY, JLab, Cornell, and KEK, and took on the task of creating the database, updating and presenting the data. The results of this effort are a clear, objective, and publicly accessible database where the progress of cavity R&D can be tracked. At the time of this TDR the database contains information on 134 cavities. This includes both cavities produced purely for gradient R&D and ones produced for development purposes.

Great progress has been made during the GDE program in essentially all aspects of high-gradient cavity production. A strong technology transfer program from the various national labs to their respective vendors dramatically improved the cavity production techniques and associated QA programs. Diagnostic improvements have resulted in an enhanced understanding of the reasons for gradient performances that failed to meet the desired goals and help guide additional processing. The growing success of the R&D program has resulted in the cavity gradient performance being time dependent. It has improved, within the statistical scatter, throughout the 5-year program.

The most recent data for first-pass and second-pass results [68] are shown in Fig. 2.19. By way of definition, the plots only include results from a vendor/laboratory combination who have previously demonstrated the ability to fabricate and process a cavity that achieves at least 35 MV/m in a vertical test. The plots only include cavities that were processed according to the baseline processing cycle, which includes electropolishing. These criteria were imposed in order to facilitate an *apples to apples* comparison of the gradient results. The first-pass test results drive the second-pass re-processing. If the gradient $\geq$ 35 MV/m and $Q_0 \geq 8 \times 10^9$, the cavity is accepted. If the gradient is below 35 MV/m, the cavity will be given a second-pass reprocessing. The choice of second-pass process is made by the test laboratory based on first-pass test results, but must be one of two standard second-pass processes, i.e. a high-pressure rinse or a second standard EP.

**Figure 2.19** Cavity yield for two gradient thresholds as a function of years, based on the global ILC cavity database updated as of October 2012 [67, 68]. Numbers in parentheses refer to cavity sample size. The cavities received standard treatment and were provided by established vendors.

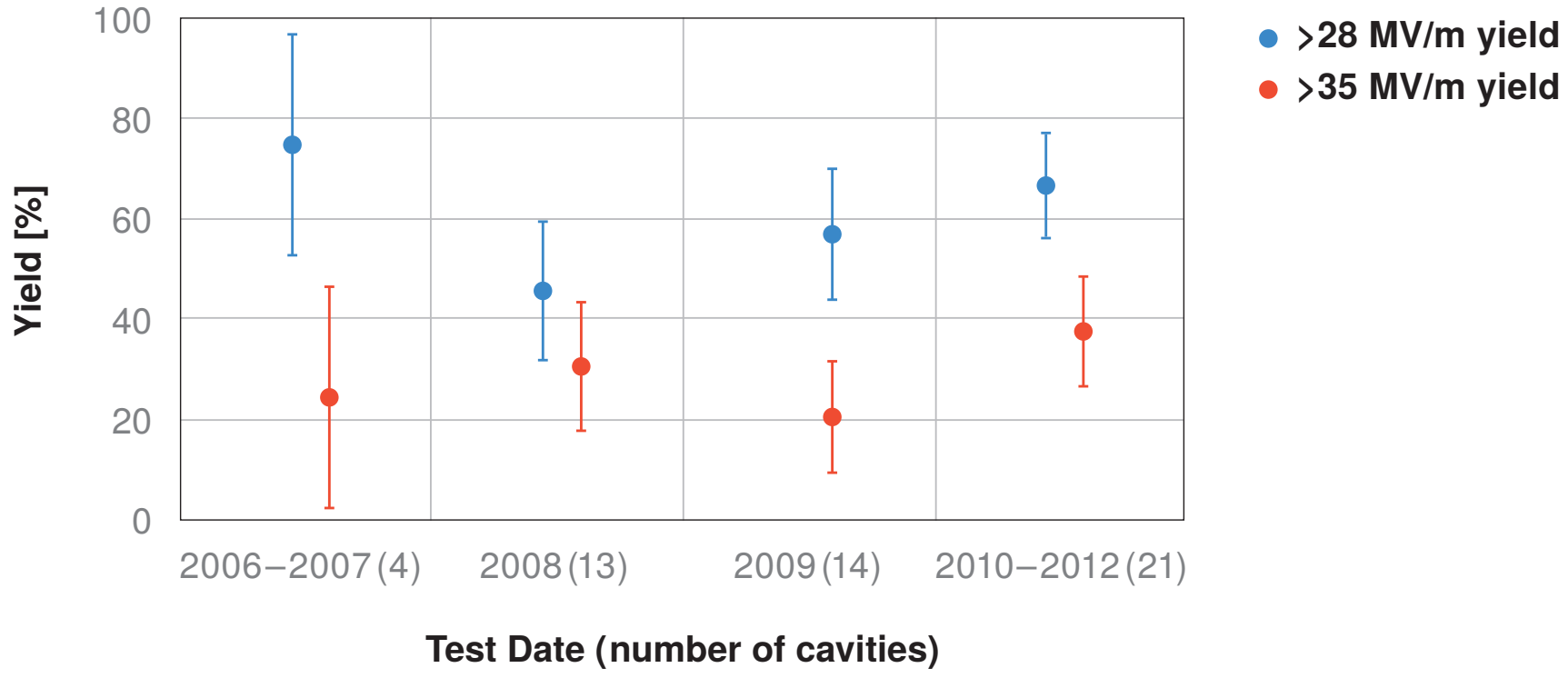

**(a)** First-pass yield

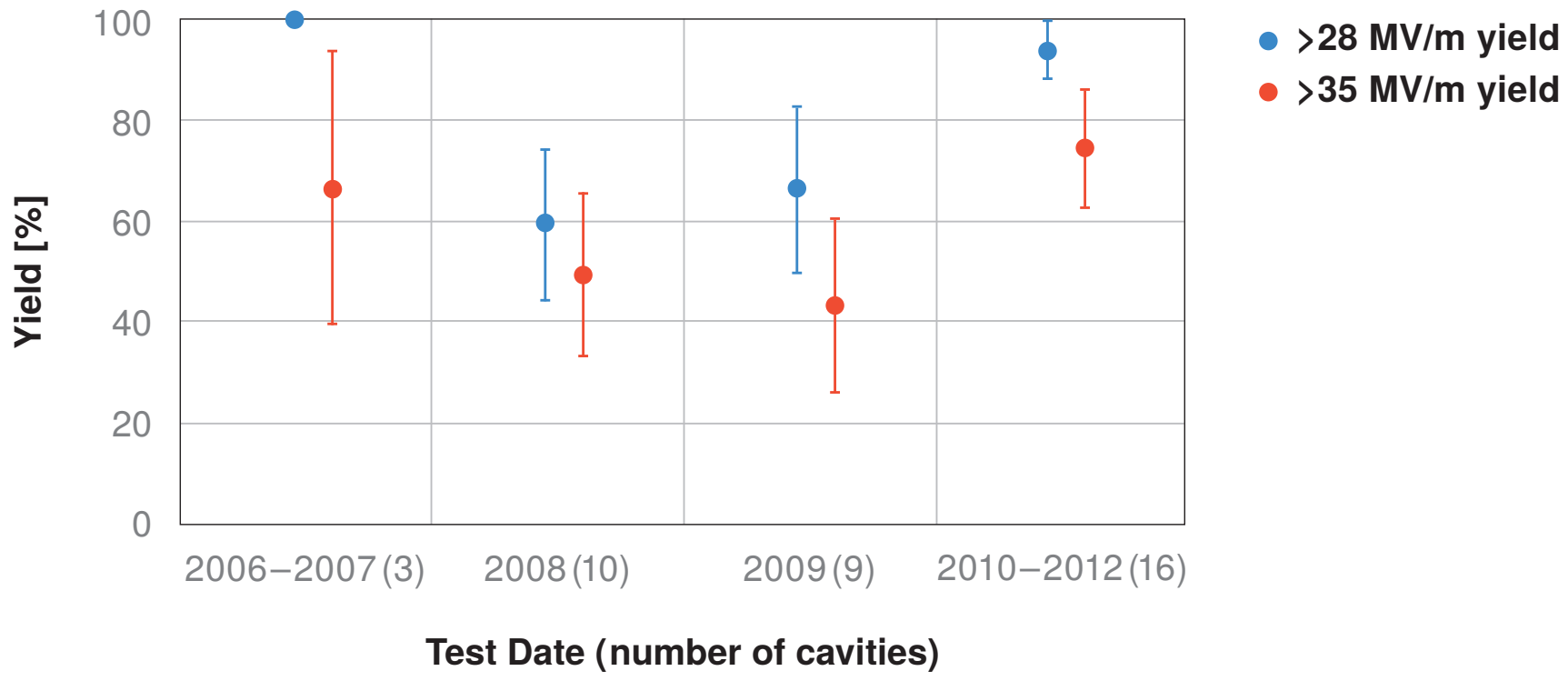

**(b)** Second-pass yield

It is apparent from the graphs that yields are improving with time, particularly for second-pass results. The latest second-pass test results from 2010-2012 show that a $(75{\pm}11)\%$ yield has been achieved at 35 MV/m and $(94{\pm}6)\%$ for $>$28 MV/m on a global basis of 16 cavities. This improvement is attributable to improved diagnostic tools that have been developed during the GDE global R&D program, to direct the second-pass process more accurately. Repeatability in EP processing, which was demonstrated in the past two years, also plays an important role in improving the second-pass yield by raising the gradient performance of cavities quench limited in the range 25–35 MV/m. It should be noted that the distribution of first-pass gradients from cavities which did not yet receive a second-pass process is statistically equivalent to those from cavities which did receive a second-pass process and appear in the plot; no bias has been introduced into the second-pass distribution based on first-pass results.

By allowing a targeted second-pass re-treatment or re-processing, a significant cost-saving benefit can be expected over simply rejecting cavities. Another important benefit is that a higher average gradient can be expected by allowing the second-pass processing, providing valuable gradient margins.

After it was decided to implement a RF power distribution system in the linac that could operate with a cavity-gradient spread of $\pm$20%, acceptance is based on cavities exceeding 28 MV/m. This then introduces the concept of average gradient of the ensembles as another important figure of merit. A 100 % yield of cavities operating at exactly 28 MV/m would exceed the yield criteria but fail to meet the operational goal of an average gradient of 35 MV/m. The average gradient for the ensemble of second-pass cavities in Fig. 2.19 meeting the $>$28 MV/m criteria is 37.1 MV/m. Thus the standard processing cycle has resulted in an average yield and gradient which meet the desired ILC goals.

As of today a small fraction ($\sim$6%) of cavities will be rejected according to this criterion. These rejected cavities may still have high enough gradient to be used for the positron source or polarised electron source.

There are two repair techniques, developed during the last few years and now in use at the national labs, which both promise to have an impact on the cavity yield: localised grinding and centrifugal barrel polishing (tumbling). The results of these repair attempts have been impressive for cavities in this category and are shown in Table 2.10. For this proof-of-principle, all fine-grain ILC-shape cavities are included, independent of whether they were fabricated by an established vendor or not; however, it should be noted that the cavities have had a wide range of surface processes before the attempted repair of recognised defects. Localised grinding, a technique implemented at KEK, is based on finding defects using an optical inspection of a cavity as described in Section 2.2.2; defects are subsequently removed with a miniature grinding tool. This inspection can take place on either a tested cavity, where the grinding may be guided to performance-limiting defects, or on an incoming cavity, for which all observed localised defects are removed. With the tumbling technique, up to four cavities may be processed in parallel in a single week-long cycle using a progressively finer abrasive until at the end a mirror-like finish is achieved. As part of the tumbling process, light electropolishing is performed. To date, two 9-cell cavities with known limiting surface defects have been tumbled to remove their defects [69]. Both procedures are being optimised. Thirteen cavities have been repaired to date for which test results are known. Substantial performance improvement is achieved in most cases, and indeed for these cases, additional repair is likely to result in further improvement. These may be viable procedures which could be incorporated into standard processing cycles.

**Table 2.10**
All fine-grain ILC cavity repairs of localised defects by local grinding or tumbling.

| cavity | repair | gradient MV/m before | after | comment |
|---|---|---|---|---|
| MHI-08 | grinding | 16 | 27 | achieved 38 MV/m in $4^{th}$ test (1 repair cycle) |
| MHI-10 | grinding | 26 | 27 | 1 repair cycle |
| MHI-14 | grinding | 13 | 37 | achieved 37 MV/m in $3^{rd}$ test (1 repair cycle) |
| MHI-15 | grinding | 23 | 36 | achieved 36 MV/m in $4^{th}$ test (3 repair cycles) |
| MHI-16 | grinding | 21 | 34 | 1 repair cycle |
| MHI-18 | grinding | 31 | 30 | 2 repair cycles; additional repair/test cycle in progress |
| MHI-19 | grinding | 26 | 37 | achieved 37 MV/m in $2^{nd}$ test (1 repair cycle) |
| MHI-20 | grinding | 9 | 35 | achieved 35 MV/m in $3^{rd}$ test (2 repair cycles) |
| MHI-22 | grinding | 32 | 36 | achieved 36 MV/m in $2^{nd}$ test (1 repair cycle) |
| AES003 | grinding | 11 | 30 | had been quench-limited below 20 MV/m after repeated EP; results shown after 1 repair cycle; 37 MV/m was reached with additional HPR |
| TB9RI026 | grinding | 20 | 36 | achieved 35 MV/m in $2^{nd}$ test (1 repair cycle) |
| HIT-02 | grinding | 35 | 41 | FE reduction in $2^{nd}$ test (1 repair cycle) |
| TOS-2 | grinding | 31 | 33 | achieved 33 MV/m in $2^{nd}$ test (1 repair cycle) |
| TB9ACC015 | tumbling | 18 | 35 | tumbling proceeded until limiting defect removed (1 repair/test cycle) |
| TB9AES006 | tumbling | 21 | 36 | tumbling proceeded until limiting defect removed (1 repair/test cycle) |

#### 2.3.3.1 Increasing Production Yield

Increasing the yield beyond the ILC goal of 90% (or ensuring the 90% target) will be most easily accomplished by incorporating a repair step into the processing cycle. As mentioned in the last section, cavity repairs (grinding and tumbling together) were successful in most cases in the 2011/12 time period. Since the standard processing cycle does not include a repair step, none of these data are reflected in the yield plots of Fig. 2.19. Indeed at this time it appears that essentially all cavities which fulfill the manufacturing specifications will meet, or can be made to meet, the gradient acceptance requirements. The decision of whether to repair or reject under-performing cavities will depend on whether a repair is cost effective. In this regard the repair scenario looks promising. Localised grinding can be performed by trained technicians at cavity vendors in about one day. A tumbling cycle at Fermilab can process four cavities in parallel in one week, and studies are ongoing to reduce or remove the chemistry steps, which may result in both reduction of acid use and touch labor. A cavity optical inspection is part of the normal acceptance process. After this optical inspection, targeted repair can proceed.

#### 2.3.3.2 Summary on cavity yield

Based on the cavity data for phase II of the R&D programme, the 90 % second-pass yield goal has been met for gradients >28 MV/m. This ensemble of cavities has an average gradient of 37.1 MV/m, slightly exceeding the 35 MV/m goal. Cavity repair appears to be a viable option to ensure or raise the yield should that prove cost effective.

### 2.3.4 ILC Cavities for the 1 TeV Upgrade

Long-term R&D addresses both the gradient and $Q_0$ need for the ILC 1 TeV-upgrade. With an improved cavity cell shape and optimised material properties, 9-cell niobium cavities should be able to reach gradients in the range of 40–60 MV/m, as indicated below.

#### 2.3.4.1 Cavity Performance Goal: Gradient and Quality Factor

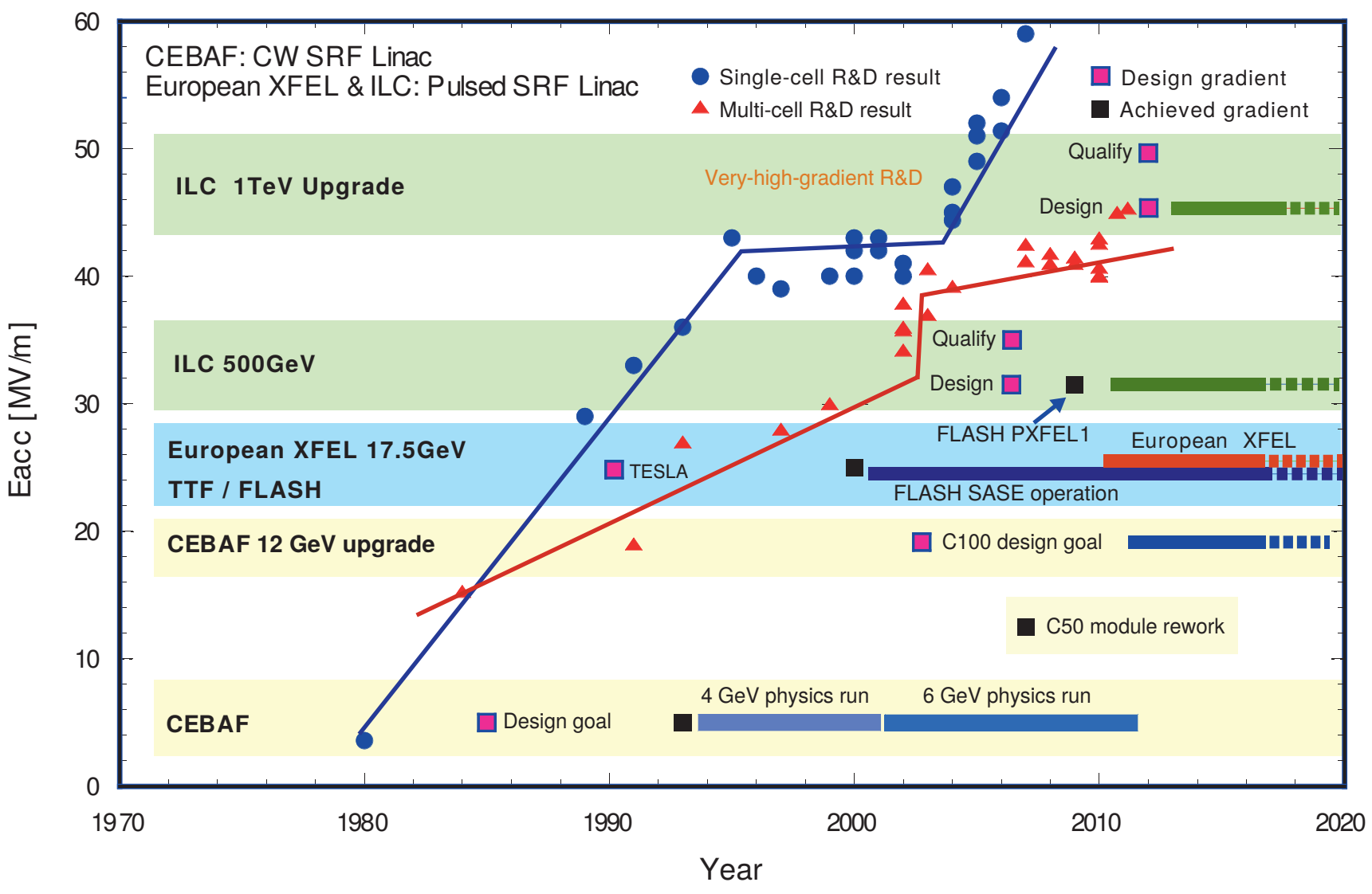

**Figure 2.20** L-band SCRF niobium-cavity-gradient envelope and gradient R&D impact on SCRF linacs.

The cavity performance goal for the 1 TeV upgrade of ILC is 45 MV/m at $Q_0$ of $2 \times 10^{10}$. A higher $Q_0$ is necessary to keep the cryogenic load at a manageable level as the cavity gradient is raised. Many 1-cell cavities have demonstrated a gradient of more than 50 MV/m in alternate-shape cavities. A 9-cell large-grain niobium cavity has achieved a record gradient of 45 MV/m [70]. Figure 2.20 illustrates the gradient envelopes of L-band SCRF niobium cavities and gradient R&D impact on SCRF linacs. Based on these results, it is believed that the SCRF cavity-gradient goal for 1 TeV ILC upgrade is reachable. The most promising path is the alternate-shape cavities processed by the current baseline recipe for the final surface processing.

In the long term, thin-film-coated cavities (such as $Nb_3Sn$) or multi-layer coated cavities offer potential of significantly increased cavity performance over the present state of the art.

### 2.3.4.2 Alternate-shape cavities

Alternative cavity shapes have a reduced ratio of the peak surface magnetic field to the acceleration gradient. There are three major alternate cavity-shape designs, namely re-entrant [71], low loss [72] and low surface field [73]. An effective increase of gradient up to 15 % can be expected as compared to the baseline shape. This is very attractive as this gain can be expected readily from the established baseline cavity processing that has been proven with the baseline cavity. In addition, these alternative shapes are more energy efficient. A reduction of dynamic heat loss of up to 25 % can be expected. There are some disadvantages to alternative shapes, for example reduced cell-to-cell coupling and increased HOM loss factors. These issues need to be further investigated in future ILC high-gradient cavity R&D.

Many 1-cell cavities have demonstrated a gradient of $> 50$ MV/m [74]. The current record gradient is 59 MV/m achieved by a 1-cell re-entrant-shape cavity [43].

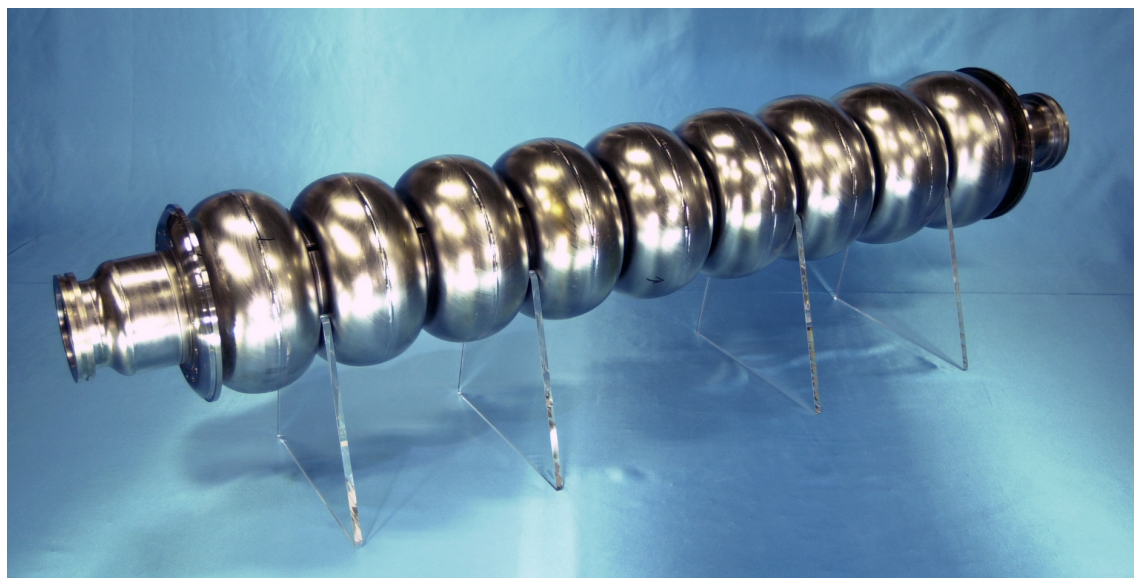

**Figure 2.21**
A 9-cell low-loss "ICHIRO" niobium cavity developed at KEK

The excellent 1-cell cavity results validate the rationale of lowering the peak surface-magnetic-field ratio to give higher acceleration gradient. However, the cleaning of the electropolished cavity surface must be improved, as the peak surface field in the real 9-cell cavities of alternate shapes are significantly (20%) higher than that in the baseline cavity. So far, most experience in alternate-shape 9-cell cavities is obtained with the ICHIRO shape developed at KEK (Fig. 2.21). The best ICHIRO 9-cell cavity has reached an accelerating gradient of 40 MV/m, limited by strong field emission [75]. Cornell University is continuing the development of 9-cell re-entrant shape cavities. JLab is developing the first prototype 9-cell cavity of the Low-surface-field (LSF) shape. The LSF shape, designed by SLAC, achieves 11% improvement in the peak surface-magnetic-field ratio without raising the peak surface electric field [73]. Table 2.11 gives a comparison of RF parameters of alternate shape cavity with the baseline cavity.

**Table 2.11**
Comparison of RF parameters of alternate-shape cavities with the baseline

| | | TESLA | Low-loss/ ICHIRO | Re-entrant | Low-surface field |
|---|---|---|---|---|---|
| frequency | GHz | 1.3 | 1.3 | 1.3 | 1.3 |
| Aperture | mm | 70 | 60 | 60 | 60 |
| $E_{peak}/E_{acc}$ | – | 1.98 | 2.36 | 2.28 | 1.98 |
| $H_{peak}/E_{acc}$ | mT/(MV/m) | 4.15 | 3.61 | 3.54 | 3.71 |
| Cell-cell coupling | % | 1.90 | 1.52 | 1.57 | 1.27 |
| G*R/Q | $\Omega^2$ | 30840 | 37970 | 41208 | 36995 |

### 2.3.4.3 Large-grain niobium cavities

The large-grain material is sliced directly from the high-purity niobium ingot. Many processing and handling steps for producing sheet niobium can be avoided. Therefore the large-grain material has potential for material cost saving. The forming and welding of large-grain half-cells are somewhat trickier. Due to considerable progress in the past few years, about twenty 9-cell large-grain niobium cavities have been fabricated and tested. Quite a number of 1-cell large-grain cavities have achieved a high gradient of > 35 MV/m with buffered chemical processing [76]. Eight industrially fabricated 9-cell large-grain niobium cavities have been processed and tested at DESY. Many of them reached a high gradient, with one of them reaching a remarkable 45 MV/m (Fig. 2.22), a record in 9-cell niobium cavities [70]. DESY's experience has shown that EP is still a necessary surface-processing step to reach high gradient in such 9-cell cavities. KEK and IHEP are also developing large-grain cavities [77, 78]

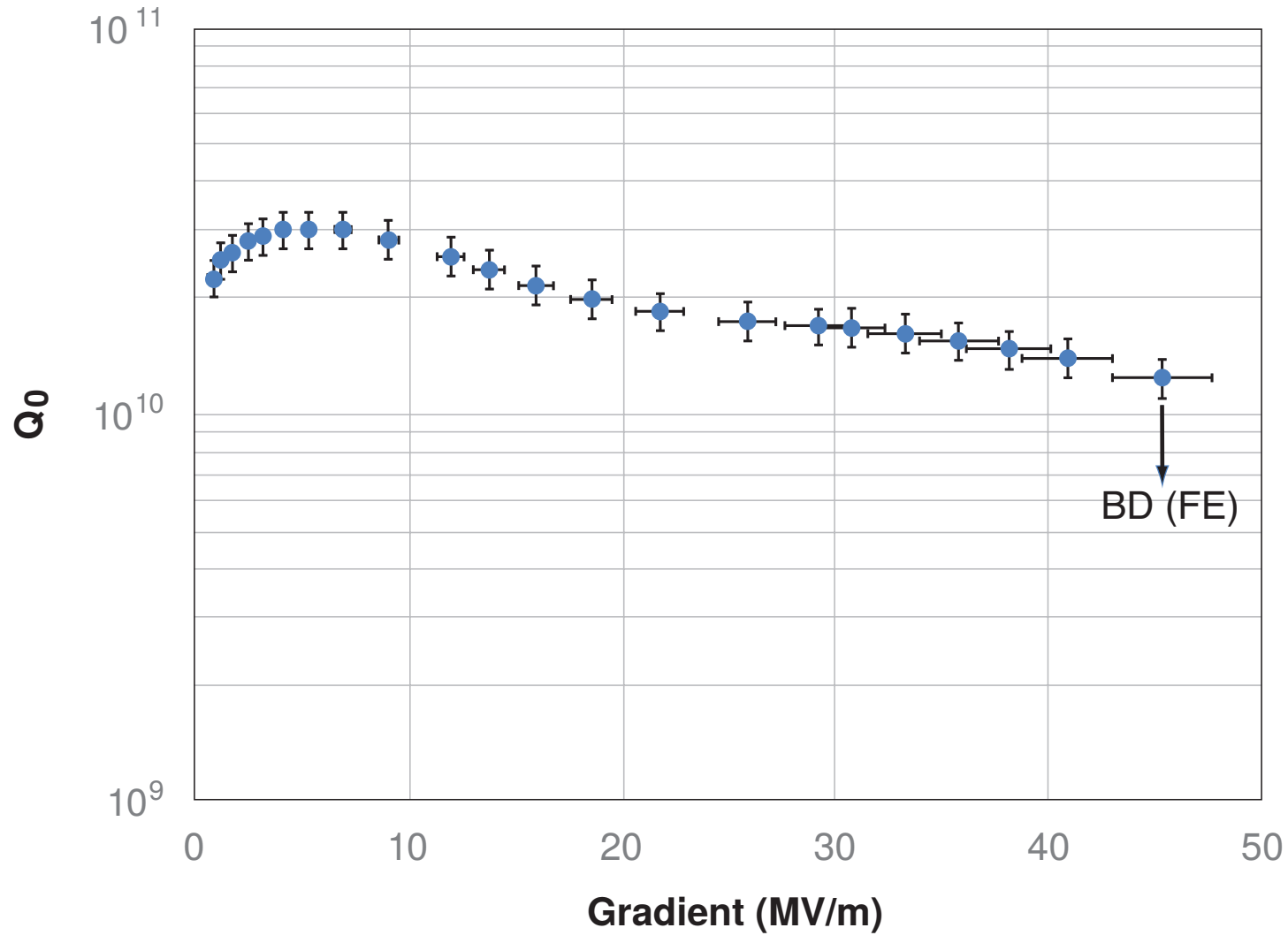

**Figure 2.22**
Demonstration of 45 MV/m by a 9-cell large-grain niobium TESLA shape cavity (AC155) at DESY.

An interesting observation is that the $Q_0$ value of a large-grain niobium cavity is higher as compared to that of a baseline fine-grain niobium cavity [79, 80]. Therefore, the large-grain material offers the added benefit of achieving higher $Q_0$ while the gradient is further pushed towards the 50 MV/m range. It is expected that large-grain material will be further evaluated for future high-gradient SCRF cavity R&D toward the 1 TeV upgrade of ILC.

### 2.3.4.4 Seamless cavities

The seamless cavity technology eliminates electron-beam welding at the equator region, where the peak surface magnetic field is high. It offers not only a possibility of cost saving in cavity fabrication but also a possibility of improved gradient yield. There are different approaches for making seamless cavities. Significant experience has been accumulated with the hydro-forming approach [81]. Many 1-cell, 2-cell and 3-cell hydro-formed seamless units have been produced and tested, showing gradients in the range comparable to that of welded cavities. More recently, several 9-cell seamless cavities have been fabricated by welding three 3-cell units together at the iris. The best result achieved so far is 34 MV/m, limited by quench [82].

To facilitate seamless-cavity fabrication, improved seamless tubes have been recently developed [83]. The first 9-cell cavity using these tubes (made by joining three 3-cell units through electron beam welding at iris) has been completed and is in the process of processing at testing at JLab [84].

The seamless-cavity fabrication technique is most suitable for building 9-cell cavities by using niobium-copper (Nb-Cu)-clad material [85, 86]. Due to the reduced use of niobium, the Nb-Cu clad material cavity has the potential for cost saving. The excellent thermal conductivity of copper at cryogenic temperatures provides good thermal stabilisation against local heating due to the presence of defects and hence is expected to improve gradient yield.

#### 2.3.4.5 Mechanical polishing

Mechanical polishing has been developed for a long time at KEK [87, 88]. It effectively eliminates surface defects originated from cavity fabrication steps. Therefore, the tolerance for fabrication flaws can be expected to increase. The mechanical removal of the damaged layer also reduces the amount of bulk surface removal by EP. These lead to improved fabrication yield and possibly some cost saving in fabrication and processing of ILC cavities. More recently, the mechanical polishing has been further optimised at FNAL and a mirror-finish inner surface has been achieved [49]. The FNAL efforts demonstrated effective cavity-performance improvement by using mechanical polishing to remove known quench-causing defects (see Fig. 2.23) [69]. Work is continuing at FNAL in assessing the suitability of implementing the mechanical polishing into the baseline cavity-processing steps.

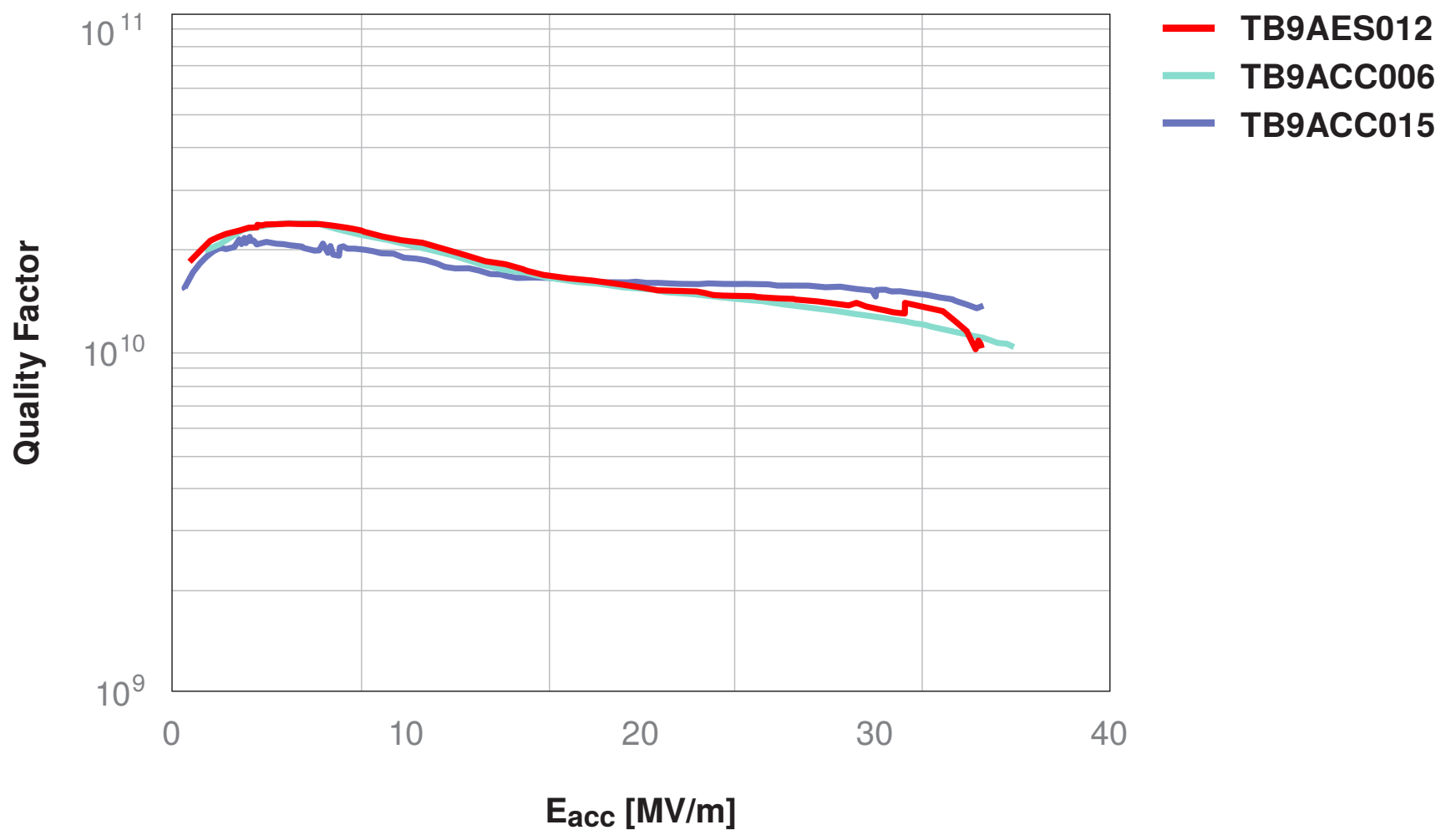

**Figure 2.23** Q($E_{acc}$) of 9-cell niobium cavities mechanically polished at FNAL followed by standard ILC final surface processing.

#### 2.3.4.6 Challenges and Perspectives

There are two major challenges toward the realisation of the SCRF cavity performance goal for the 1 TeV upgrade of ILC. The first one is to reduce field emission in the regime of very high surface field (100–120 MV/m). Although there is no known intrinsic limit to the sustainable electric field on the superconducting niobium, there is no experience yet with 9-cell cavities at this peak surface field. Fundamental studies of the nature of field emitters on a niobium surface are necessary. Advanced surface cleaning and in-situ processing procedures need to be developed. A quality factor of $2 \times 10^{10}$ at 45 MV/m has not been demonstrated even in a 1-cell cavity. Fundamental studies of medium-field Q-slope and high-field Q-slope are essential. Optimised heat treatment and surface coating need to be explored. A strong material R&D including RF characterisation (surface resistance and RF critical field) of small samples must be an integral part of the post-TDR SCRF R&D portfolio.

## 2.4 Cavity Integration

The cavity package in the cryomodule is an integrated system which consists of the niobium cavity proper, the input-power coupler, the titanium helium vessel, the frequency tuner, and the magnetic shield. The R&D performed on the frequency tuners, the input-power couplers, and the magnetic shields is summarised in this section.

### 2.4.1 Frequency Tuners

The frequency tuners play an important role in the efficient operation of a cavity. Both the fundamental frequency of the cavity can be adjusted and the detuning of the cavity under the Lorentz force (LFD) in a long RF pulse can be counteracted. All tuners apply a longitudinal force to change the length of the cavity. Typically a motor-driven adjustment applies the force to adjust the length of the cavity to the fundamental mode. A piezo-driven fine-adjustment applies a force to counteract the effect of the Lorentz force. Sophisticated feedback systems are required to control the action of the tuners during pulse operation. Achieving high reliability of the tuners (both mechanical and piezo) is a key requirement, and careful design of the stepper-motor drivers and associated gearing is crucial in this respect. This is particularly important for those designs that have the motor drive mounted directly at the tuner (in the cold) and which are not accessible outside the module.

The so-called Saclay/DESY tuner (Section 2.4.1.2) is the latest version of the tuner in operation at the TTF/FLASH facility, where 56 such tuners are routinely operated in the 7 installed cryomodules. (The same tuner will be used for the 800 cavities in the European XFEL). However, the Saclay tuner is mounted at the end of the cavity, extending the beam pipe on that end by 3.5 cm. To save this distance and increase the packing factor of the cavities in the ILC modules, it was considered attractive to include the tuner directly into the cavity helium tank, thus increasing the effective accelerating gradient. This has resulted in the design of the blade tuner which has been adopted as the baseline solution.

#### 2.4.1.1 Blade Tuner

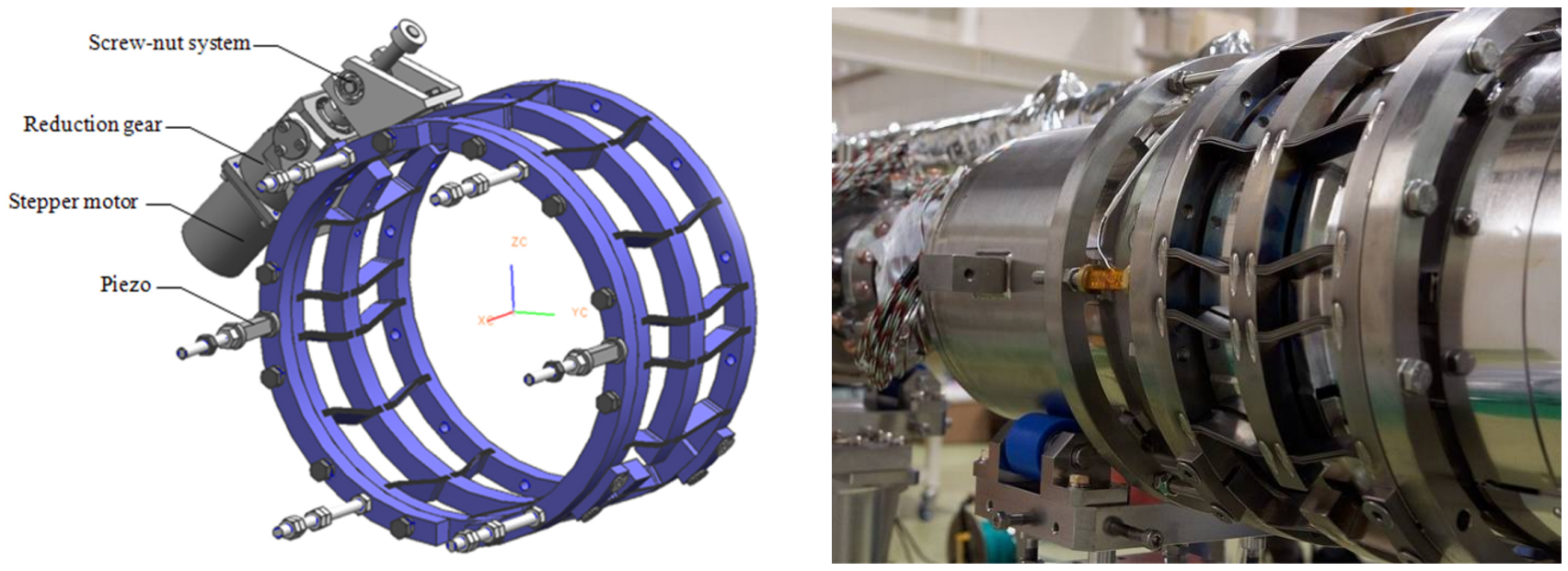

(a) (b)

**Figure 2.24** Blade tuner. (a) Drawing. (b) One of the two mid-plane mounted piezoelectric actuators is shown.

The blade tuner has been developed by INFN Milan as a coaxial and light-weight tuning solution for TESLA-type cavities, starting from the original model built for DESY. As shown in Fig. 2.24a and Fig. 2.24b, it features the concept of elastic blades capable of transforming a rotational movement of the central ring into a longitudinal cavity strain with zero backlash. The tuner mechanics as well as the blades are made from titanium, utilising its mechanical strength and also in order to minimise the effect of differential thermal contraction. The slow-tuning action is generated by a cold drive unit based on a stepper motor and coupled to a mechanical reduction gear. A CuBe-threaded shaft is used as a screw-nut system that generates the azimuthal motion of the central rings. The fast-tuning action is carried out by two piezoelectric ceramic actuators that are installed on the longitudinal plane on opposite sides of the helium jacket; their action is in series to that of the tuner in order to achieve the highest possible efficiency in transferring their stroke to the cavity. The entire mechanical load acting on the tuning system is generated by the cavity elasticity. Upon installation, an initial pre-load is set via a calibrated stretching of the cavity in order to achieve the optimal working condition for the piezo actuators. The cavity tension is large enough to accommodate the mechanical tuner range, thus guaranteeing that the piezos are always operated under compression. Details of the tuner and the actuators installed are also reported in Table 2.12. The blade-tuner cavity unit has been produced by a joint collaboration between FNAL and INFN in order to produce CM-2 for the NML/ASTA facility at Fermilab (see Section 3.4). The coaxial tuner is installed on the outside of the helium vessel that itself is split in two halves by a bellows to allow longitudinal adjustability. Moving the tuning bellow to the central position allowed for a simplification of the end cones region compared to the original TTF design.

**Table 2.12**
Various tuners investigated in the Technical Design Phase.

| | Blade tuner | Saclay/DESY tuner | Slide-jack tuner |
|---|---|---|---|
| Type | Coaxial | Lateral-Pick-up side | Coaxial and lateral coupler side |
| Tuner stiffness (design) | 30 kN/mm | 40 kN/mm | 290 kN/mm |
| Drive unit | Inside vessel, Stepper motor + Harmonic Drive | Inside vessel, Stepper motor + Harmonic Drive | Outside vessel, both manual or stepper motor actuation |
| Nominal frequency | 1.3 GHz | 1.3 GHz | 1.3 GHz |
| Nominal tunable range | 600 kHz | 500 kHz | 900 kHz |
| Nominal sensitivity | 1.5 Hz/step | 1 Hz/step | 3 Hz/step |
| Piezo | 2, thin-layer (0.1 mm), dim. $10\times10\times40\,mm^3$ | 2, thin-layer (0.1 mm), dim. $10\times10\times40\,mm^3$ | 1, thick-layer (2 mm), dim. diameter $35\times78\,mm^2$ |
| Piezo Voltage | 200 V | 200 V | 1000 V, operated at 500 V |
| Nominal piezo stroke at R.T. | 55 μm | 55 μm | 40 μm |
| Nominal piezo capacitance at R.T. | 8 μF | 8 μF | 0.9 μF |

## 2.4.1.2 Saclay/DESY tuner

**Figure 2.25**
Saclay/DESY tuner pictorial view

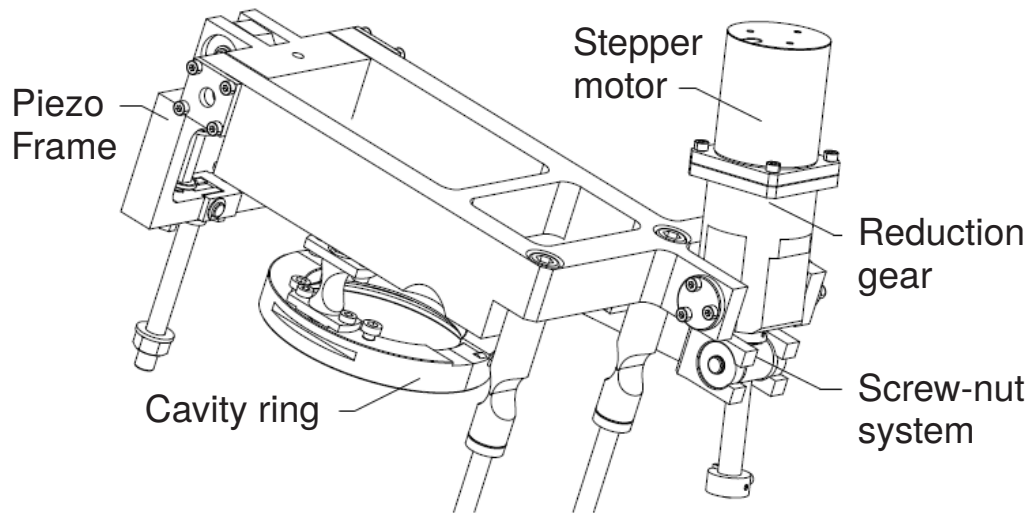

**(a)** Tuner assembly

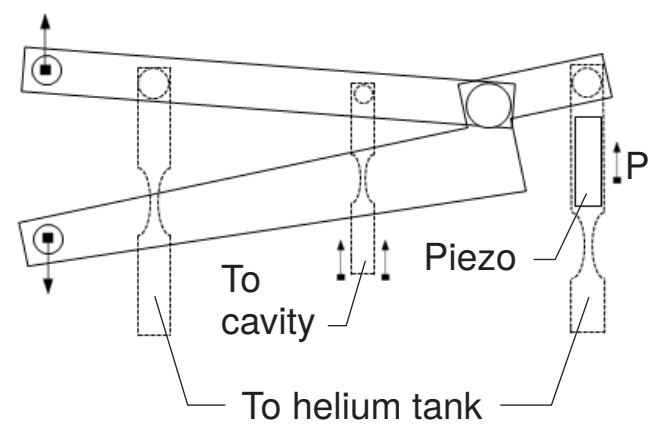

**(b)** Lever system for stepper-motor and piezos

The Saclay/DESY tuner has been further developed at DESY based on long operational experience. The Saclay/DESY tuning mechanics, shown in Fig. 2.25 and Fig. 2.26, is composed of a compact double lever installed at the pick-up side of the helium jacket made of stainless steel. The lever is

**Figure 2.26**
Saclay/DESY Tuner

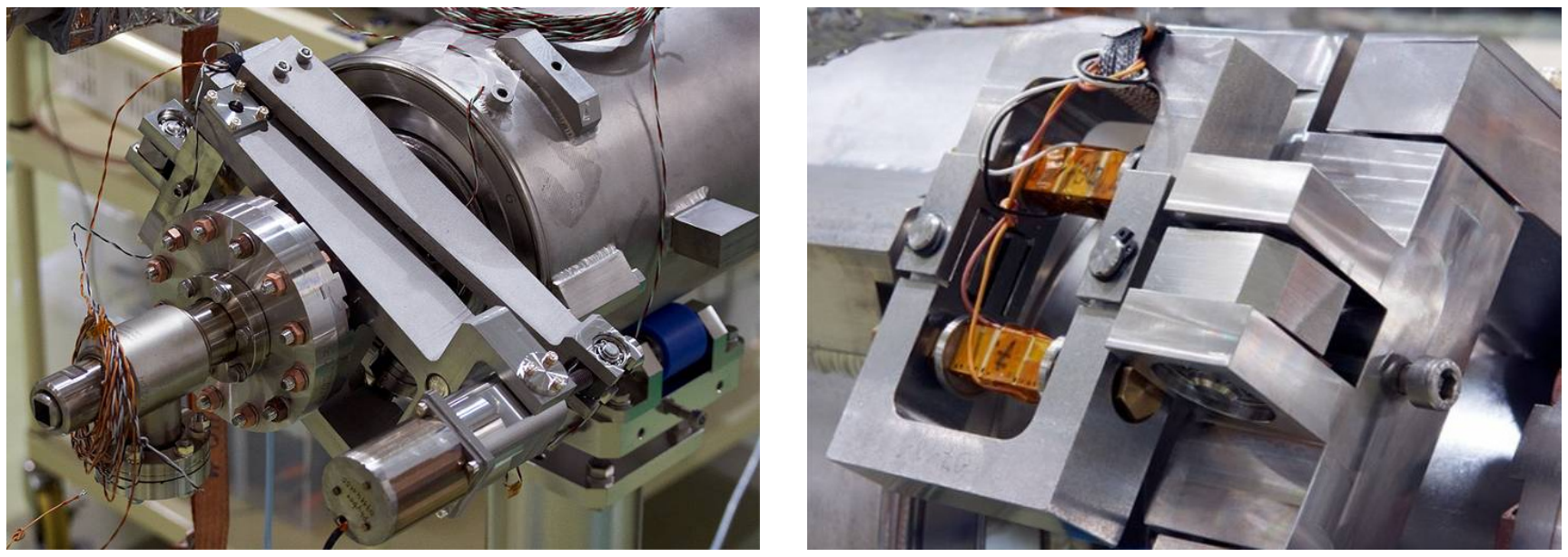

**(a)** Saclay/DESY tuner mounted on the cavity jacket

**(b)** Piezo pre-loading frame of the Saclay/DESY tuner

designed for a 1:25 reduction ratio and transfers to the cavity ring the tuning action generated by a cold drive unit based on a stepper motor equipped with a reduction gear and a screw-nut system. Two piezo actuators are installed to provide dynamic fast-tuning capabilities; they are both contained in a single preloading frame mounted at one side that acts on the cavity through the same lever mechanics. The piezo frame provides the required mechanical load on the actuators (almost independently of the load supplied by the tensioned cavity). The entire tuning system is installed outside of the helium tank (the tank itself can then be a relatively simply cylindrical design). A bellow with reduced radius is provided at the end opposite the power coupler, and fits in between the tuner joints, in order to accommodate the tuning strain. The double piezo configuration is, in this case, mainly chosen for redundancy reasons and the parallel action of the two actuators is not leading to a net doubling of the force. Moreover, the double piezo configuration, as in both blade and Saclay/DESY tuners, allows for the possibility of using the spare actuator as a passive sensing element. The salient characteristics of the Saclay/DESY tuner are summarised in Table 2.12.

### 2.4.1.3 Slide-jack Tuner

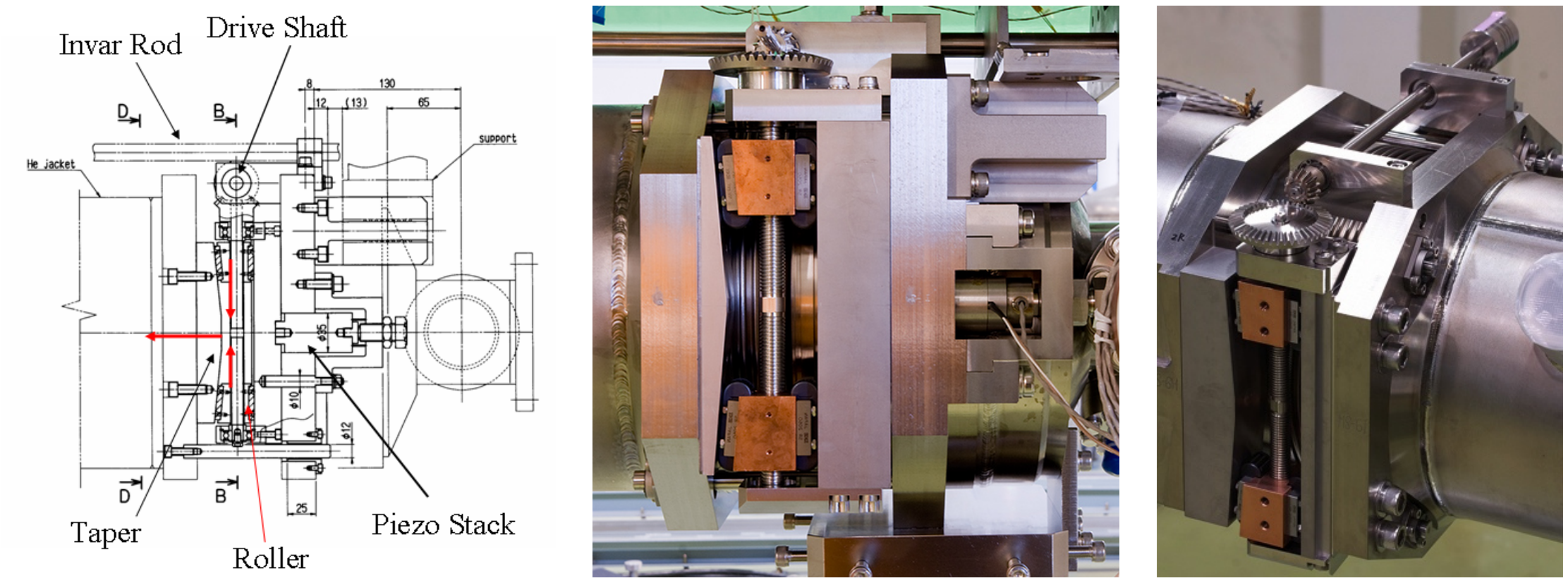

**(a)** Schematic drawing **(b)** Photograph **(c)** Driver

**Figure 2.27.** Slide-jack tuner

The slide-jack tuner has been specifically designed for the TESLA-like cavities developed at KEK for the STF project. Here the strategy of minimising the Lorentz force detuning has been pursued from the start. The design thus corresponds to a stiffer layout of the cavity and its constraints. The geometry of the resonator and the end cone regions have been optimised, which resulted in a longitudinal spring constant significantly higher than the original TESLA-type system. The coaxial tuning mechanism has been designed accordingly and is shown in Fig. 2.27.

The slow-tuning action is generated through rolling elements sliding on a sloping surface; this mechanism transforms the rotation of a driving shaft into a longitudinal cavity strain. In order to maximise reliability and ease of maintenance, no cold drive unit is foreseen in the design; the main driving shaft passes through the cryomodule vacuum vessel and the actual drive unit is placed outside. Either a stepper motor or a manual driver can be used. A single high-voltage, multiple thick-layer piezo actuator is installed at the coupler side of the jacket to provide fast-tuning capabilities. In the baseline design of this coaxial tuner, the slide-jack mechanics is installed between the pads, thus acting as a middle tuner. Similarly to the blade-tuner design, the helium jacket is split into two halves separated by a bellow. This slide-jack-tuner geometry will be referred to as *central* slide-jack. In order to compare the effects of different geometries for the tuning system, an additional cavity package has been developed which places the same slide-jack coaxial mechanical system outside the brackets on the helium tank on the coupler side, thus effectively acting as a *lateral* slide-jack tuner. The individual characteristics of the slide-jack tuners are summarised in Table 2.12.

#### 2.4.1.4 Measurement of Tuning Action

Comparative tests of the four tuner variants were performed at KEK as part of the S1-Global study program (Section 2.6). The full tuning ranges were successfully made on all four tuner variants in the cooled state: 300 kHz for the blade tuners, 400 kHz for the Saclay/DESY tuners, and 800 kHz for both KEK tuner variants. The resolution of the frequency control was also measured. Tuner performance using the piezo actuators was also measured: both DC and transient responses where successfully measured on each tuner type. Further details of the test results can be found in Section 2.6 and in the S1-Global report [7]. A key outcome from these tests was the successful validation of the technical performance of all four tuner design variants.

### 2.4.2 Input power couplers

Two types of input couplers were explicitly examined and evaluated in the Technical Design Phase, at the S1-Global experiment (see Section 2.6). Their key parameters are summarised in Table 2.13.

**Table 2.13**
Main characteristics of input couplers

| | TTF-III coupler | STF-2 coupler |
|---|---|---|
| Frequency | 1.3 GHz | 1.3 GHz |
| Pulse width | 1.5 ms | 1.5 ms |
| Repetition rate | 5 Hz | 5 Hz |
| Beam current | 9 mA | 9 mA |
| Required RF power | 350 kW | 350 kW |
| Range of external Q value | $1 \times 10^6$–$10 \times 10^6$ | $2 \times 10^6$–$4 \times 10^6$ |

While the power requirements are fulfilled by both couplers it should be noted that the range of supported external Q-values is not yet sufficient for the STF-2 coupler.

#### 2.4.2.1 TTF-III power couplers

**Figure 2.28**
TTF-III input coupler.

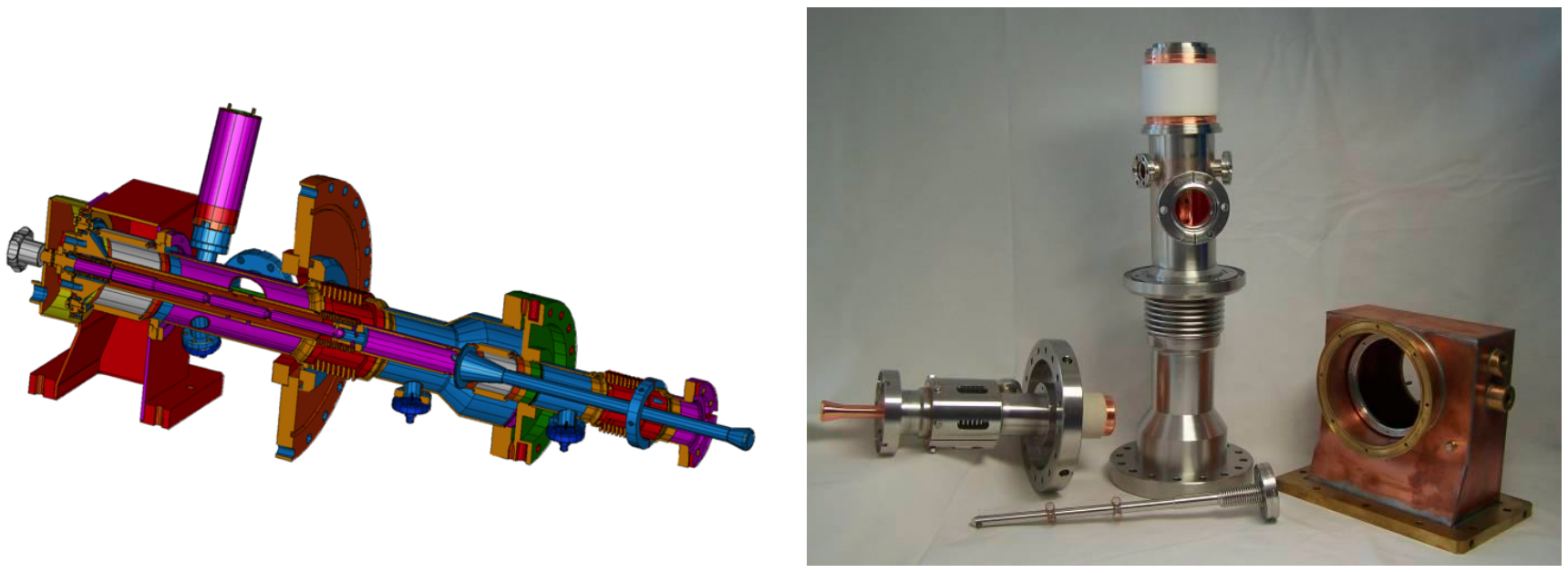

**(a)** Drawing **(b)** Assembly components

The TTF-III input power coupler is a mature design having now over a decade of R&D, and has been used extensively for the cryomodules developed at DESY for FLASH (56 couplers total), as well as a slightly modified design for the required 800 couplers for the European XFEL. The TTF-III coupler uses a double window design (warm and cold). Both the warm and cold RF windows consist of cylindrical ceramics made of $Al_2O_3$. The vacuum surface of the ceramic windows is coated with titanium-nitride (few nm) in order to prevent multipacting. The presence of bellows in the outer conductor allows the antenna protrusion into the cavity beam tube to be adjusted for variable power coupling. The inner conductor of the coupler can be DC biased in order to suppress multipacting. The input coupler has three electron current pick-up probes in order to measure electronic activity inside the coupler. The coupler and some of its components are shown in Fig. 2.28.

**Figure 2.29**
Set-up of high-power test stand at LAL.

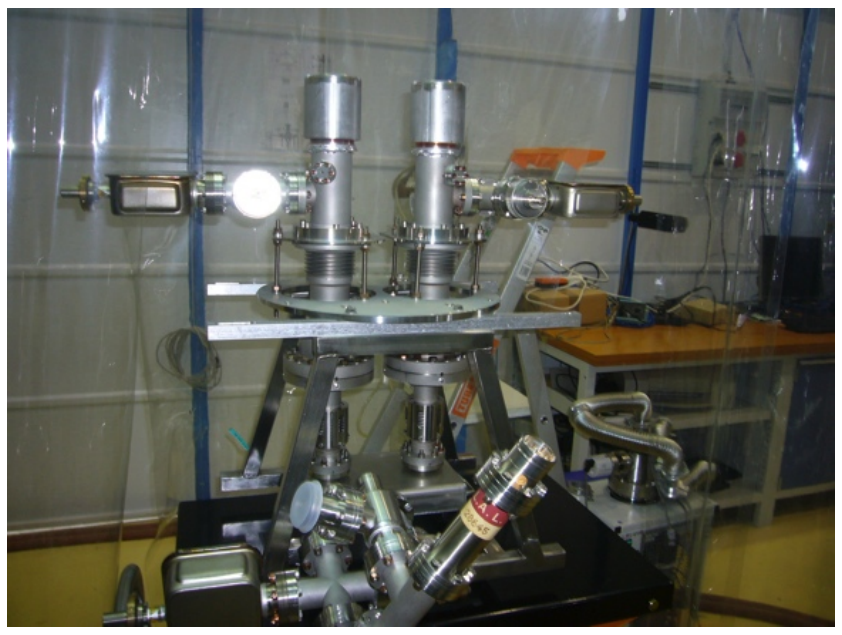

**(a)** Two couplers being conditioned

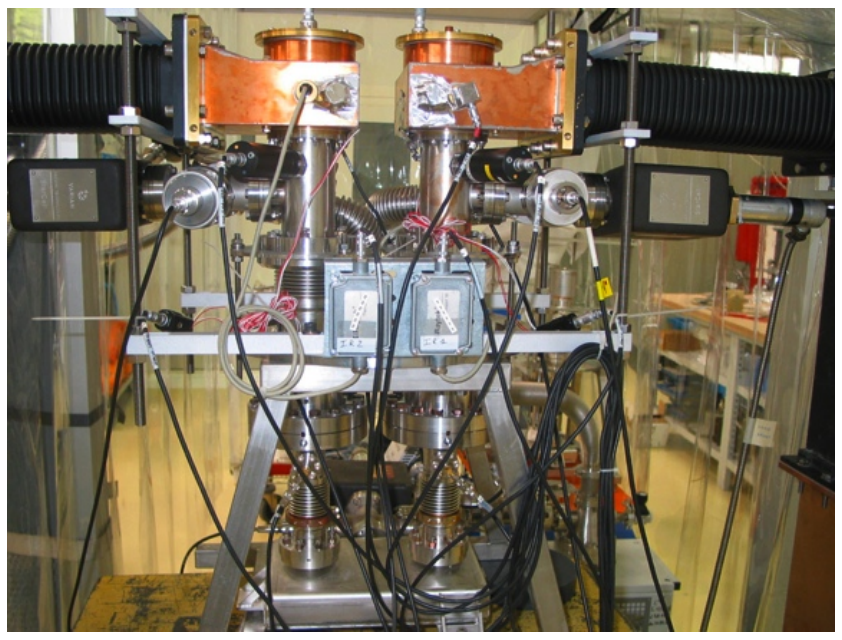

**(b)** Coupler test infrastructure

Four Type-III couplers were used for the S1-Global experiment (see Section 2.6). The two couplers for the DESY cavities were successfully conditioned at DESY and LAL respectively, while the remaining two couplers for the FNAL cavities were conditioned by SLAC. Figure 2.29 shows the processing set-up at LAL, and a similar system is installed at SLAC.

The RF conditioning of four TTF-III input couplers in the S1-global cryomodule was carried out with a high-power RF source connected to a 5 MW pulsed klystron. A monitoring system for electron activity using electron pick-up probes, and the interlock system for coupler protection, was supplied from DESY; both were attached to the TTF-III input couplers. The conditioning of the input couplers at room temperature before cooling had been carried out at up to 500 kW with pulse width of 0.5 ms and up to 200 kW with pulse width of 1.5 ms under total-reflection condition. The average conditioning time in four TTF-III input couplers was about 21 h. The vacuum interlock level was set at $2 \times 10^{-4}$ Pa.

#### 2.4.2.2 STF-2 power couplers

The input coupler, which consists of cold and warm sections, has two TRISTAN-type coaxial-disk RF windows as shown in Fig. 2.30. A bellow at the outer conductor in the 5 K part was eliminated to facilitate cavity assembly. The STF-1 input couplers used for the STF phase-1.0 had a simple structure with no variable coupling. Several improvements were made in the STF-2 input couplers for the S1-Global cryomodule. The schematic drawing and the completed STF-2 input couplers are shown in Fig. 2.31. The main features of the STF-2 coupler are summarised as follows:

1. Bellows were attached at the antenna tip of the inner conductor, so that a variable coupling of $\pm 30\%$ can be achieved.
2. The characteristic RF impedance is 41.5 Ω after enlarging the diameter of the inner conductor to insert a mechanism for variable coupling inside the inner conductor.
3. Thermal anchors at 5 K and 80 K were improved to suppress heat losses efficiently.
4. The design of the doorknob transition was modified to reduce the total length of the connection to the waveguide system.

**Figure 2.30**
Ceramics windows for the STF coupler.

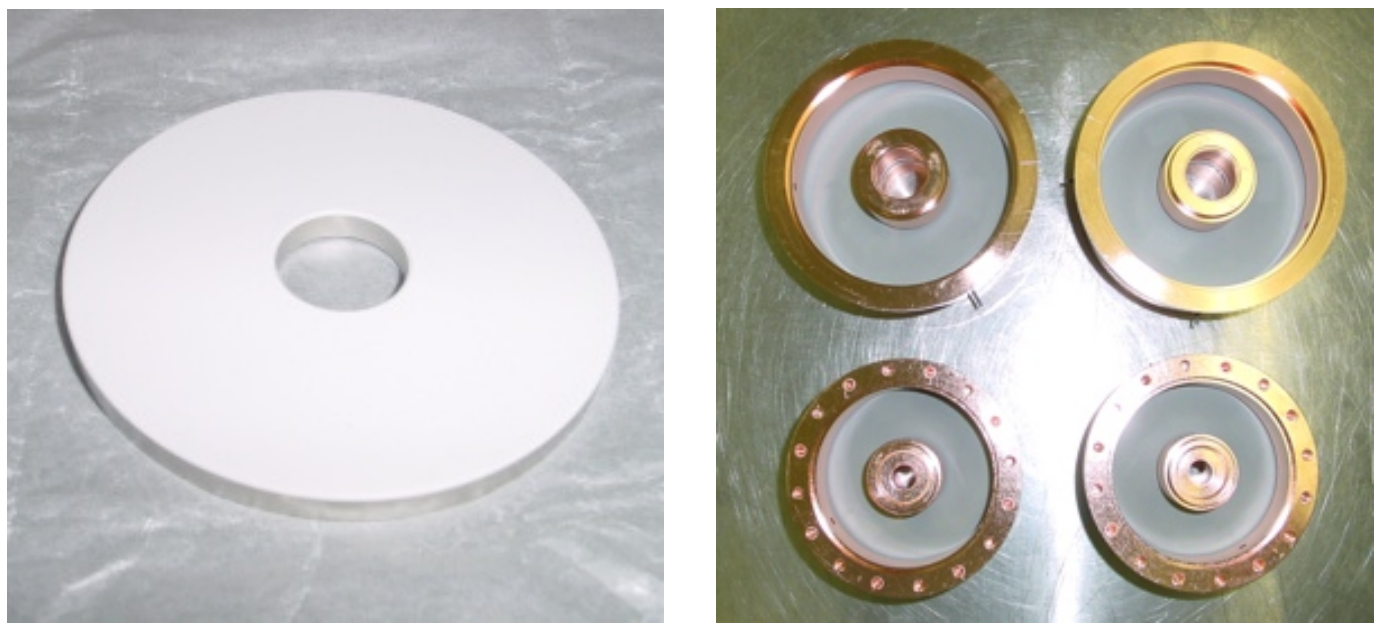

**(a)** Tristan-type coaxial ceramic disks

**(b)** RF windows after first brazing

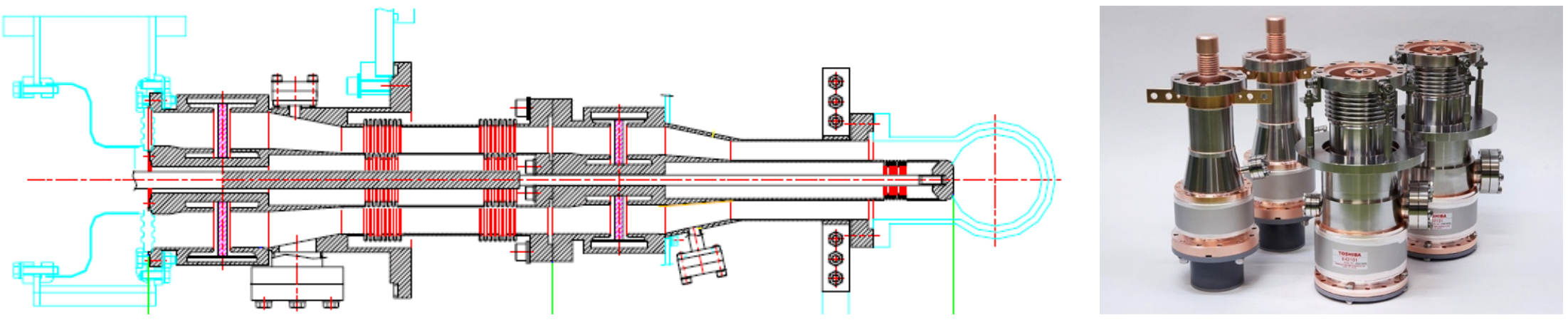

**(a)** Schematic drawing of the STF-2 input coupler

**(b)** Two sets of completed STF-2 input couplers

**Figure 2.31.** STF-2 input coupler

The static loss of 1.1 W at 5 K is five times larger than the dynamic loss of 0.2 W at an input RF power of 350 kW. The elimination of the thin bellows at the outer conductor in the 5 K part might be one of the causes for the high static loss at 5 K.

Prior to the assembly of the 9-cell cavities, the input couplers were conditioned using a high-power test stand, as shown in Fig. 2.32. The RF input power required for the conditioning in the travelling-wave mode at the test stand was determined to be four times higher than that in the standing-wave mode in the cryomodule. The conditioning was initially started using short pulses of 0.01 or 0.1 ms and the RF power level was increased very carefully. Finally, conditioning up to 1.0 MW in pulsed operation with 1.5 ms pulse width at 5 Hz was successfully performed on the four input couplers. The conditioning time at the test stand was approximately 60-120 h, as shown in Table 2.14.

**Figure 2.32**
Couplers in high-power test stand at KEK.

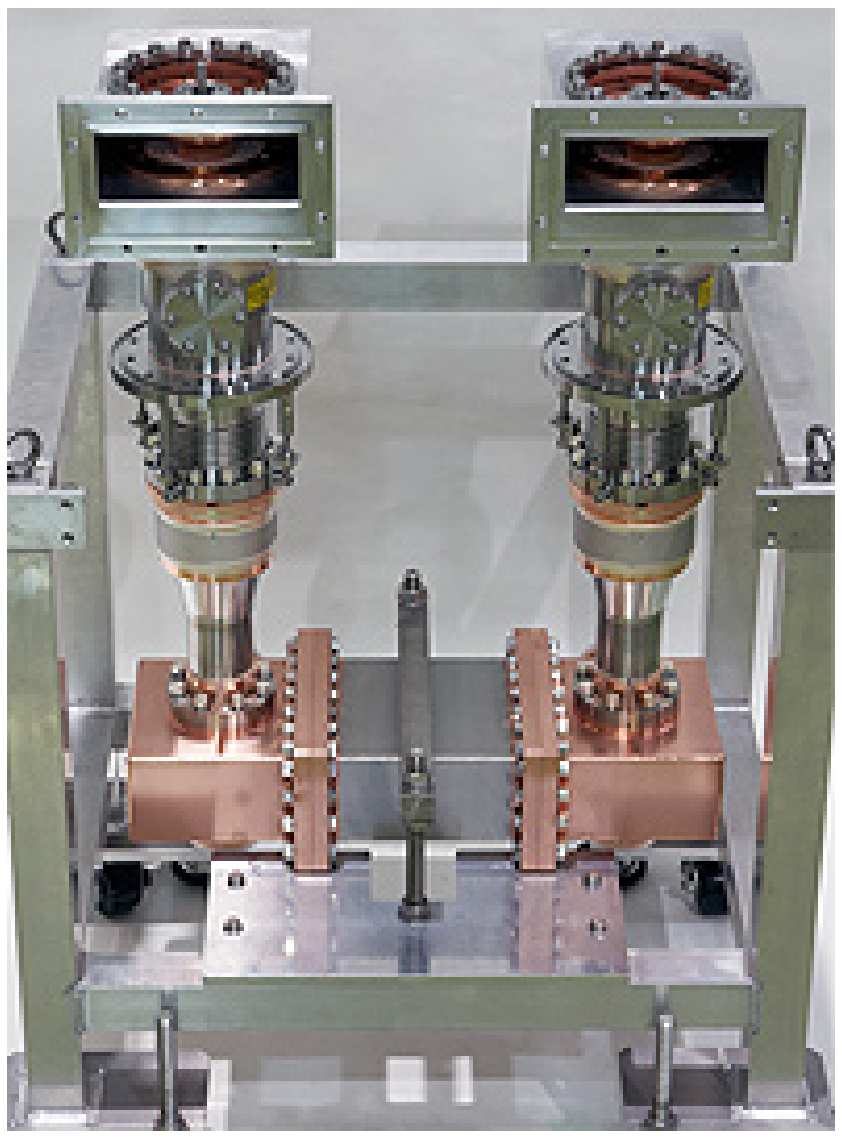

**Table 2.14**
RF conditioning of STF-II input couplers

| Pulsed operation | Couplers #1 and #2 max. $P_{rf}$, time | Couplers #3 and #4 max. $P_{rf}$, time |
|---|---|---|
| 5 ms, 1-5 Hz | 1500 kW, 33 h | – |
| 10 ms, 1-5 Hz | 1230 kW, 47 h | 1080 kW, 33 h |
| 20 ms, 5 Hz | – | 1140 kW, 1 h |
| 30 ms, 5 Hz | 1010 kW, 5 h | 1120 kW, 2 h |
| 60 ms, 5 Hz | 1060 kW, 4 h | – |
| 0.1 ms, 5 Hz | 950 kW, 6 h | 1050 kW, 7 h |
| 0.2 ms, 5 Hz | 880 kW, 6 h | – |
| 0.5 ms, 5 Hz | 820 kW, 4 h | 800 kW, 14 h |
| 1.0 ms, 5 Hz | 810 kW, 6 h | – |
| 1.5 ms, 5 Hz | 750 kW, 8 h | 270 kW, 6 h |
| Total time | 119 hours | 63 hours |

After installation of the cryomodule in the STF tunnel, the warm and cold parts of the coupler were joined inside a special booth to maintain a clean environment. In-situ baking of cold RF windows inside the cryomodule was carried out at 85 °C for 15 h. The baking of cold RF windows prior to conditioning is very effective for reducing the conditioning time. The conditioning of the input couplers at room temperature before cooling was carried out at up to 500 kW with a pulse width of 0.5 ms and up to 200 kW for pulse widths of 1.5 ms under total-reflection conditions. The average conditioning time of the four STF-2 input couplers was about 15 h - shorter than for the TTF-III input couplers.

## 2.4.3 Magnetic shields

The magnetic shield for the TESLA-style cavities is well established; no further R&D has been carried out recently. The shield design however had to accommodate the blade tuner.

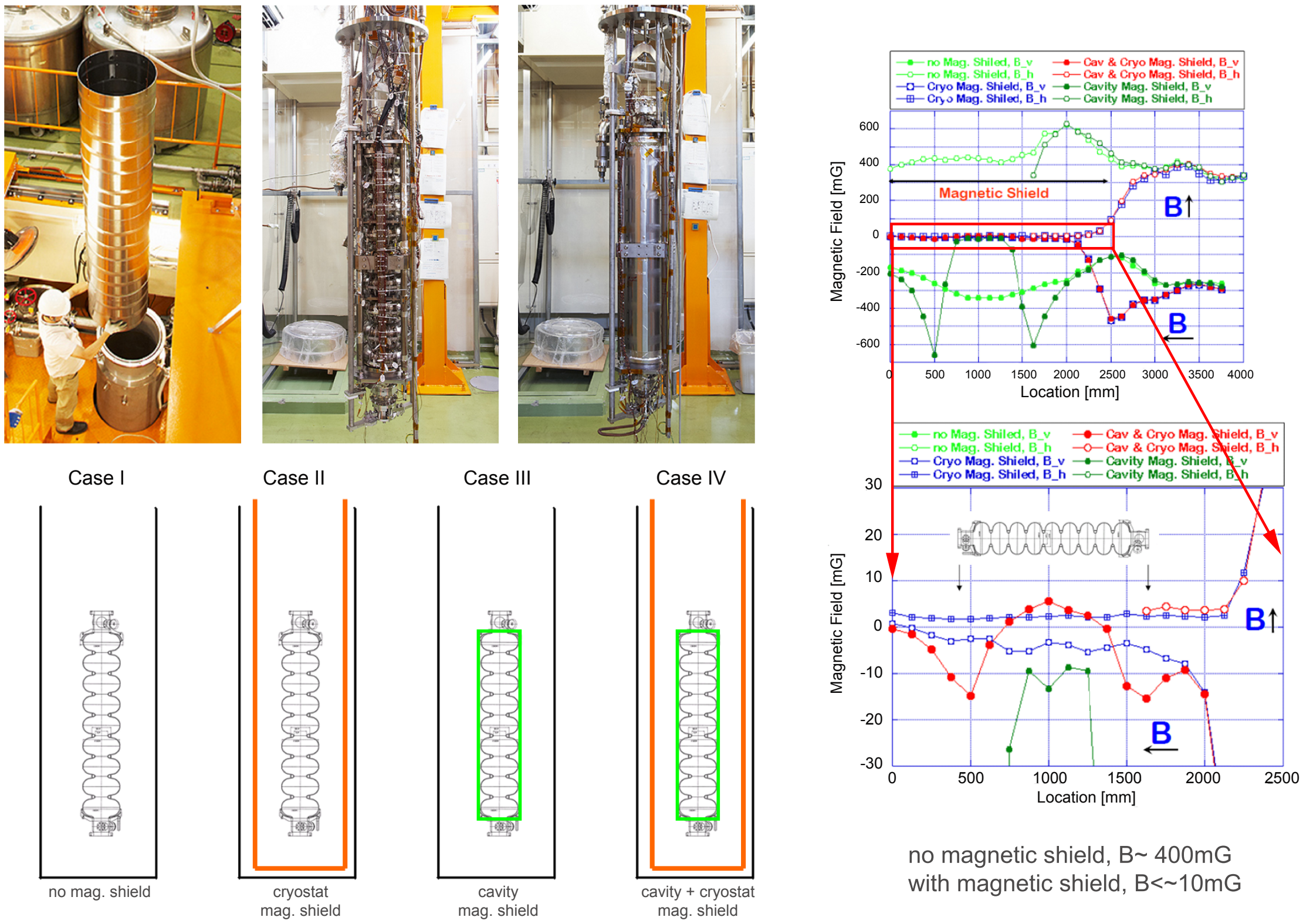

**Figure 2.33.** Four setups for the magnetic shield study in the vertical cryostat test and residual magnetic field inside the vertical cryostat.

In contrast, it was decided to test the design of the magnetic shield for the KEK cavities from the ground up. The design comprises the assembly of the cylindrical magnetic shield inside the titanium helium vessel with two conical shields between the end plates and the end cells. This configuration of the magnetic shield inside the vessel simplifies the cavity package installation into the cryomodule. The effectiveness of the magnetic shielding could be systematically studied in four configurations in the vertical test setup in STF (see Fig. 2.33); the configurations range from no shielding at all via shielding of cryostat or cavity to shielding of both cryostat and cavity. The configurations are detailed in Table 2.15, which includes the measurement of the residual resistance derived from the $Q_0$-measurement of the cavity in each configuration (Fig. 2.34).

**Table 2.15**
Magnetic shielding configurations in the vertical test

| Case | Shielding of | | $R_{res}$ |
|---|---|---|---|
| | Cryostat | Cavity | |
| 1 | – | – | 126.0 nΩ |
| 2 | yes | – | 10.3 nΩ |
| 3 | – | yes | 13.2 nΩ |
| 4 | yes | yes | 8.3 nΩ |

The measurements show that the cavity shielding (case 3) is almost as effective as the full cryostat shielding (case 2). The difference of 3 nΩ is small and has to account for the contribution from the beam-pipe port. That contribution will be further reduced once the cavity is mounted in the steel cryostat of the cryomodule, which provides extra shielding.

The magnetic shielding for the KEK developed cavities has thus been qualified and is adequate. The magnetic shield could be placed inside LHe tank allowing for a simplified the shield design with a single cylinder and the added convenience of unobscured mounting of the blade tuner outside the LHe tank.

**Figure 2.34**
The vertical-test results for the four setups of the magnetic-shield study.

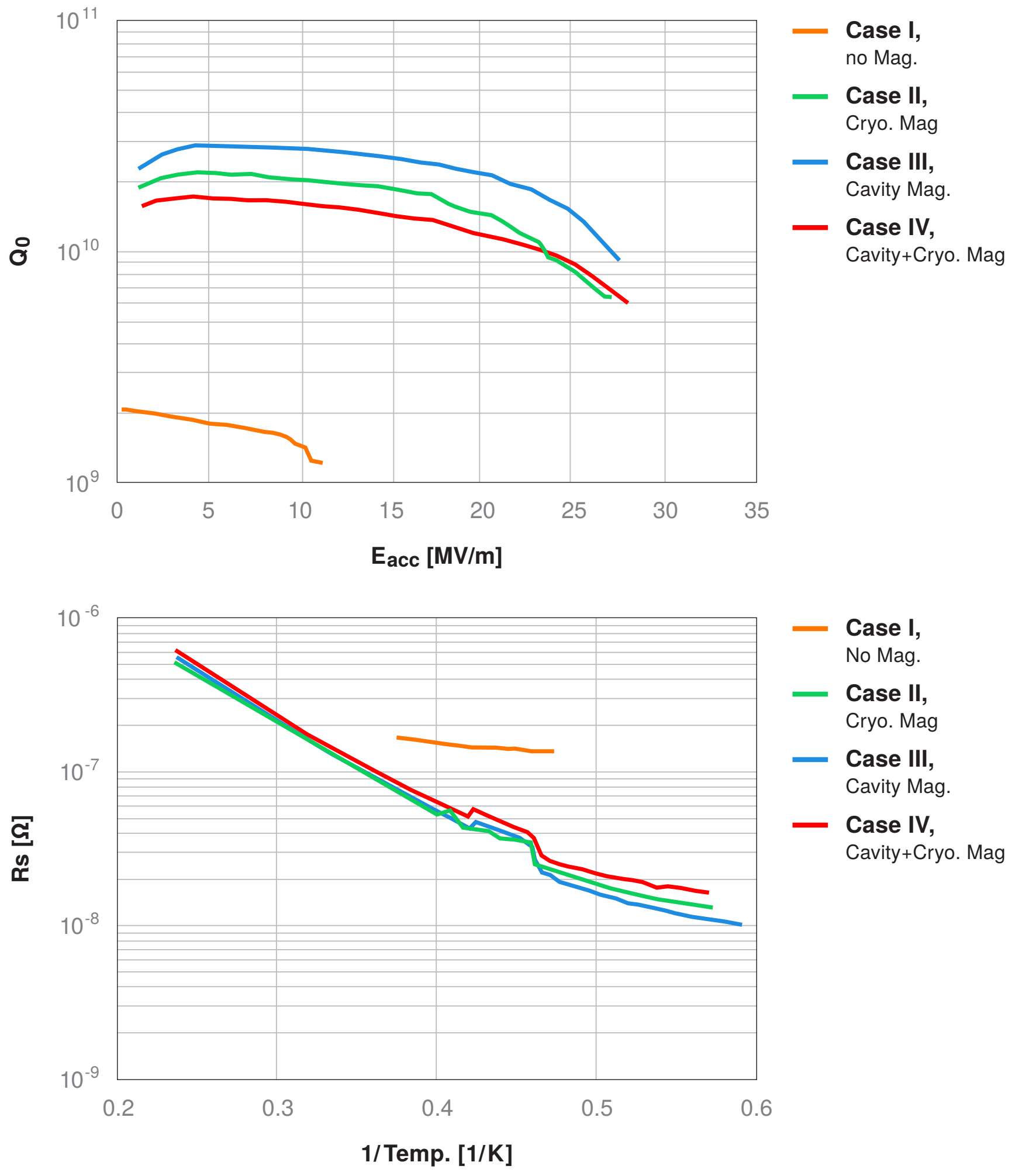

### 2.4.4 Summary of Cavity Integration R&D

Three frequency tuner designs have been successfully evaluated: blade tuner, DESY/Saclay tuner; and slide-jack tuner. All three designs met the ILC requirements under the same cryomodule-operation conditions, both in low-power and high-RF-power tests.

The two coupler types (TTF Type-III and STF2-coupler) have been compared with respect to installation procedure, RF processing, power capabilities and thermal performance. The Type-III coupler is by far the most mature design (with several years of operational experience), and has repeatedly achieved all necessary specifications for the ILC. However, the STF-2 coupler allows for a simpler installation procedure since it has no flexible bellows on the outer cylinder of the cold coupler part, and has a simpler design overall. The power capability is within requirements and the conditioning time was similar for both coupler types. The STF-2 coupler showed excess heat load around the cold part, with the first model installed in the S1-Global cryomodule. It has since been improved and is now consistent with the required design value after a test in the Quantum Beam program at KEK. In all, it is hoped that with further R&D, the STF coupler design can be made to be a more cost-effective alternative to the current TTF Type-III baseline design.

Finally, two approaches to the magnetic shield for the cavity have been evaluated, namely one external to and one internal to the helium tank, the former being the more common approach adopted by DESY and the European XFEL. The advantages of the internal tank are a simpler design (simple cylindrical shield with conical end parts), and easier assembly of the tuner during module assembly (cost factor). The internal design tested at S1-Global showed a tolerable 3 nΩ increase of the residual resistance compared to the external design. Although the internal design is currently the ILC baseline, the final design of the shield compatible with a cost-effective approach for the helium tank and tuner still needs to be made.

## 2.5 Anticipated Benefits from the European XFEL

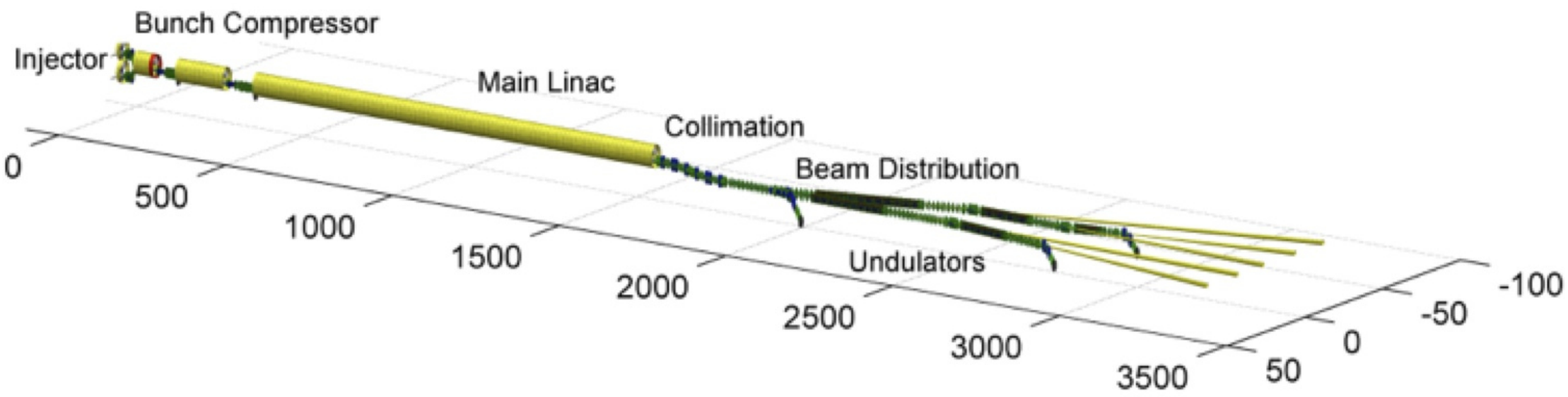

**Figure 2.35.** The layout of the European XFEL

The internationally organised European X-ray Free Electron Laser is currently under construction at the DESY and will begin operation in 2015 [6]. The layout is shown schematically in Fig. 2.35. Both the European XFEL and ILC use fundamentally the same pulsed SCRF technology originally developed for the TESLA linear collider [11]; both projects have shared in the worldwide R&D over the last twenty years. With 100 cryomodules containing 800 1.3 GHz superconducting cavities, the European XFEL is the first large-scale deployment of this technology. The strong synergy between the two projects provides unique and critically important benefits for the ILC development, ranging from fundamental R&D, through to integration and mass production and ultimately cost. As an international collaboration based on in-kind contributions, the European XFEL has also provided invaluable insights for the governance aspects of any future large-scale internationally funded project. Once operational, the European XFEL will provide unique operations and commissioning experience of a large-scale linac (7 % of an ILC linac). Table 2.16 lists the key parameters for the European

XFEL compared to the current ILC baseline.

**Table 2.16**
Comparison of some key parameters for the European XFEL and ILC Main Linac. For the ILC, the positron linac has been taken as an example. Where there is a difference between the regional variants, the values for mountainous/flat topographies are given. (The explicit numbers for the flat-topography design using KCS are average values.)

| | | European XFEL | ILC |
|---|---|---|---|
| Maximum beam energy | GeV | 17.5 | 250 |
| Accelerating gradient | MV/m | 23.6 | 31.5 |
| Charge per bunch | nC | 1 | 3.2 |
| Number of bunches per pulse | | 3250 | 1312 |
| Repetition rate | Hz | 10 | 5 |
| Bunch spacing | ns | 200 | 554 |
| Beam current | mA | 5 | 5.8 |
| Beam pulse length | μs | 650 | 727 |
| Matched loaded Q | | $4.7 \times 10^6$ | $5.5 \times 10^6$ |
| Fill time | μs | 803 | 927 |
| RF pulse length | ms | 1.45 | 1.65 |
| | | | |
| Number of klystrons | | 29 | 188/205 |
| Number of cavities | | 800 | 7332 |
| Number of cryomodules | | 100 | 846 |
| Cavities per klystron | | 32 | 39/36 |
| Average beam power per klystron | MW | 3.92 | 7.37/6.80 |
| | | | |
| Normalised Emittance (linac exit) | $\times 10^{-6}$m | 1.4 | 0.003 |
| BPM resolution | μm | 50 | 1 |
| Quadrupole lattice spacing | m | 12 | 38 |

The 800 superconducting cavities will be operated at an average gradient of 23.6 MV/m to accelerate the electron beam to an energy of up to 17.5 GeV. The 800 cavities — complete with input couplers, HOM couplers and mechanical tuners —will be installed into 100 cryomodules, each containing a string of eight cavities plus one superconducting quadrupole package (including dipole correctors and a beam position monitor). A total of 29 RF stations will supply the necessary RF power of typically 4 MW of beam power per four accelerator modules. The relatively low gradient (as compared to ILC) is considered conservative. However, it is important to note that the approach to cavity production for the European XFEL follows the same basic recipe that has essentially demonstrated the ILC yield (see Section 2.3.1.2), and there is every expectation that the average achievable gradient will exceed the specification, facilitating the possibility to upgrade to shorter wavelengths (higher beam energy). For this reason, the beam-line sections downstream of the linac are designed to accommodate a maximum beam energy of 20 GeV, corresponding to an average gradient of 28 MV/m.The construction of the cold linac is a common effort of many institutes. The overall coordination is with DESY as chair of the European XFEL Accelerator Consortium. Table 2.17 summarises the major contributions.

**Table 2.17**
Contributions to the European XFEL Cold Linac

| Institute | Component 7 Task |
|---|---|
| CEA Saclay / IRFU, France | Cavity string and module assembly; cold beam position monitors |
| CNRS / LAL Orsay, France | RF main input coupler incl. RF conditioning |
| DESY, Germany | Cavities & cryostats; contributions to string & module assembly; coupler interlock; frequency tuner; cold-vacuum system; integration of superconducting magnets; cold beam-position monitors |
| INFN Milano, Italy | Cavities & cryostats |
| Soltan Inst., Poland | Higher-order-mode coupler & absorber |
| CIEMAT, Spain | Superconducting magnets |
| IFJ PAN Cracow, Poland | RF cavity and cryomodule testing |
| BINP, Russia | Cold vacuum components |

The following sections briefly describe the status of the major sub-systems of the SCRF linac, emphasising the synergy with the ILC.

### 2.5.1 Cavities

The European XFEL cavities, high-power input and HOM couplers use the same modified TESLA design on which the ILC design is based (see Section 2.3.1). The mechanical design of the cavity itself differs slightly in the length of the long-end beam pipe, which has been reduced by ∼3 cm for the ILC cavity design to increase the packing factor of the much larger machine. As with the ILC, the complete European XFEL cavity package also includes the helium tank, magnetic shield and slow mechanical tuner and a fast piezo tuner, although here the actual baseline designs differ (Section 2.4), in part to accommodate the shorter inter-cavity spacing in the ILC design.

Two European vendors have been successfully qualified for the European XFEL industrial production of the required 800 cavities over a two-year period. The vendors are responsible for both the mechanical fabrication of the cavity as well as the surface preparation. The niobium and niobium-titanium material used for the production has been procured by DESY, which provides the necessary strict quality controls before delivery to the cavity vendors. The finished complete cavities are delivered to DESY ready for performance testing in a vertical-test cryostat. No performance guarantee has been specified for the industrial production; however the mechanical fabrication and surface preparation must precisely follow detailed specifications, which include the exact definition of infrastructure to be used. These specifications developed by DESY for the cavity production process (and also used during the tendering) have since been made generally available to the SCRF community, and have been used by the GDE as the basis for the industrial quotes and studies for the mass production of the 18,000 cavities required for the ILC.

The European XFEL cavity production benefits from the ∼20 years of linear collider R&D efforts at DESY and elsewhere. The long collaboration within the TESLA Technology Collaboration led to the set-up of infrastructure at DESY, mostly for cavity preparation, assembly and testing. The European XFEL cavity contract is based on improved infrastructure, both at DESY and at the two cavity vendors. Quality control of the required Niobium material — altogether twelve different lots of sheets, tubes, and rods — is done in a new dedicated DESY infrastructure. In addition optical inspection of the mirror-like inner surface of the finished cavities has been improved. DESY has developed a fully-automated optical system based on the KEK/Kyoto University high-resolution camera system (see Section 2.2.2) which is playing an important role during qualification of the first pre-series cavities. Any change in the production process (e.g. modification of welding parameters) will require optical inspection. The specifications for the surface preparation has equally benefited from the advances due to worldwide R&D. As a result, the European XFEL treatment will follow the same steps as the ILC (see Section 2.3.1.2), with the possible exception that buffered chemical polishing (BCP) is allowed as an alternative to electro-polishing (EP) for the final 'fine' polishing step. It is expected that at least one-half of the total of 800 cavities for European XFEL will follow the ILC recipe exactly (i.e. using final EP), and will be tested to their maximum performance: hence the European XFEL production will provide a large and unparalleled data sample for ILC-like cavity production using the now standard recipe for achieving high gradients.

As with the ILC, all European XFEL cavities will be tested at 2 K in a low-power vertical cryostat. The European XFEL production rate of 8 cavities per week requires two vertical test cryostats housing four cavities each, which are then RF tested in one cool-down / warm-up cycle. Development of automated systems and procedures for these tests are directly applicable to future ILC production.

## 2.5.2 Cryomodule

The cryostat design for the European XFEL is based on the first cryostats developed and built by INFN Milano within the TESLA R&D effort. Although the ILC module has developed several differences, it remains very close to the European XFEL design. Figure 2.36 shows a comparison of the ILC and XFEL cryomodule designs.

From the perspective of production, the cryomodule can be separated into three parts:

- the string assembly which comprises of the eight cavities and their associated auxiliary components (high-power input coupler, HOM couplers, helium tank, mechanical tuner etc.), the superconducting quadrupole package including a beam position monitor and a HOM absorber;
- the so-called cold-mass of the cryostat, which includes the 300 mm gas-return pipe, support fixtures (for the cavity string), thermal shields, cryogenic piping etc.;
- the outer (insulating) vacuum vessel.

The original type-I cryostat design was improved for FLASH at DESY (type-II); the further improved type-III design was finally shared with the worldwide community and forms the basis of the ILC type-IV module. Only minor modifications have been made to the type-III for the European XFEL. The procurement of the 100 cryostats (cold-mass and outer vessel) is organised by DESY; in total four vendors were qualified, and two finally contracted for the series production.

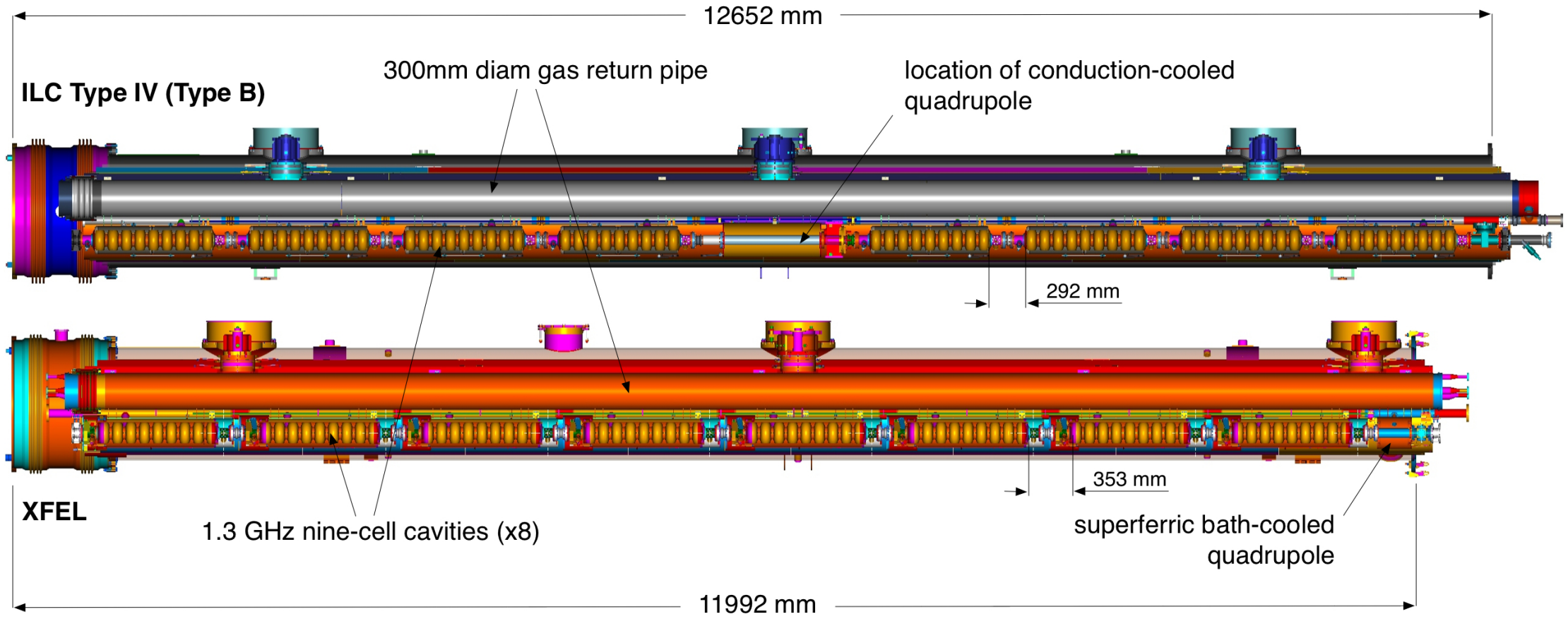

**Figure 2.36.** A comparison of the ILC (top) and XFEL (bottom) cryomodules. For the ILC the Type-IV module design with 8 cavities and one quadrupole package is shown. The most obvious difference is the longer length of the ILC module, driven primarily by the larger centrally located quadrupole (the longer quadrupole is required for the higher beam energy). The XFEL uses a superferric bath-cooled quadrupole located at the end of the module, while the ILC baseline locates the conduction-cooled magnet at the more stable central location. The reduced inter-cavity spacing is also indicated (ILC being 6 cm less than XFEL).

The cavity string and module assembly for the European XFEL is the responsibility of CEA Saclay / IRFU (France). A new dedicated infrastructure has been set-up for this purpose. Construction of the infrastructure started in 2009 and major parts were commissioned in 2010. An intense two-phase training period was used to transfer the assembly procedures from DESY to the IRFU supervisors and then to a sub-contracted company who will provide the approximately 30 personnel required to assemble the cryomodules at a rate of one per week. All major components required for the assembly are supplied by European XFEL collaboration partners as shown in Table 2.17. The work done at IRFU starts with the reception of already tested components (cavities, couplers, etc.) and ends with the shipping of completed accelerator modules ready for testing at DESY. As with the cavity production, the detailed specifications for the module assembly have been made available to the GDE and have been used as the basis for the industrial studies for ILC module assembly and cost estimation (see Section 2.9). Furthermore, the similarity in cryomodule designs has provided important technical information such as cryogenic heat loads, which can easily be scaled to the ILC specifications.

The first XFEL prototype module (PXFEL-1) achieved an average gradient on the module test stand of 32.5 MV/m when each cavity was driven independently (see Fig. 2.37), and represents the best performing module to date. The module average reflects a 10 % performance drop from the average of the individual cavity measurements achieved in the low-powered vertical tests (36.1 MV/m), mainly due to the large degradation observed in the last two cavities in the string (cavities No. 7 and 8). Such degradation in one or two cavities is not atypical and is an indication of contamination during string assembly. The XFEL series production will provide significantly larger statistics to help mitigate such assembly errors. PXFEL-1 is now installed in the FLASH linac, where it routinely accelerates beam, albeit at a reduced average gradient due to the limitations of the RF power distribution system. With its high-gradient cavities, it has been a focus of the ILC experimental programme at FLASH (see Section 3.2).

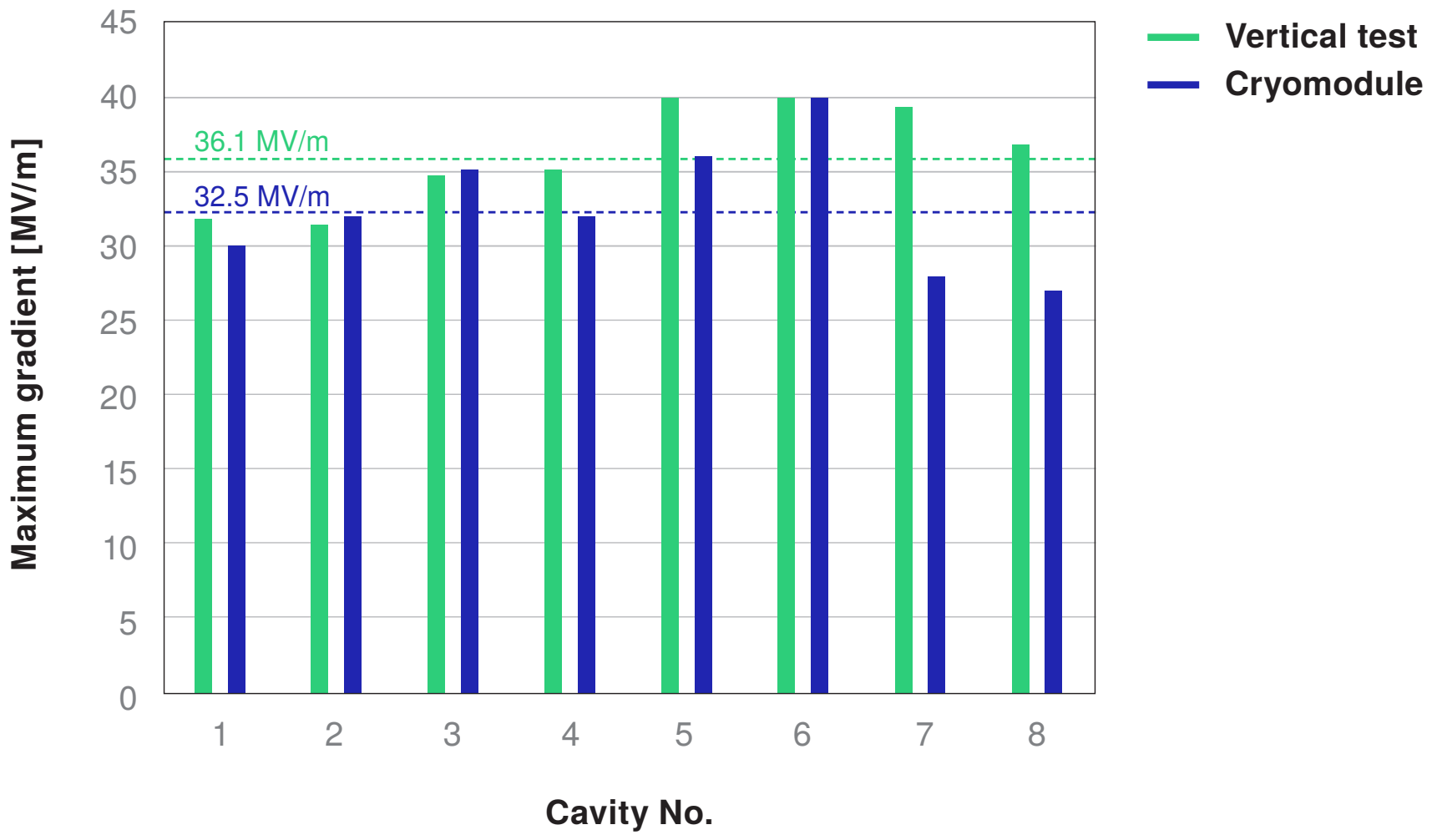

**Figure 2.37** Maximum gradient results for the XFEL prototype module PXFEL-1, where an average of 32.5 MV/m was successfully demonstrated (without beam) [89]. The individual cavity performance results from the low-power vertical tests are also shown for comparison. The observed 10 % degradation is primarily due to cavities 7 and 8.

## 2.5.3 RF Power System

The RF system includes all components required to convert AC line power to pulsed RF power and to distribute it to the superconducting cavities of the accelerator. The main components have been developed, constructed and tested for several years since the early 1990s, when DESY started to host the TESLA test facility (later converted to FLASH). The RF power source has now had nearly two decades of operational experience at TTF/FLASH, and more recently at the RF gun test-stand PITZ at DESY, Zeuthen. The European XFEL will use altogether 29 complete RF stations. The stations are split into the modulator installed outside the accelerator tunnel, and the pulse transformer / 10 MW multi-beam-klystron units inside the tunnel, close to the accelerator modules. Pulse cables of a length up to approximately 1 km connect both parts of the RF station. The 10 MW multi-beam klystron technology is shared by the European XFEL and ILC (Section 2.8). The many years of development of this klystron have lead to a mature design and the qualification of three vendors, from which two were contracted for the production of European XFEL klystrons. Modulator development started with the Fermilab bouncer-type modulator originally provided for the TESLA Test Facility in the mid 1990ies. A second generation of bouncer-type modulator was built by a German company. After further R&D, the European XFEL has finally adopted a pulse-step modulator — a solid-state modulator not unlike the ILC baseline Marx modulator. Beyond the klystron and modulator, further development of the RF system was required for the European XFEL. Due to the limited space inside the European XFEL tunnel, a compact waveguide distribution system has been developed. The waveguide distribution is based on a binary cell which consists of two circulators connected to a shunt tee with integrated phase shifters. Four binary cells are combined by three asymmetric pre-tunable shunt tees. The asymmetric shunt tees allow adjustment of the RF power to each pair of cavities to maximise the available cryomodule voltage. One MBK will drive 32 cavities (four cryomodules) via the distribution system. This is directly comparable to the approach adopted by ILC, although the details of the waveguide distribution systems are different; in particular the ILC distribution system can remotely adjust power to individual cavities.

### 2.5.4 Accelerator Module Test Facility

The primary performance testing of both cavities (vertical test) and the final complete cryomodules delivered from CEA Saclay will be made at DESY. A testing rate of 1 module and 8 cavities per week is foreseen. A new dedicated infrastructure has been set up for these tests —the Accelerator Module Test Facility (AMTF) — which offers not only the vertical tests for the cavities delivered from industry but includes also three complete test benches for assembled accelerator modules, which can be operated in parallel. AMTF also provides a test and assembly area for the waveguide distribution system, which is mounted to the module before installation in the tunnel. A test cryostat for the superconducting magnets is located in a neighbouring hall. All tests are performed by a dedicated team of some 20 people from IFJ PAN Cracow, Poland (as an in-kind contribution).

### 2.5.5 Electronic Documentation System

In addition to the technical specifications for the production process (from material QA/QC through to the final testing), the documentation process is also critically important. DESY has developed a set of comprehensive electronic documents which support every step of the manufacturing and testing process. Personnel at both DESY and in industry have been trained to use the central DESY engineering data-management system (EDMS), which will provide complete and traceable documentation for every component in the machine. This is particularly important for the cavity production, not least for the adherence to the pressure vessel regulations. Every single part can be tracked through the successive cavity production, testing, and ultimately installation in the accelerator tunnel. DESY is providing the same EDMS support for the GDE, facilitating a central database for the technical design documentation for the ILC baseline. The development of the EDMS system for European XFEL and ILC by the DESY team continued in parallel, to the mutual benefit of both projects.

### 2.5.6 European XFEL as a Beam Test Linac for ILC

As noted above, the European XFEL is due to begin operation with beam in 2015, effectively providing the ILC with a demonstration of an approximate 1 km of superconducting linac based on the same technology. Both through early commissioning and development and finally to routine user operation, the European XFEL will provide unprecedented experience in the operation of such a linac. Although much of the development work and “proof of principle” operations issues have been successfully dealt with at FLASH, the European XFEL will offer the unique advantages of understanding the issues pertaining to a much larger-scale deployment. In particular:

- operational experience with a long cryogenic string, in which half the cryogenic heat load is due to the pulsed RF;
- control system development — especially for the LLRF controls – including large-scale automation of tuning required for European XFEL (and scalable to ILC);
- Machine-protection philosophy, including trip-recovery, quench protection etc.;
- Energy management (dealing with failed RF stations);
- Short and long-term stability, including statistics on cavity detuning, calibration drifts etc.;

- Beam-dynamics issues associated with a long linac, including emittance preservation and HOM wakefield effects[4].

Until construction of the ILC, the European XFEL will represent a unique facility in the world. The direct synergy with ILC will allow much of the operations experience and control system development to be directly applicable to the larger machine. Together with FLASH and the more directly dedicated beam-test facilities at KEK and Fermilab, it is expected that the European XFEL will prove invaluable in understanding how to effectively operate a superconducting linac of the scale of the ILC.

## 2.6 The S1-Global Experiment

The S1-Global program was proposed in early 2008 with the aim of carrying out a string test of superconducting RF cavities during the ILC-GDE Technical Design Phase. The proposal was to coordinate a global collaboration involving various institutions around the world that would bring together eight 9-cell L-band cavities and associated hardware components, install them in common cryostat modules (cryomodules), and demonstrate their operation. This proposal by the GDE received strong support from the collaborating laboratories and was completed in a three-year period, ending in 2011 [7].

Two cavities with blade tuners were provided by Fermilab, two with Saclay/DESY-tuners were provided by DESY, and four with slide-jack tuners were provided by KEK. Information on the cavities is summarised in Table 2.18. Prior to being brought together for assembly into cryomodules, each of these cavities was processed and individually cold tested by the contributing laboratory. The cryomodules were provided by KEK (Hitachi) and INFN (Zanon). Input-power couplers for the cavities were contributed by SLAC, DESY, and KEK. The RF wave-guide components were provided by KEK and SLAC.

**Table 2.18**
Cavities used in the S1-Global experiment

| Cryomodule position | Cavity name | Company | Institute | Tuner | Coupler |
|---|---|---|---|---|---|
| C1 | TB9AES004 | AES | FNAL | Blade | TTF-III |
| C2 | TB9ACC011 | ACCEL | FNAL | Blade | TTF-III |
| C3 | Z108 | Zanon | DESY | DESY | TTF-III |
| C4 | Z109 | Zanon | DESY | DESY | TTF-III |
| A1 | MHI-05 | MHI | KEK | S-J central | KEK-STF |
| A2 | MHI-06 | MHI | KEK | S-J central | KEK-STF |
| A3 | MHI-07 | MHI | KEK | S-J lateral | KEK-STF |
| A4 | MHI-09 | MHI | KEK | S-J lateral | KEK-STF |

The programme successfully addressed such critical issues as plug-compatibility of hardware components (i.e. that parts must be compatible but not necessarily identical), as well as single- and multiple-cavity operation with pulsed microwave power and associated LLRF controls. While the S1-Global program did not involve beam operations, it covered all the essential steps required prior to beam acceleration in a superconducting linac.

The configuration and setup are shown in Fig. 2.38 and Fig. 2.39. The participating institutions contributed their hardware and human resources on an equal footing, and closely shared the experience of assembling a complex superconducting linac whose component designs and manufacturing were remotely coordinated. The STF site at KEK was selected as the host site, so as to take advantage of a one-year time slot that was available at the STF.

[4] while important experience can be gained from the European XFEL, the direct application to ILC is limited due to the factor of 500 difference in (vertical) emittance.

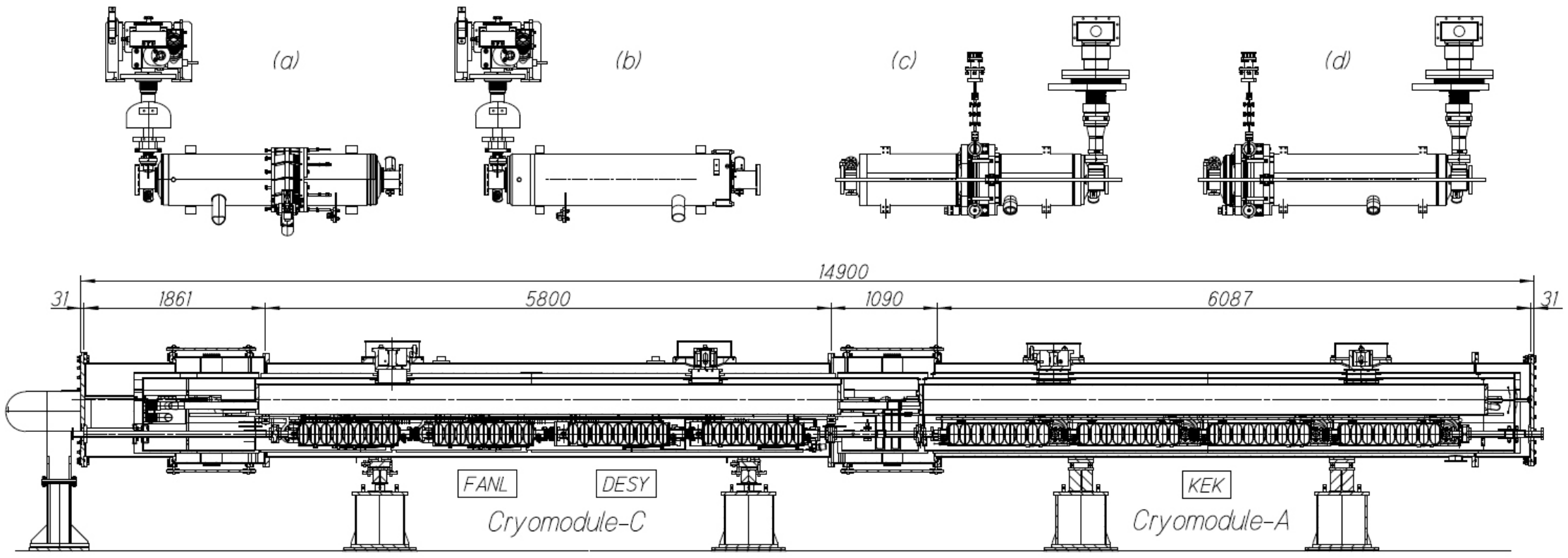

**Figure 2.38.** Layout of cavities in the S1-Global cryomodule string.

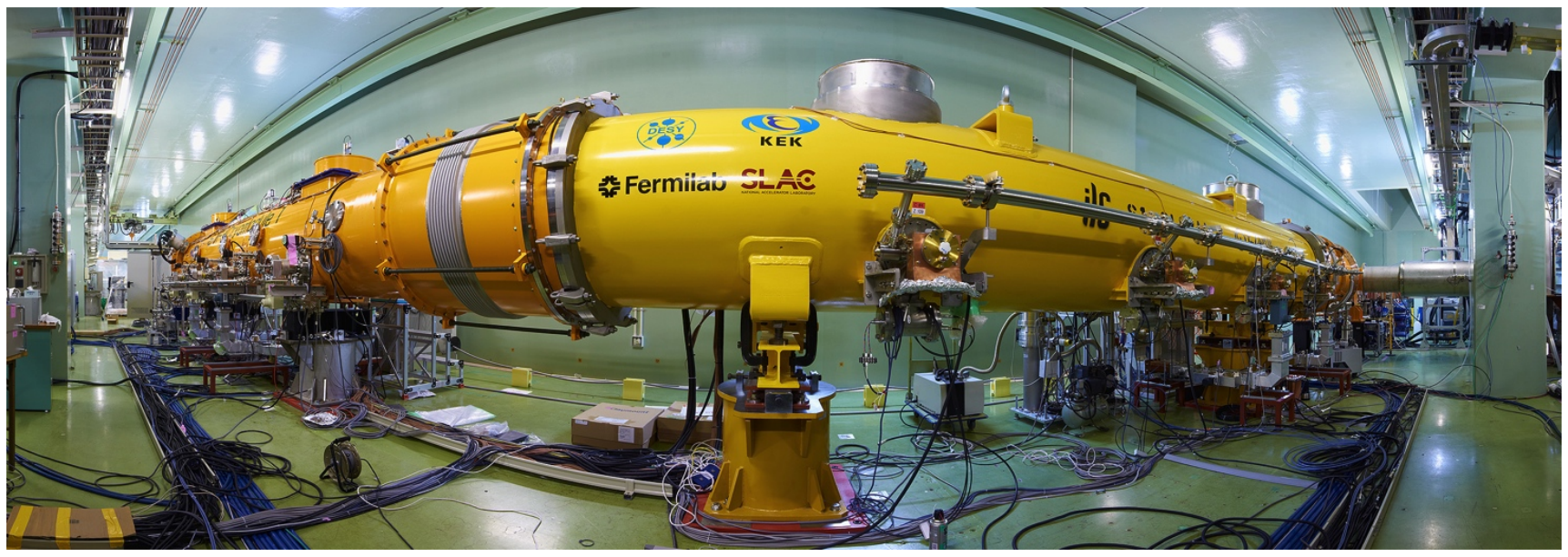

**Figure 2.39** Fish-eye view of the S1-Global cryomodule from the side of the input coupler.

## 2.6.1 Cryomodules and Cryogenics

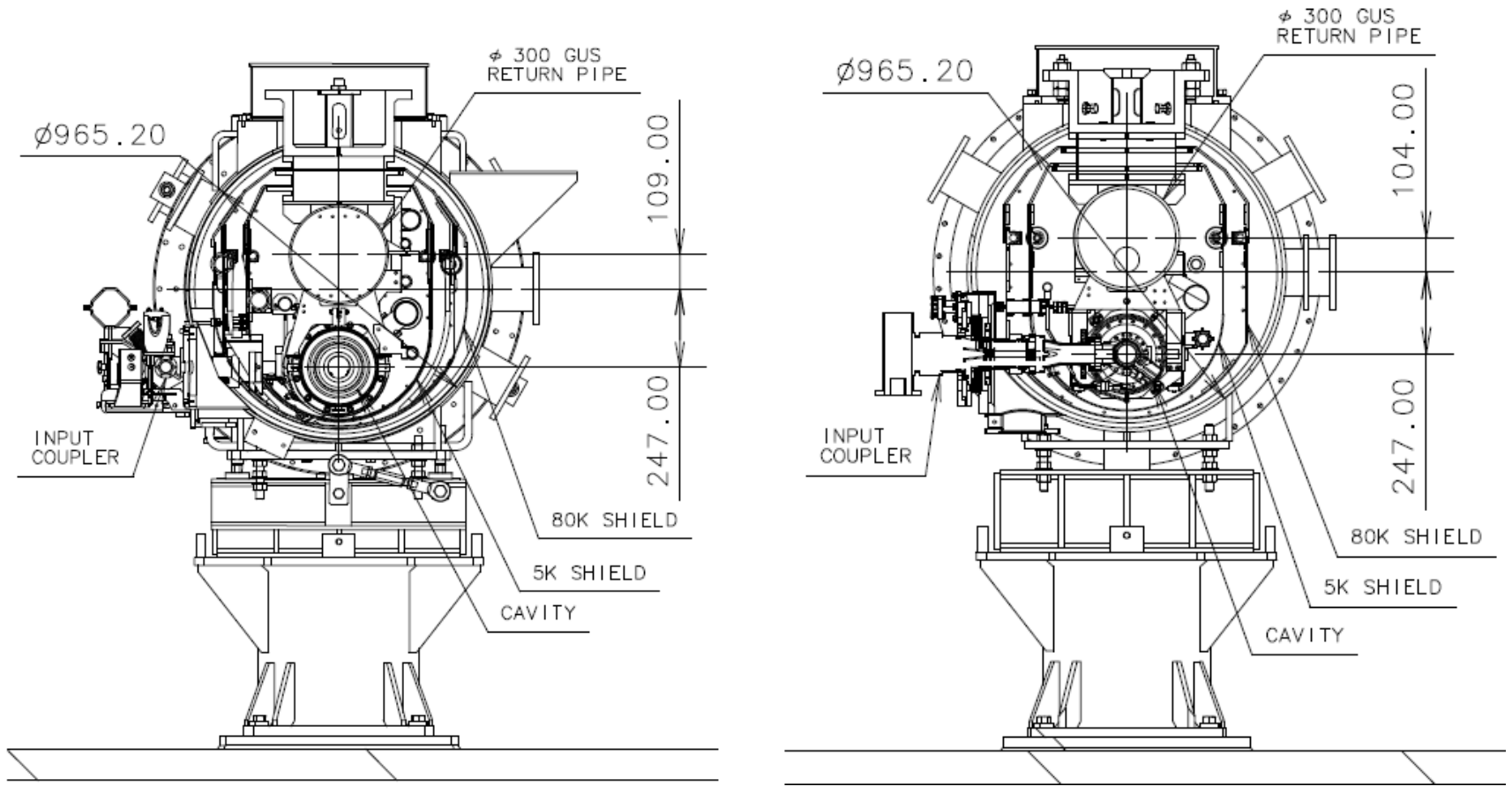

**(a)** Module-C for FNAL cavity

**(b)** Module-A for KEK cavity

**Figure 2.40** Cross section of Module-C for FNAL cavity (b) Cross section of Module-A for KEK cavity

Two cryomodules were used at the S1-Global experiment, cryomodules A and C. The cross section of Cryomodule-C is almost the same as the TTF-type III cryomodule, as shown in Fig. 2.40a. The cryomodule consists of the cavity packages, input couplers, the gas return pipe (GRP), magnetic shields, two sets of thermal shields, cooling pipes and the vacuum vessel. The distance between the input couplers is the same as for the TTF-type III and European XFEL design. The design of the cavity LHe jackets differs because of the different types of frequency tuners, namely the blade tuner and the Saclay/DESY-type tuner, as shown in Fig. 2.38. However, the cavity-support lugs have the same distance between two lugs along their axis, and a fixed distance from the input coupler axis, independent of the type of cavity package design and as a result, the cavity package supports to the GRP are identical, i.e. *plug compatible*. The frequency tuners of these cavities were driven by a cold motor and the piezo element. The cold motors were installed inside the 5 K shield. The DESY and FNAL cavities were designed to be enclosed with the magnetic shields. The cold mass at 2 K was enclosed with two thermal shields at 5 K and 80 K which were cooled with LHe and $LN_2$, respectively.

Cryomodule-A, which was used for the cold tests at STF phase-1, has been arranged to accommodate components of different dimensions and assembly. The KEK tuners have a drive motor outside the vacuum vessel and the ports for the shafts are welded on the vacuum vessel. The frequency tuners of these cavities were driven by the warm motor, while the piezo element is on the cold side. The cryomodule parameters are summarised in Table 2.19.

**Table 2.19**
S1-Global Cryomodule Parameters.

| | Module-A | Module-C |
|---|---|---|
| Vacuum-vessel length | 6087 mm | 5800 mm |
| Vacuum-vessel O.D. | 965.2 mm | 965.2 mm |
| Gas-return-pipe length | 5830 mm | 6000 mm |
| Gas-return-pipe O.D. | 318.5 mm | 312.0 mm |
| 2K LHe-supply pipe O.D. | 76.3 mm | 76.1 mm |
| 5K-shield pipe O.D. [F/R] | 30/30 mm | 60/60.3 mm |
| 80K-shield pipe O.D. [F/R] | 30/30 mm | 60/60.3 mm |
| Cool-down pipe O.D. | 27.2 mm | 42.2 mm |
| Distance between couplers | 1337.0 mm | 1383.6 mm |
| Cavity package | KEK-a/KEK-b | FNAL/DESY |
| Cavity type | TESLA-like | TESLA-type |
| Tuner type | Slide-jack | Blade / Saclay/DESY |
| Input-coupler type | STF-2 | TTF-III |
| Magnetic shield | Inside jacket | Outside jacket |
| Package length | 1247.6 mm | 1247.4/1283.4 mm |

The 2 K cryogenic system consists of a helium liquefier/refrigerator, a liquefied-helium storage vessel, a 2 K refrigerator cold box, two cryomodules, a helium-gas pumping system and high-performance transfer lines. Liquid helium is produced with a helium liquefier/refrigerator and collected in a storage vessel. The 2 K refrigerator cold box contains a He I pot, a He II pot, a heat exchanger and a Joule-Thomson (J-T) valve. After filling up the He II pot and the cryomodules with liquid helium, the system was pumped down to 3.2 kPa (the saturation pressure of helium at 2 K).

A more complete description of the S1-Global hardware is included in the full S1-Global report [7].

### 2.6.2 High-level and Low-level RF

Two high-level RF (HLRF) systems were used for S1-Global; stations 1 and 2. A TH2104C klystron was used for station No. 1 and a newly procured TH2104A for station No. 2. Both klystrons were capable of 5 MW output power at 1.3 GHz with a pulse width of 1.5 ms and a repetition rate of 5 Hz. The modulators incorporated a bouncer circuit and pulse droop was successfully compensated. The flat-top performance of the bouncer circuit in both modulators was better than $\pm$ 0.8%. The power-distribution system (PDS) comprised of two designs; the linear PDS and the tree-type PDS using 3 dB power dividers. Both systems were evaluated in the S1-Global test, as shown in Fig. 2.41. A tree-type power-dividing scheme using variable 3 dB hybrids was used with cryomodule-C. In this system, variable power dividers, i.e. variable tap-offs (VTO) supplied by SLAC, were utilised. A linear distribution system (TESLA-type PDS) was used in cryomodule-A. In order to protect the klystron from large reflected rf power, a 5 MW circulator was installed filled with $SF_6$ gas. In each coupler port, a 400 kW circulator was installed to eliminate any reflected power from the cavity. In addition, a variable loaded Q system ($Q_L$) using a reflector and a phase shifter was introduced in each coupler port.

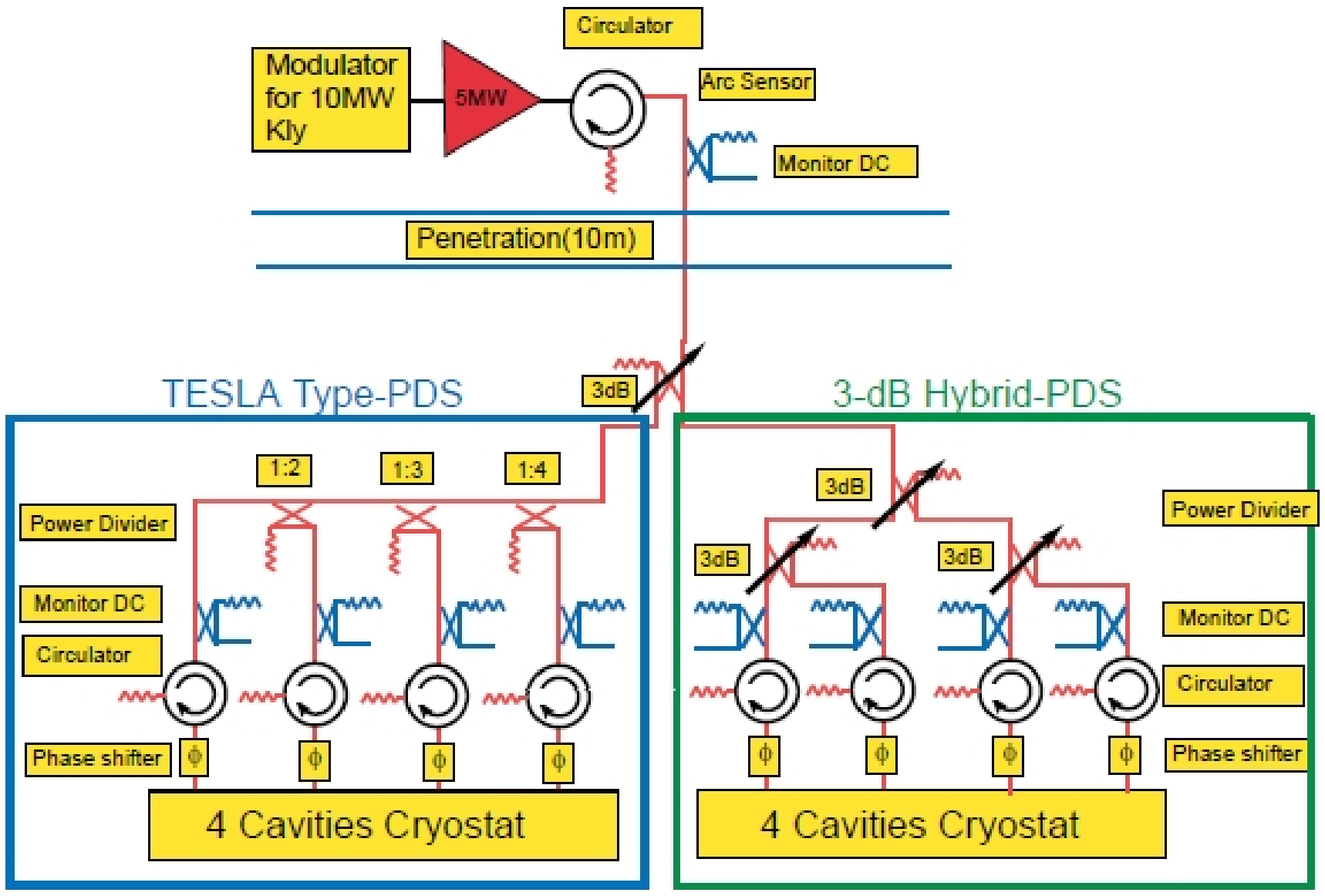

**Figure 2.41** Power-distribution system used in the S1-Global test.

## 2.6.3 Performance Test of Individual Cavities

The cryomodules were cooled down three times in total during the period June 2010 to February 2011. The first cool-down test was carried out from June to July 2010 to measure the thermal and mechanical performance of the cryomodules, together with low-power RF tests of all cavities [90, 91]. The second test was done from September to December 2011 to verify the cavity performance, Lorentz-force detuning (LFD) measurements and compensation by piezo tuners, long-term operation, and dynamic loss measurements under high RF power. The distributed RF system (DRFS) was tested in the third cold test [92] from Jan to Feb 2011 by using two 800 kW klystrons. In the low-power RF tests [90] by the INFN/FNAL/KEK team in the first cool-down, it was found that tuners attached in TB9ACC011 and MHI-09 did not function properly. Therefore, simultaneous operation of multiple cavities was limited to 7 cavities rather than 8. All of the operating tuners provided an acceptable tuning range. The results of the low-power tuner tests are shown in Fig. 2.42.

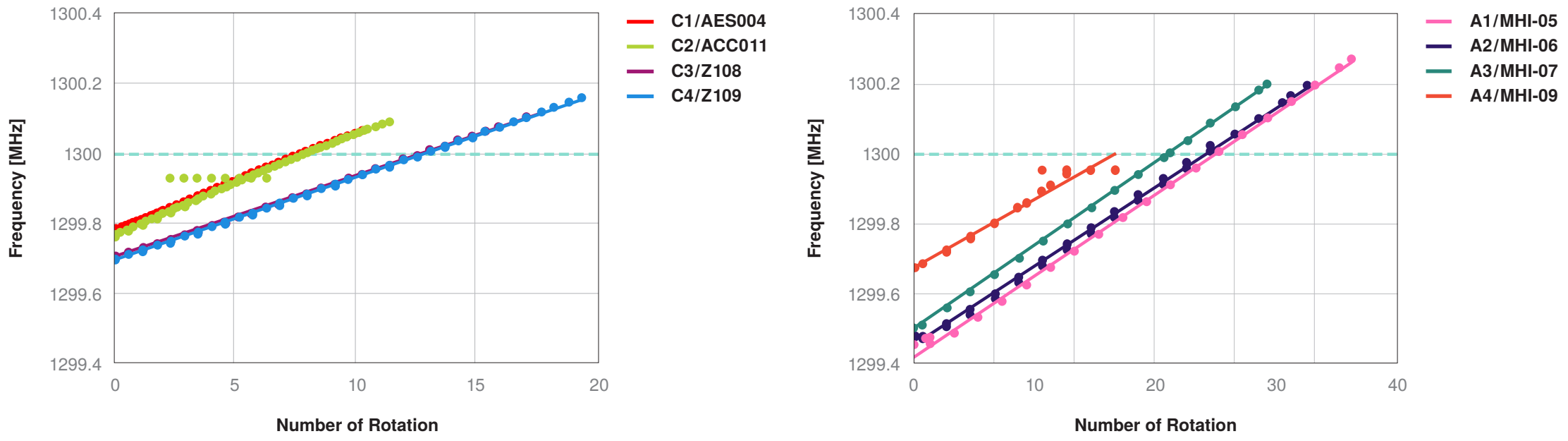

**Figure 2.42.** The result of the drive test for the motor tuner at low power. The tuner of TB9ACC011 did not work at 2 K, and MHI-09 could not be set to 1.3 GHz.

Adjustment of the variable coupling could be performed for all power couplers, and they were set to the optimum coupling of $2.4 \times 10^6$, resulting in a pulse rise time of 540 µs, as shown in Fig. 2.43.

The problems with the tuners and actuators were investigated and were the result of mechanical problems and are meanwhile well understood. One tuner failure was caused by the working loose of two screws in the shaft, presumably from mechanical vibrations during operation. The other failure related to a flange bending under the welding operation. The actuator failure was caused by misalignment of the brass caps holding the actuator stack and is hence related to mechanical tolerances during assembly.

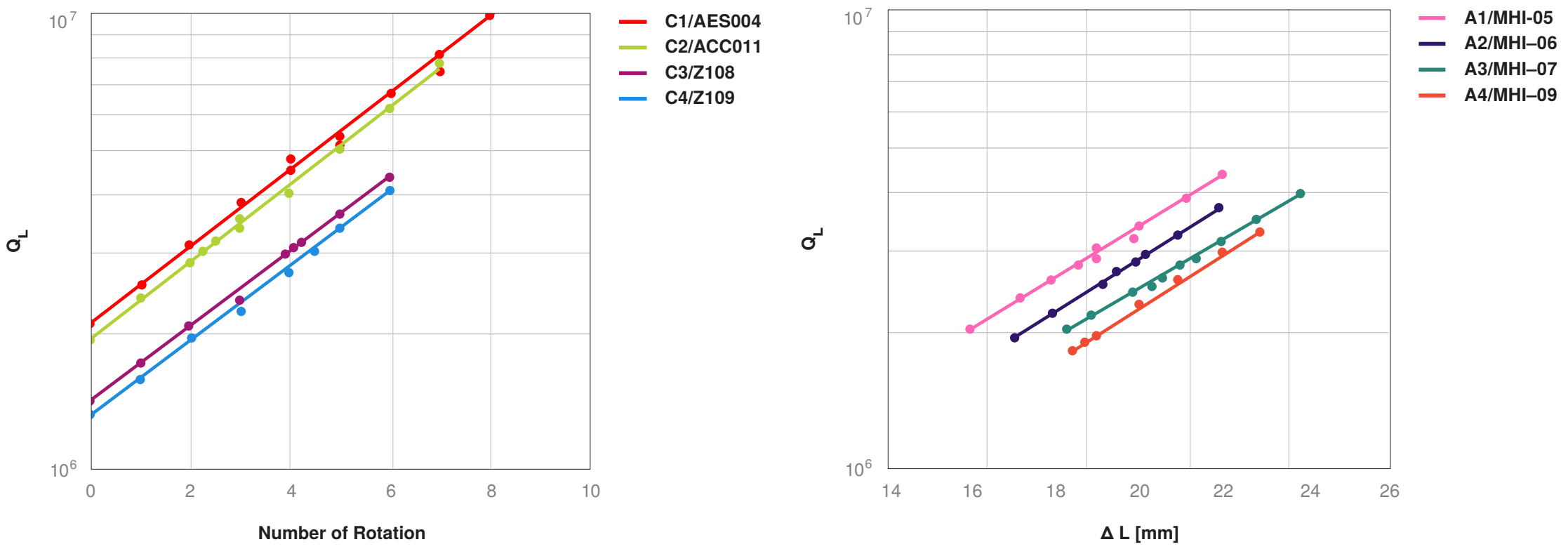

**Figure 2.43.** The results of the drive test for the variable coupling of the power couplers.

## 2.6.4 Cavity Performance in the S1-Global Experiment

After the second cool-down, the cavities were conditioned at high power. The achieved gradient values in vertical and cryomodule tests are shown in Fig. 2.44. The maximum average gradient was 30.0 MV/m in the vertical test, 27.7 MV/m for single cavity operation and 26.0 MV/m for simultaneous operation of seven cavities in the cryomodule test.

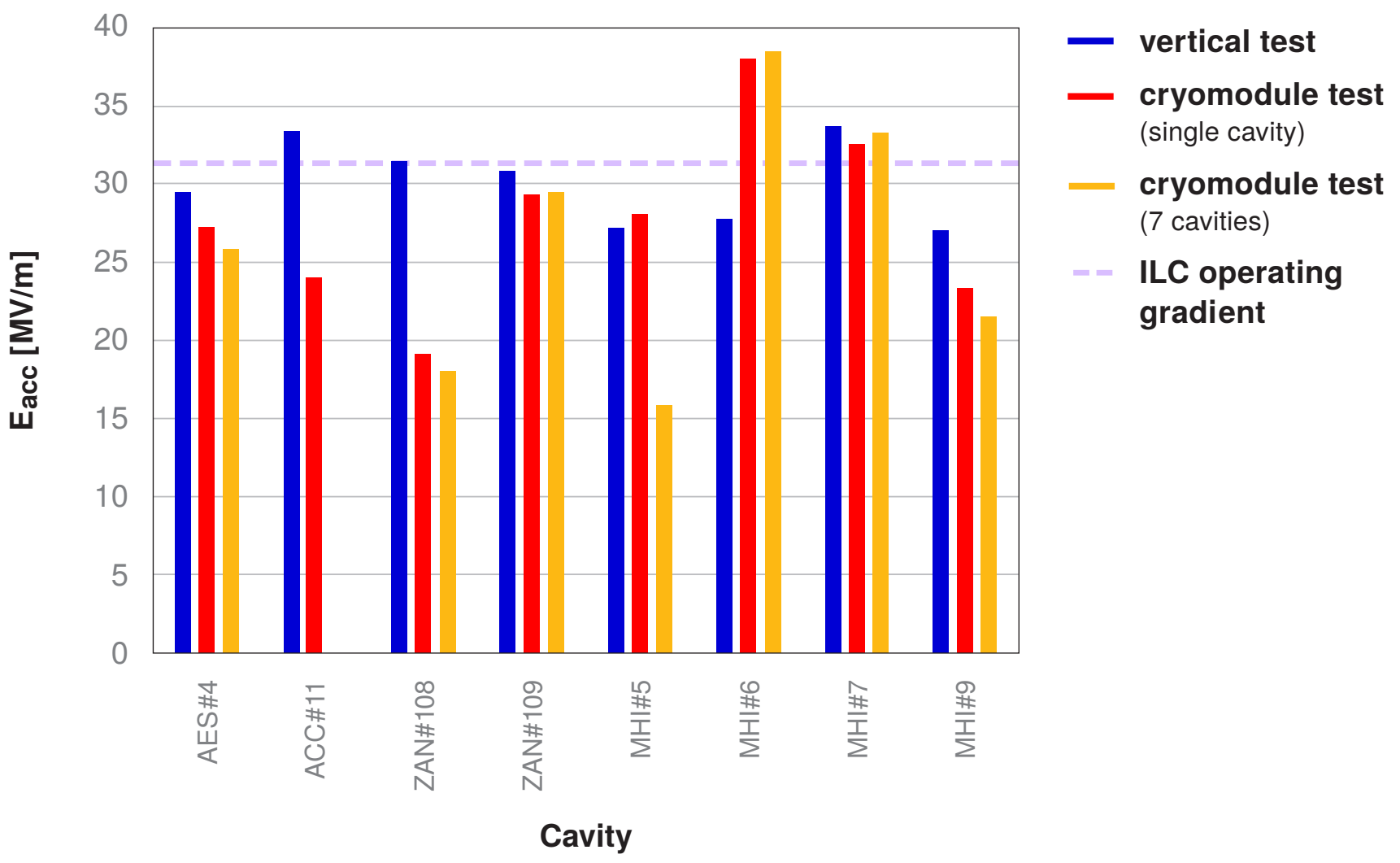

**Figure 2.44**
The achieved gradient values for eight cavities in vertical and cryomodule tests. The maximum average gradient is 30.0 MV/m at vertical test, 27.7 MV/m for single cavity operation and 26.0 MV/m for simultaneous operation of seven cavities at cryomodule test. The purple dotted line shows the ILC specification of 31.5 MV/m.

The performance of MHI-06 was much improved over the vertical test, reaching a gradient of 38 MV/m. On the other hand, the gradients of the two cavities (TB9ACC011 and Z108) were significantly reduced. The reasons for this are still unclear. While the assembly processes or the transport to KEK are possible candidates, the correct explanation will only come from a "post-mortem" study. The performance of MHI-05 (actually, the performance of the input coupler) also degraded following operation at high power levels.

## 2.6.5 Power-coupler Conditioning

All power couplers were RF conditioned at room temperature. During conditioning, the standing wave persists in the power coupler due to total reflection. The achievable power was 500 kW for a pulse width of 500 μs and 200 kW for 1500 μs at a repetition rate of 5 Hz. The average conditioning time was 21 hours for TTF-III-type couplers and 13 hours for STF-2-type, respectively. The difference can probably be attributed to the structure of the RF window. None of the RF windows had a vacuum leak during the experiments.

## 2.6.6 Lorentz-force Detuning

The magnitude of the Lorentz-force detuning (LFD) was measured by stepping the pulse length in the so-called *pulse-shortening* method (see Fig. 2.45). The detuning frequency for the three phases – rise, flat top, and full-pulse – was evaluated to compare the stiffness of all the cavities. The slopes of linear fitting between the detuning frequency for the three periods and the square of the gradient gives the cavity stiffness, as shown in Fig. 2.46. The MHI cavities were found to be more rigid than the others. The effect is noticeable in the flat-top period, while the difference is smaller in the rise period.

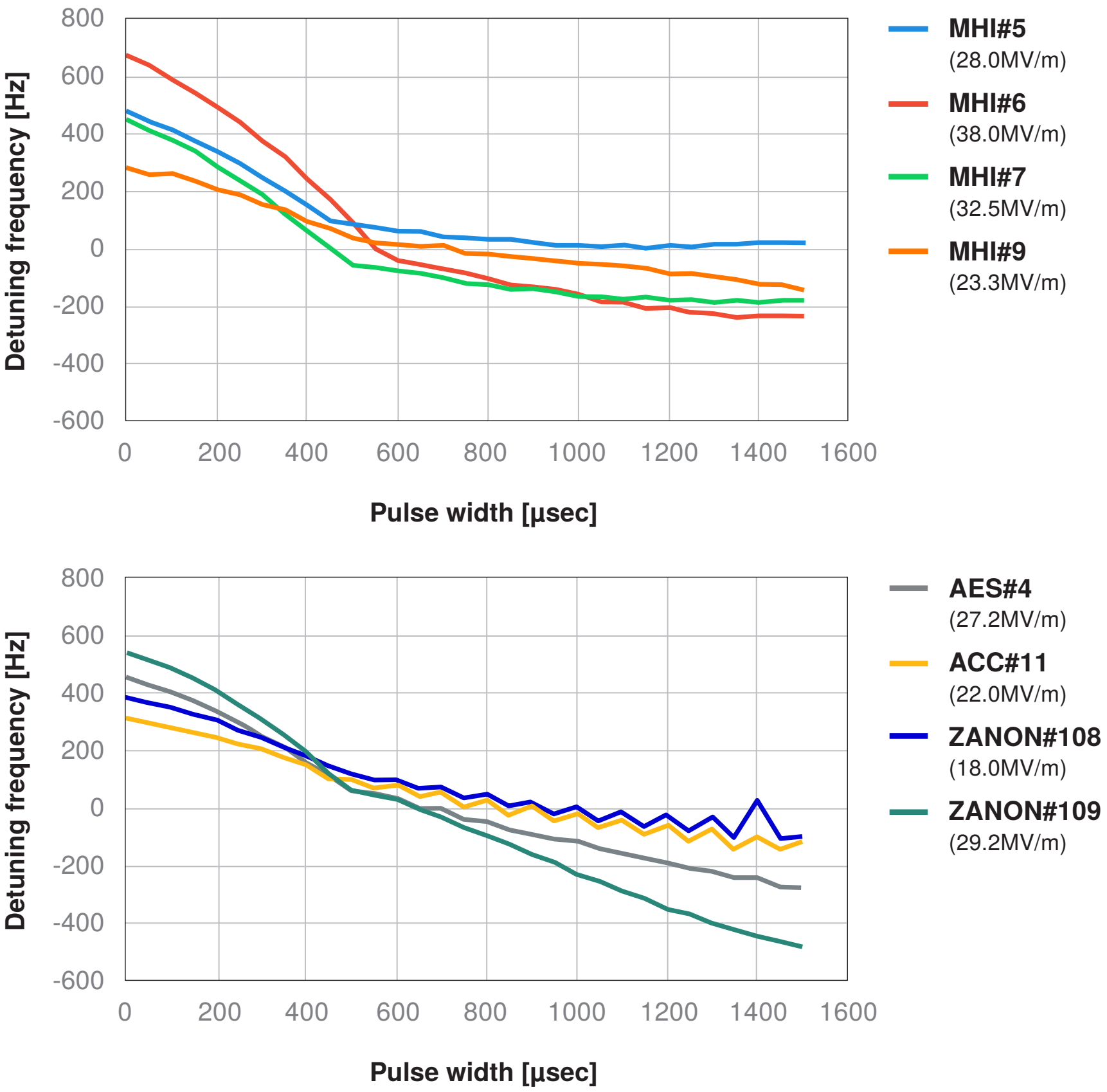

**Figure 2.45** The result of LFD measurement at the maximum gradient for every cavity. (top) Cryomodule-A, (bottom) Cryomodule-C

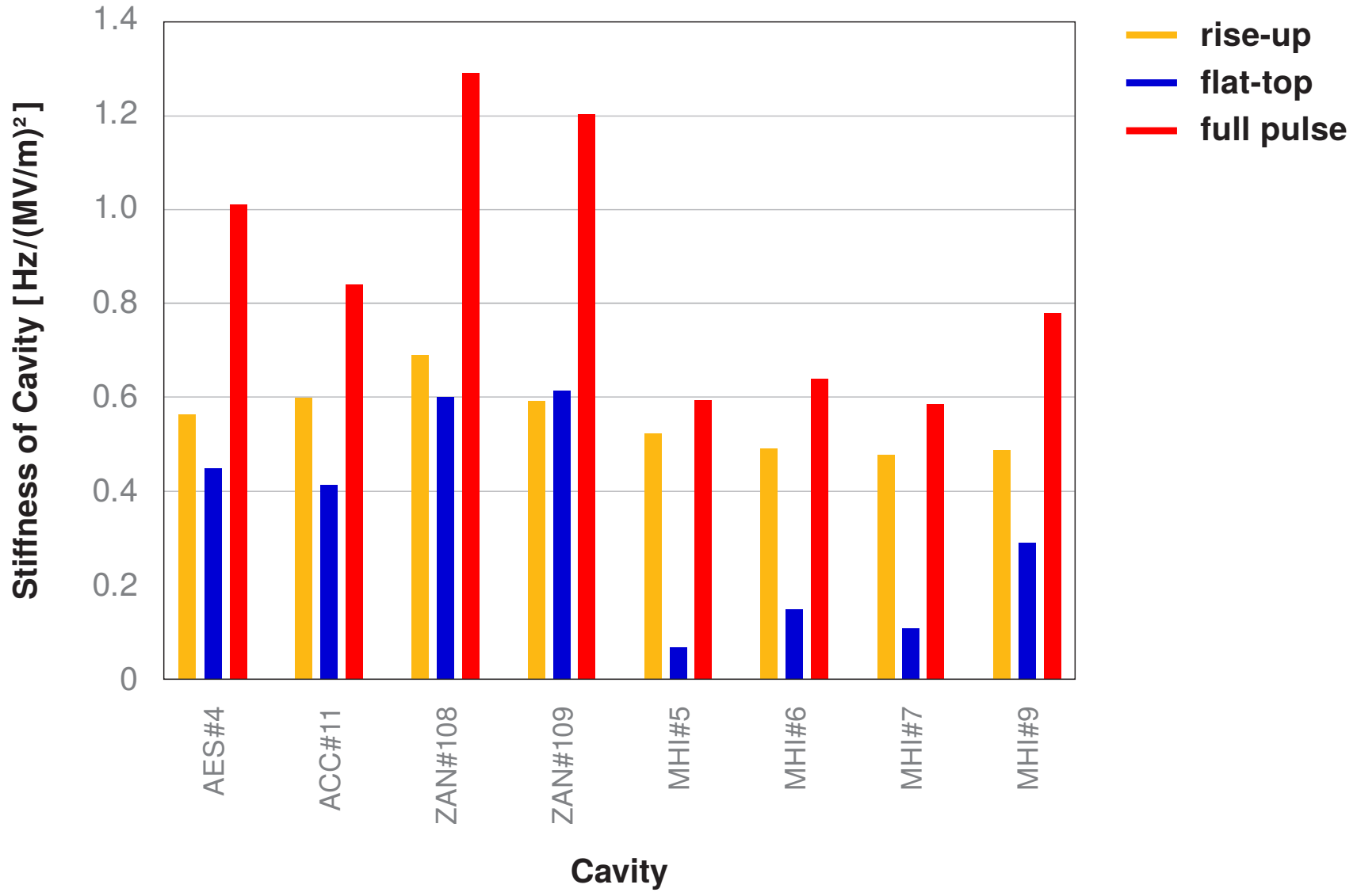

**Figure 2.46** Comparison of the slopes from rise, flat top and full-pulse.

Two methods of LFD compensation were successfully tried: *single-pulse compensation* by a half-sine waveform, which was used at STF Phase-1 [93] and a new method known as *adaptive* *compensation* [94].

To compensate the LFD, a pulse corresponding to a half-period sine-wave is applied to the piezo tuner before the start of the RF pulse. There are four adjustable parameters, namely drive frequency, delay time, pulse height, and pulse offset. The excursion peak-to-peak of the detuning frequency at the flat top of the pulse was introduced as a measure of the compensation [93]. Figure 2.47 shows the correlation between the peak-to-peak of the detuning frequency and the gradient for the best three results of the compensation by the pulse-shortening method. After the compensation, the MHI cavity still has a smaller peak-to-peak of the detuning frequency at the flat-top period, which is expected from the stiffer structure of the MHI cavity package.

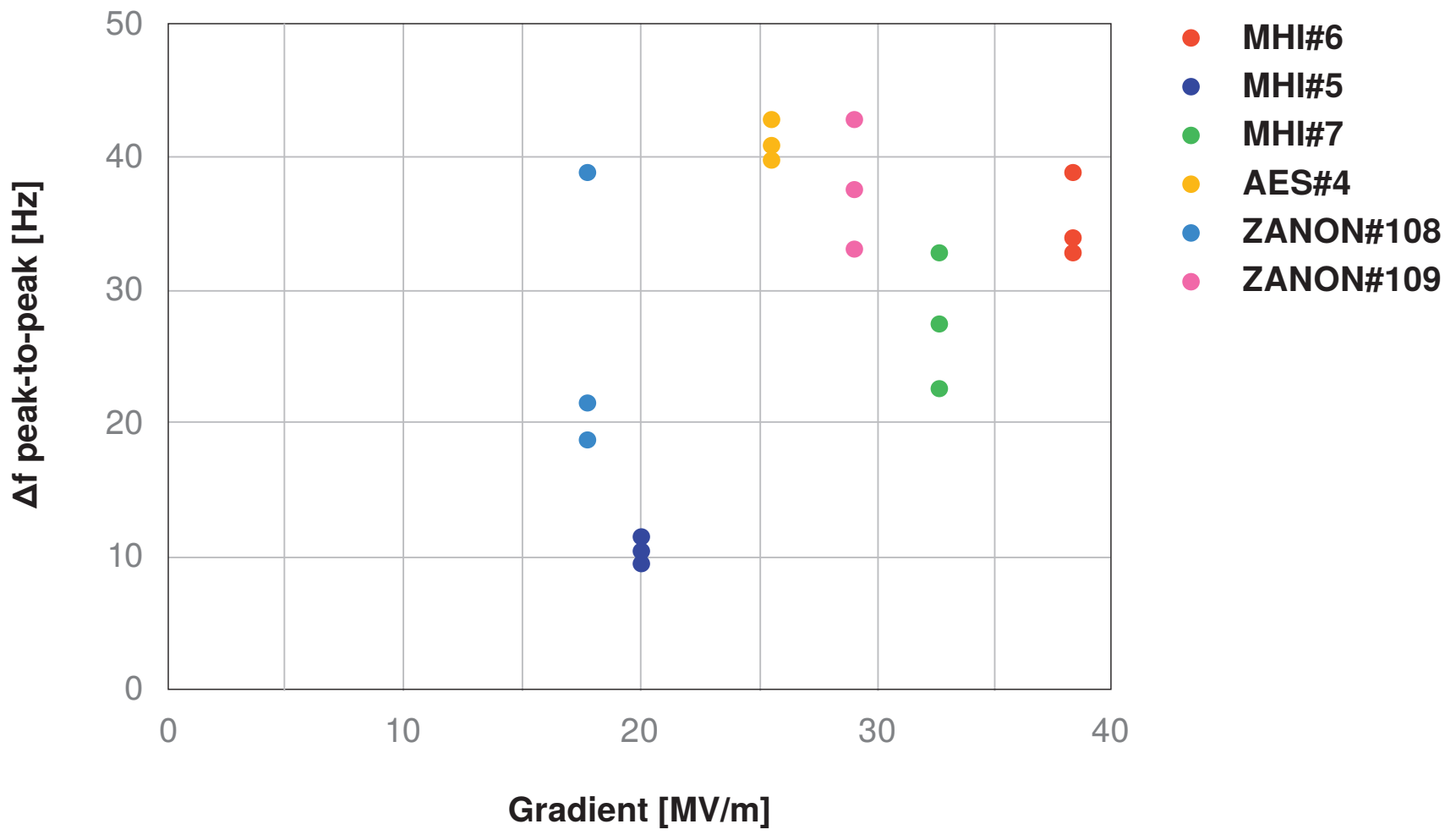

**Figure 2.47**
The correlation between the excursion peak-to-peak of the detuning frequency at flat top and field. The best three results are shown.

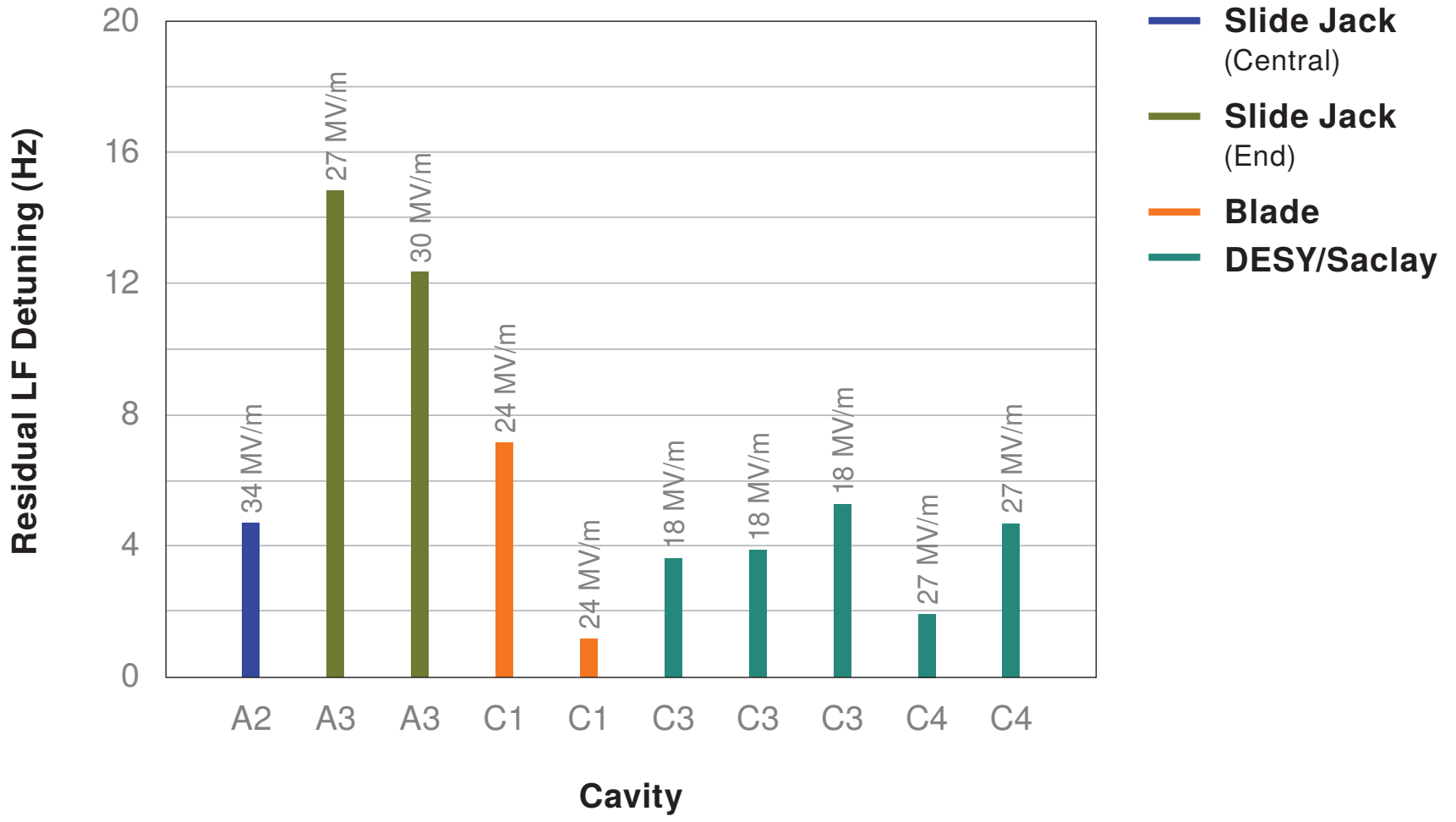

**Figure 2.48**
A comparison of the average residual detuning during the flat top in different S1-Global cavity designs following adaptive feed-forward compensation.

In addition to the standard method of LFD compensation, an adaptive feed-forward method developed at FNAL for CM1 was employed to compensate detuning in six of the eight S1-Global cavities [94]. The adaptive feed-forward procedure first measured the detuning response of each individual cavity to a series of small-amplitude, short-duration piezo stimulus pulses and to changes in the DC bias applied to the piezo. The matrices of stimulus and response data were then inverted using least squares to obtain the small-signal cavity detuning response for any arbitrary piezo waveform and to determine the piezo waveform required to cancel any arbitrary detuning profile. The feed-forward method allows the detuning compensation process to be fully automated; it can track changes in cavity gradient without the need for operator intervention; and it can automatically adjust the piezo bias to correct for changes in the cavity resonance frequency caused by changes in the pressure of the surrounding helium bath and any other slowly changing sources of detuning. The residual

RMS detuning following compensation was less than 15 Hz in all cavities, regardless of design or gradient, as shown in Fig. 2.48. With the exception of the end slide-jack design, which may not have received an adequate piezo drive signal, the residual detuning ranged between 2–8 Hz. At this level of compensation, residual detuning will have no significant impact on the design or operation of the ILC.

## 2.6.7 Operation of Seven Cavities with Vector-sum Feedback Control

In the final stage of the experiment, one 5 MW klystron fed its power to all seven cavities and the performance of operation with vector-sum feedback control was evaluated [95]. During this operation, one cavity (C-2) could not be used for the vector-sum control because of the mechanical problem with the tuner. The input power to the cavities was adjusted by the tuneable hybrids and the variable tap-offs [96] to yield the maximum gradient for each cavity. As described in the previous section, the detuning of cavities during the RF pulse flat-top (1 ms) was effectively eliminated using the piezo tuners.

The result of vector-sum feedback control operation for seven cavities is shown in Fig. 2.49. The average gradient of the seven cavities after input-power optimisation for each cavity was close to 26 MV/m. Since the average of each cavity's quench limit was 26.7 MV/m, the operation near quench limit of each cavity was achieved. The amplitude and phase stabilities were 0.005 % rms and 0.015° rms, respectively, which satisfy the ILC requirements of 0.07 % rms in amplitude and 0.24° rms in phase [3].

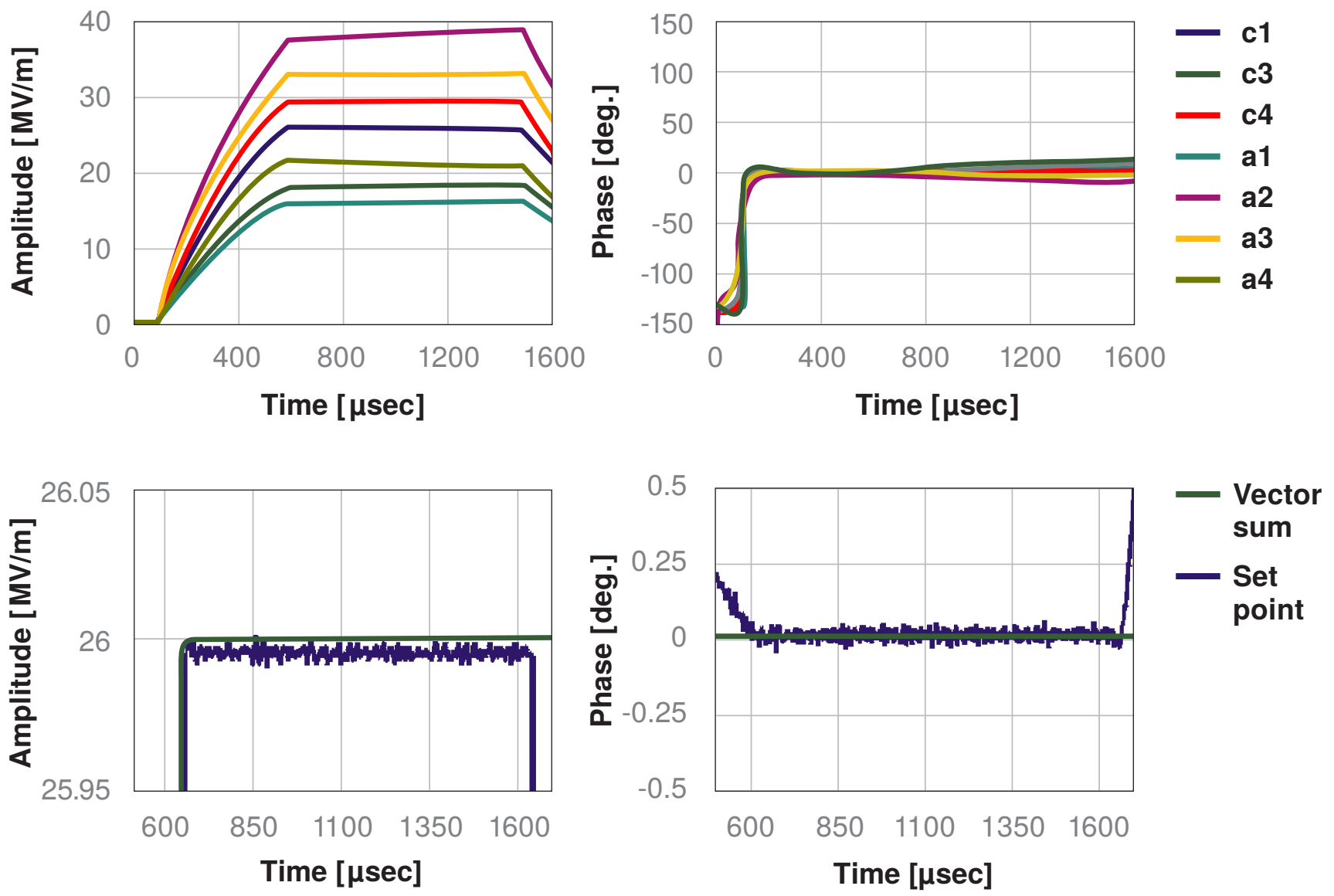

**Figure 2.49** Vector-sum operation. Left: each cavity's gradient and vector-sum gradient, right: each cavity's phase and vector-sum phase.

## 2.6.8 Summary of S1-Global Experiment

The S1-global program successfully achieved the GDE R&D goal of the fabrication and test of a cavity string. The cavity ensemble achieved an average gradient of 30.0 MV/m before installation, 27.7 MV/m for single cavity operation after installation, and 26.0 MV/m for 7 cavity simultaneous operation in a cryomodule. The ability to operate (and control) cavities in a cryomodule close to their nominal quench thresholds with a tunable PDS was also shown. Compensation of Lorentz-force detuning maintained the cavity frequencies well within the required tolerances. Although all of the tuner designs achieved the necessary frequency range, mechanical failures in two of these tuners impacted the operation in two of the cavities. These failures are not systematic and can be attributed to the lack of maturity of the cryomodule mechanical-assembly procedures and the associated quality controls.

The *plug-compatibility concept* was verified by building one cryomodule from cavities and couplers supplied from several national laboratories. A good example is the half-size cryomodule-C which was built from an INFN cryostat, DESY cavities and couplers, and FNAL cavities and couplers in cooperation with SLAC. The cryomodule power systems is another good example. The ability to incorporate and test several different component designs within a single integrated test set-up was an important aspect of the R&D program.

A less obvious but also important aspect of this program was the fact that it was the manifestation of a global engineering effort under the GDE organisation. Distributed design and fabrication requires the adoption and utilisation of common CAD design tools and data bases together with an unprecedented level of communication among the participants. This degree of detailed collaboration has rarely, if ever, been attempted across major research laboratories. Not only did the GDE learn how to build cryomodules but also how to organise the efforts to accomplish this. This bodes well for a construction project which is envisaged to embody significant international co-operation.

## 2.7 Cryomodule and Quadrupole R&D

The overall ILC cryogenic scheme evolved from the TESLA proposal [11]. The TESLA cryomodule design was conceived in the late 90s to meet the project requirements of high-filling factors, moderate capital and operation costs and effective alignment and assembly procedures [97]. This module design has been employed for the realisation of the TESLA Test Facility (TTF) linac, now operating as the FLASH FEL facility in DESY, and was improved in several iterations during the deployment of the facility [98]. The successful TTF design, with minimal modifications, was later adopted for the linac of the European XFEL, which is composed of 100 modules of this type, and evolved with few variations, into the baseline for the ILC [3, 50].

The energy reach and beam requirements of the ILC require

- a *high filling factor* (the ratio between the total length of the linac and the length in which acceleration takes place) to reduce the machine footprint;
- *moderate capital and operation cost* per unit length - a simple design, based on proven and reliable technologies, readily available in the industrial contexts, will tend to minimise costs, while operational cost is reduced by a careful thermal design aimed at minimisation of all spurious static losses to the 2 K environment;
- effective *assembly and alignment procedures* for beam-line components - cost-effective module assembly procedures compatible with the cleanliness needed by high-gradient bulk-Nb cavity technology provide the former, while preservation, reproducibility and stability of beam-line components is crucial for the achievement of beam parameters.

The high filling-factor concept favours long cryomodules (containing several cavities) and their connection in long cryo-units, separated by short interconnections. This aspect, combined with the requirement for low cryogenic loss, led to the integration of the cryogenic-cooling circuits into the cryostat. Thus in this scheme the cryomodule, besides providing the usual functions of mechanical-support structure for the beam-line elements and isolation from the warm-temperature environment, becomes a substantial part of the cryogenic system, since it represents the region where nearly all losses occur and also takes care of the fluid distribution in the linac.

The TTF module has been developed in order to meet the requirements mentioned above and has been extensively characterised, thoroughly tested and operated for years in a linac driving the high-availability user facility FLASH. From this design all the DESY FLASH and European XFEL linac modules, the FNAL-type-IV modules for NML and the S1-Global modules at KEK have been derived, with small variations, and manufactured by several companies worldwide.

The ILC-module concept therefore builds on a reliable and mature technology, currently available in different laboratories worldwide as a result of the ILC S1 and S2 efforts. The already large experience gained at FLASH and at the facilities based on this design will be soon consolidated by the large amount of information coming from the European XFEL, which will deploy a 100-module linac at a tunnel installation rate of 1 module per week. The experience so accumulated will be particularly relevant for the further evolution and the finalisation of the ILC-module concept. The following subsections briefly summarise the recent status of R&D activities worldwide in this area..

### 2.7.1 Thermal Performance of the S1-Global Modules

The S1-Global (Section 2.6) collaboration launched a strong effort dedicated to the experimental assessment of the thermal performance of the different cryomodule and subcomponent variants [99]. Such an activity is motivated by:

- the study of design variants for a possible cost reduction of the mass-produced cryomodules;
- the opportunity given by the experimental activities at KEK to provide a comparison for the thermal performance of the different variants integrated in the S1-Global setup;
- the need to benchmark with experimental measurements the heat-load calculations based on engineering calculations and FEM models.

#### 2.7.1.1 Simplification Studies of the 5 K-Shield

One of the original STF modules was used to explore the feasibility and the consequences of removing the inner 5 K shield of the cryomodules, as a possible cost-reduction measure, leading to a decrease in fabrication costs and envisaged simplifications of the assembly operations. This concept has been tested at KEK by heat-load measurement with and without the shield. Heat loads were measured by means of the evaporated mass flow for the 2 K region and of the shield-temperature increase after stopping the coolant for the shield circuits [99].

In the ILC cryomodule both thermal shields provide a manifold for the thermalisation of direct conduction paths to the many penetrations reaching the 2 K environment (couplers, HOM, current leads, supports, cables etc.). Thus a thermal sinking of these conduction paths must be provided by a circuit at approximately 5 K in order to limit direct heat flow by conduction at 2 K; a complete elimination of the 5 K shield is therefore not feasible.

The possibility of partial removal of the heat shield remained. This was investigated at KEK by retaining the top shield parts while removing the bottom shield portions. Heat loads were measured in these two setups (with and without bottom shield portions) to support and validate heat-load estimates from FEM modelling. The removal of the bottom shields led to an increase of static loads to the 2 K region of approximately 0.8 W in the 6 m-long S1-Global module, as a consequence of the thermal radiation from the 70 K shield now impinging on the 2 K surfaces. This data allowed the FEM-model parameters (e.g. surface emissivities to account for multilayer insulation) to be tuned in order to explore mitigating actions by means of alternate cooling schemes. Further FEM studies originating from the analysis of these measurements showed that the radiation-load increase can be avoided by implementing an alternate scheme for cryogenic-coolant flow. The decrease of the average shield temperature obtained by reversing the coolant-flow direction allows a reduction of the thermal static load to the coldest temperature level. Thus it is in principle possible to remove the lower 5 K shield while keeping the overall static consumption of an ILC module to the same level as the nominal RDR design with a complete 5 K shield. However, the proper implementation of a reversal of the cryogenic flow would necessitate a complete redesign of the module transverse cross section, with redistribution of the cryogenic piping. A complete module redesign, even when retaining nominal heat-load performance, could severely compromise the consolidated experience gained from the evolution of these modules which will be further supported in the near future by the European

XFEL modules. It was therefore decided to retain the nominal cooling scheme as foreseen by the RDR (matching the module layout for the European XFEL) and leave the possibility for reconsidering the 5K removal/flow inversion possibility at a later engineering stage of the project, together with the finalisation and final optimisation of the temperature stages of the cryosystems.

#### 2.7.1.2 Analytic Heat-Load Measurements

The characterisation of the thermal performance of the design variants for the components is important in order to include heat-load estimations for the plug-compatible components of the ILC cryomodules that do not exceed the overall heat-load budget available for the cryomodule unit. Such tests were carried out; details of the configuration are given in Section 2.6.

The cryomodules and STF cryogenic infrastructure have been instrumented with a large number of sensors to monitor component temperatures and cryogenic flow conditions. This allows the assessment of heat loads at the different circuit levels and a comparison with the evaluations. Experimental procedures have been devised to assess individual contributions from the components. Table 2.20 compares the static estimations and measurements. As for the case of the 5 K shield studies previously reported, the overall heat loads on the 2 K environment have been derived from the mass flows of the evaporated LHe by the cavity vessels, and the static losses on the 5 K and 80 K circuits have been derived from the temperature (i.e. enthalpy) rises of the average shield temperatures after stopping the coolant flow. A dozen temperature sensors were installed on each thermal shield of the module in order to perform these measurements. Temperature sensors were also placed at thermal sinks in order to analyse the heat-load contributions of individual conduction paths.

**Table 2.20**
Comparison of static S1-Global module heat-load estimations with measurements.

| Temperature | Component | Module-A [W] | Module-C [W] |
|---|---|---|---|
| 2 K | Thermal radiation | ≈0.0 | ≈ 0.0 |
| | Input couplers (4×) | 0.29 | 0.08 |
| | HOM RF, Piezo cables | 2.1 | 0.71 |
| | Tuner driving shafts (4×) | 0.48 | NA |
| | Temperature sensor wires | 0.18 | 0.18 |
| | WPM, pin diodes wires | 1.72 | 0.82 |
| | WPM connection pipe | 0.17 | ≈0.0 |
| | Support posts (2×) | 0.25 | 0.25 |
| | Beam Pipe | 0.02 | <0.01 |
| **Total Estimated 2 K load at module** | | 5.2 | 2.1 |
| **Total Estimated 2 K load (both modules)** | | 7.3 (6.8 without support posts) | |
| **Total Measured 2 K load (both modules)** | | 7.2 | |
| 5 K | Thermal radiation | 0.66 | 0.68 |
| | Input couplers (4×) | 4.00 | 0.92 |
| | Support posts (2×) | 1.54 | 1.54 |
| | Sensor wires | 0.9 | 0.9 |
| | Beam Pipe | 0.1 | 0.05 |
| **Total Estimated 5 K load at module** | | 7.2 | 4.1 |
| **Total Measured 5 K load at module** | | 7.3 | 5.3 |
| 80 K | Thermal radiation | 16.6 | 15.9 |
| | Input couplers (4×) | 9.60 | 7.28 |
| | Support posts (2×) | 10.78 | 10.78 |
| | RF cables | 6.88 | 1.30 |
| | Sensor wires | 0.08 | 0.08 |
| | Beam Pipe | 0.37 | 0.10 |
| **Total Estimated 80 K load at module** | | 44.4 | 35.3 |
| **Total Measured 80 K load at module** | | 48.7 | 34.4 |

The good agreement between estimates and measurements in the S1-Global experiment, which has been made possible by the large amount of diagnostics implemented during the test cool-downs by the KEK team, is another confirmation that the module design is well understood in terms of its thermal behaviour. Thus design options of subcomponents can be explored by means of suitably detailed FEM models in order to derive their thermal performance with a reasonable error margin.

Other studies have been performed in the past also to benchmark transient behaviour during cooldown, with similarly good agreement [100].

Table 2.20 is also a useful example to illustrate the opportunity for value engineering and overall optimisation for a complex system such as the ILC cryomodule. For example, Module-A was equipped with better-grade RF cables for the HOM and Wire Position Monitor (WPM) sensors, resulting in an increased heat leak to the 2 K environment. The use of external motor drives for the tuner action to allow replacement without dismounting the module (as a measure to reduce maintenance times) introduces additional heat loads to the cavity environment. The different coupler designs also have different thermal performance due to different design choices (e.g. the STF-2 coupler design, contrary to the TTF-III, has no bellows between the 5 K and 70 K intercepts to simplify clean-room assembly) , which should be factored in when considering plug-compatible component design.

A third important activity carried by the S1-Global collaboration has been the assessment of the cavity and coupler dynamic loads during module testing. This has been achieved by a sequence of four measurements of static and dynamic loads of each cavity on resonance and in detuning conditions, evaluated from the flow rate of evaporated LHe. All cavities were driven to high field values (25–38 MV/m) with a $Q_0$ estimated from this analysis in the range $\approx$ 4-9 $\times 10^9$. These measurements were then repeated by driving all four cavities simultaneously in each of the modules, leading to the heat-load values reported in Table 2.21. The extensive temperature diagnostics implemented in the modules showed that the different behaviour in the coupler dynamic loads in the two modules was associated with a substantial temperature increase at the interconnection flange between the STF-2 couplers and the cavity port. The temperature rise originated at the 3 µm-thick Cu layer on the inner surface of the outer conductor. The measurements taken at S1-Global were instrumental in a redesign of the Quantum-Beam cryomodule couplers at KEK to use a different coating procedure and more performant thermal anchoring to reduce the static and dynamic heat losses.

**Table 2.21**
S1-Global RF dynamic heat-load measurements.

| | Module C 4 cavities | | Module A 4 cavities | |
|---|---|---|---|---|
| | resonant | detuned | resonant | detuned |
| Gradient MV/m | 20.0 | 32.0 | 26.9 | 32.0 |
| Total RF load, W | 2.7 | – | 6.9 | – |
| Coupler dynamic load, W | 0.2 | 0.5 | 2.5 | 4.6 |
| Cavity RF load, W | 2.5 | – | 4.4 | – |

## 2.7.2 Thermal Performance of the European XFEL Prototype Module

The European XFEL has started the procurement of 100 cryomodules. All modules will be thoroughly tested for their RF and cryogenic performance at the Accelerator Module Test Facility (AMTF) in DESY before tunnel installation, and will yield a copious amount of data of interest to the ILC.

An update of all static heat-load evaluations for the European XFEL module design, integrating the minor changes introduced with respect to the Type-III TTF modules, has been performed at DESY [101] and compared to the thermal performance of the European XFEL prototype modules, assessed experimentally at the DESY CryoModule Test Facility (CMTB).

**Table 2.22**
European XFEL updated evaluation of static heat load.

| Temperature level | Static heat load (W) Calculated | Measured |
|---|---|---|
| 2 K | 2.1 | 3.5-6 |
| 5-8 K | 6-12 | 6-11 |
| 40/80 K | 100-120 | 100-120 |

Thermal loads have been assessed and measured several times, under different operating conditions, as a function of the outer-shield temperature (which varies along the cryostring). The heat loads in the 2 K region are evaluated from the mass-flow rate of the evaporated helium during the tests, while the loads on the 5 K and 80 K circuits are evaluated from the helium-coolant-flow conditions and cross-checked with the enthalpy response of the cold mass. A separate test with a dummy module was performed to separate the heat loads of the end/feed caps of the CMTB from that of the module.

Table 2.22 lists the updated heat-transfer evaluations (for current leads, power couplers, support posts, thermal radiation etc.) and the comparison with the measured values of the three prototytpes, which are also shown in Fig. 2.50, along with the budget values foreseen by the refrigerator.

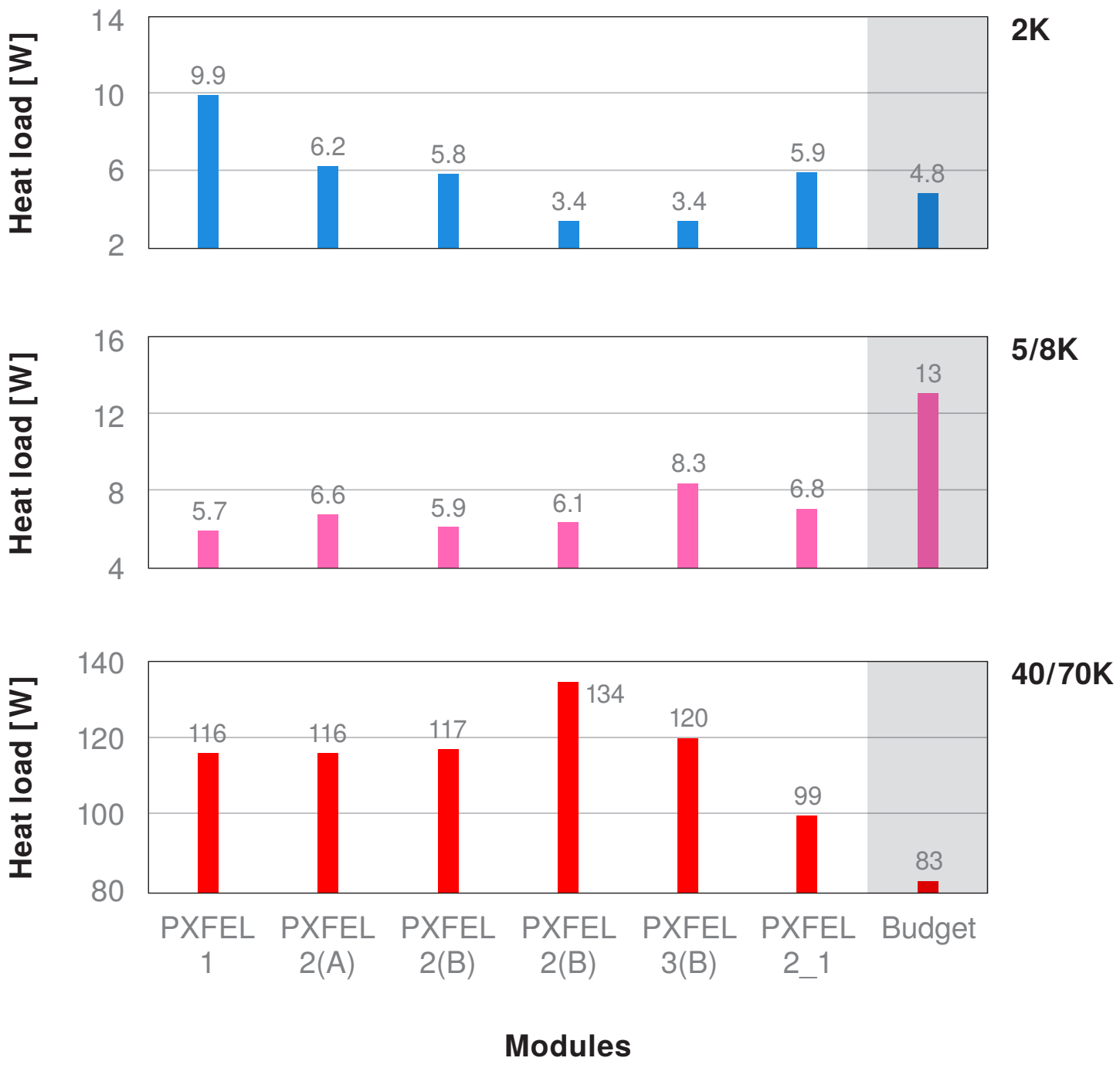

**Figure 2.50** European XFEL prototype module heat-load measurements at CMTB, compared to the budget values.

Generally speaking, measured values agree with calculated values for the 5/8 K and 40/80 K circuits. The increase in heat loads in the outer shields during the PXFEL3 (A) test has been traced to an improper tuner-assembly operation, which caused a thermal contact. The lower 40/80 K heat loads shown by the PXFEL2_1 measurements seem to suggest a better performance of the new MLI blankets that have been used for this test.

The measured values for the 2 K environment deviate from the evaluations due to an underestimation of the cabling heat loads and, furthermore, varied due to inconsistencies in the installation. The high static loads of PXFEL1 have been traced to an incorrect layout of a prototype current lead.

As for the S1-Global measurements, the European XFEL experience shows that it is possible to predict thermal performance of cryomodules with reasonable error margins, but great attention needs to be paid to the details of the models, which need to include all heat-transfer mechanisms. Figure 2.50 clearly show also a training effect for the most sensitive 2 K environment, showing a decrease in time of measured heat loads. One of the crucial outcomes of the European XFEL AMTF test for ILC would be the confirmation of this trend for the testing of the series modules.

### 2.7.3 Cryogenic Thermal Balance

The activity in the Technical Design Phase has confirmed that the thermal behaviour of the module and its subcomponents can be reliably estimated with numerical models to explore design variants and simplifications so that design complexity and thermal performance can be balanced and to allow reliability-driven options to be explored. The European XFEL will also provide substantial experience from the AMTF operation with the series module testing in the coming years.

The baseline ILC TDR cryomodule is therefore almost unchanged from the description provided in the RDR, with a unique slot length of 12.652 m for the two main variants (nine-cavity module and eight-cavity-plus-quadrupole module) . The present baseline relies also on the use of a conduction-cooled split quadrupole, as discussed in the next section.

### 2.7.4 R&D on the Split Quadrupole

Superconducting linear accelerators need a number of cryomodules with superconducting quadrupole magnets for beam focusing and steering. Various superconducting magnet designs have been investigated. Nearly all are bath cooled by LHe and they need to be assembled with the string of superconducting RF (SCRF) cavities inside a clean room, leading to the risk of particle contamination of the cavity surfaces and to increased chances of cavity-performance deterioration. In contrast a splittable magnet can be assembled around the beam pipe after all SCRF cavities are installed and sealed inside the clean room. In this case the magnet will never enter the clean room and its installation will not lead to the risk of contaminating the SCRF cavity inner surfaces. The splittable quadrupole was designed and built at Fermilab and tested in a 4.4 K helium bath at the FNAL Magnet Test Facility (MTF).

**Table 2.23**
Specifications and parameters for the splittable quadrupole magnet.

| Parameter | Value | Unit |
|---|---|---|
| Peak gradient | 54 | T/m |
| Integrated gradient | 36 | T |
| Peak operating quadrupole current | 100 | A |
| Magnet total length | 680 | mm |
| SC-wire diameter | 0.5 | mm |
| NbTi-filament size (vendor value) | 3.7 | µm |
| Cu:SC volume ratio | 1.5 | |
| Superconductor Critical current ( 5 T and 4.2 K) | 200 | A |
| Coil maximum field at 100 A current | 3.3 | T |
| Magnetic-field stored energy | 40 | kJ |
| Quadrupole inductance | 3.9 | H |
| Number of turns/pole in quadrupole coil | 900 | |
| Yoke outer diameter | 280 | mm |

The main issue for the ILC quadrupole is to provide magnetic axis stability of 5 µm over a $\pm$ 20% field range. This requirement arises from the Beam-Based Alignment (BBA) technique. The magnetic and mechanical effects which correlate with the magnetic axis stability must be minimised.

Table 2.23 shows the specification and design parameters for the splittable quadrupole shown in Fig. 2.51a. The quadrupole has a vertical split plane and is assembled from two half cores having racetrack superconducting coils on magnet poles. The magnet halves are tightened to each other by stainless-steel bandage rings. This assembly is surrounded by Al thermal leads which have a good thermal contact with the cryomodule LHe supply line. The LHe line provides the cooling by conduction to this cryogen-free magnet [102].

The fabrication and test of a splittable quadrupole confirmed the design concept. Figure 2.51b shows the quadrupole on the vertical test insert ready for measurement in pool-bath mode at FNAL.

After training up to 110 A, the magnet reached 20% current margin, as shown in Fig. 2.52a; the specified peak gradient of 54 T/m was reached at 90 A [103].

**Figure 2.51**
Split quadrupole.

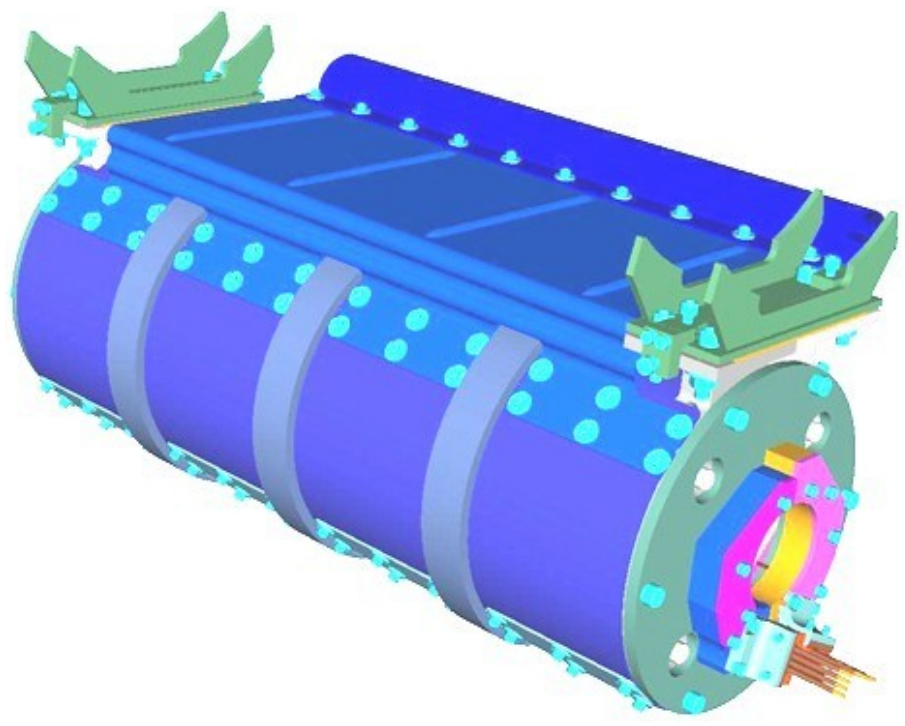

**(a)** Final assembly of the split quadrupole.

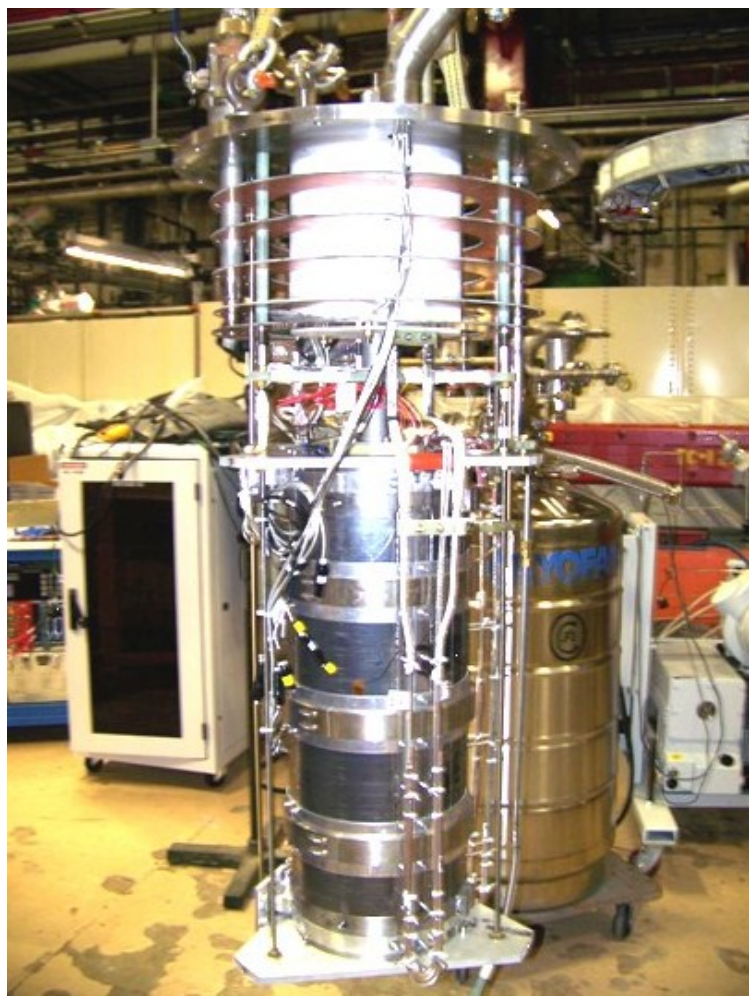

**(b)** The split quadrupole mounted on the insert of the cryostat.

**Figure 2.52**
Tests of the quadrupole.

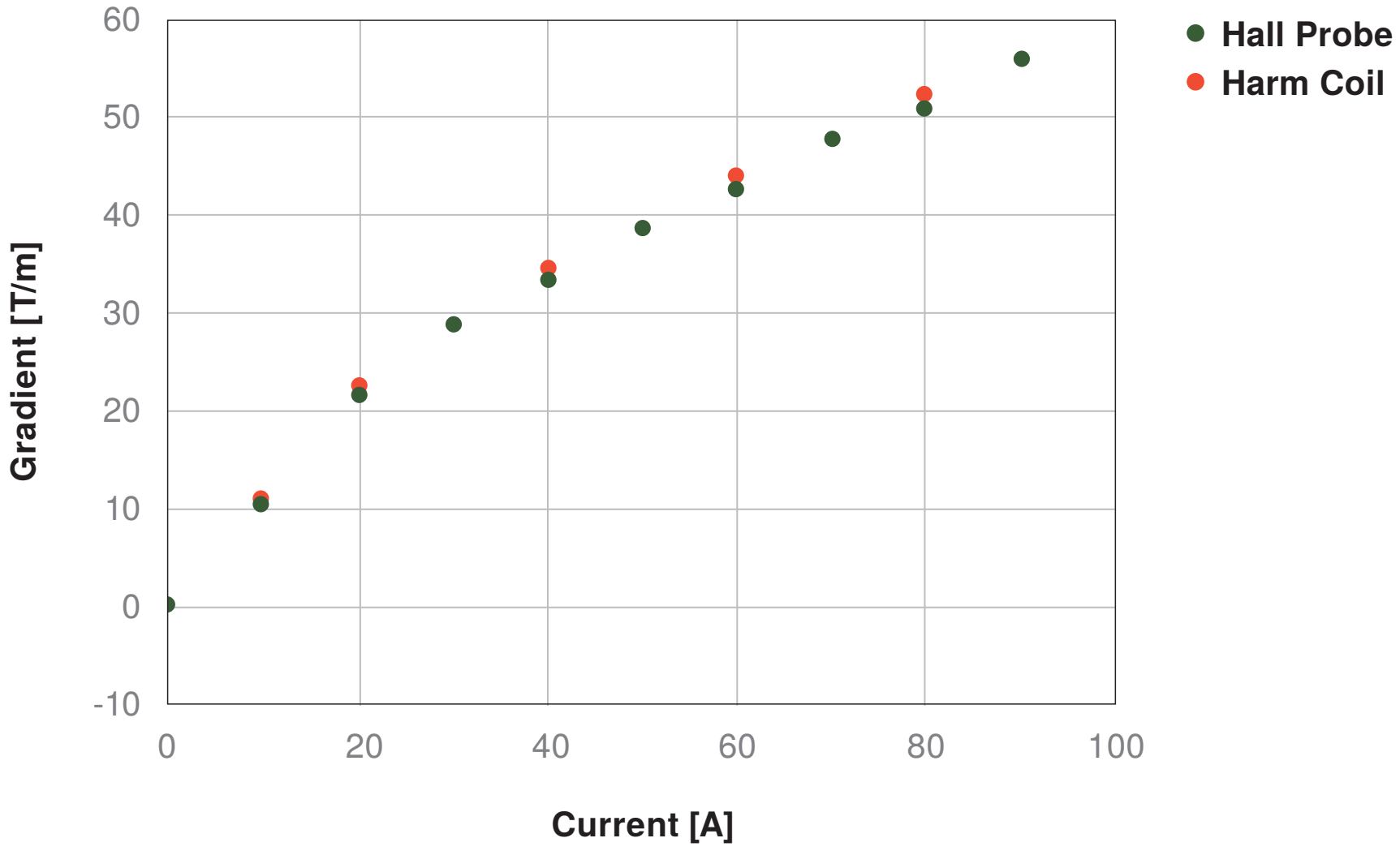

**(a)** The quadrupole gradient as a function of current.

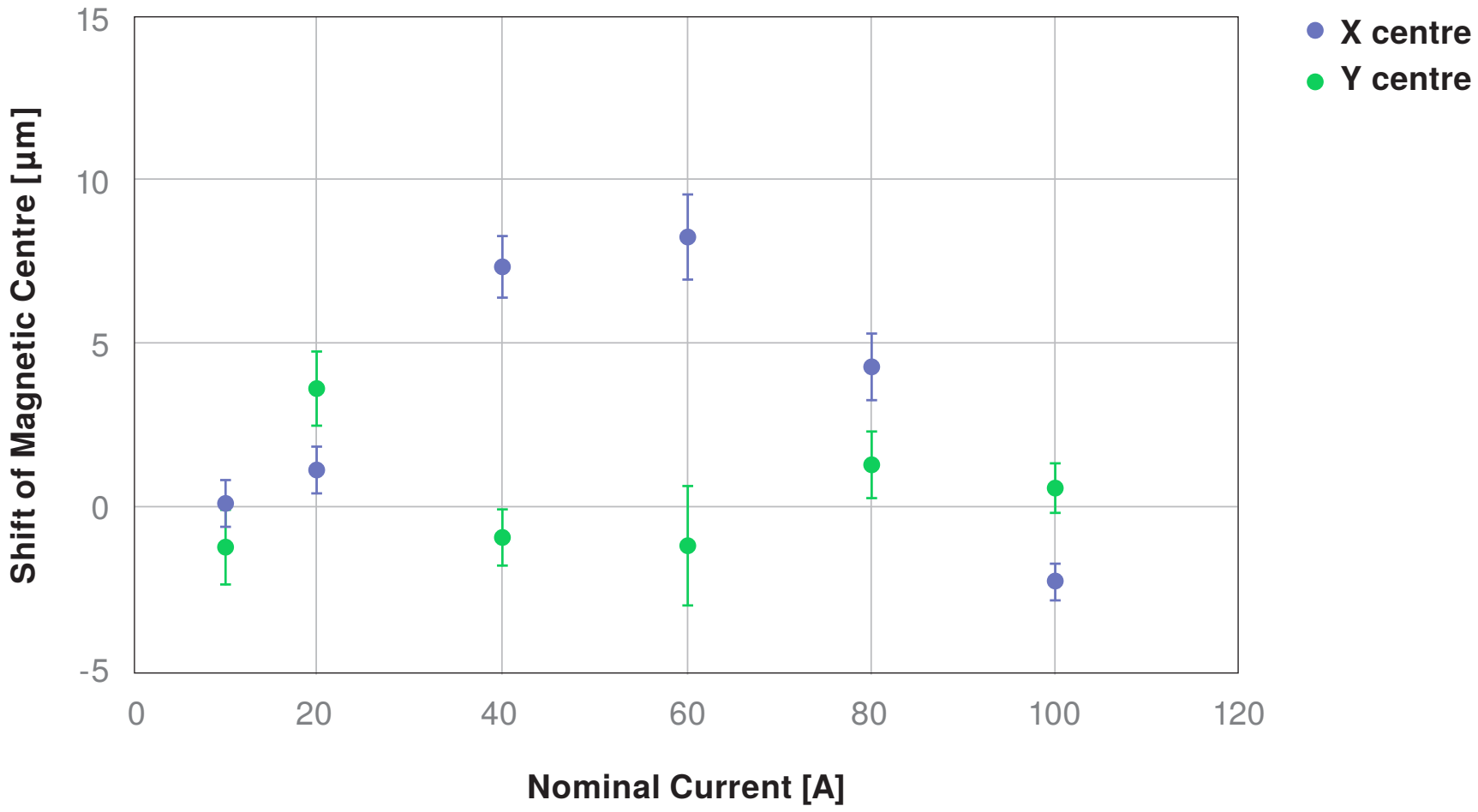

**(b)** The shift of the magnetic centre as a function of current.

The shift in the centre position of the quadrupole over a 20 % gradient change is shown in Fig. 2.52b as a function of the nominal operating current; it only slightly exceeds the specifications of 5 µm on the horizontal plane at currents in the 30–70 A range. More careful control of the yoke gap size and uniformity may improve this. Future plans are to improve the magnet split-plane flatness to eliminate small gaps, and test again in a conduction-cooling mode.

## 2.8 RF Power Generation and Distribution

### 2.8.1 Overview of HLRF R&D

#### 2.8.1.1 RDR RF System Design and TESLA R&D

The High Level RF (HLRF) system as described in the 2006 ILC RDR [3] is very similar to that developed for the TESLA linear collider design that was tested starting in 1997 at the DESY TESLA Test Facility (TTF), which is now called FLASH. The RF system features a 10 MW, high efficiency (65%), Multiple-Beam Klystron (MBK) that produces 1.6 ms pulses at a rate up to 10 Hz. For this application, three vendors developed tubes of somewhat different designs, of which two were successful to the extent that they have been adopted for use in the European XFEL Linac where they will run nominally at 5 MW but have 10 MW capability. Two horizontal-oriented MBKs are shown in Fig. 2.53. At TTF, the klystrons (both single and multiple beam) were powered with Pulse-Transformer style, 120 kV modulators with solid-state switches and 'bouncer' circuits for droop compensation (i.e., to offset the voltage decrease as the storage capacitors discharge). Finally, for the distribution of the RF power to the cavities at TTF, the waveguide components (e.g. power dividers, isolators and loads) were developed with industry [11].

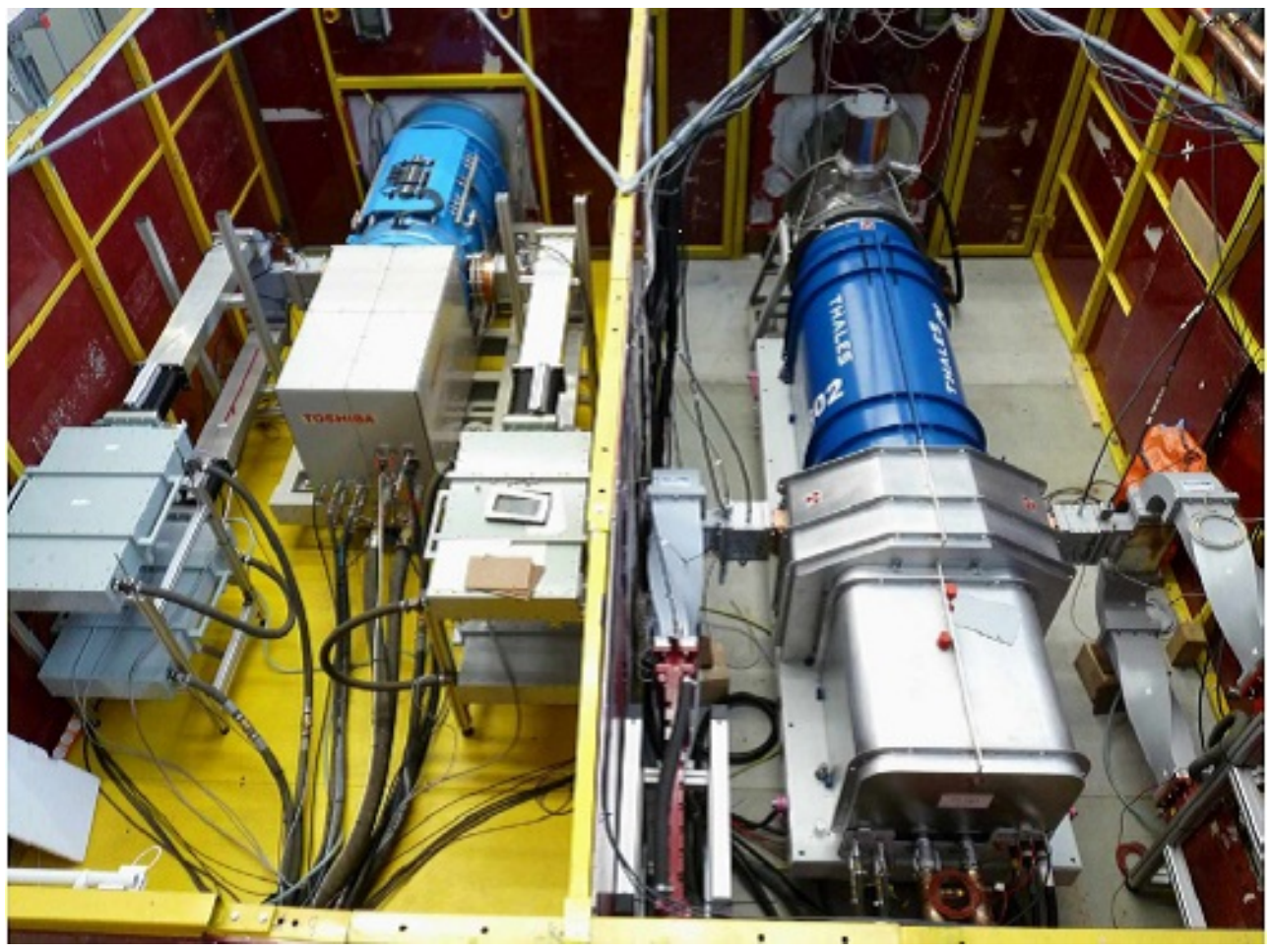

**Figure 2.53**
Horizontally mounted MBKs; (left) Toshiba E3736, (right) ThalesTH1801.

#### 2.8.1.2 ILC RF System Evolution

After the ITRP decision in 2004, KEK and SLAC/FNAL embarked on setting up their own HLRF test stations to support the SCRF infrastructure for STF at KEK and for NML at FNAL, respectively. SLAC also started a program to improve the modulator, klystron and RF-distribution designs to make them more versatile, more efficient and less costly.

During the Technical Design Phase after the RDR completion, two RF-system layouts in a single tunnel for the Main Linacs were proposed as cost-saving measures and presented in SB2009 [104]. The evolution and development of these proposals up to 2011 is described in the Interim Report on the ILC R&D Program [50]; the basic concepts are summarised below.

The Distributed RF System (DRFS) has all RF-system components located in a single tunnel, which has advantages for a mountainous site. Instead of trying to fit the RDR RF system in a single tunnel along with the cryomodules, smaller, lower power (800 kW) klystrons are used to feed two (for high-power upgrade) or four cavities (for low-power baseline) locally. The lower-power tubes should have a higher reliability to offset the much longer time to repair resulting from access to the tunnel now being limited to periods when the beams were off. The DRFS klystron features an anode electrode that is voltage modulated to switch the tube to a 70 kV, droop-compensated supply. Assessment of availability, operability and maintainability were based on performance of similar types of tubes. The DRFS design and related R&D are described in more detail in Section 2.8.5. Though DRFS has many advantages for use in mountainous regions, it was estimated to cost more than the RDR solution. After the introduction of the large Kamaboko-style tunnel for a mountainous site, DRFS was abandoned, as this tunnel includes a thick partition wall that allows access for repair to the RF system during operation. Instead, the Distributed Klystron System (DKS) was adopted, which is similar to the RDR RF-system layout (i.e. with 10 MW MBKs locally driving multiple cavities), as described in Section 3.8.

The Klystron Cluster System (KCS) has all klystrons and associated power supplies located in a series of ten surface buildings along the Main Linacs. In each building there are two sets of typically 19 klystrons (for the low-power baseline) or 29 klystrons (for the high-power upgrade) whose power is combined in a 0.5 m diameter, pressurised aluminum circular waveguide. These two waveguides extend down and along the accelerator tunnel (one upstream and one downstream) where power is tapped off periodically to feed groups of 26 cavities (3 cryomodules). KCS is particularly suitable for a flat site. The KCS concept and related R&D are described in Section 2.8.6.

### 2.8.2 European XFEL RF-System Evolution

The RF system for the European XFEL has also evolved since the original version that was to be part of the TESLA Linear Collider. A single tunnel was always assumed for the European XFEL with the RF system split between the tunnel and surface buildings. That is, 10 kV pulse generators and droop compensation circuits for the modulators are located in clusters in surface buildings, which allows easy access for repair. They connect through cables (four per unit) that extend along the linac tunnel to 1:12 pulse transformers that drive 10 MW MBKs. The RF power from each klystron is then distributed to 32 cavities.

After the ITRP decision, DESY and collaborators continued to develop the pulsed transformer modulators, multi-beam klystrons and power-distribution components with both the European XFEL and ILC in mind. In particular, the Modulator Test Facility (MTF) was constructed at PITZ and RF system testing has been on-going there since 2006. However, two major changes have since been incorporated in the European XFEL RF-system design. One was to use a Pulse Step Modulator in which several switching modules are connected in series to generate the 10 kV pulses that drive the transformers in the tunnel. This design allows better control of the HV waveform shape and costs less than the Pulse-Transformer approach. Figure 2.54 shows photos of both types of modulators.

**Figure 2.54**
Modulator development for the European XFEL.

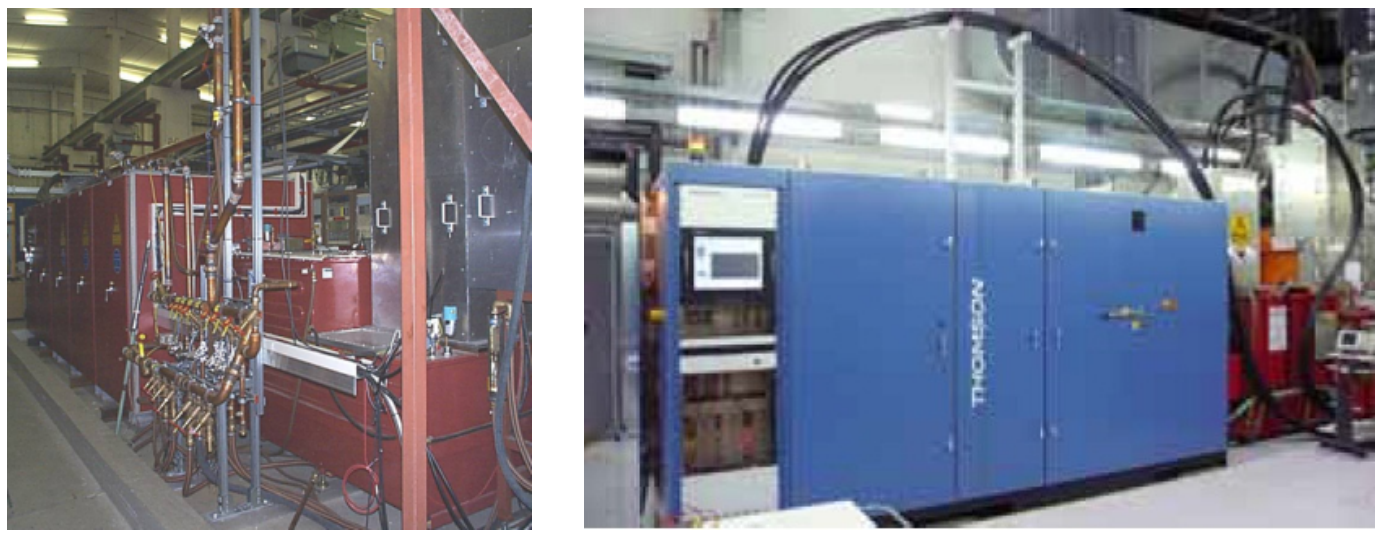

**(a)** Bouncer-type modulator

**(b)** Thomson pulse-step modulator in the MTF hall

The other change addresses the power distribution to the cavities. For TTF/ FLASH, the power-distribution system includes back-terminated power splitters and 3-stub tuners to control the cavity phase and loaded Q. As this system is fairly bulky, has no provisions for changing the power split to the cavities and the cavity phase is hard to tune, a more versatile and streamlined waveguide distribution system that can come attached to the cryomodules during installationwas adopted for the European XFEL. It features Asymmetric Shunt Tee (ASTs) which are non-back-terminated power dividers that can be readily customised before installation to achieve the desired power division based on the cavity performance measured in the test stands. Compact phase shifters are also used instead of the 3-stub tuners (the loaded Q can still be controlled through the coupler antenna position). Figure 2.55 shows such a distribution system attached to a cryomodule [105]. Commissioning of the European XFEL linac is expected to start in 2015.

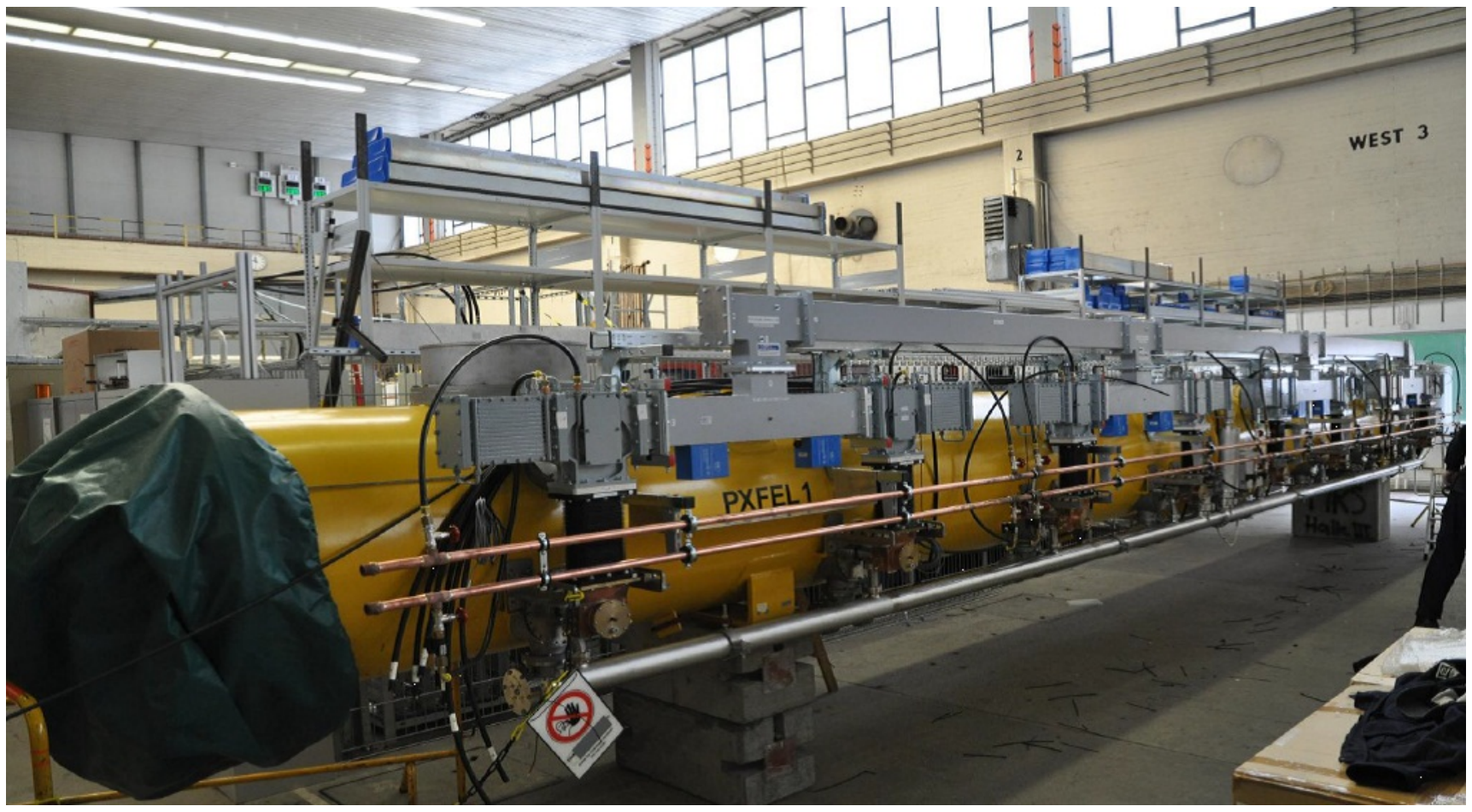

**Figure 2.55** European XFEL-type waveguide distribution tested on the cryomodule ACC7 at FLASH.

## 2.8.3 Radiofrequency System R&D for the ILC

This section describes the various RF-system R&D that has been carried out during the past eight years for the ILC, and the facilities where the RF systems have been used as part of a larger SCRF linac R&D program. These efforts include the development of 120 kV Marx modulators, 10 MW sheet-beam klystrons, 800 kW, high-availability klystrons, a variety of variable power dividers and a clustered klystron scheme for delivering RF power from surface buildings to the tunnel below. Overall, the programme has gone a long way to demonstrating that the RF system requirements for the ILC are achievable.

### 2.8.3.1 RF Systems at ILC Test Facilities

At KEK, the Superconducting Test Facility (STF) was started in 2005 to test SCRF technology including RF systems for the ILC. It has seen various successful development stages, namely STF-0, STF-1 and the S1-Global experiment; the Quantum-Beam project and STF-II will follow and include beam acceleration. Three RF stations were developed at STF that all use the Pulse-Transformer-style modulators. Two klystrons, a 5 MW Thales TH2104A/C, and a horizontally mounted 10 MW Toshiba Multi-Beam Klystron (MBK), were procured and tested in this station [106], see Fig. 2.56.

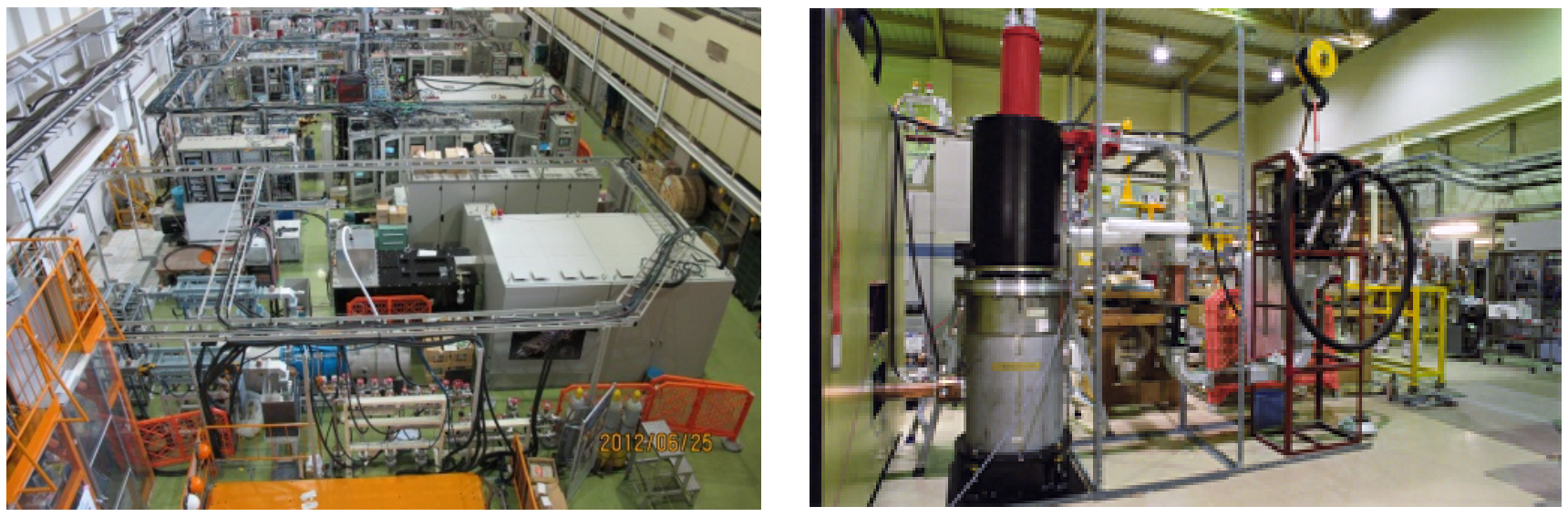

**(a)** Hall **(b)** Klystron

**Figure 2.56** STF test station in KEK.

At the New Muon Lab (NML) at FNAL, a SCRF test area was also constructed for a similar purpose. It includes older style RF-system components (Pulse-Transformer modulators and commercial 5 MW klystrons and waveguide components) and R&D components provided by SLAC (8-cavity cryomodule RF-distribution systems and eventually a Marx Modulator). Recently, FNAL also acquired a horizontally mounted 10 MW Multi-Beam Klystron from CPI.

At SLAC, two L-band RF stations were built in End Station B for the ILC RF system program. The first uses a SNS High-Voltage-Converter Modulator to drive a 5 MW Thales TH2104C klystron. It has been used for various components tests (waveguides, couplers, windows, a 5-cell cavity, RF distribution systems and initial tests of the klystron cluster scheme). The second system is meant to be as ILC-like as possible and includes a state-of-the art control system. As discussed below, it is currently configured with the P1 Marx Modulator driving a vertically mounted 10 MW Toshiba Multi-Beam Klystron. Recently, the power from one arm of the MBK has been used to test a 40 m-long circulator waveguide and bend for the KCS program. It will be upgraded with the P2 Marx modulator during 2013 and another MBK will be delivered.

#### 2.8.3.2 Marx-Modulator R&D

To drive the 10 MW klystrons that had been developed for the TESLA linear-collider design, a Pulse-Transformer-style modulator was developed by FNAL and industrialised by DESY. They are specified to produce 120 kV, 140 A, 1.6 ms pulses at up to 10 Hz. For these modulators, storage capacitors are charged to 11 kV between pulses and then connected to the primary of a 12:1 transformer via solid state switches (IGCTs initially and later IGBTs) to produce the HV that drives the klystron. A 1 kV resonant 'bouncer' circuit was included in series in the primary circuit to offset the voltage droop of the storage capacitors as they discharge. While this design had proven robust (10 had been built by 2005), they required a large oil-filled transformer and the pulse shape could not be finely controlled.

With the emergence of affordable, low-loss, high-power (MW) transistor (IBGT) switches in the late 1990's and early 2000's, various adder-modulator topologies became viable that eliminated the need for traditional transformers. During this period, SLAC developed inductive adder modulators for the NLC klystrons that consisted of stacks of ferrite cores that were each driven in parallel via IGBT switches. A common, multi-turn secondary then added the induced voltages to drive high-voltage (500 kV), short-pulse (1.5 μs) klystrons. A capacitive adder (aka Marx Modulator) program was started at SLAC as well for the NLC, and with the ITRP decision in 2004, the program was redirected to designing modulators for the ILC where the inductive approach is not feasible given the long (1.6 ms) pulses required.

The operational modes and basic Marx circuit without droop compensation are illustrated in Fig. 2.57. One advantage of this scheme is the modular design, which both simplifies fabrication and improves reliability as spare modules are included that automatically turn on if others fail. Also, these modules, which are pulsed up to 120 kV, can be supported and connected in ways that does not require immersion in oil to suppress arcing. For cooling, a closed forced-air system suffices. The biggest advantage is that no large, oil-filled transformer is required (such a transformer would in any case become prohibitive in size if the pulse length needed to be significantly increased).

As part of the SLAC ILC RF program, two Marx Modulators (P1 and P2) were successfully built and tested during the past eight years, as discussed below. More recently, two Marx Modulators were funded as part the US SBIR program: one from ISA Corporation, which was never fully completed, and one from Diversified Technology Inc. (DTI), which was built, tested briefly at DTI at low average power and is currently at KEK awaiting full evaluation. The main design features of these four modulators are listed in Table 2.24.

The P1 Marx Modulator [107] consists of sixteen, 11 kV modules and a single Vernier Marx module in a cantilevered arrangement of corona-shielded circuit boards that connect to a G10 backplane, which runs along the centre (see Fig. 2.58 – the far end is grounded and a 120 kV output cable connects to the near end). The triggering sequence is designed to promptly turn-on eleven modules, then stagger turn on the remaining five modules to coarsely compensate the storage-capacitor droop. The Vernier Marx module (with sub-modules charged to 1 kV) staggers the sub-module turn-on and turn-off to further regulate the output to the specified 0.5%. The waveforms in Fig. 2.59a illustrate this droop compensation (also shown is output power versus modulator-voltage measurements over a running period of several thousand hours).

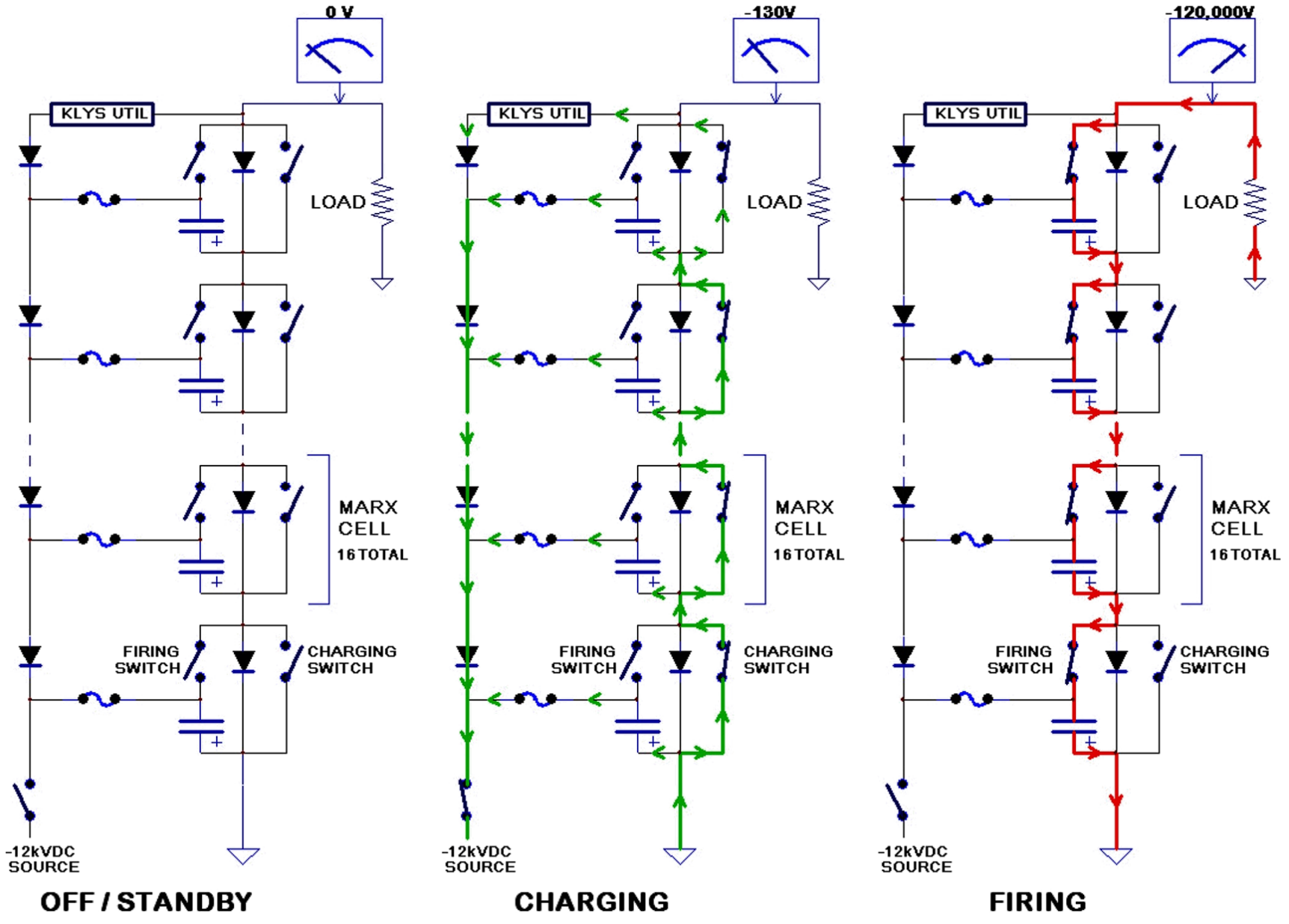

**Figure 2.57** Block diagram of Marx Modulator with illustration of the basic mode of operation. The modes off/standby, charging and firing are illustrated schematically (from left to right)

**Table 2.24** Variants of the US DOE SBIR-funded Marx-Modulator research

| | SLAC P1 | DTI | ISA Corp. | SLAC P2 |
|---|---|---|---|---|
| Cell voltage | 11 kV | 6 kV | 3.5 kV | 3.75 kV |
| Number of cells | 16 | 20 | 42 (7 delay) | 32 |
| Redundancy | Vernier (16) 1.2 kV + delay | Correction cells (16) 0.9 kV | Vernier (16) 0.5 kV + delay | Regulated cell (PWM correction) |
| Regulation | N+1 | N+3 | N+1 | N+2 |
| Status | full test completed | SLAC/KEK for MTBF test | Voltage test completed | full test completed |

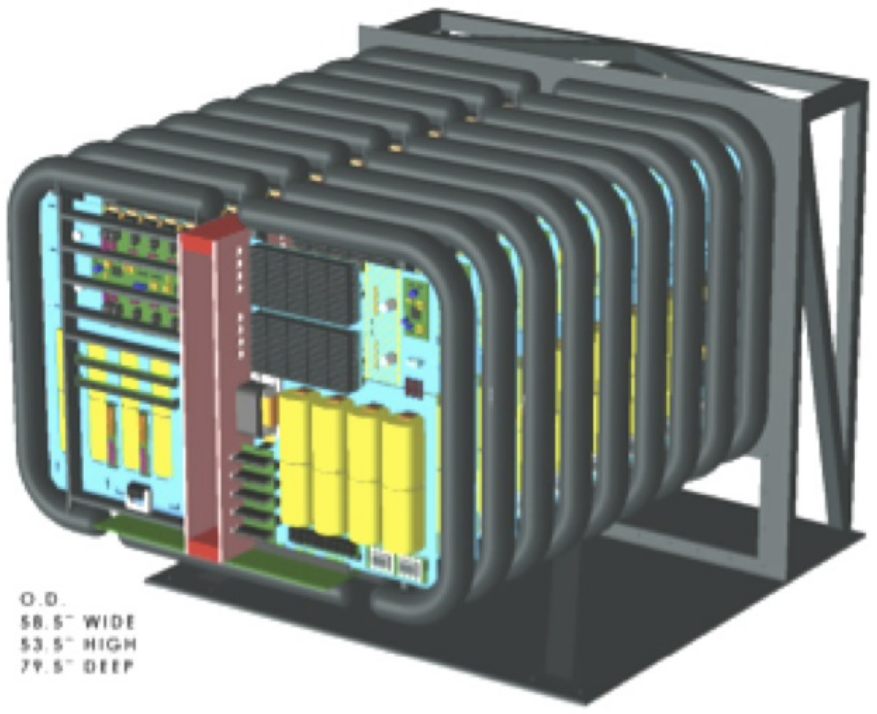

**(a)** SLAC P1 Marx-Modulator cell arrangement.

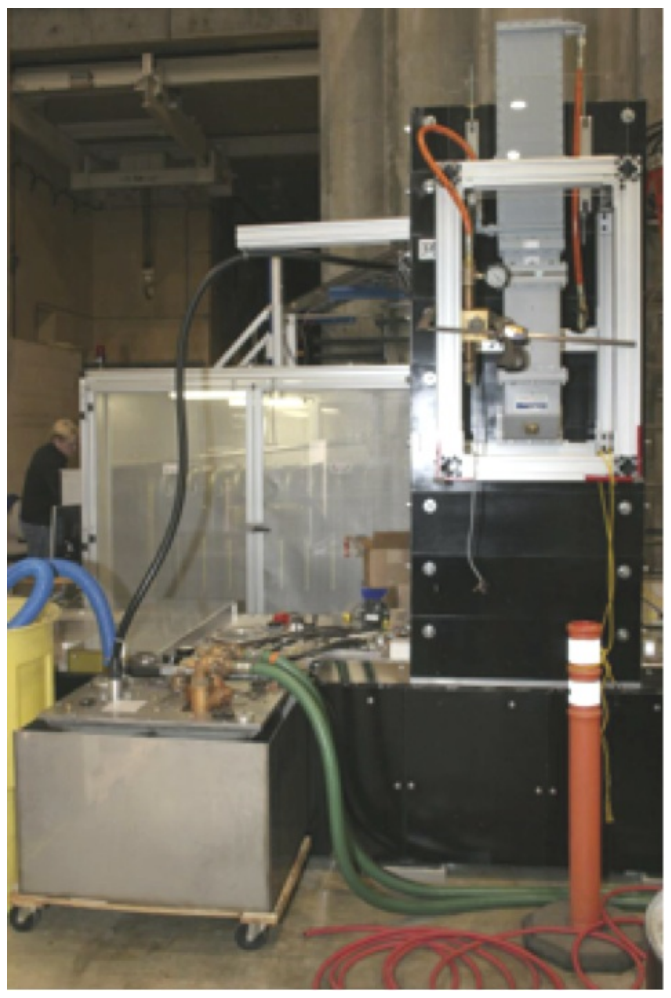

**(b)** P1 test stand

**Figure 2.58** SLAC Marx Modulator and test stand in which the modulator (in a electrically shielded enclosure in the background) drives either a water load (cart in foreground in which the modulator HV is shown connected through the black cable) or a 10 MW Toshiba Multi-Beam Klystron (located in the black lead-lined 'tower' that sits on top of an oil tank).

Until recently, the P1 had been undergoing lifetime testing, driving a 10 MW Multiple-Beam Klystron (MBK) as shown in Fig. 2.58. The modulator has been operated for nearly 6000 h, although about half of this time at half pulse width after it was discovered that the aluminised polyethylene

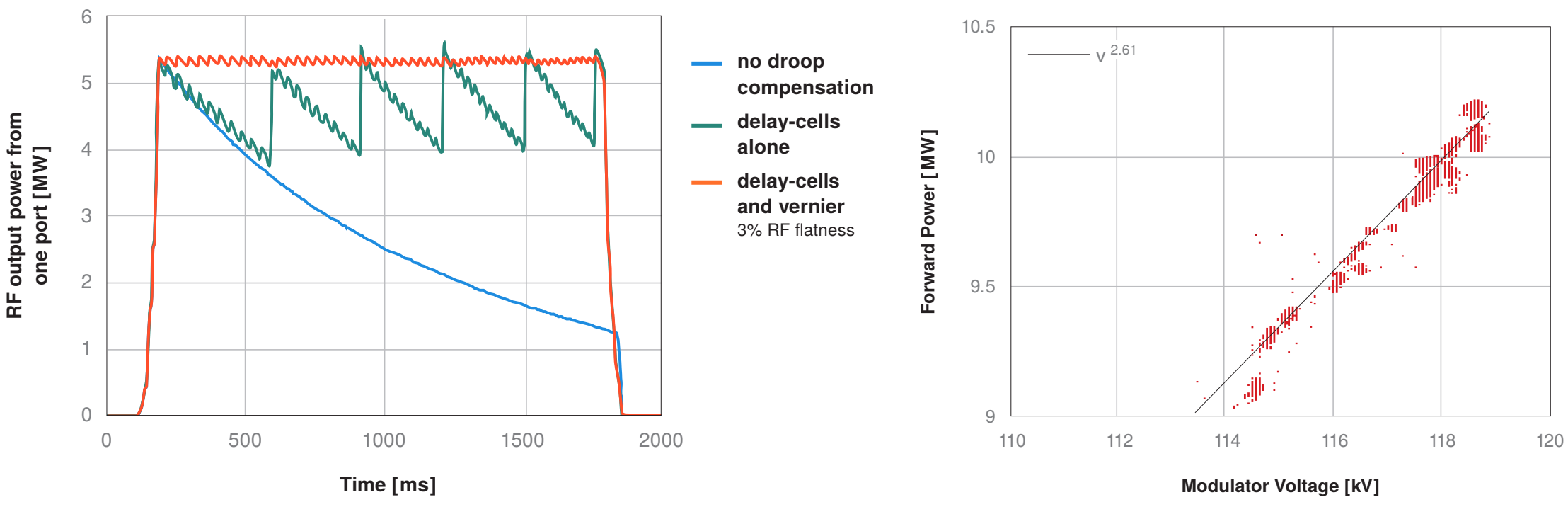

**(a)** Droop compensation. **(b)** Output power vs voltage of the SLAC P1.

**Figure 2.59.** SLAC P1 Marx-Modulator tests.

storage capacitors degraded with full-pulse but not half-pulse operation. The capacitors were replaced recently with more robust zinc-based versions but they have not run long enough at full pulse width to verify that the issue has been fully resolved (bench tests were not conclusive). Damage from corona arcing on the P1 backplane was also discovered and a fix was implemented. For the klystron, there have only been a few faults (window and gun arcs) at full power (10 MW) operation.

Building upon the experience with the P1 Marx, the SLAC P2 Marx was designed to include several improvements [107]. It has 32 modules 3.75 kV where each module individually regulates its output via a buck converter, including compensating the capacitor-voltage droop. If any two modules become inoperable, they can be bypassed by increasing the applied charging voltage or turning on spare modules, allowing the modulator operation to continue. In addition, the modular design allows better utilisation of high-volume manufacturing techniques, and there is no arraying of solid-state switches within a module, simplifying the control and protection schemes. Finally, the module layout is much different (see Section 3.6.1).

The SLAC P2 Marx utilises a hierarchical control system with a system-level application manager and module-level hardware manager [108]. Communication is accomplished using Gigabit fibre-optic Ethernet through a commercial switch. The application manager coordinates the state control of the modulator, handles external interlocks, interfaces with DC chargers, archives diagnostics, processes prognostic routines, coordinates module timings and implements the regulation scheme. The module hardware managers control the state of the modules, implement module IGBT switching routines, gather and transmit diagnostics, and provide fault protection. Each hardware manager incorporates twelve 12-bit, 1 MS/s ADCs, which monitor module voltages, currents, and temperatures. A feedforward algorithm is used to generate very flat pulses – a 0.05% flatness has been achieved when operated with a water load (see Section 3.6.1).

The P2s ability to survive an arc-down at full output power without damage to components has been successfully demonstrated, along with the requirement for klystron protection, namely that less than 20 J be deposited in the load. Since its storage capacitors discharge one-half as much as those in the P1, the lack of capacitor damage seen when running the P1 at half pulse width would indicate that the P2 capacitors should not degrade. The P2 has not yet been tested at full average power but will be run long term to full specification (driving the 10 MW MBK) starting in 2013. It has been adopted for the ILC Linacs to replace the Pulse-Transformer modulator as it is less expensive, has better performance ($< 0.1\,\%$ pulse flatness) and has no large transformers.

#### 2.8.3.3 Sheet-beam Klystron

As a plug-compatible alternative to a multi-beam klystron, the development of a L-band sheet-beam klystron (SBK) was pursued at SLAC, with the same 10 MW, 1.6 ms, 5 Hz specifications, operating at the same voltage and beam current. The goal was a less expensive and more compact RF power source with potentially greater efficiency [109]. Figure 2.60 shows the considerable progress made in the design, which uses a periodic-cusp magnet-focusing scheme and a 40-to-1 beam aspect ratio. Unfortunately, after discovery through 3-D simulations of strong trapped-mode-driven beam instabilities [110], it was realised that a stronger solenoidal focusing scheme would be needed to produce stable operation. Since this would be make the tube less competitive as an alternative to the MBK, the program was discontinued after a gun with an elliptically-shaped cathode was tested.

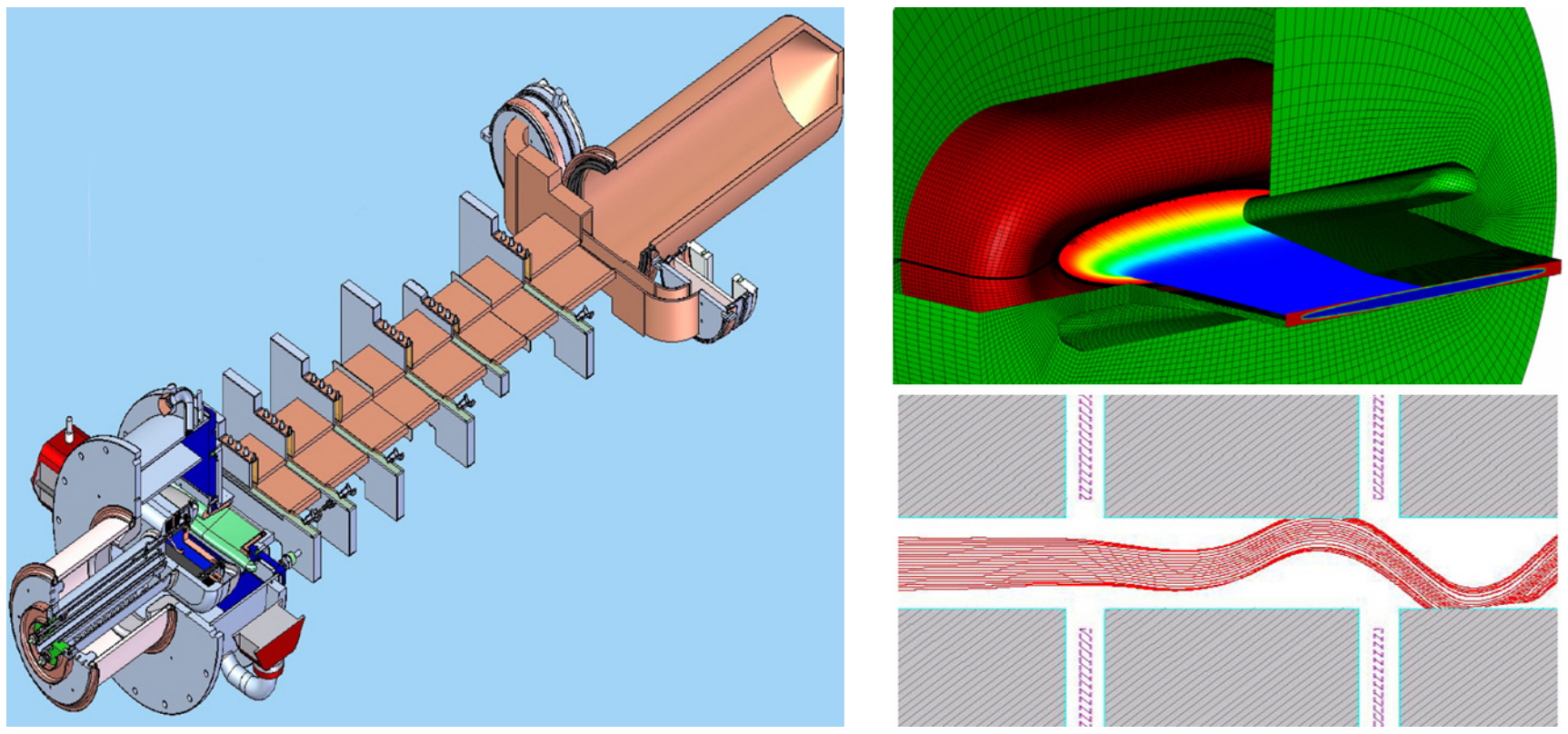

**Figure 2.60** Sheet-beam Klystron design and MICHELLE tracking-code simulations

**(a)** Sheet-beam klystron design

**(b)** Simulations of the focused sheet beam leaving the cathode, and of the beam hitting the wall due to instability oscillations

### 2.8.4 Waveguide Components R&D

#### 2.8.4.1 Waveguide Components

The 1.3 GHz L-band rf power needed by the cavities to accelerate the beam is delivered to their fundamental power-coupler inputs through nitrogen-filled aluminium WR650 (16.51 cm $\times$ 8.255 cm) rectangular waveguides. Simple pieces such as straights, bends, Ts and semi-flexible bellows can generally be customised by waveguide vendors. However, a number of functions need to be accomplished in the waveguide circuit that require special components. Some of these were developed for FLASH/European XFEL and could be directly used for the ILC design. These include isolators, loads, directional couplers and in-line phase shifters, all available from specific vendors. Even for these, some modifications and variations were pursued, while for other functions, particularly power division, new ILC components had to be developed.

The isolator assumed in the RDR is based on a design, still used by DESY, that can handle 400 kW peak power with full reflection at any phase and 8 kW average power, with fairly low-loss —approximately 0.1 dB [3]. In particular, a version was developed that incorporates directional coupler pickups in the input and load ports, eliminating the need for a separate component. In order to explore cost efficiency, a comparable isolator was developed by a Japanese company and tested in the S1-Global test.

In the RDR design, RF phase was adjusted via a three-stub tuner. This was superseded by a DESY phase-shifter design using a motorised movable side wall inside a short WR650 section to vary the guide wavelength. This gives a range of more than 100° with a reflection below −23 dB. At KEK, a variant phase shifter was developed that uses a movable floating pontoon to change the effective phase length, also allowing a wide range of cavity-phase adjustment.

The klystron output windows require pressurisation, whereas the cavity-coupler input boxes are not pressurisable. Therefore a waveguide pressure window is needed somewhere in between. Although a commercial pillbox window with rubber seals is available, SLAC developed a simpler, more compact, radiation-hard version. Its main component is a block of ceramic with transverse dimensions equal to that of the waveguide and with a thickness (few centimeters) to achieve an RF match. It is electro-plated around its perimeter with a centimetre-thick layer of copper and, after machining, fits into an aluminum flange frame.

All pieces of waveguide upstream of the window, in which the power is higher, are pressurised to 300 kPa with dry nitrogen, which requires that the waveguide have thicker walls and a more robust construction. To join WR650 waveguides, whether pressurised or not, aluminium gaskets are used at SLAC and FNAL. They have partially recessed rubber rings on each side to provide a pressure seal and the inner edges have raised, dimpled surfaces to assure an RF seal. At KEK, rubber inserts are placed in flange grooves to seal pressure and the flanges are machined flat at the 25–50 µm-level to achieve good electrical contact.

### 2.8.4.2 Power-Split Adjustment

The one area that called for a focused R&D effort was the development of RF power dividers, which are required for distributing RF from a given source to multiple cavities. For the RDR, it was assumed that all cavities would get equal power; fixed waveguide splitters of various increasing couplings were to be used in linear distribution lines, as they had been at DESY. Also, 50 % power dividers (i.e. 3 dB splitters) were required to feed pairs of cavities in the RF distribution systems that SLAC built for FNAL, where the goal was to achieve better isolation of the output ports than that typical from commercial 3 dB splitters. For this purpose, SLAC designed and built a four-port 3 dB H-hybrid [111] via dip-brazing with highly isolated output ports, as seen in Fig. 2.61a.

**Figure 2.61**
Power-distribution systems developed at SLAC.

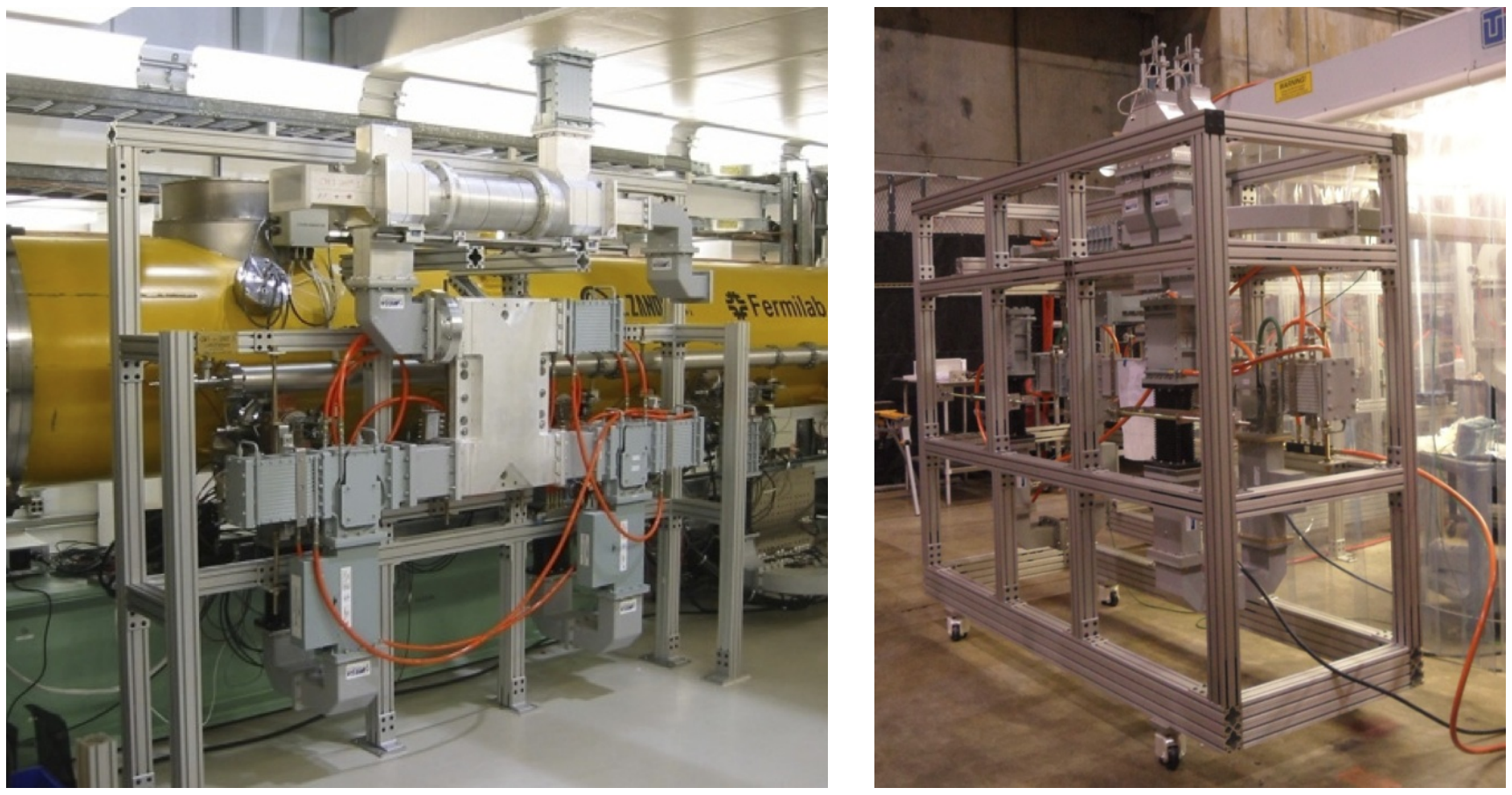

**(a)** A two-feed unit of PDS1, incorporating a VTO and H-hybrid, installed on a cryomodule at NML.

**(b)** A two-feed unit of PDS2, incorporating a VPD, ready for high-power testing at SLAC.

When it was decided to allow a range of cavity operating gradients to be accepted. increasing the cavity yield, R&D was done on tailoring the power distribution so each cavity could reach its maximum sustainable gradient. As noted above, the European XFEL project also added provisions for this purpose. For the ILC, three 4-port devices were developed, each of which operates with a load on the unused port at the input end to back terminate it. They are summarised below – the latter two have been adopted for use in the local RF-power-distribution system for the ILC.

The variable tap-off, or VTO [112], shown in Fig. 2.62, was developed first to deal with the

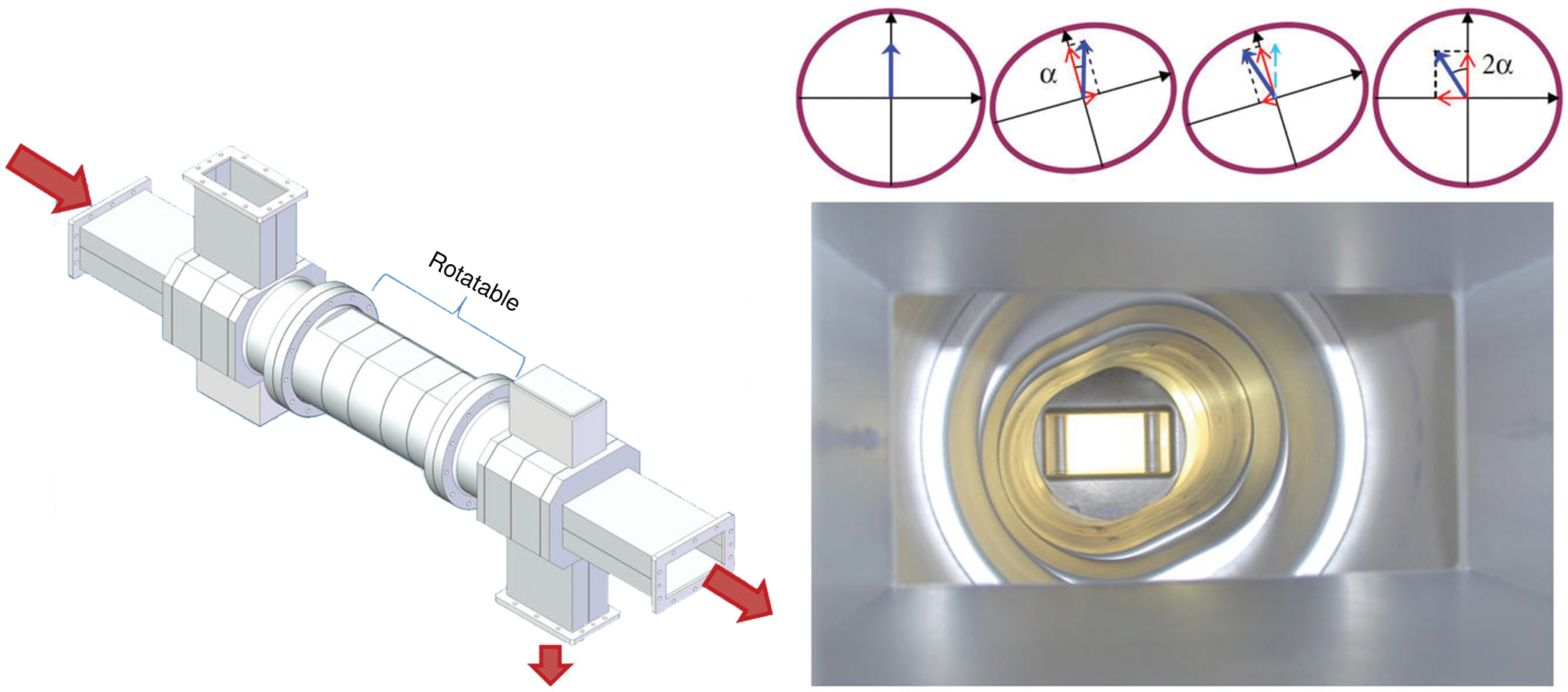

**(a)** VTO Assembly. The rotatable central section allows adjustment for any coupling ratio

**(b)** Above: polarisation rotation along the VTO; one component slips in phase by 180 degrees relative to the other by the time circular symmetry is restored. Below: photograph looking through the VTO gradient spreads in the FNAL cryomodules. This 4-port device has three pieces: each end consists of a 3-port section whose two orthogonal rectangular WR650 ports each couple into a different polarization of $TE_{11}$ in the circular third port; the central section has round ends step-tapered to an oval cross-section at its centre. The latter serves basically as a mode rotator; its orientation determines how power entering a given port at one end of the VTO is divided between the ports at the other end. It can be adjusted to provide any desired split without affecting the output phases. However, since the adjustment requires opening the waveguide and rotating the centre section, it would require an access to the accelerator tunnel if used for the ILC.

**Figure 2.62.** Variable Tap-off power divider (VTO)

**Figure 2.63**
a) the VPD (Variable Power Divider), composed of folded magic-Ts and U-bend phase shifters (with offset spacers), and b) the U-bend phase shifter, with motorised feed-through-supported trombone-like inner waveguide.

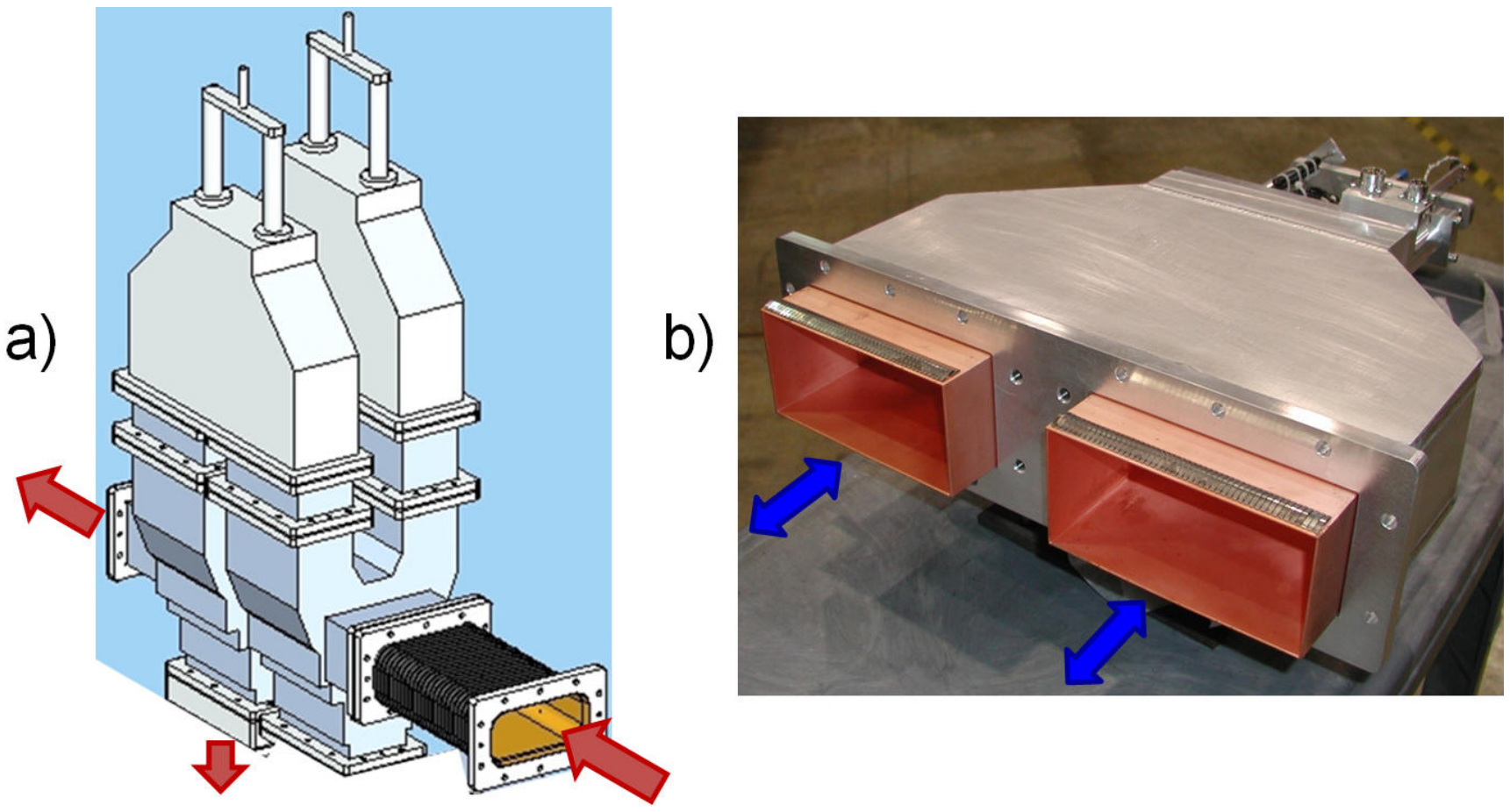

The variable power divider (VPD) was developed to provide the same functionality as the VTO but with remote adjustability. Illustrated in Fig. 2.63, it consists of a pair of folded magic-Ts connected by trombone-like U-bend phase shifters. The latter component, also pictured, was developed at SLAC for this application. It contains a thin-walled mitred H-plane U-bend in copper-plated stainless steel that fits inside WR650, with springy finger-stock on the outer broad-wall edges for electrical contact. Housed in an aluminium outer shell for pressurisation, this bend can be moved by motorised feed-throughs attached to its back. The RF phase shift through the bend is thus adjusted by a change of path length, rather than a change in guide wavelength. Tests of a prototype showed $\sim 0.02$ dB loss and reflections of $-51$–$-36$ dB over a phase range of 120° . Moving the two U-bend phase shifters in opposite directions keeps the combined VPD output phases fixed, but allows a full range of amplitude variation. The VPD and VTO have been tested to power levels of 3–4 MW with 1.2 ms pulses without breakdown.

Finally, the variable H-hybrid was developed in parallel at KEK. This four-port device is based on the SLAC H-hybrid, with a pair of movable pontoons introduced in the dual-moded interaction region to provide coupling variation by affecting phase lengths, the same innovative mechanism used in the KEK phase shifter. This scheme provides a wide range of adjustment in the power division ratio, but the output phases are also affected as the amplitudes are changed. The variable H-hybrid is a relatively simple and cost-effective device. A schematic drawing and simulation results are shown in Fig. 2.64.

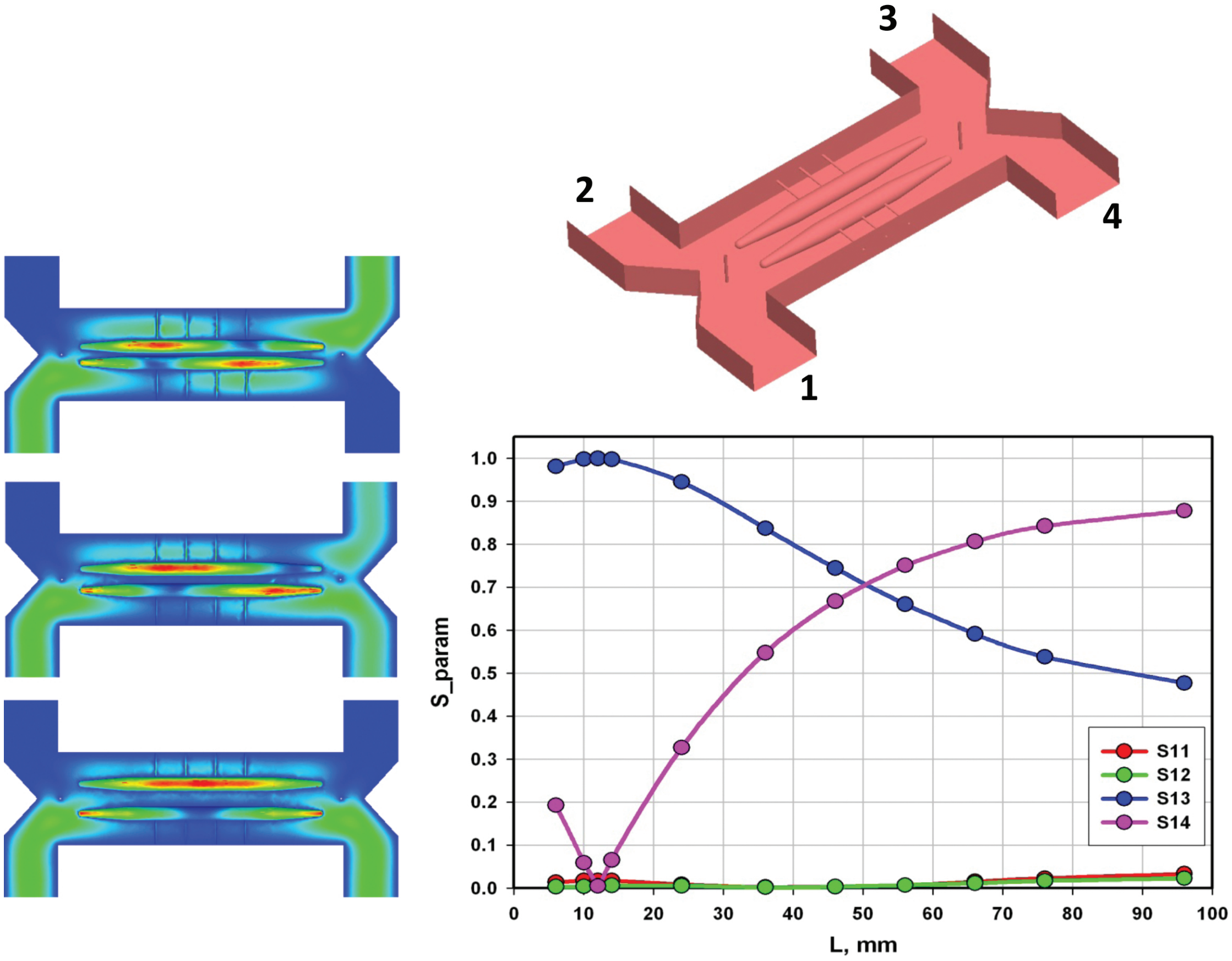

**(a)** Simulated electric field plots illustrating power split variation mechanism in the Variable H-Hybrid.

**(b)** Model and scattering parameters as functions of pontoon position for the physically shorter not-full-range design.

**Figure 2.64.** Variable H-hybrid.

### 2.8.4.3 RF Power-Distribution Systems

Two full-cryomodule (8-cavity) power-distribution systems have been built, high-power tested and delivered by SLAC to FNAL for use in the NML SCRF facility. The first uses the VTO to allow the power to be adjusted to pairs of cavities. The power out of each VTO was split through the SLAC 3 dB H-hybrid with the idea that tests without isolators on each of the cavity inputs could be done. In this case, the reflected power from the two cavities goes to the load on the input side of the hybrid (such operation was subsequently shown to work reasonably well at STF, see Section 2.8.5). Additional components in the system include isolators, bends, loads, air-to-air windows, phase-shifters, directional couplers, and semi-flex guides to connect to the cavity-coupler boxes. Also a 5 MW load is included for the unused power that passes through the four VTOs. A second full-cryomodule system followed with the VTO replaced by the VPD, providing remote control of the power division albeit still by pairs. It also substituted a T-waveguide for the H-hybrid to reduce cost (The ILC eventually decided the cavities would be feed individually, not in pairs, so the idea of eliminating the isolators was abandoned). Figure 2.61 shows a 2-cavity unit of each system on its support frame.

At KEK, two types of RF-power-distribution systems were built and tested, a RDR-style linear type and a *3 dB hybrid tournament* tree type, each of which power a 4-cavity cryomodule. These were used for the S1-Global test in 2011 [7] at the Superconducting Test Facility (STF). With the latter distribution method, the isolators connected to each superconducting cavity could be eliminated as discussed above. Some results of the tests are described in Section 2.8.5. A schematic drawing of the power-distribution system and a drawing of the layout are shown in Fig. 2.65.

**Figure 2.65**
STF Power-distribution system in the S1-Global test at KEK.

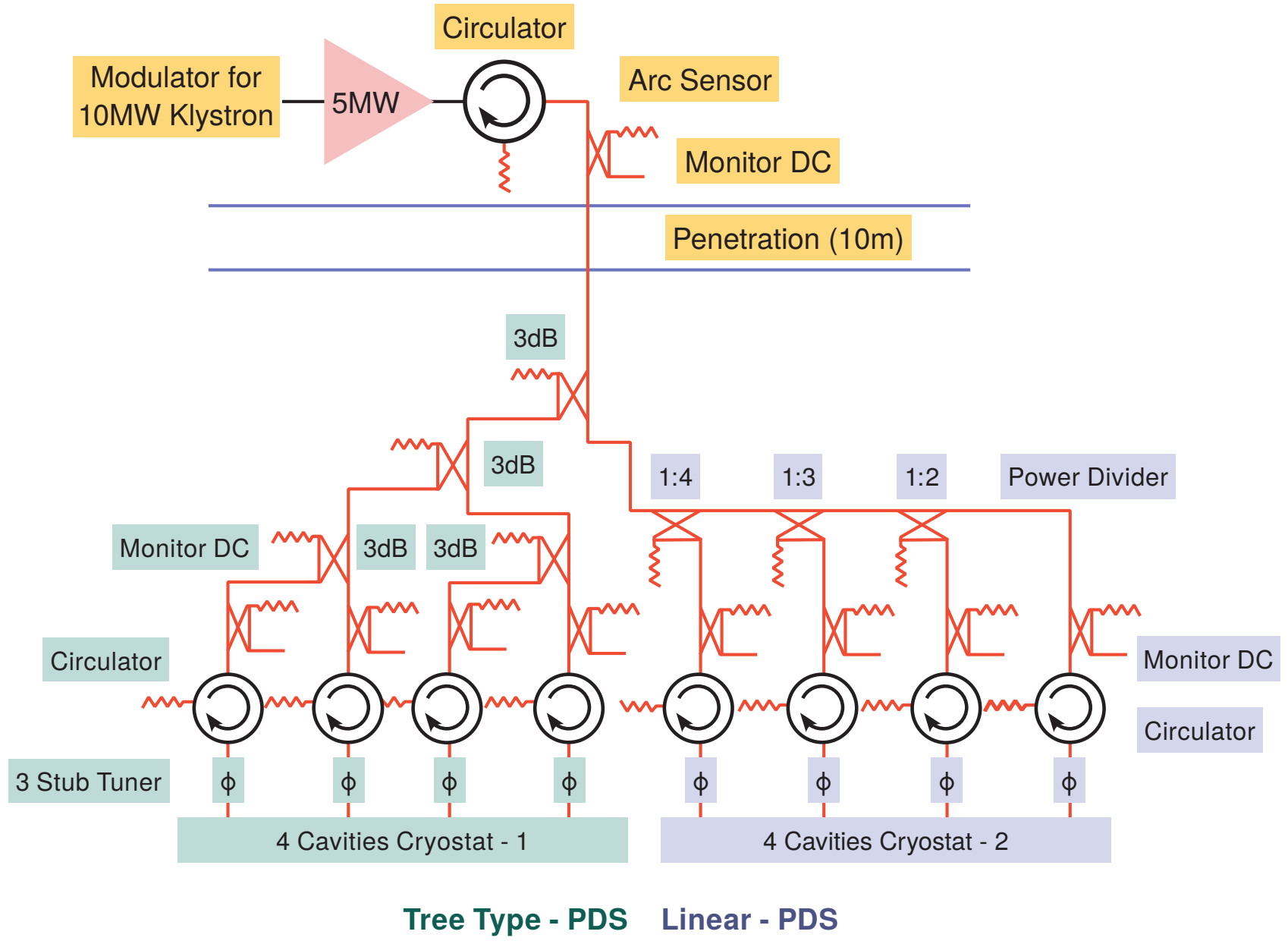

**(a)** Schematic

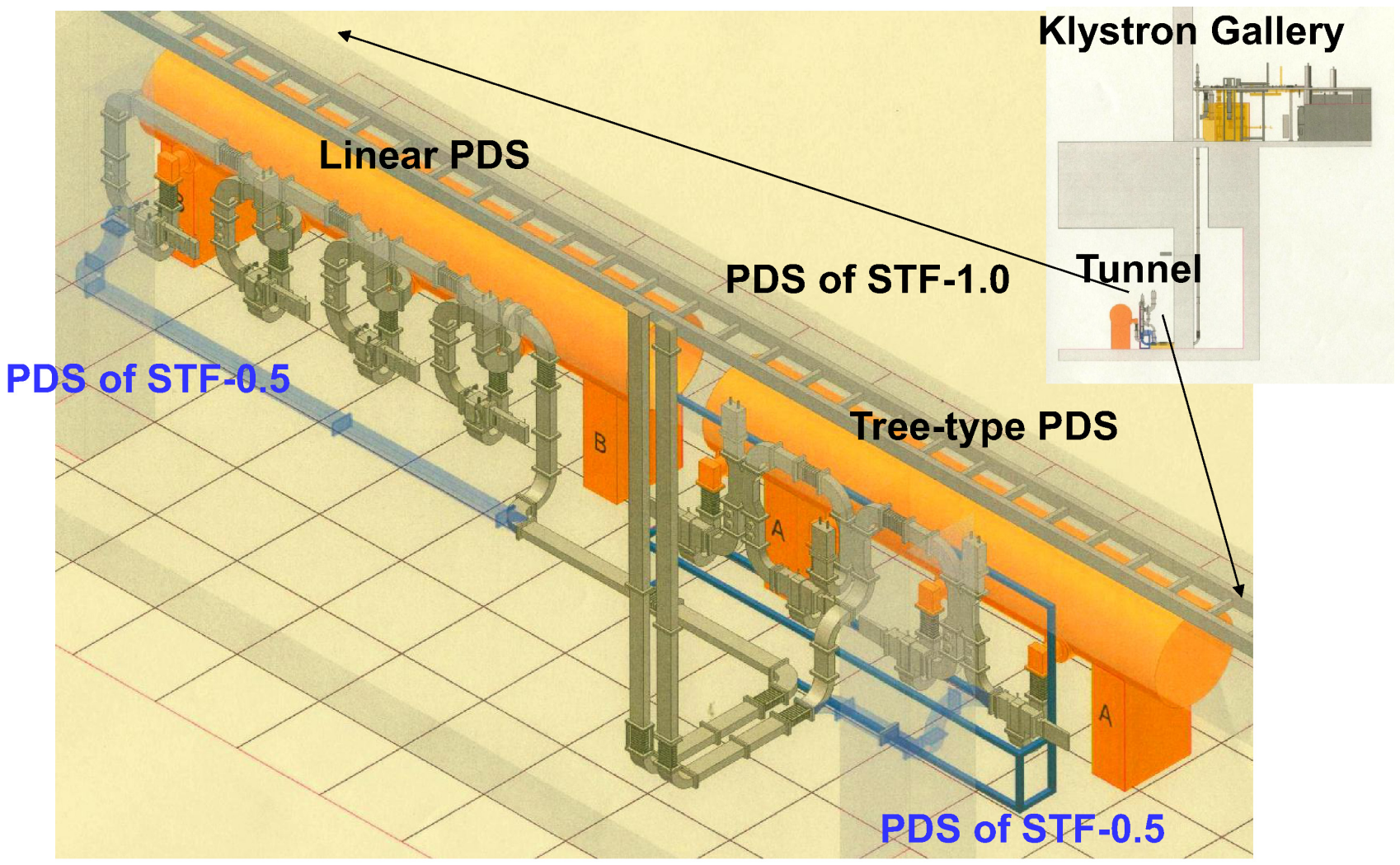

**(b)** Layout

### 2.8.5 R&D of the Distributed RF Scheme (DRFS)

A Distributed RF System (DRFS) suitable for a single-tunnel main linac has been designed and studied [50, 104]. The key idea was to produce lower-power, higher-reliability klystrons so availability would remain high given the limited access to the klystrons in a single-tunnel configuration. However, the klystron and modulator produce heat dissipation in the accelerator tunnel and protect the control devices from radiation damage.

For the DRFS design, an 800 kW klystron would power two cavities with the high-power upgrade and four cavities for the low-power baseline design. To reduce cost and save space, the possibility of eliminating the isolators was also considered. At KEK, the smallest functional unit of a DRFS device, including modulator and klystron, were manufactured and tested in the S1-Global experiment [7]. The results and the remaining R&D task for the DRFS klystron are described in the following.

The DRFS klystron is a modulated-anode-type klystron which has an output capability of 800 kW at an applied voltage of 67 kV, pulse width of 1.5 ms and a repetition rate of 5 Hz. Four DRFS klystrons were manufactured and two klystrons were successfully operated in the S1-Global experiment. Figure 2.66 and Fig. 2.67 show the performance of the DRFS klystron and a picture of the S1-Global operation, respectively. to achieve high availability, the DRFS klystron uses permanent-magnet focusing (Fig. 2.67b), which allowed for the elimination of the power supply and cooling system [113]. Ferrite material has been used for economic reasons; the test was carried out in the summer of 2012.

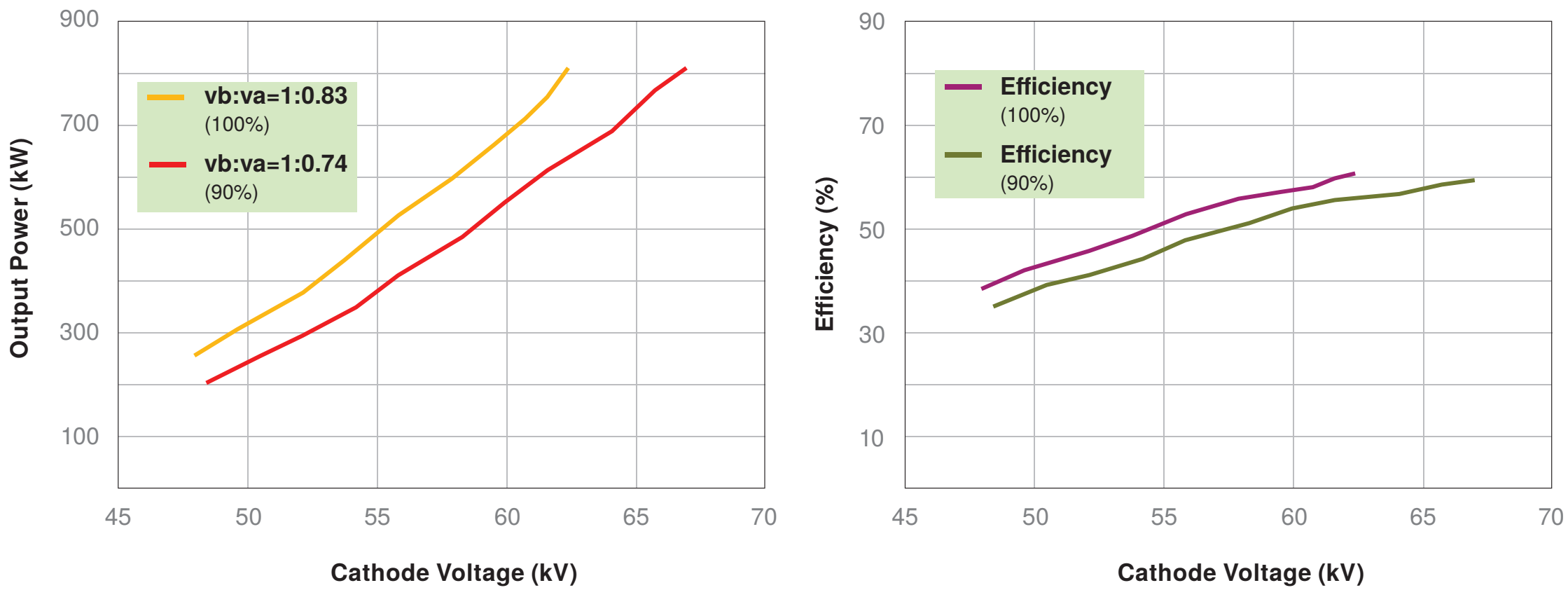

**(a)** Measured output power as a function of cathode voltage

**(b)** Measured efficiency as a function of cathode voltage

**Figure 2.66.** Performance of Toshiba DRFS Klystron powered by Marx Modulator

A prototype pulse modulator for the DRFS klystron has been manufactured at KEK, which consists of a DC power supply and a modulation-anode pulse modulator. Its function was tested in the S1-Global experiment. Since the prototype only supported a minimum set of functions, several components remain to be developed and tested. This includes development of a cost effective crowbar circuit and a gap switch, development of the sag-compensation circuit other than bouncer-type so that Marx-type circuits can be applied and development of a reliable HV relay to detach the klystron from the HV line in case of klystron failure. These developments are no longer pursued after the decision favouring another RF distribution scheme.

The power distribution scheme (PDS) of the DRFS is inherently simple. The power from the klystron is divided into two (two feed) or four (four feed) fractions. When the power load for cavities is paired (by selection) the expensive isolator can be eliminated. Reflection from the cavities and interference between the waveguide branches raises concerns for the LLRF feedback system in this case. Nonetheless the approach without isolators was tested and evaluated in the S1-Global experiment. A stable vector-sum operation was successfully established and the $Q_L$ diagnostics at the cavity worked

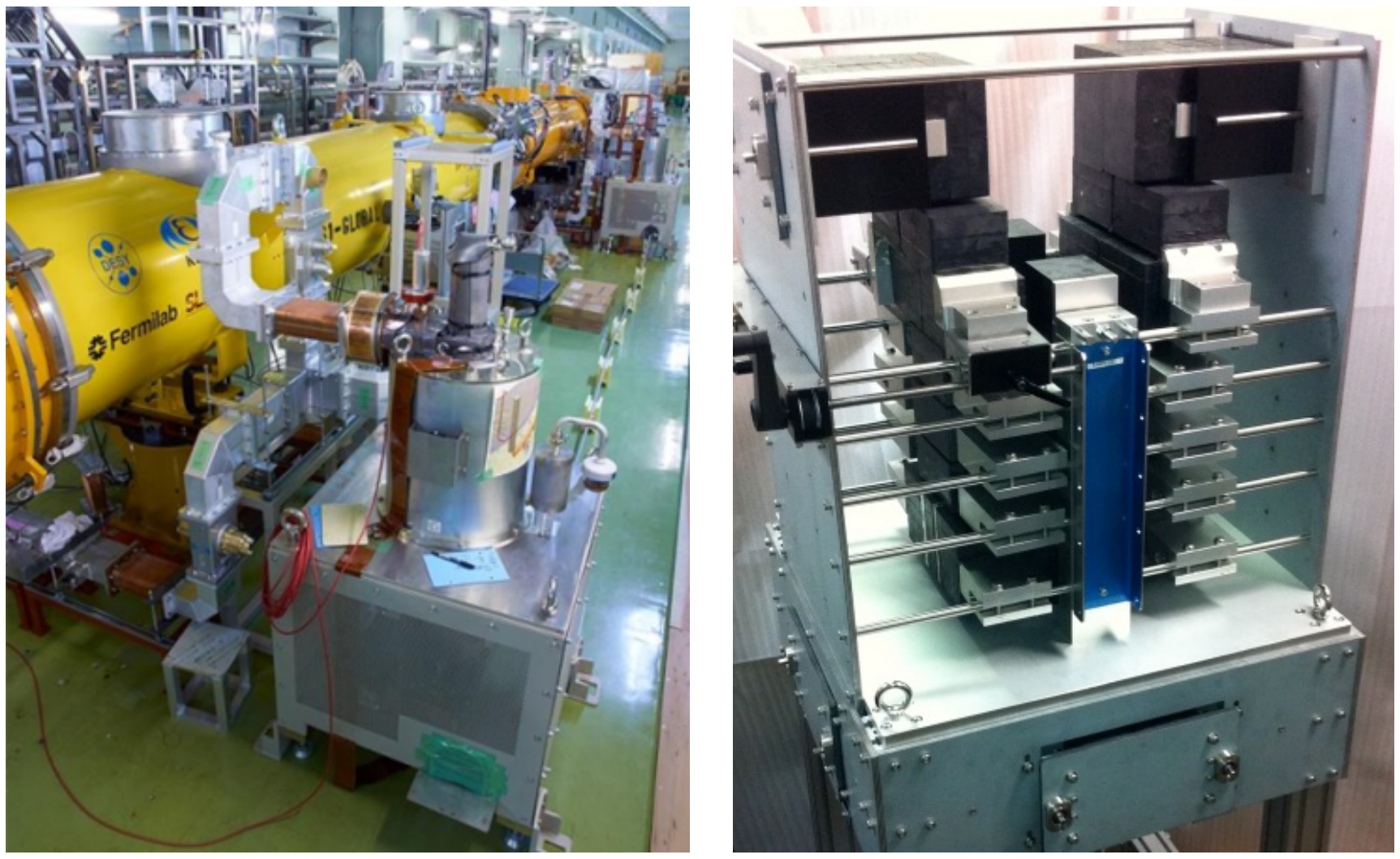

**(a)** DRFS klystron in the S1-Global experiment

**(b)** Permanent magnet for beam focusing in the DRFS klystron.

**Figure 2.67**
Picture of Toshiba DRFS Klystron powered by Marx Modulator

well in the two cases of low reflection (VSWR $\sim 1.1$) and large reflection (VSWR $\sim 3$) under varying detuning. Figure 2.68 shows the characteristics of time vs. field gradient and $Q_L$ diagnostics for operation without circulators.

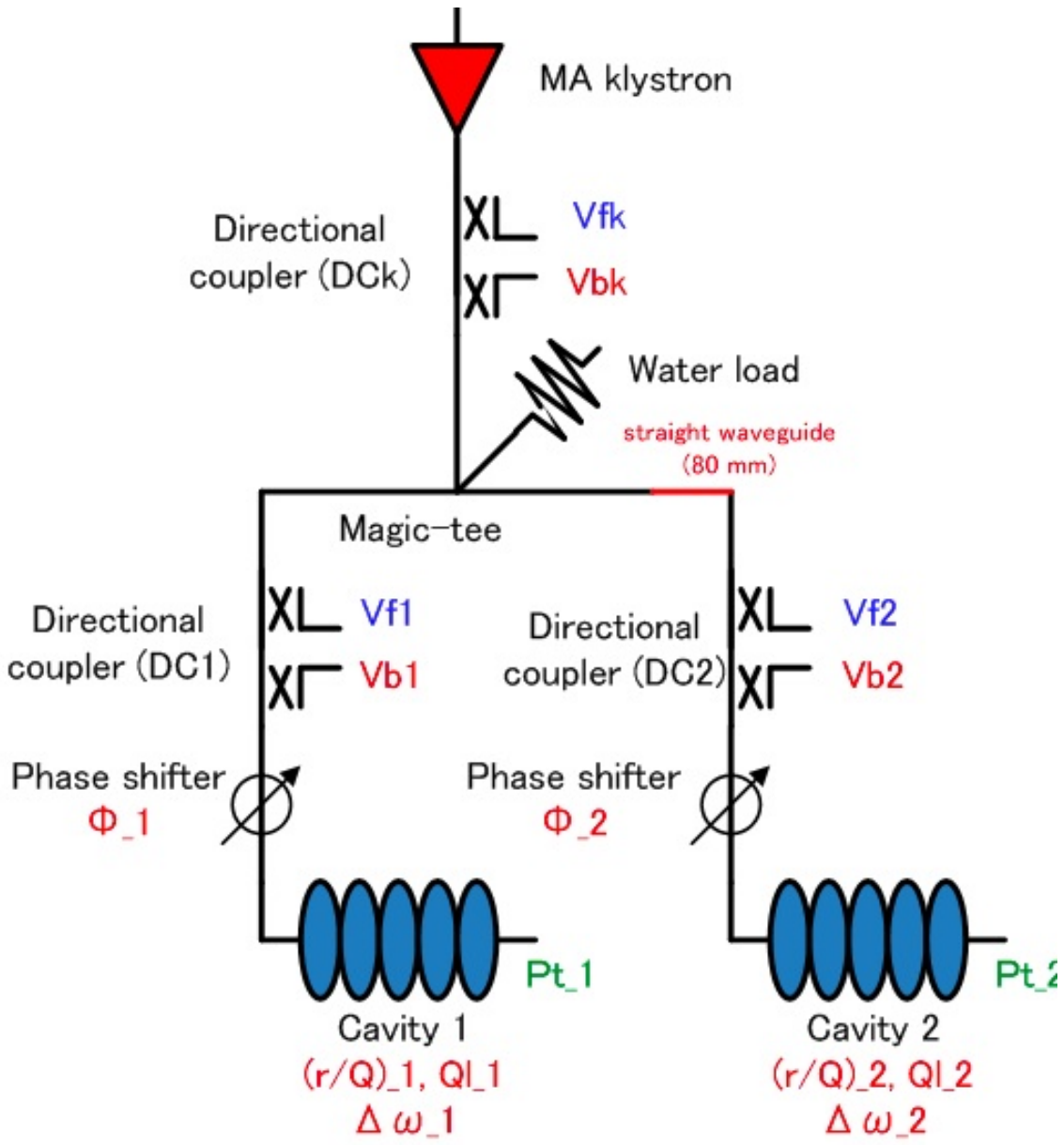

**Figure 2.68**
Result for operation without a circulator in the S1-global test.

## 2.8.6 R&D of the Klystron Cluster RF Scheme (KCS)

In the wake of the post-RDR decision to eliminate the parallel utility tunnel that was to have housed the powering equipment, one of several single-tunnel solutions that emerged was the Klystron Cluster Scheme (KCS) [96], in which RF production is moved to the surface. Unlike in the RDR and current DKS option, where it is brought down as AC power, or the European XFEL, where it is brought down as DC power to underground klystrons, with KCS, the power for accelerating the beam is transported between the surface and the tunnel as RF. This approach follows the example of the SLAC linac, which served the only previous linear collider, the SLC. The differences arise both from having to accommodate a deep-bored (as opposed to cut-and-cover) tunnel and from the need to minimise surface impact over a much larger footprint. Thus the idea of clustering was adopted. Power from groups of approximately 20-30 klystrons is combined into a single low-loss, over-moded waveguide and transported down to and along the tunnel to power approximately a kilometre of linac. At $\sim$ 38 m main linac unit intervals, partial power is siphoned from this main waveguide in 10 MW decrements, to be distributed to 26 cavities, as if from a local klystron.

### 2.8.6.1 Main Waveguide, Bends and Tap-offs

For low transmission loss and robustness against RF breakdown, an over-moded, 0.480 m-diameter circular waveguide (WC1890), operated in the $TE_{01}$ mode, was chosen as the main high-power RF conduit. This azimuthally symmetric mode has the lowest loss and no electric fields terminating on the wall. Pressurisation with dry nitrogen further suppresses breakdown, so that peak powers of up to 300 MW might be transmitted. Minimising mode conversion requires millimetre-scale tolerances on radius, roundness and alignment, and straightness tolerances of half a degree. A flange joint or bellows is needed that can provide sufficient longitudinal flexibility to take up thermal expansion locally while maintaining concentricity and straightness.

Tapping partial RF power into and out from the circular $TE_{01}$ mode transmission waveguide in a compact way, without breaking the internal symmetry, presented a special challenge. For this, a novel waveguide component was invented, the coaxial tap-off, or CTO [96]. In this device (see Fig. 2.69), a diameter step mixes $TE_{01}$ with $TE_{02}$. Reintroduction of a wall at the original diameter divides the interior from a coaxial region. Because the two modes beat, the longitudinal distance between the step and the inner wall determines the power-split ratio, allowing the multiple couplings needed with minor design change. A small ridge at the input suffices to cancel any reflection. Power in the inner region continues on as $TE_{01}$; the tapped-off coaxial $TE_{01}$ power in the shorted outer region is extracted through several apertures into a rectangular waveguide encircling it and thence through two radial WR650 ports in a wrap-around configuration [114].

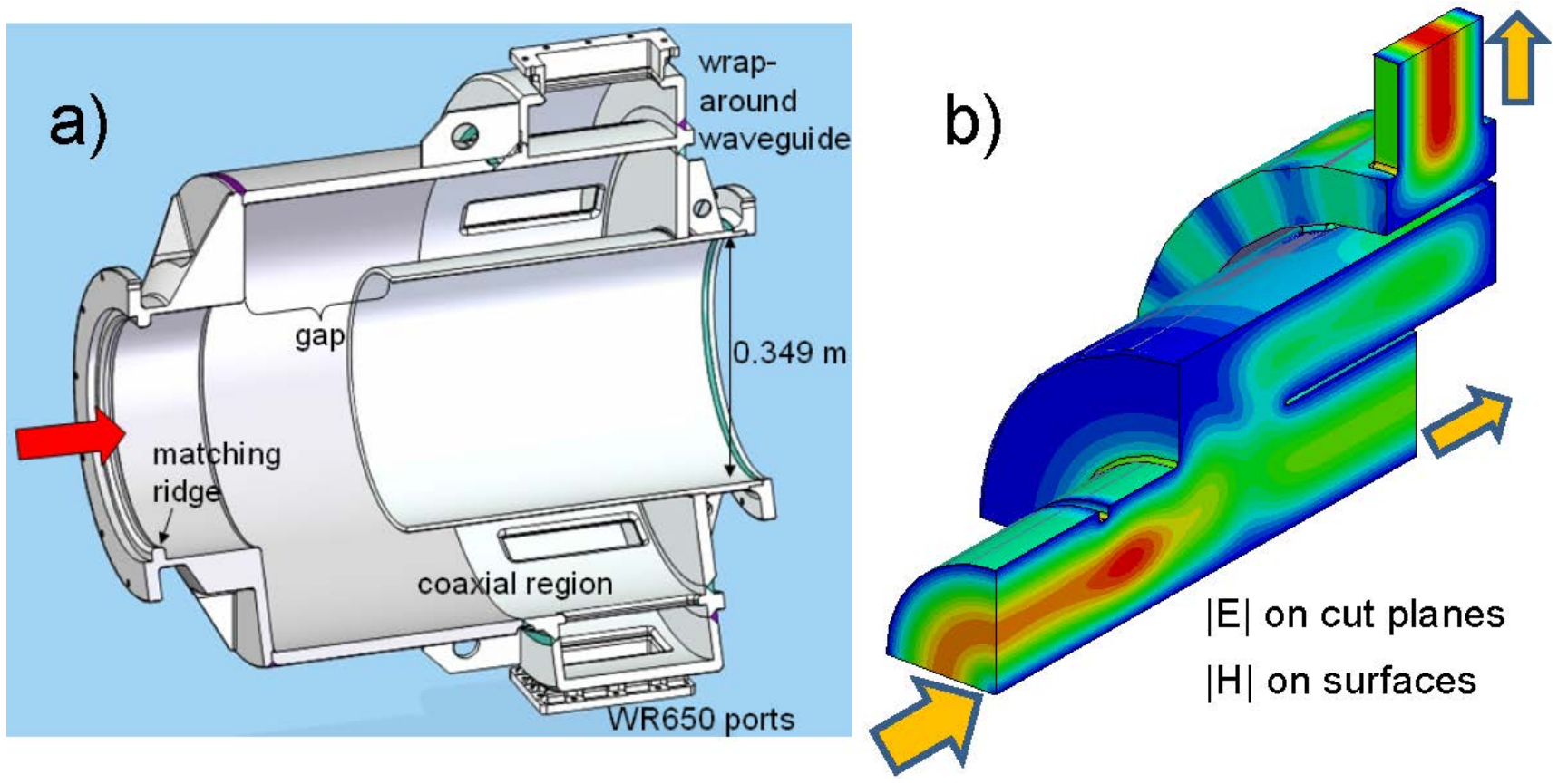

**Figure 2.69**
a) Cutaway and b) Simulated field patterns of a Coaxial Tap-Off (CTO), designed to extract (inject) fractional RF power from (into) a flowing $TE_{01}$ wave in the circular KCS main waveguide.

To bring the RF down to the tunnel, bends are required. Bending in an over-moded waveguide is non-trivial, as geometrical changes tend to scatter power into parasitic modes. A special L-band $TE_{01}$ bend, illustrated in Fig. 2.70 has been developed at SLAC. It includes two stepped transitions between a circular and an over-moded rectangular cross section, designed to map circular $TE_{01}$ into a pure $TE_{20}$ rectangular mode [115]; the actual bending is accomplished with a swept arc in the rectangular guide that restores the incoming mode after mixing with $TE_{10}$.

Both the bend and the CTO have 0.349 m diameter ports, somewhat smaller than the main waveguide. They interface with the latter through a simple matched double-step taper to 0.49 m.

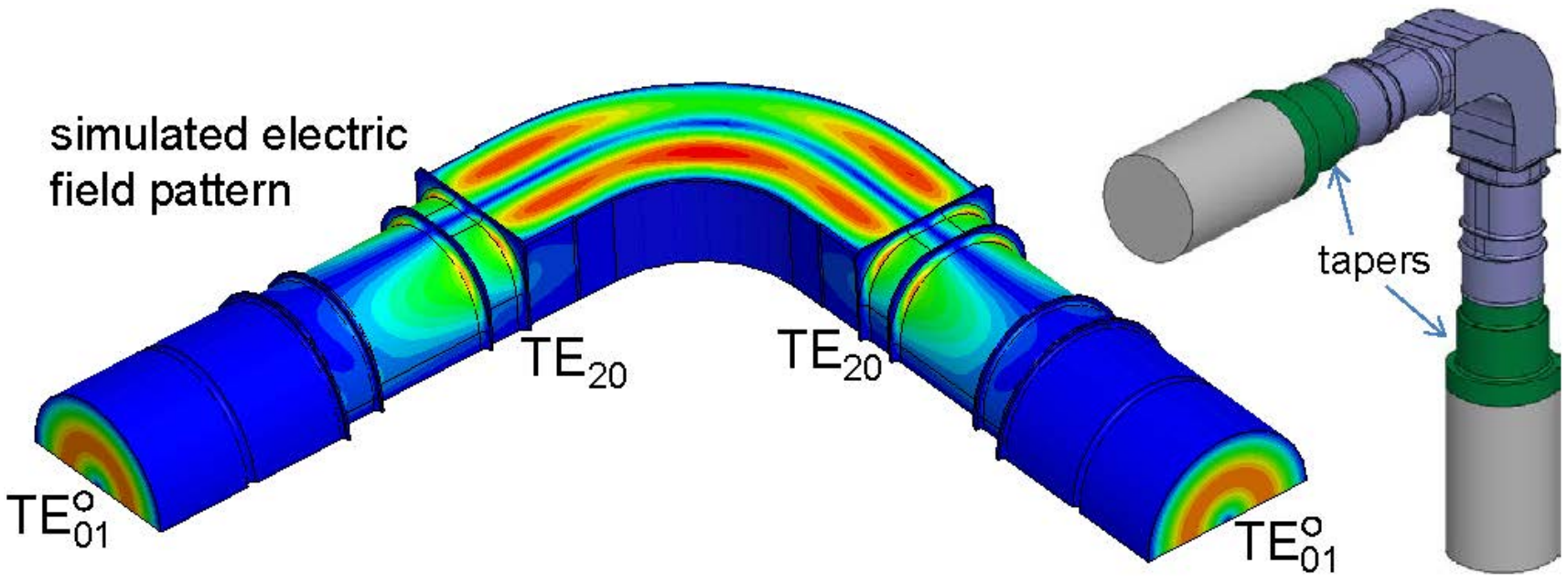

**Figure 2.70** Overmoded bend for the circular $TE_{01}$ mode KCS main waveguide. Mode-converting sections allow the actual bending to be done in the rectangular $TE_{20}$ mode.

## 2.8.6.2 Experimental Program

A pair of 3 dB CTOs and step tapers were fabricated and a $\sim$ 10 m run of the KCS circular waveguide, in four sections, was constructed at SLAC. With proper depth shorting caps, the CTOs could be used as launchers for a successful transmission test using a Thales 5 MW klystron. Since a full-power transmission test is not feasible, comparable field levels were achieved by resonating the big pipe with one end blanked off and the shorting cap of the remaining CTO adjusted for near-critical coupling. Building up forward and backward waves to the 75 MW level gave peak field values equivalent to a single 300 MW wave.

Recently, a longer, 40 m run of WC1890 has been assembled (see Fig. 2.71) and a prototype of the $TE_{01}$ bend has been built. The latter, unavoidably having high surface electric fields (2–3 MV/m), is the bottleneck of the whole scheme. Resonant testing was performed using a SLAC Marx Modulator and a Toshiba MBK. With the longer waveguide, more stored energy is available to test for damage from breakdowns. This system can be pressurised up to 210 kPa allowing the required pressurisation for tolerable break-down rate to be verified. After resonant testing of the pipe alone, the bend was installed at the end, and the combination tested. Breakdowns occurred during processing, but over 100 hours continuous operation without breakdown at field levels equivalent at the anti-nodes to over 250 MW travelling-wave has been observed. Testing will continue but the KCS main waveguide and components do appear to work as designed and to be capable of handling the ILC power levels.

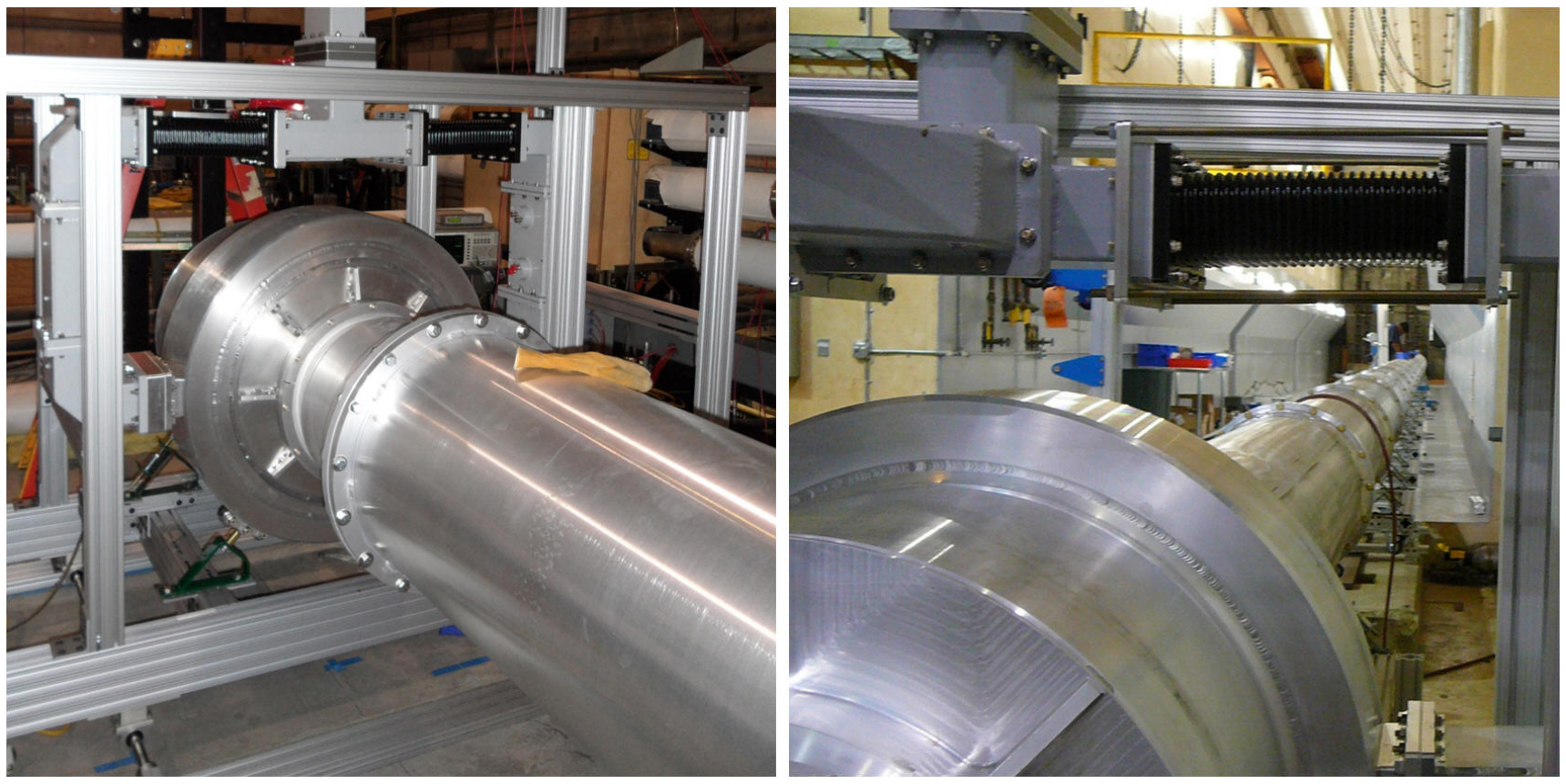

**Figure 2.71** Views of a CTO used as a launcher or coupler in high-power tests of KCS WC1890 main waveguide runs.

## 2.9 R&D towards Mass-production and Design for Manufacture

The ILC has commissioned studies at experienced industries in all three regions to understand the fabrication costs for the cryomodules and components in the quantities required for the ILC. The focus of these studies has been to create a more robust cost estimate for the TDR, so companies were asked to consider changes to production methods that could result in savings for the ILC. In addition to the commissioned studies, two industrialisation workshops were hosted, in conjunction with IPAC 2010 and SRF2011, where industry and laboratory experience was solicited and shared.

One of the important changes since the RDR has been the exploration of different production models: at the time of the RDR, a single-vendor model was used for the cost estimate. Although a final production model has not yet been chosen, more-distributed models dividing production of key components among multiple vendors and the possible roles of laboratories in the production model have been explored. Figure 2.72 shows one such model. The dashed circle in the figure indicates technical coordination links between regional laboratories, designated as 'hub-laboratories' in the model. The arrows in the figure highlight the central role world-wide industry has as 'build-to-print' manufacturors.

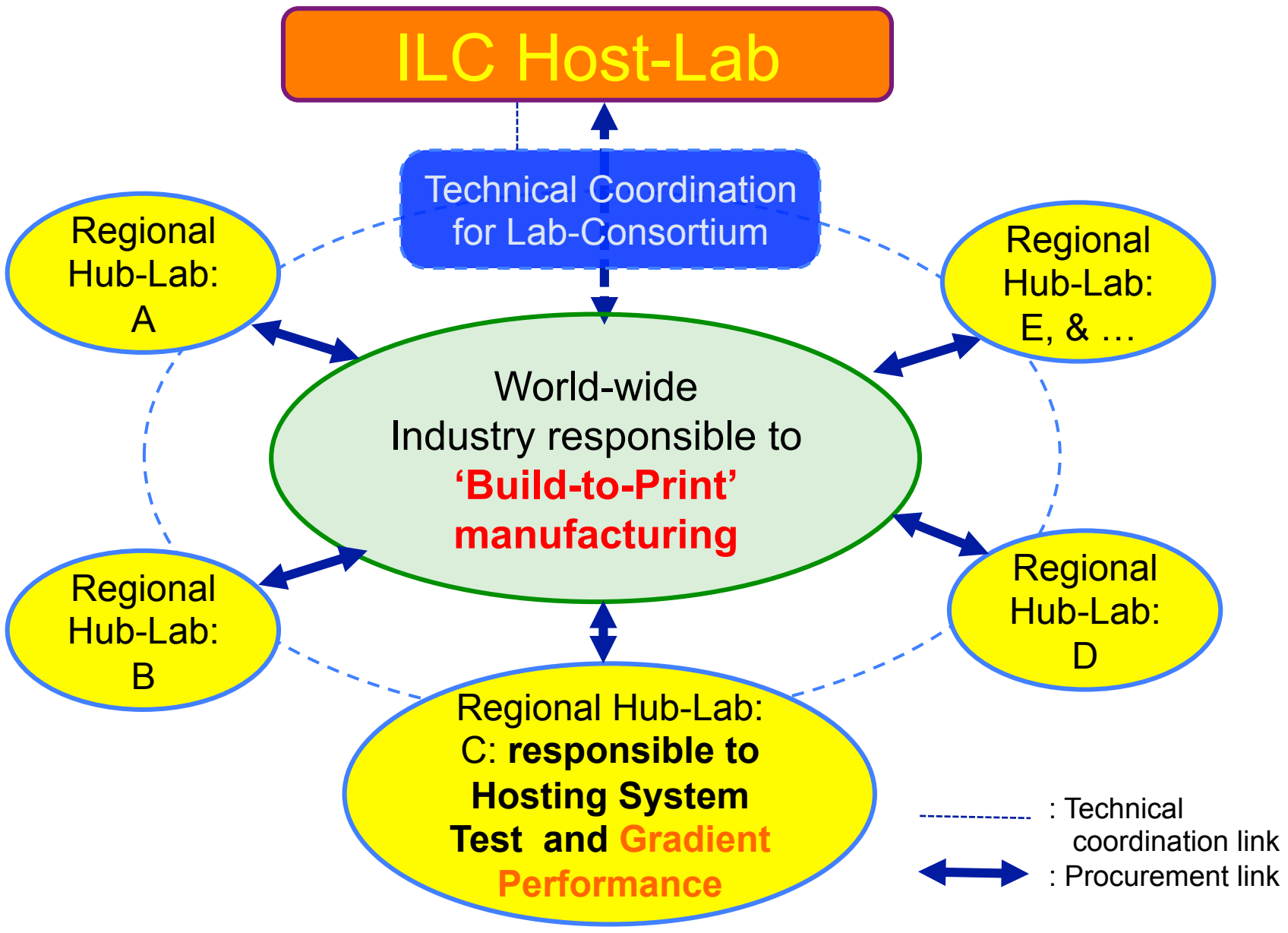

**Figure 2.72**
The industrialisation model.

### 2.9.1 R&D and Studies towards Cavity Industrialisation

The most direct source of information on the production of cavities in quantity will be the European XFEL production of 800 cavities, by two companies, over the course of 2 years. This is equivalent to about 5 % of the total required ILC production, and is the largest production of such cavities. This production and contractual experience will be invaluable.

To expand on this experience, and better understand the regional differences that might come into play during ILC production, three studies have been commissioned. In Europe, a study evaluating the needs and costs associated with the production of all 18,000 cavities, by a single vendor, over the course of 3 years, and the facility required for such a run, has been requested. Given the economies of scale, this study should present a best-case scenario for the cost per cavity. In the Americas, the study considered a 20 % production (3600 cavities) over the course of 6 years, and the plant associated with such numbers. Finally in Asia, a study considering full utilisation of the KEK pilot plant was requested, and resulted in a study considering 540 cavities per year. The last two scenarios are consistent with a distributed model where 5 or 6 vendors worldwide share the production.

In addition to the baseline process studies, sensitivity and bottleneck analyses were included in the studies, on items like the assumed first- and second-pass yields, learning curves, the number of e-beam welding machines, and an extra study looking at alternative processing techniques.

Common threads through the studies include the role of the laboratories in providing test facilities and assuming risk in situations where industry does not have the expertise. Even for the smaller quantity studies, the use of a single supplier for a sub component, where existing expertise exists, is identified as a viable strategy, even if the larger components are assembled in a distributed manner. A distributed model may be more consistent with existing infrastructure and size of companies, and more robust and sustainable with respect to economic upsets, but still use sole sources for subcomponents.

## 2.9.2 R&D and Studies towards Cryomodule Industrialisation

As with the cavities, in the short term the most direct and applicable source of information for the cryomodule sub-components and assembly will be from the European XFEL, where the 100 cryomodules represent about 5 % of the 1950 similar devices for the ILC. The current experience is further limited by the low number of cryomodules built and tested worldwide to date, and that the majority of this work has been done in a laboratory setting rather than industry. Furthermore, when looking at subassemblies, such as tuners, the amount of cold mass, operational testing for some of the designs is limited. However the data at hand has been shared with companies familiar with cryomodule work. The LHC-dipole cryostat assembly provides a valuable benchmark, which has been utilised in the ILC studies. Figure 2.73 shows one such analysis where the volume of production is shown versus the number of variants required at the LHC.

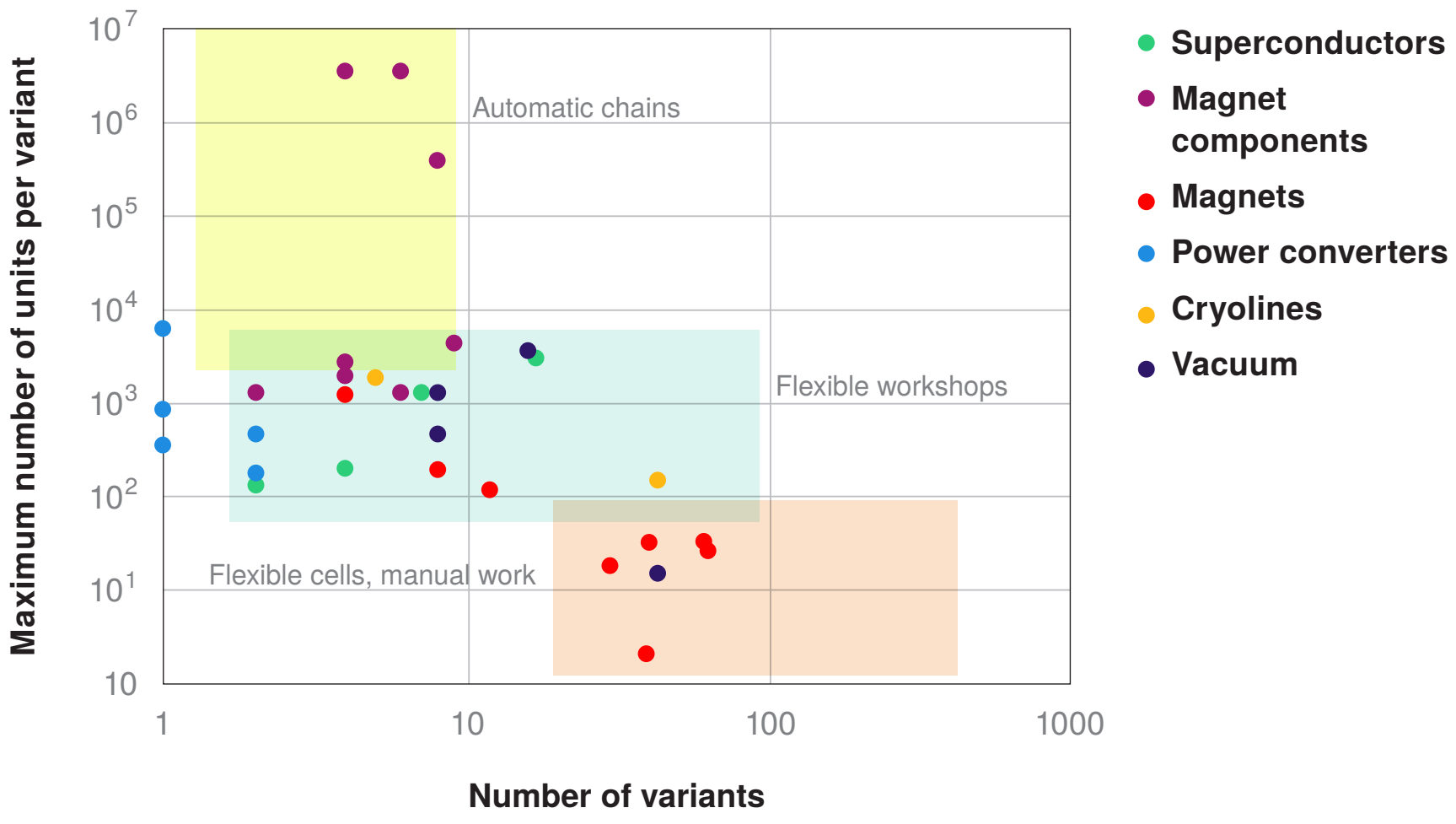

**Figure 2.73** Units produced and variants required for major LHC components [116].

The commissioned studies can be separated into two types: one evaluating the costs of subcomponents in the cryomodule, the other looking at the labour and plant requirements for cryomdule assembly. As with the cavity, for these assembly studies all cold test facilities are assumed to be at the laboratories; subcomponents are fully qualified when delivered to the assembly facility and only incoming inspection is required.

In Europe, the assembly study covered full production of 1950 cryomodules over 4 years, or a subset of 650 (30 %) over the same timeframe; in the US, the study assumed 450 cryomodules over 6 years, consistent with the cavity production study; and in Japan the study looked at 390 (20 %), 950 (50 %), or 1950 (100 %) cryomodules over 6 years.

For parts and subcomponents, the studies looked at the needs consistent with the same quantities and durations listed above; parts costs were evaluated using current experience modified with learning curves to the quantities needed based on industrial best judgement. An additional specific study, looking solely at the split quadrupole design, has been commissioned from Toshiba, who are an expert and experienced superconducting-magnet company and are currently assisting KEK in the re-assembly of a split quadrupole originally fabricated by Fermilab.

The guiding philosophy behind the cryomodule production model is that industry will be used whenever possible for component production. Final integration into cryomodules with involve both industry and the national labs. Cryomodule testing and shipping will take place in the hub laboratories, i.e. the regional laboratories that concentrate the regional efforts in component production in a particular region. LHC experience indicates that all high-technology components produced in industry will require engineering support from the national labs both during the initial set-up and pre-production phase as well as during the production cycle itself. It is likely that the ILC experience will be similar to the LHC in this regard.

Dressed cavities, couplers, tuners, quadrupoles, cryomodule vessels and other components will all be procured in volume from industry. Delivered cavities will be processed and cold-tested vertically for gradient performance at a hub laboratory. All cavity assemblies will be vertically cold tested. If necessary, cavity repair and a second-pass processing cycle will be performed at the hub lab. After cavity performance is certified, the final component integration into a complete cryomodule will be performed in or near the hub laboratory with industrial labour and laboratory oversight in a similar fashion to the LHC dipoles or the XFEL cryomodules. The final integration in or near a hub lab minimises the handling and transportation of the delicate cavity assemblies after processing and testing. The completed cryomodules will be cold tested at the hub laboratories. It is expected that all cryomodules in the pre-production phase will undergo cold testing; in full production mode only a fraction ($\approx 20\,\%$) will be cold tested before shipping. The overall scheme for cryomodule production is shown in Fig. 2.74.

**Figure 2.74**
Scheme of manufacture for cavities and cryomodules.

| Step hosted | Industry | Industry/Laboratory | Hub-laboratory | ILC Host-laboratory |
|---|---|---|---|---|
| Regional constraint | no | yes | yes | yes |
| Accelerator<br>- Integration, Commissioning | | | | Accelerator sys. Integ. |
| SCRF Cryomodule<br>- Performance Test | | | Cold, gradient test | As partly as hub-lab |
| Cryomodule/Cavity<br>- Assembly | | Coupler, tuner, cav-string/cryomoduleassmbly work | | As partly as hub-lab |
| Cryomodule component<br>- Manufacturing | V. vessel, cold-mass ... | | | |
| 9-cell Cavity<br>- Performance Test | | | Cold, gradient test | As partly as hub-lab |
| 9-cell Cavity<br>- Manufacturing | 9-cell, end-group assembly<br>Chem-process, He-Jacketing | | | |
| Sub-comp/material<br>- Production/Procurement | Nb, Ti, specific comp. ... | | Procurement | |

# Chapter 3
# Beam Test Facilities

## 3.1 Overview

Beam Test Facilities are required for critical technical demonstrations, including accelerating gradient, beam dynamics, and precision beam handling. The scope of test-facility activity needed to mitigate critical technical risks was assessed during the development of the Reference Design with the intention of modifying the design in accordance with the results obtained. Purpose-built test facilities were then either constructed by collaborative teams or provided through adaptation of existing facilities. With two exceptions, high-gradient beam operation (SCRF – main linac) and beam-size tuning (BDS), these test facilities have met or exceeded goals established for the Technical Design.

Primary beam test facility goals are:

- demonstration of ILC linac performance and evaluation of realistic cavity performance with beam acceleration;
- demonstration of a number of cavities operated in an accelerator showing repeatable performance and providing an estimate of reliability;
- studies of instabilities, such as electron cloud, and mitigation techniques;
- demonstrations of the generation and handling of low-emittance beams using precision optics and stabilisation tools.

It is especially important that each region deploy a full superconducting linac system, including cryomodules, beam generation and handling, and RF power source and distribution systems, to integrate the accelerator technology and gain sufficient experience in that region. The strategy for accomplishing this depends on infrastructure limitations and schedule constraints at each of the test facilities, due in part to institutional commitments to non-ILC related projects. As shown in Table 3.1, the aggregate effort is evenly distributed among the three regions; for two of the regions, a separate, specialised beam facility has been built in addition to an SCRF linac facility.

**Table 3.1** Table of Beam Test Facilities. Note that the Main Linac test facilities in USA and Japan will not be fully operational until after the Technical Design Phase.

| | Test Facility | Purpose | Host lab |
|---|---|---|---|
| 1 | TESLA Test Facility(TTF)/ Free-Electron Laser Hamburg (FLASH) | Main Linac | DESY, Germany |
| 2 | SCRF Test Facility (STF) (2014) | Main Linac | KEK, Japan |
| 3 | New Muon Lab (2013) | Main Linac | FNAL, USA |
| 4 | Accelerator Test Facility (ATF) | Damping Ring | KEK, Japan |
| 5 | Cornell Test Accelerator | Damping Ring | Cornell, USA |
| 6 | Beam-Delivery Test Facility (ATF2) | Beam-Delivery System | KEK, Japan |

Test facilities also serve to train scientific and engineering staff and regional industry. For example, ATF2, the final-focus test facility at KEK, was constructed through in-kind contributions and has been commissioned and operated by an international team of researchers. It is considered a possible model for a future ILC collaborative effort. A key component, present in both the test facility and the much larger-scale project, is good advance planning which takes into account the diverse and complementary skills and resources of large accelerator laboratories and smaller university groups alike.

Beam tests done during the Technical Design Phase at other facilities are also important. For example, beam-coupling corrections applied at the Australian Synchrotron [117], have demonstrated beam emittance of 1.2 pm-rad, well below that needed for ILC.

## 3.2 FLASH 9 mA experiment

### 3.2.1 Introduction

The TESLA Test Facility (TTF) was constructed by the TESLA collaboration [118] to demonstrate that a linear collider based on superconducting accelerating cavities would be feasible and cost competitive with one based on conventional copper structures. Technical feasibility of ILC-type superconducting accelerating cavities was demonstrated in 2000 when an 800 microsecond-long 8-mA beam was accelerated through a single cryomodule to 168 MeV. The TTF was renamed FLASH (Free electron LASer in Hamburg) and became a user facility operating as a soft X-ray free-electron laser in 2005. The FLASH linac is a 1.25 GeV linac based on Tesla-type technology and operates 5000 hours per year on average. The '9-mA' program was proposed by the GDE in 2008 with the goals of demonstrating reliable operation of the TTF/FLASH linac with ILC-like bunch-trains and to characterise the limits of operation of gradient and RF power. Typical beam properties for FEL user operation (charge, number of bunches, average beam power) are far lower than those required for the 9mA studies. For DESY, however, these studies have been important for integration and operational issues associated with running long bunch trains and high bunch charge, both for FLASH itself and for the European XFEL (see Section 2.5).

The ILC main linac will accelerate a 5.8 mA (upgradeable to 9 mA) 726 microsecond beam pulse to 250 GeV with 0.1 % rms energy stability at a pulse-repetition rate of 5 Hz. The beam energy must be stabilised over two timescales: long-term pulse-to-pulse stability over minutes and hours; and energy stability within a bunch train.The ILC main linac also requires precision control of high-gradient SRF in the presence of heavy beam loading. Since gradient performance has the greatest cost impact, each cavity in the ILC will be set to a stable voltage near its gradient quench-limit. To leverage the most cost-effective performance, low-level RF controls are used to push the cavities to achieve the maximum practical gradient, expected to be within 5 % of the nominal maximum. Effects such as beam loss, beam turn-on, beam-current fluctuation, Lorentz-force detuning and errors in power input coupling should be properly managed and disturbances minimised in order to maintain stable operation. In addition, studies were made of the required high-level RF-power overhead needed for reliable operation.

The above scientific programme attracted strong interest and participation of low-level RF and machine experts from DESY and also internationally, from Argonne, Fermilab and KEK. The studies were performed over a total of about 5 separate week-long runs between September 2008 and September 2012.

Table 3.2 compares beam parameters for the 9 mA Studies with the ILC Main Linac design parameters, and those of the European XFEL.

**Table 3.2** Parameters for the 9 mA Experiment in context

| | units | TDR Baseline | TDR Upgrade | European XFEL | FLASH 9 mA Expt. |
|---|---|---|---|---|---|
| Number of bunches per pulse | | 1312 | 2625 | 3250 | 2400 |
| Bunch repetition rate | MHz | 1.8 | 2.73 | 5 | 3 |
| Beam pulse length | μs | 727 | 960 | 650 | 800 |
| Bunch Charge | nC | 1.9 | 3 | 1 | 3 |
| Beam current | mA | 5.8 | 9 | 5 | 9 |

### 3.2.2 FLASH Main Linac

The main elements of the FLASH linac are shown in Fig. 3.1 before and after the energy upgrade in 2009/2010, when the number of accelerating cavities was increased from 48 to 56, increasing the maximum operating energy to 1.25 GeV. The injector comprises a 5 MeV laser-driven photo-cathode RF gun and a two-stage bunch compressor. The RF system of the first-stage compressor has eight cavities in a single cryomodule (module ACC1), while that of the second bunch compressor has 16 cavities in two cryomodules (modules ACC2 and ACC3). The main linac comprises modules ACC4 onwards. Prior to 2010, the main linac comprised modules ACC4, ACC5, ACC6 (24 cavities), all fed from a single klystron and regulated using vector-sum control.

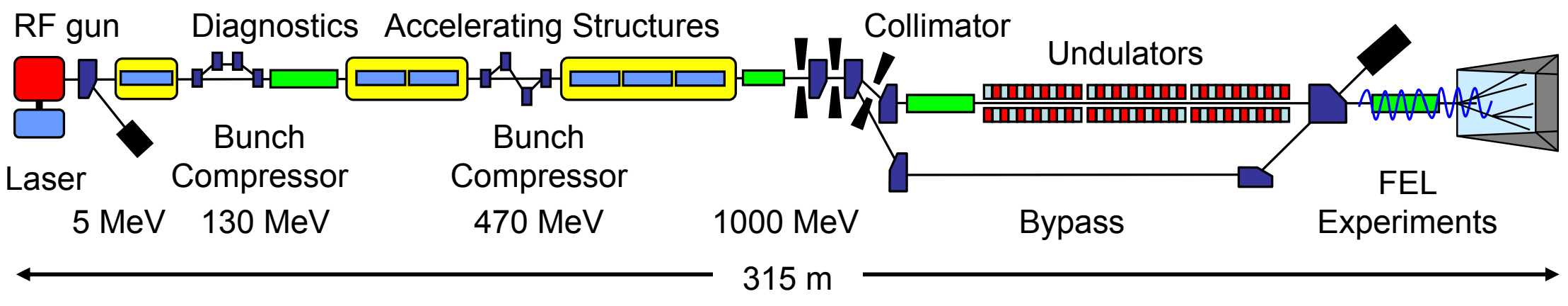

**(a)** 2008-2009 prior to energy upgrade

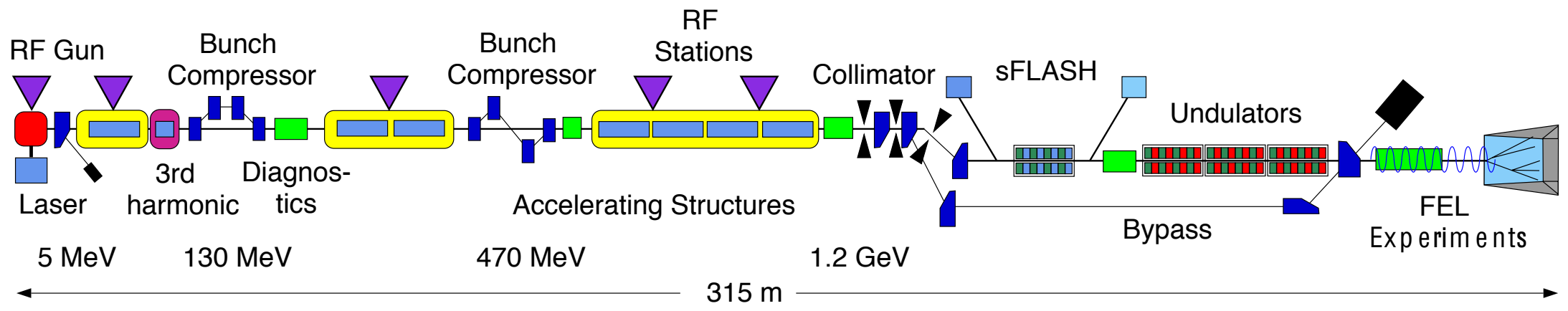

**(b)** 2010-2012, following upgrade to 1.25 GeV

**Figure 3.1.** Layout of the FLASH linac

The FLASH high-level RF systems are the basis of the HLRF system design for both the ILC and the European XFEL. The RF power is distributed from individual multi-beam klystrons of either 5 or 10 MW through a series of power-dividing elements to either one, two, or three groups of eight cavities. The relative power to each cryomodule (group of eight cavities) can be adjusted remotely. Pairs of adjacent cavities receive some fraction of the total RF power, with the power ratios having been set during fabrication based on measured quench limits of the cavities. These power-dividing ratios are not adjustable. Prior to 2010, the RF unit comprising the 24 cavities in ACC4,5,6 was of most interest for the 9 mA studies. In 2010, an additional cryomodule (ACC7) was added and the HLRF systems were reconfigured into two groups of 16 cavities each fed from it's own klystron (ACC4+5 and ACC6+7). Subsequent studies were focused on operation of ACC6 + 7. At the end of the linac, the beam is directed through a series of undulators for SASE FEL operation, or alternatively to a bypass line and then to the beam dump.

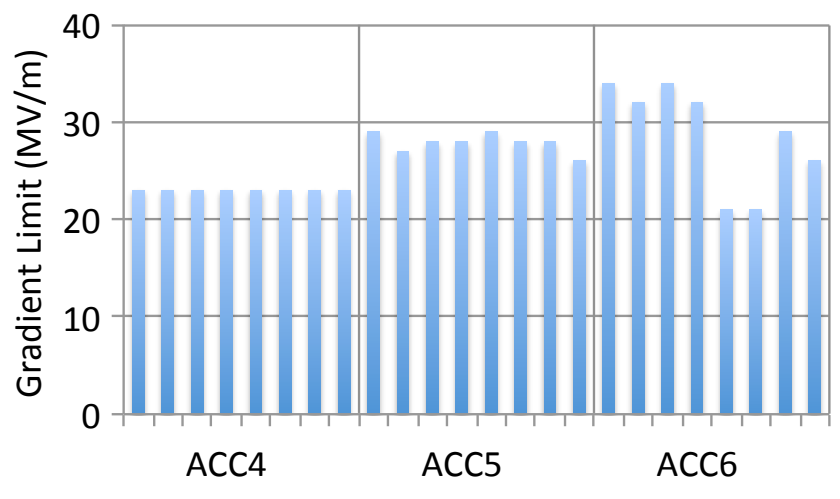

**(a)** Modules ACC4,5,6 (2008-2009)

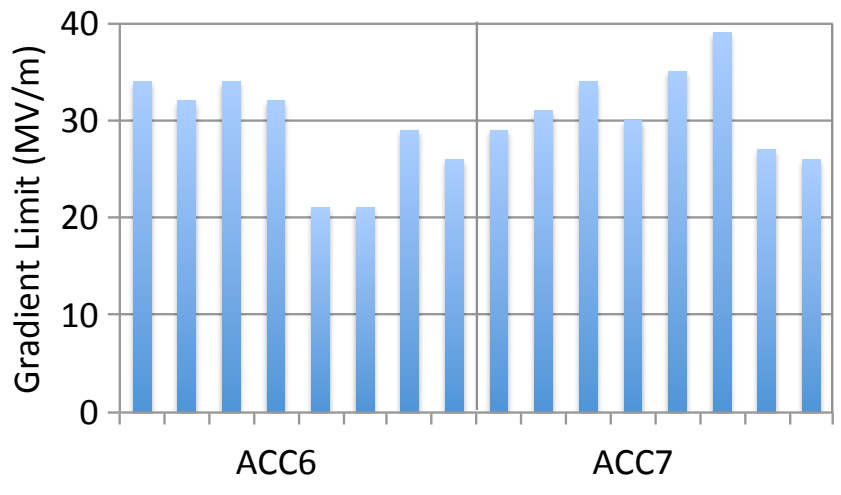

**(b)** Modules ACC6,7 (2010-2012)

**Figure 3.2** Measured cavity-gradient limits for the cavities in the FLASH Accelerator cryomodules for the 9 mA studies

The strong similarity of the ILC HLRF design to that at FLASH is a very important motivation for the 9 mA experiment. The most significant functional difference between the FLASH and ILC power distribution systems is the absence of remote adjustability of the power-divider ratios at FLASH, the impact of which will be discussed in Section 3.2.8.

Maximum operating gradients for the FLASH cavities for the two linac configurations are shown in Fig. 3.2.

### 3.2.3 Low-Level RF Control

DESY has been a pioneer in the field of digital LLRF control for pulsed superconducting linacs. The LLRF systems currently implemented at FLASH use the third generation of digital LLRF controllers developed by DESY ("Simcon-DSP"), and will shortly be upgraded to a fourth-generation system that has been developed with the European XFEL in mind. The main elements of the FLASH LLRF system are shown in Fig. 3.3 [119].

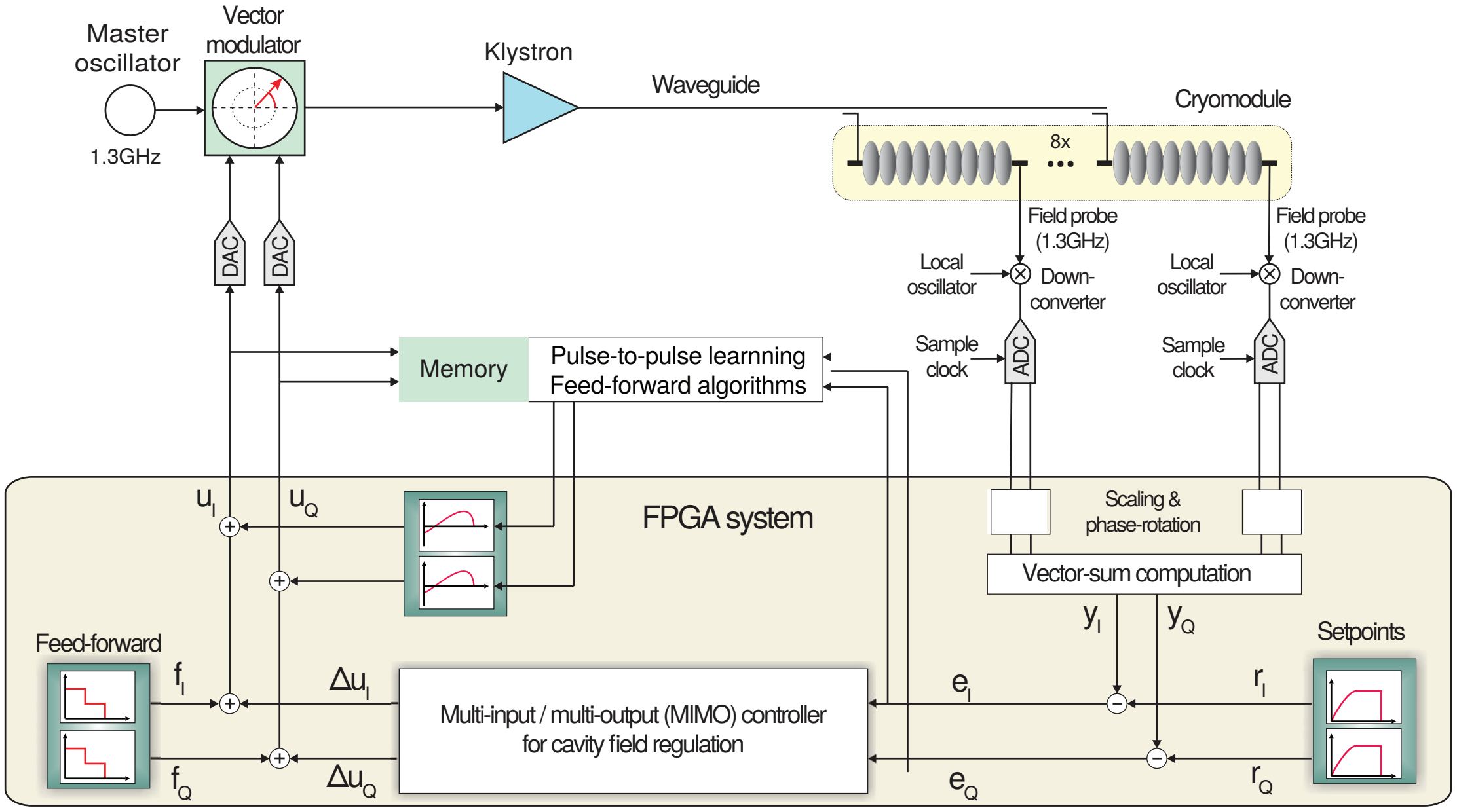

**Figure 3.3.** Block diagram of the FLASH digital LLRF control system [119]

The LLRF controller output to the klystron comprises two components, a feed-forward term to generate the cavity-field profile and handle the repetitive pulse-to- pulse artefacts while a dynamical intra-pulse feedback controller compensates for unpredictable pulse-to-pulse variations. A learning feedforward algorithm iteratively updates the shape of the feed-forward waveform in order to minimise the repetitive pulse-to-pulse errors and reduce the effort demanded of the feedback regulator.

Over the course of the 9 mA studies program, DESY has implemented many performance and functionality improvements on the LLRF systems: hardware and firmware upgrades and incremental development and refinement of high-level applications. Improvements in pulse-to-pulse jitter, intra-train regulation, repeatability, robustness, and long-term drift have all significantly improved over the course of the studies. Beam-based feedback has been incorporated into the RF systems of the two bunch compressors in order stabilise bunch-arrival time and bunch compression; beam-loading compensation is based on measured bunch charge. There have also been substantial benefits from the increasing degree of automation. FLASH FEL operations has also seen significant benefit from the improved performance of the LLRF systems.

### 3.2.4 FLASH Control System

The FLASH control system is implemented in DOOCS [120], which provides a powerful environment for control and monitoring of technical equipment and a framework for implementing integrated high-level applications. Hooks into the control system and data-acquisition system are provided for Matlab [121] and Octave [122], which are extensively used for developing high-level integrated applications.

Extensive use has also been made of the FLASH Data Acquisition System (DAQ), which provides access to several thousand signals that are pulse-to-pulse synchronous, including 1 MHz sample-synchronous and bunch-synchronous waveforms from the LLRF system and beam diagnostics [123]. The DAQ also serves to provide the data for pulse-to-pulse synchronous high-level applications, including computation of final beam energy from beam orbits and LLRF iterative learning feed-forward vector-sum control and piezo-tuner waveform optimisation for compensation of Lorentz-force detuning.

### 3.2.5 High-power long-pulse studies

The principal high-power long-pulse study goals were achieved during a two-week period in September 2009, having operated for several hours at 9 mA and pulse lengths of 500-600 µs (1500-1800 bunches). There were many additional hours of operation with 800 µs pulses (2400 bunches) and beam currents up to 6 mA. Figure 3.4 shows example energy profiles with the long bunch-trains at beam currents of 0.3 mA and 7.5 mA.

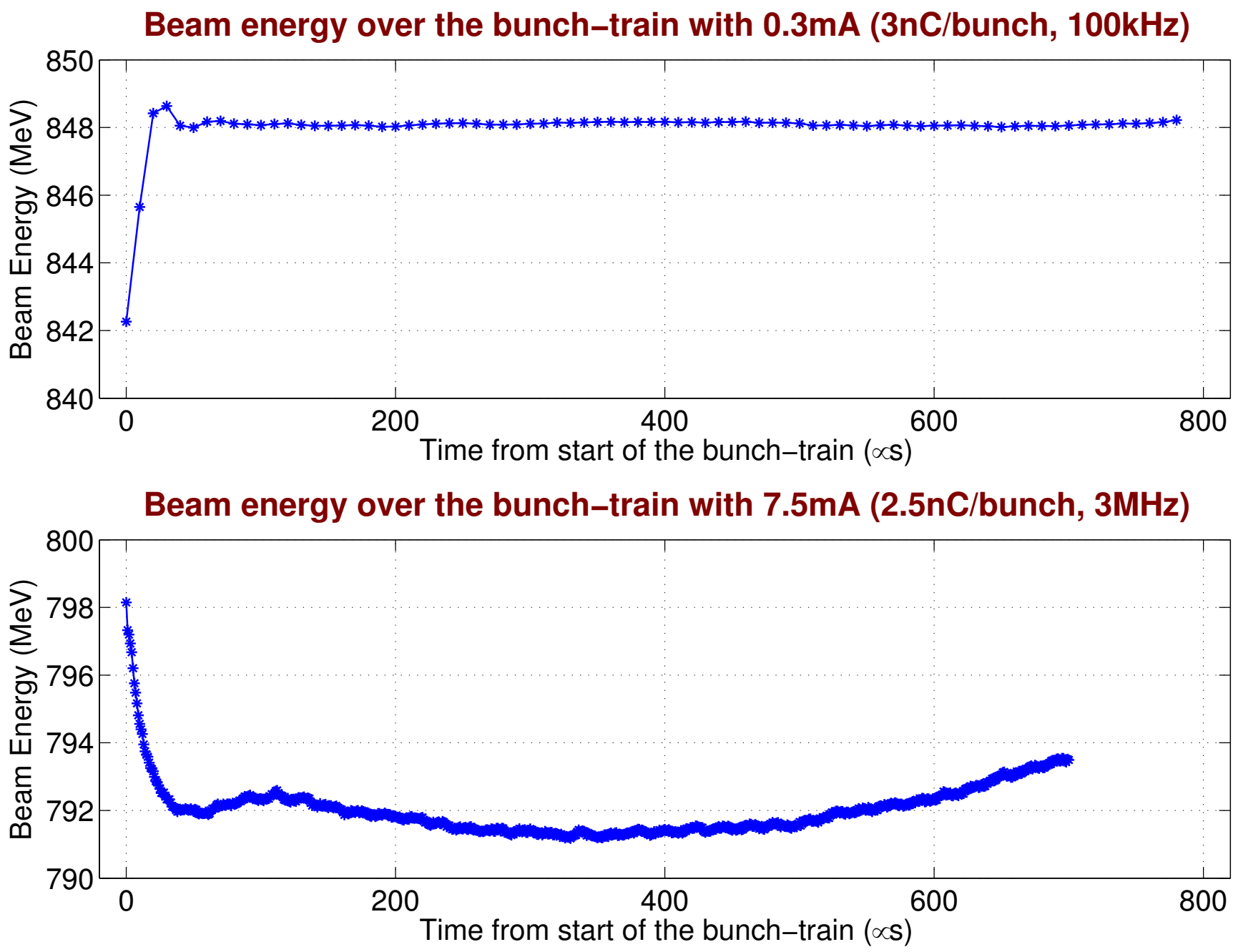

**Figure 3.4** Linac energy profiles from 2009 study at 0.3 mA (top) and at 7.5 mA (bottom)

Operation at these high beam currents and long bunch trains required increasing the total charge per pulse by two orders of magnitude over typical FEL operation parameters. A significant side effect of this was the increased sensitive to changes in machine parameters that occur over periods of hundreds of microseconds, such as pulse-heating effects in RF components and drive-laser-beam transport that resulted in energy variations or orbit changes over the bunch train. An additional important consequence of the high total charge per pulse was that the per-bunch losses had to be very low in order to avoid reaching thresholds for integrated beam-loss per pulse.

A history of the number of bunches over the week of studies leading up to the 9 mA and full-length bunch trains operation during the September 2009 studies is shown in Fig. 3.5. Of particular note are the rapid recovery following a tunnel access on 19th September (less than one hour), followed by the 15-hr stable run at 3 mA with the full 800 μs bunch-train on 19 September before reconfiguring the laser for 3 MHz operation and ramp-up to 9 mA on 20th September.

**Figure 3.5**
Number of bunches and bunch charge over 7 days of studies.

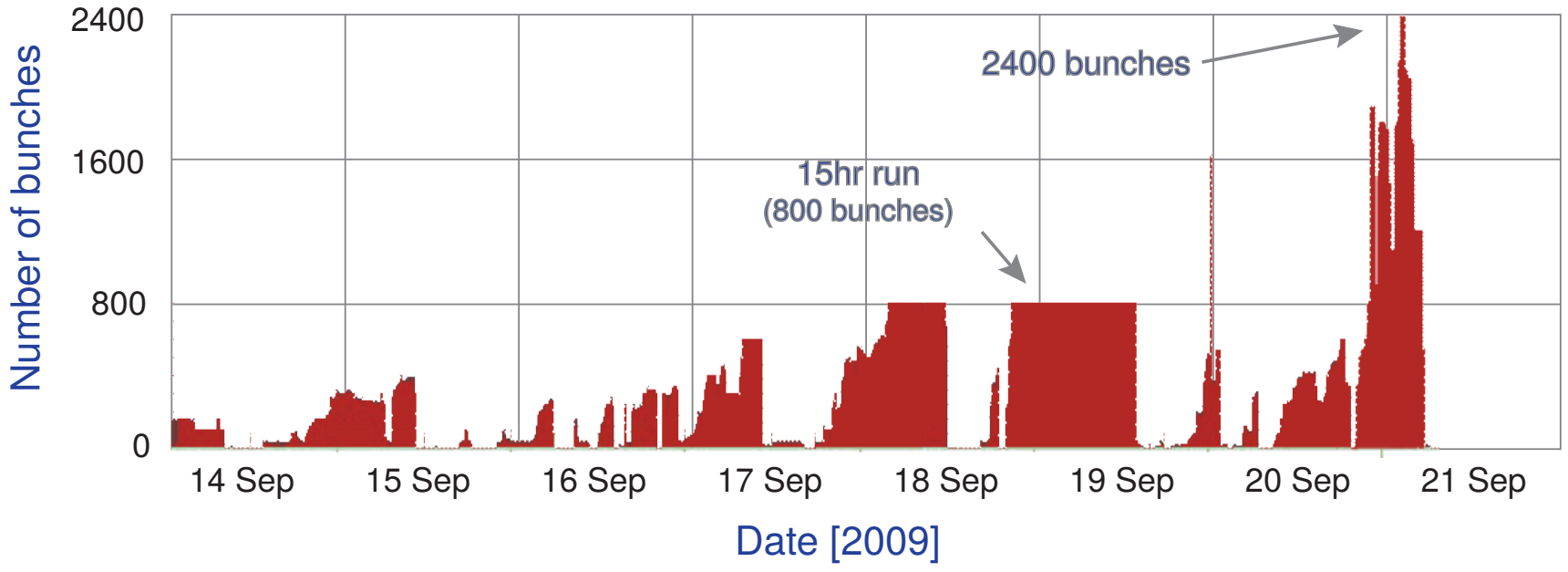

## 3.2.6 RF-power overhead studies

The ILC baseline design assumes that the main linac klystrons can be operated reliably within 10 % of their 10 MW maximum output, with some klystrons operating within 7 % of the maximum power. This has two consequences for RF control: there is limited power overhead available for vector-sum regulation; and the LLRF controller must contend with a power-dependent gain in the klystron as it approaches a saturation limit at the maximum output power. Figure 3.6 shows representative curves of the output power and gain at different anode voltages for the Toshiba 10 MW multi-beam klystron.

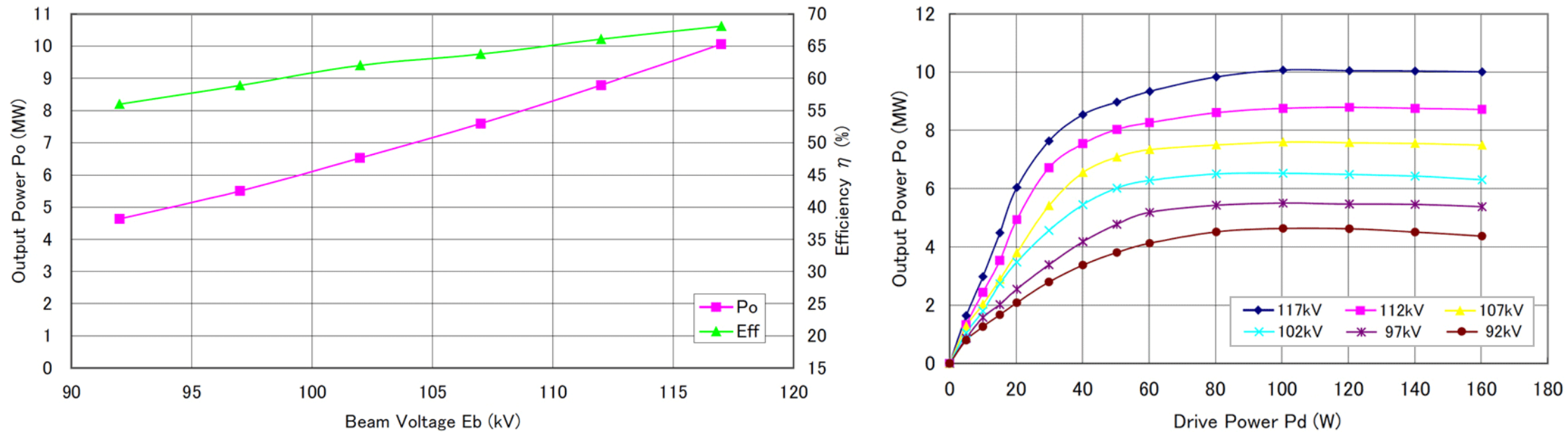

(a) (b)

**Figure 3.6.** Representative RF power and gain curves for Toshiba multibeam klystron

**Figure 3.7**
Klystron gain curve and working point at 86.5 kV anode voltage

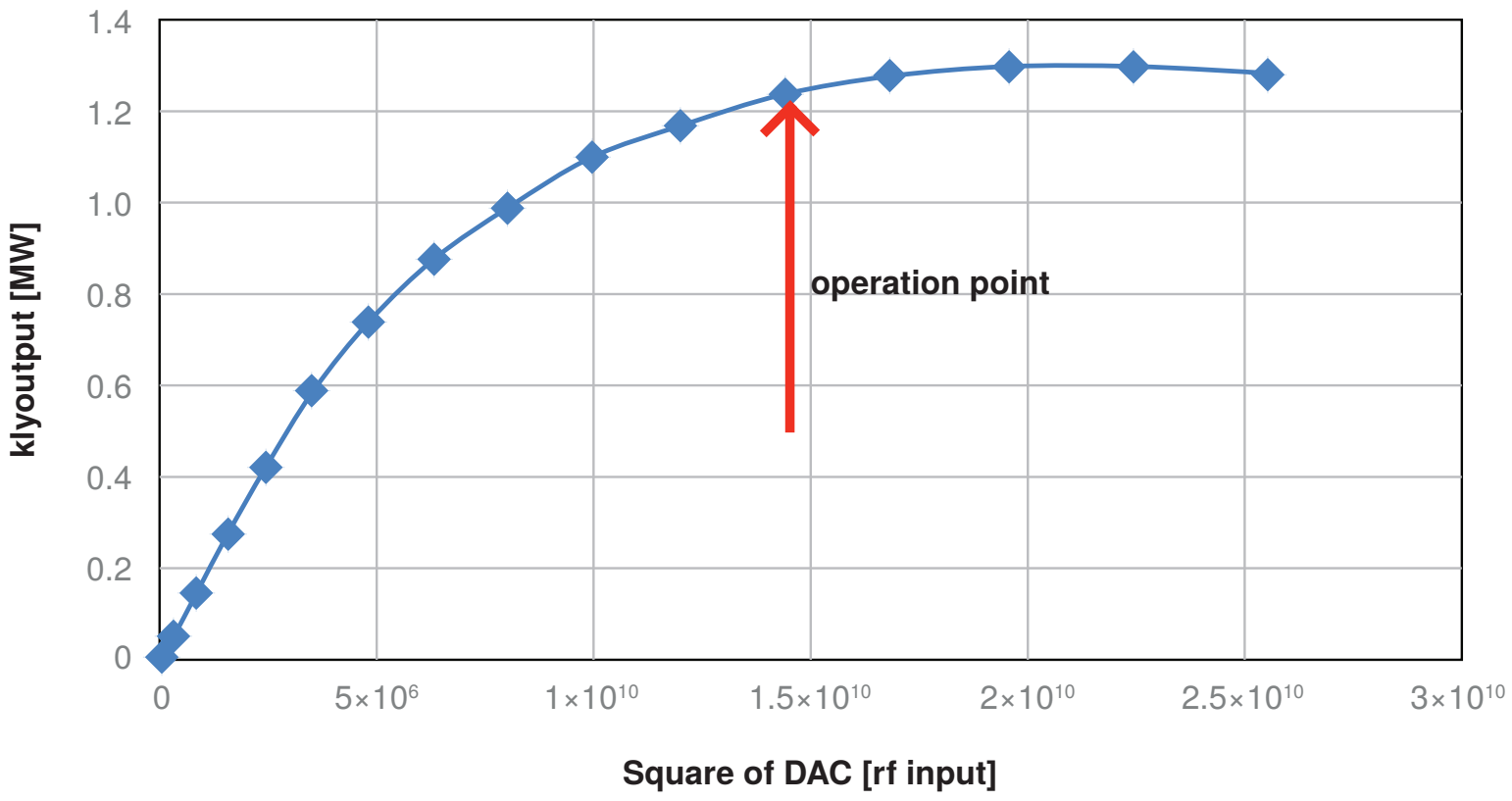

First tests of beam operation with klystrons running within a few percent of saturation were performed in February 2012. The studies was performed on the RF unit comprising ACC6/ACC7 with a beam current of 4.5 mA, which required approximately 1.2 MW of forward power from the klystron, which is nominally rated at 10 MW. The saturation point of the klystron was reduced to just above 1.2 MW by dropping the anode voltage. Figure 3.7 shows the measured klystron output power as a function of drive at an anode voltage of 96.5 kV. It shows a factor five change in gain from the linear region to the operation point for the test and also that the output power starts to roll over if the klystron is driven too hard.

In order to observe the effects of klystron saturation during the beam pulse, a notch was introduced into the vector sum setpoint table. An example of the tracking response to the notch is shown in Fig. 3.8.

The effect of klystron saturation can be seen by comparing the speed of response on the vector sum at the start and end of the notch. The lower trace shows the negative- and positive-going blips in klystron forward power at the start and end of the notch as the LLRF controller attempts to make the vector sum follow. On the step up, the klystron output power demanded by the regulator is clipped by saturation, slowing the recovery time compared with the initial step down.

**Figure 3.8**
Klystron forward power and vector-sum response to a notch in the vector-sum setpoint

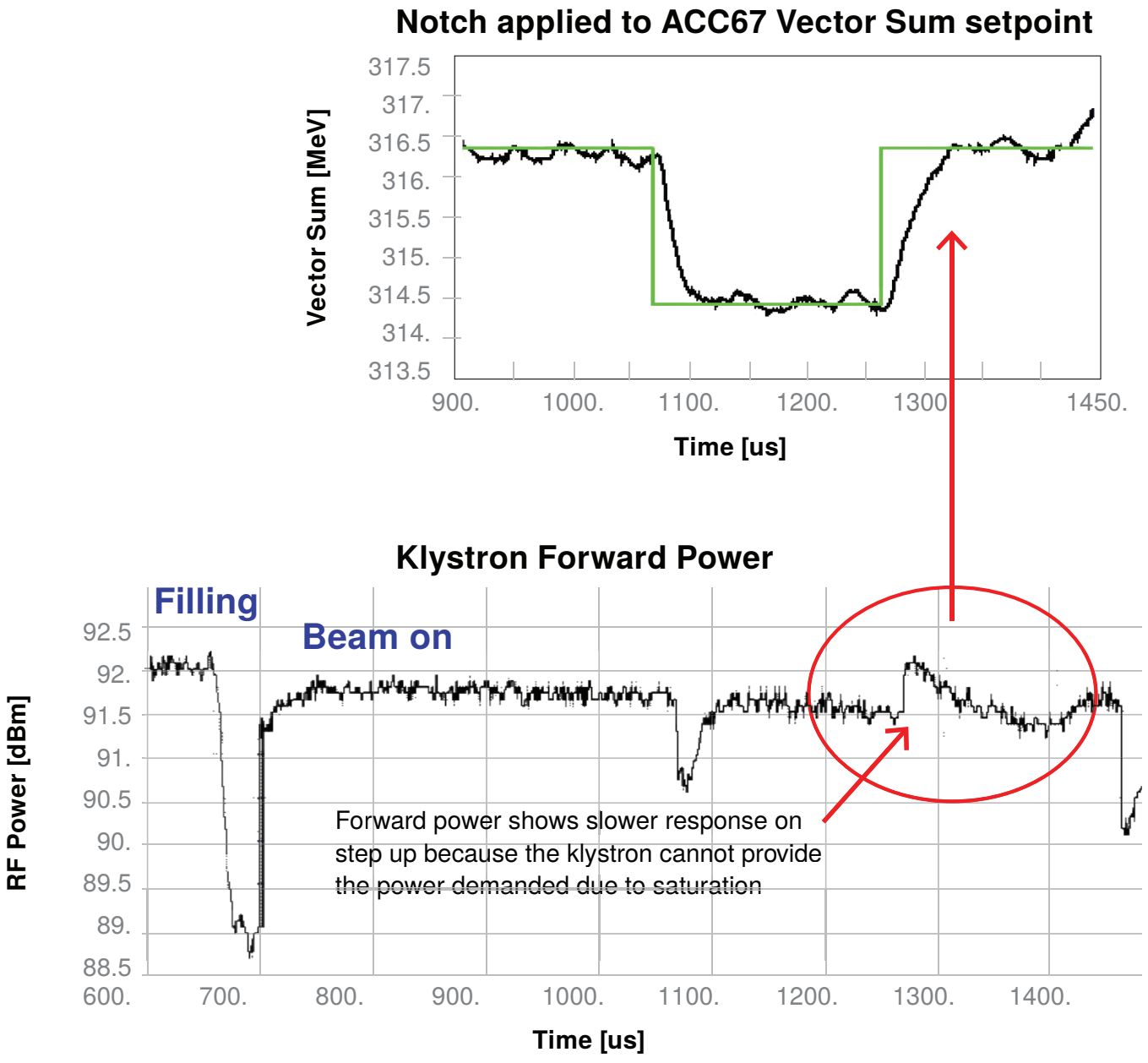

First beam tests with a klystron linearisation function were successfully performed in September 2012. This was implemented as a look-up table in the LLRF-controller output signal chain that has the inverse characteristic to the measured klystron saturation curves. Although the nonlinearity and saturation is a function of output power amplitude, linearisation tables must be applied to both amplitude and phase. Results from these first linearisation tests are shown in Fig. 3.9, where the input-output characteristics of the pre-amplifer/klystron drive chain are shown with and without linearisation.

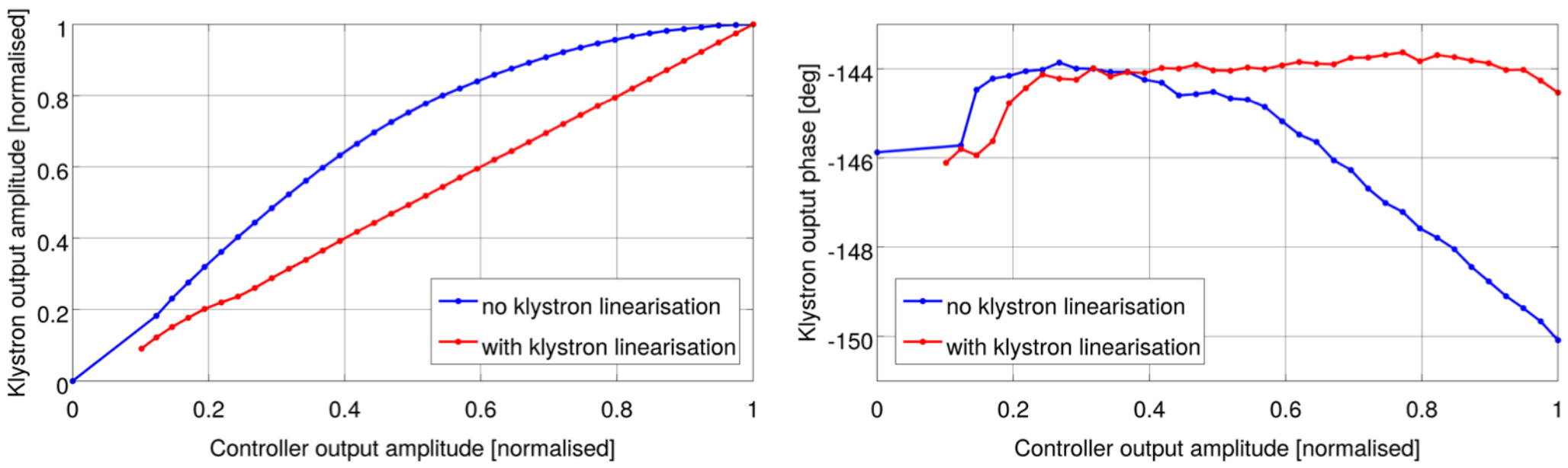

(a) Amplitude response

(b) Phase response

**Figure 3.9.** Klystron output as a function of LLRF Controller output with and without linearisation

### 3.2.7 Compensation Studies for Lorentz-force detuning

Lorentz forces cause the resonant frequency of the cavity to change linearly as a function of time over the RF pulse. At 35 MV/m, the change in detuning over the beam-on period can be 600 Hz or more. This Lorentz-force detuning must be compensated to the level of a few hertz in order to avoid an RF power penalty for driving a detuned cavity and to avoid cavity gradient deviations that otherwise could not be removed by the vector-sum controller.

Methods for compensation of Lorentz-force detuning at FLASH have been widely reported. A pulsed excitation is applied to the piezo tuners several milliseconds before the RF pulse in order to excite mechanical resonances of the cavity structures that apply forces to the cavity during the RF pulse that counteract the Lorentz forces. The excitation is in the form of a pulsed sinusoid where the amplitude, frequency, number of periods, start-time, and DC offset are adjusted in order to null the measured detuning profile during the beam-on period.

Observed mechanical modes are typically in the range of 200-400 Hz with damping times of tens of milliseconds. Figure 3.10 shows time-domain responses over 80 ms for the eight cavities in module ACC7.

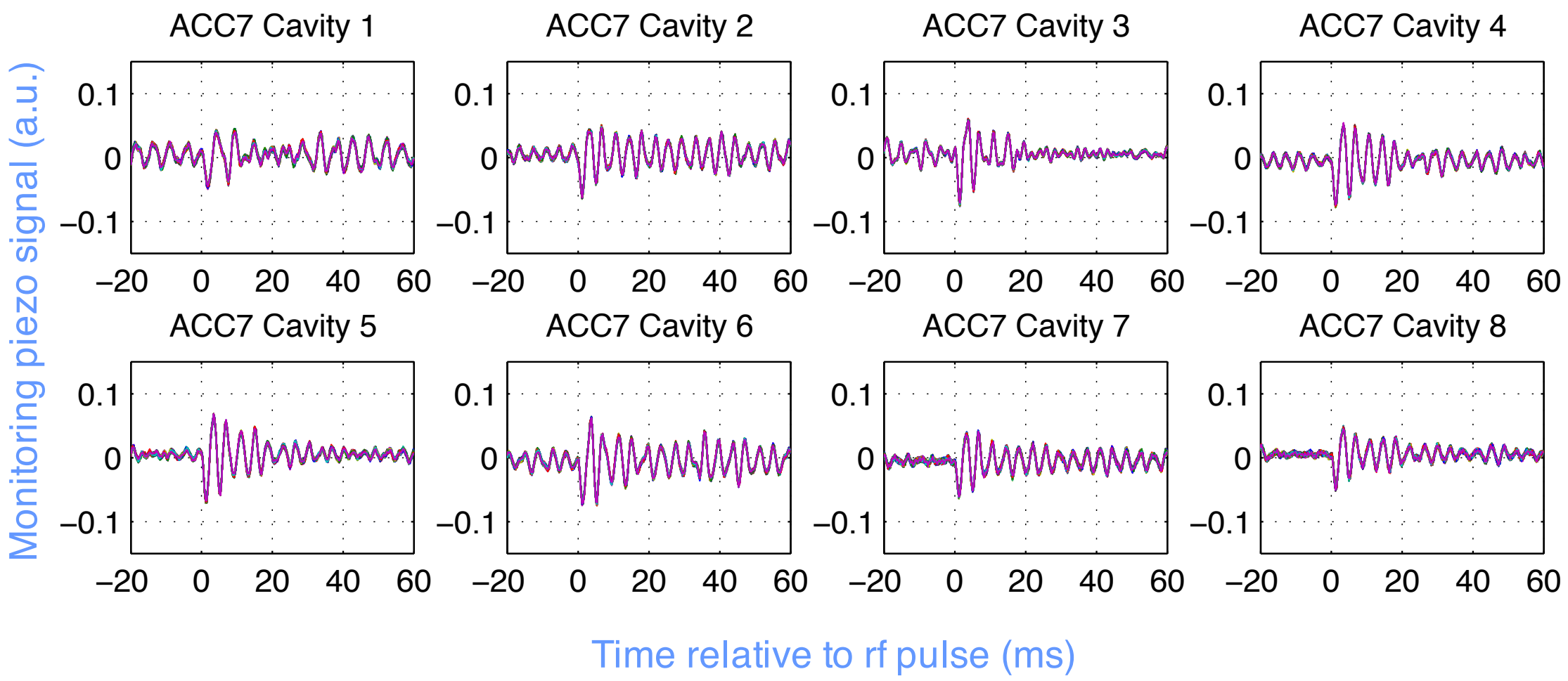

**Figure 3.10.** 'Monitoring' piezo signals from $-20\,\mathrm{ms}$ to $+60\,\mathrm{ms}$ relative to RF pulse showing cavity-to-cavity differences in damping times of the mechanical resonances.

The optimisation algorithm for the piezo tuner adjusted the amplitude, DC offset and start-time of the excitation in order to minimise the measured average ("static") detuning, change in detuning ("dynamic" detuning), and the curvature of the detuning profile over the pulse. Examples of detuning profiles showing the metrics and the result of the optimisation are illustrated in Fig. 3.11. Since each cavity operates at a different gradient, the excitation parameters must be optimised for each piezo tuner. Since the optimisation uses only the time period of the RF pulse, there is no cancellation of the actual mechanical motion, which typically takes several hundred milliseconds to decay (as was shown previously in Fig. 3.10).

The detuning profile used for piezo-tuner optimisation must be computed from the cavity probe and forward power waveforms. The computation must also take into account the beam-loading contribution to the total forward power.

Calibration of the forward power signal is complicated by the cross-contamination from the reflected power signal, and is performed by comparing forward and reflected power waveforms from knowledge that they must be equal at the start of the RF pulse, while the forward power must be zero after the RF pulse. For the studies, calibration was cross-checked by comparing the computed detuning with end-of-pulse measurements from scanning the RF-pulse length. The intra-pulse computation of detuning was also extended to include the beam-loading term. Figure 3.12 shows the computed

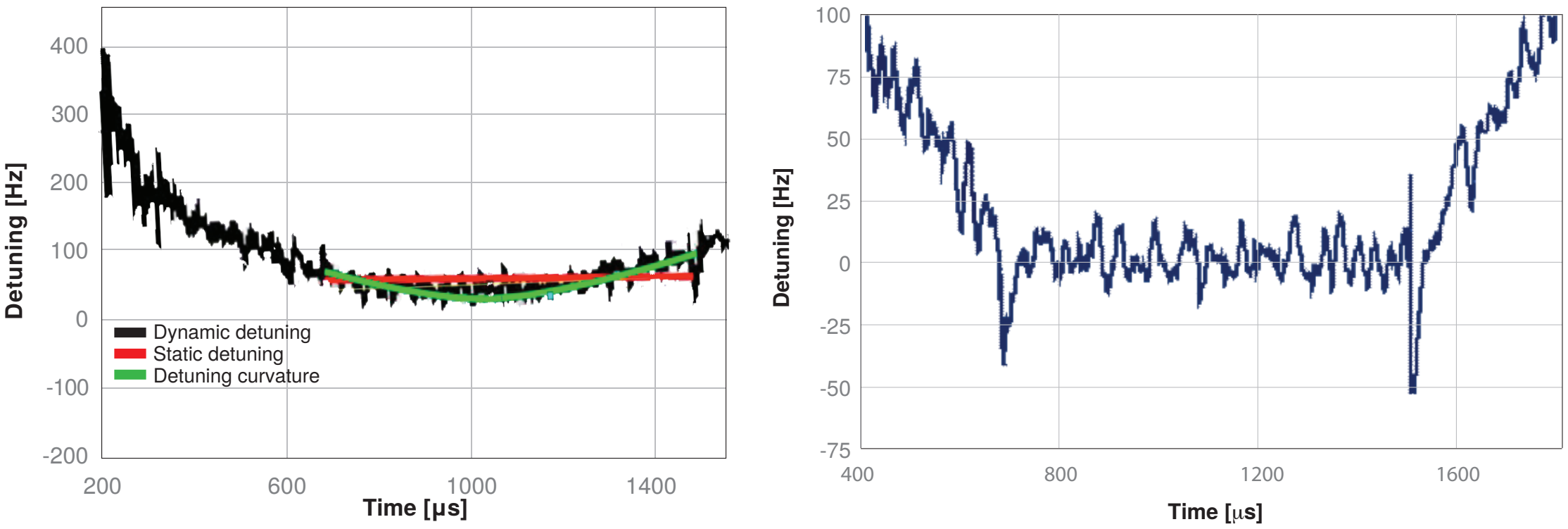

**(a)** Measures of detuning

**(b)** Residual detuning profile with compensation (0.3 Hz dynamic, -1.2 Hz static)

**Figure 3.11.** Definition of terms for detuning compensation algorithm and residual detuning for ACC6 Cavity #1

detuning with and without correction for the forward power from beam loading. Grecki *et al.* [124] contains further discussion of calibration and beam-loading issues with the detuning computations.

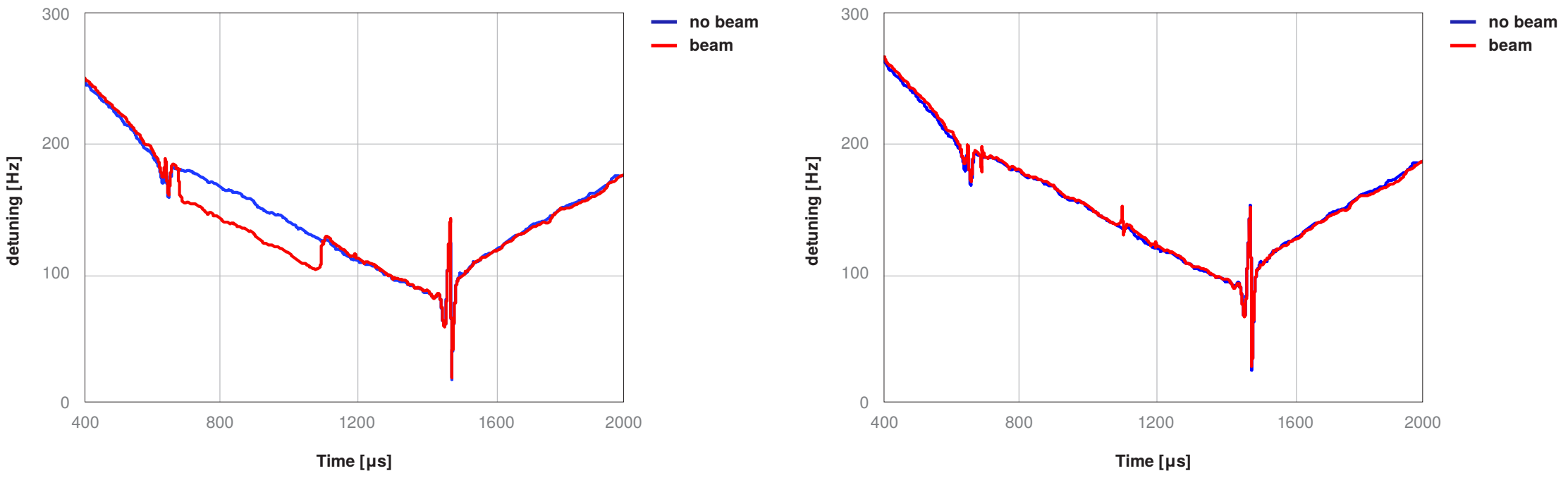

**(a)** No correction for beam loading

**(b)** With correction for beam loading

**Figure 3.12.** Comparison of computed detuning (red) with measurements from shortened pulses (blue) with and without beam loading

To date, piezo tuner studies have concentrated on minimising the measured detuning errors. In practice, however, the piezo tuners would also be used to make fine adjustments to the cavity-gradient profiles as part of the gradient-flattening algorithm. The principle benefit is that using the piezo tuners for fine adjustments avoids having to make frequent changes to the coupler motor positions.

### 3.2.8 Gradient Studies: Beam Operation close to Quench

To minimise the length of the tunnel, the main-linac design assumes that all cavities will operate close to their quench limits. Cavity strings are qualified at a nominal average of 33 MV/m with a spread from 26 MV/m to 38 MV/m, and will operate at 31.5 MV/m average. All cavities must operate within 1.5 MV/m of their quench limits. Any pulse-to-pulse gradient jitter or gradient changes along the bunch-train must be kept to a minimum in order to avoid squeezing this gradient margin even further. The primary sources of gradient changes along the bunch train are detuning (discussed in the previous section) and gradient tilts induced by beam loading.

Figure 3.13 shows examples of gradient tilts induced by beam loading from the September 2009 studies, which show that the tilts are proportional to beam current. These gradient tilts can be corrected if the individual cavity parameters (loaded Q and/or cavity forward power) are tailored to the respective cavity operating gradients and the nominal beam current. To achieve this, such that the voltage in a cavity remains constant at a specified gradient in the presence of beam loading, the forward power and loaded Q must simultaneously satisfy two constraints:

- the cavity must have reached the specified gradient at the end of the cavity fill time;
- the RF power fed to the cavity during the beam-on period must exactly balance the power being removed from the cavity by the beam.

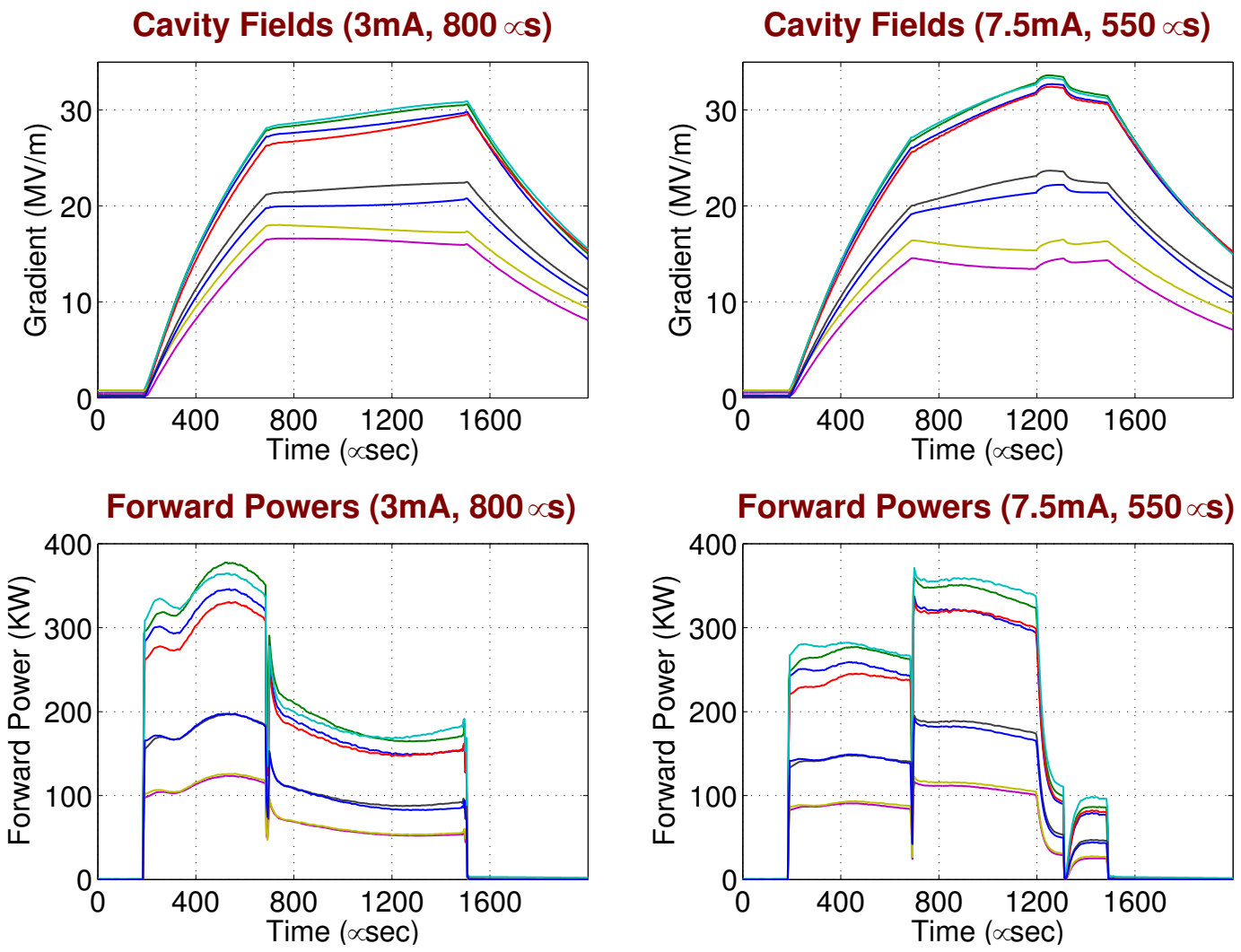

**Figure 3.13**
Gradient tilts from beam loading at 3 mA and 7.5 mA (Sept 2009)

Where there are multiple cavities fed from a single klystron, the above two constraints must be satisfied simultaneously for all cavities with the added constraint of a common fill-time. Provided the individual cavity loaded Qs and forward power ratios can be independent adjusted, a set of loaded Qs and power ratios can always be found that achieves flat gradients on all cavities simultaneously. However, if the power ratios are not adjustable, as is the case at FLASH, the only adjustments are the loaded Qs, the total forward power, and the common fill time. Solutions can still be found, but only over a restricted range of beam currents and operating gradients. The greater the spread in cavity operating gradients, the more restrictive is the range of beam currents and operating gradients over which solutions can be found. This is illustrated in Fig. 3.14, which shows the spread of loaded Q expanding exponentially with beam current to the point where there are no longer any valid solutions. The two outlier curves correspond to the two lowest-gradient cavities. Without those cavities, the spread of required Qls would be significantly less and solutions could be found over a wider range of currents.

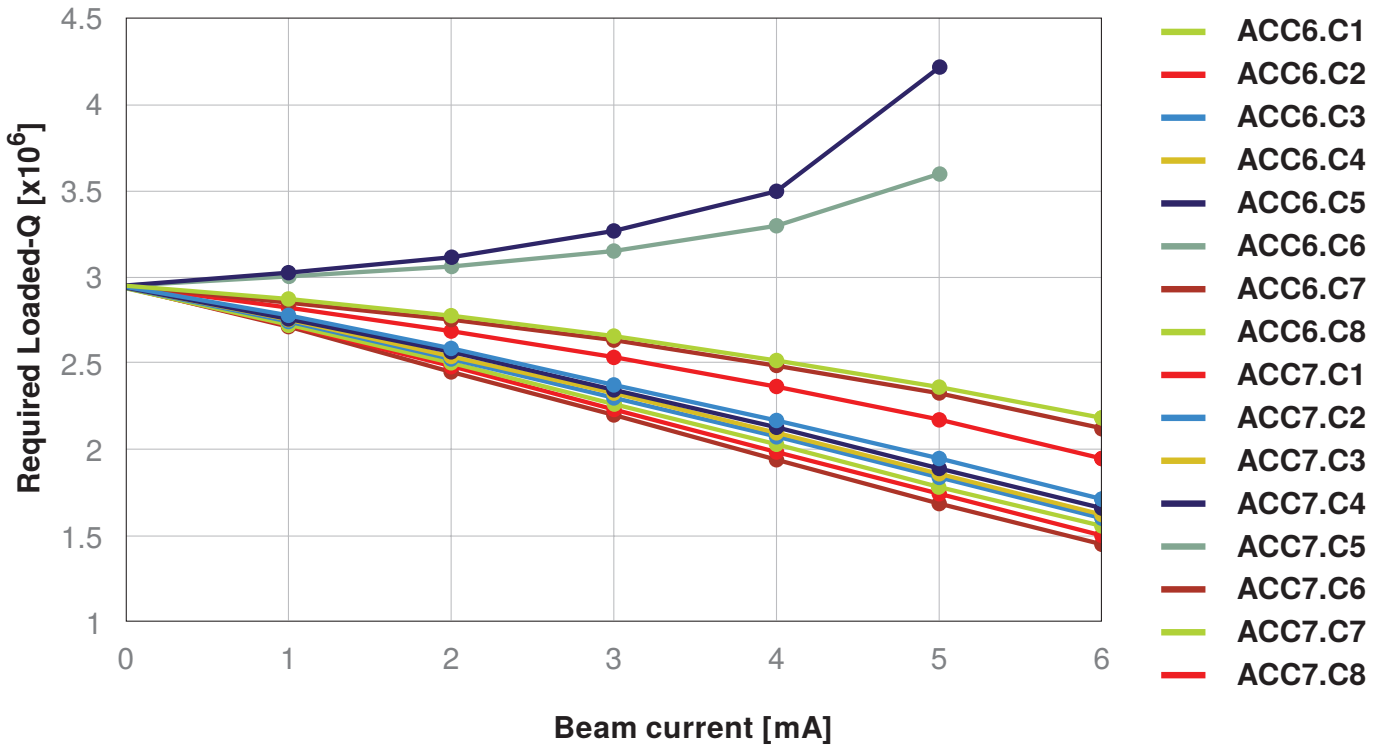

**Figure 3.14**
Calculated sets of loaded Qs that achieve flat gradients as a function of beam current, showning the divergence in the maximum-to-minimum loaded Qs as the beam current or the spread in gradients increases

The method of establishing flat gradients using both adjustable power ratios and loaded Qs has been referred to as 'Pk/Ql control.' For the FLASH studies, where only the loaded Qs are adjustable, the term 'Pseudo-Pk/Ql control' has been used. This method has been successfully applied for a range of operating conditions and beam currents up to 5 mA. Because of the limitation the range of currents and gradients where there were solutions, most of the studies were performed with the two lowest gradient cavities were completely detuned in order effectively to remove them from the system. An automated iterative optimisation algorithm is described in Section 3.2.12. The approach to finding the optimum set of loaded Qs that produced flat gradients was to start with values that were determined analytically and then manually optimise the values iteratively using a model-based cavity simulator. The quality of the solution was evaluated by sweeping the beam current above and below the nominal value and observing the tilts. For each cavity, there will be one particular beam current at which the gradient is flat, as shown in Fig. 3.15a.

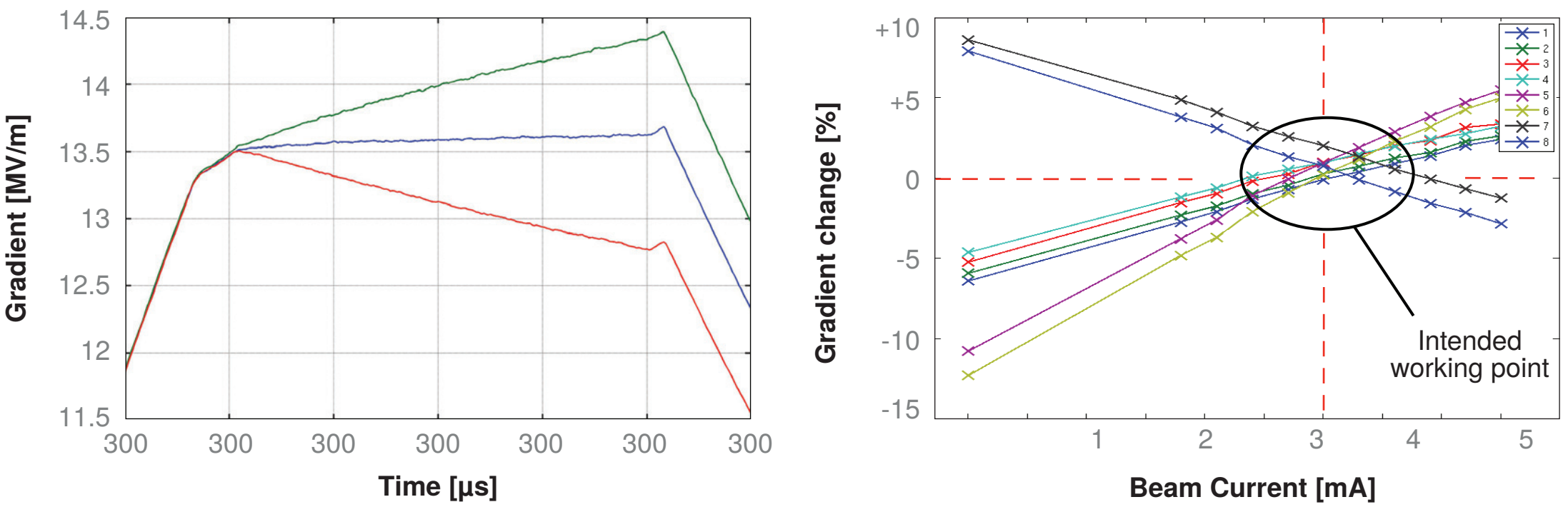

**(a)** Gradient profiles for one cavity at 1.8 mA, 3 mA, and 4.5 mA

**(b)** Gradient tilts as a function of beam current for all cavities

**Figure 3.15.** Evaluation of loaded-Q optimisation by scanning the beam current around the 3 mA nominal working point

Plotting the gradient tilts as a function of beam current for all the cavities will yield a straight line that goes through zero at a particular beam current. Shown in Fig. 3.15b are results of such a beam-current scan for all cavities where the loaded Qs have been set up for flat gradients at 3 mA beam current. Since not all cavities have zero tilts at the same beam current, the quality of the solution is judged by the spread in tilts at the nominal 3 mA.

### 3.2.9 Ramp-up to Full Beam Power and Maximum Gradients

An important operational issue associated with the beam-loading-induced tilts is how to avoid causing large gradient excursions when ramping from a cold start to nominal beam current and full pulse length. One approach would be to start at the nominal bunch charge and a short beam pulse. The loaded Qs (and power dividers if available) would be optimised to achieve flat gradients with the short beam pulse before increasing the pulse length. Given that the loaded-Q optimisation should be independent of the beam-pulse length, if the cavity gradients are flat for a short pulse, they should remain flat if the length of the bunch-train is increased. This approach was tested during the gradient flattening studies with 4 mA beam current; the results are shown in Fig. 3.16. The loaded Qs were initially optimised for flat gradients with a 400 µs beam pulse. The pulse length was then increased, first to 600 µs and then to 800 µs. The gradients did indeed remain flat, with changes in the loaded Qs being necessary to keep the gradients flat over the longer beam pulse.

In this experiment, the full RF pulse length was retained throughout in order to observe the gradient tilts after the end of the beam pulse. The large tilts occurred after the beam turned off because the loaded Qs were optimised for 4 mA and not for zero current. Clearly, if the gradients were close to quench, then the large tilts after the beam-on period would cause cavities to quench. Two options for avoiding the quenches have been tested: first, the RF pulse as well as the beam pulse can be shortened; secondly, gradient limiters can be used to prevent the gradients from reaching quench. This latter approach is discussed in Section 3.2.10.

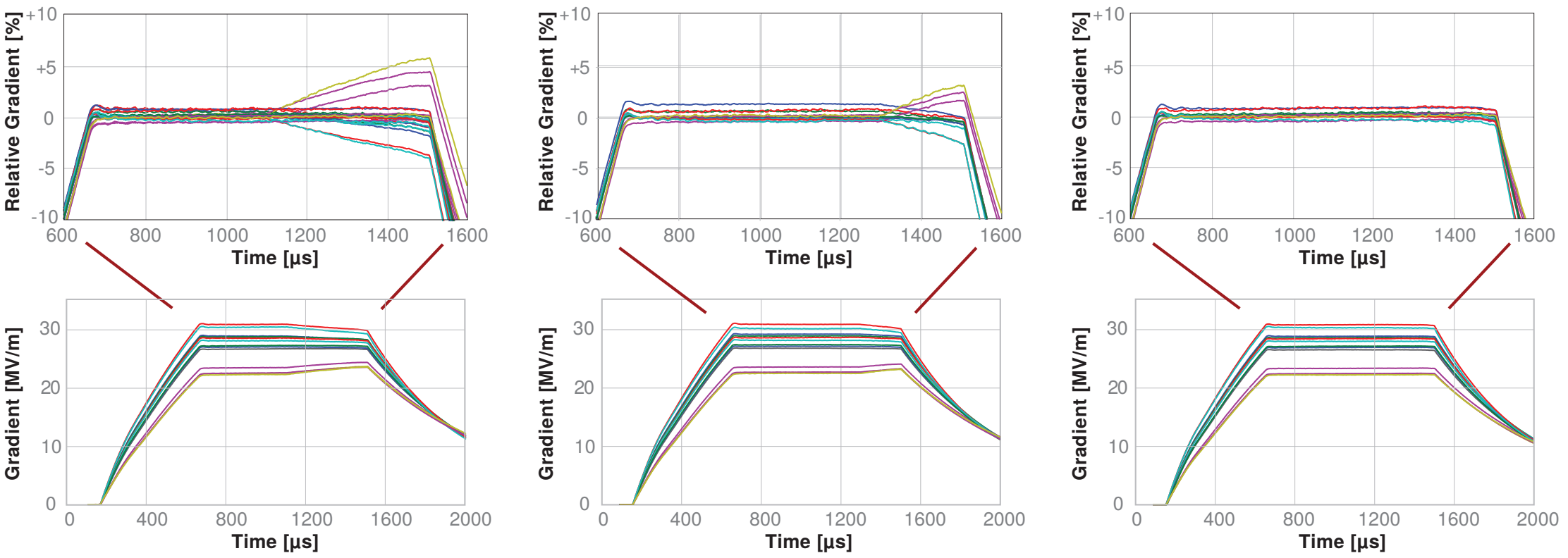

**(a)** 400 µs beam pulse

**(b)** 600 µs beam pulse

**(c)** 800 µs beam pulse

**Figure 3.16.** Cavity gradients during ramp-up of beam pulse length. Each cavity loaded Q was optimised for flat gradients during the beam-on period with 400 µs pulses (left); when the pulse length was extended to 600 µs (centre) and then to 800 µs (right), the gradients became flat for the full duration of the beam pulse without needing further loaded Q adjustments.

## 3.2.10 Management of Operational Gradient Limits

Suitable quench-protection mechanisms will be an essential requirement for routine operation close to cavity gradient limits. The prevention of quench events that require a long cryoplant recovery cycle and machine downtime is of primary importance. It is also vital to avoid false triggers from quench-detection mechanisms that are either too sensitive or have poor discrimination of quench signatures from other transient conditions.

Although this study is in its early days, two detection methods have been pursued: early detection of the onset of a quench by looking for a sudden drop in the loaded Q; and preemptive detection of a potential quench condition from a real-time comparison of cavity gradients with gradient-alarm thresholds based on knowledge of the quench limits. This latter method relies on there being consistent and repeatable quench limits.

A sudden drop in loaded Q is detected from the decay-time of the cavity field at the end of the RF pulse, where a sudden large drop in the loaded Q from pulse to pulse indicates that the cavity had started to quench during the pulse. The basic algorithm worked successfully when operating parameters were stable but is vulnerable to false triggers when parameters change, for example shortening the RF pulse or beam pulse, or when the loaded Qs are being changed by the gradient-flattening algorithm.

In the case of the gradient-alarm levels, each cavity field is continually compared with a predefined alarm threshold, allowing action to be taken before a quench begins. The RF pulse is terminated if any one of the cavities in the vector sum reaches its threshold. The most recent studies have used two gradient-alarm thresholds for each cavity. The upper threshold acts as a 'hard limiter,' turning off the RF pulse if any cavity exceeds its respective alarm threshold. The second threshold, set somewhat lower, is used as a 'pre-limiter', where instead of turning off the RF pulse, the vector-sum setpoint is dynamically reduced until all cavity gradients are again below their respective limiter thresholds. The original setpoint table is restored for the following pulse. The action of the pre-limiter is shown in Fig. 3.17.

The benefit of this pre-limiter is that it can keep all gradients below defined thresholds, pre-emptively avoid a quench without having to terminate the RF pulse. This makes it possible for beam operation at gradients right up to the pre-limiter thresholds and is particularly beneficial during machine tuning when frequent early pulse terminations could otherwise change the machine working point from changes in pulse-heating effects.

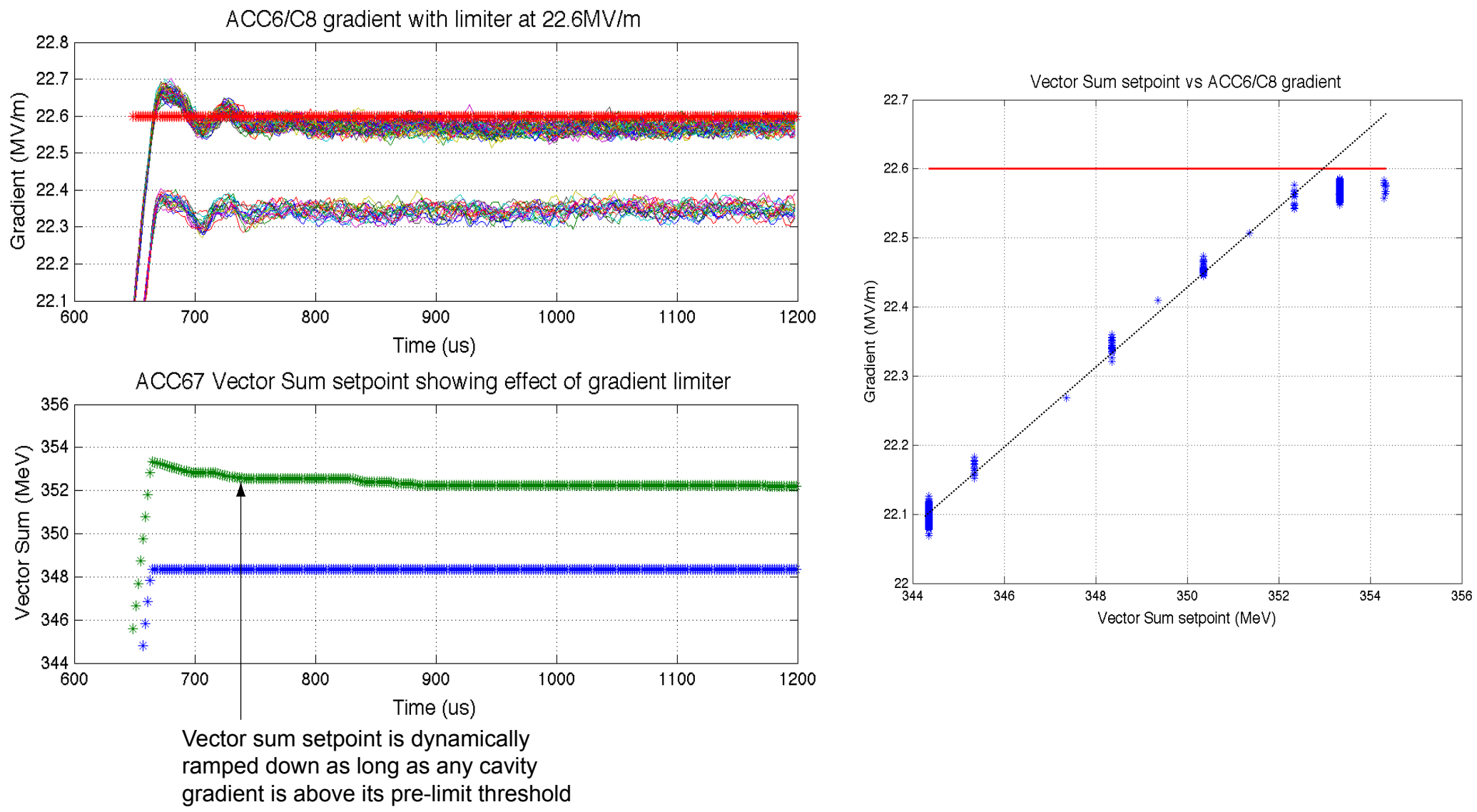

**Figure 3.17.** Cavity-gradient 'pre-limiter' action: dynamic ramping down of the vector-sum setpoint to keep the gradient below the defined threshold (left); clamping action on the vector sum (right)

## 3.2.11 Energy Stability

Requirements for energy stability encompass pulse-to-pulse energy jitter, intra-pulse jitter from bunch to bunch, and the mean energy profile over the bunch train, all of which have been monitored routinely throughout the FLASH studies. Examples of energy profiles from the high-current study were shown previously in Fig. 3.4.

Between 2009 and 2011, upgrades to the low-level RF systems, largely motivated by FLASH FEL user requirements, have significantly improved the energy stability, especially in the pulse-to-pulse jitter. These improvements are illustrated in Fig. 3.18, which overlay energy profiles from many pulses from the 2009 and 2011 studies. While the energy stability achieved in 2009 was about at ILC requirements, energy stability in 2011 is now significantly better than ILC requirements for both the main linac and bunch compressors. Representative long-term stability over a three-hour period during the 2011 studies is shown in Fig. 3.19.

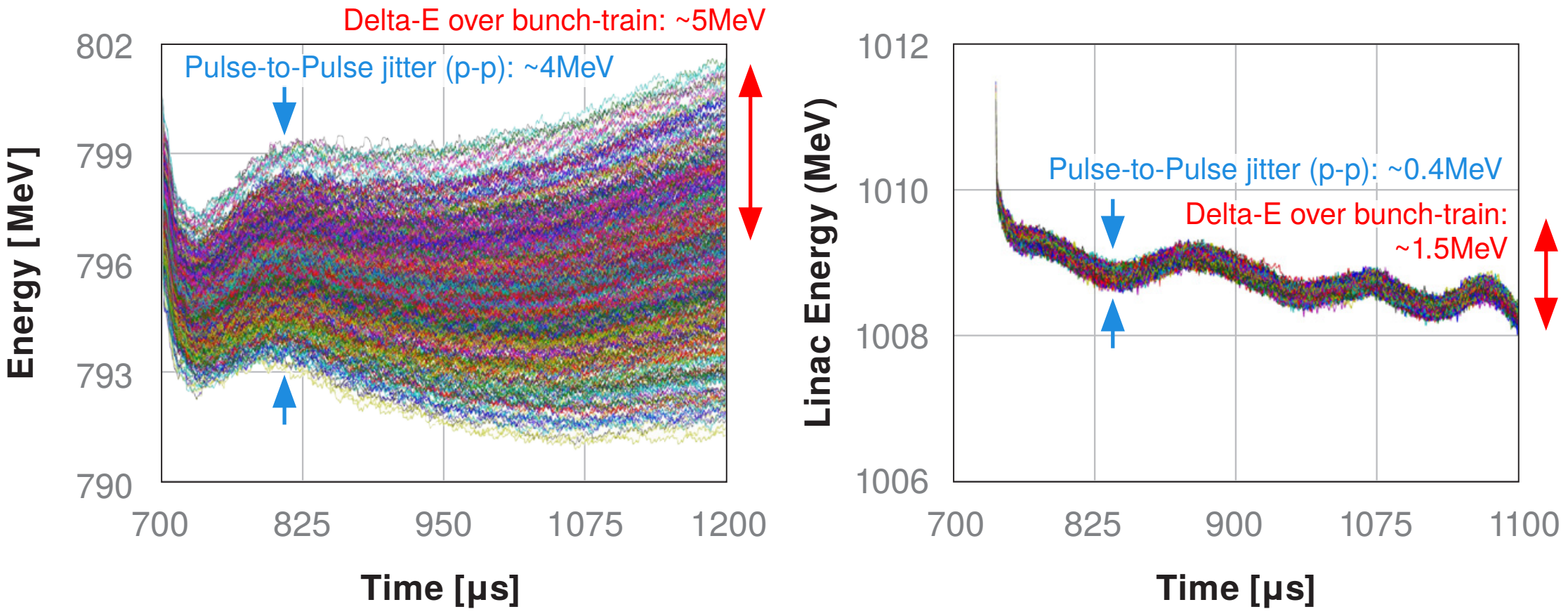

**Figure 3.18.** Comparison of pulse-to-pulse (left) and intra-pulse (right) energy stability between 2009 and 2011 4.5 mA

Beside stabilisation of the cavity vector sum, low-level RF controllers have been upgraded to make use of beam-based measurements in order to improve the pulse-to-pulse repeatability and to compensate for pulse-to-pulse jitter. All LLRF systems now compensate beam loading using bunch-to-bunch charge measurements. Additionally, the amplitude and phase of the RF systems upstream of the bunch compressors are dynamically regulated in order to stabilise the downstream bunch-to-bunch energy and compression angle. This compensates for arrival-time jitter at the start of the bunch-train and any slewing over the length of the train. Figure 3.20 shows the pulse-to-pulse arrival-time jitter as a function of bunch-number. Using bunch-to-bunch feedback, the arrival-time jitter at the start of the bunch-train was attenuated by a factor five by the tenth bunch. Similar beam-based feedback that operates on the damping ring output kicker is planned for the ILC RTML turn-around.

**Figure 3.19**
Pulse-to-pulse energy stability over a three-hour period with 4.5 mA and 400 µs bunch trains (from 2011 studies)

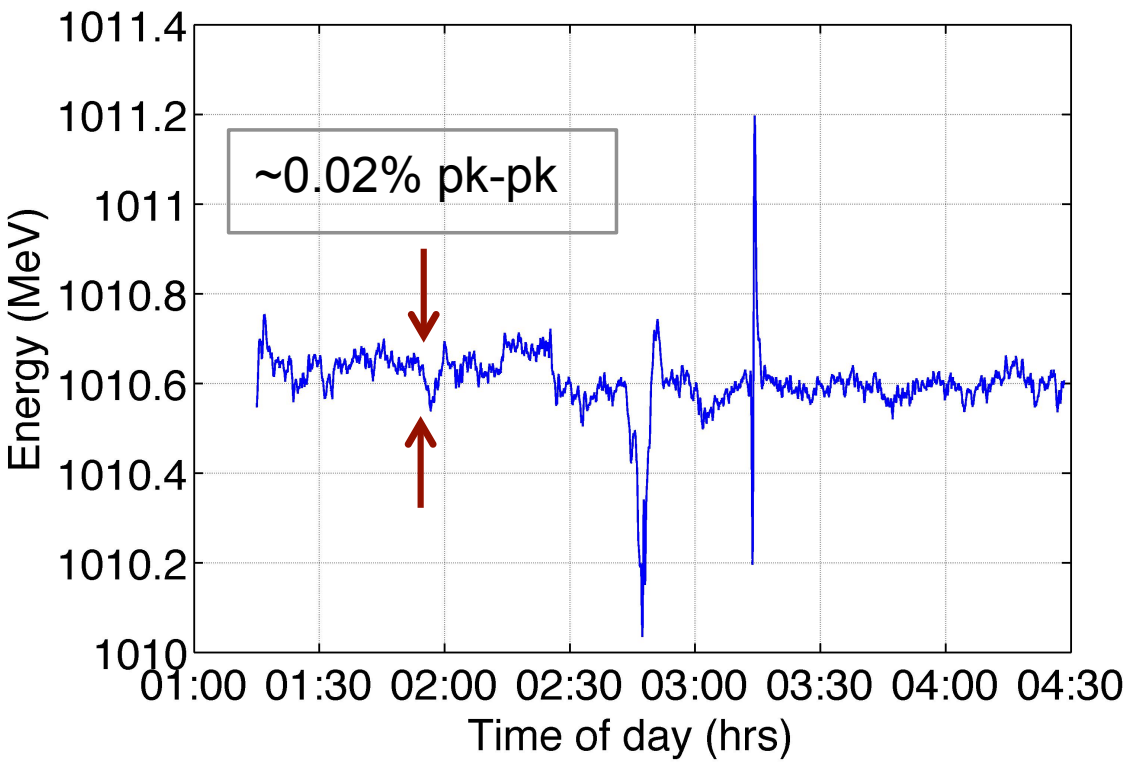

**Figure 3.20**
Arrival jitter at end of the linac with beam-based LLRF feedback [119]

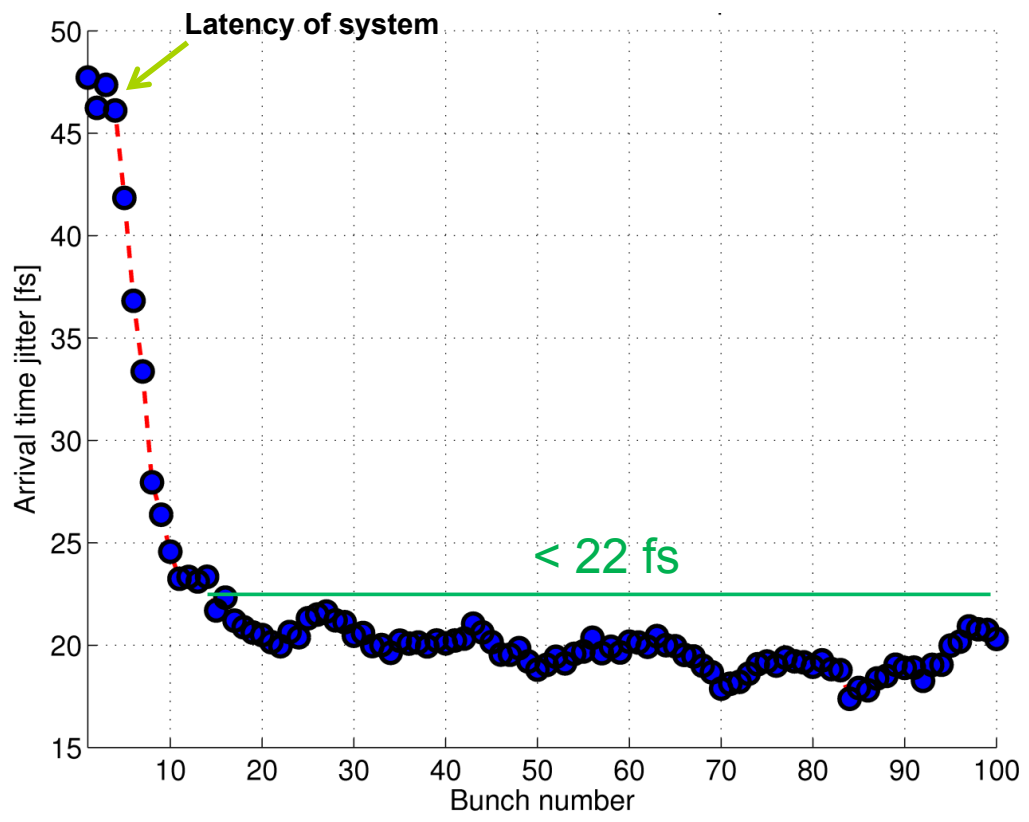

## 3.2.12 Automation

The level of automation of FLASH operations has increased significantly over the course of the 9 mA studies program, with much of the focus being on automating LLRF operations:

- use of state machines to automate procedures such as linac startup, recovery following interlock trips, and restoring standard setup files [125];
- feedback control using high level applications implemented in DOOCS servers [120]: examples include orbit feedback [126], cavity resonance control using piezo tuners [127], pulse-to-pulse iterative learning feedforward for klystron forward power [119].

Automation has also been developed specifically for the 9 mA studies, including the loaded Q optimisation for establishing and maintaining flat gradients and tracking changes in beam current to allow ramp-up of the machine energy and beam current [128].

Automation developed at FLASH will provide important lessons and a starting point for automating XFEL operations. Similarly, the ILC will benefit significantly from lessons learnt in automating XFEL.

### 3.2.13 Conclusions

The principle goals of the 9 mA Experiment at FLASH were to establish beam operation at ILC-like parameters and to study operation of an ILC-like RF unit at the limits of gradient and klystron power. These goals have been largely achieved, as summarised in Table 3.3.

**Table 3.3**
Results of high-power beam studies

| Metric | Goal | Achieved |
| --- | --- | --- |
| Pulse length & current | 800 µs and 9 mA | ▪ 800 µs pulses and current up to 6 mA<br>▪ 9 mA current and pulse lengths up to 600 µs |
| Charge per pulse | 7200 nC | ▪ 5400 nC (600 µs, 9 mA) |
| Average power | 36 kW (7200 nC, 5 Hz, 1 GeV) | ▪ 22 kW (5400 nC, 5 Hz, 800 MeV) |
| Operating gradients with beam loading | 31.5 MV/m nominal average | ▪ Several cavities above 30 MV/m<br>▪ 13 cavities totalling 380 MV/m with 13 cavities |
| Gradient flatness | 2 % $\Delta V/V$ (800 µs, 5.8 mA) | ▪ $<$ 0.3 % $\Delta V/V$ (800 µs, 4.5 mA) (800 µs, 9 mA)<br>▪ Automated the cavity-gradient flattening algorithm |
| Gradient margin | All cavities operating within 3 % of quench | ▪ Some cavities within $\sim$ 5 % of quench (800 µs, 4.5 mA)<br>▪ First tests of operations strategies for gradients close to quench |
| Energy Stability | $<$ 0.1 % rms at 250 GeV | ▪ $<$ 0.15 % p-p (400 µs pulses $<$ 0.02 % rms 5 Hz) |
| RF power overhead | Stable operation at ILC design parameters | ▪ First tests of operation within 5 % of klystron saturation with 800 µs pulse lengths and 4 mA<br>▪ First tests of klystron linearisation close to saturation |
| Linac operations | ▪ 15 hrs continuous running with 3 mA and 800 µs pulses<br>▪ Several hours operation close to 9 mA with bunch trains of 500-600 µs<br>▪ Energy deviations within long bunch trains: less than 0.5 % pulse-pulse at 7 mA<br>▪ Energy jitter pulse-to-pulse with long bunch trains: $\sim$ 0.13 % rms. at 7 mA<br>▪ Recovery to 2400 bunches and 4.5 mA on the first pulse after a beam-inhibiting cryo event | |

No fundamental technology issues with operating a superconducting linac at the ILC Technical Design baseline parameters were encountered. The operation of FLASH has also benefitted from the 9 mA experiment, most particularly with regard to providing stable routine operation with long beam pulses for FEL users. There remains, however, much to be learnt at FLASH. The European XFEL currently under construction offers an even greater opportunity for gaining invaluable experience and lessons with constructing, commissioning, and operating a large-scale superconducting high-power linac and for developing necessary tools and techniques.

## 3.3 STF beam test facility at KEK

### 3.3.1 Introduction

The STF facility at KEK has been designed from the outset eventually to house a test accelerator consisting of two ILC cryomodules, accelerating ILC-like beam from a laser-driven photocathode RF gun. Figure 3.21 shows the envisaged layout. A staged approach has been adopted in developing this test accelerator:

- for STF phase 1 (STF-1, 2007–2011), the facility was used as a module RF test set-up for the early STF-1 cryomodules and for the S1-Global programme; (Section 2.6).
- In 2011–2012, in part in preparation for the STF-2 beam test accelerator shown in Fig. 3.21, a separate project called "Quantum-Beam Accelerator" [129] was installed and tested. Primarily designed as a demonstration for a compact X-ray source using Inverse Compton backscattering of laser photons, the laser-driven photocathode RF gun and two-cavity capture cryomodule which provided the electron beam are the first stage of STF-2. The X-ray-generation experiment was performed in October 2012, after which the X-Ray-generation beam line and Compton laser system was decommissioned and replaced with the first ILC eight-cavity cryomodule in preparation for the STF Phase-2 ILC beam test facility.

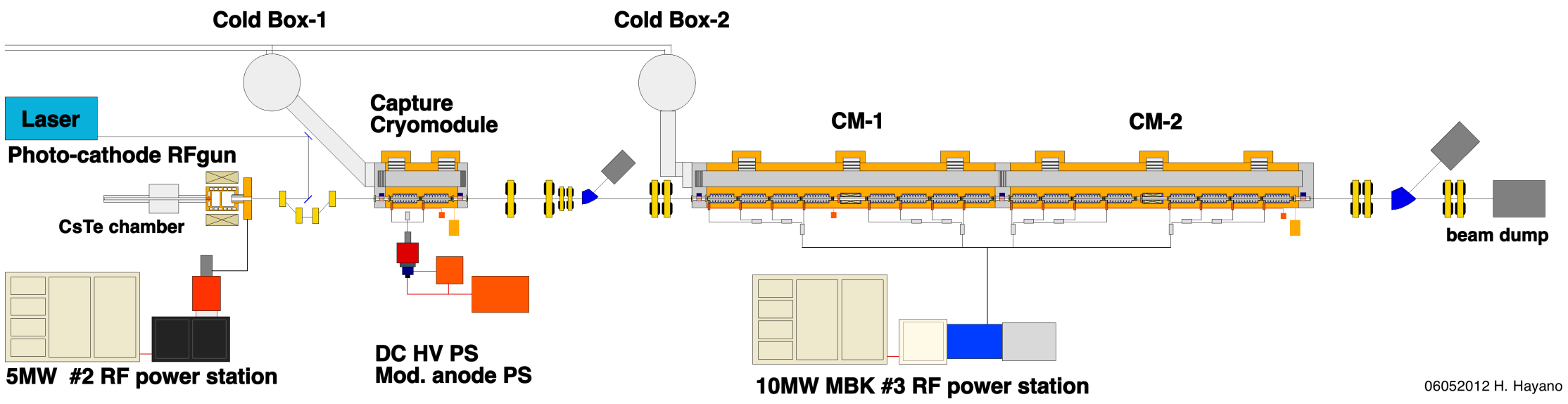

**Figure 3.21.** The STF Phase-2 accelerator, which will be constructed in the STF tunnel.

### 3.3.2 Quantum-Beam Accelerator as an injector for STF-2

**Figure 3.22**
The Quantum-Beam Accelerator. From the left to right, $Cs_2Te$ photocathode preparation chamber, 1.3 GHz normal conducting RF gun, injection beam line, capture cryomodule with DRFS klystron power system, the focusing beam line, and beam dump.

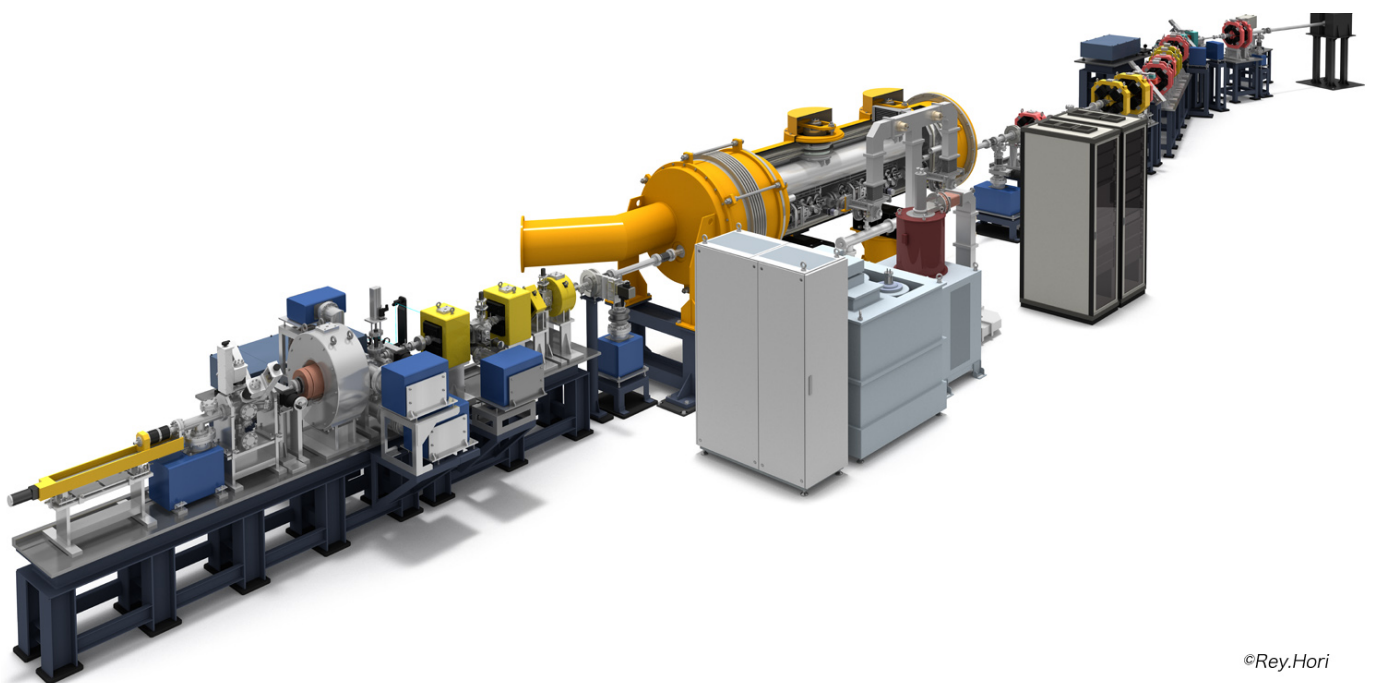

The Quantum-Beam accelerator is illustrated in Fig. 3.22; a photograph is shown in Fig. 3.23. The design parameters for the Quantum-Beam Accelerator and the STF phase-2 accelerator are listed in Table 3.4. The main difference in the electron beam requirements is the bunch spacing and bunch charge. The Quantum-Beam accelerator uses 162.5 MHz bunch-repetition frequency (6.15 ns spacing) with 62 pC/bunch, resulting in a peak pulse current of 10 mA. By comparison, the STF phase-2 accelerator requires a 2.708 MHz bunch rate (369.27 ns spacing) with 3.2 nC bunches, corresponding to a peak current of 8.7 mA (the ILC beam parameters), which will require a different RF gun laser system. STF-2 also requires a lower beam energy after the two-cavity capture accelerator of 21.5 MeV compared to 40 MeV for the Quantum-Beam Accelerator.

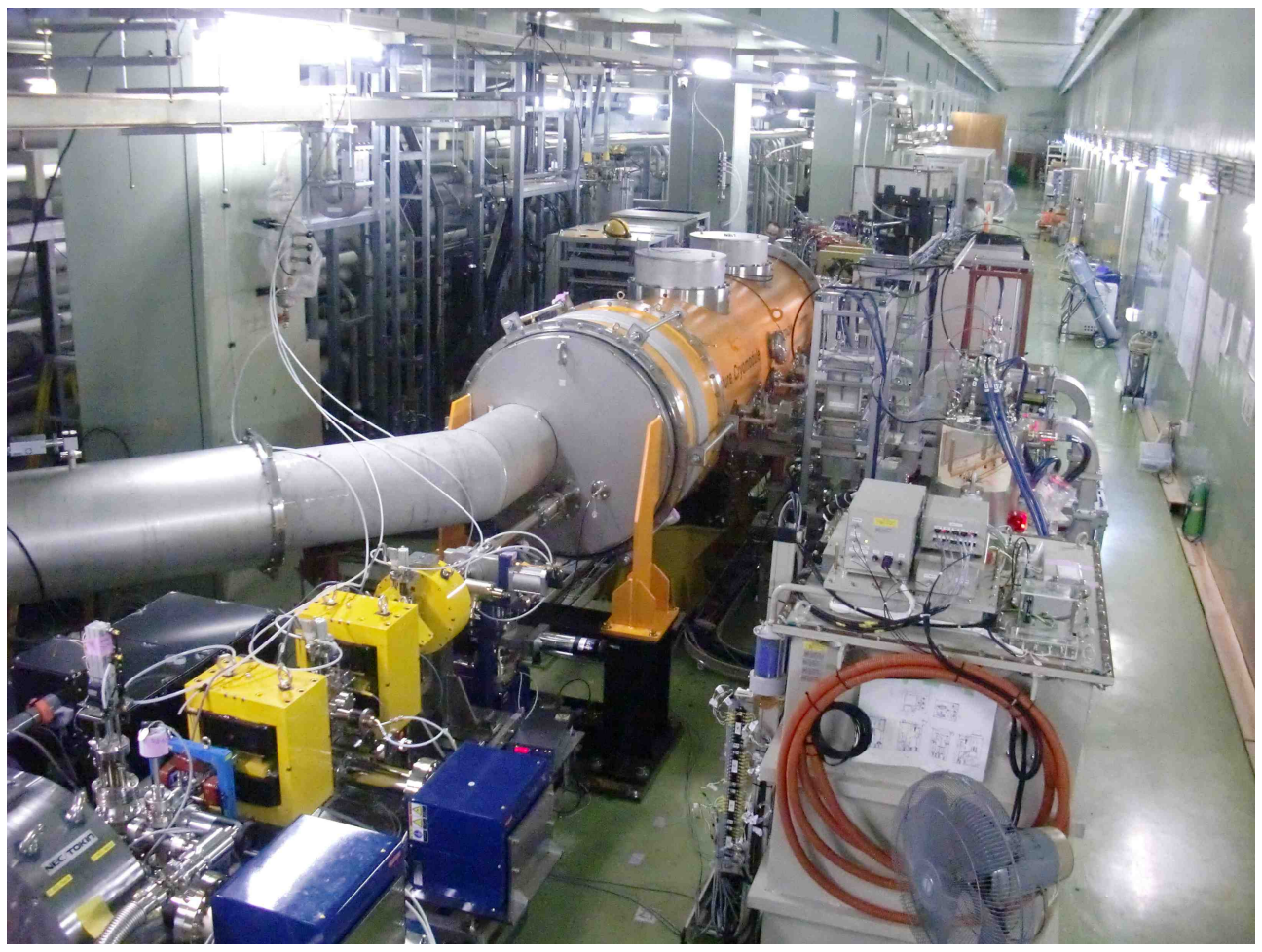

**Figure 3.23**
The injector for the STF-2 accelerator (Quantum Beam Accelerator), showing the 1.3 GHz normal-conducting RF gun, injection beam line, and SCRF capture cryomodule.

**Table 3.4**
Parameters for the Quantum-Beam Accelerator compared to those for the STF-2 injector

| | | Quantum Beam Accelerator | STF-2 Accelerator |
|---|---|---|---|
| Pulse length | ms | 1.0 | 0.9 |
| Repetition rate | Hz | 5 | 5 |
| Bunch spacing | ns | 6.15 | 329.27 |
| Bunch frequency | MHz | 162.5 | 2.708 |
| Bunches/pulse | | 162.500 | 2.437 |
| Bunch charge | pC | 62 | 3.200 |
| Total charge /pulse | nC | 10.000 | 7.798 |
| Beam current | mA | 10.0 | 8.7 |
| Bunch length (laser FWHM) | ps | 12 | 10 |
| Max. beam energy | MeV | 40.0 | 21.5 |
| Beam power | kW | 2.0 | 0.8 |

Before installation in the capture cryomodule, the two nine-cell cavities were tested and successfully reached gradients of up to 40 MV/m and 32 MV/m respectively. After installation in the cryomodule, the two cavities were connected to the 800 kW DRFS klystron power system Section 2.8.5. RF-gun commissioning began in February 2012, and acceleration of the ~1 ms beam train was successfully achieved by June 2012.

#### 3.3.2.1 Photocathode RF-gun commissioning

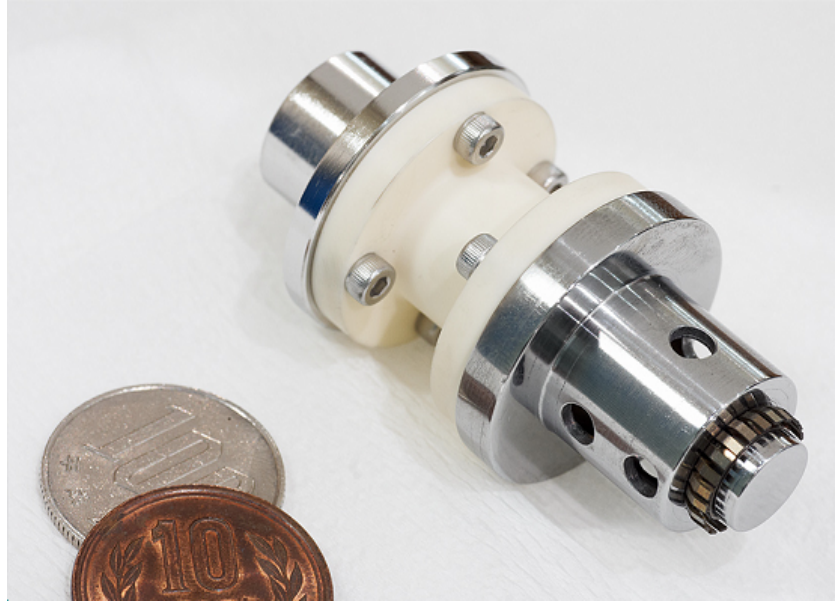

**Figure 3.24**
Photograph of the Molybdenum photocathode block. The head is thermally insulated by the ceramic interface to avoid the heat flow during surface heat cleaning. The heater is embedded in the blockhead. The Inconel crown contactor is used for the RF current contact between the block and the cavity endplate.

The design of the normal-conducting 1.3 GHz photocathode RF gun used was originally developed by DESY for FLASH and the European XFEL [130]. The Gun cavity was fabricated by FNAL and has a design peak field at the cathode of 50 MV/m at 4.5 MW RF power. To generate the long multi-bunch beam, a $Cs_2Te$ photocathode is employed, which is prepared by evaporating a thin film on to a Molybdenum cathode block under vacuum. The Molybdenum cathode block is shown in Fig. 3.24. A special preparation chamber has been designed, allowing direct transfer of the cathode to the gun cavity under vacuum. In the preparation chamber, the surface of the Molybdenum plug is first baked to remove any impurities, followed by tellurium and caesium evaporation. To date, a quantum efficiency (QE) of 7 % has been achieved. After an ethanol treatment followed by RF processing, the observed dark current was 247 µA with 46.7 MV/m (4 MW input power) for the bare Molybdenum surface cathode, and 688 µA with the $Cs_2Te$ coating.

The drive laser for the cathode consists of a 162.5 MHz CW oscillator, Pockels cell, two-stage optical-burst amplifier and wavelength converter. The oscillator generates 1064 nm infrared pulses with 12 ps (FWHM) pulse width. The Pockels cell cuts out a pulse train of 1 ms with 5 Hz repetition from the CW oscillator output. The burst amplifiers boost the energy up to 10 µJ per pulse. The pulse train is converted to the 4th harmonic (266 nm) of the fundamental mode by LBO and BBO crystals. Figure 3.25 shows the 1 ms beam pulse successfully extracted by the gun in March 2012. The RF power was approximately 2.6 MW, and the cathode field gradient was approximately 37.5 MV/m, with 162,450 bunches at $\sim$ 30 pC per bunch.

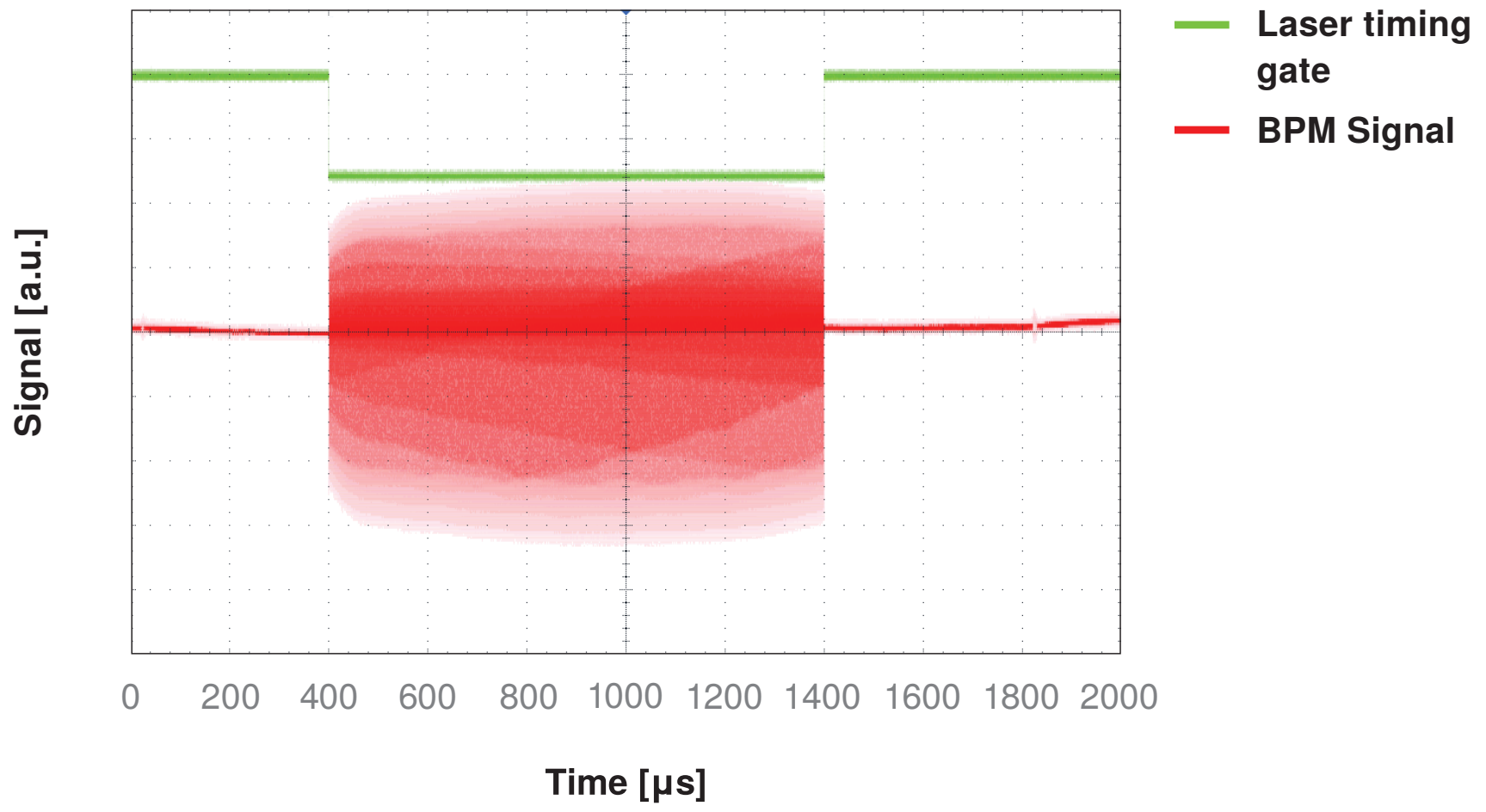

**Figure 3.25**
Plot of the 1 ms bunch train extracted from the photocathode RF gun. The lower (red) trace is the signal of the BPM; the upper (green) trace is the gate signal for the extracted laser pulse.

### 3.3.2.2 Capture-accelerator commissioning

The capture-cryomodule cool-down and adjustment of the superconducting cavities was started in April 2012. Both cavities were operated at 20 MV/m. For the initial commissioning, a short pulse of 28 bunches at 41 pC per bunch was used (5 Hz repetition rate). After successful commissioning of the digital feedback amplitude and phase control for both the RF gun cavity and superconducting cavities, a 1 ms-long beam train was successfully accelerated up to 40 MeV, with 15 pC bunches — 25 % of the target intensity — as shown in Fig. 3.26.

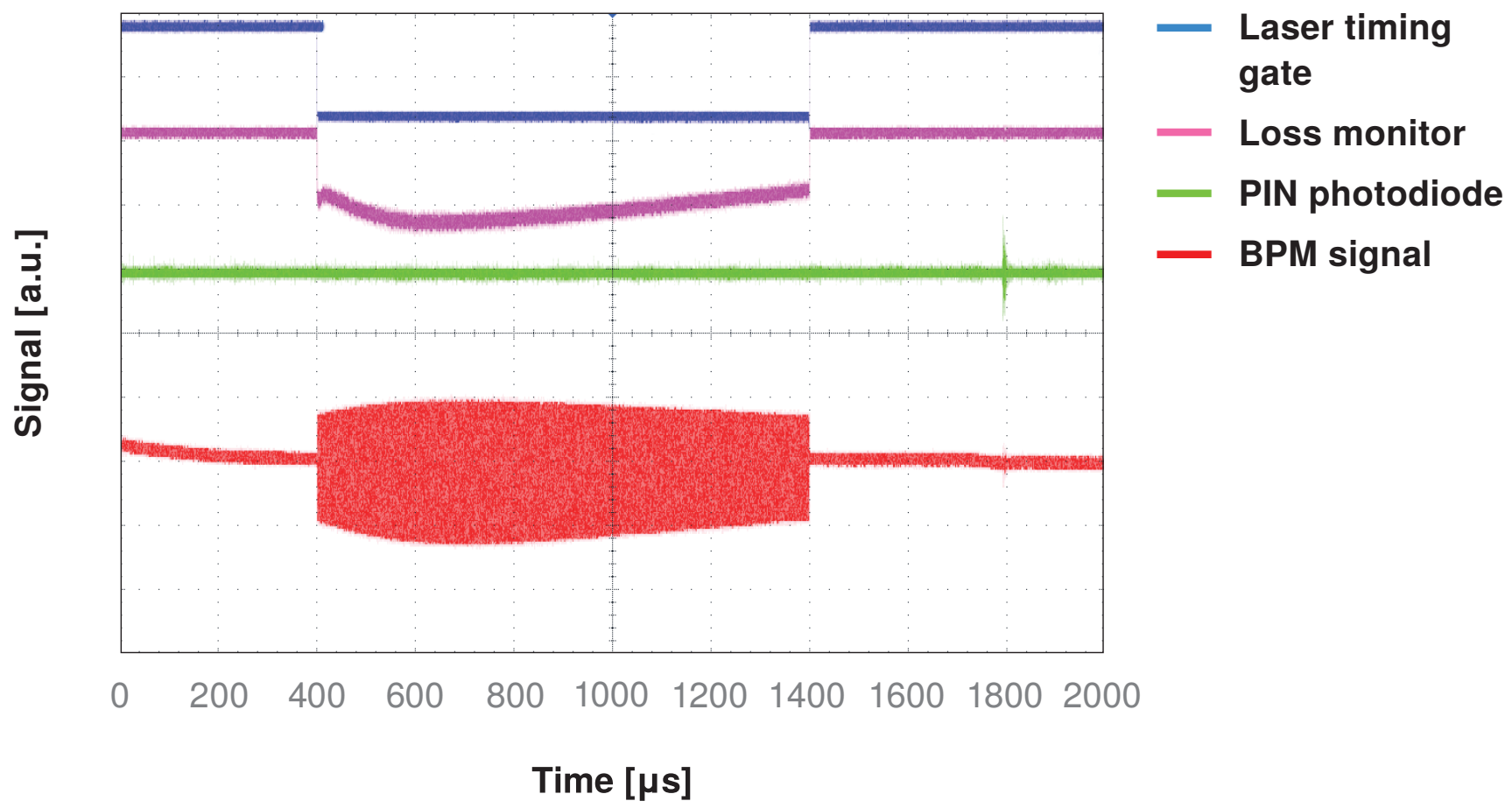

**Figure 3.26**
Plots showing 1 ms bunch train after acceleration to 40 MeV. The upper (blue) trace is the gate signal for the extracted laser pulse, the second (magenta) trace is the signal from the loss monitor, the third (green) trace is from the PIN photodiode signal used to measure beam loss at the Inverse Compton interaction point, and the lowest (red) trace is the signal from the bpm

### 3.3.3 The STF-2 accelerator

The STF-2 accelerator will consist of several ILC-type cryomodules. An existing laser capable of producing the required 1 ms pulse at 2.7 MHz (2437 bunches) will be used to drive the RF gun. The existing two-cavity capture-cryomodule will be used to accelerate the beam to 20 MeV before injection into the main ILC-type cryomodule accelerator. The first phase-2 cryomodule (CM-1) will consist of eight cavities, with a same-weight mock-up of an SC quadrupole installed at the centre of module. The quadrupole mock-up will also contain a BPM. Four cavities will have the slide-jack tuner mounted at the mid-point of the cavities (helium vessel), while the other four will have the tuner located at the end of the helium tank. The tuner motor will be placed outside the cryomodule vessel. The expected beam energy at the exit of CM-1 is 272 MeV. By November 2012, nine cavities have already been fabricated, and nearly all have completed their initial vertical test and have achieved gradients of 35 MV/m. The detail design and fabrication of CM-1 should be completed by the beginning of 2014. The originally foreseen second module (CM-2) is still under discussion. The RF power for the two ILC cryomodules (a total of 16 cavities) will be provided by a single 10 MW multi-beam klystron (Part II Section 3.6.3) driven by a Marx modulator (Part II Section 3.6.2). Beam operation of the STF-2 accelerator is planned for early 2015.

## 3.4 Fermilab Cryomodule 1 Test

### 3.4.1 Introduction/Goals

Fermilab's Cryomodule 1 (CM-1) is a Tesla Type III+ 8-cavity Superconducting RF module (Fig. 3.27). It came to Fermilab as a 'kit' from the DESY laboratory. CM-1 was assembled at Fermilab by its technical staff with assistance from colleagues from DESY and LASA/INFN, Milano. The goal of installing and operating CM-1 at NML was for Fermilab to gain expertise in assembling and operating a complete ILC-type Cryomodule as well as to demonstrate its capability. Successful cooldown and RF powering of all cavities simultaneously was the milestone for successful operation. Once installed, CM-1 was operated for fifteen months, ending in early March 2012. Concurrently with bringing CM-1 into operation, all necessary subsystems including cryogenics, high- and low-level RF, vacuum, protection systems, controls, safety systems etc., had to be commissioned and integrated. This operating period was invaluable both in terms of gaining experience in commissioning and operating an SCRF system but also in identifying improvements for future cryomodules.

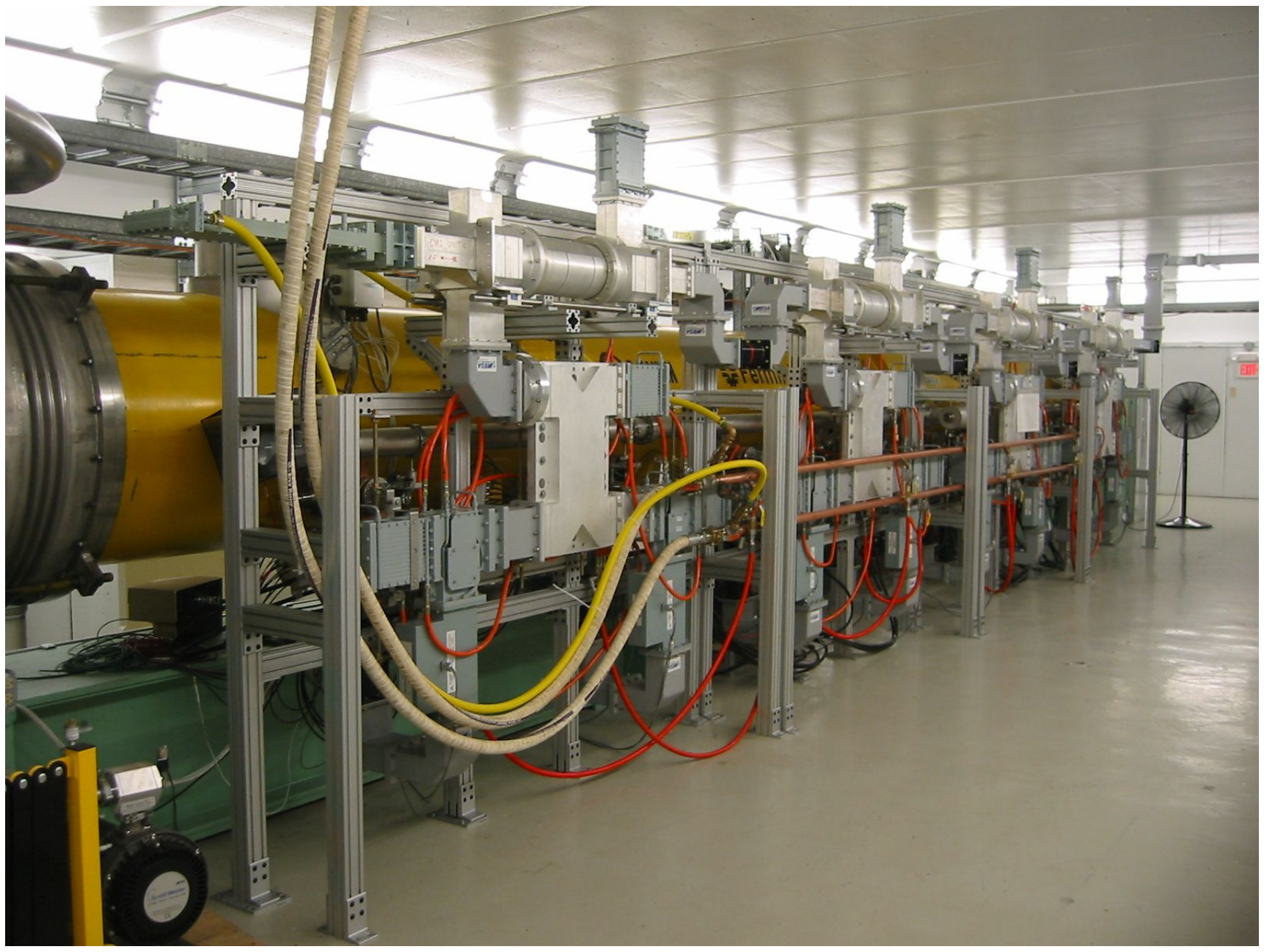

**Figure 3.27** Cryomodule 1 installed for operation at Fermilab's NML building. The waveguide distribution system, which feeds RF power to all 8 cavities simultaneously, is in the foreground.

### 3.4.2 Commissioning & Testing Protocol

CM-1 commissioning and testing approximately followed the standard protocol developed worldwide for bringing cryomodules into full operation. The steps include:

- RF cable calibration;
- technical sensor/interlock check;
- RF/waveguide check;
- warm coupler conditioning (off resonance);
- cooldown to 2 K;
- frequency spectra measurements;
- cavity tuning to nominal cold frequency (1300.00 MHz) via motorised slow tuner;
- $Q_L$ adjustment (to $3 \times 10^6$ per ILC specification);
- RF system calibrations;
- cold coupler conditioning (on resonance);
- performance evaluation including:
  - maximum gradient;
  - dynamic heat load ($Q_0$ vs. $E_{ACC}$);
  - dark current and field emission (X-rays vs. $E_{ACC}$).
- full cryomodule powering and evaluation.

As these steps were completed, additional measurements were made particularly with the LLRF and resonance control systems. Time was also made available to conduct a series of 'long pulse' 9 ms pulse-length studies on the two best cavities. As the end of the planned period of operation drew near, an additional step, namely thermally cycling CM-1 to room temperature and back to 2 K, was undertaken to determine its effect on cavity performance and provide data on longer-term operation.

### 3.4.3 Cold-Coupler Conditioning and Performance

CM-1 was installed in its final position and aligned in January of 2010 and final RF, cryogenics and vacuum connections were made thereafter. Since this was the first cryomodule of this type to be operated in this facility, commissioning entailed both the module itself as well as the RF, controls, protection system, vacuum, cryogenics, etc.

Warm off-resonance conditioning was performed with one cavity at a time connected to the output of the 5 MW klystron; it took anywhere from two weeks (for the first cavity) to four days to complete. Off resonance implies that the input RF was fully reflected back towards the load and no RF was directed into the cavity. The decrease in time was due to growing familiarity with the systems. The conditioning itself entailed applying successively higher amounts of RF power in successively longer pulse widths beginning with 20 µs to a maximum of the full 1.3 ms pulse. The peak power was as high as 1.1 MW for short pulses (20 µs – 400 µs) and 600 kW for 800 and 1300 µs pulse widths. An automated sequence was implemented which controlled the power and pulse width in a prescribed manner but responded to faults sensed by instrumentation monitoring excessive arcing, field emission, coupler temperature, and vacuum activity. In most cases, field emission coupled with vacuum activity from one of the three FEP sensors was the limiting factor in conditioning.

Once all eight couplers were conditioned, final vacuum work was completed leading to permission and initiation of cool down to 2 K. The module was cooled from room temperature to 4 K in slightly more than 2 days. The final cool-down to 2 K (23 Torr) required 2-1/2 hours. The CM-1 module first reached operating temperature on 22 November 2010. The 1.3 GHz cryomodule design requires that its individual circuits be cooled at predetermined rates in order to limit thermal stresses. One such circuit is the helium-gas return pipe (GRP) which is a 300 mm pipe that acts as the rigid support (strong back) from which all of the cavities are suspended. Based on a computer simulation of the cryomodule cool-down and operating experience with similar design modules at DESY, the maximum vertical gradient along this circuit is limited to less than 15 K. The longitudinal gradient is to be maintained at 50 K or less and lastly the overall cool-down rate is to be 10 K per hour or less over the range of temperatures from approximately 300 K to 100 K. The thermal shield circuits have identical longitudinal cool-down-rate constraints.

Continuing with the sequence of commissioning steps, each cavity was then powered on resonance to complete coupler conditioning and determine cavity performance limitations. Again each cavity was powered singly. All cavities were characterised by June 11, 2011. Individual performance characteristics and limiting factors are summarised in Table 3.5. Cavities are numbered sequentially from the upstream end of the cryomodule (where upstream is defined as the end closest to the photo-injector gun), i.e. low-energy front end.

**Table 3.5**
CM-1 individual cavity performance characteristics.

| Cavity | Peak Eacc (MV/m) | Estimated maximum $Q_0$ (E09) | Limitation/Comments |
|---|---|---|---|
| 1/Z89 | 20.2 | 11 | 'soft' quench/heat load |
| 2/AC75 | 22.5 | 12 | Quench |
| 3/AC73 | 23.2 | 0.43 | 'soft' quench/heat load |
| 4/Z106 | 24* | 2.3 | *RF-limited |
| 5/Z107 | 28.2 | 39 | Quench |
| 6/Z98 | 24.5 | 5.1 | Quench |
| 7/Z91 | 22.3 | 4.7 | 'soft' quench/heat load |
| 8/S33 | 25 | 18 | Resonant frequency at 1300.240 MHz; tuner motor malfunction |

Three cavities exhibited unexpectedly high heat loads as evidenced by cryogenic activity – change in liquid level followed by an increase in system pressure, indicating heating and a drop in the Loaded Q ($Q_L$) at relatively low gradient, <20 MV/m. Significant field emission did not accompany this behaviour. It was found that the onset of the Q drop could be delayed by shortening the flat-top pulse length from the nominal 620 μs to 100 s. No source of this excessive heating has been definitely identified. Such behaviour has been observed elsewhere, most recently on a cavity under test at Fermilab's Horizontal Test Stand (HTS). With CM-1 now removed and disassembly started, plans are being made to disassemble the cavity string and re-evaluate these sub-performing components one at a time in the HTS facility. Reprocessing will be performed if deemed necessary. Additionally the 'slow' tuner on cavity #8 ceased functioning after approximately 20 minutes of cold operation leaving the cavity +240 kHz from the nominal 1.3 GHz resonant frequency. 'On-resonance' tests for this cavity were possible by adjusting the LLRF master oscillator, but this precluded inclusion of Cavity #8 in full module tests.

Comparison of individual cavity performance in CM-1 with that of previous tests is of interest. Figure 3.28 compares the peak gradient of each cavity to the values determined in vertical and horizontal tests at DESY. Table 3.6 provides a numerical comparison. A degradation of gradient of order 15 % or less is noted, perhaps not too surprising, considering that the cavities were vented to atmosphere and pumped back down a few times while at Fermilab. Differences in system calibration could also introduce disagreement in values.

Measurements of $Q_0$ were made on all cavities by measuring the static and dynamic heat load at discrete power levels. While maintaining each cavity at a fixed gradient for one hour the helium mass flow as well as incoming and outgoing temperatures and pressures were recorded. From this, a dynamic heat load and thus $Q_0$ could be determined. The static heat load was measured before and after a powered run. Dynamic heat-load measurements were also made during operation of the entire cryomodule. Figure 3.29 shows a summary of $Q_0$ vs E for all cavities powered simultaneously. The static heat load was of order 22 W, consistent with independent measurements.

**Figure 3.28** Comparison of cavity peak gradients – DESY single cavity vertical test results (blue), horizontal test results (Chechia) (red), and the complete CM-1 at Fermilab (green).

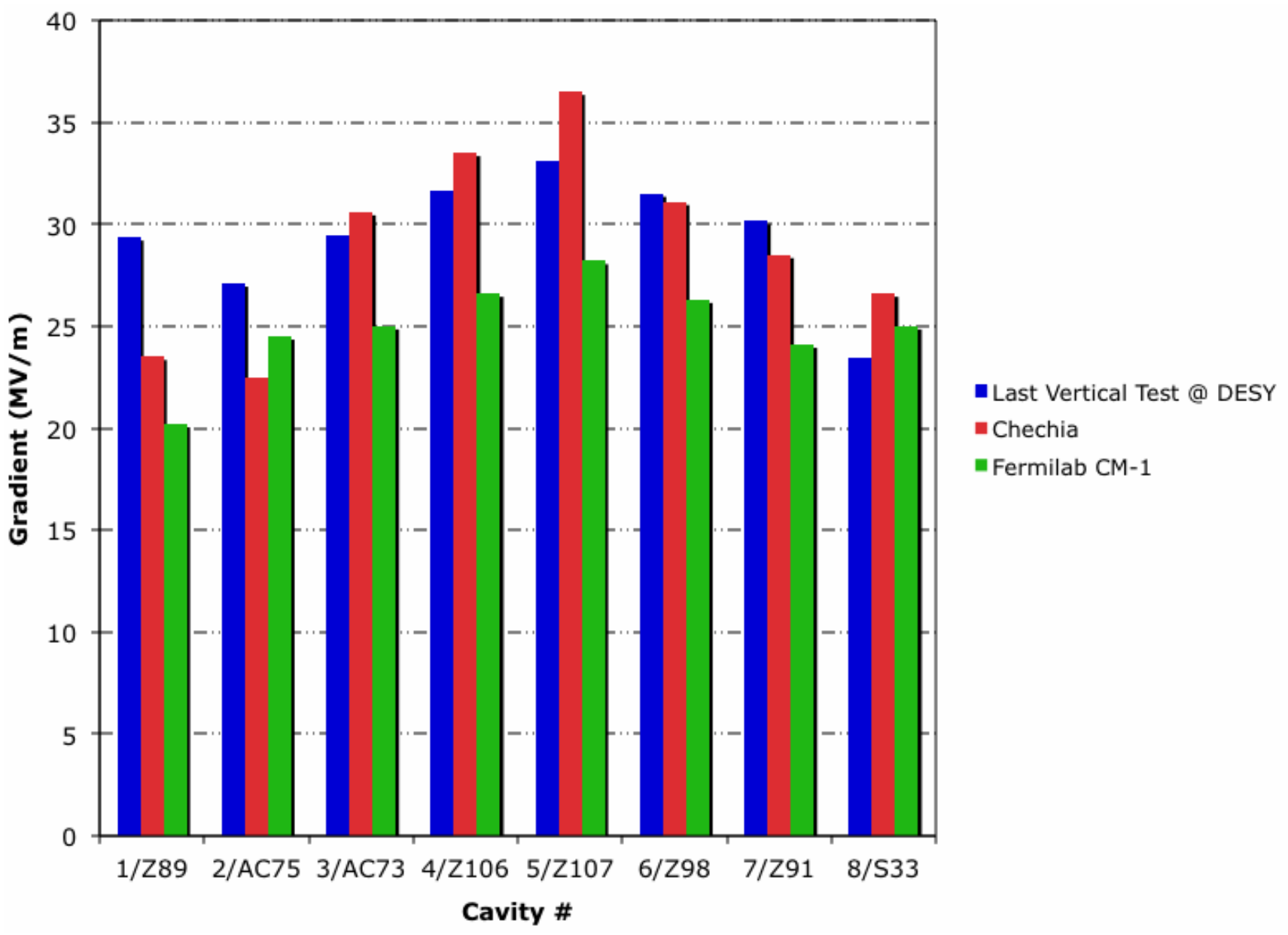

**Table 3.6** Numerical comparison of CM-1 individual cavity performance

| | Cavity | | | | | | | | | |
|---|---|---|---|---|---|---|---|---|---|---|
| | 1/ Z89 | 2/ AC75 | 3/ AC73 | 4/ Z106 | 5/ Z107 | 6/ Z98 | 7/ Z91 | 8/ S33 | Avg. | Sum |
| Last Vertical Test at DESY | 29.36 | 27.10 | 29.49 | 31.65 | 33.07 | 31.51 | 30.16 | 23.44 | 29.47 | 235.78 |
| Chechia | 23.5 | 22.5 | 30.6 | 33.5 | 36.5 | 31.1 | 28.5 | 26.6 | 29.1 | 232.8 |
| Fermilab CM-1 Final | 20.2 | 24.5 | 25 | 26.6 | 28.2 | 26.3 | 24.1 | 25 | 24.99 | 199.9 |
| Fermilab/ Vertical ratio | .688 | .904 | .848 | .840 | .853 | .835 | .799 | 1.07 | .848 | .848 |
| Fermilab/ Chechia ratio | .860 | 1.09 | .817 | .794 | .773 | .846 | .846 | .940 | .859 | .859 |

Following completion of single-cavity testing on 11 June 2011, the waveguide distribution system provided by SLAC, which allows for independent amplitude and phase control of adjacent pairs of cavities, was installed. Variable Tap Offs (VTO's) were set based upon the gradient limits identified during cavity characterisation.

Full-module testing was initiated on 6 July 2011 and largely continued until the end of the year. The bulk of the time spent powering the entire module was devoted to low-level RF (LLRF) optimisation, refining the Lorentz-force detuning compensation system and trying to understand the source of the 'soft quenching' cavities. Operation was largely reliable and there were periods of overnight operation. CM-1 ran for as many as 65.5 hours continuously at moderate gradient, average 16.5 MV/m per cavity, without a trip. Figure 3.30 shows seven cavities operating at the same frequency being powered simultaneously. The variation in peak gradient was due to individual cavity limitations and the resulting setting of the VTO's. Pairs of cavities were thus operating at similar gradients.

Development work on the low-level RF system was a constant throughout the running period. By the end of the run it was possible to control the RF amplitude and phase over 50 pulses to a RMS magnitude error of $2.5 \times 10^{-4}$ and RMS error of 0.005°.

Significant time was also spent refining adaptive Lorentz-force detuning compensation. Gradients in the cavities ranged between approximately 8 and 27 MV/m. Without compensation, the highest gradient cavities would detune by up to 400 Hz during the flattop. With adaptive compensation on, the detuning was negligible in all the cavities except for Cavity 1, which detuned by approximately 10 Hz during the RF pulse.

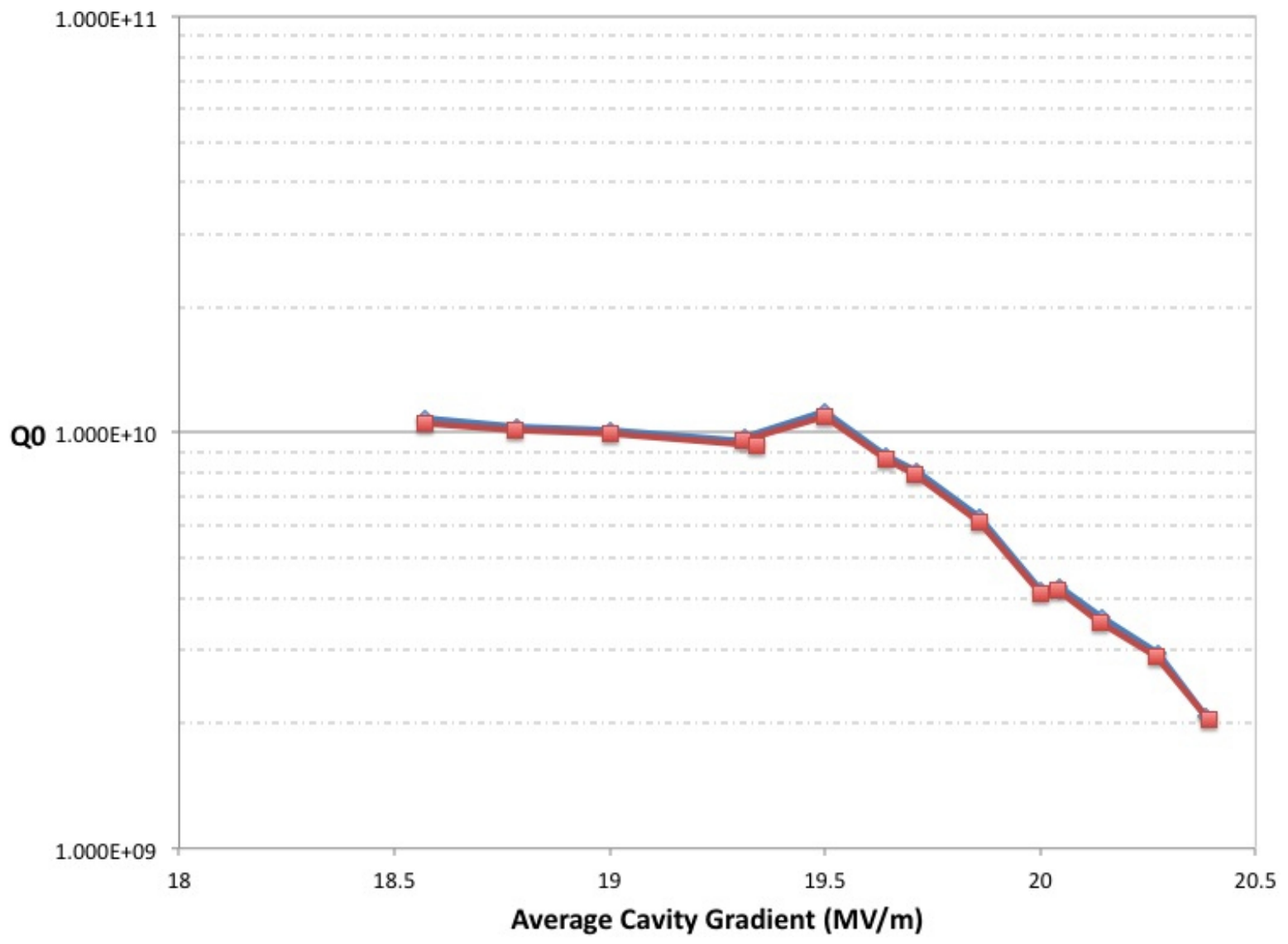

**Figure 3.29**
$Q_0$ vs $E_{Acc}$ for CM-1 cavities

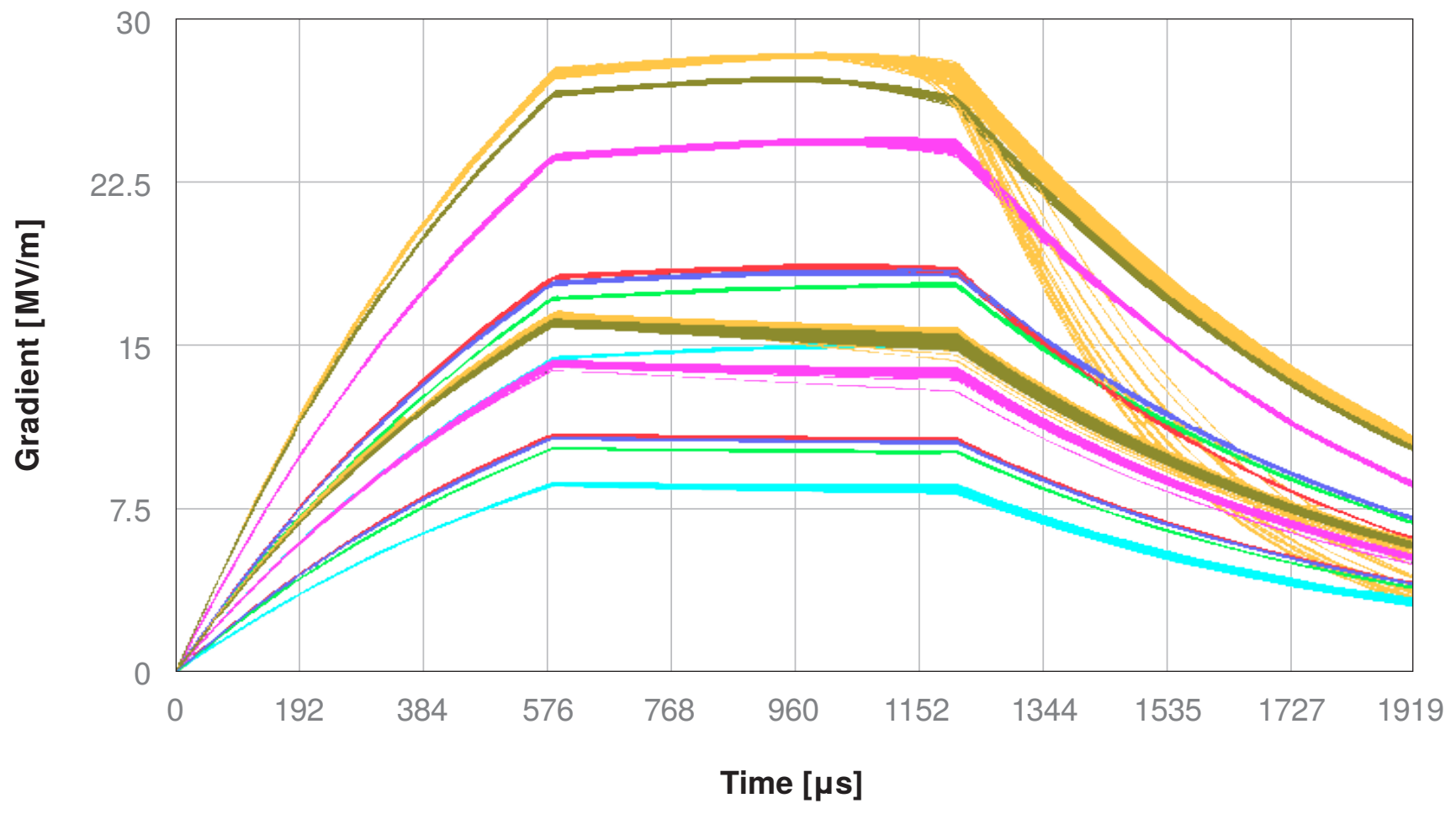

**Figure 3.30**
Transmitted power waveforms of all cavities pulsing at the nominal conditions: 5 Hz, 1.3 ms pulse length (580 µs fill, 620 µs flattop) with LLRF in closed loop and LFDC disabled just below the stable operating point. Some quenching by the highest-performing cavity and indications of Lorentz-force detuning can be observed.

Even for cavities operating at the highest gradients, Lorentz-force detuning could be compensated using relatively modest voltages of $\sim \pm 20$ V to drive the piezo tuner. The waveforms do not exhibit fast transients that could potentially damage the piezo tuners. Figure 3.31 shows a screen capture of the resonance-control system during normal CM1 operation with the adaptive compensation on.

Once all planned tests were completed, the cryomodule was thermally cycled to room temperature and back to 2 K to determine if such action might have an effect on cavity performance. Warm-up proceeded uneventfully, however upon reaching 132 K during cooldown, vacuum activity was observed in the insulating vacuum space, the magnitude of which automatically closed the gate valves between the module and the feed and end cans. A nitrogen peak was observed. Upon opening the upstream insulating vacuum space, a leaking feedthrough for a nitrogen-line temperature sensor was identified as the source of the leak. The feedthrough was replaced with a blank and an alternate surface-mount sensor installed. Cooldown proceeded and operation resumed. Although this was an unfortunate occurrence, it did prove to be a valuable experience in terms of learning how to identify the location of leaks remotely and make in situ repairs, as well as assess vulnerabilities in types of thermometry. Upon the resumption of cold operation only minor changes in operating performance were noted although some multi-pactoring was observed as some cavities approached their peak values. This was rapidly processed away.

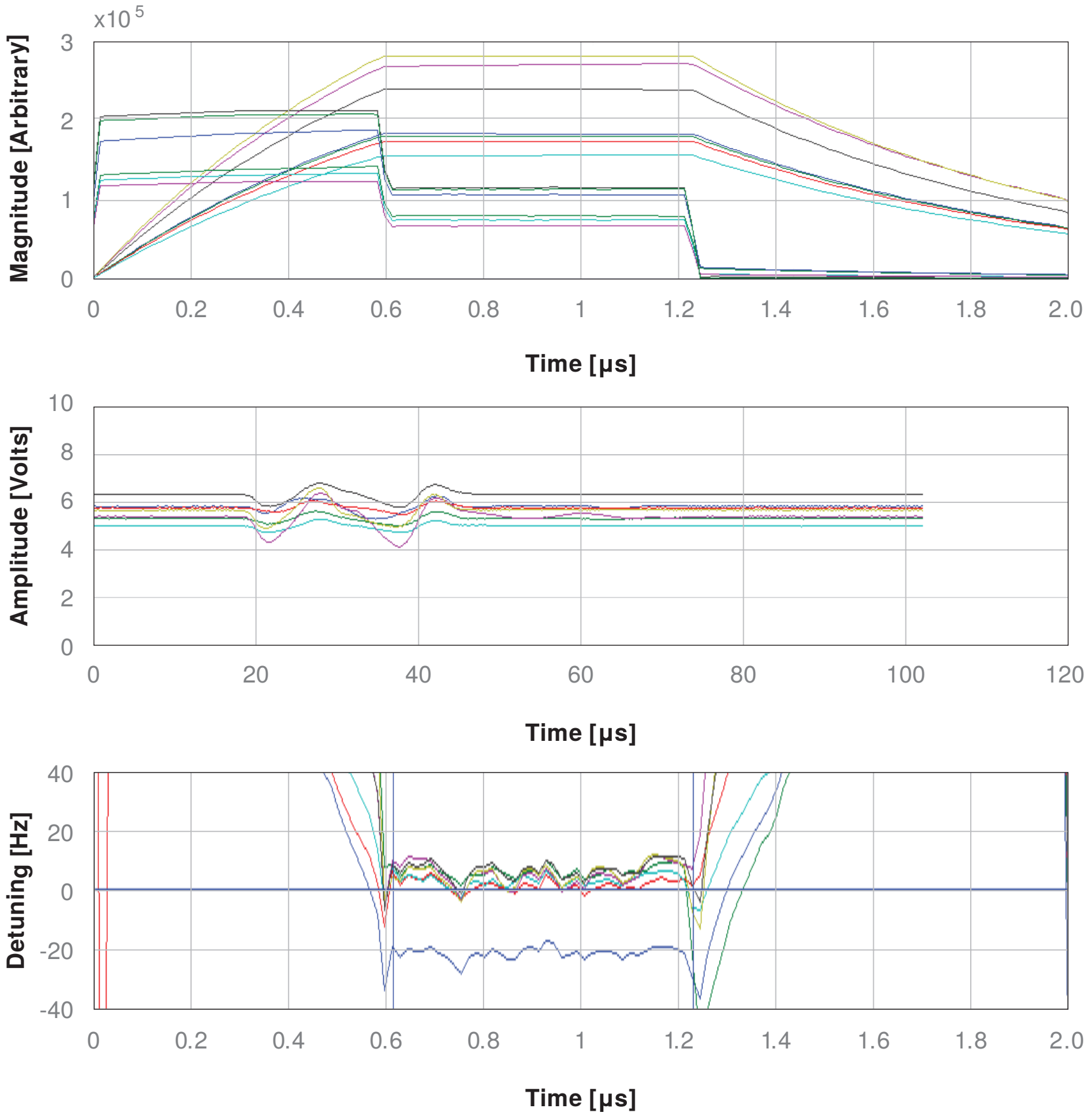

**Figure 3.31** CM1 resonance-control online display during adaptive feed-forward compensation. The top section shows the magnitude of the RF waveforms of the cavities; the middle action shows the piezo monitor waveform; the 10 V full vertical scale of the piezo monitor signal corresponds to a piezo drive of 200 V. The bottom section shows the cavity detuning calculated from the RF waveforms.

## 3.4.4 Ancillary Systems and Findings

All of the necessary subsystems for operating and protecting the cryomodule performed largely as expected. The protection system appropriately limited power or triggered system shutdowns as conditions warranted. The cryogenics plant suffered only minor interruptions, usually due to external factors.

Diagnostics were in place to monitor field emission as well as the presence of dark current emanating from the cavities. These proved to be measurable, but not limiting factors in performance.

## 3.4.5 9-millisecond-Pulse Tests

Once the Cryomodule was well characterised, a 'long pulse' study was carried out in support of an R&D program for a proposed continuous-wave SRF accelerator at Fermilab. This effort, a proof or principle, entailed operating the two best performing cavities at a pulse length of up to 9 ms and determining the LLRF and resonance-control capabilities at various input powers and QL values. Planned operating parameters were:

- RF power limitations: 80 kW; 100 kW, 120 kW per two cavities;
- external $Q_L$: $3 \times 10^6$; $6 \times 10^6$; $1 \times 10^7$;
- gradient: 15 MV/m; 20 MV/m; 25 MV/m;
- 9 ms pulse width.

This study proved to be a good first-pass test. Lorentz-force detuning (LFD) Compensation at the nominal parameters Q=$1 \times 10^7$ and 25 MV/m was demonstrated. The LLRF feedback worked with phase stability good to $\pm$ 4°. The total range of detuning was of order $\pm$10 Hz peak-to-peak with comparable contribution from microphonics and residual LFD. The residual LFD's shape of detuning was repeatable from pulse to pulse with major harmonics $\sim$ 1 kHz. It is expected that LFD compensation can be improved by cancelling out this component. The measured microphonics level was in range 2-4 Hz, similar to what was measured for short 1 ms pulses. Although the behaviour of cavities #5 and #6 were different, the results of LFD compensation were comparable.

Areas requiring further work were uncovered. Input power limitation for the low Q case ($\sim 3 \times 10^6$) limited the peak gradients to $<$ 20 MV/m. The system also proved to be highly sensitive under dynamic conditions. Nearly constant attention was required to maintain stable operation, especially when adjusting the power. Spontaneous quenching resulted from power adjustments without also adjusting LLRF system gains.

### 3.4.6 Future Prospects

The CM-1 module is the first of a series of cryomodules planned to be installed and tested at Fermilab. The second cryomodule is also a TTC Type III+, containing cavities that all achieved gradients of 35 MV/m when tested vertically and 31.5 MV/m when 'dressed' and tested at Fermilab's Horizontal Test Stand. Cooldown and first operation of this cryomodule, designed to meet ILC gradient specifications, is expected in late 2012. Most components are in hand and some assembly at Fermilab has begun on a third cryomodule, which will be a high-gradient TTC Type-IV cryomodule.

### 3.4.7 Summary

The CM-1 cryomodule was the first to be assembled at Fermilab, and as such was also the first cryomodule tested at the recently constructed New Muon Lab (NML) facility. The overall goal of demonstrating facility operations and cryomodule test capability was achieved successfully. The CM-1 module was operated stably near the peak achievable gradients with both LLRF feedback and LFDC enabled for an extended time. RF amplitude and phase were controlled within specification and detuning was successfully compensated over the full gradient range. Valuable experience was gained in both the commissioning and control of multi-cavity SCRF modules. A problem associated with a tuner prevented one cavity from operating on-resonance. A thermal cycle resulted in a vacuum leak in a feedthru. After repair and resuming cold operation the cryomodule showed only minor operating changes.

The next cryomodule, CM-2, is under fabrication and will be a complete ILC design with cavities that all reached the GDE gradient goals. Testing is anticipated in 2013.

## 3.5 CesrTA and Electron-cloud R&D

### 3.5.1 Introduction to the Electron-cloud R&D Program

One of the principal R&D issues for the positron damping ring of the ILC is to ensure that the build-up of the electron cloud (EC)) in the vacuum chambers can be kept below the levels at which EC-induced emittance growth and beam instabilities occur. During Phase I (2008-2010) of the ILC Technical Design Phase (TDP), a focused effort to study methods of suppressing the EC as well as measuring its impact on ultra-low emittance beams was undertaken at the Cornell Electron-Positron Storage Ring Test Accelerator (CesrTA). In addition, a complementary R&D program has continued at various laboratories around the world to develop better techniques to mitigate the build-up of the electron cloud. Section 3.5.2 describes the research effort that has been carried out at Cornell University by the CesrTA collaboration [131, 132], while Section 3.5.3 describes the work that has taken place at other laboratories around the world. As part of this coordinated global programme, a major emphasis has been placed on developing and benchmarking simulation tools as well as measurement techniques.

In order to incorporate the research results into the ILC damping ring (DR) design, an ECLOUD Working Group was formed whose main objective has been to provide recommendations on the EC mitigation techniques to apply to the DR design based on the results of the R&D programme [133, 134]. A dedicated Working Group meeting [135] was held during the ECLOUD10 Workshop [136], with a significant level of participation by experts attending the workshop. The recommendations, which have since been implemented in the ILC DR vacuum-system conceptual design [137], are summarised in Section 3.5.4.

## 3.5.2 The CesrTA R&D Programme

The CesrTA research programme was approved in late 2007 to carry out electron-cloud R&D in support of the ILC technical design. The first dedicated experiments using the Cornell Electron-Positron Storage Ring (CESR) began in March 2008 at the conclusion of 28 years of colliding beam operations for the CLEO experiment [138]. Two principal goals were specified for the programme. The first was to characterise the build-up of the electron cloud in each of the key magnetic-field regions of the accelerator, particularly in the dipoles and wigglers, and to study the most effective methods of suppressing it in each of these regions. This required the design and installation of detectors to study the local build-up of the cloud in each of these environments as well as a supporting simulation programme to fully characterise and understand the results. The second goal was to study the impact of the electron cloud on ultra-low-emittance beams. The ILC damping-ring design targets a geometric vertical emittance of 2 pm rad; no positron ring has yet been operated in this emittance regime. By benchmarking electron-cloud instability and emittance-growth simulations in a regime closer to that specified for the damping ring, confidence in projections of the final damping-ring performance could be significantly improved. This in turn helped to determine how much further R&D was required to achieve the necessary design specifications. In order to carry out these measurements, CESR had to be reconfigured as a damping ring and upgraded with the necessary beam instrumentation for low-emittance optics correction and characterisation of the resulting beams.

### 3.5.2.1 Conversion of CESR to a Damping-Ring Test-Accelerator Configuration

Modification of CESR into a damping-ring configuration involved three main thrusts:

1. Relocation of six of the twelve CESR-c damping wigglers [139, 140], to the L0 straight section in CESR to enable ultra-low-emittance CesrTA operation [138];
2. Upgraded beam instrumentation to achieve and characterise ultra-low-emittance beams, including deployment of a system for high-resolution beam-position monitoring [141] and X-ray beam-size monitors for both positron and electron beams [142];
3. Addition of vacuum-system diagnostics for characterisation of local electron-cloud growth in a range of vacuum chambers, including retarding-field analysers [143, 144], transverse-electric-wave transmission hardware [145] and shielded pickups for time-resolved measurements [146].

Table 3.7 shows the CesrTA lattice parameters for operation at 2 GeV and 5 GeV. At 2 GeV, $\sim 90\,\%$ of the synchrotron-radiation power is provided by the twelve damping wigglers and a horizontal design emittance of 2.6 nm rad is obtained [147]. During Phase I of the CesrTA programme, a vertical-emittance target of less than 20 pm rad (ten times the ILC damping-ring vertical-emittance target) was specified. A key element of the R&D programme has been the flexibility of CESR operation. CESR allows operation between 1.8 and 5.3 GeV with both positron and electron beams. The ability to operate over a wide range of energies, bunch spacings and bunch intensities has enabled systematic studies of primary photoelectron and secondary electron contributions to electron-cloud build-up in the vacuum chambers, which are not feasible at any other facility.

**Table 3.7**
2 GeV and 5 GeV lattice parameters for CesrTA.

| Parameter | Symbol | Unit | 2 GeV | 5 GeV |
|---|---|---|---|---|
| Energy | $E_{\mathrm{beam}}$ | GeV | 2.085 | 5.0 |
| Number of wigglers | | | 12 | 6 |
| Wiggler peak field | | T | 1.9 | 1.9 |
| Horizontal tune | $Q_{\mathrm{x}}$ | | 14.57 | |
| Vertical tune | $Q_{\mathrm{y}}$ | | 9.6 | |
| Longitudinal tune | $Q_{\mathrm{z}}$ | | 0.075 | 0.043 |
| RF voltage | $V_{\mathrm{RF}}$ | MV | 8.1 | 8 |
| Horizontal emittance | $\epsilon_{\mathrm{x}}$ | nm rad | 2.6 | 35 |
| Damping time constant | $\tau_{\mathrm{x,y}}$ | ms | 57 | 20 |
| Momentum compaction | $\alpha_{\mathrm{p}}$ | | $6.76 \times 10^{-3}$ | $6.23 \times 10^{-3}$ |
| Bunch length | $\sigma_{\mathrm{l}}$ | mm | 9.2 | 15.6 |
| Relative energy spread | $\sigma_{\mathrm{E}}/E$ | % | 0.81 | 0.93 |
| Bunch spacing | $t_{\mathrm{b}}$ | ns | $\geq 4$, steps of 2 | |

A novel element of the CesrTA upgrade has been the development of a high-resolution X-ray beam-size monitor capable of single-pass measurements of each bunch in a train. Figure 3.32 shows one of the indium-gallium-arsenide detectors wire-bonded to its circuit board along with a single-pass fit of data acquired using pinhole imaging with a 1 mA bunch. In addition to pinhole imaging, coded aperture and Fresnel-zone-plate optics have also been installed in both the positron and electron beam lines. These detectors are the principal tools for verifying the vertical beam size in the ultra-low-emittance machine optics.

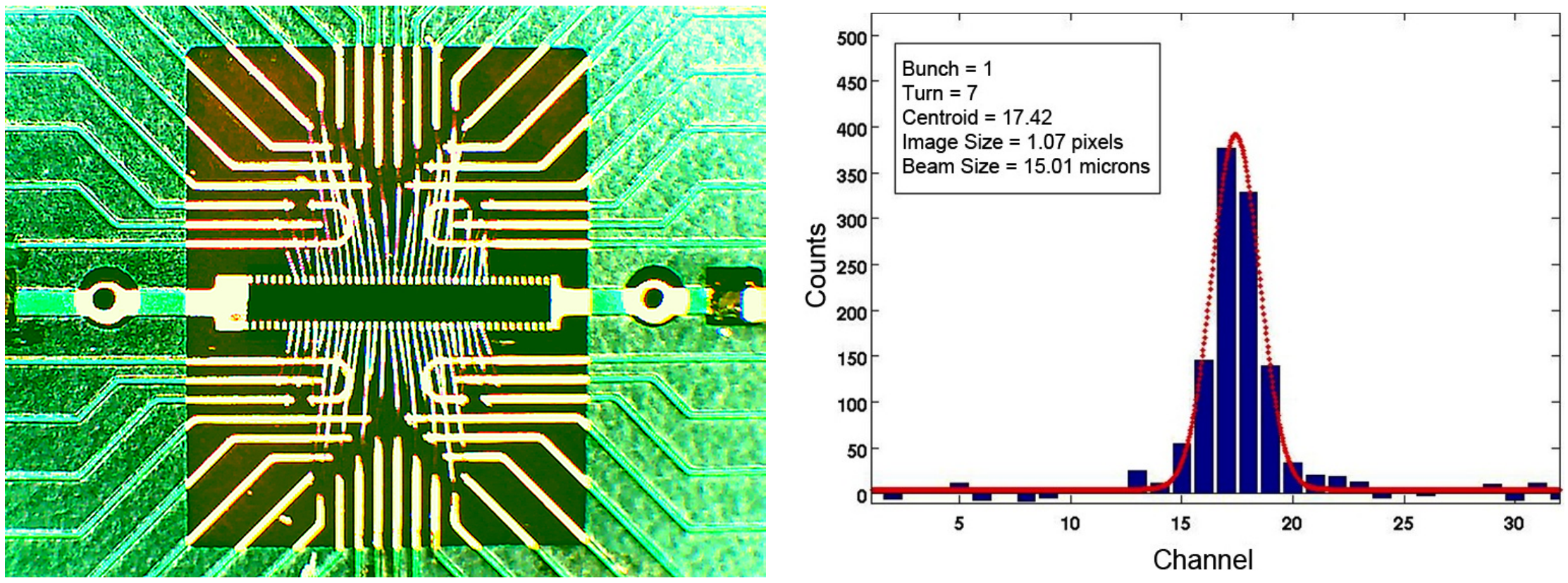

(a) (b)

**Figure 3.32.** (a) An X-ray beam-size monitor detector, an indium-gallium-arsenide diode array, mounted on its circuit board. Each detector has 32 diodes of 400 μm width and 50 μm pitch. (b) A single-turn fit to data acquired from a bunch with $0.8 \times 10^{10}$ particles (at 2.1 GeV beam energy) using a heavy-metal slit as the X-ray imaging optic.

Figure 3.33 shows the layout of the CESR L0 straight section after installation of the wiggler string. This region is one of four dedicated CesrTA electron-cloud experimental areas. It is equipped with extensive diagnostics to study the growth and mitigation of the electron cloud in wigglers. A second electron-cloud experimental region was installed on the opposite side of CESR in the L3 straight section. Figure 3.34 shows the layout of the L3 region. It supports four electron-cloud experiments: a large-bore quadrupole housing a test chamber; the Positron Electron Project (PEP) II chicane for dipole-chamber tests, which was relocated from SLAC after the early termination of PEP II operations; a drift-chamber test section currently configured for testing titanium-zirconium-vanadium (TiZrV) non-evaporable getter (NEG) test chambers; and an in-situ secondary-electron-yield measurement station, which supports studies of the processing rates and equilibrium properties of secondary-electron yield of various technical surfaces. In addition to the L0 and L3 experimental regions, two arc sections were configured for flexible installation of experimental drift chambers to study the performance of various mitigations in the photon environment of the CESR arcs.

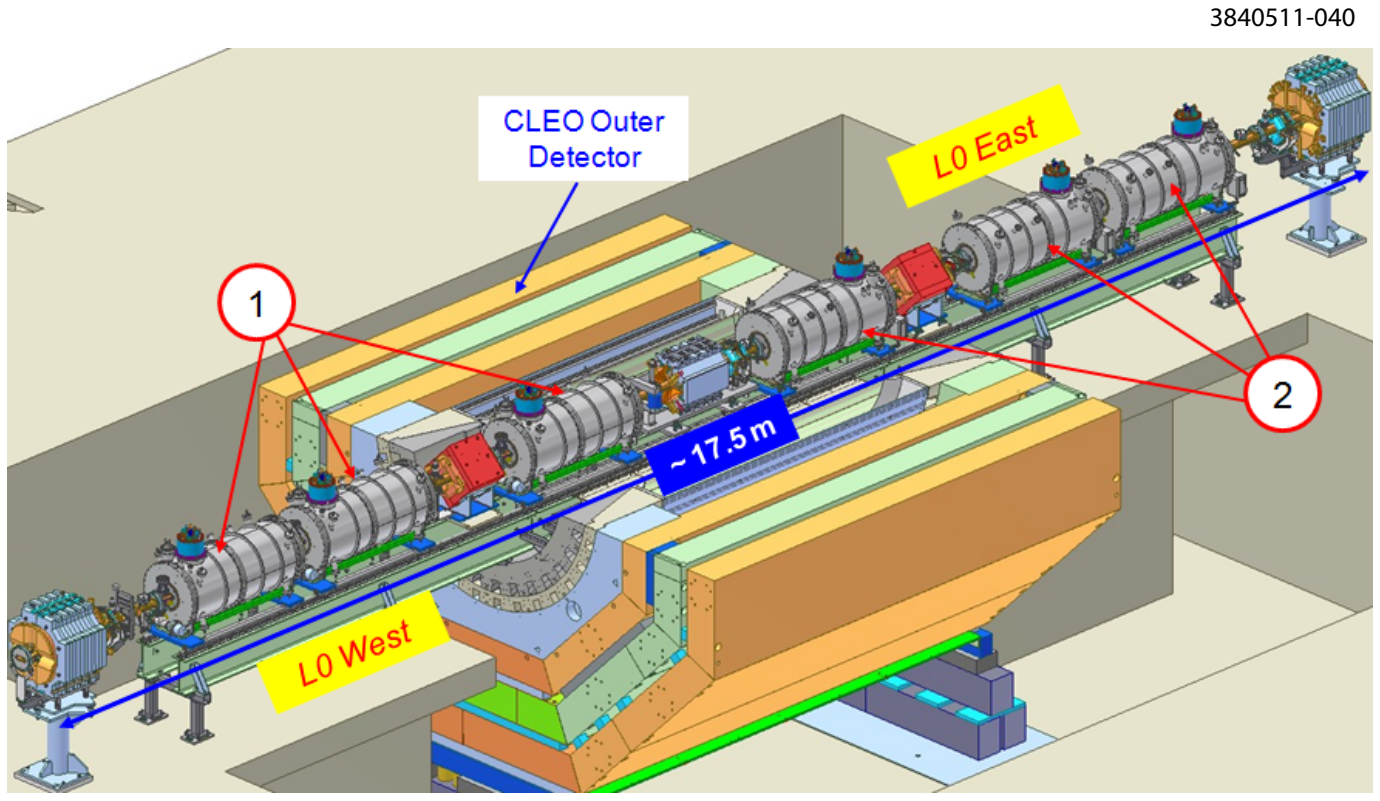

**Figure 3.33**
Layout of the CESR L0 wiggler straight and electron cloud experimental region with a cutaway view of the CLEO detector. Six superconducting CESR-c type wigglers are deployed in the straight, which is configured for zero dispersion operation. Wigglers labeled (1) are downstream wigglers, with respect to the positron beam direction, which are instrumented with retarding field analyzers. Wigglers labeled (2) are unmodified CESR-c type units. The straight section includes extensive vacuum diagnostics: retarding field analysers, residual gas analyser, and transverse electric wave measurement hardware.

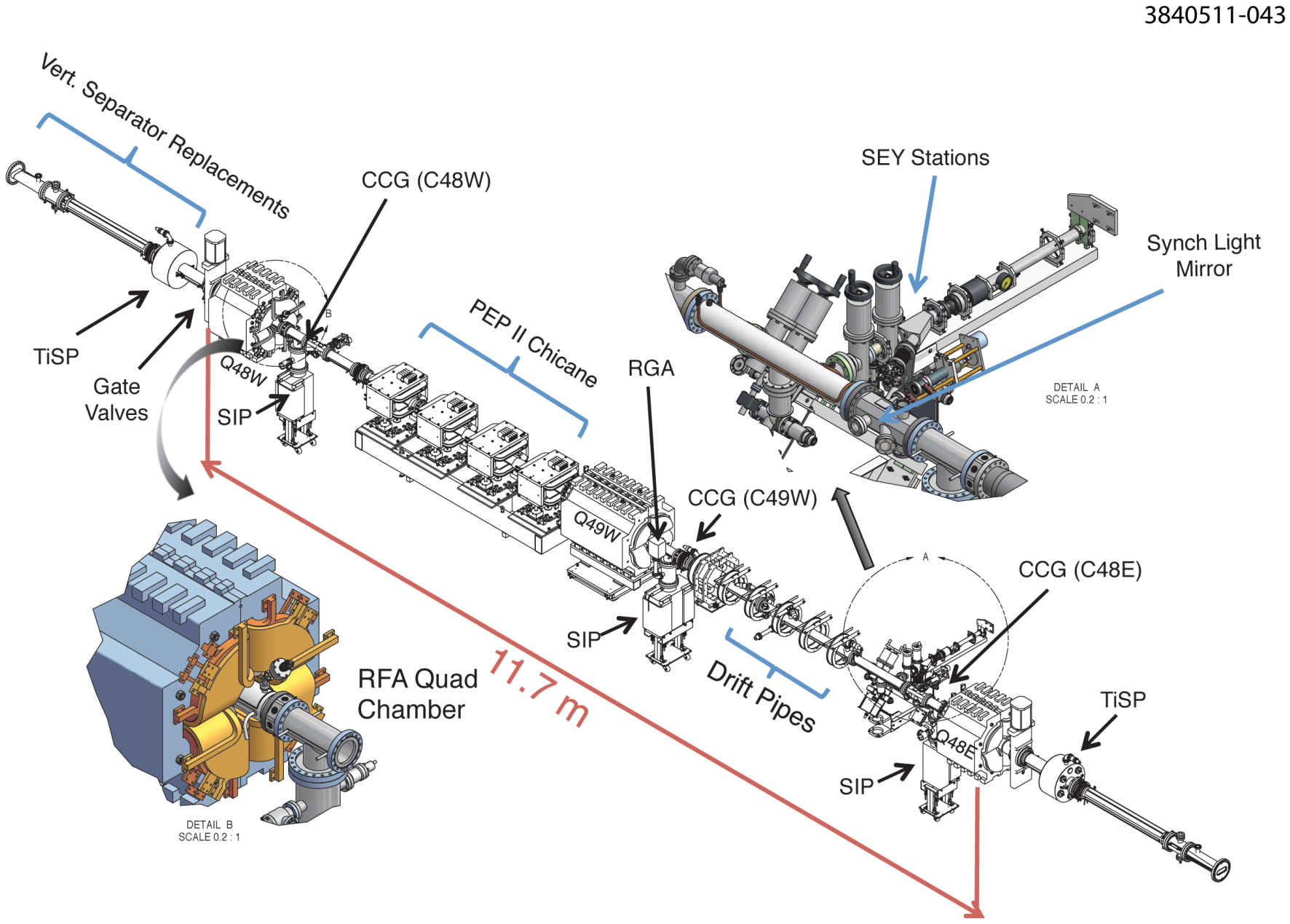

**Figure 3.34**
Layout of the CESR L3 straight and electron-cloud experimental region. Tests of electron-cloud mitigations in drift, dipole and quadrupole chambers are possible in this region. Additionally, an in-situ secondary-electron-yield station is also installed, which allows characterisation of the rate of processing and properties of equilibrium secondary-electron yield of various vacuum-system technical surfaces.

### 3.5.2.2 Electron-Cloud Build-up and Mitigation Studies

Retarding-field analysers deployed at approximately 30 locations around CESR have enabled the detailed study of local cloud build-up in a variety of vacuum chambers under a range of experimental conditions [148, 149]. The analysers provide a time-averaged current readout at each location. The majority of deployed retarding-field analysers utilise a segmented design to provide geometric information about the cloud build-up around the azimuth of the vacuum chamber. Analyser data taken in vacuum chambers fabricated with cloud-mitigation measures provide the foundation for comparison of the efficacy of different methods of electron-cloud mitigation. An active effort continues to model this analyser data in order to constrain experimentally the secondary-electron-yield and photoelectron-yield parameters of the vacuum chambers treated with mitigations in an operating accelerator environment [149–153]. In addition to the retarding-field analyser studies, transverse-electric-wave transmission methods [154–157] are also being used to characterise the build-up around the ring; a significant simulation effort is underway to take full advantage of these results [158–161]. A final method to study local cloud build-up is shielded pickup measurements [162–165], which are providing additional constraints on the vacuum-chamber surface parameters for the chambers in which they are installed. Table 3.8 summarises the range of chamber surfaces and mitigation methods that were prepared for testing during Phase I of the CesrTA R&D programme.

Figure 3.35 shows a comparison of the performance of various chamber surfaces in a dipole field along with a plot of the evolution of the transverse distribution of the electron cloud that develops in the dipole chamber as a function of the beam current. While coating with a material with low secondary-electron yield, such as titanium nitride, significantly reduces the growth of the cloud in this environment, the use of a grooved surface with titanium-nitride coating is clearly superior.

**Table 3.8**
Vacuum chambers fabricated for testing during the CesrTA R&D programme. Mitigation studies have been conducted in drift, dipole, quadrupole, and wiggler magnetic-field regions. Checks indicate chambers for which data has been acquired. CU stands for Cornell University.

| Mitigation | Drift | Quad | Dip | Wig | Contributing Institutions |
|---|---|---|---|---|---|
| Al | X | X | X | | CU, SLAC |
| Cu | X | | | X | CU, KEK, LBNL, SLAC |
| TiN on Al | X | X | X | | CU, SLAC |
| TiN on Cu | X | | | X | CU, KEK, LBNL, SLAC |
| C on Al | X | | | | CERN, CU |
| Diamond-like C on Al | X | | | | CU, KEK |
| NEG on SS | X | | | | CU |
| Solenoid Windings | X | | | | CU |
| Fins w/TiN on Al | X | | | | CU, SLAC |
| *Triangular Grooves:* | | | | | |
| On Cu | | | | X | CU, KEK, LBNL, SLAC |
| With TiN on Al | | | X | | CU, SLAC |
| With TiN on Cu | | | | X | CU, KEK, LBNL, SLAC |
| Clearing Electrode | | | | X | CU, KEK, LBNL, SLAC |

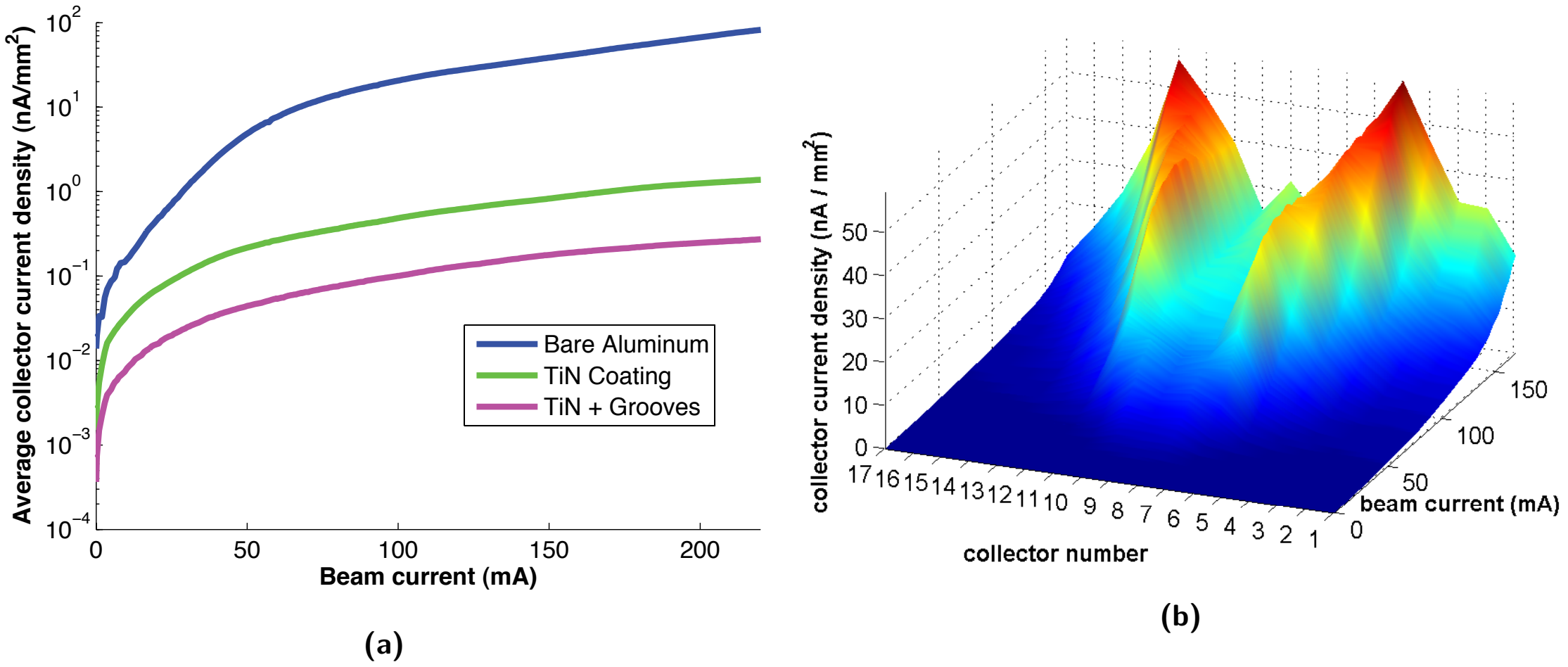

**Figure 3.35.** (a) The measured retarding field analyser current in a dipole versus beam current with a 20-bunch positron train for a bare aluminium surface, titanium nitride-coated surface and a grooved surface with titanium nitride coating. The efficacy of the grooved surface for suppressing the electron cloud is clearly evident. (b) The transverse shape of the electron cloud signal in the dipole retarding-field analyser (aluminium chamber surface) as a function of beam current.

Figure 3.36 shows two of the mitigation methods that have been tested in the CesrTA high-field damping wigglers: triangular grooves and a clearing electrode. The clearing electrode is a very thin structure developed at KEK [166] that offers very good thermal contact with the vacuum chamber and minimal impact on the chamber aperture (see also Section 3.5.3). A bare copper surface and a titanium-nitride-coated copper surface have also been tested. Figure 3.37a shows a comparison of the electron-cloud growth as a function of beam current with each of these surfaces. The data indicate that the best cloud suppression in the wiggler region is obtained with the clearing electrode. Figure 3.37b shows the transverse distribution of the cloud present in the vertical field region of the wiggler (copper surface) as a function of the retarding grid voltage, which probes the energy spectrum of the electron cloud.

Studies of the electron-cloud build-up in drift and quadrupole regions have also yielded important results. Drift measurements have been used to compare the performance of various coatings. A new coating of significant interest is amorphous carbon developed at CERN [167] for use in the Super Proton Synchrotron (SPS). Tests at CesrTA have afforded the opportunity to study the performance of this coating in the presence of synchrotron radiation. Initial studies show that the electron-cloud

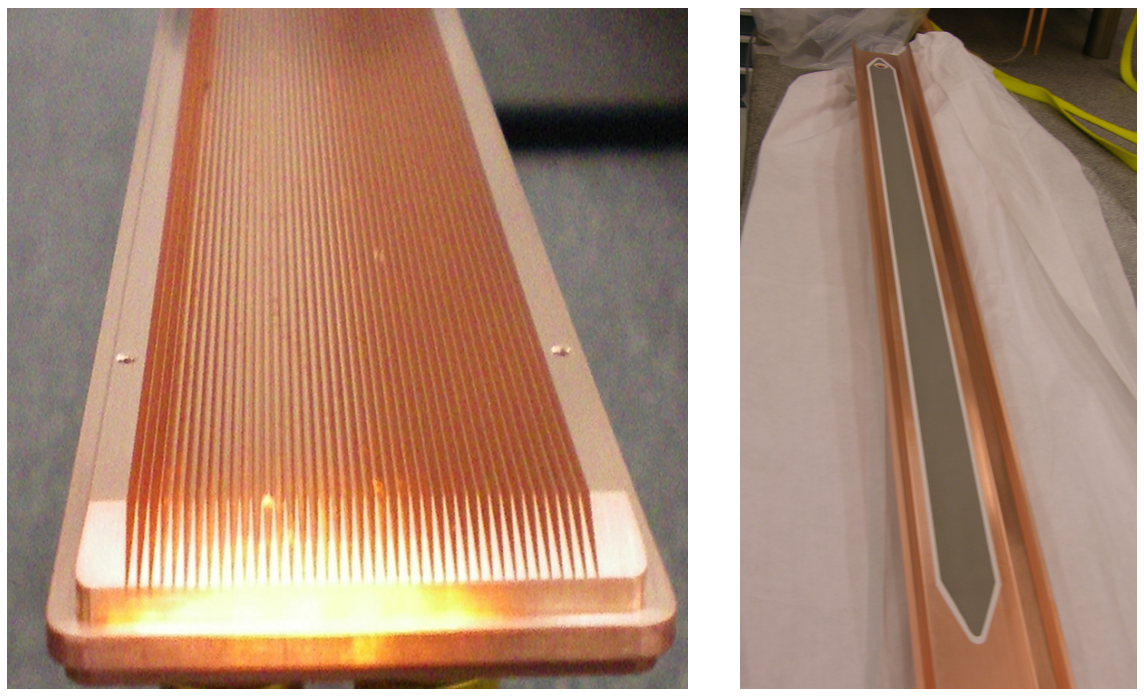

(a) (b)

**Figure 3.36**
(a) A grooved copper insert with 21.8° triangular grooves having 1 mm pitch for testing in a CesrTA wiggler. (b) A thin clearing electrode applied with a thermal-spray method to the bottom half of another CesrTA experimental wiggler chamber.

mitigation performance of carbon is quite comparable to that of titanium nitride and that its vacuum performance is quite reasonable in an environment with significant photon flux. Continued testing will provide information about the long-term durability of this very promising coating. Vacuum chambers in quadrupole magnetic fields can show quite significant cloud build-up. Concerns about long-term trapping of the cloud in quadrupole fields [168] require that cloud mitigation be incorporated into the ILC damping-ring quadrupole vacuum chambers. Tests in CesrTA have demonstrated the effectiveness of titanium-nitride coating in this region.

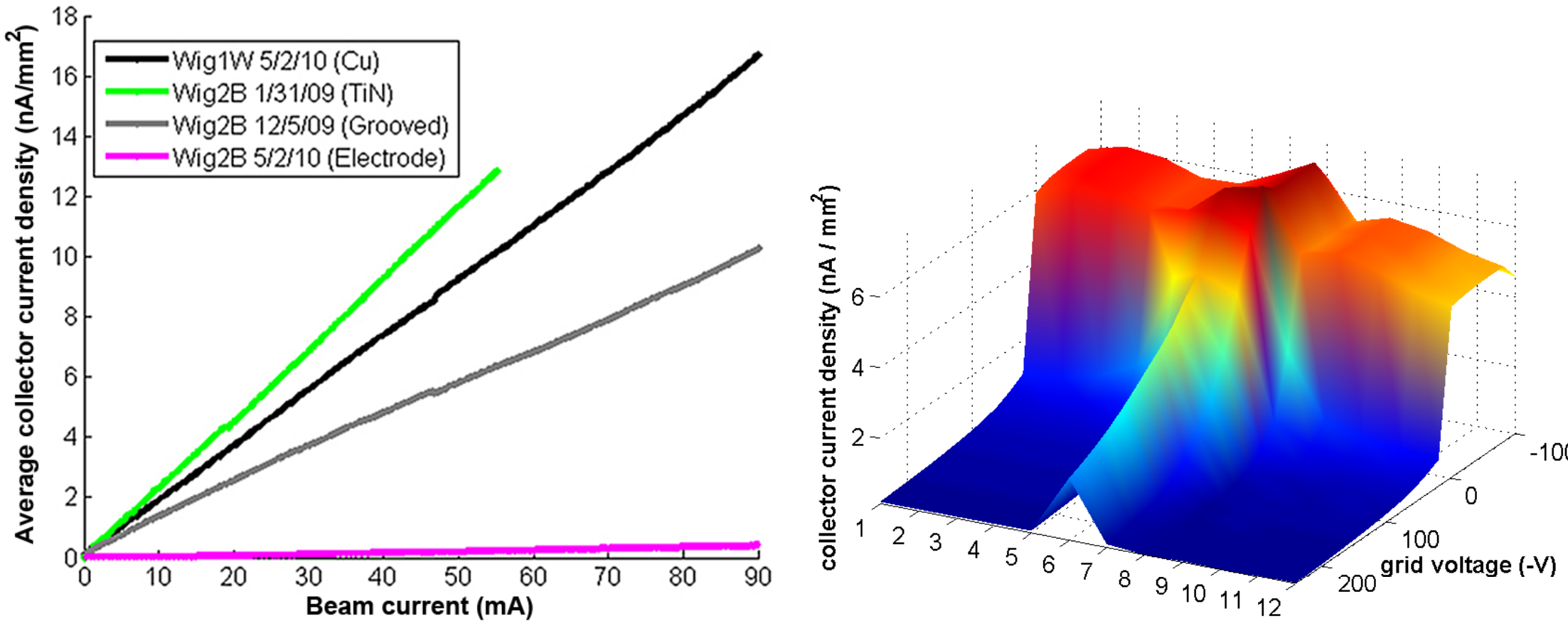

(a) (b)

**Figure 3.37.** (a) The measured retarding-field analyser current in a wiggler versus beam current with a 20-bunch positron train for a bare copper surface, a titanium-nitride-coated copper surface, a grooved copper surface and a clearing electrode. The efficacy of the clearing electrode for suppressing the electron cloud is clearly evident. (b) The transverse shape of the electron-cloud signal in the wiggler retarding-field analyser as a function of retarding voltage.

### 3.5.2.3 Studies of Electron-cloud-induced Beam Dynamics at Low Emittance

The CesrTA low-emittance tuning effort provides the basis for studying the emittance-diluting effects of the electron cloud in a regime approaching that of the ILC damping rings. As of early 2010, the low-emittance tuning programme had resulted in reliable operation at or below the CesrTA Phase I vertical emittance of 20 pm rad [169] for both single- and multi-bunch operation as confirmed by X-ray beam-size-monitor measurements of the vertical beam size [170]. As of the end of 2010, vertical emittances less than 10 pm-rad had been achieved.

A number of beam-dynamics studies have been conducted in order fully to characterise the impact of the electron cloud on beams in CESR. As the electron cloud builds up along a bunch train, the focusing effect of the cloud on the beam causes the natural frequency of oscillation of each bunch (i.e. the horizontal and vertical betatron tunes) to shift with respect to the preceding bunch.

Measurements of this electron-cloud-induced coherent tune shift [171–173] for trains of electron and positron bunches, as well as for witness bunches at various positions behind a leading train, have provided an important probe of the integrated effect of the cloud around the ring. Systematic measurements over a wide range of beam conditions (varying beam energy, emittance, bunch current, bunch spacing and train lengths) have been used to validate electron-cloud models more thoroughly. An important feature of these measurements has been the need to model accurately the photon transport, reflection and absorption around the ring in order to describe the data adequately. This has led to the implementation of a new ring photon propagation package, SYNRAD3D [174], which has most recently been applied to the design of the ILC damping-ring vacuum system and has resulted in important refinements of that design [137].

A principal deliverable of the CesrTA programme is the characterisation of instability thresholds and emittance-diluting effects in the regime of ultra-low vertical emittance [175–179]. Figure 3.38 shows the observed beam-motion spectrum for each bunch along a train obtained in these conditions. As described in the preceding paragraph, the development of the horizontal and vertical tune lines, denoted by $F_h$ and $F_v$, along the bunch train provides information about the electron-cloud density experienced by each bunch. For a positron train, the attractive force of the bunch pinches the cloud into the bunch and can lead to the development of an oscillation of the bunch tail with respect to the head. This head-tail instability is expected to induce characteristic sidebands in the bunch-motion spectrum. In Fig. 3.38, the onset of the spectral lines denoted by $F_v \pm F_s$ part way along the bunch train indicate where the cloud density build-up has become sufficient for the onset of the instability.

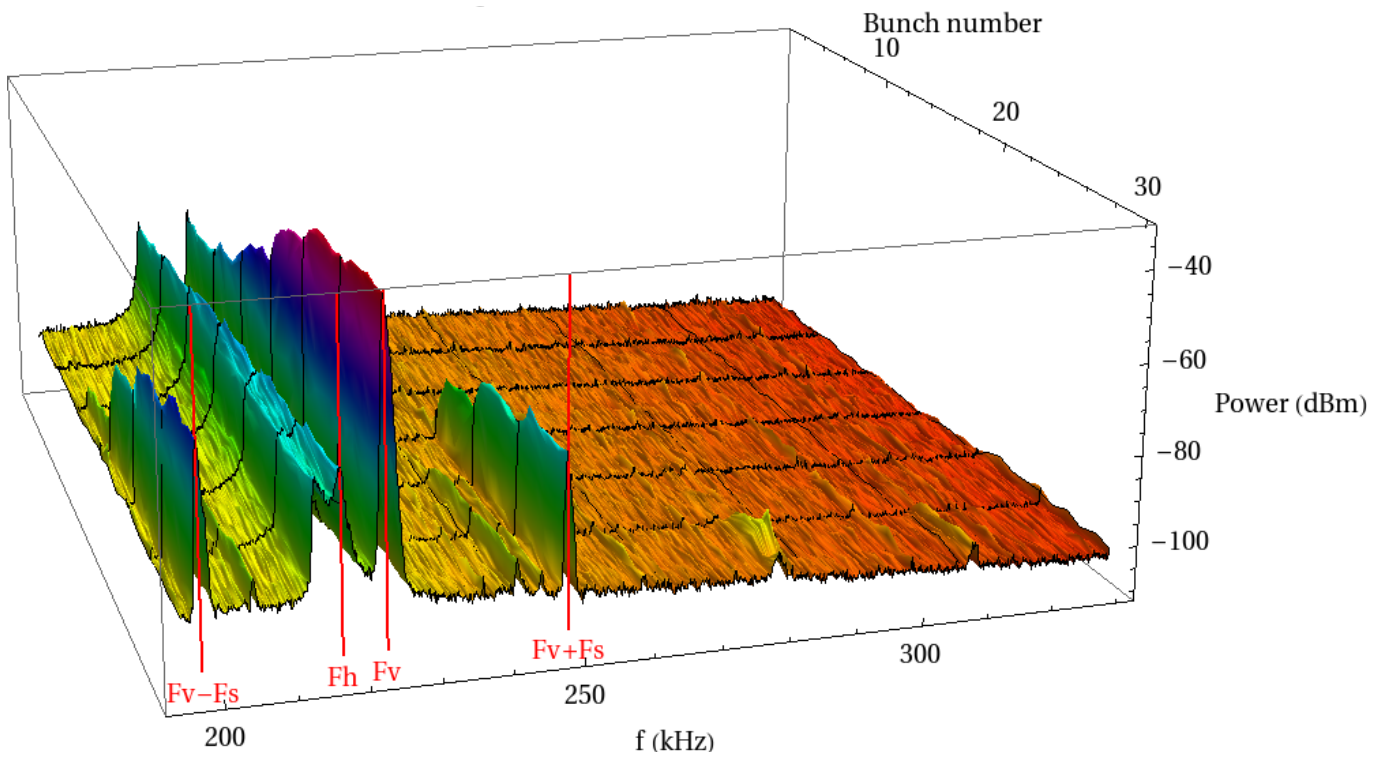

**Figure 3.38**
Bunch-by-bunch power spectrum for a positron train with a nominal bunch current of 0.75 mA/bunch. The horizontal (Fh) and vertical (Fv) tunes are clearly visible for all bunches. The onset of the sidebands labelled as $F_v \pm F_s$ are consistent with the onset of a head-tail instability around bunch number 15 in the train.

A second observable associated with this instability is a growth in the vertical beam size as measured along the train. Figure 3.39 shows bunch-by-bunch beam-size development along bunch trains with three different intensities. As the bunch currents are increased, the bunch number in the train at which beam-size blow-up occurs moves earlier in the train due to the more rapid build-up of the electron cloud. By studying both the spectral and beam-size information as a function of various parameters (bunch intensity, vertical emittance, bunch spacing, chromaticity, feedback conditions, and beam energy) and comparing with simulation [180–183] the simulations in a regime approaching that of the ILC damping ring have been validated, ensuring that the projections of the expected performance of the positron damping ring are accurate.

Analytical estimates of the head-tail instability thresholds, following the calculations described in [184, 185], have been made for CesrTA as well as for the ILC damping ring [8]. Table 3.9 summarises the key parameters and instability thresholds, based on the parameters given in Table 3.10. The observed instability thresholds observed in CesrTA low-emittance experimental conditions are in good agreement with these calculations when the EC density near the beam is determined by tune-shift measurements along the train. This applies both to the appearance of synchrobetatron side-bands in the tune measurements as well as rapid beam-size growth as observed with the xBSM. The good agreement between data and experiment provides confidence in estimates for the ILC DR.

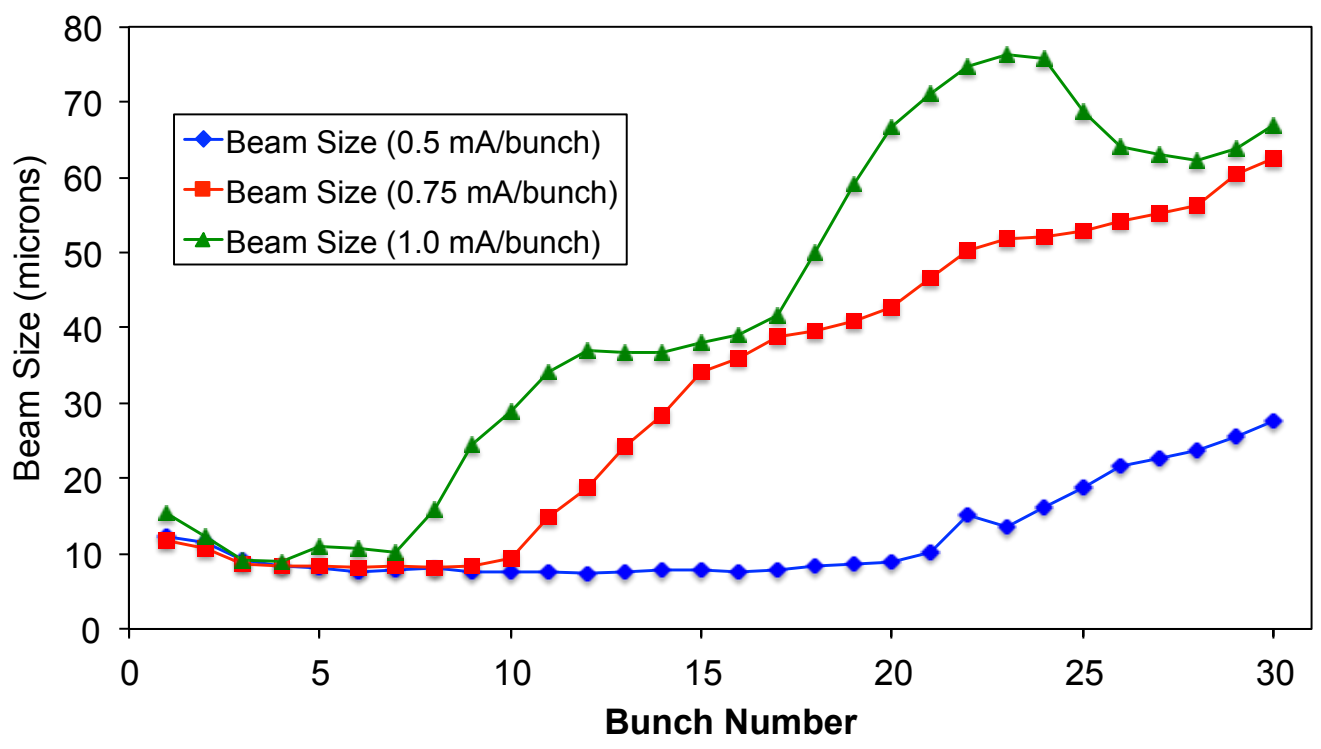

**Figure 3.39**
Bunch-by-bunch beam sizes based on turn-by-turn fits for each bunch for 30 bunch trains of varying current (0.8, 1.2, and 1.6 $\times$ $10^{10}$ particles/bunch). As the bunch currents are increased, the point in the train at which the electron cloud density is high enough to cause emittance and beam-size growth moves to earlier points in the train.

There is some evidence in the beam-size data that there may be sub-threshold emittance growth before the full onset of the head-tail instability. In the CesrTA data, such effects can occur at EC densities which are a few times smaller than the calculated instability thresholds. This suggests that some additional margin may be required in our estimates of the safe EC densities for stable operation of a damping ring and argues for an aggressive approach to the choice of EC mitigation techniques. This is consistent with the recommendations described in Section 3.5.4. Work continues on the simulations and measurements required to clarify this issue.

**Table 3.9**
Analytic estimate of the fast head-tail instability threshold for CesrTA and the ILC damping ring.

| | CesrTA (2 GeV) | CesrTA (5 GeV) | ILC DR |
|---|---|---|---|
| Bunch Population $N_+$ ($\times$ $10^{10}$) | 2 | 2 | 2 |
| Bunch Spacing $\ell_{sp}$ (ns) | 4 | 4 | 6 |
| Average Vertical Beta Function $\beta_y$ (m) | 16 | 16 | 24 |
| Electron Frequency $\omega_e/2\pi$ (GHz) | 35 | 11 | 111 |
| Phase Angle $\chi$ | 8.9 | 3.7 | 14.0 |
| Threshold Density $\rho_{e,th}$ ($\times 10^{12}$ m$^{-3}$) | 0.82 | 3.22 | 0.23 |
| Tune Shift at Threshold $\Delta\nu_{x+y}$ ($\times$ $10^{-3}$) | 9 | 14 | 5 |

**Table 3.10**
Parameters of CesrTA and the ILC damping ring used for instability threshold estimates.

| | CesrTA (2 GeV) | CesrTA (5 GeV) | ILC DR |
|---|---|---|---|
| Circumference $L$ (m) | 768 | 768 | 3245 |
| Energy $E$ (GeV) | 2.1 | 5.0 | 5.0 |
| Bunch Population $N_+$ ($\times$ $10^{10}$) | 2 | 2 | 2 |
| Emittance $\varepsilon_x$ (nm) | 2.6 | 40 | 0.45 |
| Momentum Compaction $\alpha$ ($\times$ $10^{-4}$) | 67.6 | 62.0 | 3.3 |
| RMS Bunch Length $\sigma_z$ (mm) | 12.2 | 15.7 | 6 |
| RMS Energy Spread $\sigma_E/E$ ($\times$ $10^{-3}$) | 0.80 | 0.94 | 1.09 |
| Horizontal Betatron Tune $\nu_x$ | 14.57 | 14.57 | 47.37 |
| Vertical Betatron Tune $\nu_y$ | 9.62 | 9.62 | 28.18 |
| Synchrotron Tune $\nu_s$ | 0.055 | 0.0454 | 0.031 |
| Damping Time $\tau_{x,y}$ (ms) | 56.4 | 19.5 | 24 |

#### 3.5.2.4 CesrTA Inputs to the ILC DR Technical Design

The results from the first three years (Phase I) of the CesrTA R&D programme have been incorporated into the design of the ILC damping-ring vacuum chamber [137]; the findings of the programme are documented in the CesrTA Phase I Report [8]. In particular, the observed efficacy of grooved chamber surfaces in the dipoles as well as that of the clearing electrode in the high-field wigglers provide confidence that practical electron-cloud mitigation measures can be prepared for the arc and wiggler straight regions of the ILC positron damping ring. The importance of cloud mitigation in the damping-ring quadrupole chambers has also been demonstrated. New coating technologies to suppress the secondary-electron yield show great promise. However, there is still the issue of studying the long-term performance and durability of these coatings. This will be a subject of study during Phase II of the CesrTA programme. Perhaps most importantly, the flexibility of CESR operations has enabled a systematic programme of electron-cloud build-up and electron-cloud-induced beam-dynamics studies. By benchmarking physics models and simulations against these studies, confidence in being able to make valid projections of the expected ILC positron-damping performance has been significantly enhanced.

### 3.5.3 Electron-cloud R&D at Other Laboratories

During 2007 and 2008 in the Positron Low Energy Ring of the PEP II accelerator, a magnetic chicane and special vacuum chambers were installed to study electron-cloud effects in an accelerator beamline [186, 187]. A special chamber was used to monitor the secondary-electron yield of titanium-nitride and NEG coatings, copper, stainless steel and aluminium under the effect of electron and photon conditioning in situ in the beam line. A drastic reduction of the secondary-electron yield to approximately 0.95 for titanium nitride and a still-high value for aluminium of greater than 2.0 after exposure in the accelerator beam line was measured. Other vacuum-chamber materials including NEG-coated samples were also measured. In magnetic-field-free regions, chambers were installed with rectangular groove profiles meant to reduce the secondary-electron generation at the surface. The electron signals in the grooved chambers, when compared to signals in smooth chambers, were significantly reduced. Two important results in dipoles were reported from the electron-cloud chicane tests : 1) the titanium nitride coating reduces the cloud density by several orders of magnitude with respect to a bare aluminium surface; and 2) a new resonance phenomenon has been observed that results in the modulation of the electron wall flux, and hence, presumably, of the electron-cloud density. After the PEP II shutdown the magnetic chicane and the test chambers were installed in the CesrTA ring (see Section 3.5.2.1) to continue the cloud-mitigation studies.

Tests of coated chambers, grooves and clearing electrodes have been carried out at KEK in order to mitigate the electron-cloud instability in an intense positron ring [166, 188, 189]. Aiming for application in a dipole-type magnetic field, various shapes of triangular grooved surfaces have been studied. In a laboratory, the secondary-electron yields of small test pieces were measured using an electron beam in the absence of magnetic fields. The grooved surfaces clearly had low secondary-electron yield compared to flat surfaces of the same materials. The grooves with sharper vertices had smaller secondary electron yield. A test chamber installed in a wiggler magnet of the KEKB positron ring was used to investigate the efficacy of the grooved surface in a strong magnetic field. In the chamber, a remarkable reduction in the electron density around the beam orbit was observed compared to the case of a flat surface with titanium-nitride coating.

An electron-clearing electrode with an ultra-thin structure has been developed. The electrode was tested with a positron beam in KEKB. A drastic reduction in the electron density around the beam was demonstrated in a wiggler magnet with a dipole-type magnetic field of 0.78 T. No discharge or extra heating of the electrodes and feedthroughs was observed after using the latest connection structure. The same type of electrode was also successfully tested in a CesrTA wiggler (see Section 3.5.2.1). The clearing electrode has also been applied to a copper beam-pipe with antechambers in preparation for its application in the wiggler section of Super-KEKB. Simulations indicate a small impedance for the thin structure of this electrode design.

At the INFN Frascati National Laboratories in Italy, clearing electrodes to mitigate the electron-cloud instability have been installed in all the dipole and wiggler chambers of the DAΦNE positron ring, covering approximately 16 % of the circumference [190]. The electrodes have proven very effective in combatting the electron cloud and minimising its impact on the positron beam dynamics [191].

At CERN, carbon thin films have been applied to the liners in the electron-cloud monitors and to vacuum chambers of three dipole magnets in the SPS [167]. The electron cloud is completely suppressed for LHC-type beams in the liners even after three months of air venting, and no performance deterioration is observed after one year of SPS operation. Following the positive preliminary results obtained at the SPS it was decided to test these types of coatings in an environment with high synchrotron radiation in a lepton machine at CesrTA (see Section 3.5.2.1).

### 3.5.4 Recommendations for Electron-cloud Mitigation

A working group has been set up to evaluate the electron-cloud effect and instability issues for the ILC positron damping ring and to recommend mitigation solutions. The collaborating institutions are Argonne National Laboratory, CERN, Cornell University, INFN, KEK, Lawrence Berkeley National Laboratory and SLAC. The first task of the working group was to compare the electron-cloud effect for two different damping-ring designs with 6.4 km and 3.2 km circumferences, respectively, and to investigate the feasibility of the shorter damping ring with respect to the electron-cloud build-up and related beam instabilities. The working group compared the instability thresholds and the electron-cloud formation assuming 6 ns bunch spacing in both configurations, that is, in the same beam current. Both ring configurations were found to exhibit very similar performance. The risk associated with the adoption of the 3.2 km damping-ring design, while maintaining the same bunch spacing, was deemed low and the 3.2 km ring was found to be an acceptable baseline design choice [134].

**Table 3.11** Summary of the *Recommendations for Electron-Cloud Mitigation* for the ILC positron damping ring.

| Field Region | Mitigation Recommendation | | Alternatives for Further Study |
|---|---|---|---|
| | *Primary* | *Secondary* | |
| Drift† | TiN Coating | Solenoid Windings | NEG Coating |
| Dipole | Triangular Grooves with TiN Coating | Antechambers for synchrotron radiation power loads and photoelectron control | R&D into the use of clearing electrodes |
| Quad† | TiN Coating | | R&D into the use of clearing electrodes or grooves with TiN coating |
| Wiggler | Clearing Electrode | Antechambers for synchrotron radiation power loads and photoelectron control | Grooves with TiN coating |

† Where drift and quadrupole chambers are in arc or wiggler straight regions of the machine, the chambers will incorporate features of those sections, antechambers for power loads and photoelectron control.

The mitigation recommendations for the ILC damping rings were prepared during a meeting of the working group at Cornell University on 13 October 2010 as a satellite meeting to the ECLOUD10 Workshop. The input from the workshop participants was included in the evaluation. The results of the evaluation were presented at the IWLC10 Workshop at CERN [192], and the recommendations are summarised in Table 3.11 [193]. It should be noted that the choices of mitigation methods in Table 3.11 are nearly identical to the choices that have been made for the construction of the SuperKEKB [194] vacuum system. Thus operation of the SuperKEKB positron ring will serve as a crucial performance test that will further improve understanding of the anticipated performance of the ILC DR.

### 3.5.5 CesrTA Phase II Plans

Specification of the electron-cloud mitigation, magnet alignment tolerances, emittance-tuning procedure, and beam instrumentation for the damping ring is to a large extent based on the CesrTA findings. There remain, however, a number of outstanding questions regarding beam physics peculiar to low-emittance rings that CesrTA is well equipped to address. In particular, as regards growth of the electron cloud, while global measurements of cloud lifetime with tune-shift data, as well as local measurements with shielded pickups, suggest that the cloud decays on a timescale of hundreds of nanoseconds, there is a hint of a long-lived cloud with lifetime of many microseconds, possibly trapped in quadrupole and/or wiggler fields. With respect to beam-cloud dynamics, whereas emitttance growth due to the electron cloud as well as cloud-induced head-tail instability has been observed, the former by direct measurement of beam size and the latter from the turn-by-turn bunch-position spectra, it must yet to be determined to what extent the two phenomena are related. The CesrTA team is developing both new instrumentation and computational tools to explore these and other aspects of electron-cloud physics.

As ever smaller emittance and higher charge density are achieved, the sensitivity to single particle as well as collective effects that can limit vertical emittance will be enhanced. Investigation of intra-beam scattering at CesrTA takes advantage of the instrumentation that allows simultaneous measurement of horizontal, vertical, and longitudinal bunch dimensions. Initial results show reasonable agreement with theory, but with some remaining discrepancies. Discriminating emittance blow-up due to intra-beam scattering from other collective effects will require more extensive study, including measurement of the current dependence of equilibrium emittance at different beam energies, and complementary measurements with both positron and electron beams. CesrTA is an excellent laboratory for investigating ion effects in electron beams, and in particular the fast ion instability in bunch trains in the ultra-low vertical-emittance regime. The same tools used to explore electron-cloud effects with positron beams are suitable for investigation of ion effects with electron beams.

Finally the work to collect RFA and SPU data during operation of CESR as an x-ray source for CHESS will be continued. Analysis of the data will inform models of electron-cloud growth and questions of the durability of mitigations and the long-term effects of beam processing.

## 3.6 ATF2 Final-Focus Experiment

### 3.6.1 Introduction

The challenge of colliding nanometre-sized beams at the interaction point (IP) involves three distinct issues:

- creating small emittance beams;
- preserving the emittance during acceleration and transport, and finally;
- focusing the beams to nanometers before colliding them.

**Table 3.12**
Main design parameters for ATF2 compared to ILC. The ATF2 37 nm (at the IP) includes residual effects from uncorrected higher-order optical aberrations.

| Parameter | | Unit | ATF2 | ILC |
|---|---|---|---|---|
| Beam energy | $E$ | GeV | 1.3 | 250 |
| Effective focal length | $L^*$ | m | 1 | 3.5 - 4.5 |
| Horizontal emittance | $\epsilon_x$ | nm | 2 | 1.0 (damping ring) |
| Vertical emittance | $\epsilon_y$ | pm | 12 | 2 (damping ring) |
| Horizontal IP $\beta$ function | $\beta_x^*$ | mm | 4 | 21 |
| Vertical IP $\beta$ function | $\beta_y^*$ | mm | 0.1 | 0.4 |
| Horizontal IP dispersion divergence | $\eta_x'$ | | 0.14 | 0.0094 |
| Relative energy spread | $\sigma_E$ | % | $\sim 0.1$ | $\sim 0.1$ |
| Vertical chromaticity | $\xi_y$ | | $\sim 10^4$ | $\sim 10^4$ |
| RMS horizontal beam size | $\sigma_x^*$ | µm | 2.8 | 0.655 |
| RMS vertical beam size | $\sigma_y^*$ | nm | 37 | 5.7 |

The Accelerator Test Facility (ATF) at KEK is a prototype damping ring to deal with the first issue and has succeeded in obtaining the emittances that almost satisfy ILC requirements [195, 196]. ATF is now used as an injector for the ATF2 final-focus test beam line, which was constructed in 2008 to study the third issue. ATF2 is a follow-up to the Final-Focus Test Beam (FFTB) experiment at SLAC [197], but with different beam-line optics based on a scheme of local chromaticity correction [198]. The beam line is shorter than that originally tested at FFTB and promises better performance and greater extendibility to higher energies. As with FFTB, the value of $\beta_y^*$ and hence the vertical beam size at the optical focal point are chosen to have a similar demagnification and yield a similar chromaticity as in the ILC final focus. The primary goals for ATF2 are [199, 200]:

1. achieving a 37 nm vertical beam size at the IP;

2. stabilising the beam at that point at the nanometer level.

The main parameters of ATF2 are given in Table 3.12 together with the corresponding values for the ILC.

The layout of the ATF/ATF2 facility and the design optical functions of the ATF2 beam line are displayed in Fig. 3.40 and Fig. 3.41 , respectively. The optics is a scaled-down version of the ILC design.

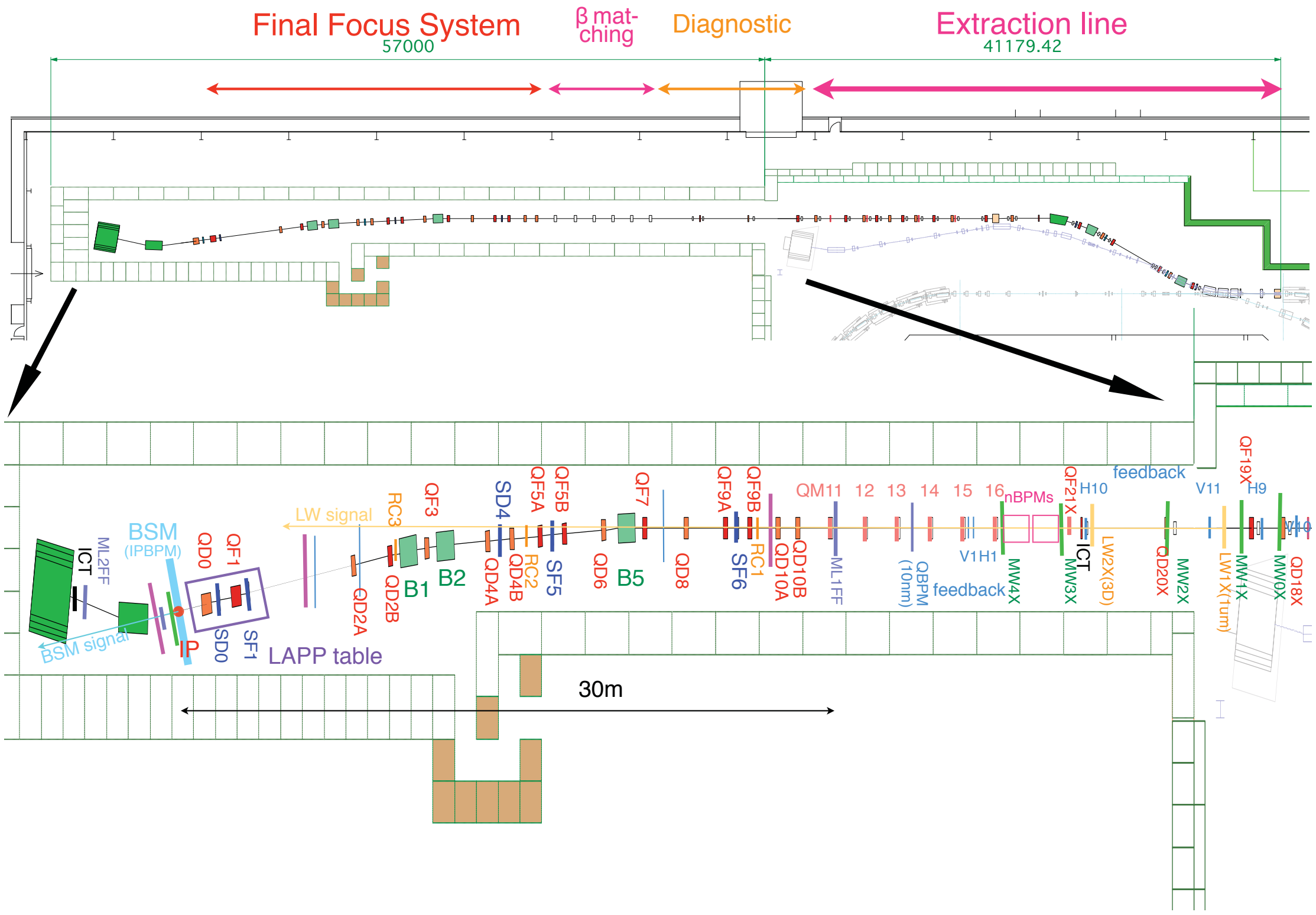

**Figure 3.40.** Top: the ATF2 beam line. Bottom: an enlargement of the final focus system.

**Figure 3.41**
Top: Optics of the ILC beam-delivery system from the exit of the main linac on the right to the interaction point on the left. Bottom: ATF2 optics from the ATF damping-ring extraction point on the right to the interaction point on the left.

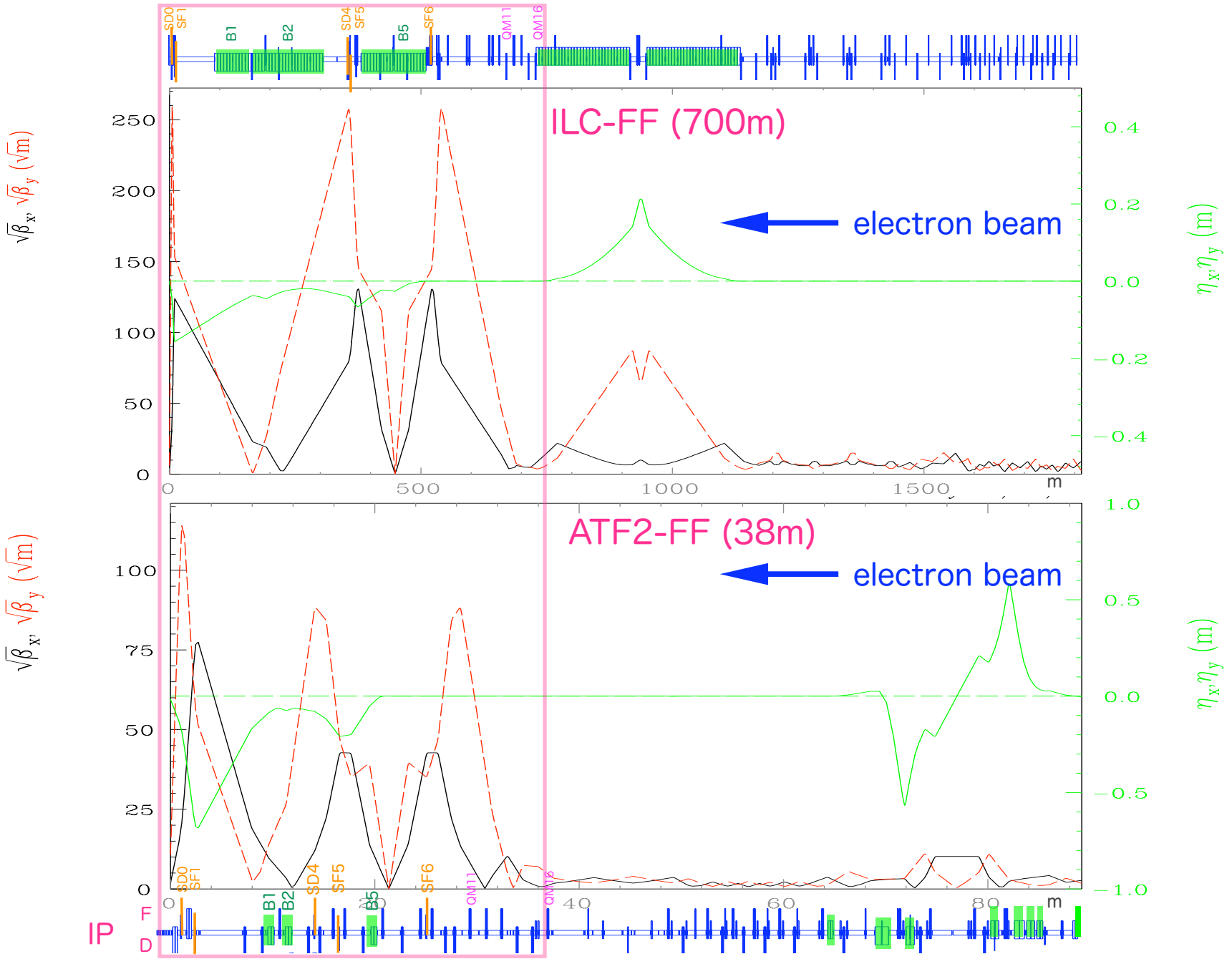

## 3.6.2 Status of ATF2 Systems

### 3.6.2.1 Magnets and Magnet Mover

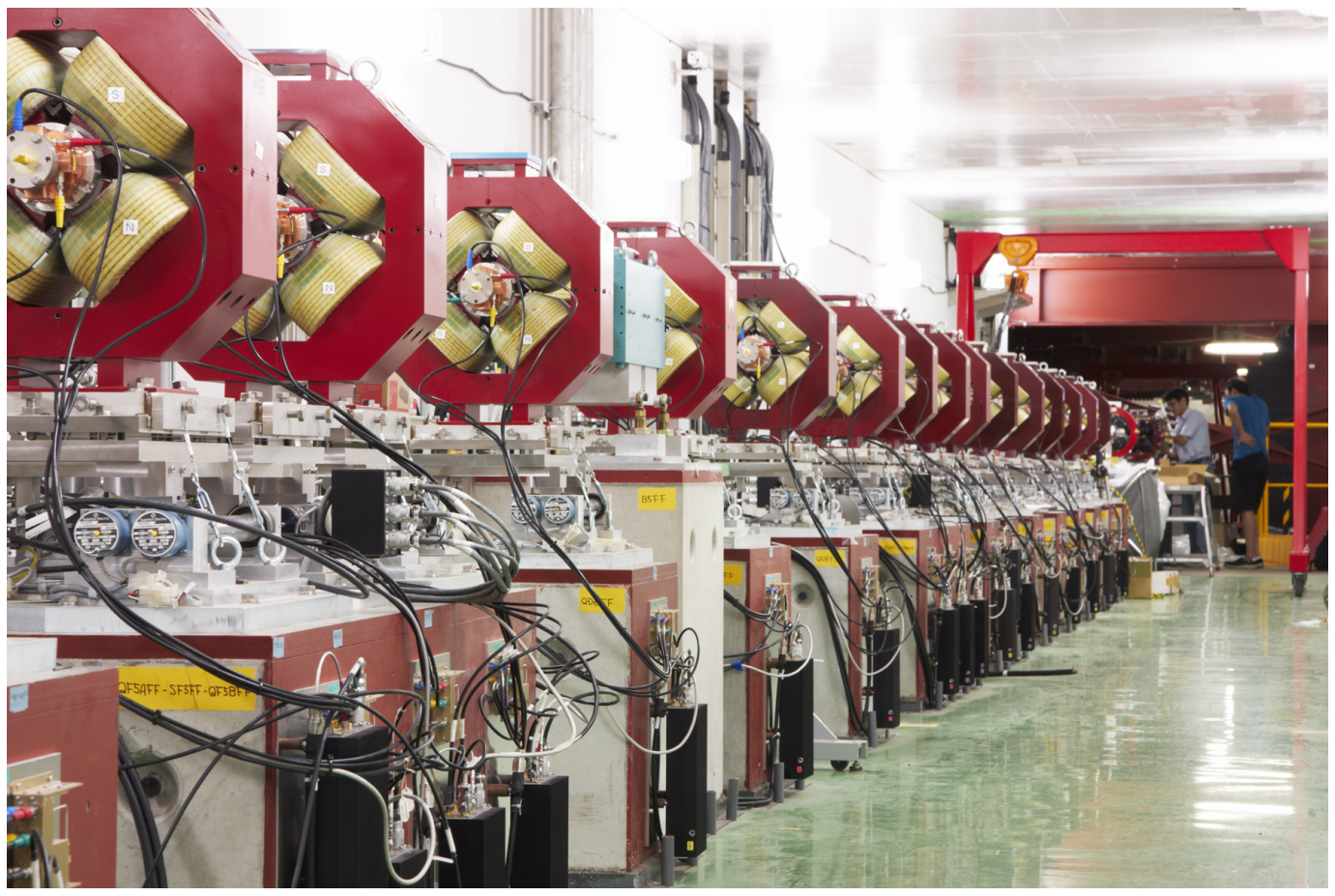

**Figure 3.42**
View looking downstream along the final focus section of the ATF2 beamline.

The ATF2 beam line extends over about 90 meters from the beam extraction point in the ATF DR to the IP (see Fig. 3.40 and Fig. 3.42). Many quadrupoles and some dipoles were fabricated for ATF2, while others were reused from the old ATF extraction beam line and from the Final Focus Test Beam (FFTB) at SLAC [197]. Among the latter were the two quadrupole and two sextupole magnets composing the strong-focusing final-doublet (FD) system just before the IP. The apertures of the FD quadrupole magnets needed to be increased to accommodate the larger $\beta$ function values at these magnets in the ATF2 optics design.

Anticipating gradual movements of supports and magnets due to thermal variations or slow ground motion, twenty quadrupole and five sextupole magnets in the final focus were mounted on remote-controlled three-axis movers recycled from the FFTB experiment. The movers have a precision of 1–2 µm for transverse motion (horizontal and vertical), and 3–5 µrad for rotations about the beam axis.

Overall alignment precisions of 0.1 mm (displacement) and 0.1 mrad (rotations) have been achieved using conventional alignment/metrology techniques. The final alignment of the magnets is achieved via beam-based alignment (BBA) techniques.

### 3.6.2.2 Final Doublet (FD)

The FD is composed of two quadrupole and two sextupole magnets (QD0, QF1, SD0, SF1 in Fig. 3.40). These magnets must be supported in a way that ensures their vertical vibration amplitude relative to the IP is smaller than 7 nm rms above 0.1 Hz, in order to limit effects on the measured beam size at the IP to less than 5 %. Below 0.1 Hz, beam-based feedback methods can be used (1/10th of the beam repetition rate of 1 Hz). A rigid support was chosen since the coherence length at ATF2 (about 4 m in this frequency range) exceeds the distance between the FD and IP, which strongly suppresses their relative motion. Vibration measurements with the table fixed to the floor and with all magnets and movers installed were performed in the laboratory for prior validation, including checking for potential effects from cooling water flowing in the magnets. Additional measurements after final installation of the FD confirmed that the residual motions of the magnets relative to the IP were within tolerance. The whole FD system is shown in Fig. 3.43 .

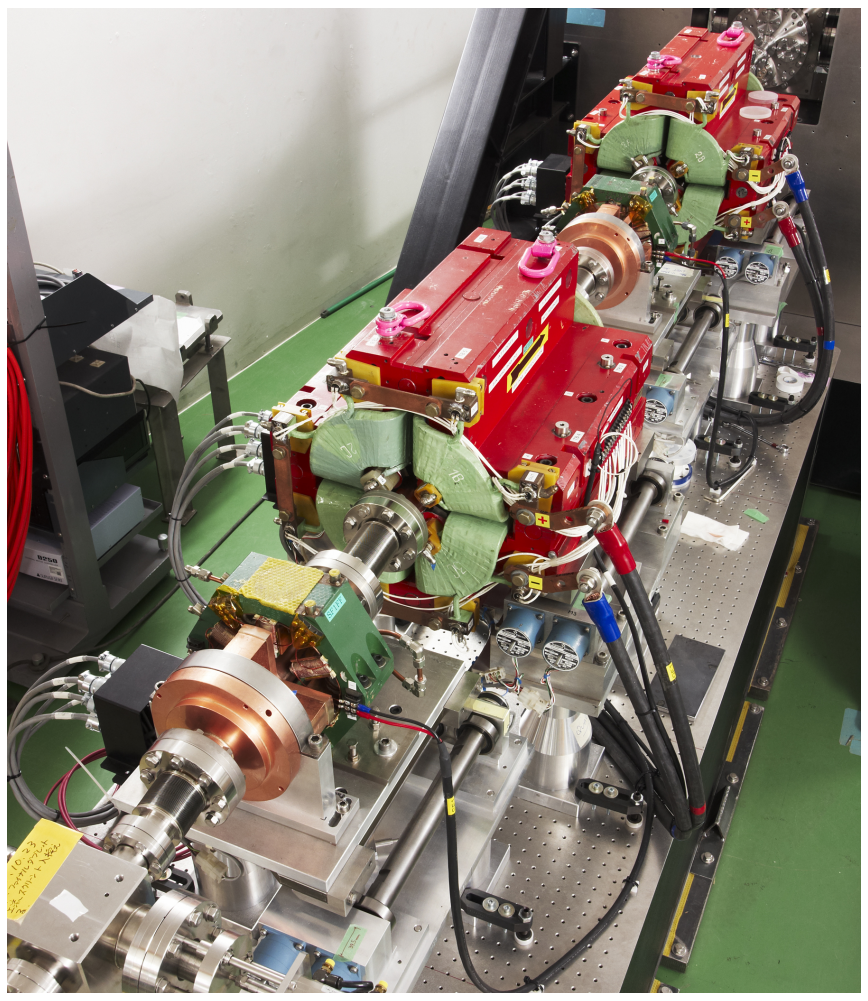

**Figure 3.43**
View of the final doublet installed on its rigid mechanical support system.

### 3.6.2.3 Cavity Beam-position Monitors

The ATF2 beam line is instrumented with 32 C-band (6.5 GHz) and four S-band (2.8 GHz) high-resolution cavity beam-position monitors (BPM). There are also four C-band and one S-band reference cavities to monitor beam charge and beam arrival phase. In the diagnostics and final-focus sections, every quadrupole and sextupole magnet is instrumented with a cavity BPM. The FD magnets use S-band BPMs, while the remaining quadupoles are equipped with C-band BPMs. The usable measurement range of the cavity BPMs was found to exceed the mechanical range of quadrupole movers ($\pm$1.5 mm). A resolution of 200–400 nm for the C-band BPMs has been demonstrated [201].

### 3.6.2.4 IP Beam-size Monitor (IPBSM)

Measuring transverse beam sizes of tens of nanometers at the IP requires specialised beam instrumentation, in particular a beam-size monitor based on laser interferometry (IPBSM, also referred to as a Shintake monitor [202]). The IPBSM is based on inverse Compton scattering between the electron beam and a laser interference fringe pattern. For the ATF2 beam energy, the energy of the generated gamma rays is typically rather low compared to that of bremsstrahlung photons, emitted when beam-tail electrons interact with apertures and start showering, which are the main detector background. In the monitor designed for ATF2, the signal is separated from this high-energy background by analysing the longitudinal shower profile measured with a multilayered detector located a few meters after the IP, downstream of a dipole magnet. The laser wavelength used is 532 nm, the 2nd harmonic of the Nd:YAG laser, which provides a suitable fringe pitch to measure the target vertical size of 37 nm. The monitor is designed to have three measurement modes by changing the crossing angle between the overlapping lasers, which give sensitives to beam sizes of 5000-350 nm (2-8 deg. mode), 100-350 nm (30 deg. mode), and 20-100 nm (174 deg. mode). In addition, a single “laser-wire” mode can be used for horizontal beam-size measurements [203].

The method to set up the electron and laser beams correctly was developed experimentally during early commissioning using diagnostics and instrumentation available for both electron and laser beams. It consists of five main steps: (i) carefully tuning the electron beam trajectory to reduce backgrounds; (ii) aligning the photon detector onto the electron beam axis at the IP; (iii) checking the synchronisation of both beams; (iv) scanning the laser beam horizontally to overlap its waist with that of the electron beam; and (v) scanning the two laser beams longitudinally across the electron beam to maximize the modulation.

#### 3.6.2.5 Other Beam-line Instrumentation

The instrumentation from the old ATF extraction line – strip line BPMs, Integrated Current Transformers (ICTs), optical transition radiation (OTR), screen profile monitors, and wire scanners – is reused in the reconfigured beam line. There are five wire scanners with tungsten and carbon wires of 10 and 7 μm diameter, respectively, located in the diagnostic section upstream of the final-focus section (see Fig. 3.40). They are used to measure the horizontal and vertical beam emittances after extraction from the DR. An additional wire scanner is installed just downstream of the IP for beam-size tuning and has tungsten and carbon wires of 10 μm and 5 μm diameter, respectively. Screen monitors are located right after the extraction, in the middle of the beam line, and before and after the FD. An optical-fibre beam-loss monitor is installed all along the beam line to localise and quantify beam losses in a relative sense. Four OTR monitors close to the wire scanners with an improved resolution of 2 μm has been installed in the extraction line. They permit the measurement of the beam sizes as well as a fast emittance measurement, with high statistics giving a low error and a good understanding of the emittance jitter.

### 3.6.3 Tuning Status Towards Achievement of Goal 1

#### 3.6.3.1 Summary of ATF2 Commissioning and IP Tuning Activities

The commissioning of ATF2 began in December 2008. During 2009, the various hardware diagnostic systems were commissioned whilst the beam line was commissioned with a relaxed optics configuration (10mm vertical beta function at the IP). The relaxed optics configuration was intended to allow for simplified checkout of the linear optics as the reduced chromaticity of the FFS at this level of focusing did not require the chromaticity compensation sextupoles to be switched on [204, 205]. It also provided a small-enough beam (∼1 μm vertical IP size) to commission the 2–8 degree modes of the IPBSM.

During 2010, a more aggressive optics (1 mm vertical IP beta function) was used, the FFS sextupoles were turned on on and the beam size tuned below 1 μm for the first time, with a smallest recorded spot size as measured by the IPBSM of 310 nm.

In the autumn runs of 2010 and at the start of 2011, the nominal vertical focusing optics was used, whilst keeping the horizontal beta function at 2.5 times the nominal design for reasons explained below.

The tuning work towards obtaining the nominal vertical IP spot size was expected to start during the spring of 2011. However, due to a fire in the modulator supplying power to the klystron generating RF for the gun, the program was delayed. The ATF was then damaged in the great Eastern Japan earthquake (M9.0), March 11 in 2011.

The ATF conventional facilities and accelerator systems were repaired during the summer and autumn of 2011, with the resumption of beam operations through ATF2 in the winter period. During the winter and spring periods of 2012, different optics configurations were applied while gaining experience on tuning the beam size and measuring in the 30-degree mode of operation of the IPBSM. The decision was made to relax the horizontal beta function at the IP to ten times that of the initial design (to 40 mm). This reduced the observed backgrounds in the IPBSM detectors and was also considered a prudent measure to lower the horizontal beam sizes throughout the FFS in order to reduce as much as possible the sensitivity to undesired multipole components in the FFS magnets. As the backgrounds detected in the IPBSM detectors were gradually reduced through alignment and beam-pipe enlargement activities, the vertical beta function at the IP was reduced, in a staged fashion, down to the design 0.1 mm. The IPBSM was ultimately found to function well in the 30-degree mode of operation with these settings. During the operation period February through June 2012, the beam was frequently tuned down to the 200nm level and below. Through careful online monitoring of both beam conditions and internal conditions of the IPBSM itself, these beam sizes could be maintained.

It was realised that, in order to achieve the beam sizes of Goal 1, a tuning program that runs continuously over periods of many days, perhaps weeks, was required. During the April - June beam operation periods in 2012, a student training program was set up, involving post-graduate-level students and postdocs from the collaborating institutions across the world that make up ATF2. The aim of this program was to produce a core team of people capable of maintaining operational efficiency of ATF whilst progressing the ATF2 tuning program.

Thus, during the November and December operations periods, i.e. the last four weeks, ATF2 ran continuously in "ATF2 tuning mode". The 174-degree mode of operation of the IPBSM was successfully commissioned. At the end of the runs, the beam was tuned down to about 70nm as shown in Fig. 3.44, where the 70nm beam size could be maintained for about 7 hours at least. It is an achievement of important milestone towards full verification of the local chromaticity correction scheme at the ILC final focus system. However, it could be performed only at very low beam intensity of $1 \sim 2 \times 10^9$/bunch. We suspected the wakefield effect at the large beta function region at the final focus system for the major cause. We will prepare the mitigation plan to achieve the 37nm beam size in coming runs.

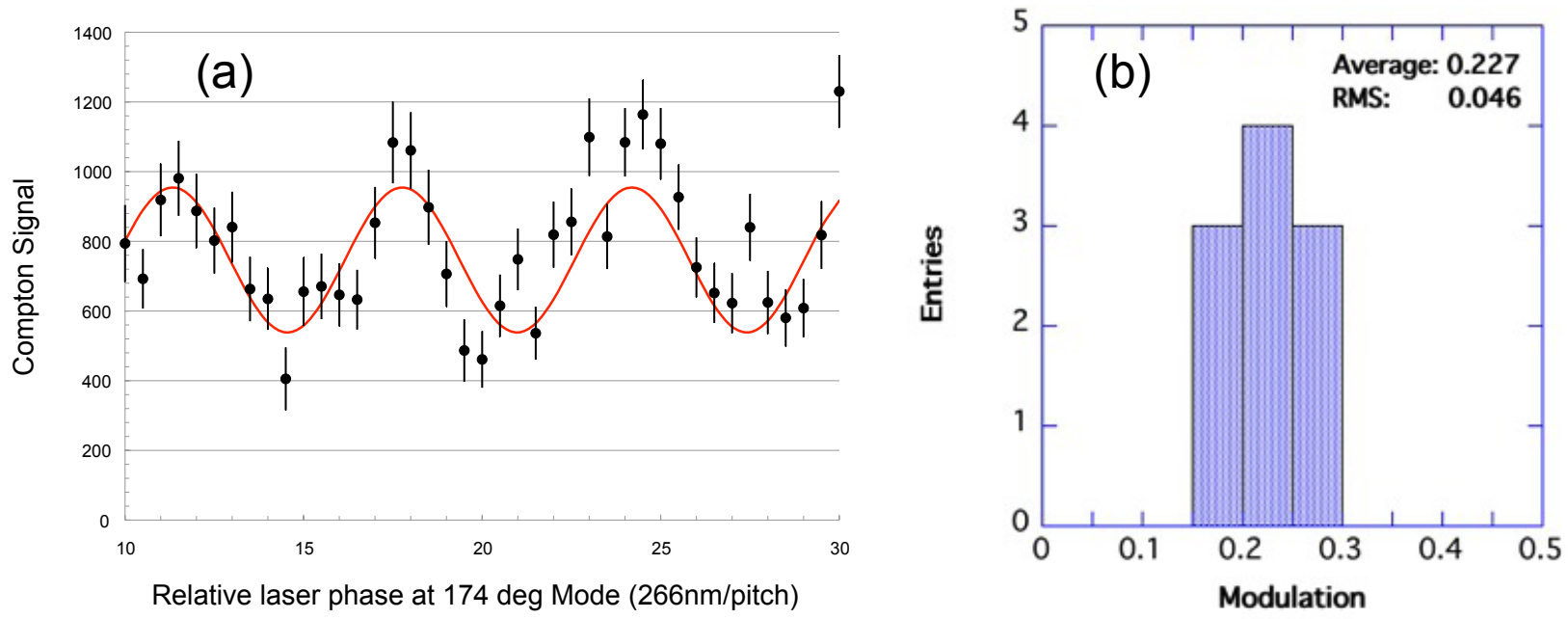

**Figure 3.44** (a) The best modulation (M) measured to date by the IPBSM beam-size monitor with the 174° crossing angle, M=0.28±0.3, taken in December 2012. The measurement corresponds to a beam size of 68±3 (statistical) nm. (b) Histogram of modulations measured 10 times at the 174 degree mode, 21 December 2012. The average beam size corresponds to 73nm with an RMS of 5nm.

## 3.6.3.2 Software for Optics Analysis and Re-Design

Due to the non-linear nature of the FFS optics, multi macro-particle tracking software has been used to test the performance of the optics, to develop beam-tuning algorithms and to test their robustness under realistic error conditions. This has been done independently with multiple codes, which provides a useful cross-check of the methods used and the tracking software in the different codes themselves. Codes used for these studies are Lucretia, SAD [206], MADX (MAPCLASS) [207] and PLACET [208].

## 3.6.3.3 Optics Preparation and Beam-dynamics Simulation

In addition to the original FFS tuning procedure developed for ILC, three different analyses environments were constructed based on MADX, SAD and Lucretia. These are able to re-match the FFS optics and study effects of errors using tracking through the model lattice together with the inclusion of all envisaged error sources.

One such error source was discovered after construction was complete: the higher-order multipolar components of many of the quadrupole and dipole magnets in the EXT and FFS were large enough to generate aberrations at the IP which noticeably increase the expected vertical spot size. Measurements of the normal and skew multipole components of these magnets were made at IHEP, KEK and at SLAC and inserted into the models. It was found that multipole components up to octupole, and in the case of the final doublet, up to 12-pole were important. To try and recover the beam size closer to the design goal, a study was performed to try and re-match the optics to mitigate the effects induced by the measured multipoles. The result of this process is shown in Fig. 3.45. By increasing the horizontal beta function at the IP by 2.5–10 mm, and thus reducing the horizontal beam size in the QF1FF aperture and other sensitive apertures, the effect of the various skew multipoles was somewhat mitigated.

**Figure 3.45**
Re-tuned IP vertical beam-spot size as a function of input horizontal emittance, with 2.5 times nominal IP horizontal beta function. Each curve corresponds to different maximum multipoles included in the simulation.

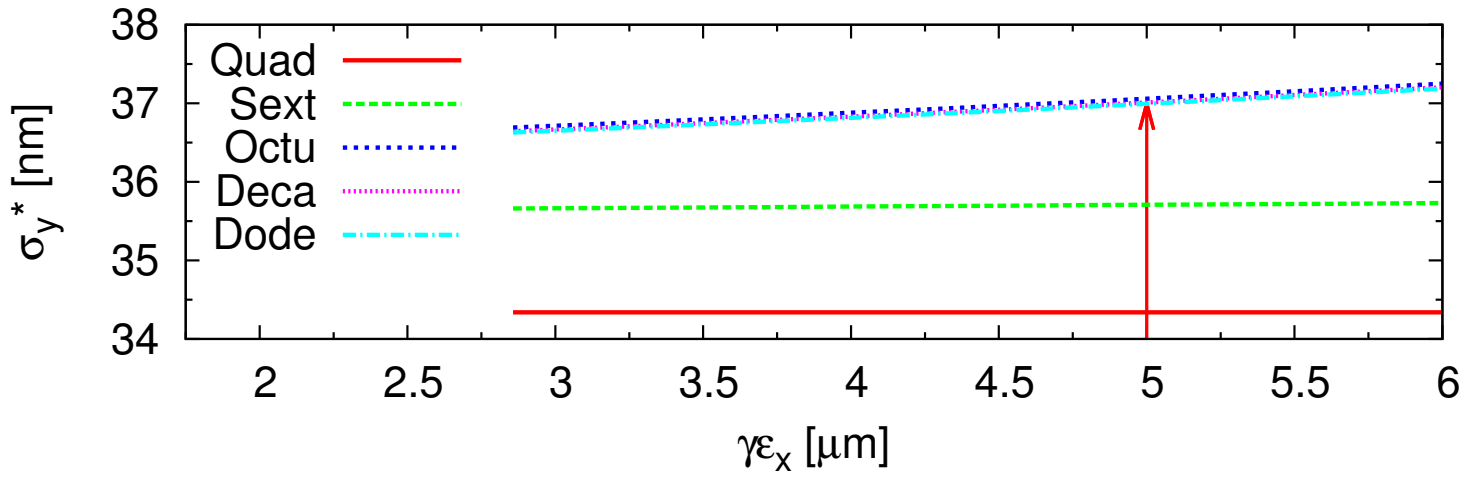

Due to the sampling of more of the skew multipole fields at larger horizontal beam sizes at the high beta points, the vertical beam size scales noticeably with horizontal emittance. The normal horizontal emittance achieved is about 5 µm rad (normalised), but has been measured to be lower at lower bunch charge due to collective effects in the damping ring.

This retuning method has been repeated for the different operational optics used as discussed in Section 3.6.3.1. The results have also been cross-checked using MADX, Lucretia and SAD.

In addition to the retuning procedure described above, it has been shown that, through the addition of four skew-sextupole magnets into the FFS, the tolerances to the sextupole multipolar errors in some of the FFS magnets are considerably relaxed. One such skew-sextupole magnet has already been installed in the beam line, resulting in some effect on beam size. Three more were installed over the summer 2012 shutdown period. This arrows "multi knobs" to be formed, allowing these particular aberrations to be tuned out. In addition, the QF1FF was replaced with a PEP-II LER quadrupole magnet with negligibly small 12 pole component just before the continuos runs for last 4 weeks. This replacement is expected to recover the horizontal beta function to the design value at IP.

After generating the desired lattice as discussed above, the expected performance of the accelerator in the presence of all expected error conditions (e.g. survey alignment tolerances, magnet field strength errors, BPM errors, ground motion, beam size measurement errors etc) has been simulated using a Monte Carlo approach in which 100 random distributions of error conditions were generated and the full tuning procedure applied to each seed. The mean and spread in the final results represents the best guess of the expected performance of the machine and an uncertainty estimate in the results based on the level of knowledge of the expected error sources. These simulations also provide the basis for developing and testing different tuning algorithms and techniques. The simulations show the initial beam sizes are expected to be of the order of about 0.5–3 µm and can be tuned down to within 20 % above the nominal beam size in about 12 steps. At each step, the range of dominant aberrations at the IP (up to third order) are examined. The most common aberration at each step is then targeted across all Monte Carlo seeds. This forms the basis for the tuning procedure to be applied to the real accelerator. Further iteration of the tuning knobs produces a final beam-spot size within about 10 % above the design goal after several hours of tuning. This simulation has been performed additionally in MADX and PLACET with similar results.

#### 3.6.3.4 Beam-tuning Procedure

The required procedures to tune the ATF2 beam are briefly outlined below. Not covered here is the process by which the beam is accelerated, transported into the ATF damping ring and the damping ring itself tuned (COD, dispersion and coupling). This is typically done once before a set of ATF2 tuning attempts, and occasionally during ATF2 tuning if the damping-ring conditions change, for example, due to a change in the ring circumference caused by temperature variations.

A notable problem that has presented itself since initial commissioning of the ATF2 is prompt emittance growth during damping-ring extraction. It is assumed that high-order magnetic fields experienced by the beam as it passes close to kicker and septa devices cause emittance growth via coupling effects as the beam is extracted. This manifests itself as a higher emittance as measured by the OTR system in the EXT than that measured in the damping ring. This effect has been observed to be highly sensitive to the extracted beam orbit, and has led to extracted emittances that vary from one operation period to the next. Typically the emittance is 2 nm $\times$ 12 pm in the DR and 2-3 nm $\times$ 20-30 pm in the EXT after coupling correction. Figure 3.46 shows vertical emittances measured at the DR and EXT in 2011 and 2012.

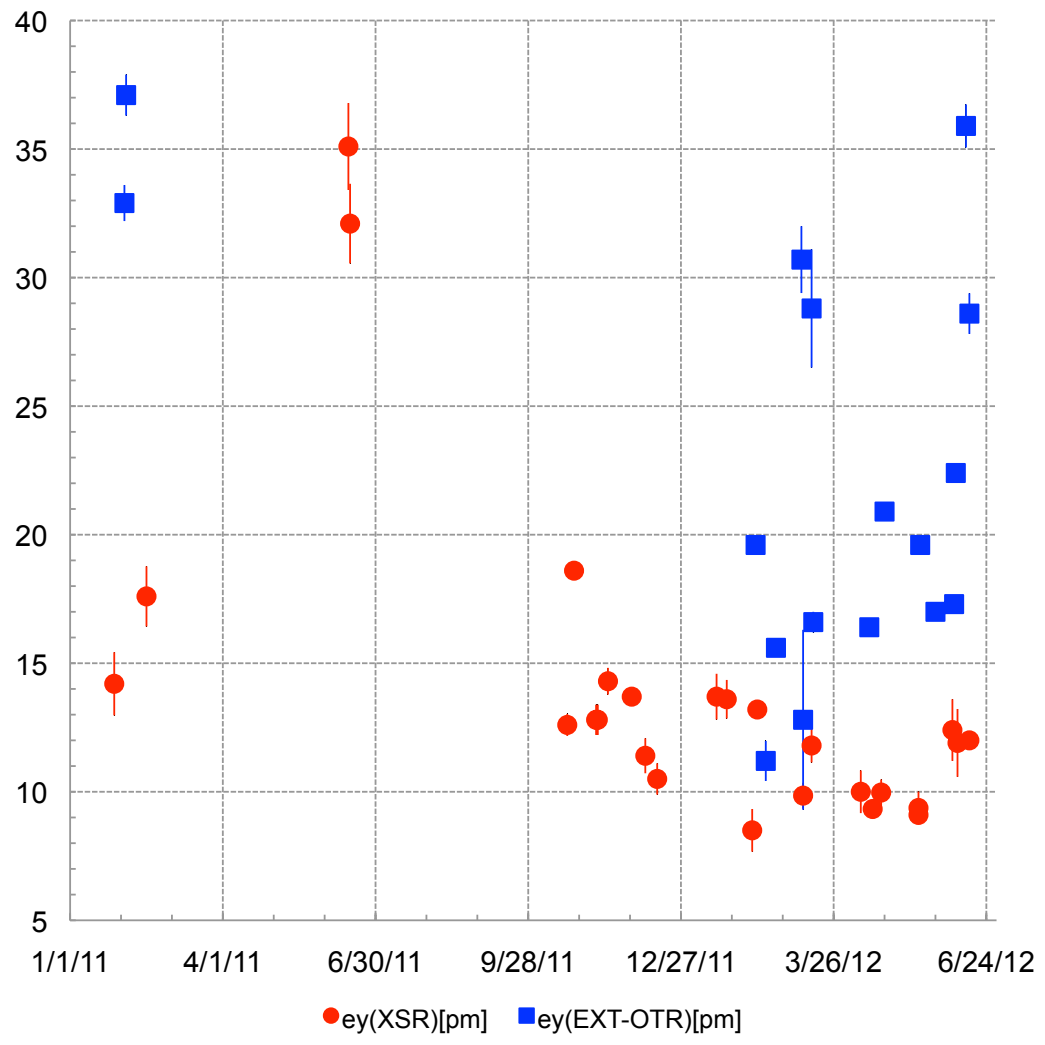

**Figure 3.46**
Vertical emittances measured by the X-ray synchrotron-radiation monitor(XSR) in the damping ring (red circles) and the optical transition-radiation monitor system (EXT-OTR) in the extraction line (blue squares), in 2011 and 2012.

Usually, beam tuning starts from steering the beam upstream of the extraction line. Next, the BPM readouts are aligned to the field centres of the quadrupole magnets by a "quad-shunting" procedure whereby the beam is steered through the magnet in question at various offsets at different quadrupole activation strengths. Then, orbit feedbacks are applied to maintain the "golden" beam orbit at both DR and EXT/FFS. It is especially critical to maintain the desired orbit through the FFS sextupole magnets to reduce drift in the beam size at the IP during tuning.

Horizontal and vertical dispersions are corrected by pairs of normal/skew quadrupole magnets at dispersive locations. The residual vertical dispersion coming from the damping ring can be corrected by rotation of QD0. Coupling is primarily corrected by four skew-quadrupole magnets in the EXT. One phase of coupling ($< x'y >$ at the IP phase) generated at the FFS would be corrected by the IP $< x'y >$knobs based on five sextupole magnets at the FFS.

The principal aberrations present at the IP are expected to be $< x'y >$ coupling, vertical waist shift and vertical dispersion. To cancel these, orthogonal knobs have been developed based on the five FFS sextupole magnets, which are deliberately offset to generate the required correction terms at the IP. Also, so-called non-linear knobs have been developed in order to correct high-order terms, predominantly T322, T326 and T324.

Figure 3.47 shows the simulation results of the tuning process (upper and lower one-sigma bounds from 100 Monte Carlo seeds) and the results from ATF2 tuning shifts.

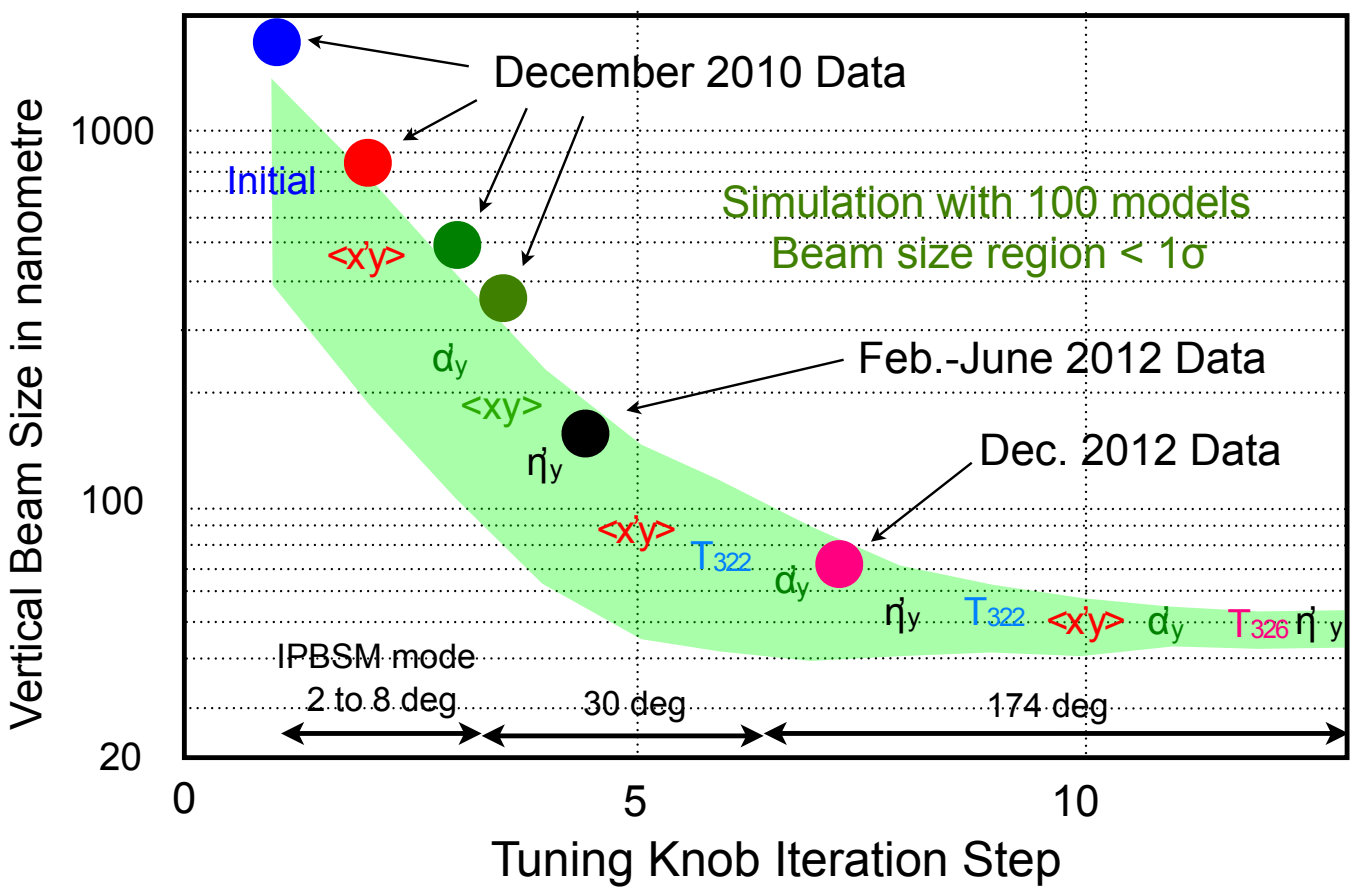

**Figure 3.47**
IP vertical beam-size tuning: experimental results and simulation. Shown are the data points from 2010 & 2012, the ±1-sigma curves from the Monte Carlo model; the tuning steps are indicated in text. The required operation modes of the IPBSM are also indicated for each relevant section of the plot.

## 3.6.4 Stability

The second goal of ATF2 is to achieve nanometer-level stability. Several R&D activities are currently being actively pursued:

- feedback on a nanosecond time scale (FONT);
- nanometer resolution IP-BPM [209];
- fast nanosecond-rise-time kicker [210];
- cavity BPM optimised to monitor angular variations of the beam near the IP with high accuracy;
- development of robust laser-wire diagnostics [211].

The most recent FONT results are shown in Fig. 3.48, where measurement of the beam offset at the first of a three-bunch train is used to correct the subsequent two bunches. Bunch separation is 151.2 ns. The data clearly indicates a reduction of the beam jitter by a factor of 5 from the first to the second bunch. The achieved 2.1 µm rms scales to 2.6 nm at the IP, given the demagnification of the optics.

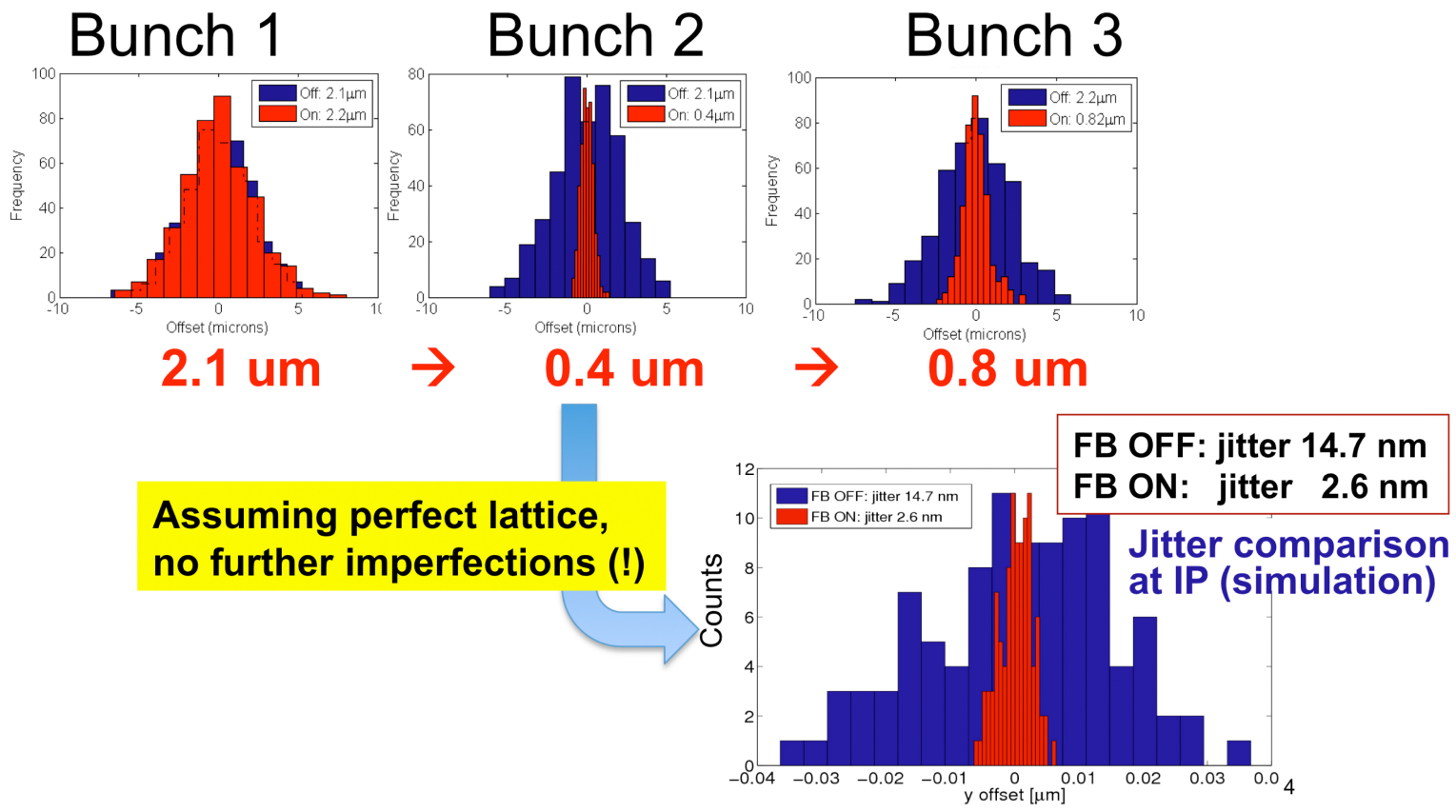

**Figure 3.48**
Recent results of FONT, the intra-train fast feedback. The three plots above are experimental results, while the bottom one is a simulated one to demonstrate the nanometer stabilisation at the IP assuming a perfect lattice for the final-focus beam line [212].

#### 3.6.4.1 Post TDR and Future prospects

Post TDR, there will be a concerted effort to achieve the goal of 37 nm vertical beam size in parallel with studies related to achieving the stability goals.

Plans to upgrade the performance of ATF2 on the time scale of a few years, after the main goals of ATF2 have been achieved, are also under consideration. In particular, optical configurations with ultralow $\beta^*$ values (2–4 times smaller than the current nominal in the horizontal and vertical planes), relevant to both the CLIC design and to some of the alternative ILC beam-parameter sets, are actively under study. An R&D program to develop a tuneable permanent magnet suitable for the FD is pursued in parallel, with an initial goal to construct a prototype for beam testing in the upstream part of the ATF2 beam line.

# Chapter 4
# Accelerator Systems R&D

## 4.1 Overview

In the R&D phase, work on accelerator systems has concentrated on alleviating technical rather than cost risk, since the cost of specific accelerator-system components is small compared to the costs associated with the main linac; however, the failure or sub-optimal operation of many accelerator components can prevent running of the ILC or severely limit its performance. The electron source photo-cathode gun, positron-source undulator, target and capture system, damping-ring kicker magnet system, and beam-delivery dumps are examples of such systems; work on these systems is described here. For much of this work, a strong collaborative team of experts that included both GDE members and collaborators on unrelated projects was an important factor in their success.

Technical risk is judged by assessing the state of the art for a given component against expected performance criteria. The strategy to mitigate technical risk in a long-term project development and construction cycle must balance potential gain against the time and resources to achieve it. The development of new ideas and beam tests of full systems have the longest lead times and should hereford be given priority.

The photo-cathode gun laser and in-vacuum high-voltage DC gun are the most critical electron-source components so these were chosen for study during the Technical Design Phase. Studies of the basic physical properties of the semiconductor photo-cathode itself were deemed successful and mature enough not to require further work. As is typical with the accelerator systems discussed here, the goal of work on the laser and the gun was focused on applying specialised high-technology solutions in an innovative manner to extend significantly the performance of off-the-shelf products. Wavelength tuneability and high average-power performance are a concern for the laser and the GDE teams are working to apply cryogenic technology to the lasing medium. Extending DC high-voltage capability without increasing field emission or dark current is the main concern for the gun. It is well known that beam quality will improve as the voltage is increased but this has to be balanced against the photo-cathode poisoning caused by excessive dark current.

The positron-source component studies are the most important effort related to accelerator systems because of the high level of integration required and the difficulty of performing a full-power system test. It is important to note that because the rate of positron creation is well beyond that achieved in existing facilities, an Alternate Configuration based on a normal-conducting linac with a high repetition rate is under active study. Both are described in this chapter. Furthermore, baseline performance goals include an option to provide 50 to 60 % positron polarisation, a performance characteristic also unmatched at any existing facility.

The undulator-based scheme has been adopted for the baseline design because the photon beam requires a thinner target and therefore dissipates less energy than an electron beam and because it provides polarised positrons. The use of a circularly polarised photon beam produced by the passage of high-energy electrons through a helical undulator to create polarised positrons is a well understood process that was demonstrated early in the ILC design phase [213]. The achievable positron polarisation depends critically on the collimation scheme used to separate the core forward–directed gamma rays from the surrounding halo. Key R&D topics of concern here are the helical undulator and the collimator. Collimator design studies are presented in this chapter.

The undulator field period and characteristic strength 'K' determine both the incoming electron-beam threshold energy required to produce the positrons and the total undulator length needed. Two full-scale baseline prototype superconducting undulators were constructed and successfully tested during the design phase; results are reported here. The greatest overall flexibility for the collider physics program is provided with low minimum threshold energy which in turn corresponds to a small undulator period; development is focused on reducing the period from 11.5 to below 10 mm. The baseline design threshold energy is 150 GeV.

Power deposition in the positron target during the 730 microsecond-long train must be smeared out to avoid damage. Simulation work has identified average-power deposition, 'shock-wave' stress, thermal stress and eddy-current heating (from interaction with an externally applied axial magnetic field), as the most important concerns. It has not been practical during the design phase to test the power-handling capability of the 14 mm-thick titanium target. With the baseline 1 m diameter target rotating with a speed of 100 m/s at the rim, the incident beam energy is spread along a 75 mm long by 1.4 mm (rms) wide stripe. Additional target heating, of almost equivalent amplitude to the average beam heating, is caused by eddy currents. Applying a high-vacuum seal to the rotating shaft and providing cooling water through the target rim are the support-engineering tasks of greatest concern. Reports on studies of these topics using a full-scale test model are included here.

Two key elements of the capture section that follows the target are the high-field tapered solenoid, located 2 mm from the target rim, and the normal-conducting RF accelerator section. Design and prototyping of the high-field tapered solenoid, also referred to as an 'optical matching device' (OMD), and the high-current pulse modulator to power it is underway. The design uses a stack of single-turn flux-concentrating disks that are transformer-coupled to large – diameter 10- to 25-turn coils. Design studies of other kinds of OMD, such as a liquid-lithium lens, have also been done. A full power normal-conducting RF accelerator capture section has been built and fully tested.

R&D on an alternate positron production scheme based on a smaller peak-power incident beam on target is underway and is also reported in this chapter. A short, high-power, high-repetition-rate normal-conducting linac reduces concerns over thermo-acoustic shock-wave damage to the target. A state-of-the-art conventional fixed target can be used, removing the need for a system rotating at high speed.

The most serious performance concern raised for the damping ring is the potential impact of electron-cloud instabilities, notably in the positron damping ring. These instabilities have been studied extensively at the purpose-built test facility 'CesrTA,' the results of which are reported in Section 3.5. The technological aspects of the mitigation strategies, including vacuum-chamber electrodes and coatings, are also covered in that chapter.

The damping ring, as the source of low-emittance beams, and the beam-delivery system, as the demagnification section, must both be tested to validate the expected performance of beam-line components and tuning-strategy convergence. This topic is covered in Section 3.6.

The primary technical concern for the damping rings is the injection and extraction ultra-fast pulse kicker system. Roughly 40 pulser-kicker pairs are required for injection and extraction of the beam bunches from the damping ring. The ring minimum inter-bunch spacing, (and ultimately therefore the size of the ring) is determined by the pulse rise and fall time of these devices. A two-pair system has been built and tested successfully using the ATF.

The primary R&D goals for the BDS are associated with the beam demagnification and stability tests at the ATF2. However, there are several other key R&D programmes not directly related to this. The GDE team is pursuing design and prototyping of the ILC superconducting final doublet, with most of the relevant ILC features reproduced. This prototype is not intended for a beam test but rather for laboratory studies with various instruments. A second focus is the engineering design work associated with the machine-detector interface, in support of the detector push-pull option. Finally, simulation work on the main beam dump has been done [214].

## 4.2 Polarised electron source

The R&D for the ILC polarised electron source has been focused on two aspects: first, to build a prototype of the source laser system and to use the laser system to generate an electron beam with ILC beam parameters (Section 4.2.1); secondly, R&D on the electron gun itself, the goal being to achieve the ILC specification for a gun voltage of 200 kV while maintaining a low dark current to ensure a long cathode lifetime. This development has been done at the Jefferson Laboratory's injector facility. Both the prototype of the ILC source laser and the high-voltage gun have been demonstrated.

### 4.2.1 Beam Parameters

The key beam parameters for the electron source are listed in Table 4.1.

**Table 4.1**
Electron Source system parameters.

| Parameter | Symbol | Value | Units |
|---|---|---|---|
| Electrons per bunch (at gun exit) | $N_-$ | $3 \times 10^{10}$ | |
| Electrons per bunch (at DR injection) | $N_-$ | $2 \times 10^{10}$ | |
| Number of bunches | $n_b$ | 1312 | |
| Bunch repetition rate | $f_b$ | 1.8 | MHz |
| Bunch-train repetition rate | $f_{rep}$ | 5 | Hz |
| FW Bunch length at source | $\Delta t$ | 1 | ns |
| Peak current in bunch at source | $I_{avg}$ | 3.2 | A |
| Energy stability | $\sigma_E/E$ | <5 | % rms |
| Polarisation | $P_e$ | 80 (min) | % |
| Photocathode Quantum Efficiency | QE | 0.5 | % |
| Drive laser wavelength | $\lambda$ | 790±20 (tunable) | nm |
| Single-bunch laser energy | $u_b$ | 5 | μJ |

### 4.2.2 System Description

Results from previous photocathode R&D projects have demonstrated that materials are available that can provide the ILC beam charge and polarisation. The ILC source will use a strained gallium-arsenide phosphide (GaAsP) highly doped photocathode. Figure 4.1 illustrates the performance of such a cathode. With bunch spacing of ~500 ns, the surface-charge limit is not expected to be an issue. The optimum doping level remains to be determined. The prototype laser system could be used to help clarify this issue.

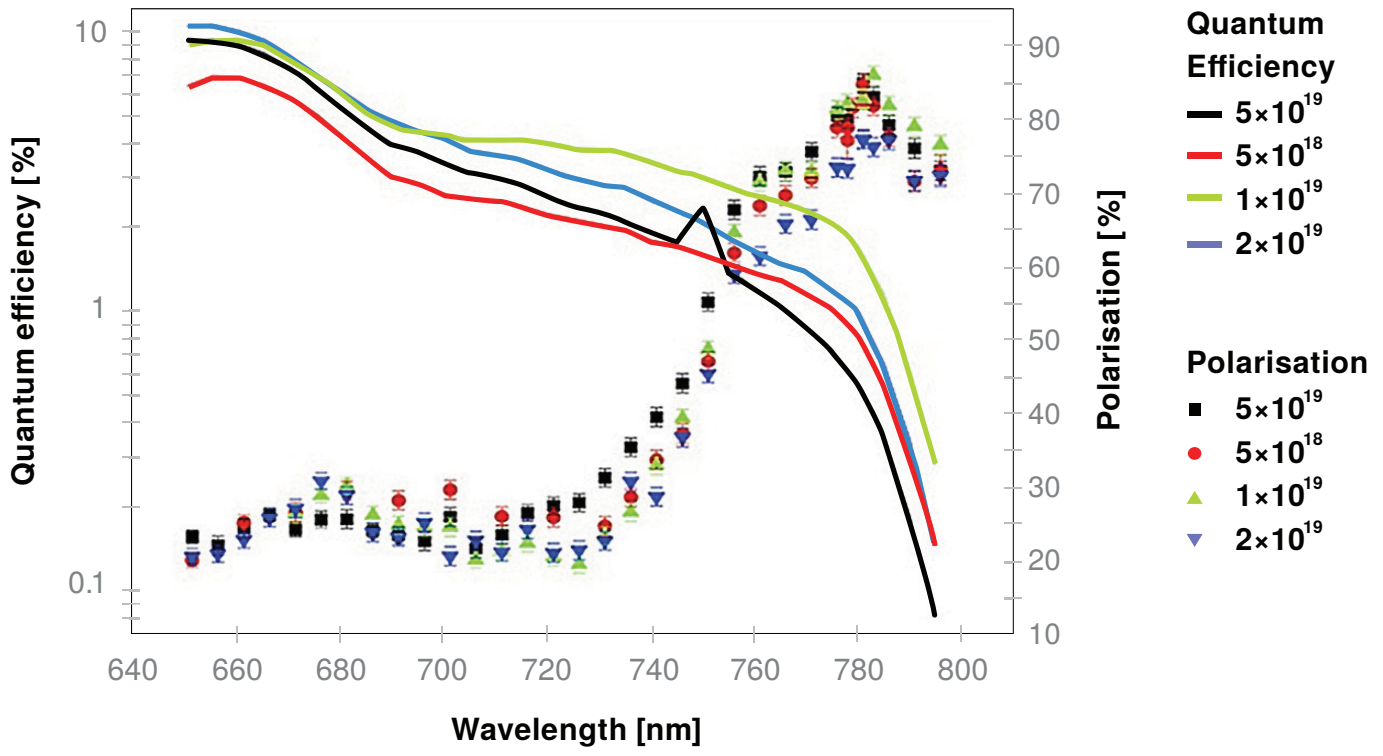

**Figure 4.1**
Performance of strained layers of GaAsP photocathodes at different doping levels.

#### 4.2.2.1 Laser system development

A laser system has been developed for the ILC polarised injector that is capable of generating the ILC bunch train. This system was developed in the US by Kapteyn-Murnane Laboratories (KM Labs) of Boulder, Colorado.

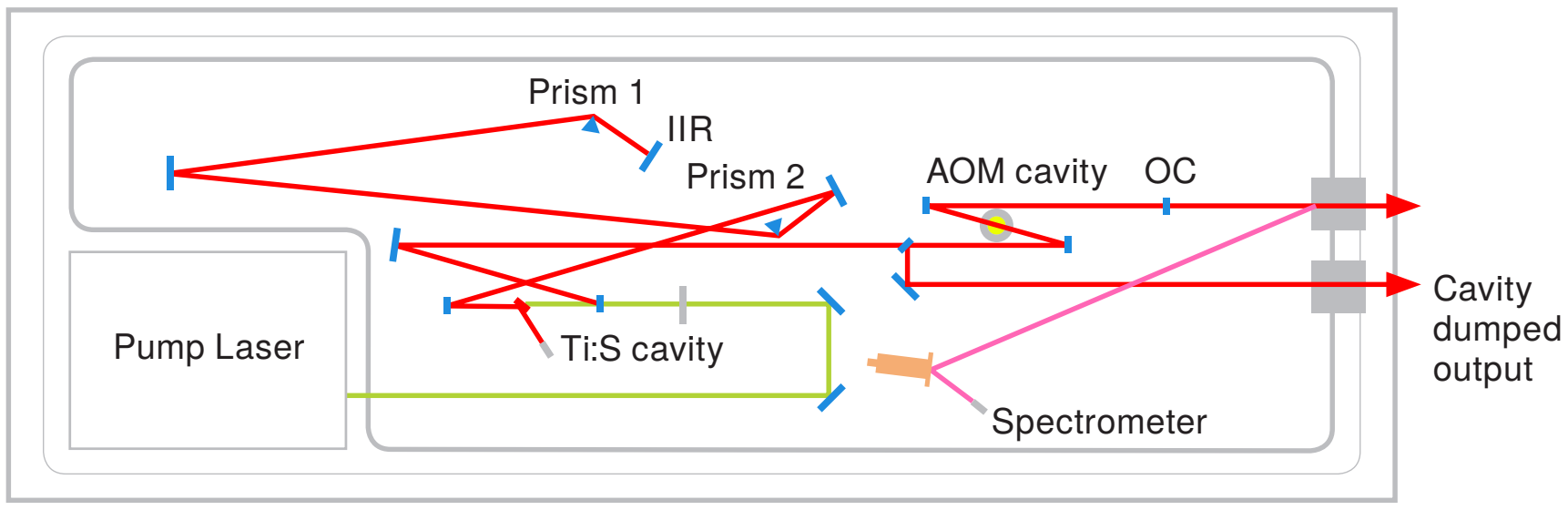

**Figure 4.2** Optical layout of the KM Labs mode-locked cavity dumped ps oscillator

The laser wavelength must match the band gap of the cathode material. For GaAsP, a wavelength of approximately 790 nm is necessary. The laser system must provide the time structure of the ILC pulse train. One basic component of this laser system is the mode-locked oscillator that operates at a harmonic frequency of the micro-bunch repetition rate, which can be locked to an external reference frequency (see Fig. 4.2). A regenerative amplifier is used to amplify the pulse train (see Fig. 4.3). A key component of this amplifier is the cryogenic cell containing a titanium-sapphire crystal. Cryogenic technology allows large pump power and efficient amplification, minimising the effects of thermal lensing in the amplifying medium. The micro-bunch structure (1.8 MHz pulse train) is formed in the oscillator using a cavity-dumping acousto-optic modulator. The macro-bunch structure (5 to 10 Hz) is generated by electro-optical switching of the amplified beam. The KM Labs laser was successfully demonstrated in November 2010 at the KM Labs facilities. Due the design constraints and limited amplifier pump power, the laser output was limited to 1.5 MHz with pulse energies of ∼3 µJ. These laser systems are presently being installed at the SLAC ILC Injector Test Facility with stronger pump lasers. The prototype laser performance at SLAC is expected to exceed 5 µJ per pulse at 1.5 MHz.

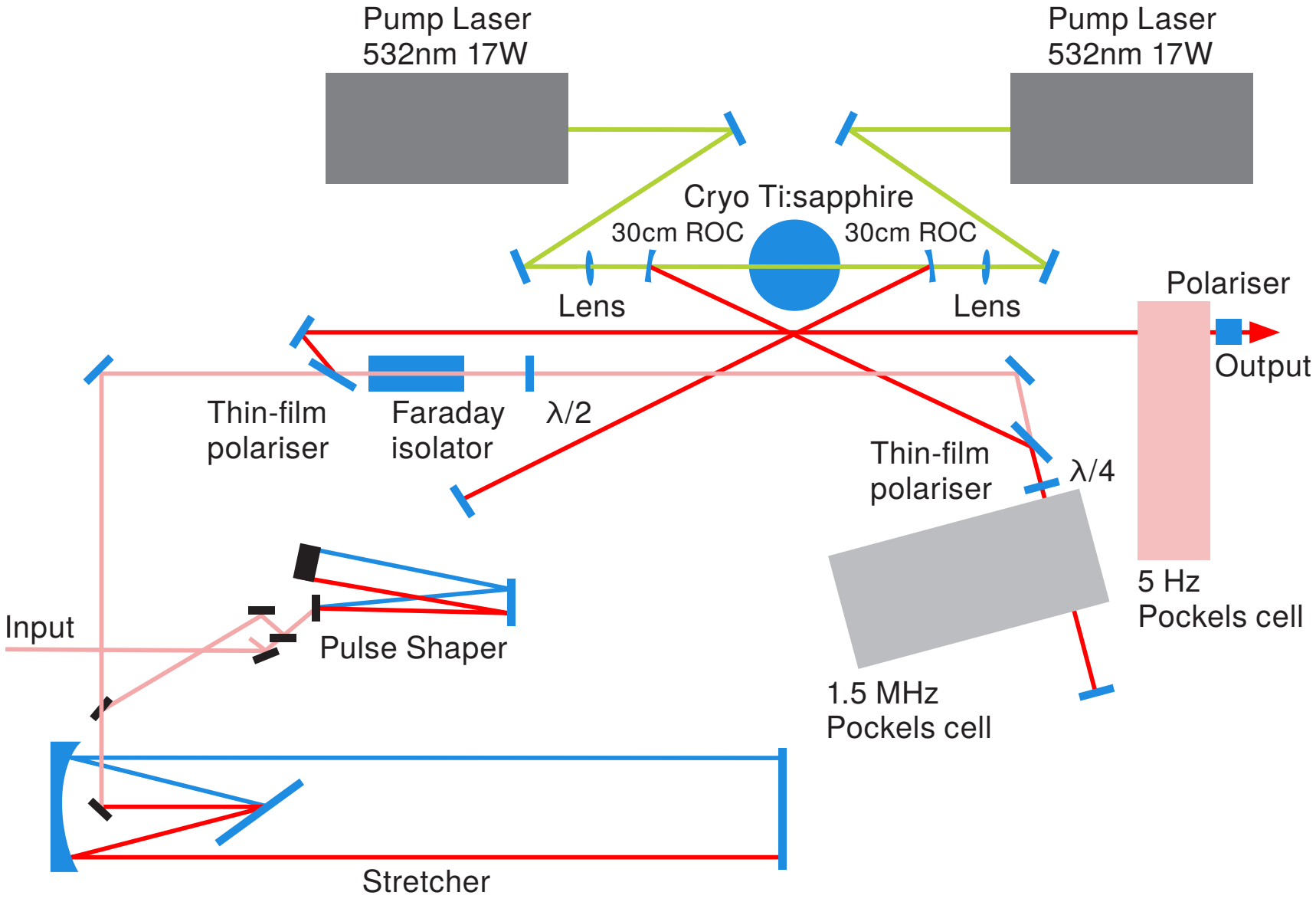

**Figure 4.3** Optical layout of the KM Labs pulse stretcher, pulse shaper, and regenerative amplifier systems

#### 4.2.2.2 Direct-current gun

The main goal of R&D towards a direct-current gun for polarised-electron generation is to increase the high-voltage capability while maintaining or reducing the dark current. A higher voltage is desirable to reduce the space-charge forces that the electrons experience at low energy before further acceleration. The reduction of space-charge forces is desirable to lower the transverse and longitudinal emittances of the generated electron bunches. A low dark current is necessary to maintain the negative electron-affinity properties of the photocathode, thereby increasing the lifetime of the electron source. The most important issue is to reduce field emission within the gun, which is the fundamental source of dark current.

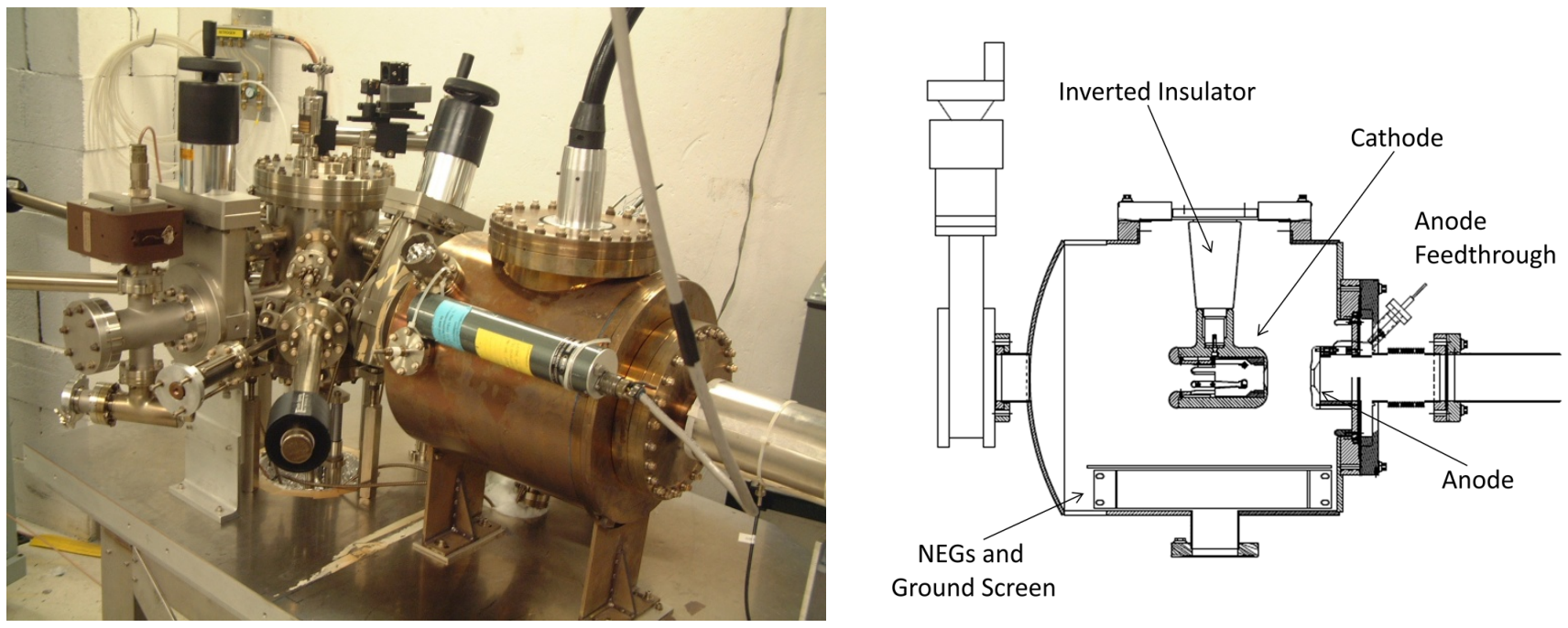

(a) (b)

**Figure 4.4**
The chamber of the 200 kV DC high voltage photogun (a) and its schematic view (b).

A 200 kV DC high voltage photogun (Fig. 4.4) was constructed at Jefferson Lab using a compact inverted insulator (Fig. 4.5) and with a vacuum load-lock that supports relatively quick photocathode replacement [215]. The photogun employs a cathode electrode manufactured from large grain niobium that was demonstrated to reach higher voltages and field strengths compared to stainless steel electrodes that were prepared using traditional diamond-paste polishing. This photogun has undergone extensive testing, demonstrating reliable beam delivery from strained-superlattice GaAs/GaAsP photocathodes at average current up to 4 mA. A second load-locked photogun with an inverted insulator was constructed for CEBAF [216]. It employs a stainless steel cathode electrode biased at 130 kV. It has operated reliably for over two years, delivering more than 200 μA average beam current for month-long periods without interruption and with electron beam polarization $> 85\,\%$.

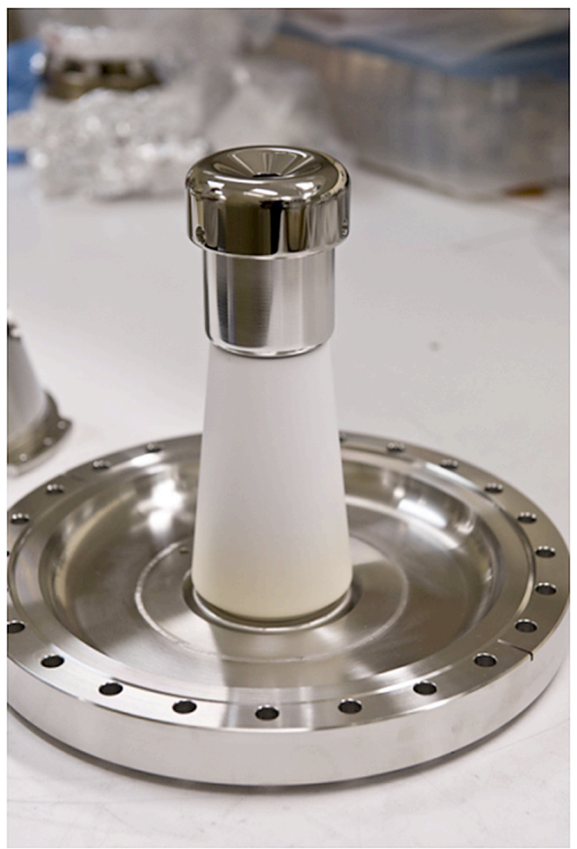

**Figure 4.5**
Compact inverted insulator with test electrode used to evaluate electrode materials, polishing techniques and inert gas processing.

High voltage processing in the presence of inert gas (He and Kr) was demonstrated to significantly improve the performance of stainless steel and niobium cathode electrodes, eliminating field emission ($<$ 10 pA) at voltages to 225 kV and field strengths $>$ 18 MV/m [217] (see Fig. 4.6).

The vacuum chambers and many internal components were baked at 400 °C prior to final construction which served to reduce the outgassing rate by a factor of $\sim$20 and resulted in the lowest observed static vacuum of all the Jefferson Lab photoguns to date. The pressure registered by a Leybold extractor gauge was $2 \times 10^{-12}$ Torr (nitrogen equivalent), which is very close to the X-ray limit of the gauge.

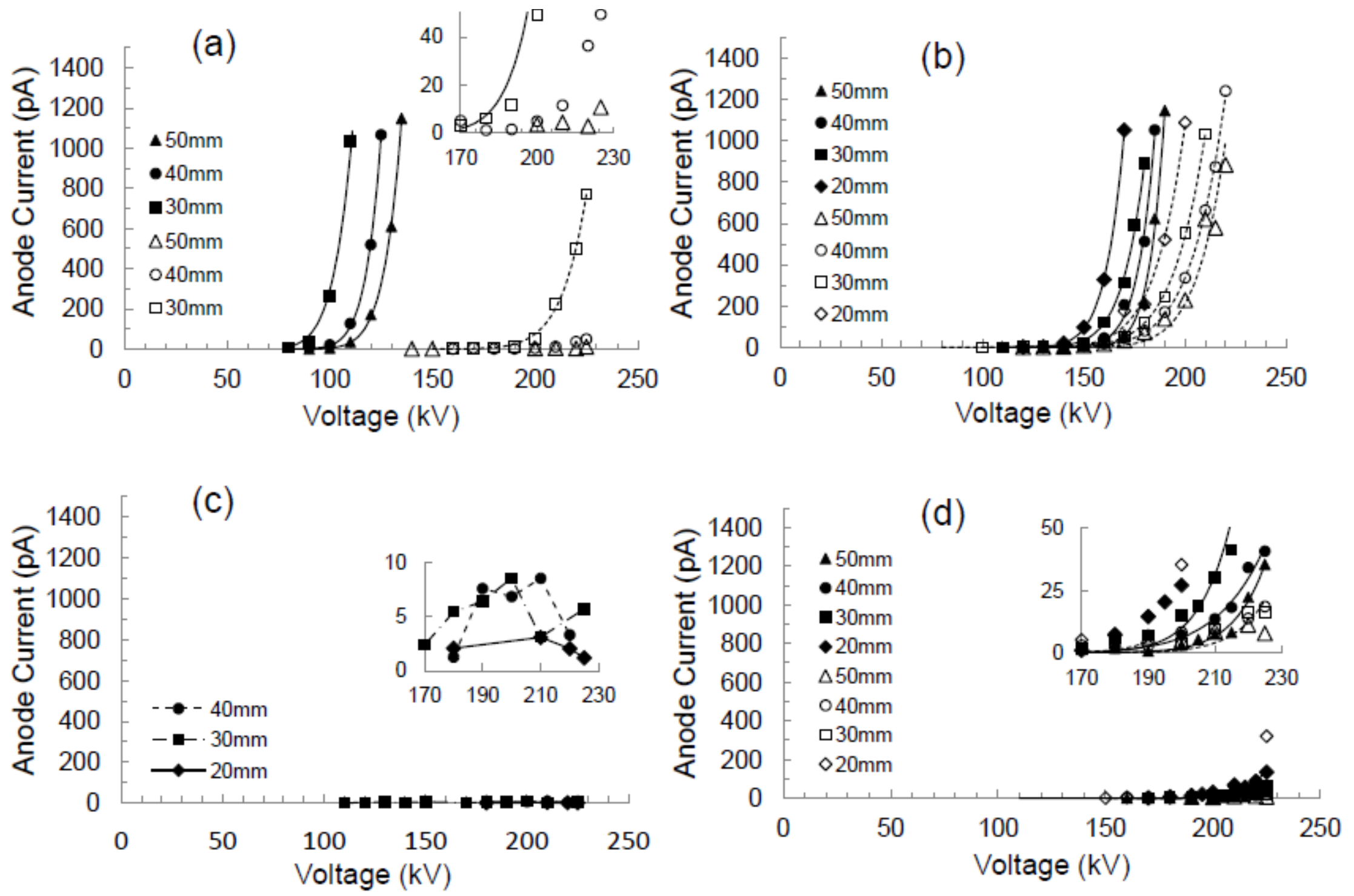

**Figure 4.6.** Results from a field emission test stand. Field emission current versus bias voltage and anode/cathode gap spacing for (a) DPP stainless steel, (b) fine-grain Nb, (c) large-grain Nb and d) single-crystal Nb. Each plot shows field emission behavior before (solid symbols) and after (open symbols) krypton processing, except for large-grain Nb which did not require krypton processing. Insets show an enhanced view of the low current data points. For all cases except large-grain Nb, the lines between data points represent Fowler-Nordheim fits.

## 4.3 Positron Source

### 4.3.1 Overview

The ILC Positron Source generates the positron beam for the ILC. To produce the positrons, the beam from the electron main linac passes through a long helical undulator to generate a multi-MeV photon beam which hits a thin metal target to generate showers of electrons and positrons. This system pushes the state of the art in many areas and there has been substantial R&D on several critical items including the undulator, positron-conversion target, optical-matching device, photon collimator, normal-conducting accelerating structures, radiation shielding and remote handling. There has also been work on the alternative conventional positron source and beam-line lattice design.

### 4.3.2 Undulator

At the time of the *Reference Design Report* (RDR), short superconducting helical-undulator prototypes using niobium-titanium superconductors had been successfully fabricated and tested by groups at Rutherford Appleton Laboratory (RAL) in the UK and at Cornell University [218, 219] in the US. This gave confidence that the undulator period and field strength selected for the ILC were feasible. Since that time the RAL group has successfully fabricated two identical long undulators, each 1.75 m in length, which have been magnetically tested and proven easily to achieve the field strength required. In fact, both exceeded the magnetic field specification by more than 30 % [220]. The quench training for the two magnets is shown in Fig. 4.7.

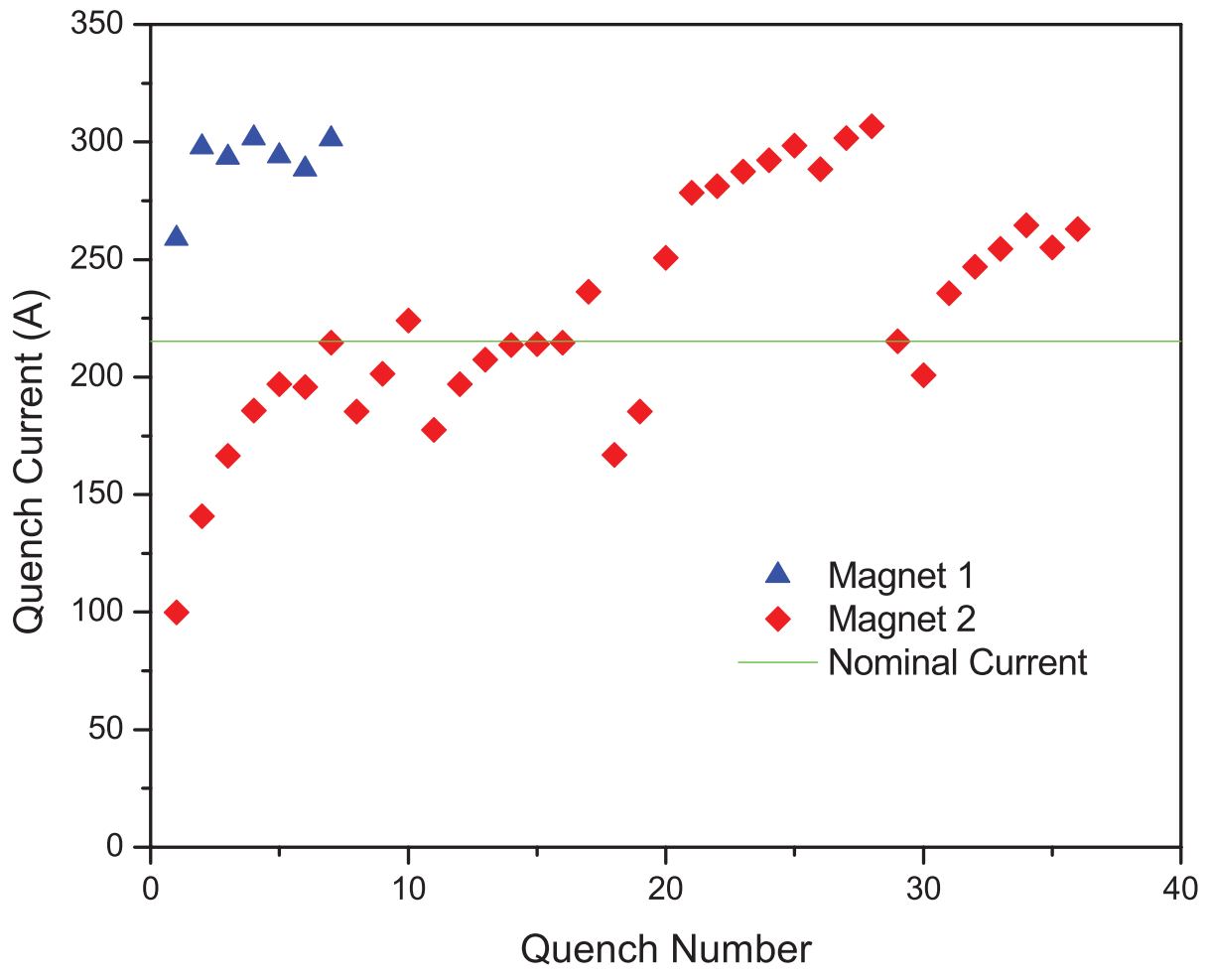

**Figure 4.7**
Quench history of the 2 of 1.8 m prototype superconducting helical undulator

In addition the subsequent analysis of the magnetic field results by staff at Daresbury Laboratory in the UK has shown that both undulators have a very high field quality, certainly more than sufficient to provide the intense source of photons that is required. The RAL team has since incorporated both of these undulators into a single 4 m-long cryogenic module (which operates at -269 °C) of the design required by the ILC, and has proven that both undulators can be powered simultaneously at the field levels required [221]. A photo of the complete undulator cryomodule is shown in Fig. 4.8. In the future it would be valuable to test the module in an electron beam to measure the properties of the light generated by the undulators. The parameters of this undulator are given in Table 4.2.

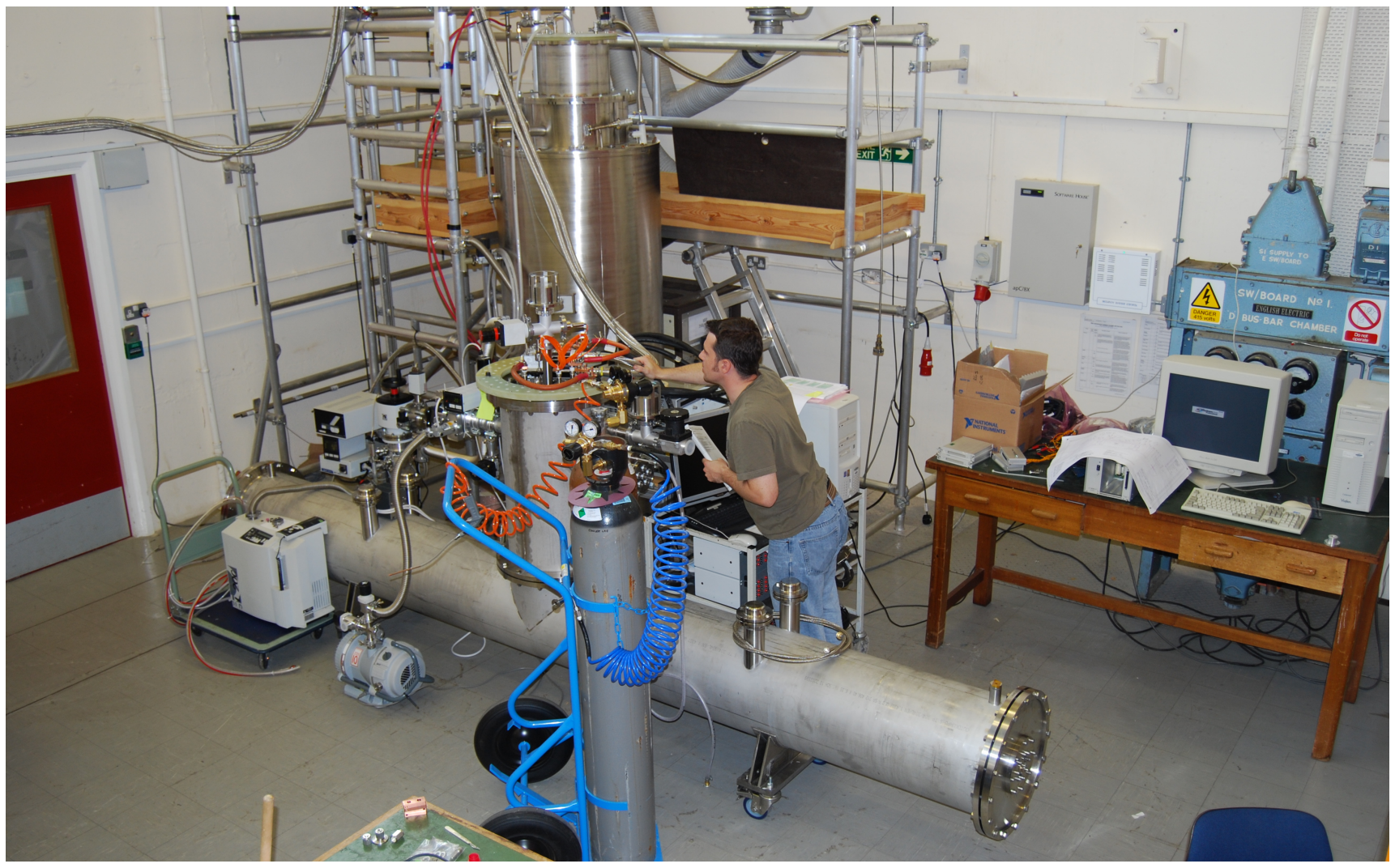

**Figure 4.8**
The 4-m prototype superconducting helical undulator under test at Rutherford Appleton Laboratory.

**Table 4.2**
Parameters of ILC undulator [222].

| | |
|---|---|
| Undulator Type | helical |
| Undulator Period | 11.5 mm |
| Undulator Strength, $K$ | 0.92 |
| Field on-axis | 0.86 T |
| Beam Aperture (diameter) | 5.85 mm |
| Winding Bore | 6.35 mm |

The Daresbury/RAL team is now investigating the use of a more advanced superconducting material, niobium tin, which should produce even higher field strengths. This would make an even shorter-period undulator, which could generate the required positron yield at lower electron drive-beam energies. Currently the team is winding short prototypes to gain experience with this technically more challenging material and also to allow a direct comparison with the other prototypes built using niobium titanium [223]. Simulations of an undulator with niobium-tin conductor predict that a higher field on axis is possible than with niobium titanium. The increased field could allow the period to be reduced to around 10 mm whilst maintaining a $K$ value of 0.92, which could reduce the active undulator length by 10 %.

### 4.3.3 Conversion target

The conversion target is a 1-m-diameter wheel of titanium alloy that rotates at 100 m/s at the rim. To increase the positron-capture efficiency, the target rim passes through a strong magnetic field. Unfortunately, this then induces unwanted eddy currents in the wheel, causing the wheel to heat up. The level of heating that can be tolerated limits the usable magnetic field. Several groups have tried to model the eddy-current heating but inconsistent results were obtained from the different simulation codes used [224–226]. Consequently a full-scale prototype target was built at the Cockcroft Institute in the UK to benchmark the simulation codes. The target wheel was fabricated from the required titanium alloy and was rotated over a range of rim velocities in a strong magnetic field (Fig. 4.9a). The results of this unique experiment have accurately quantified the eddy-current effects and have confirmed which simulations were correct [227]. Furthermore, the experiment has proven that the magnetic-field level assumed by the positron-source design at the target wheel is feasible, with the eddy-current heating being easily tolerated.

**Figure 4.9**
Rotating target.

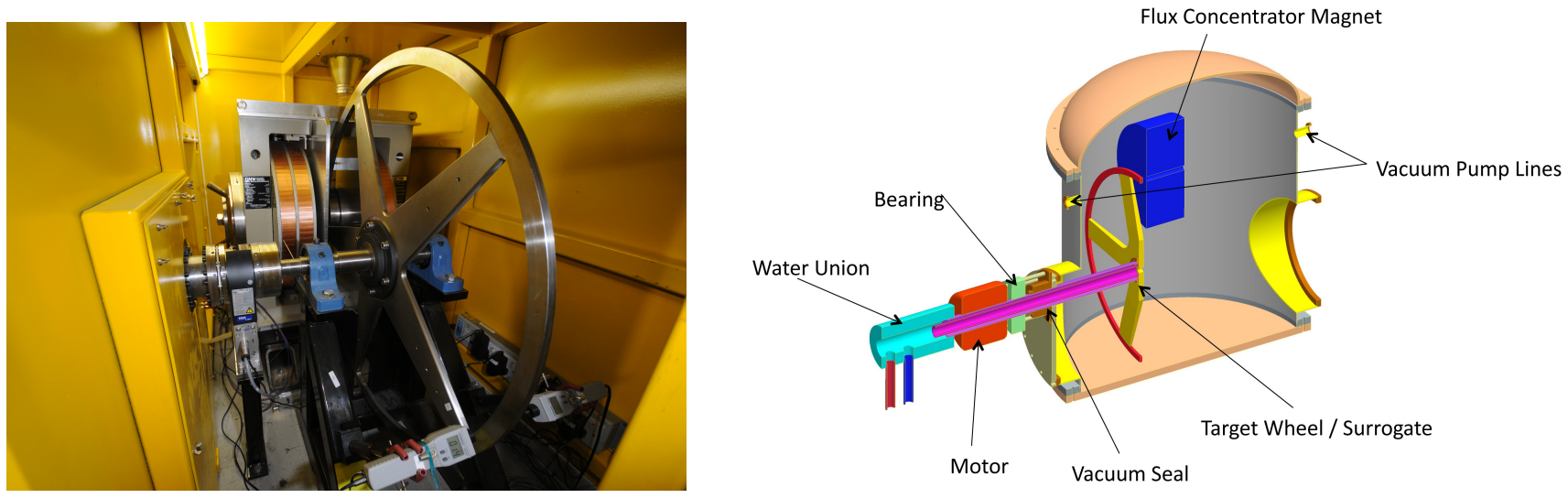

**(a)** Prototype setup for eddy current tests at Daresbury Lab.

**(b)** Schematic of the engineering design.

The target wheel also has to operate inside a vacuum chamber while the motor is in air. This means that a rotating vacuum seal is required that is capable of operating at high velocity, near a magnet and in a high radiation environment – quite a demanding challenge. The team has identified a commercial vacuum seal that appears suitable for ILC conditions. Tests of the long-term performance of the seal were performed at Lawrence Livermore National Laboratory (LLNL) in the US.

Initially, an equivalent load to the target was rotated in a vacuum at the design 2000 RPM. The performance of the seal was evaluated by monitoring the vacuum level within the chamber along with the torque, vibration and temperature of the shaft. The ongoing vacuum test looks promising although these seals have occasional pressure spikes. A differential pumping solution will be required to mitigate the effect of this feature of the ferrofluidic seal. The temperature of the ferrofluid at these speeds is near the manufacturer's limit using standard seal designs. Seals with improved cooling characteristics will be needed for the final design. The full-size target wheel used at the Cockcroft Institute is available at LLNL, and can be installed for testing under vacuum. The engineering design concept for this test is shown in Fig. 4.9b.

Another issue is the effect of the shockwave on the target as a consequence of being struck by the intense pulses of gamma photons generated by the undulator [227]. Concerns were raised over possible material damage to the target itself on a shot-by-shot basis. Simulations with a numerical code at LLNL suggested that the effect is not significant. This has since been confirmed with a detailed analytical study, carried out at Durham University in the UK [228], and at DESY and Hamburg University [226].

### 4.3.4 Optical Matching Device

The flux concentrator is the pulsed magnet that generates the strong magnetic field close to the target wheel in order to enhance the positron yield. Many of these have been used successfully in the past but the parameters of the ILC are more technically challenging. A detailed R&D study has been initiated at LLNL to confirm the feasibility of the proposed magnet and later to build a suitable prototype to demonstrate the design performance.

The team has now completed the design [229] which operates at room temperature using a water-cooled coil and disk (see Fig. 4.10). In this design, the device sits in vacuum; all power and cooling connections move to the rim; the coils are kapton wound, hollow copper and water cooled; the plates are OFHC copper with water cooling pipes soldered in; there is only metal in the high-radiation areas; plates and coils are stacked and bolted together. With the help of a pulse-forming network, the flux-concentrator B field has a much better flat top than the previous design (see Fig. 4.10b).

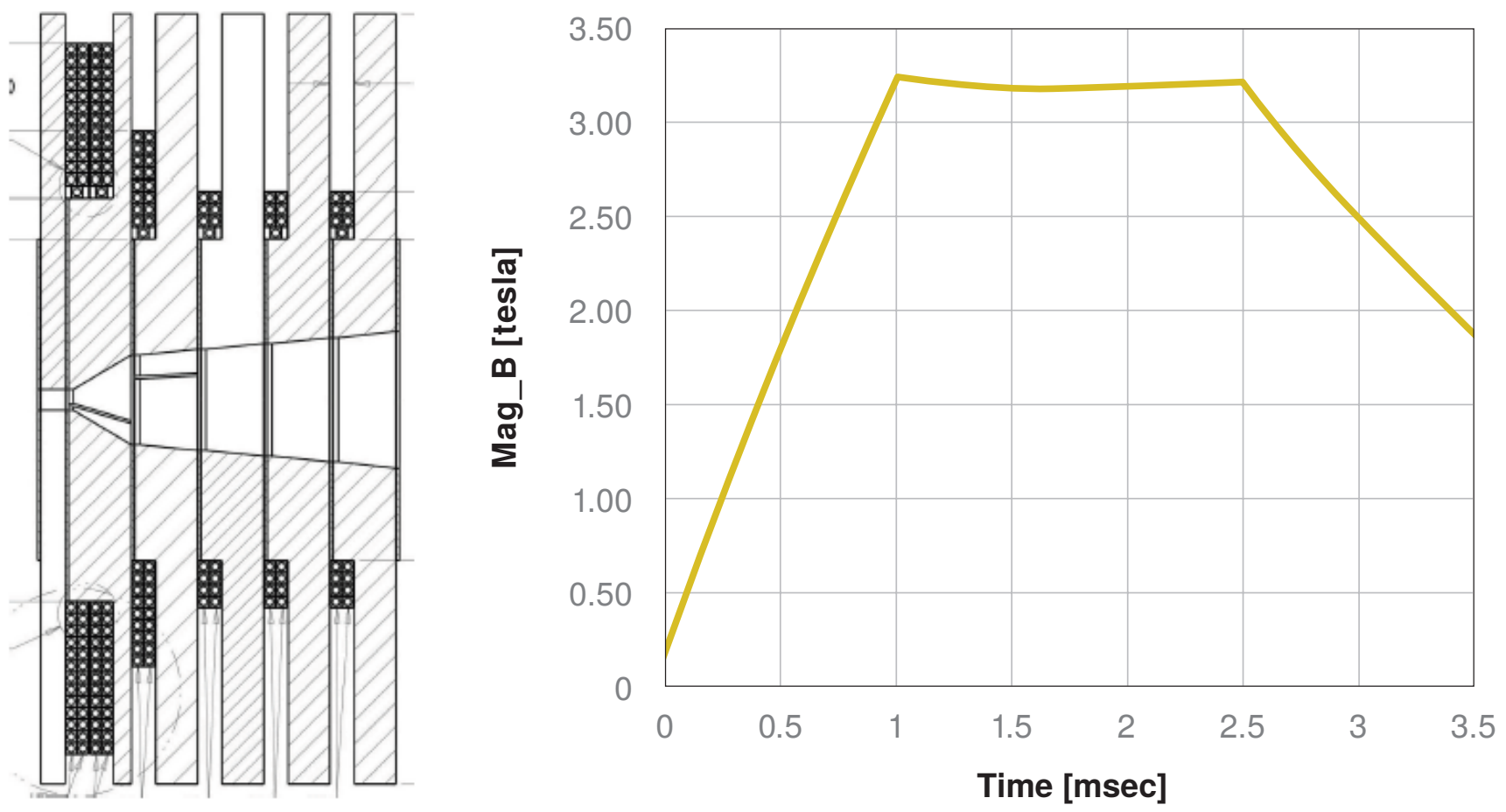

(a) (b)

**Figure 4.10**
(a) Drawing of new room-temperature flux concentrator; (b) the designed $B$ field pulse of the flux concentrator

### 4.3.5 Photon collimator

The positron-beam polarisation of $\sim$ 30 % can be increased to 50–60 % with a photon collimator upstream of the target. This reduces the number of photons, requiring compensation by a longer active undulator (a length up to 231 m is to be reserved). The collimator must withstand a large heat load as well as the deposition of a huge peak energy density [230]. Table 1 summarises the basic requirements for the photon collimator depending on the centre-of-mass energy and the undulator parameters based on a positron yield of $Y = 1.5\ \mathrm{e}^+/\mathrm{e}^-$.

**Table 4.3**
Basic requirements for the photon collimator at the positron source and resulting degree of positron polarisation. The number of bunches is 1312 with a bunch population of $2 \times 10^{10}$ positrons; the positron yield is $Y = 1.5\ \mathrm{e}^+/\mathrm{e}^-$ for the smallest aperture of the final collimator [222]

| Parameter | Unit | | | | | L upgrade |
|---|---|---|---|---|---|---|
| Centre-of-mass energy | GeV | 200-250 | 350 | 500 | 500 | 500 |
| Drive-electron-beam energy | GeV | 150 | 175 | 250 | 250 | 250 |
| Undulator $K$ value | | | | 0.92 | | |
| Undulator period | cm | | | 1.15 | | |
| Positron polarisation | % | 55 | 59 | 50 | 59 | 50 |
| Collimator-iris radius | mm | 2.0 | 1.4 | 1.0 | 0.7 | 1.0 |
| Active undulator length | m | 231 | 196 | 70 | 144 | 70 |
| Photon beam power | kW | 98.5 | 113.8 | 83 | 173 | 166 |
| Power absorbed in collimator | kW | 48.1 | 68.7 | 43.4 | 121 | 86.8 |
| Power absorbed in collimator | % | 48.8 | 60.4 | 52.3 | 70.1 | 52.3 |

The collimator parameters, in particular the radius of the collimator iris, are coupled to the drive-beam energy. Hence, instead of a universal collimator for the whole centre-of-mass-energy range, a multi-stage collimator design with decreasing iris is used. Each stage is constructed from the same materials but has different lengths to absorb the part of the photon beam with lower polarisation.

The first and longest section of the collimator is made of pyrolytic Carbon. This is followed by layers of materials with increasing $Z$, Titanium, Iron and Tungsten, which stop the shower and reduce the intensity of the absorbed photon beam after the final collimator to less than 0.1 % . Each section must be long enough to reduce the power to a value which can be handled by the next section. Conical bores in the Carbon and Titanium parts ensure a smooth heat-load distribution over a large region. The average heat load as well as the peak values of the energy deposition (PEDD) in the material are below the recommended stress limits. The collimators are water cooled; the pipes are embedded in a copper layer which encloses the carbon, titanium and iron parts.

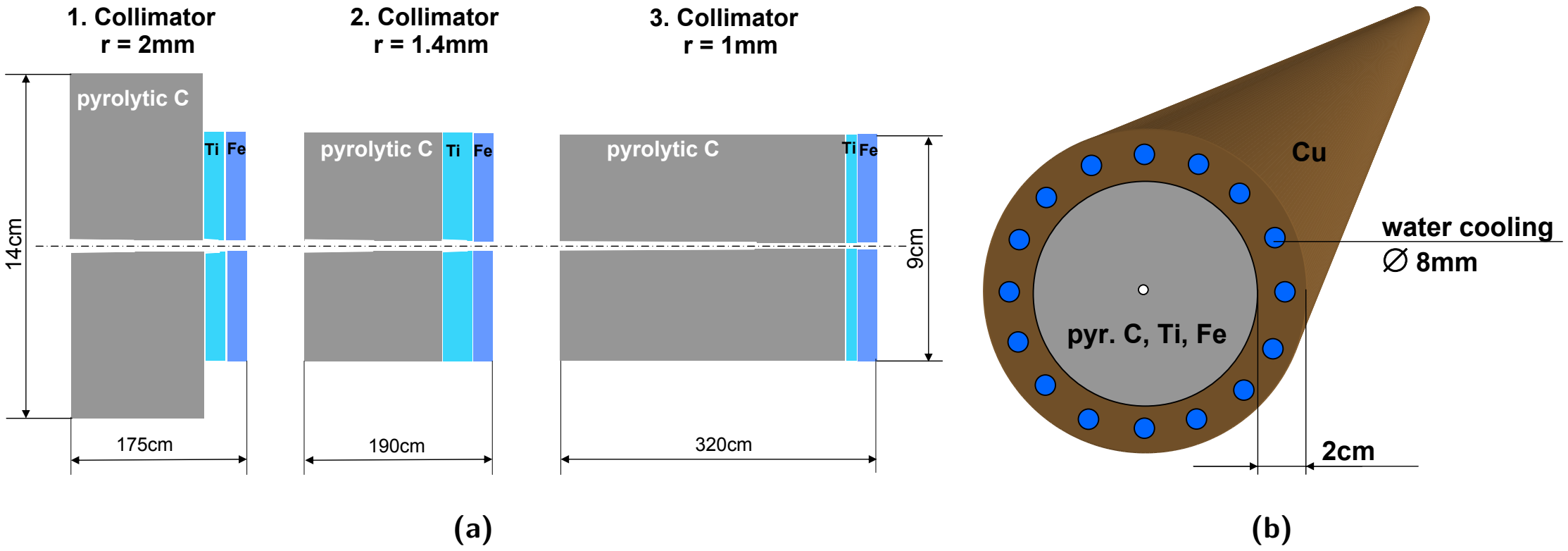

**Figure 4.11.** Basic layout of the multi-stage photon collimator at the positron source: (a) longitudinal section; (b) three-dimensional CAD model showing copper lading and water-cooling channels.

A sketch of the collimator layout is shown in Fig. 4.11. Each collimator stage can be moved onto or off the beamline. The maximum energy deposition and the average heat load in the collimator materials is summarised in Table 4.4. For the luminosity upgrade at 500 GeV, the values given in the last column have to be doubled.

**Table 4.4.** The peak power densities deposited in the components of the first (($r = 2\,\mathrm{mm}$), second ($r = 1.4\,\mathrm{mm}$) and third ($r = 1\,\mathrm{mm}$) collimator stage for different centre-of-mass energies and the corresponding degrees of polarisation. The number of bunches is 1312 with a bunch population of $2 \times 10^{10}$ positrons; the positron yield is 1.5 $\mathrm{e}^+/\mathrm{e}^-$ for the smallest aperture of the final collimator. Note the 10 Hz mode is used for 200-250 GeV case.

| Parameter | Unit | | | | | | |
|---|---|---|---|---|---|---|---|
| Centre-of-mass energy $E_{\mathrm{cm}}$ | GeV | $200-250$ | 350 | | 500 | | |
| Drive-electron-beam energy | GeV | 150 | 175 | | 250 | | |
| collimator stages | | 1. | 1. | 2. | 1. | 2. | 3. |
| iris radius $r$ | mm | 2 | 2 | 1.4 | 2 | 1.4 | 1 |
| $\mathrm{e}^+$ polarisation | % | 55 | 59 | | 50 | | |
| max. energy density in | | | | | | | |
| pyrolytic C | J/g | 176.5 | 143.9 | 183 | 36.1 | 73.0 | 81.7 |
| Titan | J/g | 16.4 | 18.4 | 21.6 | 5.8 | 30.0 | 12.9 |
| Iron | J/g | 12.8 | 12.8 | 13.3 | 5.1 | 11.1 | 9.5 |
| total power deposition in | | | | | | | |
| pyrolytic C | kW | 44.8 | 36.3 | 25.9 | 7.9 | 12.9 | 17.9 |
| Titan | kW | 0.8 | 0.8 | 0.9 | 0.2 | 0.6 | 0.2 |
| Iron | kW | 0.2 | 0.3 | 0.2 | 0.1 | 0.1 | 0.1 |
| Copper (cooling) | kW | 2.3 | 1.9 | 2.4 | 0.4 | 1.2 | 1.8 |

The dimensions of the collimator stages, the average heat load in the sections and the corresponding cooling power are given in Table 4.5. The outer diameter is 20–30 cm. Due to radiation activation, remote handling will be required.

**Table 4.5.** Dimensions of the photon collimator parts. [222]

| | 1st collimator | | | 2nd collimator | | | 3rd collimator | | |
|---|---|---|---|---|---|---|---|---|---|
| iris radius $r$ | 2 mm | | | 1.4 mm | | | 1 mm | | |
| | length [cm] | out. radius [cm] | weight [kg] | length [cm] | out. radius [cm] | weight [kg] | length [cm] | out. radius [cm] | weight [kg] |
| pyr. C | 135 | 7.0 | 72.7 | 140 | 4.5 | 31.2 | 290 | 4.5 | 64.6 |
| Titan | 20 | 4.5 | 13.8 | 30 | 4.5 | 8.6 | 10 | 4.5 | 2.9 |
| Iron | 20 | 4.5 | 24.2 | 20 | 4.5 | 10 | 20 | 4.5 | 10 |
| active length | 175 | | | 190 | | | 320 | | |

### 4.3.6 Prototyping of normal-conducting RF accelerating structures and RF breakdown Study

Due to the extremely high energy deposition from positrons, electrons, photons and neutrons behind the positron target, the 1.3 GHz pre-accelerator has to use normal conducting structures up to an energy of 400 MeV. Major challenges are achieving adequate cooling with the high RF and heating from particle loss, and sustaining high accelerator gradients during millisecond-long pulses in a strong magnetic field. The positron-capture section contains both standing-wave (SW) and traveling-wave (TW) L-band accelerator structures. Detail about the parameters of these standing wave and travelling wave L-band accelerator structures can be found in Part II Section 5.5.4.

A half-length (5-cell) prototype standing-wave (SW) cavity was built at SLAC to verify that the gradient of 15 MV/m in 1.0 ms pulses can be achieved stably and without significant detuning from the RF heat load of 4 kW per cell. The cavity cross section is shown in Fig. 4.12. Figure 4.13a is a plot of the cold-test measurement of the mode frequencies (dots); the solid line is the fitted dispersion curve. The unloaded Q of the cavity is ~29000 and the operating frequency (at $\pi$ phase advance) is 1299.7 MHz. The time constant of this critically coupled cavity (0.5 $Q_o/\omega$) is 1.8 µs.

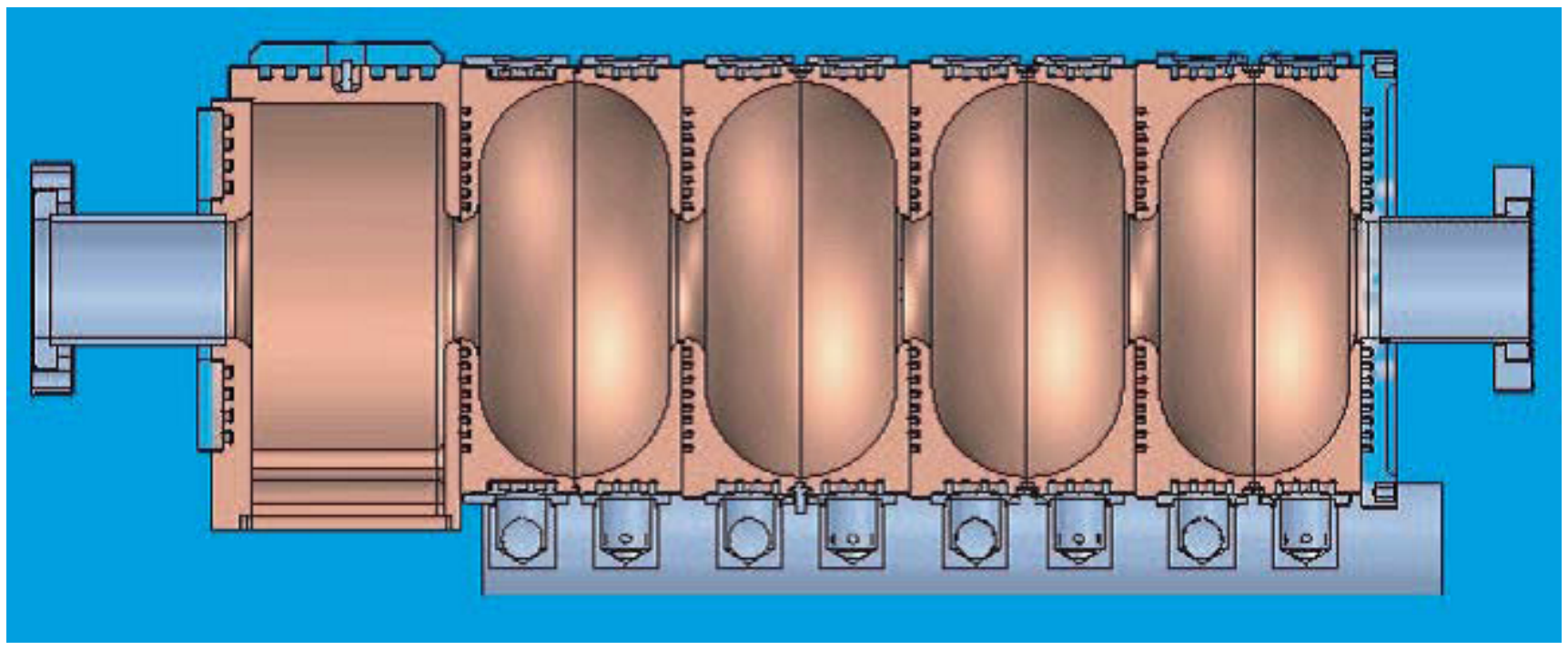

**Figure 4.12** Cross section of 5-cell standing-wave cavity with cooling channels

The cavity has been rf processed at the $\pi$ mode for about 530 h and it has incurred about 6200 breakdowns. The gradient goal of 15 MV/m with 1 ms pulses has been achieved. Figure 4.13b shows the breakdown-rate history during processing. For these data, the pulse repetition rate was 5 Hz except for 1 ms pulses where it was lowered to 1 Hz to reduce the detuning as the reflected power was causing waveguide breakdowns (the source of these breakdowns has since been eliminated and 5 Hz operation is expected to be possible in the future). The cooling system was designed to have about 25 % reflected power when the cavity was turned on 'cold' which then dropped to zero in steady state with full RF power dissipation (20 kW at 15 MV/m). In this way, a cavity temperature control system is not needed (at least for testing). However, the flow rate that was achieved (due to a limited supply of temperature-regulated water) was 86 gpm compared to the 140 gpm desired, which increased the cavity temperature by about 50 % (to 0.13 °C per kW dissipated). Also, the detuning of -2.7 kHz/kW was about 25 % larger than expected from simulations using the actual temperature rise and led to an overall reflection of about 50 % when cold with the appropriate choice of RF frequency to minimise the reflection at full power in steady state.

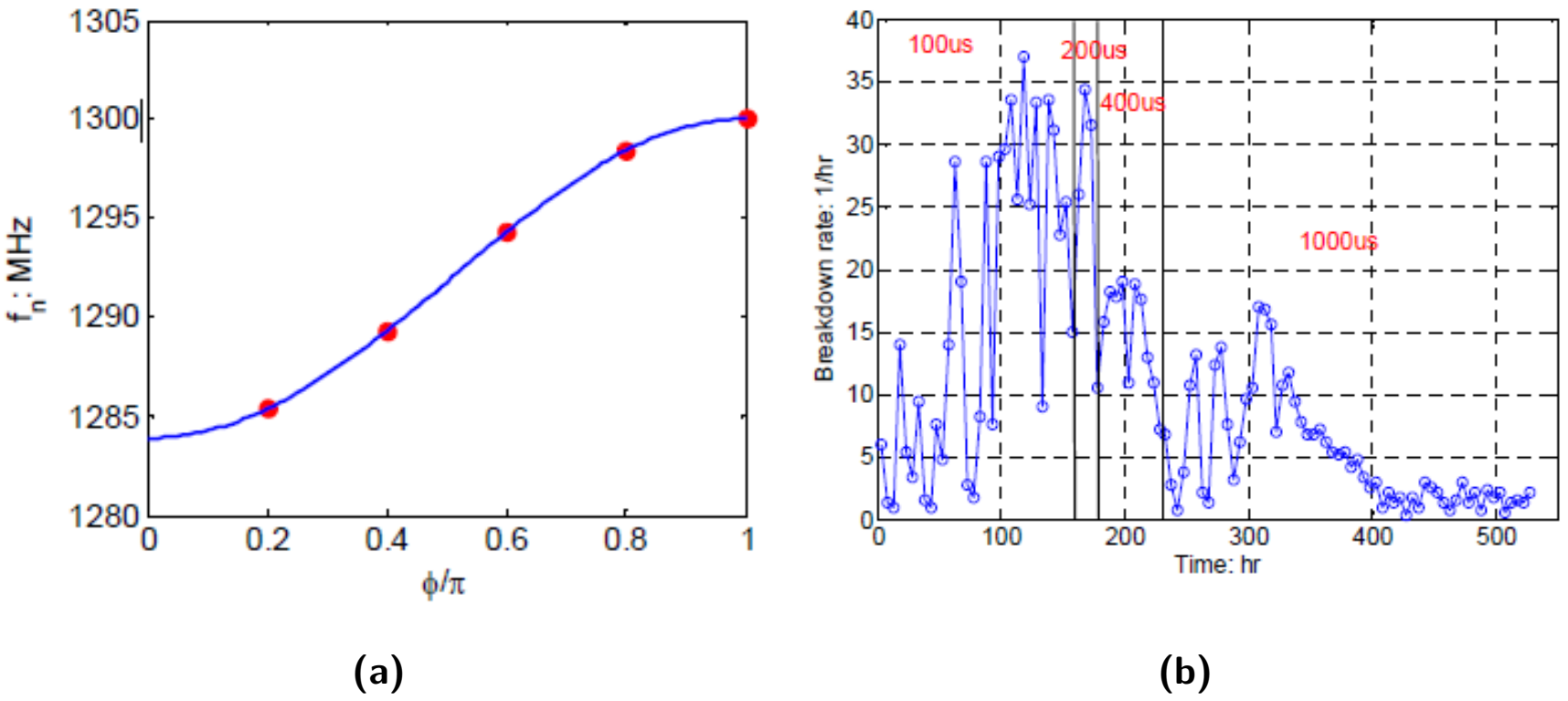

**Figure 4.13**
(a) The 5 cell standing wave linac dispersion curve. (b) Breakdown-rate history of the cavity with different pulse length.

After hundreds of hours of conditioning, the breakdown rate was measured as a function of unloaded gradient (G), as shown in Fig. 4.13b. A $\mathrm{G}^{21}$ dependence was found, which is less steep than the $\mathrm{G}^{32}$ dependence recently observed for a CERN-designed, low a/$\lambda$, TW, X-band (11.4 GHz) structure. However, the gradient exponent is within the 20–30 range measured for NLC/GLC X-band prototype structures and close to the 19.5 value measured for the FNAL 805 MHz injector linac cavities.

## 4.3.7 Performance simulation

The parameters for source subsystems and the determination of the source performance can only be quantified using complex simulations. These simulations quite often require the combination of several sophisticated computer codes. The primary performance figure of merit is the yield, defined as the number of positrons captured in the acceptance of the damping ring per electron passing through the undulator. (Ideally a yield of 1 is required, but the design goal is set at 1.5 to allow a 50 % safety factor.) The second figure of merit is the polarisation of the captured positrons, which depends on additional parameters such as the collimation aperture of the photon beam before the target.

As shown in Fig. 4.14 and in Part II Fig. 5.3, for a given undulator and positron-capturing system, the performance of the positron source strongly depends on the energy of the main electron beam. At higher energy, the undulator B field is re-optimised to restore the polarisation to 30 %. The final parameters are listed in Part II Table 5.2. Performance simulation studies have also been done to determine the target energy deposition and the impact of the undulator on the electron-beam parameters [231–233].

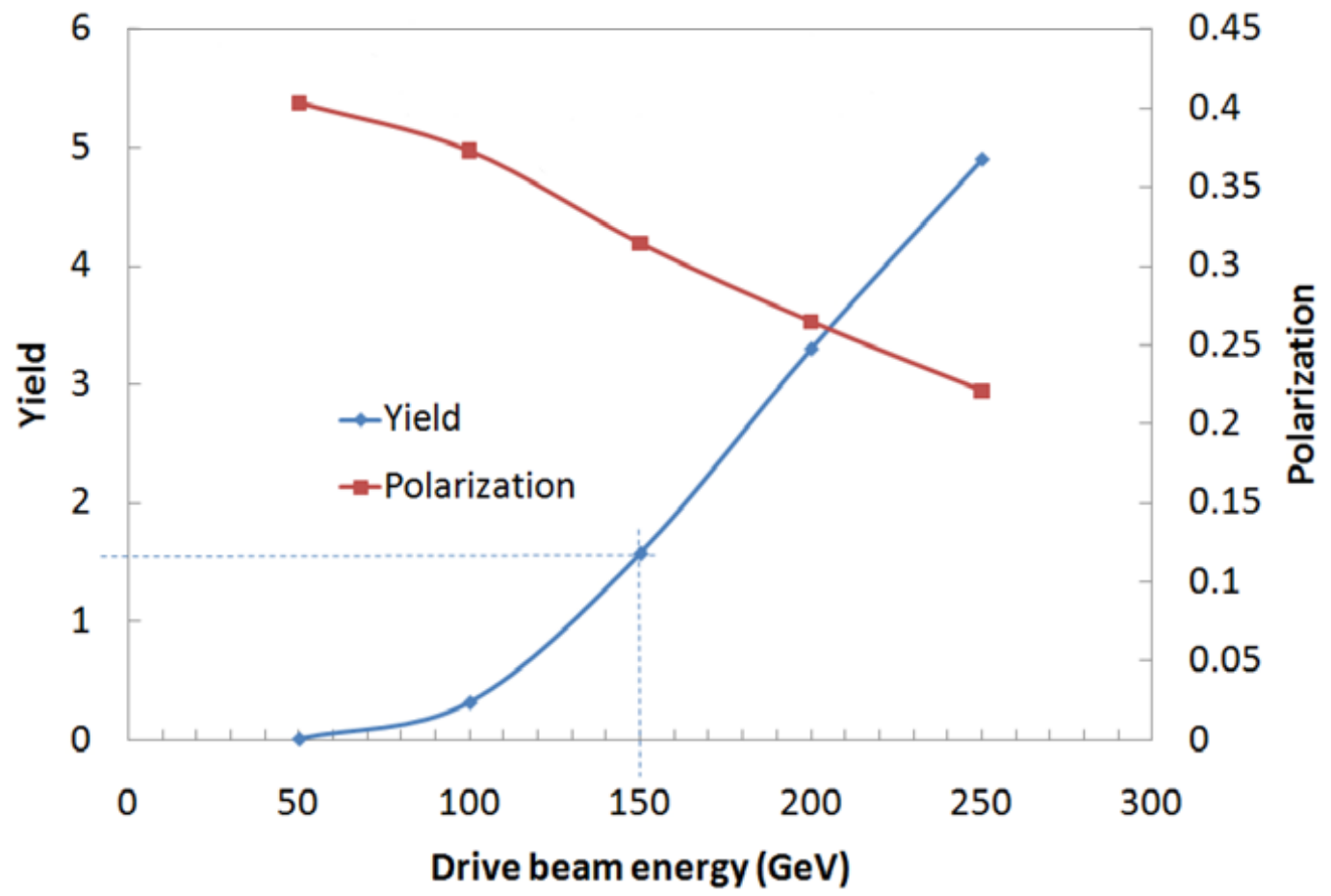

**Figure 4.14**
Simulation results of positron source yield and polarization for 231 m RDR undulator with QWT as OMD.

## 4.3.8 Lattice design

The layout of the ILC positron source beam line is shown in Part II Fig. 5.2 and the optics functions are shown in Part II Fig. 5.6. The lattice downstream of the separation chicane and through the transport line is the same as in the RDR. The quad settings of the FODO transport lattice have been re-optimised to minimise emittance growth and maximise transmission. The booster Linac lattice has been redesigned around the real geometry of cryomodules. The Linac-to-Ring lattice has also been redesigned to adapt to the central-region integration [234].

## 4.3.9 Spin flipper

The positron source based on a helical undulator provides a beam with a net polarization; its direction is defined by the helical winding of the undulator. Since the helicity of the positrons must be reversible as fast as that of the electrons, a spin flipper [235] complements the spin rotator system located in the PLTR line. The principle of the spin rotator and flipper system is shown in Fig. 4.15.

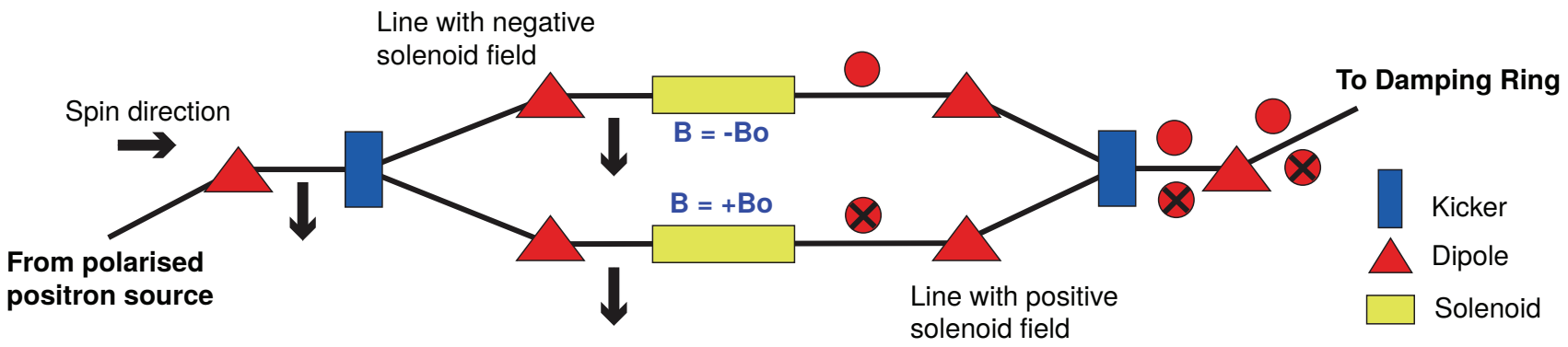

**Figure 4.15** Schematic layout of positron transport to Damping Ring with spin rotator section

The beam is kicked into one of two identical parallel transport lines to rotate the spin in either the up or down vertical direction. Horizontal bends rotate the spin by $3 \times 90°$ from the longitudinal to the transverse horizontal direction. In each of the two symmetric branches a 5 m long solenoid with an integrated field of 26.2 Tm aligns the spins parallel or anti-parallel to the B field in the damping ring. Both lines are merged using horizontal bends and matched to the PLTR lattice. The length of the splitter section is approximately 26 m. Since the position of the solenoids is shifted along the straight section a horizontal offset of 0.54 m for each branch is obtained.

## 4.3.10 Remote handling and radiation shielding

The target will be highly activated after one year of operation. The RAL group produced a conceptual design of the target remote-handling system. Recent calculations confirmed the results from the RAL studies and showed that, with the nominal 150 kW photon beam, after 5000 hours of operation and 1 week of shutdown, the equivalent dose rate at 1 m from the target wheel will be approximately 170 mSv/h. Concrete shielding 0.8 m thick around the target is thought to be sufficient fully to contain the radiation associated with the beam and with the subsequently activated materials. Additional shielding may be needed upstream and downstream of the target and around the beam dump. Figure 4.16 shows the calculation results for the target activation and required shielding.

A remote-handling system is used to replace the target, OMD and 1.3 m NC RF cavities. To minimise the time required to exchange the target, the whole target system is designed into a plug with the top shielding. Inflatable seals for the vacuum enclosure form the interfaces of the target plug with the two beam lines. The targets will be removed and replaced vertically by an overhead crane, the disconnection and reconnection of the target being done locally. The used target is placed into a shielding cask for storage until the radiation has decayed to a safe level. Figure 4.17 shows the conceptual design of such a system.

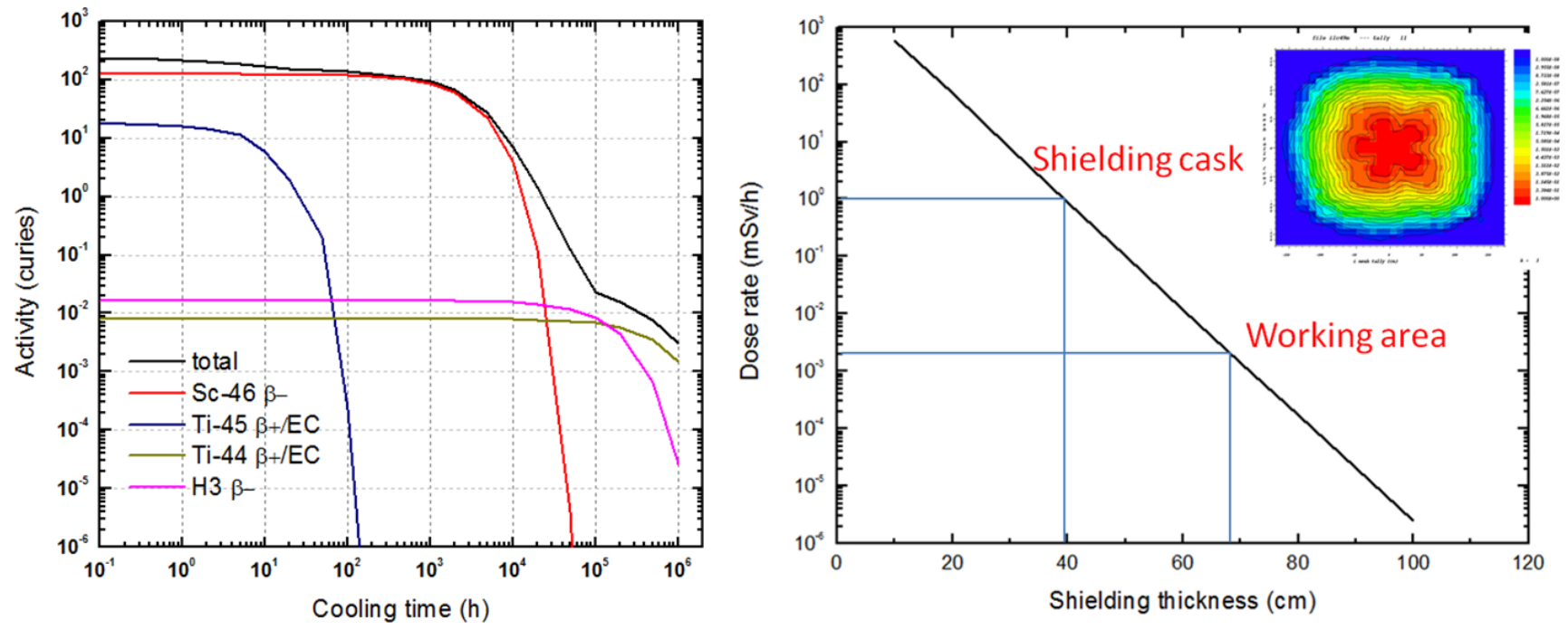

**Figure 4.16**
Shielding calculation for used target

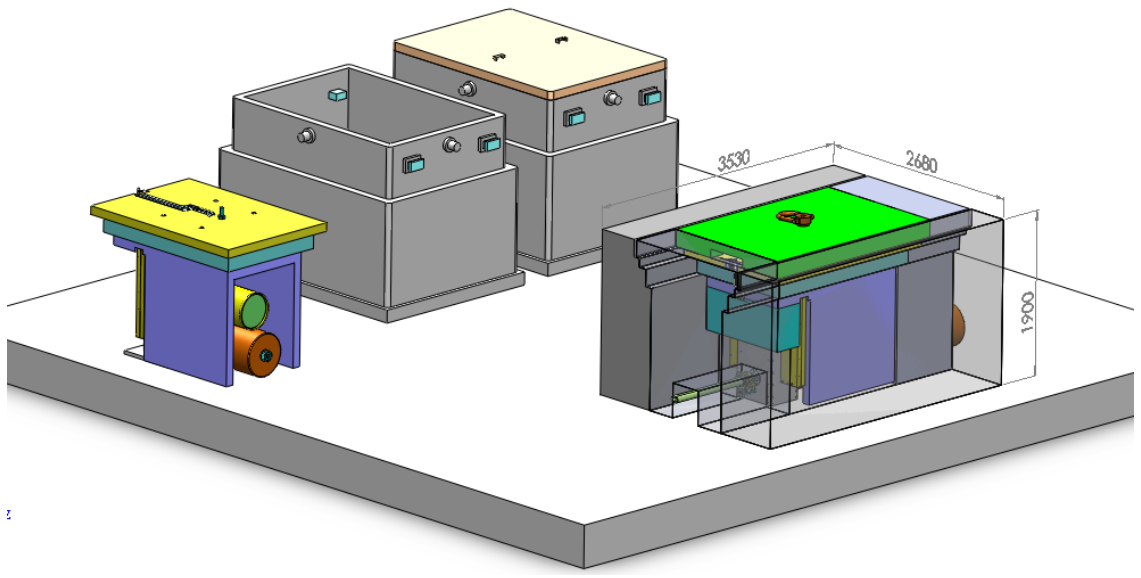

**Figure 4.17**
General layout concept for used-target remote handling.

## 4.3.11 Alternative source

As described above, the baseline design adopts a novel concept of positron generation, i.e. the undulator-based scheme. Although it can satisfy necessary requirements for the ILC positron source, some components, in particular the rotating target and the flux concentrator, are still under development for the final design. In view of this, two alternative positron sources are being investigated. First is the conventional positron source [236], which makes use of a few GeV electron beam on a thick target to produce positrons. This has been studied as a backup scheme for the baseline. The disadvantage is that it cannot create a polarised positron beam. The second option is a laser-Compton-based polarised-positron source as a possible future advanced scheme [237]. Both of these alternatives have an advantage that the positron source is independent of the electron beam to be used for collision. This would eliminate the complexity of operation, the need for an auxiliary source, the need for 10 Hz operation, and the constraint on the beam-line length.

### 4.3.11.1 Conventional source

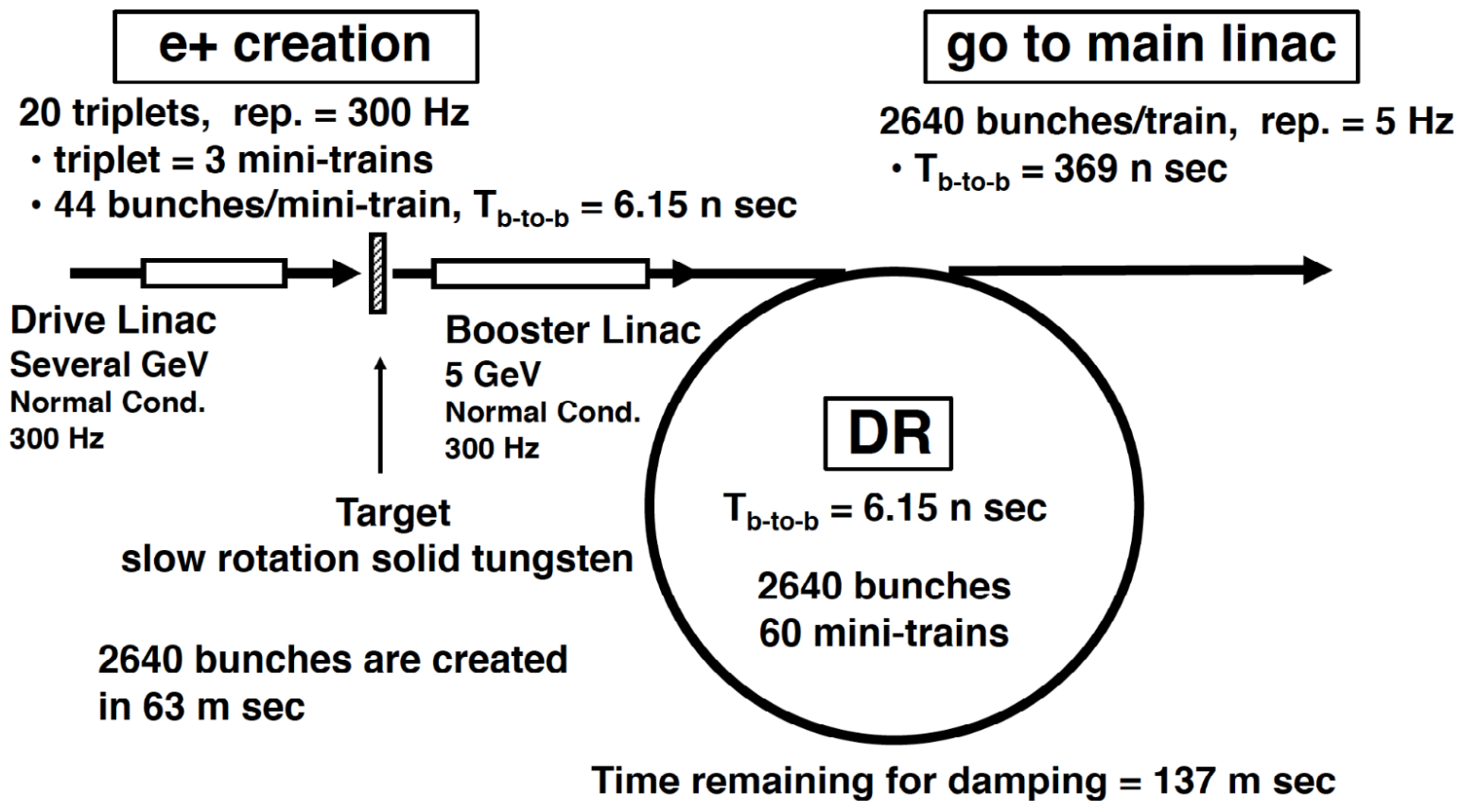

**Figure 4.18**
Schematic view of the 300 Hz positron generation scheme. The repetition rate of the drive linac and the booster linac is 300 Hz, whereas the repetition rate of the main linac is 5 Hz.

Though the concept of the electron-driven positron source is well known, ILC requires a new regime of the thermal and shockwave parameters in the target. The current design avoids these problems by stretching the pulse length of a bunch train. The schematic view of the 300 Hz scheme is shown in Fig. 4.18. (The parameters for 2640 bunches per pulse are given, which are close to those at full power.)

The repetition rate of the ILC main linac is 5 Hz, giving an interval of 200 ms between two pulses, which is adequate for pulse stretching. We employ a normal conducting 300 Hz electron linac to create positrons. The pulse-to-pulse separation of the linac is 3.3 ms. Each pulse of the linac creates about 130 bunches, so 20 linac pulses create about 2600 bunches of positron in about 63 ms. This means 137 ms is left for the damping in the damping ring.

In the linac, a bunch-to-bunch separation of 6.15 ns is chosen, which is identical to that in the damping ring. To match the mini-train beam structure in the damping ring, one RF pulse of the linac accelerates three mini-trains with inter-mini-train gaps. This package of three mini-trains is called a triplet. Each mini-train contains 44 bunches. With 6.15 ns bunch-to-bunch separation and 44 bunches per mini-train, the length of a mini-train is 264 ns. Since a triplet contains three mini-trains, it consists of 132 bunches, thereby requiring 20 triplets to form 2640 bunches. There are gaps of about 100 ns between the mini-trains in a triplet. They are necessary for suppressing the instability caused by electron clouds in the damping ring.

In addition to employing 300 Hz generation, it is important to optimise the drive beam and target parameters to reduce target thermal issues. An intense simulation study concluded that a single production target made of tungsten alloy can be employed in the conventional source if appropriate parameters are used. The typical choice of the drive-beam parameters are: bunch charge of 3.2 nC, energy of 6 GeV, and R.M.S. spot size on the target of 4 mm. With these drive-beam parameters, the optimal target thickness is 14 mm. The slow rotation (tangential speed 0.5-1 m/s) is enough to deal with the thermal load. This is much slower than that of the undulator-based positron source.

Downstream of the target is a positron-capture system consisting of an Adiabatic Matching Device (AMD) and an L-band RF-section. A flux concentrator forms the AMD. The required pulse length of the flux concentrator is about 1 μs, which is the length of the triplet. This length is similar to that of existing flux concentrators. Moreover, the pulse length is short enough to use a high accelerating gradient in the RF section. After exiting the capture section, the positron energy is boosted to 5 GeV in a 300 Hz normal-conducting linac. Then a kicker with a pulse length of about 1 μs and repetition rate 300 Hz is employed to inject triplets of positrons into the DR.

#### 4.3.11.2 Compton Sources

A disadvantage of the conventional scheme is that it cannot provide polarised positrons. To overcome this problem an upgrade to a polarised source in a later stage is considered by making use of the inverse-Compton scattering between electrons and a laser. Three possible schemes are under investigation: Ring-Compton; ERL-Compton; and Linac-Compton. The first two are similar: both use a 1-2 GeV electron beam, 1 micron laser light, and high-finesse laser stacking cavities in multiple laser-electron collision points. Both assume stacking of generated positrons in a storage ring. The Linac-Compton is rather different. It uses a 5-6 GeV electron beam and $CO_2$ laser light. Regenerating amplification in low-finesse laser cavities are assumed at the collision points. Positron stacking is not assumed.

The Compton source must match the 300 Hz normal-conducting linac for a smooth upgrade from non-polarised to a highly polarised positron source. Linac-Compton will smoothly upgrade the conventional source. The 6 GeV 300 Hz drive linac of the conventional scheme is reused. At the end of the linac, Compton IPs are placed, followed by the positron-production target. The 5 GeV booster linac is also reused. On the other hand, the Ring- and ERL-Compton will need relatively large modification of the system. What must be added are: a storage ring or an ERL as the electron driver for the Compton scattering; a low-energy storage ring for positron stacking; a long pulse linac (most probably superconducting) between positron production and the stacking ring. Figure 4.19 shows a typical schematic view of the Ring- and ERL-Compton scheme as upgrade options from the conventional source.

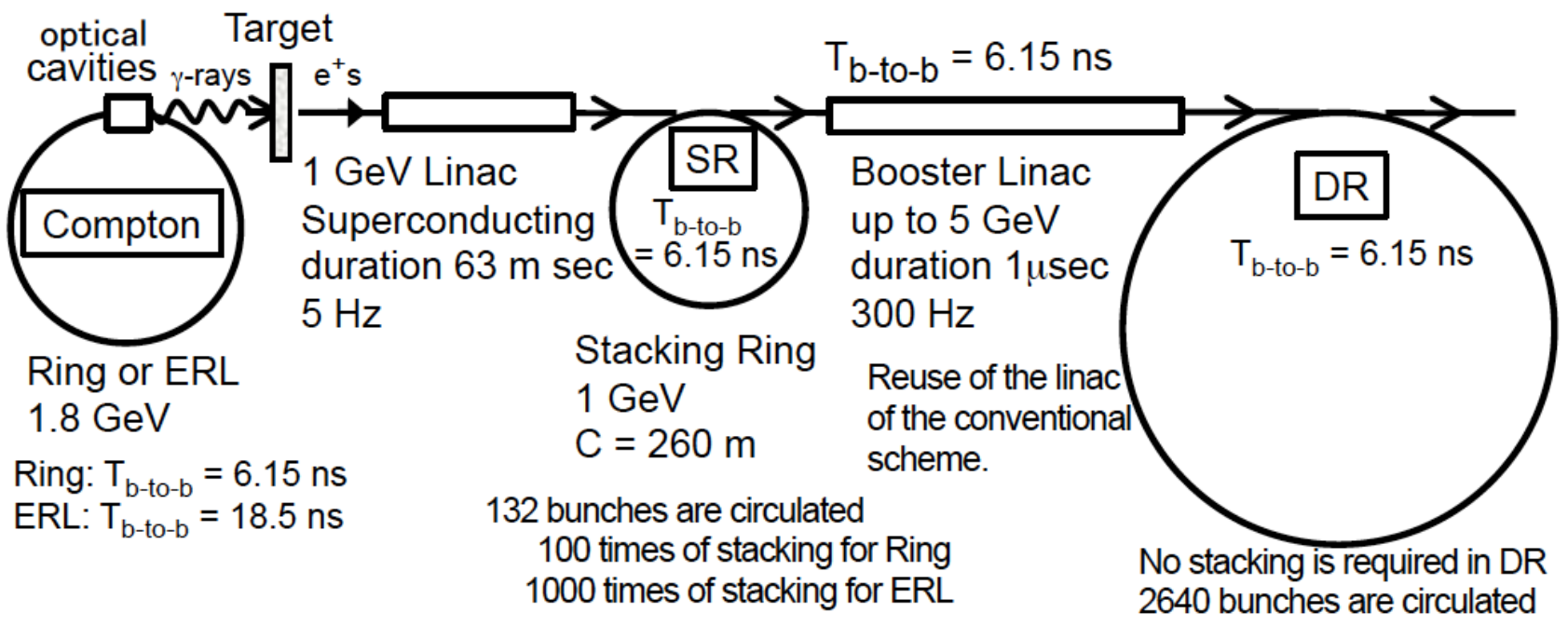

**Figure 4.19** The conceptual design of Ring-Compton and ERL-Compton for the future upgrade from the 300 Hz conventional positron source.

Several critical issues remain with the Compton scheme.

For the Linac-Compton scheme, reliable $CO_2$ laser technology with a regenerating laser cavity and the heavy beam-loading compensation scheme of the linac are essential. For the Ring- and ERL-Compton schemes there are many common issues to be addressed. Firstly, a high-finesse laser-light stacking cavity is necessary. Both schemes assume a cavity with the enhancement factor $10^4$, while the currently achieved value is 1200. Further improvement of the feedback system is needed to achieve this goal. Secondly, the large stored power in the optical cavity is also an issue. It raises heat and destruction issues for the cavity mirrors. To deal with the issue of the laser-light power, it is necessary to develop very good mirror-coating with the surface loss less than 1 ppm. A larger laser spot size on the mirror is also required to deal with the high power. It requires a longer cavity (longer mirror-to-mirror distance). The length of the current optical 4-mirror cavity is 0.42 m (round-trip path length 1.64 m). Development cavities with twice the lenght is now under way at KEK. Reduction of the crossing angle is also one of the critical R&D issues. It significantly increases the luminosity of laser-electron collision. A longer cavity makes a smaller crossing angle possible. Head-on collision is optimal for luminosity production. A test of head-on collision with two bending magnets upstream and downstream of the Compton collision point is under way at KEK. If head-on collisions are employed, the generated gamma-rays pass through the downstream mirror. Studies of possible mirror damage by the gamma-rays is necessary. Employing a crab-crossing is another way to obtain head-on collisions. A detailed feasibility study will be necessary. In the ERL-Compton scheme, the effect of finite crossing angle is much smaller than that in the Ring-Compton scheme because a bunch length significantly shorter than that in the storage ring can be achieved.

The positron stacking in a storage ring is another challenge of the Ring- and ERL-Compton schemes. The simulation studies performed at CERN showed that a positron-stacking efficiency around 95 % can be achieved. This is significant progress but a still-higher efficiency is necessary to avoid radiation activation of the stacking ring.

## 4.4 Damping ring

The ILC R&D programme identified the key areas for work during the Technical Design Phase:

- developing methods to suppress the electron cloud instability
- demonstration of operation at ultra-low vertical emittance (vertical emittance of 2 pm)
- demonstration of fast injection/extraction kickers performance.

Two dedicated test facilities were identified for this effort: CesrTA at Cornell University and ATF at KEK. Both programmes have been carried out by large collaborations, with contributors from institutions worldwide working on simulation, experiment and design.

### 4.4.1 Electron-cloud mitigation

The R&D effort on electron-cloud mitigation involves the large international collaboration participating in the CesrTA programme plus the effort that is in progress at other laboratories.

Results of the damping ring R&D program are described in Section 3.5 on the CesrTA programme.

### 4.4.2 Ultra-low-emittance operation

The demonstration of ultra-low emittance was carried out in the framework of the CesrTA and ATF collaborations, but important results have also come from the synchrotron-light-source community.

#### 4.4.2.1 Diagnostics for low-emittance beams at ATF

The ATF damping ring achieved a vertical emittance as low as 4 pm before the publication of the RDR and has supported a wide range of important research for many years: low-emittance tuning and intra-beam scattering studies, studies of the fast-ion effect and fast-kicker tests. Currently, the main focus of activity at the damping ring is the production of an extracted beam with the required characteristics for the ATF2 programme (see Section 3.6) and the development and test of diagnostics for low-emittance beams. Instrumentation development includes laser wire, optical transition radiation, optical diffraction radiation, and a high-resolution X-ray monitor [196, 238, 239].

#### 4.4.2.2 Diagnostics and tuning algorithms at CesrTA

The low-emittance-tuning effort provides the foundation for studies of the emittance-dilution effects of the electron cloud in a regime approaching that of the ILC damping rings. The vertical-emittance goal for the initial phase of the CesrTA programme is less than 20 pm. Low-emittance-tuning efforts have focused on the systematic elimination of optical and alignment errors that are the sources of vertical-emittance degradation [169]. Techniques have been developed to eliminate systematic errors in the beam-position monitors, measuring gain variation among the four button electrodes on each beam-position monitor, and to centre the monitors with respect to the adjacent quadrupole. Work has also been carried out to optimise the sextupole design, thus minimising sources of emittance coupling. During the most recent experimental run, this effort resulted in measurements of the vertical emittance consistent with having achieved the target vertical emittance of 20 pm in both single-bunch and multi-bunch operations.

An X-ray beam size monitor has been developed and successfully demonstrated at CesrTA. It is able to measure both integrated and single-bunch turn-by-turn beam sizes at positions for monitoring the progress of the low-emittance tuning of the machine and for beam dynamics related to instabilities driven by the electron cloud [170, 176].

#### 4.4.2.3 Demonstration of vertical emittance below 2 pm at synchrotron light sources

A step forward in the demonstration of very low vertical emittance has been achieved at some synchrotron light sources, which operate low-emittance storage rings with characteristics very similar to the ILC damping ring and have developed alignment procedures, machine modelling, tuning algorithms, and orbit stabilisation for coupling correction and low-vertical-emittance tuning [240]. In particular, the Diamond Light Source in the UK, the Swiss Light Source and the Australian Synchrotron storage ring have achieved betatron-coupling correction down to 0.1 % and vertical emittances below 2 pm [241–243]. Significant progress has been made in the development of diagnostic systems for the measurement of such small vertical emittances [244–246]. Two low-emittance-ring workshops, LER10, at CERN in January 2010 and LOW$\epsilon$RING 2011, in Crete in October 2011, were organised by the joint ILC-CLIC working group on damping rings. They were very successful in strengthening the collaboration within the two teams designing the damping rings and with the rest of the low-emittance-rings community, including synchrotron-light sources and B factories.

### 4.4.3 Performance of fast injection/extraction kickers

#### 4.4.3.1 ILC-like multi-bunch extraction at ATF

The injection/extraction kickers act as the bunch-by-bunch beam manipulator to compress and decompress the bunch spacing into and from the damping ring. The kickers require high-repetition rate, 3 MHz, and very fast rise and fall times of the kicker field: 6 ns for the nominal configuration and 3 ns for the proposed luminosity upgrade. The tolerance on horizontal beam jitter of the extracted beam is approximately 10 % of the beam size, which requires the relative stability of the extraction kicker amplitude to be below $7 \times 10^{-4}$.

A rise and fall time of 3 ns has been already demonstrated in the ATF using a 30 cm-long strip-line kicker together with a semiconductor source of high-voltage pulses [247]. The time response of the strip-line kicker was observed by measuring the resulting betatron-oscillation amplitude of the stored electron beam.

An ILC-type beam-extraction experiment using two strip-line kickers has been carried out at ATF [248]. The length of the strip lines is 60 cm and the gaps of the two electrodes are 9 mm and 11 mm. Two pairs of pulsers with a peak amplitude of 10 kV, a rise time of 1.5 ns and a repetition rate of 3.3 MHz are used to drive the strip lines. The strip-line kicker system produced a 3 mrad total kick angle for the 1.3 GeV beam. The rise time of the kick field is less than 5 ns.

The multi-bunch beam stored in the damping ring with 5.6 ns bunch separation was successfully extracted with 308 ns bunch spacing in the extraction line (see Fig. 4.20). No deterioration of the extracted vertical beam size was observed (as measured with the laser wire). The resynchronisation circuit used for precise timing adjustment worked stably. The relative angle jitter of the single-bunch beam extraction was $3.5 \times 10^{-4}$ rms, which is better than the requirements for ILC damping-ring extraction. For multi-bunch beam extraction a trigger timing circuit is needed to compensate the time drift of the pulser. Very recently, 30-bunch extraction with an rms angle jitter $10^{-3}$ has been achieved. This value can be further reduced by precise tuning of the timing system or by using a feed-forward system.

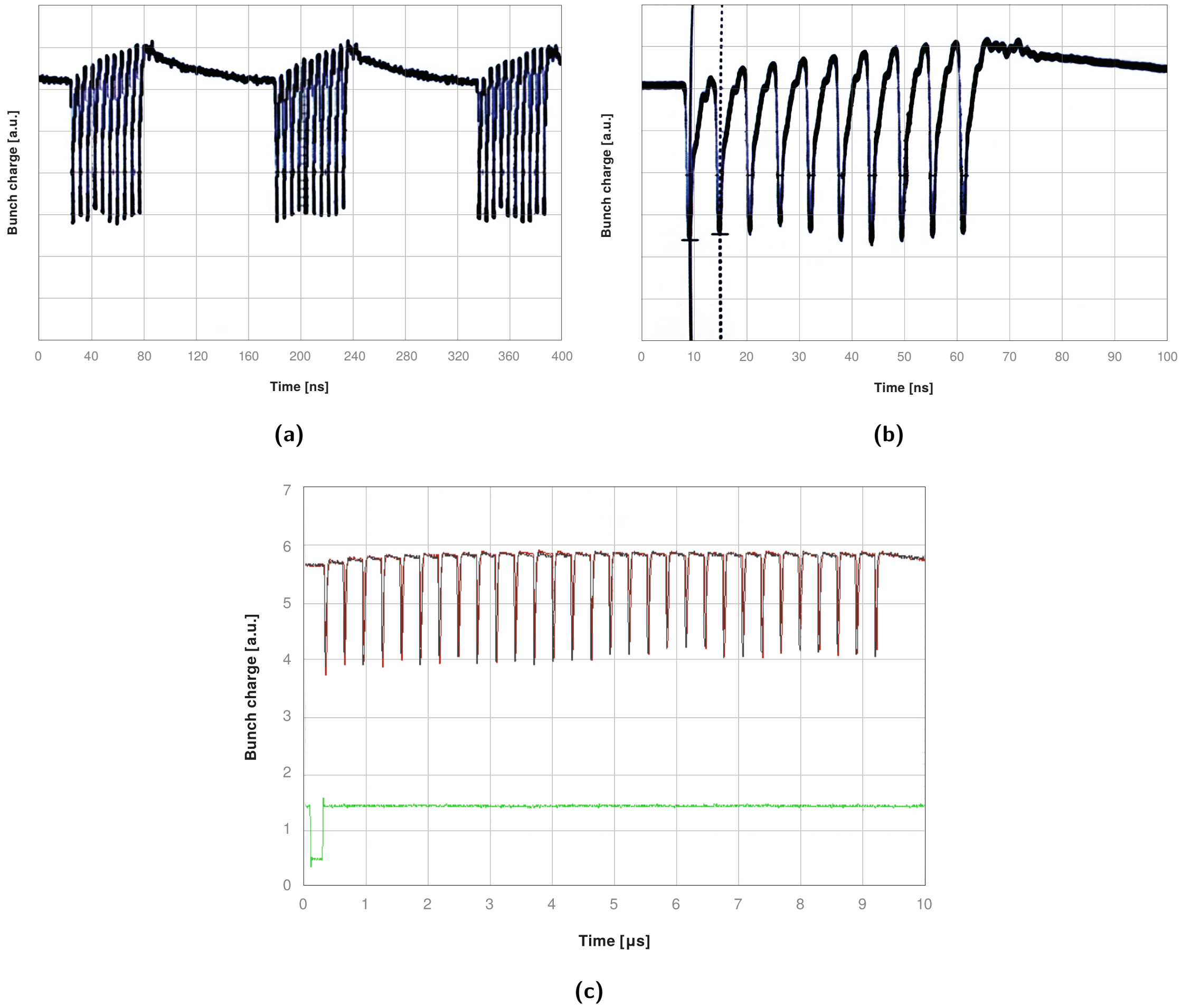

**Figure 4.20.** Bunch current monitor signals from the ATF damping ring and damping ring extraction line: (a) 3 trains of 10 bunches with 5.6 ns spacing stored in the damping ring; (b) close-up of one train in (a); (c) 1 train of 30 bunches with 308 ns spacing in the extraction line.

### 4.4.3.2 Strip-line-kicker design at DAΦNE

The design of the new, fast strip-line kickers for the injection upgrade of DAΦNE is based on strip-line tapering to obtain a device with low beam impedance and an excellent uniformity of the deflecting field in the transverse plane (see Fig. 4.21) [249]. These characteristics are essential also for the ILC damping ring; the experience gained with the new DAΦNE injection system will be applied to the design of the ILC damping-ring injection system. The rise and fall times of the kickers are all less than 6 ns, corresponding to the ILC damping-ring requirement for the nominal configuration.

**Figure 4.21**
DAΦNE strip-line kicker.

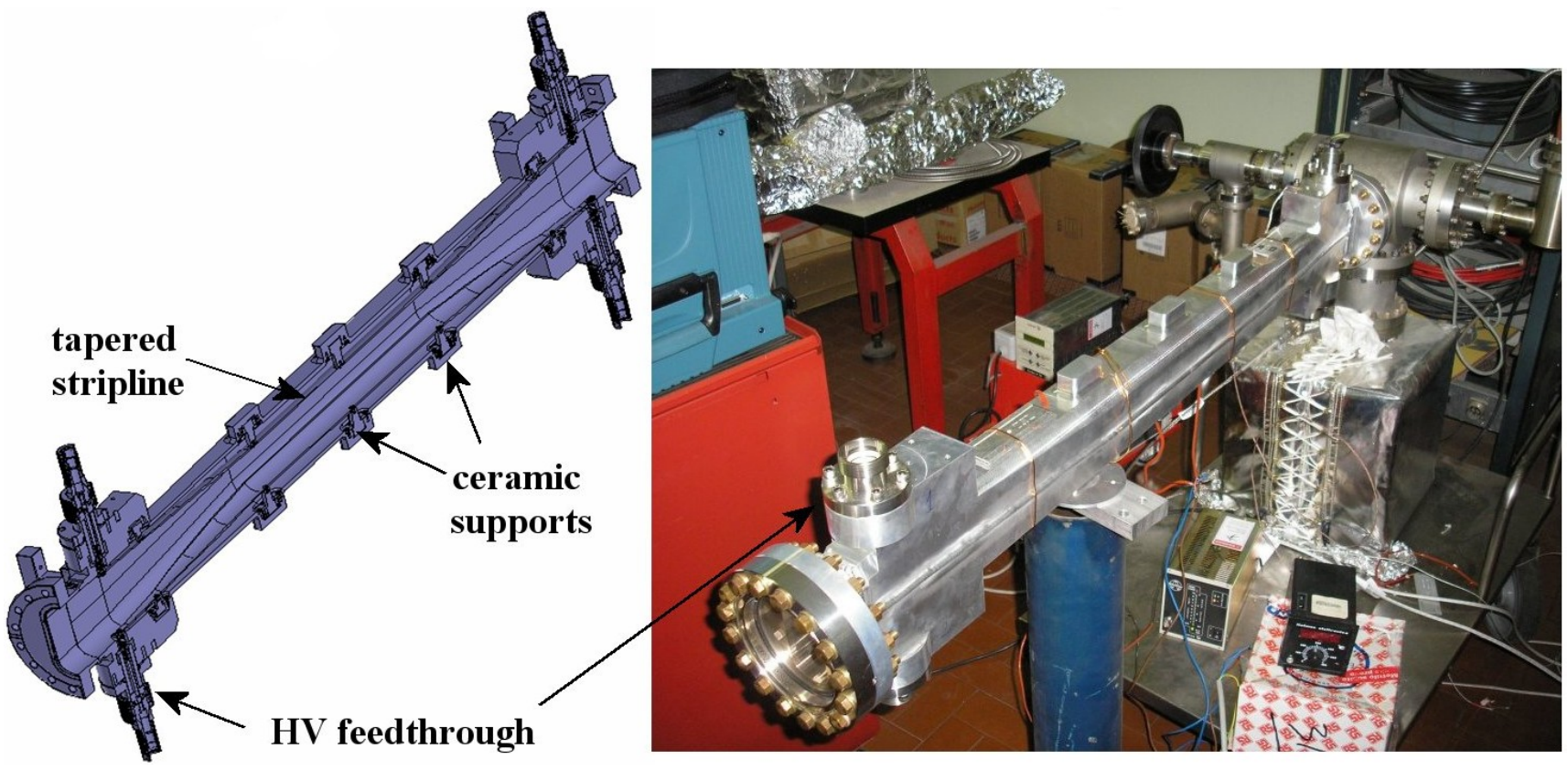

The coupling impedance measurements and simulations have pointed out the absence of trapped higher-order modes in the longitudinal and horizontal planes when at least two ports are loaded by 50 W [250]. In the vertical plane, only four trapped higher-order modes were found. The instability growth rates of these resonances (in the worst case) were well below the damping rates provided by the DAΦNE feedback systems. After installing the injection system, no instability effects due to the kickers were observed and the DAΦNE broadband impedance arising from this and other vacuum-chamber modifications made at the same time was reduced by about 50 % [251].

#### 4.4.3.3 SLAC pulser modulator

At SLAC, two related paths to meet the ILC kicker driver requirements were studied: a transmission-line adder topology, which combines the output of an array of ultra-fast MOSFET switches and a drift step-recovery diode (DSRD) approach.

For the adder topology, an ultra-fast hybrid MOSFET/driver, recently developed at SLAC, has achieved 1.2 ns switching of 33 A at 1000 V with a single power MOSFET die [252]. A transmission-line adder has been designed based on the ultra-fast hybrid MOSFET/driver switching module. The initial test demonstrated that the adder can combine pulses with 1.4 ns switching time without any degradation [253]. An improved switching module (Fig. 4.22) has been tested in 6 MHz burst operation [254].

**Figure 4.22**
Photo of improved HSM the insert shows the reverse side.

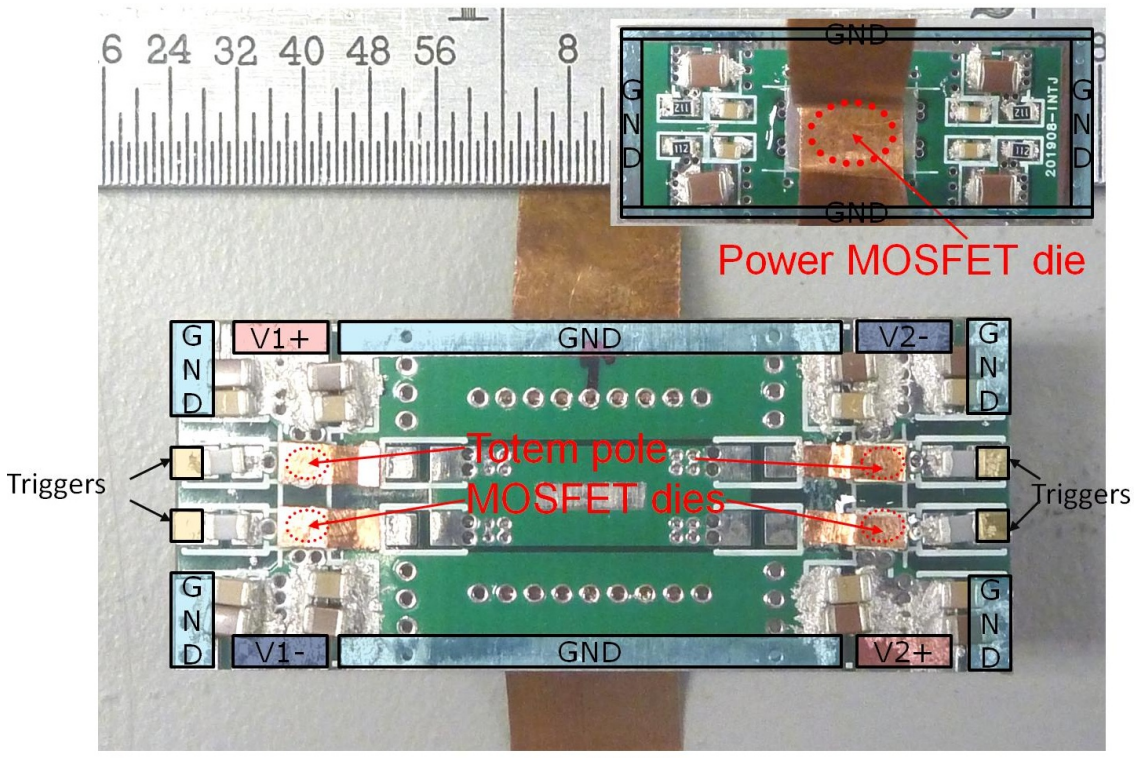

For the DSRD programme, development of a fully capable DSRD kicker driver is proceeding well, with excellent results obtained from the first commercially produced DSRDs, and from a refined circuit for the MOSFET driver [255]. A prototype with a 4 ns pulse length (as required for ATF) and 3 MHz pulse train has been demonstrated (see Fig. 4.23). A recent success was to eliminate the post pulse, which is unacceptable for the ILC kicker driver since it affects the bunches adjacent to the kicked bunch [256].

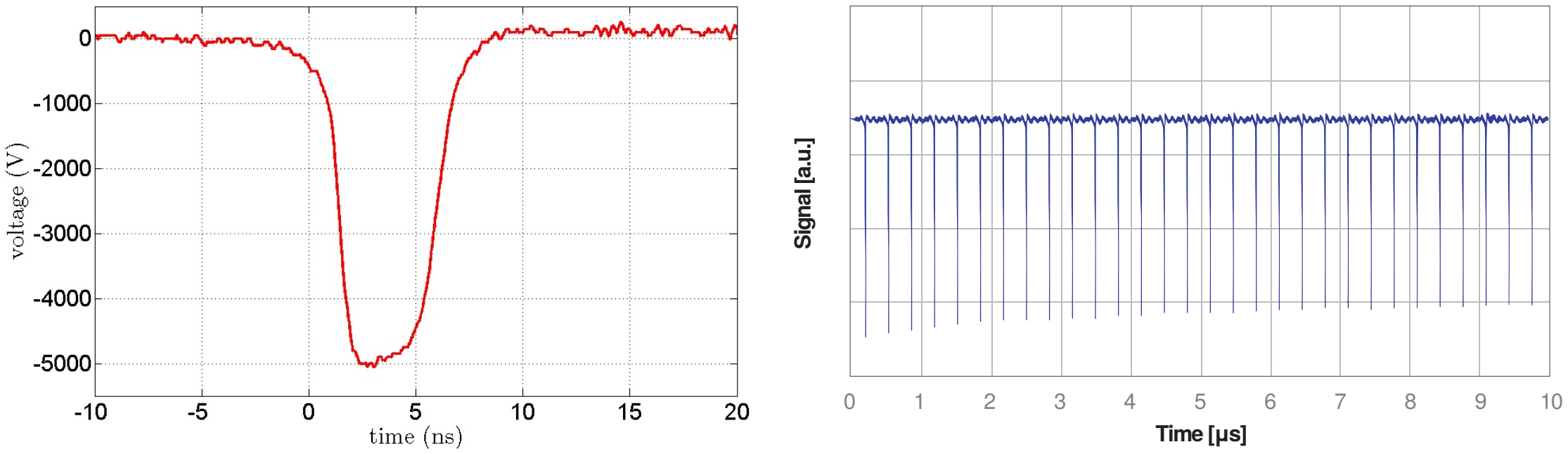

(a) (b)

**Figure 4.23.** DSRD-based driver output; single pulse detail, 3 MHz 30-pulse pulse-train [256].

Figure 4.24a shows the output voltage of a kicker pulser (FPG10-3000KN, built by FID GmbH)

and tested at ATF [210]. Figure 4.24b shows the time dependence of the estimated kick field when applied to the KEK-ATF 60 cm strip line. In the ATF experiment, two 60 cm strip lines were installed to extract single bunches from trains of ten bunches circulating in the damping ring. The bunch spacing was 5.6 ns. The measured angle jitter of the extracted beam in the KEK-ATF experiment is $3.5 \times 10^{-4}$ of the kick angle.

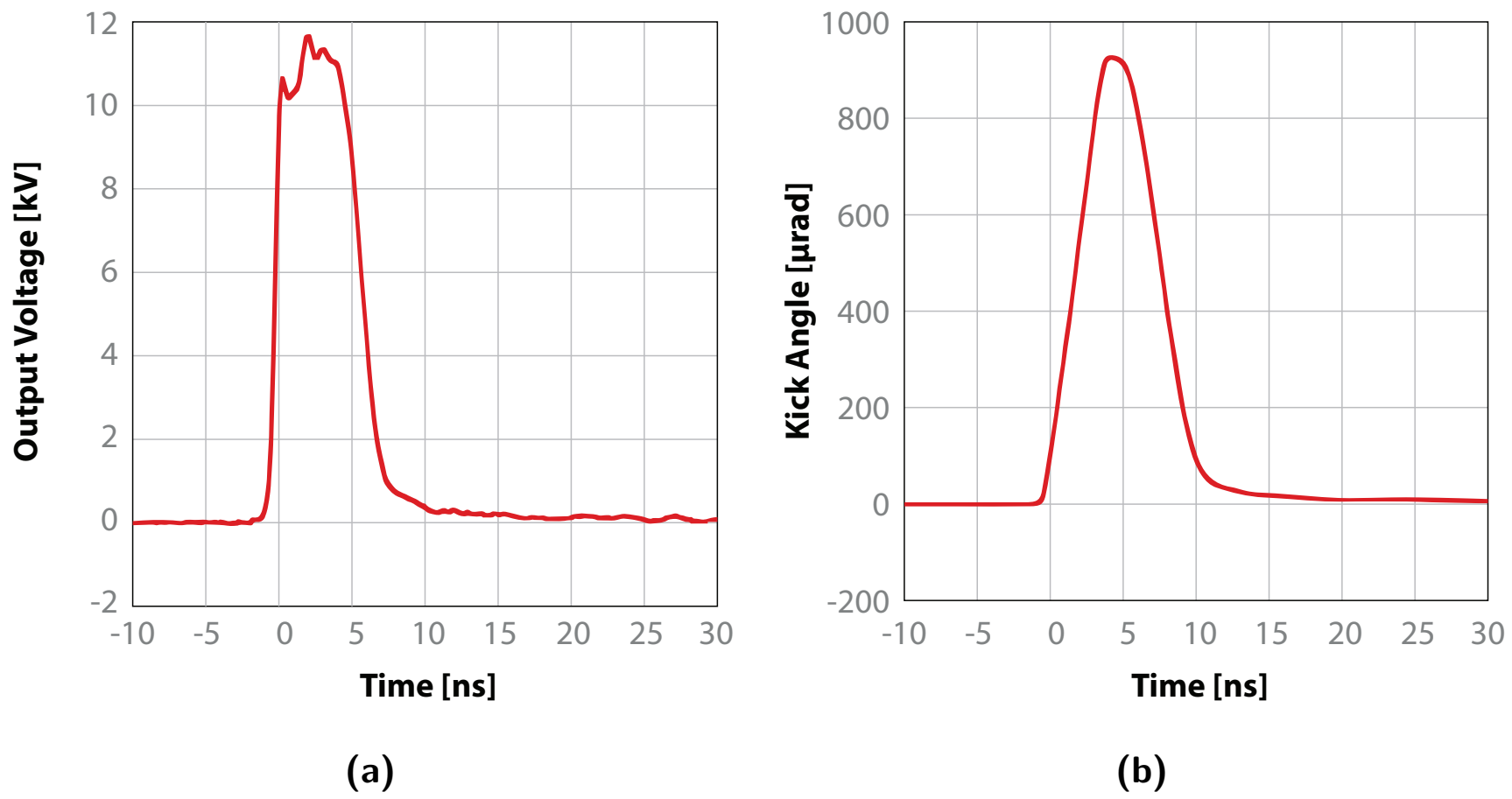

**Figure 4.24**
(a) Pulse waveform of the FID pulser; and (b) estimated kick obtained using the 60 cm stripline kicker in the KEK-ATF beamline.

## 4.5 Beam Delivery System and MDI

### 4.5.1 Introduction

The beam-delivery system and the interface between the accelerator and detectors near the IP constitute some of the foremost challenges to be faced at any linear collider. This is a direct consequence of the extremely small beam size required for a "one-pass" collider to attain a luminosity competitive with that of a storage ring in order to observe the rare physics processes of interest. In many areas, the requirements of the ILC are beyond the current state of the art. This section describes the R&D carried out in several of these areas: the design of the final-focus magnets, including both superconducting and permanent-magnet options; maintaining the stability of the beams so that luminosity during collisions in maximised using a fast intra-train feedback system based on digital processing; the design of a high-accuracy but relatively rapid "push-pull" system that can exchange two massive particle-physics detectors with a minimal loss of luminosity; and a precise and rapid alignment system to ensure that all parts of the detector are aligned both internally and with respect to the final-focus machine elements.

### 4.5.2 Cold QD0 Design

The ILC IR final-focus (FF) magnets must satisfy many conflicting requirements; before the start of QD0 Prototype R&D work it was not clear that construction of such magnets would be feasible. The ILC FF system brings the incoming $e^+e^-$ beams to nanometre-scale focus with optical demagnifications much greater than previously attempted but at the same time it must catch and refocus the disrupted outgoing beams with sufficient aperture to avoid particle loss and generation of detector backgrounds. The FF has strong quadrupoles with close side-by-side spacing, sextupole magnets to correct for higher-order chromatic effects and octupole magnets to reduce halo-particle backgrounds. The need for precision energy scans and low-energy running for detector calibration demands tuneable magnets making a compact superconducting system advantageous. The presence of strong detector-solenoid fields overlapping the magnets closest to the IP precludes magnetic yokes and requires incorporation of an active shield to reduce the unwanted external field at the extraction beam line.

The Prototype R&D programme investigated producing the FF magnets using BNL direct-wind magnet-production technology. This involves building up a multilayer coil structure with epoxy impregnation that is then fibre-compression wrapped and cured. The resulting structure handles large magnetic Lorentz forces on the coil without the need for external clamping collars. Warm field-harmonic measurements that are used to control harmonic errors are made between winding different coil layers.

Initially, short prototype quadrupole and sextupole test coils were wound and tested. These coils exceeded ILC operational requirements by reaching the conductor short-sample limit in the presence of deliberately enhanced solenoidal background fields. Subsequently an opposite-polarity outer-quadrupole coil was added outside the main quadrupole so as to demonstrate active shielding via external field cancellation. Studies showed that it is important to rotationally align the main and cancellation coils in order to achieve the best performance; with proper adjustment the external field was dramatically reduced, with only a small impact on the overall QD0 operating margin.

In parallel, prototype full-length coils suitable for FF operation were produced. A major challenge in winding the full-length coil is that the aspect ratio (20 mm ID, 3.5 m length) of the coil support tube results in droop in the middle under a combination of its own weight and pressure applied to the tube during winding. This droop causes systematic conductor mis-positioning and modulation of the coil pattern, which lead to the generation of unwanted field harmonics. However, based on earlier work, small, consistent, support-tube offset modulations can be compensated via software-based corrections during coil winding. Rolling tube supports, that move along with the coil-winding head and reduce local support-tube deflection, were hence implemented.

Despite these measurers, field harmonic variations symptomatic of significant residual tube-offset motion were observed. Unfortunately stiffening the rolling support structure increased the risk of damaging a coil already wound and, more problematically, was found to increase positioning hysteresis that in turn made software compensation even more challenging. The moving support scheme was hence abandoned in favour of a simple fixed central-tube support that was precisely aligned via a laser system with the rest of the coil-winding mechanical structure.

The above approach requires that the quadrupole coil must be split into two parts. This has the advantage that it is possible to energise the coil closer to the IP more strongly at low beam energy, while reducing the gradient in the subcoil further from the IP. This causes the effective focusing centre to be closer to the IP at low energy, and can be used to improve the FF optical performance.

**Figure 4.25**
QD0 cryostat

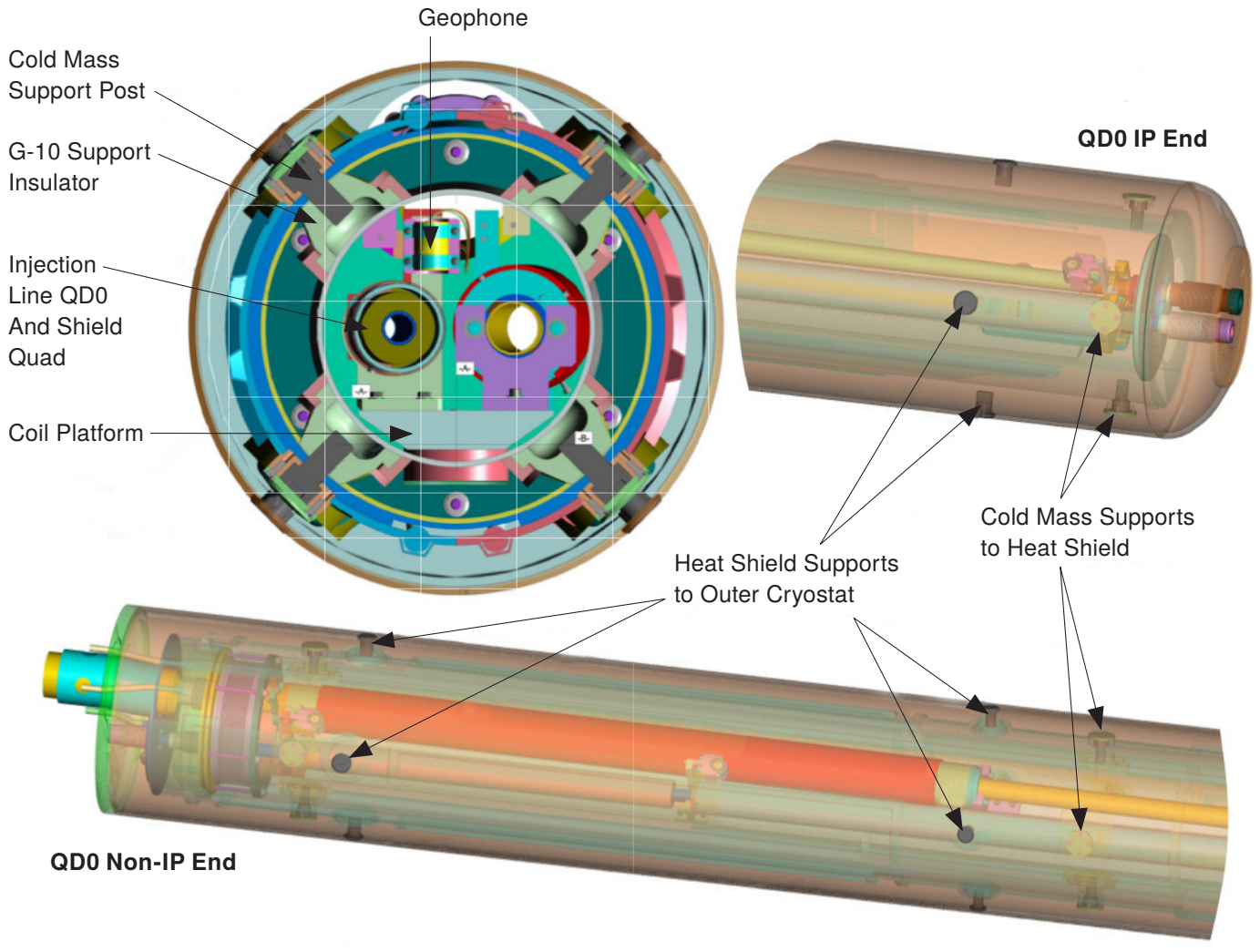

Another fundamental component of the R&D programme was to develop practical solutions for installing and operating FF magnets located inside two different particle detectors. The MDI issues are especially challenging in light of the requirement for rapid push-pull swapping of the detectors. The conclusion is that each experiment should carry its own customised pair of QD0 and cryogenic-feed cryostats. Each detector must hence interface with a pair of fixed QF1 and service cryostats that remain in place to define the IR beam-line axis. The QD0 cryostats must accommodate the incoming and extraction beam-line magnetic elements with different $L^*$ and have customised features to facilitate their installation and alignment, as well as compatibility with detector access and maintenance. A QD0 cryostat is shown in Fig. 4.25.

**Figure 4.26**
QD0 service cryostat

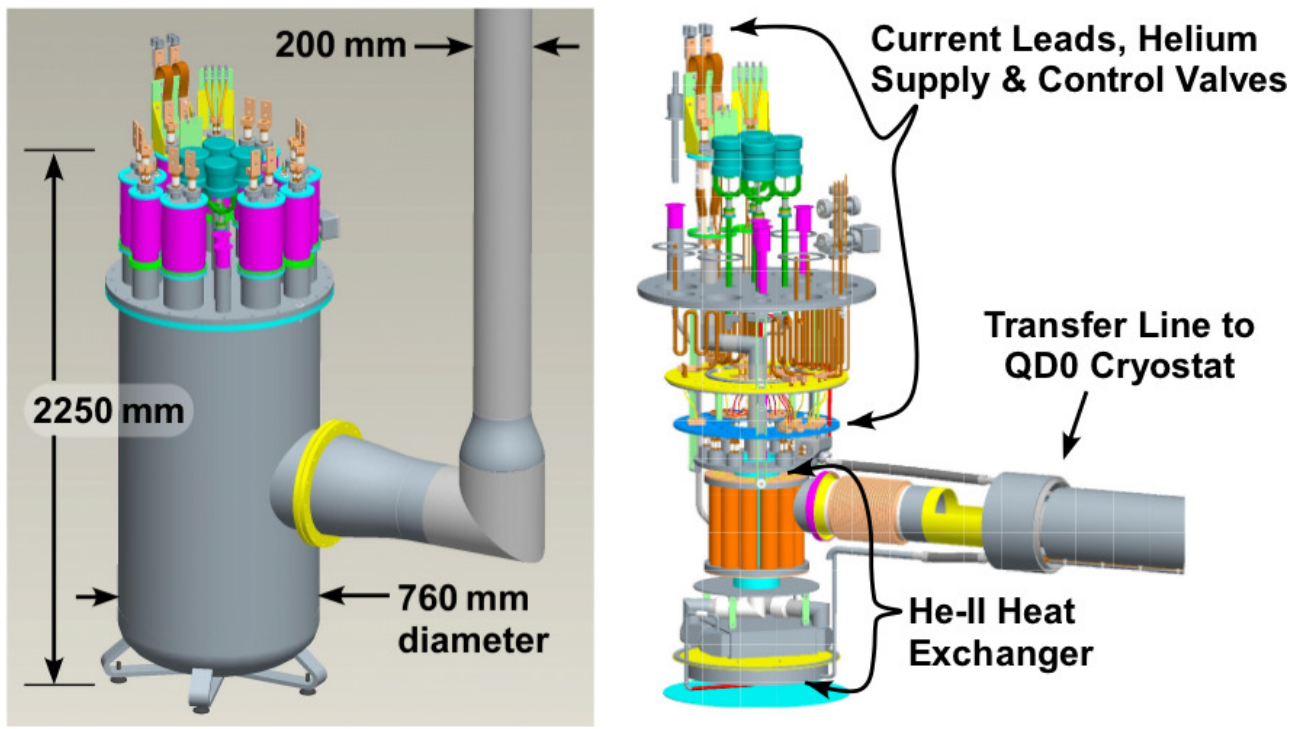

The FF magnets are designed to operate using pressurised superfluid He-II in order to avoid cryogens flowing through a long transfer supply line and the magnet cryostat structure. The interface point for the He-II supply heat exchanger and the magnet current power leads is the service cryostat shown in Fig. 4.26. Together, the service cryostat and cryogenic transfer line comprise an intricate system that should be made as stable as possible against vibration during operation. In order to explore the challenges inherent in such a system fully, the QD0 magnet cryostat is being fabricated to accept the full-length prototype coils. This system should be horizontally tested while connected via a transfer line to an ILC-style service cryostat that is also under construction. The partially assembled QD0 helium-containment vessel with its alignment supports is shown in Fig. 4.27a and the start of assembly of the service cryostat is shown in Fig. 4.27b. The QD0 prototype is designed to be compatible with vibration stability measurement via a laser doppler vibrometer system, as well as some internal accelerometers and external geophones. A conceptual design has also been developed for a stabilised pickup-coil system that could be inserted into the QD0 bore in an attempt to measure changes in the magnetic-field centre directly. Such a stabilised probe is also very useful for direct measurement of the field-centre stability of the SuperKEKB IR quadrupole magnets; its planned deployment there will gain valuable experience of direct relevance to ILC.

**Figure 4.27**
(a) Partially assembled QD0 helium-containment vessel with alignment supports; (b) the start of assembly of the service cryostat.

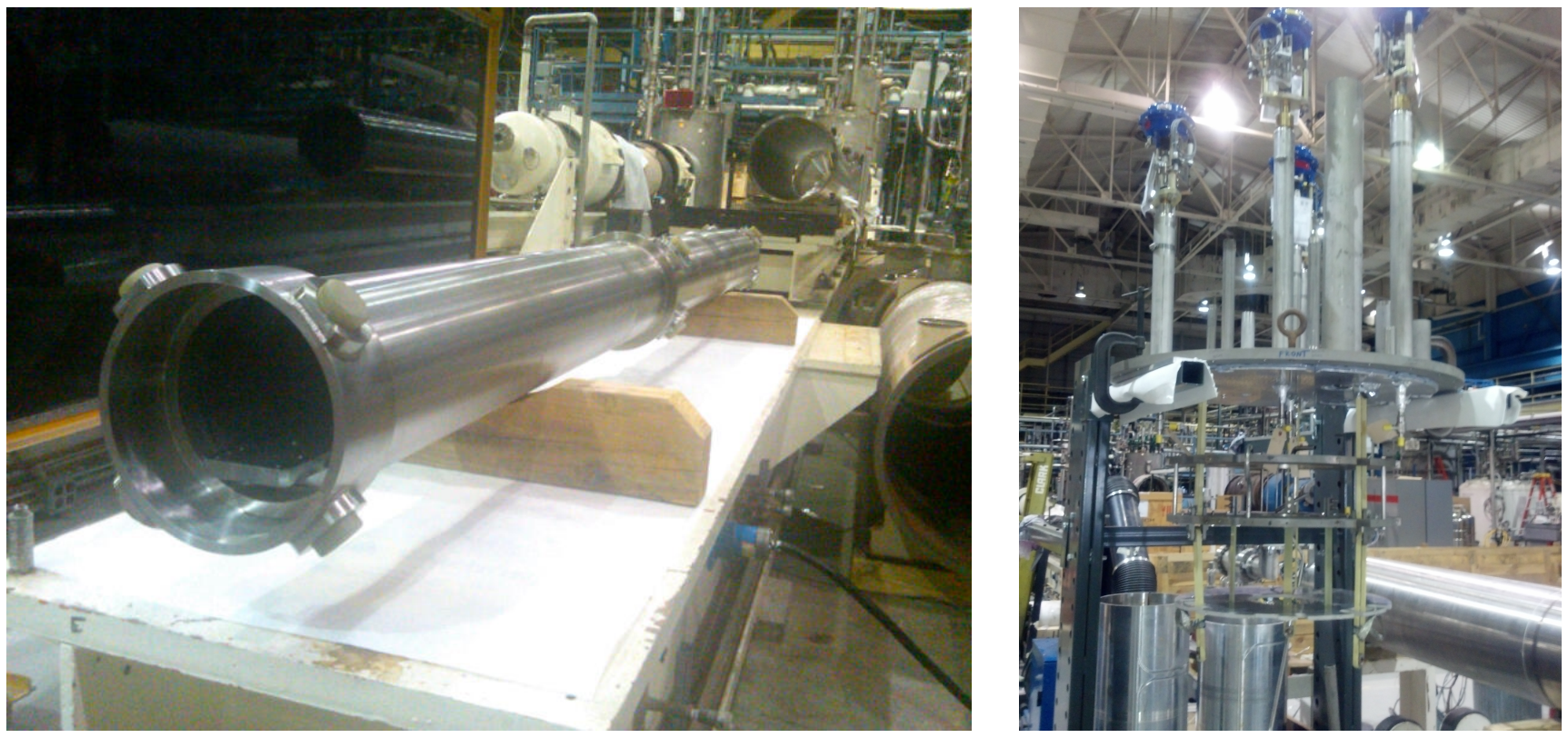

**(a)** **(b)**

## 4.5.3 Permanent QD0 design

### 4.5.3.1 Magnet Design and Test

Permanent-magnet quadrupoles (PMQ) [257–260] have been under study as an alternative technology for the final-focus magnets in a linear collider since 2002. Since the system is passive, there is no risk of introducing jitter via a cryogenic plant. However, demagnetisation is a concern, although a rough estimate shows that continuous operation for 10 years at the ILC may cause about 1 % demagnetisation on NEOMAX 32EH. Magnets made from SmCo are more durable against radiation but has lower magnetic strength.

An adjustable-strength PMQ, which is divided into five rotatable rings, is shown in Fig. 4.28. The rotation angles $\pm ø/2$ of the PMQ rings at even positions are opposite in sign compared to those at odd positions. By choosing a proper length for each magnet ring, the skew component can be eliminated up to the 5th order of rotation angle. A model magnet unit was fabricated with five magnet rings that have lengths of 20, 55, 70, 55, 20 mm, respectively. The total length is about 24 cm. Each magnet is divided into 20 pieces azimuthally and radially. The outer diameters of the magnet rings are 100 mm (without the magnet holder ring). The inner diameters are set at 55 mm for beam test at ATF2. Since the beam energy of ATF2 is about 0.5 % of that of the ILC, the gradient required is smaller. This 24 cm-length unit has nearly the same focusing strength as the currently used QD0 electromagnet in ATF2. Before the assembly of the magnet unit, the five rings have to be adjusted individually [261–263].

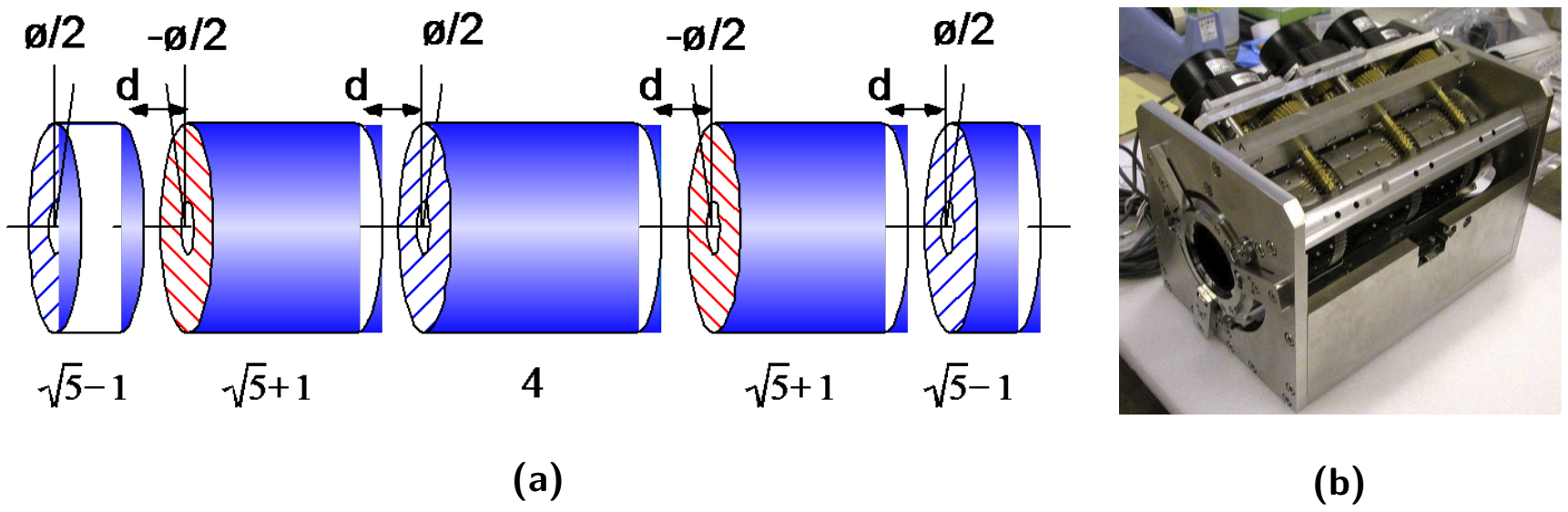

**Figure 4.28**
(a) Schematic view of the five-ring singlet; (b) the fabricated model.

Assuming a field gradient of 140 T/m and using 12 units of the five-ring-singlets, the total length becomes about 3 m. Using this design for QD0, a preliminary fine-tuning simulation was carried out with matching requirement for Twiss parameters: $\alpha_x = \alpha_y = 0$, $\beta_x = 0.021$ m, $\beta_y = 400$ µm, $\eta_x = 0$ at IP, starting with the ILC deck "ilc2006b.ilcbds1" (14 mrad version). The final rotation angle ø of the PMQ is 6.58°. Off-momentum matching was performed by re-optimising the K2 of the sextupoles by looking at the beam size at the IP [264]. The coupling between $x$ and $y$ was well suppressed and the final beam sizes at the IP are $\sigma_x/\sigma_y = 656/\ 544$ nm for $\gamma_{\mathrm{ex}}/\gamma_{\mathrm{ey}} = 9.2 \times 10^{-6}$ m/ $3.4 \times 10^{-8}$ m and $\sigma\delta = 6 \times 10^{-4}$. Further optimisation will improve these results. A rough sketch of the closest optical components is shown in Fig. 4.29. The QDEX1 and SD0 magnets can be also fabricated using permanent magnets.

**Figure 4.29**
Rough sketch of QD0 and QDEX1.

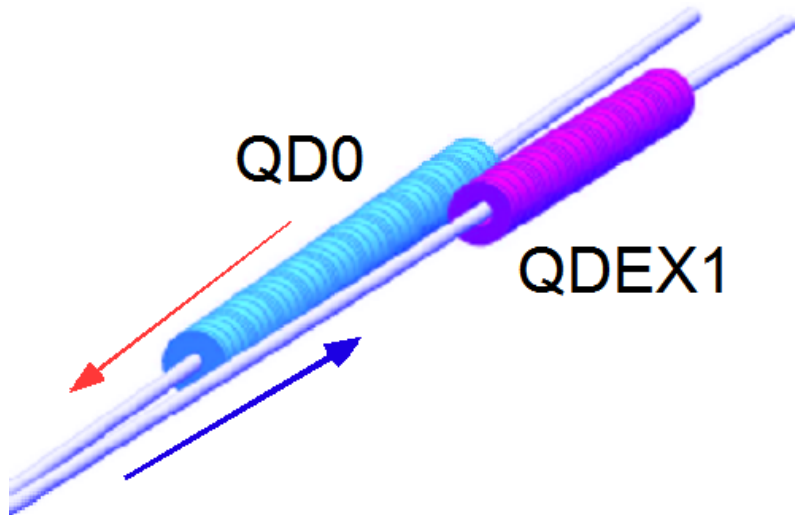

An evaluation test of the fabricated five-ring-singlet model was successfully carried out at ATF2 to obtain practical experience in handling this new device [265].

### 4.5.3.2 Anti-solenoid for Permanent QD0

Since the system has to be located close to the IP where stray magnetic field from the detector solenoid exists and the permanent magnet should not be immersed in a high magnetic field, the stray field has to be reduced. Figure 4.30 shows the partially passive hybrid anti-solenoid, where the major part of the magnetic field is cancelled by the anti-solenoid coil [266]. The residual magnetic field can be easily shielded by the iron shield pipe as long as it is not saturated. This passive scheme makes the adjustment of the anti-solenoid current less critical. The outward thrust force on the solenoid coil and inward force on the iron shield can be balanced. While the force may be cancelled out in normal operation, it should have enough mechanical strength to hold the structure against an emergency case such as a fault on the power supply of the anti-solenoid coil.

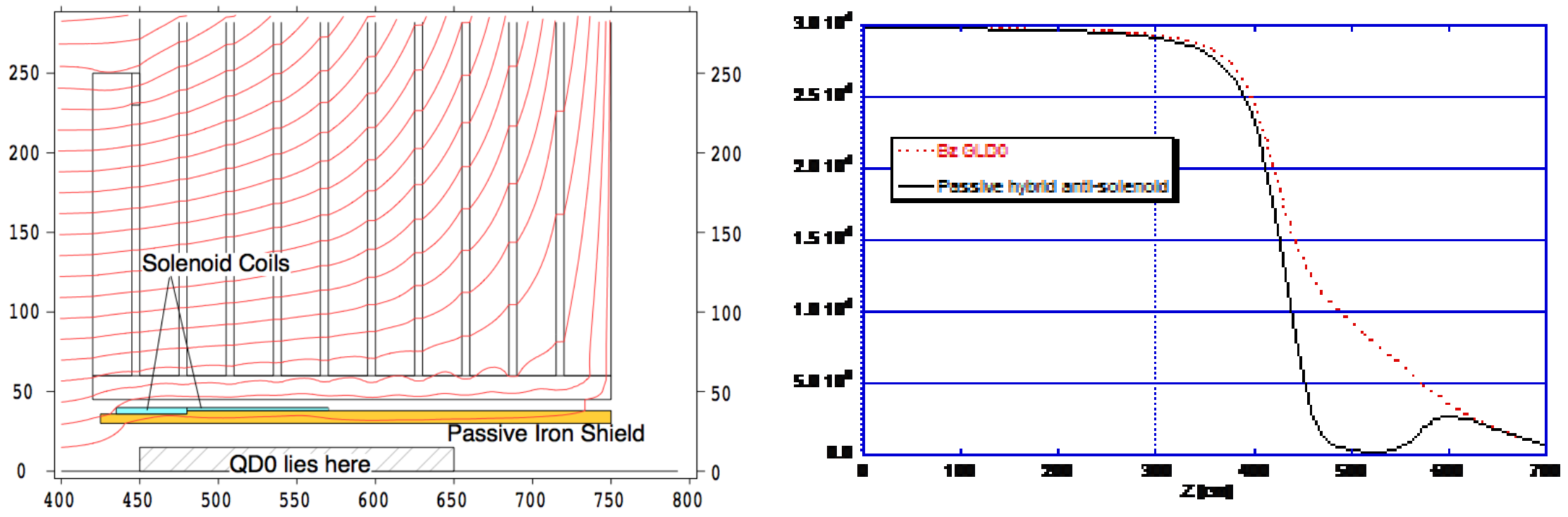

(a) (b)

**Figure 4.30.** (a) Partially passive hybrid anti-solenoid. The blue boxes denote the anti-solenoid coils wound on the passive iron magnetic-shield pipe and the QD0 lies in the hatched box area; (b) the stray field is well suppressed.

### 4.5.3.3 External Stray Field of PMQ

Assuming an $L^*$ of 4.5 m, the distance between the in-coming and out-going beam is 63 mm at $L^*$ from the IP and the bore diameter required for the outgoing beam would be 25 mm. The external stray field at the outgoing beam, which is located from $x = 50$ mm to 75 mm (see Fig. 4.31), is less than 10 G. Figure 4.32 shows the measured stray field on a horizontal line 10 cm from the beam axis. The distribution changes with the strength adjustment. An iron case as the PMQ holder will reduce the field by a factor of one hundred. The external magnetic field from the detector solenoid has to be dealt with when soft magnetic material is used; the iron case should be made of laminated iron with small enough packing factor to have less longitudinal permeability.

**Figure 4.31**
PMQ and outgoing beam.

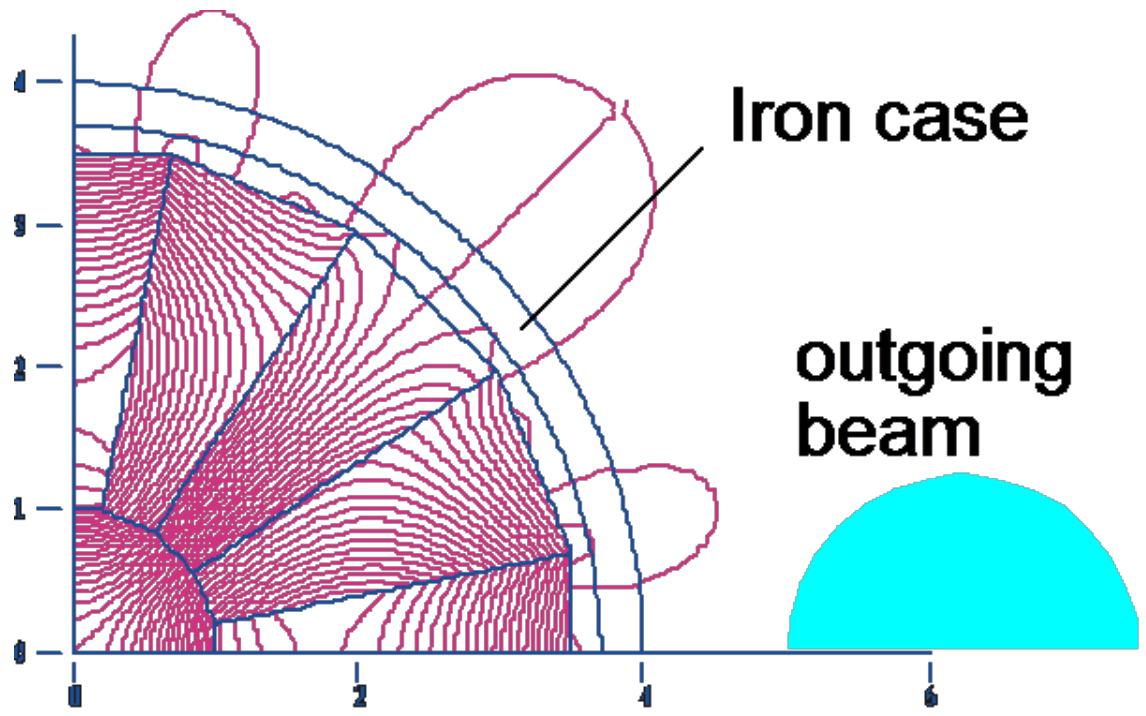

**Figure 4.32**
Measured external stray field.

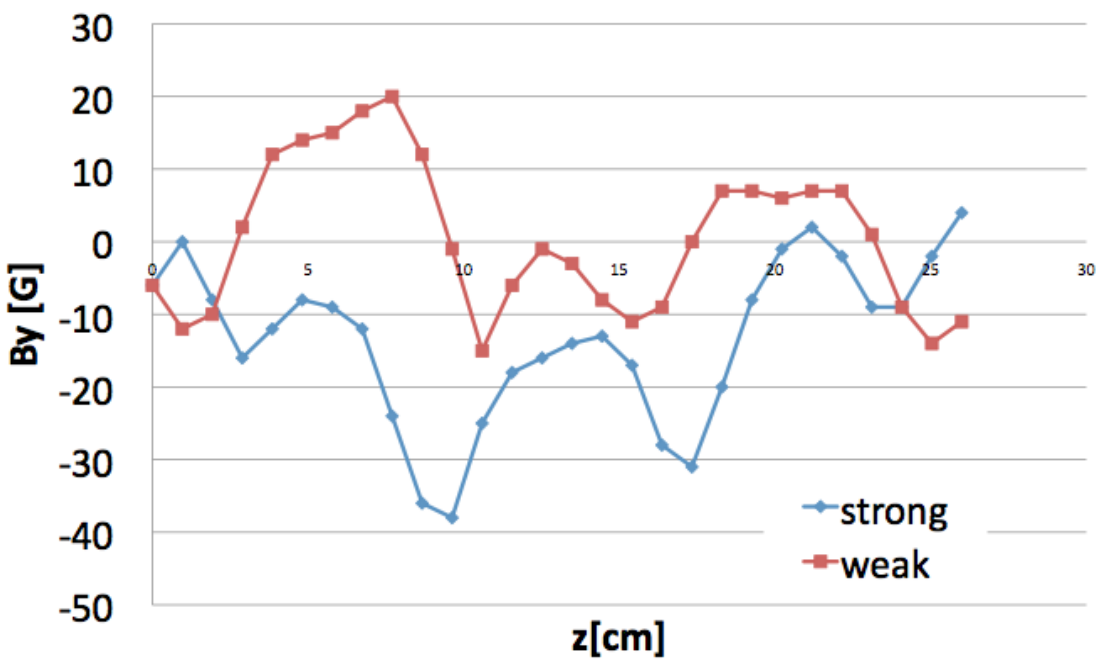

## 4.5.4 Intra-train Feedback System

A schematic of the IP intra-train beam-based feedback is shown in Fig. 4.33. Critical issues for the feedback performance include the latency of the system, as this affects the number of corrections that can be made within the duration of the bunch train, and the feedback algorithm.

**Figure 4.33**
Schematic of IP intra-train feedback system. The deflection of the outgoing beam is registered in a BPM and a correcting kick applied to the incoming other beam.

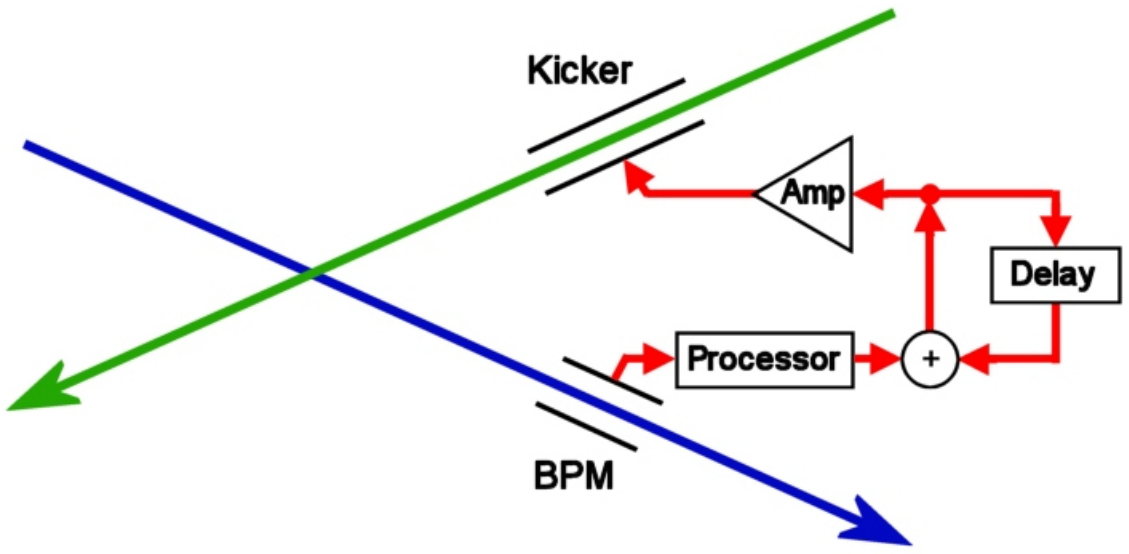

A prototype system has been designed and tested on the extraction line of the KEK Accelerator Test Facility (ATF). The FONT5 intra-train feedback system aims to stabilise the beam orbit by correcting both the position and angle jitter in the vertical plane on bunch-to-bunch time scales, providing micron-level stability at the entrance to the ATF2 final-focus system. The system comprises three stripline beam-position monitors (BPMs) and two stripline kickers, custom low-latency analogue front-end BPM processors, a custom FPGA-based digital processing board with fast ADCs, and custom kicker-drive amplifiers.

A schematic of the prototype of the FONT5 feedback system and the experimental configuration in the upgraded ATF extraction beamline, ATF2, is shown in Fig. 4.34. Two stripline BPMs (P2, P3) are used to provide vertical-beam-position inputs to the feedback. Two stripline kickers (K1, K2) are used to provide fast vertical beam corrections. A third stripline BPM (P1) is used to witness the incoming beam conditions. Upstream dipole corrector magnets (not shown) can be used to steer the beam so as to introduce a controllable vertical position offset in the BPMs. Each BPM signal is initially processed in a front-end analogue signal processor. The analogue output is then sampled, digitised and processed in the digital feedback board. Analogue output correction signals are sent to a fast amplifier that drives each kicker.

The ATF can be operated to provide an extracted train that comprises up to 3 bunches separated by an interval that is selectable in the range 140–300 ns. This provides a short ILC-like train which can be used for controlled feedback system tests. FONT5 has been designed as a bunch-by-bunch feedback with a latency goal of around 140 ns, meeting the ILC specification of c. 150 ns bunch spacing. This allows measurement of the first bunch position and correction of both the second and third ATF bunches.

The front-end BPM signal processor [267] utilises the top and bottom ($y$) stripline BPM signals, added with a resistive coupler and subtracted using a hybrid, to form a sum and difference signal respectively. The resulting signals were band-pass filtered and down-mixed with a 714 MHz local oscillator signal which was phase-locked to the beam. The resulting baseband signals are low-pass filtered. The hybrid, filters and mixer were selected to have latencies of the order of a few nanoseconds to yield a total processor latency of 10 ns [268, 269].

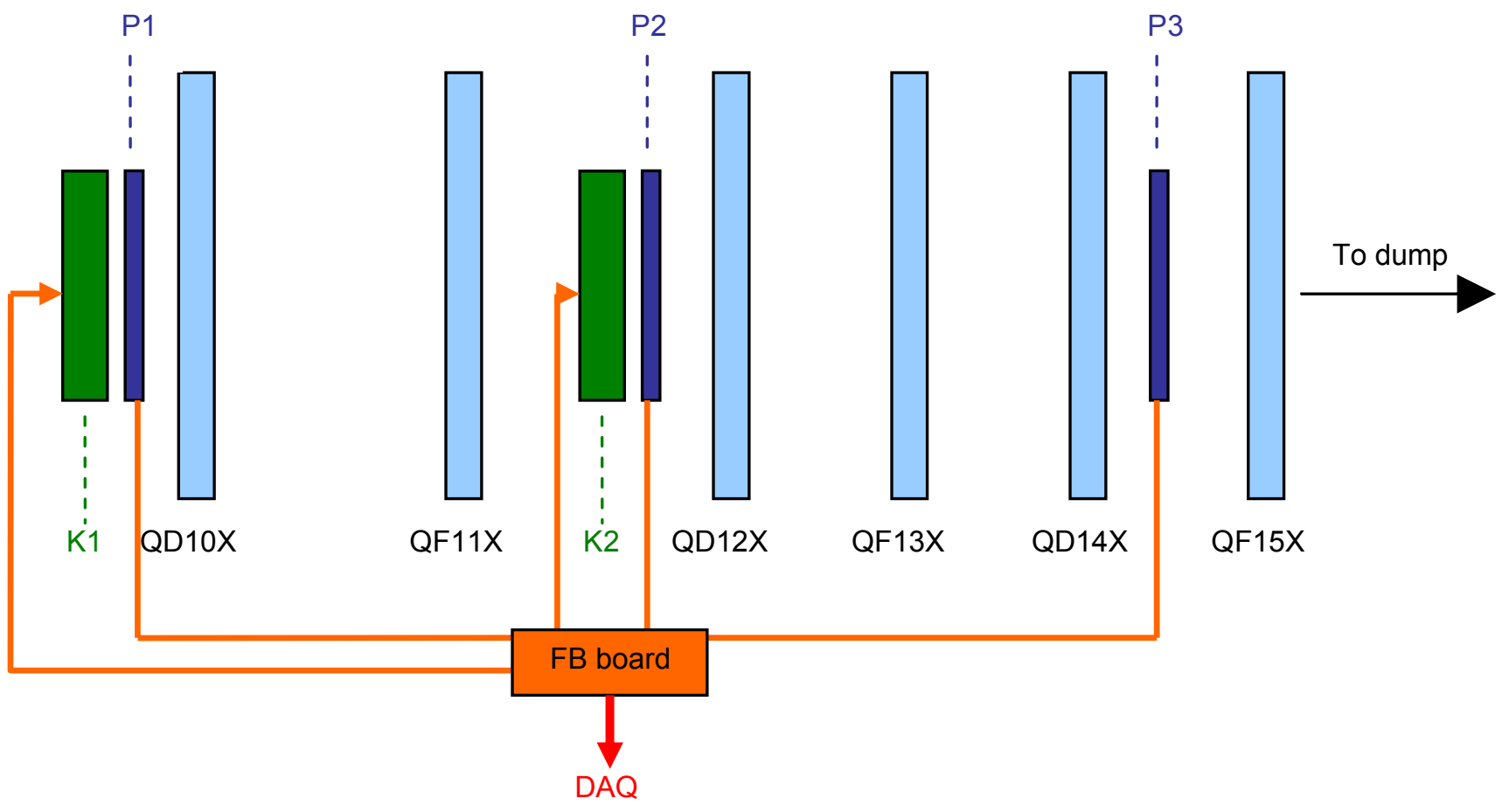

**Figure 4.34**
Schematic of FONT5 at the ATF2 extraction beam line showing the relative locations of the kickers, BPMs and the elements of the feedback system.

The custom digital feedback processor board is shown in Fig. 4.35. There are 9 analogue input channels in which digitisation is performed using ADCs with a maximum conversion rate of 400 MS/s and 2 analogue output channels formed using DACs, which can be clocked at up to 210 MHz. The digital signal processing is based on a Xilinx Virtex5 FPGA [270]. The FPGA is clocked with a 357 MHz source derived from the ATF master oscillator and hence locked to the beam. The ADCs are also clocked at 357 MHz. The analogue BPM processor output signals are sampled on peak to provide the input signals to the feedback. The gain stage is implemented via a lookup table stored in FPGA RAM, alongside the reciprocal of the sum signal for beam charge normalisation. The delay loop is implemented as an accumulator in the FPGA. The output is converted back to analogue and used as input to the driver amplifier. A pre-beam trigger signal is used to enable the amplifier drive output from the digital board.

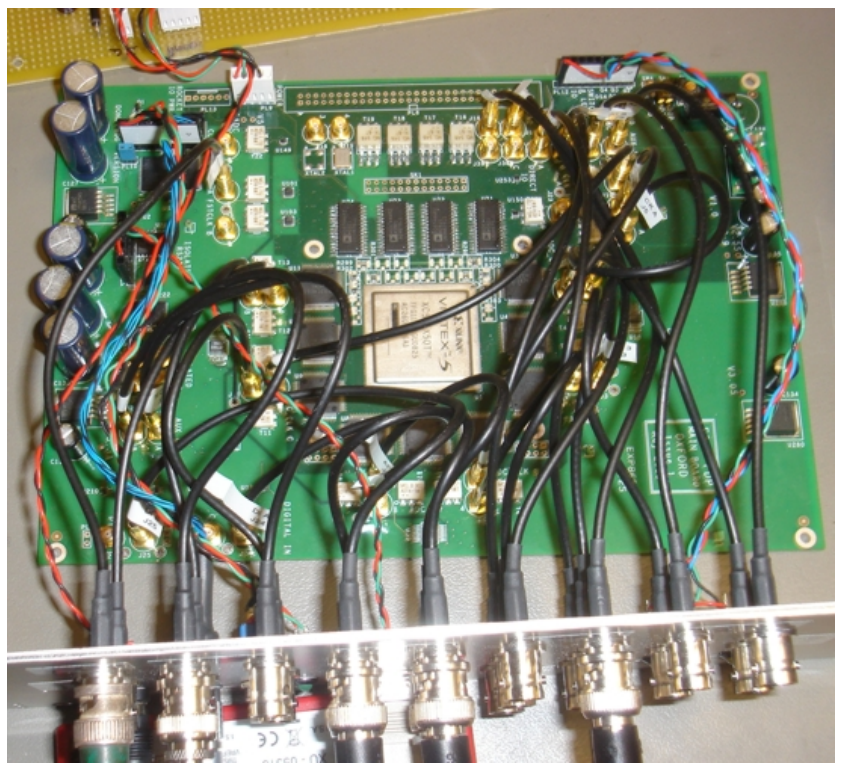

**Figure 4.35**
FONT5 digital feedback board.

The driver amplifier was manufactured by TMD Technologies [271] and provides ± 30 A of drive current into the kicker. The risetime is 35 ns from the time of the input signal to reach 90 % of peak output. The output-pulse length was specified to be up to 10 μs. The latencies were measured to be 133 ns (P2-K1) and 130 ns (P3-K2).

An example of the feedback performance is given in Fig. 4.36a and Fig. 4.36b, which show the RMS vertical beam position (the 'jitter') of bunch 2 measured at P2 and P3, respectively. With the feedback off, the incoming jitter was measured to be 3.42 μm at P2 and 3.21 μm at P3. With the feedback on, the measured jitter was 0.64 μm and 1.04 μm, respectively, representing correction

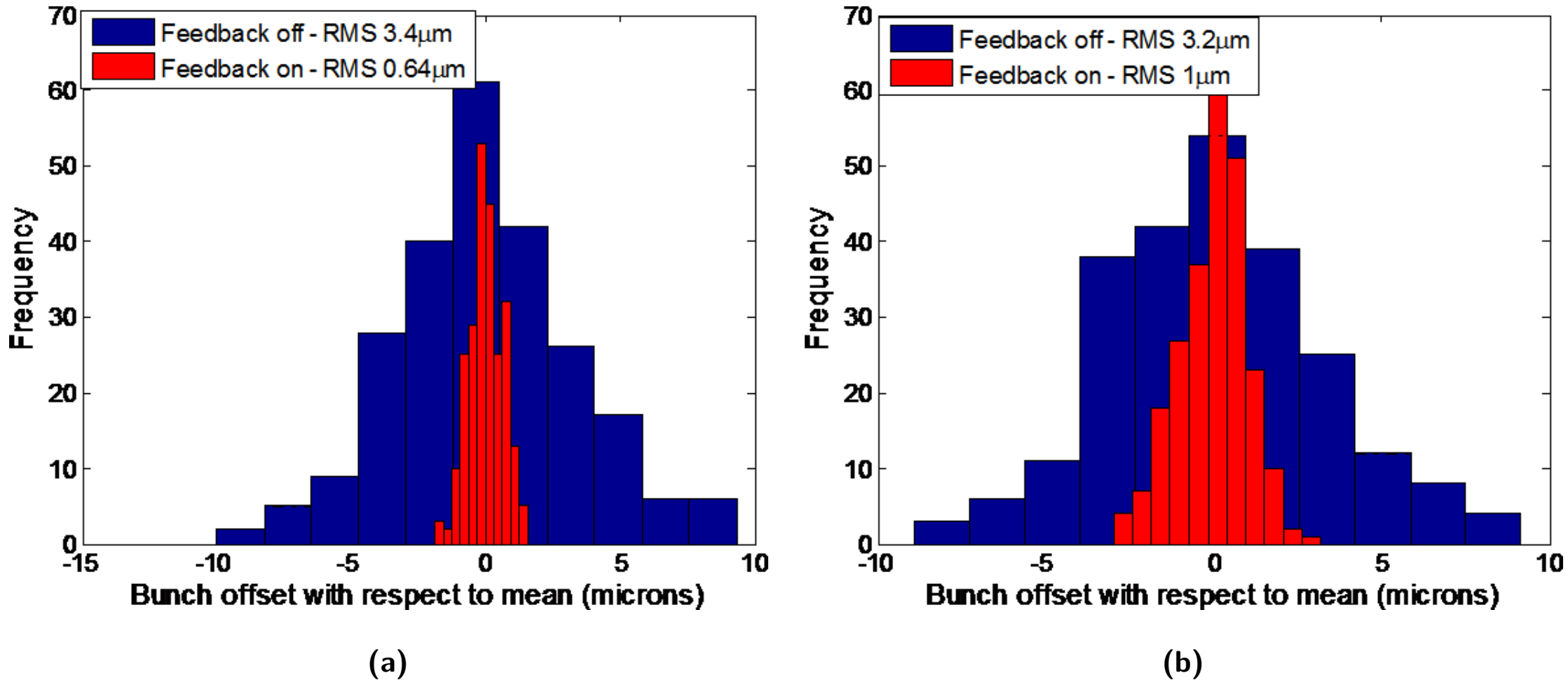

**Figure 4.36.** (a) Distribution of vertical beam position for bunch 2 at P2, without (blue) and with (red) feedback. (b) Distribution of vertical beam position for bunch 2 at P3, without (blue) and with (red) feedback.

factors of approximately 5 and 3 respectively.

The expected performance of the feedback can be calculated from the measured incoming jitter and knowledge of the bunch 1 – bunch 2 correlations. These correlations were measured to be 98 % and 97 % at P2 and P3, respectively. Using:

$$\sigma_1^{'2} = \sigma_1^2 + \sigma_2^2 - 2\sigma_1\sigma_2\rho_{12} \geq 2\sigma_r^2$$

it follows that one expects corrected beam jitters of 0.64 μm and 0.83 μm for bunch 2 at P2 and P3 respectively, in very good agreement with the measured values. The coupled-loop feedback is operating at close to the optimal performance given the degree of correlation between the two bunches.

This measured performance of the system was input into a beam-transport simulation [272] of the ATF2 beam line and the expected vertical beam position downstream of the FONT5 system was evaluated and compared with measurements. In the absence of additional jitter sources and lattice imperfections the performance is equivalent to stabilising the beam at the ATF2 IP to below the 10 nm level [272].

### 4.5.5 Push-pull system

The most efficient architectural scheme for the push-pull operation is to have both detectors installed on individual reinforced-concrete platforms (see Fig. 4.37). A detailed report describes the outcome of a platform design study, which has been carried out together with an external contractor [273] based on the following assumptions:

- Maximum Detector Weight: 15 000 tons
- Movement duration: 5 hours
- Speed : $>$ 1 mm/s
- Number of movements: 10/year
- Limit of acceleration: $0.1g$
- Maintenance allowances On Beam: 2 m
- Maintenance allowances Off Beam: 6 m
- Positioning relative to beam: $\pm$ 1 mm

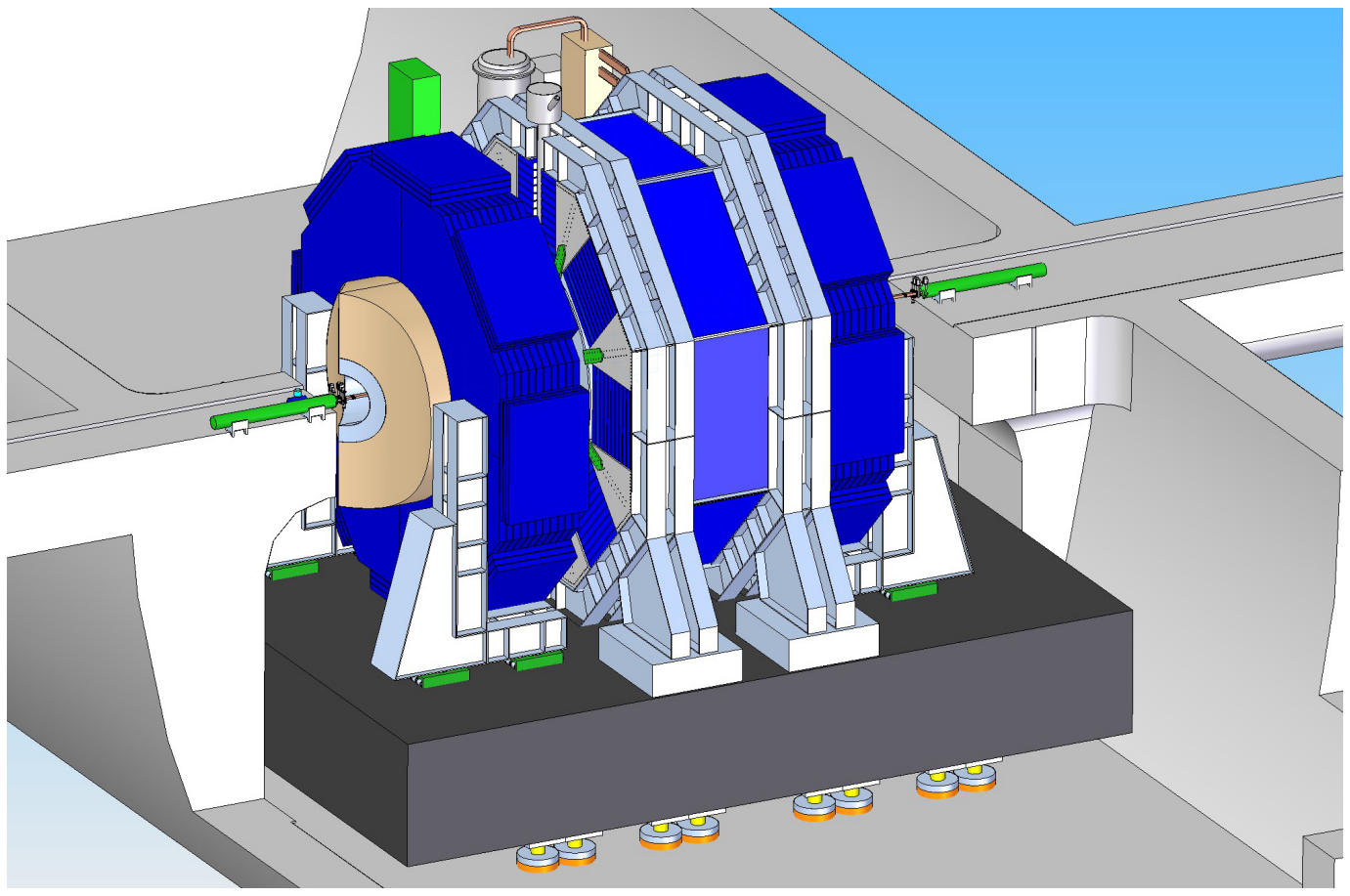

**Figure 4.37**
Illustration of detector on motion platform.

There are three essential results. First, a geological analysis indicates that special attention must be given to the design of the cavern invert, i.e. the floor where the platform motion system acts, to limit the deformations. Secondly, a reinforced-concrete platform with a maximum deformation $< 1$ mm under the detector-support points must have an approximate thickness of 2.5 m with a weight of $\sim$ 3000 t. The contact point between the platform and the floor must be carefully located under the respective contact points of the detector with the platform. Finally, two motion systems were considered, air-pads and rollers. Both can be engineered to meet the requirements with a guiding system on the floor. Fifty air pads with 370 tons capacity (maximum capacity available on the market) will be required. Only eighteen rollers of 1 ktons will be required. Air pads will need a lighter moving system because the friction factor is only 1 % while for the roller it is estimated to 3-5 %. In both case a propulsion system based on gripper jacks is possible

A detailed analysis of the propagation of the ground and technical noise has been carried out [274, 275] to check if the push-pull scheme adopted is compatible with the stability requirements. The basic results show that, with a combined transfer function of the detector plus the platform, the ground vibration propagation at QD0 is $< 50\,\mathrm{nm}$, well under the maximum budget allowed. The relative displacements of the QD0, for different ground models, is also very low (see Fig. 4.38) [275].

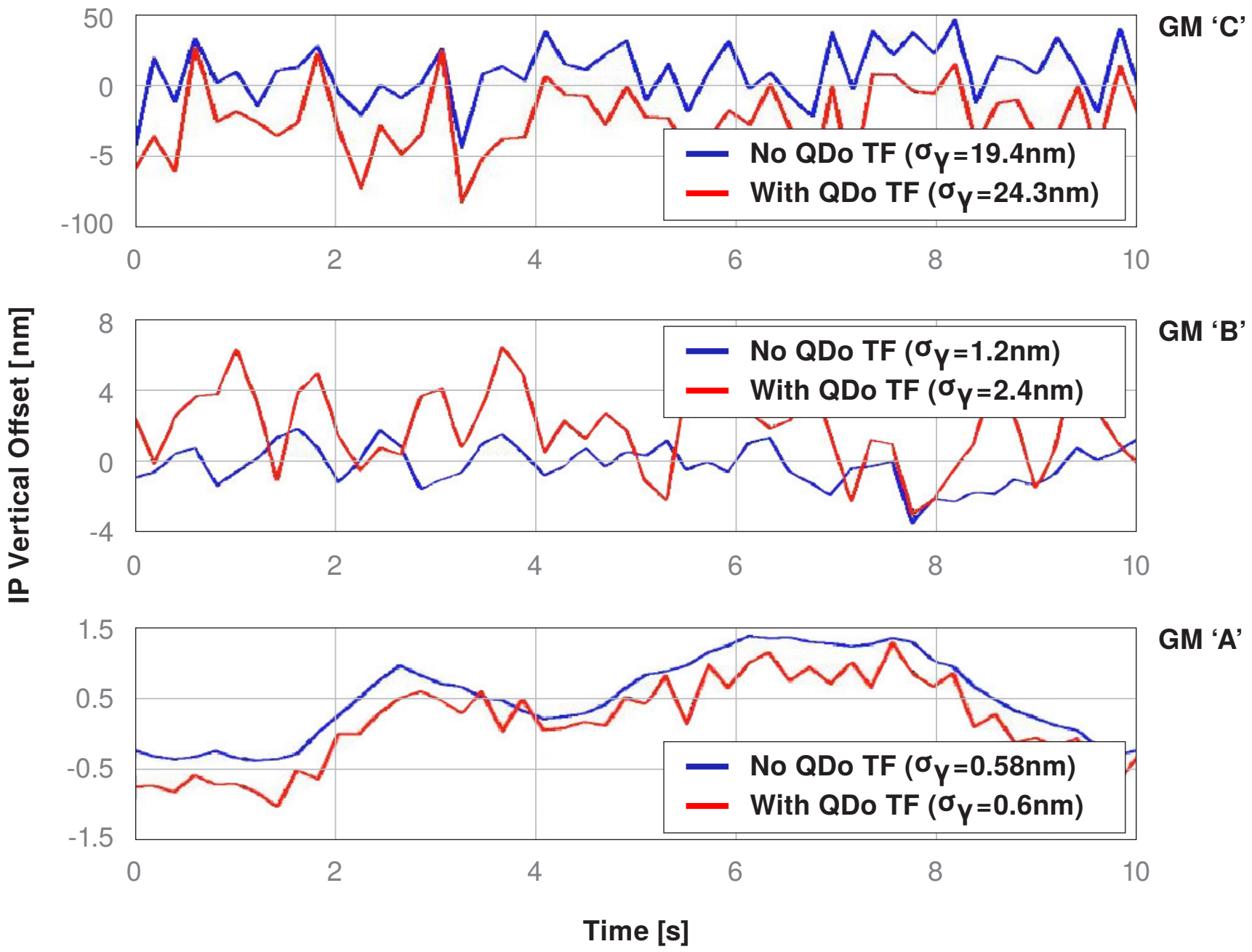

**Figure 4.38**
Vertical offset of the electron and positron beams at the IP for ground-motion models A, B and C, with and without the QD0 transfer function included.

A more refined model developed by Collette *et al.* (see Fig. 4.39) [276] shows that the technical noise developed by the detector is not harmful for the stability of a QD0 supported from the endcap.

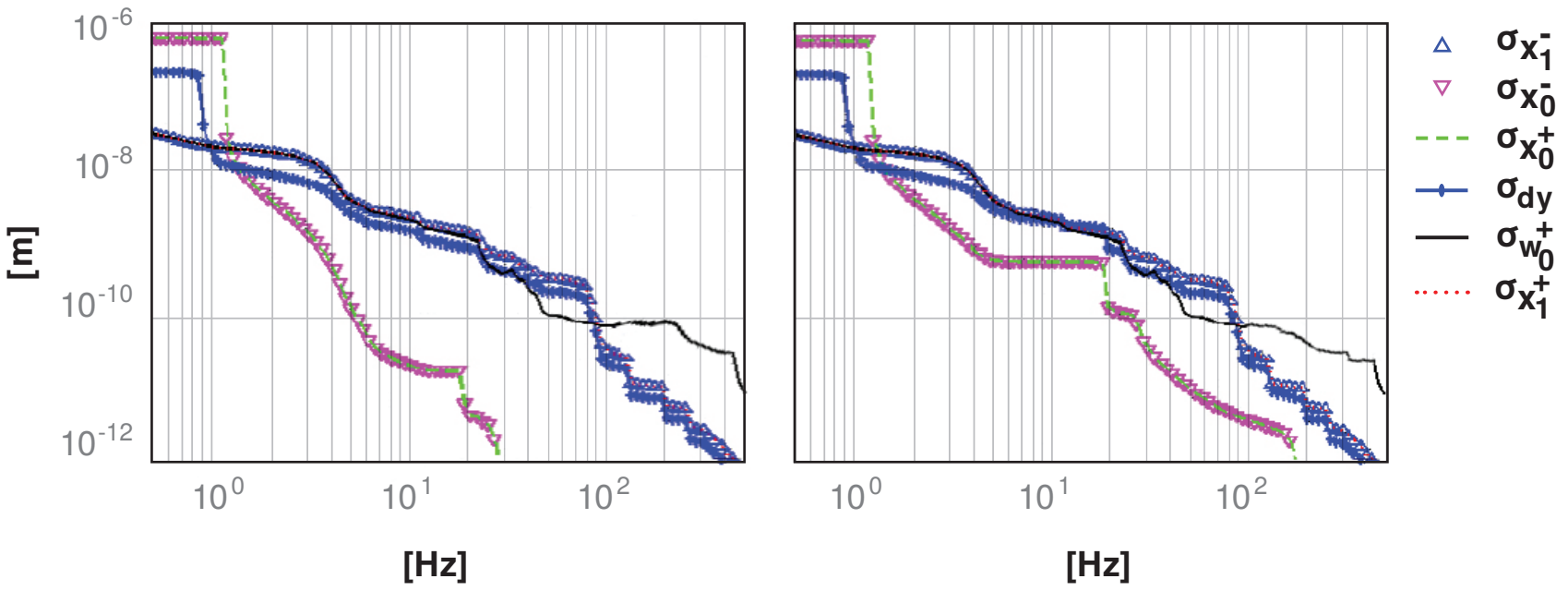

**Figure 4.39** Induced vibrations (r.m.s.) vs. frequency on QF1(x0) and QD0 (x1) with (right) and without (left) technical noise [276].

## 4.5.6 Alignment systems

The IR detector push-pull configuration places severe demands on the alignment system both for the final-focus magnets and for the tracking subdetectors of the SiD and ILD detectors. Rapid realignment of the quadrupoles will be essential to minimise the downtime for colliding-beam running. In addition, the unprecedented momentum and impact-parameter resolutions envisioned for the charged-particle trackers also argue for prompt and precise *a priori* alignment, without the need to rely on the accumulation of millions of charged particles in the detectors. Hence it is desirable to establish an integrated alignment system for the final-focus magnets and the tracking subdetectors that can: 1) measure and correct magnet misalignments that interfere with establishing colliding beams; and 2) measure any global tracker-distortion parameters that degrade tracking performance (e.g. translation, rotation, twist, sag).

An attractive solution is offered by Frequency Scanned Interferometry (FSI), an alignment method pioneered and implemented by the Oxford group for the ATLAS Experiment [277, 278]. In FSI a geodetic grid of points attached to the object to be aligned and to nearby bedrock is used to measure positions, rotations and distortions of detector and accelerator elements. The points in the grid are optically linked via a network of laser beams, using optical fibres for beam launch and collection. Small beam splitters and retroreflectors define interferometers for which a scanning of laser frequency over a known range defines absolute distance measurements to sub-micron precision. The beam lines (between beam splitters and retroreflectors) can be thought of as optical "trusses," where degrees of freedom that would be made most rigid by a corresponding mechanical truss network are those most constrained by the FSI measurements. Conversely, degrees of freedom with the greatest remaining "play" are least constrained by measurement.

Focusing here on the needs for quadrupole alignment, the preliminary design requirements for *a priori* alignment for QD0 before attempting to establish beam collisions are 50 µm in $x$ and $y$, 20 mrad in roll, and 20 µrad in pitch and yaw [279]. Simulations [280] indicate that these tolerances can easily be met, in principle, by a small network of 4 beam launchers mounted on the front face of the QF1 cryostat, each of which sends two beams to two similarly situated retroreflectors on the back end of the QD0 cryostat. Much better performance can be achieved, however, by placing (in addition) beam launchers on inner sections of the detector with retroreflectors on the QD0 front ends, provided a complete grid network that spans the IP is installed. More simulation work is needed, incorporating detailed detector line-of-sight constraints, to establish a design for the spanning task.

On the instrumental side, there has been recent R&D specific to an FSI system for SiD quadrupole and silicon-tracker alignment [281, 282]. This work, carried out on a bench at the University of Michigan, has used a multi-channel, dual-laser system to ensure robustness against systematic errors due to environmental disturbances, such as from unstable temperatures and turbulent air. Two lasers are injected and scanned over the same wavelength range but in opposite directions and with optical choppers creating alternating beam injections (a technique pioneered by the Oxford ATLAS group).

A number of significant technical complications had to be overcome in implementing the dual-laser system. The most important were: the reduction by half of the light seen by the interferometer photodiode from each chopped beam; the reduction of useful Fabry-Perot transmission peaks; and the difficulty in handling "edge effects" at chopped-beam transitions. All of these lead to increased statistical uncertainties in FSI distance determinations, despite the decrease in systematic uncertainties. With refinement of the beam-chopping method and with improved analysis software, however, these hurdles were overcome. As a result, precisions better than 200 nm have been achieved for distances up to 40 cm under highly unfavourable conditions, using the dual-laser scanning technique. These early single-channel studies used relatively large, commercial retroreflectors and commercial clamps for the beam launchers. More recent work has focused on expanding to a multi-channel system, in addition to developing small, customised retroreflectors and beam launchers.

A six-channel system has now been commissioned on a bench, allowing for successful sub-micron reconstruction of 3-dimensional retroreflector translations of O(mm), with independent sub-micron confirmation of translations in 2 dimensions and O(micron) confirmation in the 3rd dimension, based on a precision, multi-axis translation stage. At the moment, the dynamic range for successful reconstruction is limited to $\pm 3$ mm, with an eventual goal of $\pm 1$ cm.

Although material burden of FSI components is not a critical consideration in alignment of quadrupole magnets, that burden is important for the use of FSI in the inner tracking system, where multiple scattering and photon conversions degrade detector performance. Hence there have also been recent studies using much smaller (but still commercial) retroreflectors which perform nearly as well as the larger units used previously.

In addition, customised plastic cartridges have been fabricated that allow easier mounting of launch/return fibres and beam splitters, with some flexibility in defining the relative displacement and tilt of the beam splitter with respect to the launch and return fibres. These cartridges have given comparable performance to the previous clamp system.

Significant R&D remains to be done in the context of an integrated alignment system for the quadrupole magnets and tracker. Specific tasks include:

- full miniaturisation of FSI components, using lightweight materials;
- establish a detailed conceptual design for an integrated grid of FSI beams connecting the magnets and tracker to bedrock on both sides of the interaction region;
- move from current visible to infrared laser wavelengths, from which significant cost reductions may be gained by using mass-produced modulated infrared lasers and beam components, because of the prevalence of infrared devices, including scannable lasers, in the telecommunications industry;
- build a large-scale mock-up of an inner-tracker barrel layer attached to mock quadrupole magnets, to test "bootstraping" the alignment to the required precision over tens of metres, using the ends of the IR as anchors, without interfering with the hermeticity or performance of the detector.

## 4.6 Beam dynamics (simulations)

### 4.6.1 Overview

The luminosity performance of the ILC will be affected by many issues ranging from space-charge effects at the electron gun to instabilities in the damping rings to timing errors at the IP. This section addresses issues associated with the emittance preservation from the damping-ring extraction to the IP which is referred to as the Low Emittance Transport (LET). Other accelerator physics issues are addressed in the respective subsystem descriptions.

Static and dynamic imperfections in the LET impact the luminosity performance: examples are the survey errors of beam-line components and ground motion. Preserving the ultra-small emittances requires component-alignment tolerances far beyond that which can be achieved by traditional mechanical and optical alignment techniques, hence the use of beam-based alignment and tuning techniques are essential in obtaining the design luminosity. The corresponding sensitivity to ground motion and vibration mandates the use of continuous trajectory-correction feedback systems in maintaining that luminosity. The necessary procedures and specifications of the required hardware and assessment of the potential luminosity degradations are described here.

Estimation of the luminosity performance, both the static (peak luminosity) and dynamic (integrated luminosity) behaviour of the machine in realistic conditions relies on complex simulations. The performance of the ILC has been simulated for a variety of errors and procedures. Design performance was achieved in essentially all of these studies. There are no major obstacles that would prevent the ILC from reaching design performance.

### 4.6.2 Sources of Luminosity Degradation

The performance of the real machine is degraded by errors in both component alignment and field quality. For example, misaligned magnets result in beam-trajectory errors which cause emittance growth via chromatic effects (dispersion) or impedance effects (wakefields). The primary sources of emittance degradation considered are:

- dispersion - the anomalous kicks from misaligned quadrupoles, coupled with the non-zero energy spread of the beam, cause dispersive emittance growth;
- cavity tilts - the transverse component of the accelerating field causes a transverse kick on the beam, which, coupled with the non-zero energy spread of the beam, causes dispersive emittance growth;
- $x - y$ coupling - rotated quadrupoles and vertically misaligned sextupoles (for example) couple some fraction of the large horizontal emittance into the small vertical emittance leading to beam-emittance growth;.
- single-bunch wakefields - an off-axis bunch in a cavity or beam pipe generates a dipole wakefield, causing a transverse deflection of the tail of the bunch with respect to the head; the wakefields are relatively weak for the SCRF accelerating cavities, and the cavity-alignment tolerances correspondingly loose;
- multi-bunch wakefields (higher-order modes) - leading bunches kick trailing bunches, which can lead to individual bunches in a train being on different trajectories.

## 4.6.3 Impact of Static Imperfections

### 4.6.3.1 Beam-Based Alignment and Tuning

The beam emittance at damping ring extraction is $\gamma\epsilon_{\mathrm{x}} = 8\,\mu\mathrm{m}$ and $\gamma\epsilon_{\mathrm{y}} = 20\,\mathrm{nm}$. In a perfect machine, the emittance would be essentially the same at the interaction point. To allow for imperfections, the ILC parameters specify a target emittance at the IP of $\gamma\epsilon_{\mathrm{x}} = 10\,\mu\mathrm{m}$ and $\gamma\epsilon_{\mathrm{y}} = 35\,\mathrm{nm}$. Depending on the actual misalignments, the machine performance can differ significantly. The goal for the alignment and tuning procedures is to ensure that the emittance growth is within the budget with a likelihood of at least 90 %.

Similar beam-based alignment and tuning procedures are applied in the different subsystems of the LET. First, the elements are aligned in the tunnel with high precision. When the beam is established, the corrector dipoles are used to zero the readings in the Beam Position Monitors (BPMs) (so-called one-to-one steering). Even with a very good installation accuracy, the final emittance will be significantly above the target. Achieving the emittance goal requires more complex beam-based alignment (BBA) to minimise the dispersive emittance growth (the dominant source of aberration).

All BBA algorithms attempt to steer the beam in a dispersion-free path through the centres of the quadrupoles, either by physically moving the magnets (remote magnet movers) or by using corrector dipoles close to the quadrupoles. The exact details of the algorithms and their relative merits differ. The three most studied methods are:

- dispersion-free steering (DFS) – beam trajectories are measured for different beam energies by changing acceleration upstream; the final trajectory minimises the difference, thereby minimising the dispersion;
- kick minimisation (KM) – the BPM offsets with respect to the associated quadrupole magnetic centres are measured by varying the quadrupole strength and monitoring the resulting downstream beam motion; this information is used in a second step to find a solution for the beam trajectory where the total kick from quadrupoles and correctors on the beam is minimised;
- ballistic alignment (BA) – a contiguous section of quadrupoles (and in the linac the RF) is switched off and the ballistic beam is used to determine the BPM offsets with respect to a straight line. The quadrupoles/RF are then restored, and the beam is steered to match the established straight line.

All BBA techniques rely on precise measurements from the BPMs to determine a near dispersive-free trajectory. The final performance of the algorithms is determined by the resolution of the monitors.

Once BBA is complete, a final beam-based tuning either minimises the beam emittance by direct measurement of the beam size (emittance) or maximises the luminosity. Closed-trajectory bumps or specially located and powered tuning magnets are used as orthogonal knobs to generate specific aberrations, such as dispersion or $x - y$ coupling. The knobs are tuned to minimise the emittance by cancelling the remaining aberrations in the beam.

### 4.6.3.2 RTML before the Bunch Compressor

The issue of static emittance growth from misalignments and errors has been studied in detail for the section of the RTML from the turnaround to the launch into the bunch compressor. The strong focusing, strong bending, strong solenoids, and large number of betatron wavelengths in this area can potentially lead to very serious growth in the vertical emittance, despite the relatively low energy spread of the beam extracted from the damping rings.

The tolerances used in the study for the warm solid-core iron-dominated magnets are listed in Table 4.6 and were similar to those found at the SLAC Final Focus Test Beam.

BPM is attached to a quadrupole magnet and its offset error with respect to the quadrupole-field centre is critically important for preserving low emittance. The accuracy is assumed to be 7 µm, which will be achieved by a technique called "quad shunting", where the strength of every quadrupole magnet is changed one by one and the downstream orbit is measured. (There should be no orbit change when the beam goes through the centre of the magnet.)

**Table 4.6**
Tolerance for RTML section up to the bunch compressors.

| Error | RMS Value | Reference |
|---|---|---|
| Quadrupole Misalignment | 150 µm | Design Line |
| BPM Misalignment | 7 µm | Quad Center |
| Quadrupole Strength Error | 0.25% | Design Value |
| Bend Strength Error | 0.5% | Design Value |
| Quadrupole Rotation | 300 µrad | Design |
| Bend Rotation | 300 µrad | Design |
| BPM Resolution | 1 µm | |
| Beam-size monitor | 1 µm | |

**Dispersion Correction** The preferred dispersion correction method was found to be a combination of Kick Minimization (KM) and dispersion knobs, the latter consisting of pairs of dedicated skew quadrupoles located in the turnaround, where there is non-zero horizontal dispersion. The two skew quads in a pair are separated by a $-I$ transform such that exciting the quads with equal-and-opposite strengths causes the resulting betatron coupling to cancel and the dispersion coupling to add. There are two such dispersion knobs in the turnaround, which allow correction of dispersion at each betatron phase. Simulations indicate that in the absence of measurement errors, the combination of KM and dispersion knobs (DK) can eliminate dispersion as a source of emittance growth in this part of the RTML. The principal remaining source of emittance dilution is betatron coupling, which typically contributes about 7 nm of emittance growth.

**Coupling Correction** The coupling-correction section consists of four skew quads phased appropriately to control all four betatron coupling parameters of the beam. The skew quads are used to minimise the vertical beam sizes as measured in the downstream emittance-measurement station. The correction system can eliminate the betatron coupling introduced by misalignments and errors in this section of the RTML, without measurement errors.

In addition to the studies described above, the emittance preservation issues in the long transfer line from the damping ring to the turnaround have been examined. Because of the weaker focusing, the alignment tolerances are much looser than in the turnaround area, and emittance preservation is relatively straightforward [283, 284].

Emittance growth in this section is expected to occur dominantly in the turnaround, assuming 1 µm BPM resolution. Simulations show the total emittance growth from the exit of the damping ring to the entrance of the bunch-compressor section is 5.4 nmon average and 9.9 nm at 90 % confidence level [285].

### 4.6.3.3 Bunch Compressors

The RF in the bunch compressor introduces an energy spread correlated to the longitudinal position in the bunch. The long bunch from the Damping Ring (6 mm) makes the beam particularly sensitive to cavity tilts in the bunch-compressor RF. The near-zero phase crossing of the bunch induces a strong transverse kick which is also correlated to the longitudinal location in the bunch (i.e. the bunch is crabbed), and therefore also strongly correlated to the induced energy spread. The resulting transverse kick-energy correlation can effectively be compensated using downstream dispersion knobs. Wakefield-driven head-tail correlations can also be compensated in the same way. As with other sections of the LET, the other primary source of emittance dilution is dispersion due to misaligned quadrupoles.

Transverse kicks due to the input coupler and HOM couplers of an accelerating cavity have been extensively studied [286]. The couplers induce asymmetries of the electromagnetic fields in the cavity, which cause transverse components of the accelerating field (coupler RF kick) and transverse wake field (coupler wake). The effects is included in the simulations.

Simulation studies with RMS random quadrupole offsets of 0.3 mm, cavity offsets of 0.3 mm and cavity pitch of 0.3 mrad indicate that combined DFS and DK will reduce the mean emittance dilution to 1.1 nm in average and 1.5 nm at 90 % confidence level [285].

#### 4.6.3.4 Main Linac

Single-bunch emittance dilution in the main linac is dominated by chromatic (dispersive) effects and wakefield kicks arising from misaligned quadrupoles and cavities respectively. The $x-y$ coupling arising from quadrupole rotation errors also adds a small contribution to the vertical-emittance growth. The assumed installation errors are listed in Table 4.7. The tolerances for cavity offsets and quadrupole rolls can be achieved mechanically, but beam-based tuning is required for reducing the effects of the quadrupole and BPM offsets.

**Table 4.7**
Assumed installation errors in the main linac, and the emittance growth for each error assuming simple one-to-one steering. Note that with perfectly aligned BPMs, the one-to-one steering eliminates dispersive effects from quadrupole magnet offset and cavity tilt. With realistic BPM errors, the required emittance preservation can only be achieved using beam-based alignment of the magnets/BPMs.

| Error | with respect to | value | $\Delta\gamma\epsilon_y$ [nm] |
|---|---|---|---|
| Cavity offset | module | 300 µm | 0.2 |
| Cavity tilt | module | 300 µrad | <0.1 |
| BPM offset | module | 300 µm | 400 |
| Quadrupole offset | module | 300 µm | <0.1 |
| Quadrupole roll | module | 300 µrad | 2.5 |
| Module offset | perfect line | 200 µm | 150 |
| Module tilt | perfect line | 20 µrad | 0.7 |

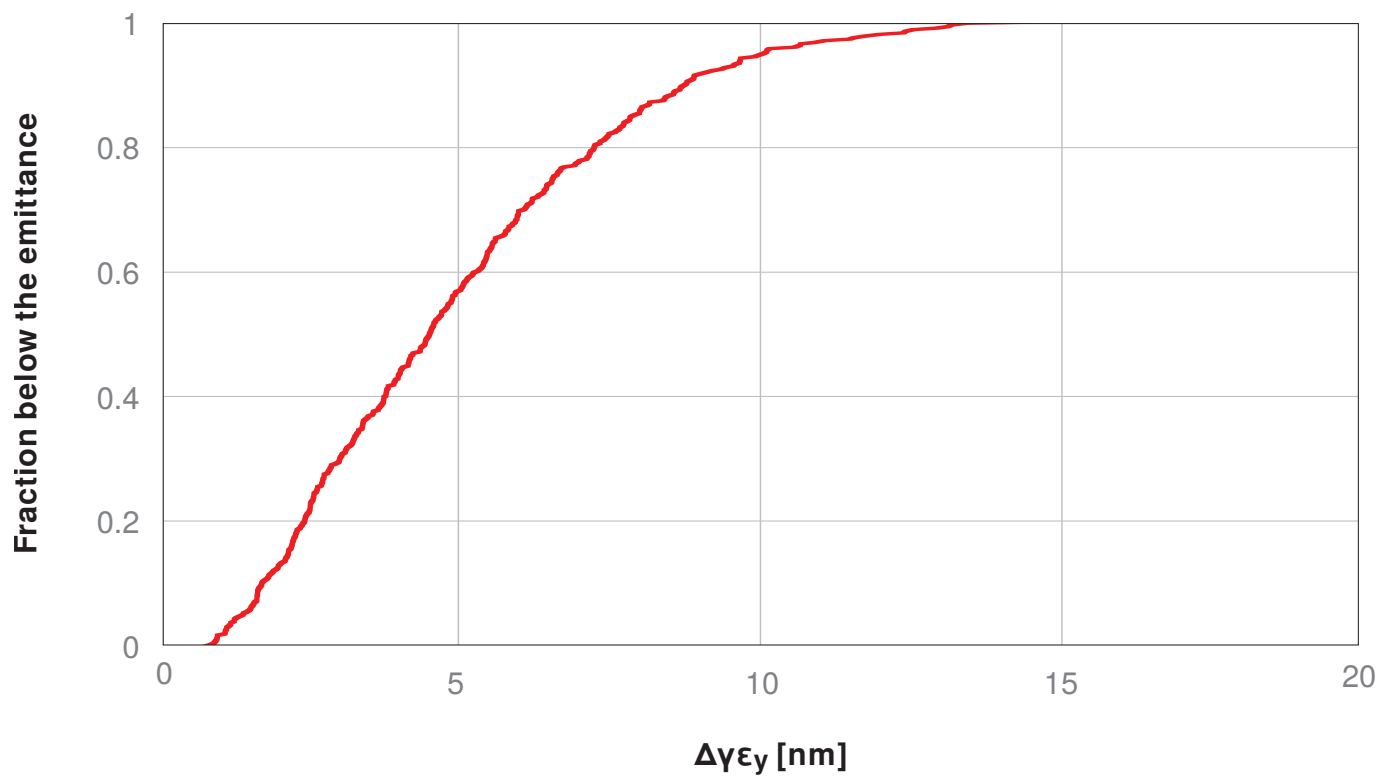

**Figure 4.40**
The fraction of simulated cases staying below the emittance-growth target for the main linac after Dispersion Free Steering.

The main linac follows the gravitational equipotential of the earth, and is therefore not laser-straight. This gentle bending in the vertical plane results in a small but non-negligible design dispersion which must be matched, and taken into consideration during beam-based alignment. A variant of dispersion-free steering, dispersion-matched steering (DMS), is used to attain the matched dispersion function along the lattice. This modified form of DFS requires well calibrated BPMs [287] to the level of 5 % with very stable readout. The method achieves the required performance in simulations [287–291]. An example simulation result is shown in Fig. 4.40. Additional tuning knobs to modify the dispersion at the beginning and end of the linac reduce the emittance growth still further [292]. Further improvement is possible with wakefield tuning knobs.

Studies of kick minimisation have shown similar performance as DMS [293, 294]. The ballistic alignment method has not been applied to the latest ILC lattice, however studies for TESLA showed that ballistic alignment and dispersion-free steering yielded comparable results (for a laser-straight machine) [295].

Multi-bunch wakefields (high-order modes, HOMs) with very high Q-values could lead to unacceptable multi-bunch emittance growth. Suppression of HOMs is achieved by random cavity detuning ($\sim$ 0.1% spread in the HOM frequencies, expected from the manufacturing process), and by damping using HOM couplers. All the modes for the baseline cavity shape have been calculated and measured at FLASH. The resulting multi-bunch emittance growth due to cavity misalignment is expected to be below 0.5 nm. If the transverse wakefield modes are rotated due to fabrication errors, they can lead to a coupling of the horizontal and vertical planes, potentially increasing the vertical emittance [296]. This effect is mitigated by using a split-tune lattice in which the vertical and horizontal beam-oscillation wavelengths are different, thus avoiding resonant coupling.

Different codes have been compared in detail for the main linac [297], finding excellent agreement for both tracking and performance predictions for a specific beam-based alignment method. This cross-benchmarking increases confidence in the results of each individual code.

#### 4.6.3.5 Undulator Section for Positron Production

At the end of the electron main linac, the beam passes through an undulator and emits hard photons for positron production. This insert has several potential consequences for emittance preservation:

- stronger focusing in the 1.2 km insert leads to additional dispersive emittance growth, which should be correctable using BBA methods;
- the undulator increases the energy spread of the beam, which increases the dispersive emittance growth in the downstream linac; this effect is estimated to be small;
- the narrow-bore vacuum chamber of the undulator is a potential source of transverse wakefields; the effect is expected to be small and can be corrected using precise alignment movers.

The total emittance growth in this insertion is estimated to be small compared to the overall emittance-growth budget.

Final beam energy will be changed as physics requires. Low-emittance preservation tends to be difficult for lower-energy operation, because of larger relative energy spread (dispersive effect) and relatively stronger transverse-wakefield effect. However, simulation studies for beam energies of 100 GeV and 250 GeV shows the difference of expected normalised emittance is smaller than 2 nm.

#### 4.6.3.6 Beam-Delivery System (BDS)

Beam-based procedures have been developed to align and tune the BDS. First, all multipole (sextupoles and more) magnets are switched off and the quadrupoles and BPMs are aligned. Second the multipoles are switched on and aligned. Finally, tuning knobs are used to correct the different beam aberrations at the interaction point. Detailed simulations have been made assuming the realistic installation alignment errors and magnetic-field errors given in Table 4.8.

**Table 4.8**
Assumed imperfections in the BDS. The assumed magnet strength errors are very tight; it is expected that more realistic larger errors mainly lead to slower convergence of the procedures.

| Error | with respect to | size |
|---|---|---|
| Quad, Sext, Oct x/y transverse alignment | perfect machine | 200 µm |
| Quad, Sext, Oct x/y roll alignment | element axis | 300 µrad |
| Initial BPM alignment | magnet center | 30 µm |
| Strength Quads, Sexts, Octs | nominal | $10^{-4}$ |
| Mover resolution (x/y) | | 50 nm |
| BPM resolutions (Quads) | | 1 µm |
| BPM resolutions (Sexts, Octs) | | 100 nm |
| Power supply resolution | | 14 bit |
| Luminosity measurement | | 1 % |

Studies have been performed using the beam-beam interaction code GUINEA-PIG to give a realistic estimate of the luminosity, assuming the accuracy of luminosity measurement is 1 %. Important results of the beam-beam interaction studies has been crosschecked with the CAIN code.

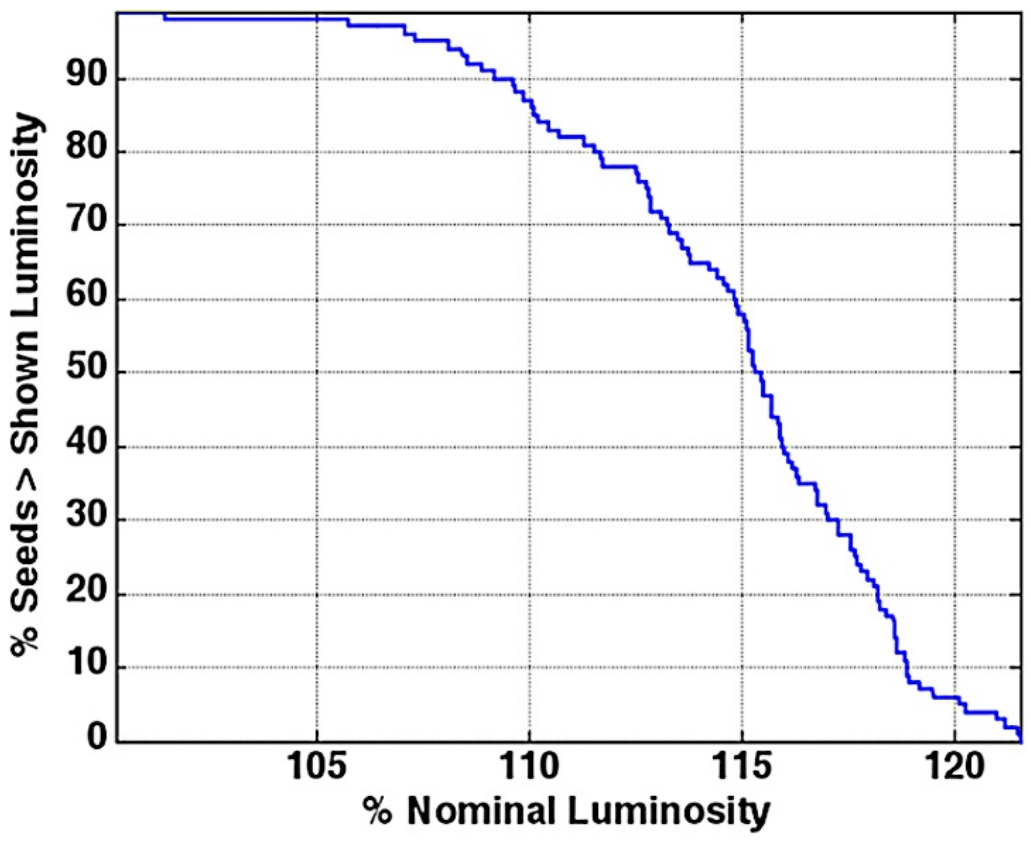

**Figure 4.41**
Results of simulation of the luminosity tuning in the beam-delivery system. The vertical axis indicates the ratio of the random seeds simulated that results in a relative luminosity greater than values in the horizontal axis [298].

Figure 4.41 shows an example of the BDS luminosity-tuning simulations for static errors. The vertical emittance at the entrance of BDS was assumed to be 34 nm, which assumes an emittance growth of 14 nm for the RTML and ML combined. The results show that for the given assumptions on static errors and input conditions, all seeds exceed the design luminosity after application of the beam-based tuning algorithms.

### 4.6.4 Dynamic Effects

The ILC relies on several different feedback systems to mitigate the impact of dynamic imperfections on the luminosity. These feedback systems act on different timescales. The long $\sim$ 1 ms pulse length and relatively large bunch spacing ($\sim$ 554 ns, $\sim$ 366 ns in update configuration) makes it possible to use bunch-to-bunch (or intra-train) feedbacks located at critical points. The most important is the beam-beam feedback at the interaction point that maintains the two beams in collision. Other feedback systems act from train to train (inter-train) at the 5 Hz pulse repetition rate of the machine. Over longer timescales (typically days or more), the beam may have to be invasively re-tuned.

Main dynamic error sources and their effects to orbit jitter and emittance growth in Main Linac are listed in Table 4.9.

**Table 4.9**
Dynamic errors and their effects in the main linac.

| Error | assumed RMS amplitude | Orbit change | emittance growth |
|---|---|---|---|
| Quad offset change (vibration) | 100 nm | 1.5 $\sigma$ | 0.2 nm |
| Magnet-strength jitter | $10^{-4}$ | $1\sigma$ | 0.1 nm |
| Cavity-tilt change | 3 µrad | $0.8\sigma$ | 0.5 nm |
| Cavity-to-cavity strength change | 1% | $0.8\sigma$ | 0.5 nm |

Important sources of dynamic imperfections are ground motion and component vibration. The ground motion depends strongly on the site location. For the ILC-TRC study, three ground motion models were developed, all based on measurements at existing sites: Model A represents a very quiet site (deep tunnel at CERN); Model B a medium site (linac tunnel at SLAC); Model C a noisy site (shallow tunnel at DESY). A fourth model (K) was later developed based on measurements at KEK and is roughly equivalent to C. These models have been used in all subsequent simulations of the dynamic behaviour of the ILC.

Another possibly important source is cavity-to-cavity strength change within the $\sim 1\,\mathrm{ms}$ pulse, with mechanical cavity tilt (static alignment error). The effect leads to trajectory errors of individual bunches in a bunch train. While these trajectory errors (which are relatively slow over the 1 ms pulse) can easily be corrected with intra-train feedback at the end of the linac, the large variation in trajectories in the ML give rise to emittance growth. To suppress this effect, voltages in the individual cavities will be corrected to within 1 % over the pulse.

One possible issue is time-varying stray fields in the long transfer line of RTML which can drive orbit oscillations. This orbit change can be cancelled downstream by the feed-forward system located across the turnaround, though this system cannot compensate possible emittance growth in the turnaround. Measurements at existing laboratories [297] indicate a reasonable estimate for the magnitude of time-dependent stray field is about 2 nT, which will not cause a problem. Even in the case that the stray fields turn out to be significant, adding an intra-pulse orbit-feedback system before the turnaround can eliminate the emittance growth of most of the bunches in a bunch train.

Tolerances of other dynamic errors are listed in Tables 4.10–4.11 and 4.12. None of these errors are expected to affect luminosity performance significantly.

**Table 4.10**
Other dynamic errors relevant to transverse motions.

| Error | assumed RMS amplitude |
|---|---|
| Offset change (vibration) of warm magnet | 10 nm |
| Strength change of warm magnet | $1 \times 10^{-5}$ |
| Strength change of cold magnet | $1 \times 10^{-4}$ |

**Table 4.11**
Bunch-compressor RF dynamic errors, which induce 2 % luminosity loss.

| Error | RMS amplitude | RMS phase |
|---|---|---|
| All klystron correlated change | 0.5 % | 0.32° |
| Klystron-to-klystron uncorrelated change | 1.6 % | 0.60° |

**Table 4.12**
ML RF dynamic errors, which induce 0.07 % beam energy change.

| Error | RMS amplitude | RMS phase |
|---|---|---|
| All klystron correlated change | 0.07 % | 0.35° |
| Klystron to klystron uncorrelated change | 1.05 % | 5.6° |

#### 4.6.4.1 Bunch-to-Bunch (Intra-Train) Feedback and Feedforward Systems

The damping-ring extraction kicker extracts each bunch individually. If this kicker does not fully achieve the required reproducibility, the beam will have bunch-to-bunch variations that cannot be removed by an intra-pulse feedback system (effective white noise). The feed-forward system in the RTML is designed to mitigate this effect. The position jitter of each bunch is measured before the turn-around and then corrected on that bunch after the turn-around.

Quadrupole vibration in the downstream bunch compressor and (predominantly) in the main linac will induce transverse beam jitter (coherent betatron oscillations). The tolerance on the amplitude of this jitter (and hence on the quadrupole vibration) from the main linac itself is relatively loose. Quadrupole vibration amplitudes of the order of 100 nm RMS lead to negligible pulse-to-pulse emittance growth. However the resulting oscillation (one- to two-sigma in the vertical plane) in the BDS could lead to significant emittance degradation from sources such as collimator wakefields. An intra-train feedback at the exit of the linac solves this problem. In addition, this feedback could correct any residual static HOM disturbance in the bunch train. If the main-linac quadrupole vibrations are significantly less than 100 nm (e.g. 30 nm RMS, as expected for a typical quiet site), then intra-train feedback at the exit of the linac may not be required.

Small relative offsets of the two colliding beams, in the range of nanometers, lead to significant luminosity loss. The offsets are particularly sensitive to transverse jitter of the quadrupoles of the final doublet. Fortunately, the strong beam-beam kick causes a large mutual deflection of the offset beams, which can be measured using BPMs just downstream of the final quadrupoles. The intra-train feedback system zeros the beam-beam kick by steering one (or both) beams using upstream fast kickers. The system typically brings the bunch trains into collision within several leading bunches (depending on the gain). The IP fast feedback and the long bunch train also affords the possibility to optimise the luminosity within a single train, using the fast pair monitor as a luminosity monitor [300].

### 4.6.4.2 Train-to-Train (5 Hz) Feedback

There are number of local feedbacks. At certain locations in the machine, a few correctors are used to steer the beam back through a few selected BPMs, thus keeping the trajectory locally fixed. These feedback systems can be used in a cascaded mode where each of the feedback anticipates the trajectory change due to the upstream feedback systems. Such a system was successfully implemented at SLC.

Since the system corrects only locally, a residual of the dynamic imperfections will remain, due to deterioration of the trajectory between the feedback locations. After longer times, this will require a complete re-steering of the machine back to the exact trajectory determined from the initial beam-based alignment (gold orbit).

There are other alternative options. One is to perform permanent re-steering with a very low gain; this method avoids the additional layer of steering but may be slower than local feedback. Another option is the use of a MICADO-type correction. In this procedure all BPMs are used to determine the beam orbit. A small number of the most effective correctors is identified after each measurement and these are used to correct the trajectory.

### 4.6.4.3 Luminosity Stabilisation

A complete and realistic simulation of the dynamic performance of the collider requires complex software models that can accurately model both the beam physics and the errors (e.g. ground motion and vibration). The problem is further complicated by the various time scales that must be considered, which span many orders of magnitude: performance of the fast intra-train feedbacks requires modelling of the detailed 10 MHz bunch train; fast mechanical vibrations at the Hz level need to be accurately modelled to test the performance of the pulse-to-pulse feedback systems; long-term slow drifts of accelerator components over many days must be studied to determine long-term stability and the mean time between invasive (re-)application of BBA. Ideally all these elements need to be integrated into a single simulation of the complete machine.

However, practically it is enough to have sets of various simulations focusing on individual aspects of the problem, with varying degrees of the feedback models. The results thus far give every indication that the ILC can achieve and maintain the desired performance.

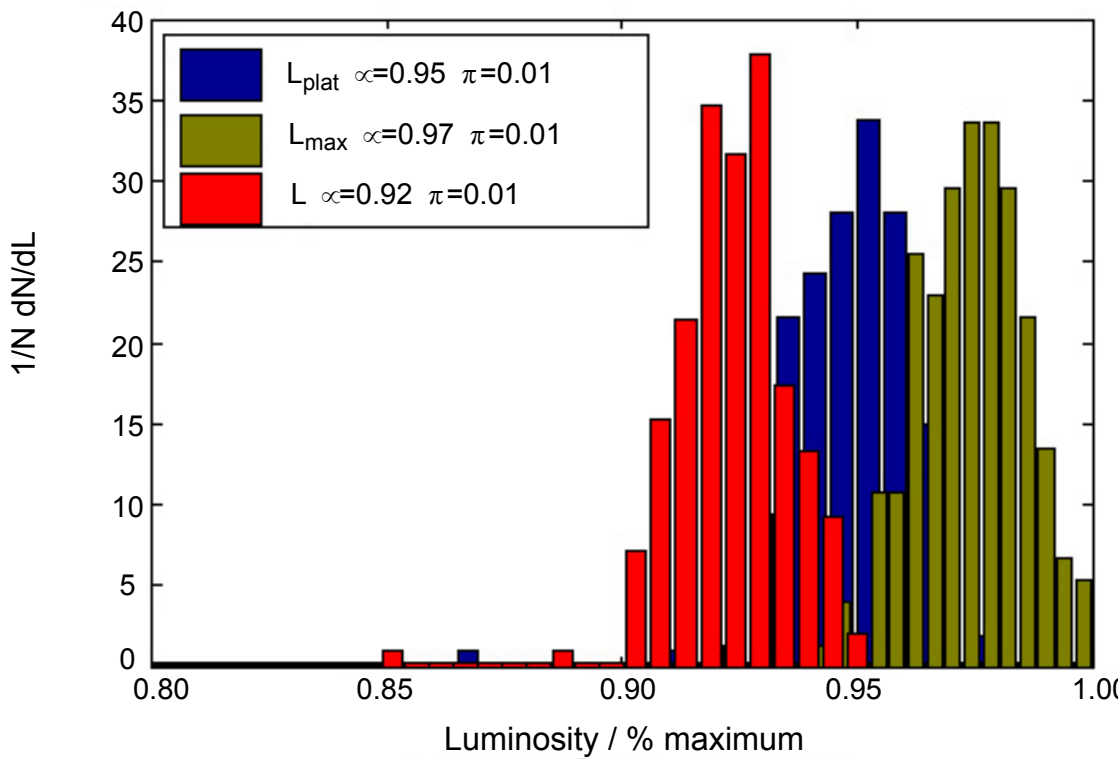

**Figure 4.42**
Example of integrated dynamic simulations, showing the performance of the beam-beam intra-train feedback system with realistic beams and beam jitter (simulated from the Main Linac and BDS). The histograms show performance over 100 seeds of random vibration motion: green - achieved luminosity for an infinitely fast beam-beam feedback and no bunch-to-bunch variations (3 % reduction from ideal); blue - performance including bunch-to-bunch variations (driven by long-range wakefields in the Main Linac); red - as blue but including a finite response time for the feedback (8 % reduction from ideal) [301, 302].

Extensive simulations have been made of the performance of the fast beam-beam (and other) intra-train feedback using a model of the main linac and BDS to generate realistic bunch trains [301, 302]. For realistic component vibration amplitudes, the results indicate that feedback can maintain the luminosity within a few percent of peak on a pulse-to-pulse timescale (5 Hz), as shown for example in Fig. 4.42. These results are in agreement with earlier studies [303, 304].

Drifts of components on the timescale of seconds to minutes have been studied [303, 304]. Simulations of 5 Hz operation with all ground-motion models, and assuming the beams are maintained in collision by the fast IP feedback, indicate a slow degradation in luminosity. This can be mitigated by pulse-to-pulse feedback, especially in the BDS, where the tolerances are tightest. Noisy sites (model C) showed the most pronounced effect, and would place most demand on the slower feedback systems.

Longer-term stability has been studied, assuming a variety of configurations for the slower pulse-to-pulse feedbacks. Studies of the main linac [305] using local distributed feedback systems indicate that the time between re-steering ranges from a few hours to a few days for ground-motion models C and B respectively. After 10/200 days (models C/B), simple re-steering does not recover the emittance, at which point a complete re-tuning would be necessary.

Dynamic studies integrating the main linac and BDS, again based on distributed local pulse-to-pulse feedback systems (including one in the BDS) and incorporating many error sources and comparing all ground-motion models have been made [306]. It was shown that luminosity reduction in the noisy sites (models C and K) comes almost entirely from the BDS.

Other examples of such simulations of luminosity performances for models of ground motion and vibration are shown in Fig. 4.43. The primary effect is a beam-beam offset at collisions, which is quickly compensated by the intra-train feedback at the interaction point. The luminosity loss will not be serious even for the models C and K, which are much noisier than expected for possible ILC sites.

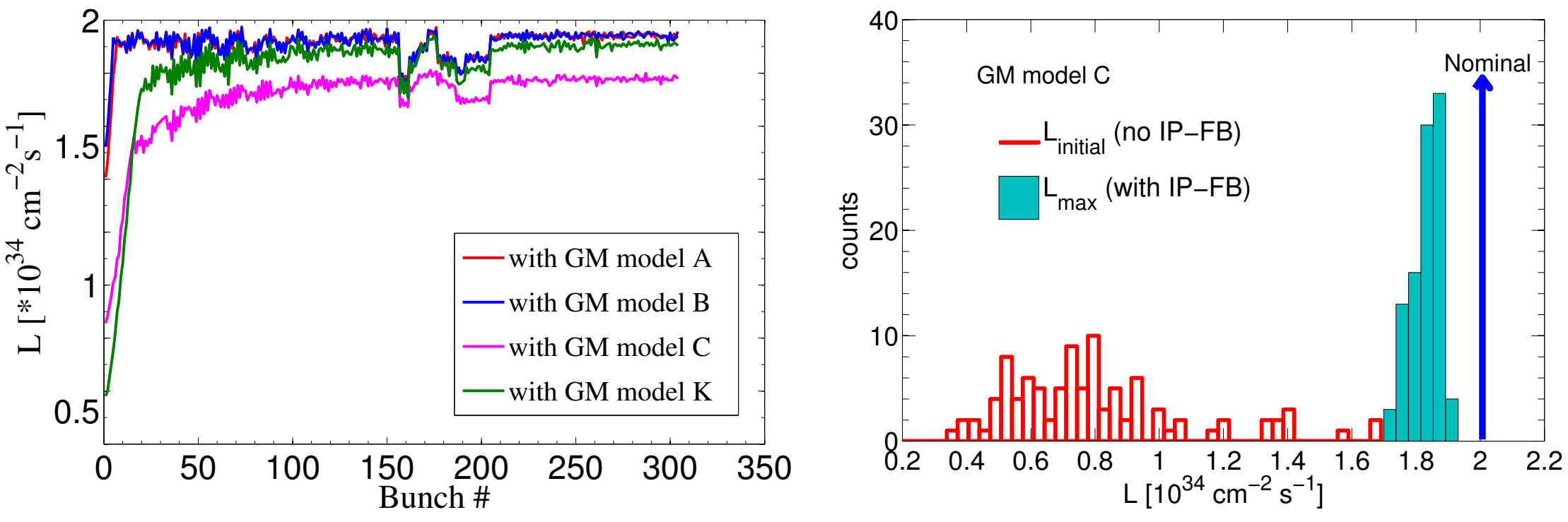

**(a)** **(b)**

**Figure 4.43.** Time-dependent luminosity modelling in the beam-delivery system. (a): the luminosity is shown as a function of bunch number for the first 300 bunches of a pulse for various ground motion (vibration) models. The luminosity is quickly recovered by the beam-beam fast feedback. (b): a histogram of the results of 100 seeds, assuming model C, as referenced in the left graph, for the ground motion, with and without intra-pulse orbit feedback [299].

A key parameter is the maximum-allowed vibration of the Final Doublet (FD). This is primarily set by the limitations of the IP fast feedback, which becomes increasingly ineffective for larger beam-beam offsets. Figure 4.44 shows the luminosity as function of RMS offset of both final-doublet cryomodules. The allowed RMS FD offset tolerances is conservatively specified as 50 nm.

#### 4.6.4.4 Requirements of Field-Ramping Speed of Main-Linac Components

The field strengths of the superconducting quadrupole and correction dipole magnets must be able to change at a sufficient speed to allow efficient commissioning and to adapt to the ground motion. The required speed of have of magnetic fields in the main linac has been roughly estimated as follows. [307]:

- Quadrupole magnet:

  0.001 T/m×m/s (0.003 %/s) for adjustment of optics with RF failure.

  0.01 T/m×m/s (0.03 %/s) for performing "quad shunting", beam-based alignment measurement, within a reasonable time.

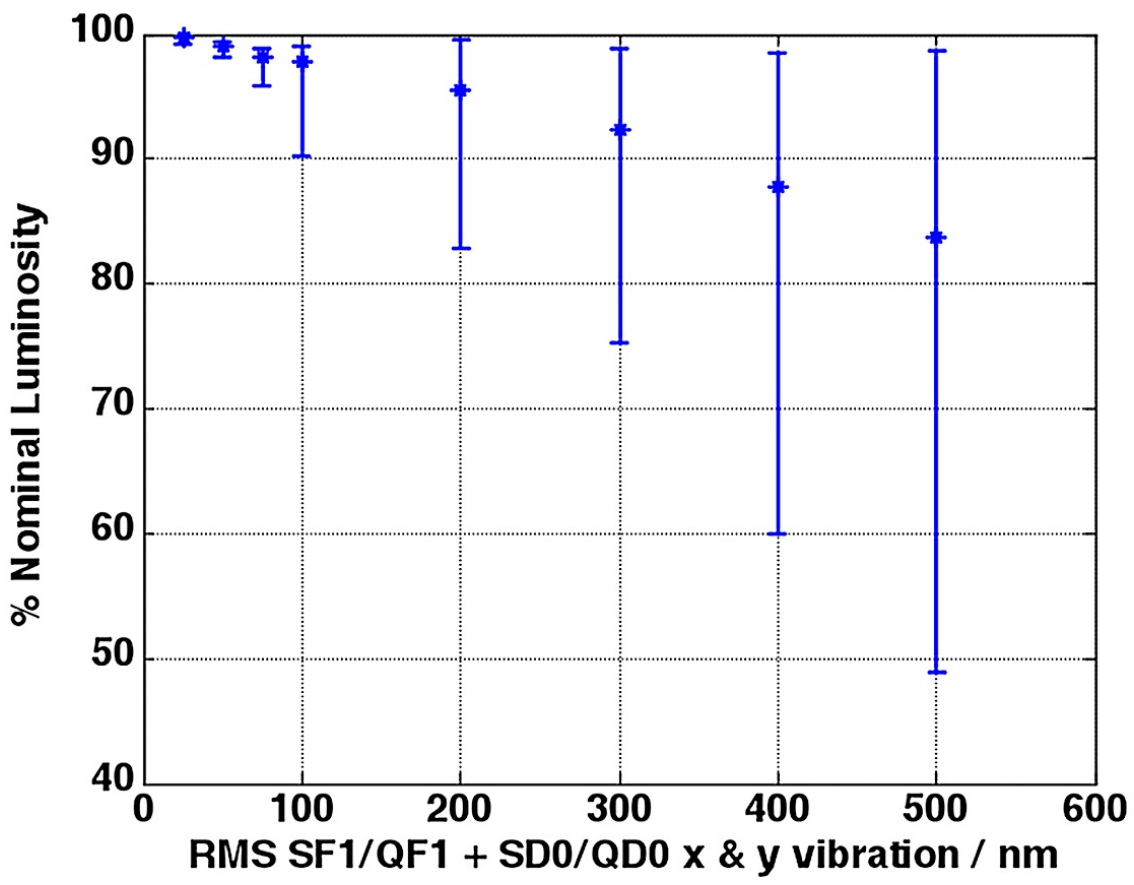

**Figure 4.44**
Simulated luminosity as a function of rms offset of both final-doublet cryomodules. Mean and standard deviation from tracking simulation are shown [298].

- Steering magnet:

  $5 \times 10^{-6}$ T×m/s (0.01 %) for following ground motion

  $3 \times 10^{-4}$ T×m/s (0.6 %/s) for orbit correction after a long shut down within a reasonable time end

For each type of magnets, the second number is the most stringent and should be adopted. These numbers are obtained assuming very pessimistic cases for the correction process under the ground-motion model C to safely define the specification of the magnets.

### 4.6.5 Optics for the Upgrade to 1 TeV

A possible scenario to upgrade ILC from the centre-of-mass energy 500 GeV to 1 TeV is described in Part II Section 12.4. Each main linac, from 15 GeV (after the bunch compressor) to 250 GeV in the baseline, will be extended to 500 GeV in the following way:

- the majority of the cryomodules of the 250 GeV linacs remain in place and become the final 225 GeV of the 500 GeV;
- the first section (from 15 GeV to 25 GeV) of the old linac will be moved to the upstream end of the new linac. This is because the magnets (quadrupole and correction dipole) in this section are shorter than others in order to avoid fields that are too low to be accurate;
- the newly constructed modules are inserted between these sections (25 GeV to 275 GeV).

The adoption of a FODO lattice in the entire new linac as in the old linac with the limited maximum strength of the quadrupole magnets, the beta-function and the vertical dispersion (induced by the beam line following the earth's curvature) would be large in the high-energy beam part of the linac. For a given lattice, the beta function is roughly proportional to the inverse of the magnet strength (normalised by the beam energy) and the dispersion to inverse square.

To preserve low emittance, DMS (Dispersion Matching Steering), in which the vertical dispersion at every BPM will be measured and adjusted to the designed non-zero value will be applied. This correction requires accurate measurement of dispersion, whose error will be proportional to the scale error of the BPMs and the designed dispersion. With large designed dispersion, the required accuracy of the BPM scale becomes too tight.

To keep the dispersion small along the main linac, a FOFODODO lattice will be used instead of a FODO lattice for the 250 GeV to 500 GeV section of the linac. In a curved linac, the typical vertical dispersion with a FOFODODO lattice is approximately half of that with FODO lattice. Simulations showed that emittance growth will be small enough in the whole linac if a FODO lattice is adopted from 15 GeV to 250 GeV and FOFODODO lattice from 250 GeV to 500 GeV [307].

# Chapter 5
# Conventional Facilities and Siting Studies

## 5.1 CFS design considerations

The Conventional Facilities and Siting Group used the Technical Design Phase to improve the maturity of the conventional design and respond to changes in the technical baseline criteria for the International Linear Collider Project. The maturity of the conventional design was influenced, in general, by two factors. First, more precise criteria were developed for the civil, mechanical and electrical design. Secondly, instead of the more generic approach taken for the design of the conventional facilities in the original reference design, regional conditions were taken into consideration and had a major impact on the design. The changes to the technical baseline were provided primarily through the Strawman Baseline 2009 (SB2009) analysis [104], which resulted in fundamental changes to the design of the conventional facilities. This chapter will describe these changes and also provide brief descriptions of the supporting consultant work and studies.

The overall maturity of the conventional facilities design was improved over the course of the TDP. Several value-engineering exercises were conducted. Optimisation of mechanical- and electrical-systems design, comparisons of tunneling techniques and configurations and more detailed criteria were pursued in all regions by the Conventional Facilities Group. This effort has provided better understanding of the impact of the design requirements and an overall increase in the level of detail. The local site conditions had a large influence on the conventional facilities design. This was especially the case with the two candidate sites in the Asian Region.

### 5.1.1 Central Region integration

In SB2009, the repositioning of the Electron Source and adjustments to the overall circumference, location and size of the Damping Ring contributed to a major revision for the central-region design and underground enclosures. Instead of encircling the Interaction Hall, the Damping Rings are now shifted to the side but with their centre still aligned with the interaction point. Moving the Positron Source to the end of the Main Linac reduced the length of transfer lines. The size of the tunnel for the Damping Rings was enlarged to allow for the possibility of accommodating three rings. Also the Electron and Positron Sources and transfer lines to and from the Damping Rings, the RTML and Beam-Delivery Systems are all co-located in the same enclosures, complicating issues of installation, life safety, personnel egress and equipment replacement. In addition, there are various improvements to the requirements and the design maturity of the Detector Hall.

## 5.1.2 Main Linac underground enclosures

Another major technical change, the development of different schemes for the High-Level RF system, introduced opportunities for adjustments to underground enclosures for the main linac. Two options for the High-Level RF system were considered: the Klystron Cluster system (KCS) and the Distributed Klystron system (DKS).

In the reference design, all sites used vertical shafts for access to the main-linac tunnel. The KCS system involved relocating all of the klystrons and supporting power-supply equipment to surface buildings at the vertical shafts. The microwave power is then transported to the cryomodules that make up the main linac through large-diameter waveguides that extend from the klystrons down the shaft and through the main-linac tunnel. The KCS approach provided the opportunity to eliminate the RDR service tunnel for the main linac since the klystrons and all supporting equipment are now located in surface buildings. This applies to both the Americas and European Sites. This single main-linac tunnel and shaft design was analysed by a Tunnel Configuration Study and a Tunnel Lining Study in the Americas Region as well as Life Safety studies in both the Americas and European Regions. The tunnel cross sections for the Americas and European regions are shown in Fig. 5.1 and Fig. 5.2, respectively.

**Figure 5.1**
Americas Region tunnel Cross Section

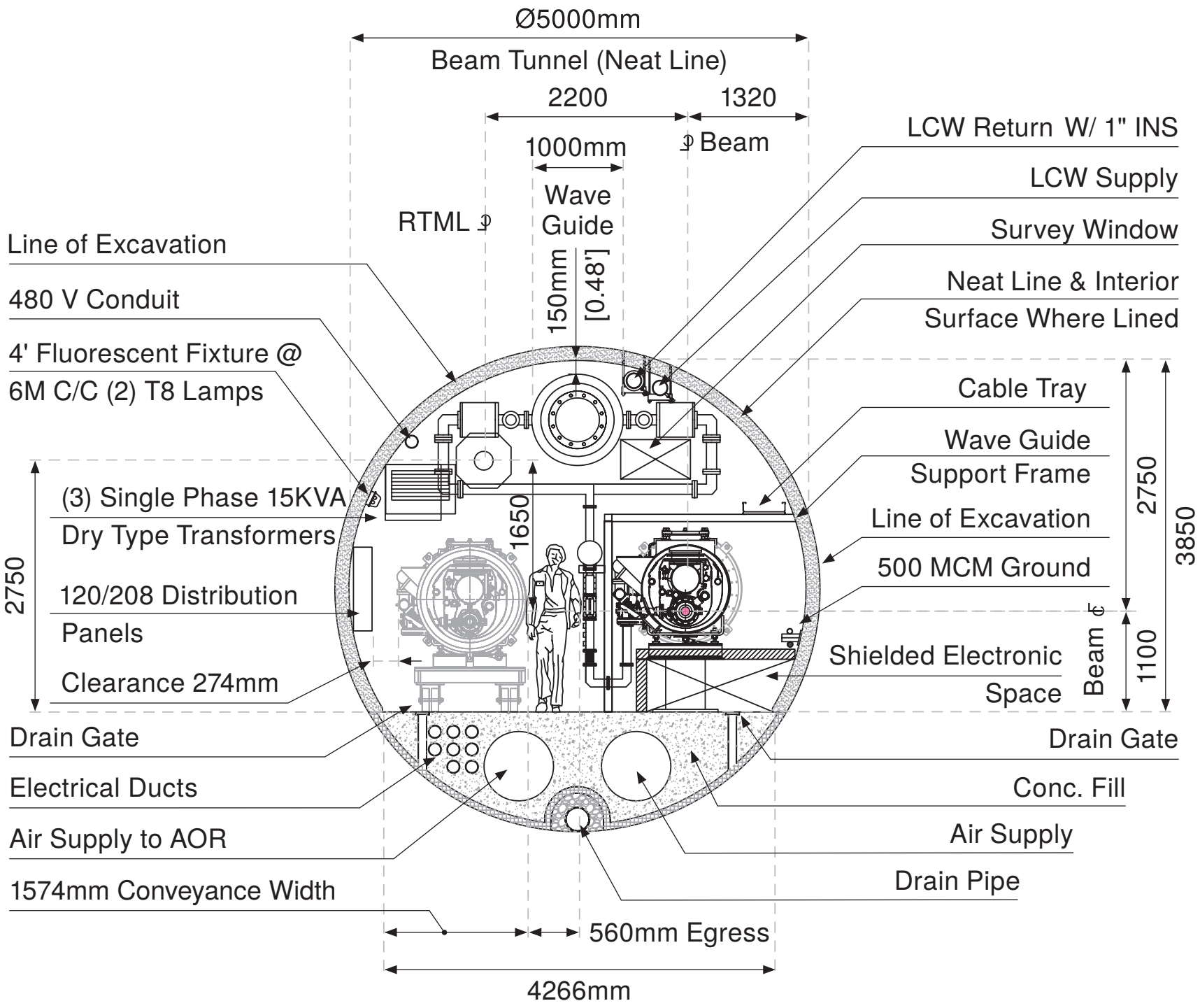

For the Asian region, another solution was required. Two candidate sites were identified in mountainous regions of Japan, resulting in vertical shafts becoming less desirable. Horizontal access tunnels, with a slight downward incline, became the preferred solution for these sites. The lengthier horizontal tunnel access posed a problem for the KCS RF system. As the conventional design progressed in the Asian Region, different tunnel cross sections were studied. Eventually it was decided that, for the Japanese candidate sites, drilled tunnels using tunnel boring machines (TBMs) were not cost effective. As an alternative, the Asian CFS Group proposed the use of the New Austrian Tunnelling Method (NATM), which is basically a drill-and-blast method of construction, as a more cost-effective approach for construction of the underground enclosures. The cross section of the main-linac tunnel is larger than the circular cross sections of the Americas and European Regions that result from using a TBM. The larger tunnel illustrated in Fig. 5.3 is wide enough to be divided into two compartments, making an alternative klystron system possible. One side will be used for the accelerator and the other for the DKS scheme, which is similar to the approach used in the ILC Reference Design. The DKS klystrons are shielded by the tunnel wall and can therefore be accessed while the beam is on. This solution works for both Asian candidate sites. The Civil design in the Asian region was also supported by a Tunnel Configuration Study and preliminary geotechnical investigation at both candidate sites.

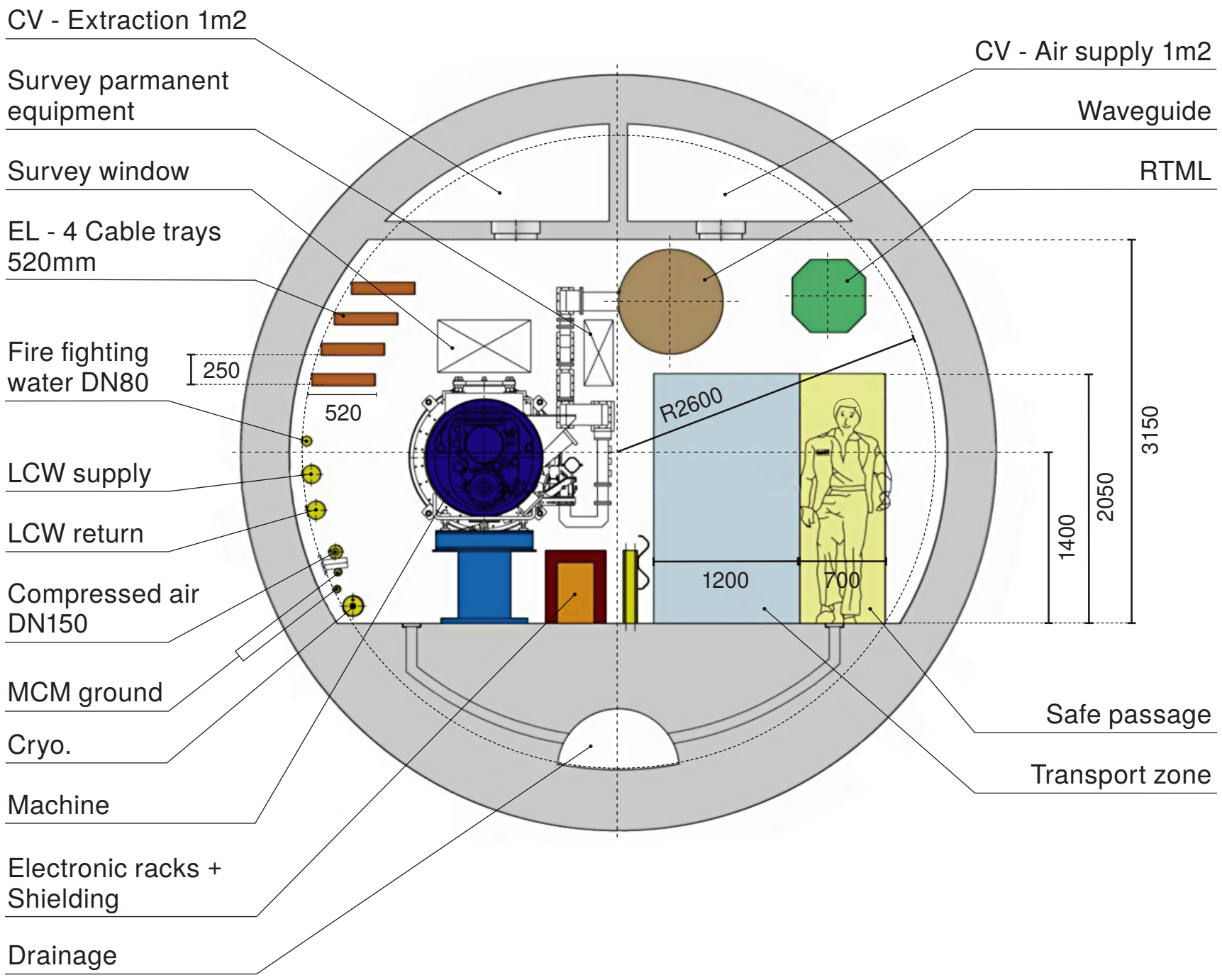

**Figure 5.2**
European Region tunnel Cross Section

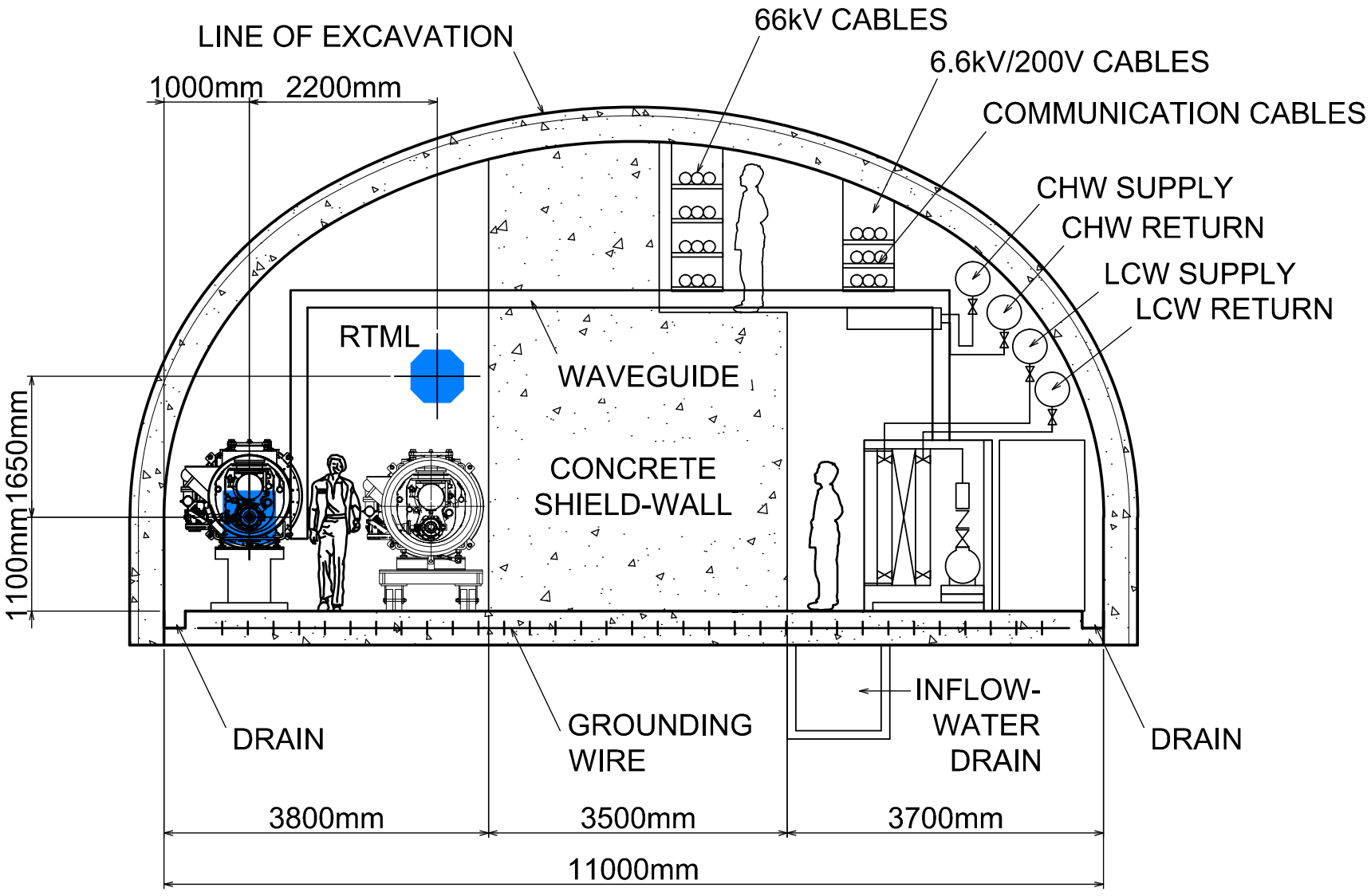

**Figure 5.3**
Asian Region tunnel Cross Section

### 5.1.3 Detector hall and detector assembly

The local site conditions also had a direct impact on the configuration of the Detector Hall and ultimately the assembly of the detectors themselves. In both the Americas and European Regions, large sections of the detectors are assembled in a surface assembly hall and lowered into the detector hall through a large diameter central shaft located directly over the eventual interaction point. Other vertical shafts are used to access the out-of-beam detector "garage" areas and for personnel access. In the Asian candidate sites, a single horizontal access tunnel provides the only access to the detector hall. Detector components must be assembled in small pieces and transported to the detector hall for final assembly. These regional differences, based on local conditions, have a significant effect on detector assembly and shaft/tunnel coordination strategies. In all regions, both detectors will be assembled on individual moveable platforms that will allow for movement to the interaction point. In the Americas Region, a consultant study was completed to provide design and movement alternatives for the detector platforms. In the European Region, a consultant study provided a 3-dimensional model to study the effects of the sample-site geology on detector hall construction.

### 5.1.4 Life-safety and egress

In all regions, life safety and egress solutions for local underground tunnel configurations were an important aspect of the technical design. The decision to eliminate the main-linac service tunnel and have only one main linac tunnel presented challenges for the Americas and European Region with respect to life safety and personnel egress. While the main-linac tunnel configuration in the Asian Region went through significant changes, the final solution produced a configuration very similar to the original twin-tunnel RDR scheme. Each regional solution for life safety and egress was shaped by local codes and regulations. It is important to note that the solutions described below are the direct result of extensive in-house analysis of regional requirements by the CFS Group, supported in part by independent consultant studies. Work to date has provided confidence that a single-tunnel solution can be constructed in all regions that will provide a safe working environment when personnel are underground performing machine installation and maintenance activities.

The Asian main-linac tunnel configuration for the two Japanese mountainous sites provides a single tunnel that is divided into two compartments separated by a concrete radiation shielding wall. This effectively creates two separate enclosures that can be used for egress. If a fire or other hazard occurs in one of the tunnels, the second tunnel can be isolated and used as the emergency escape route. The compartments are connected by labyrinths, spaced every 500 m, through the shielding wall. Each labyrinth is protected by fire doors to allow safe passage from the affected area to the other side of the tunnel for egress to the surface. At each horizontal access tunnel, a similar protected pathway is provided to the surface.

In the Americas Region, the main linac enclosure is a single tunnel. The prevailing codes require containment of those areas of the underground space that have the highest hazard potential by fire-rated walls and doors, so that the most likely hazardous areas are contained in the event of an emergency. Oil-filled electrical equipment, water pumps, motors and other utility equipment constitute the highest potential for fire. This equipment is located in the caverns at the base of the vertical access shafts located along the single main-linac tunnel enclosures. In the Americas Region solution, these local areas are isolated by fire-rated walls and doors, leaving the main linac (or damping ring) tunnel enclosure available for personnel access to the surface. Due to the overall length of the tunnels, it is required to have a fire-protected area of refuge at intervals of 1200 m along the length of the single tunnel to provide an intermediate safe area for injured personnel or to await emergency-response assistance.

In the European Region, the main-linac enclosure is also a single tunnel. However, local regulations and previously established precedents with the LHC at CERN produced a very different life-safety solution that divides the single-tunnel enclosure for the main linac and damping ring into 500 m increments or "compartments" that are separated by fire walls and automatic fire doors that can isolate areas that are involved in a fire or other hazardous incident. Another important part of this approach is the requirement for the control of air flow in an emergency. A continuous duct for air supply and extraction, located at the tunnel crown, extends along the entire length of the tunnel. In the event of fire, dampers will close in the supply duct, preventing fresh air from contributing to the fire and stopping smoke from permeating beyond the area of the fire. In this way, the unaffected areas of the tunnel can be used for personnel egress to the surface and for emergency-response personnel to access the affected area.

### 5.1.5 Electrical and mechanical utilities

Another important part in the development of a mature design for the ILC conventional facilities is the optimisation of the design for the utility systems that support the accelerator operation. After the underground civil construction, the mechanical and electrical systems represent the second and third largest cost drivers in the conventional facilities cost. As design development progressed and local conditions contributed to different regional design solutions, the mechanical and electrical designs were also adapted to local conditions.

Another factor that contributed to progress in the mechanical and electrical design was the refinement of the mechanical and electrical criteria used to develop the designs. Early in the TD phase, the Americas CFS team completed a formal value-engineering review of the process water system. The value-engineering process involves the identification of major drivers of the design and identifying alternatives for evaluation and possible inclusion into the design to improve efficiency and reduce costs. One of the most important outcomes of the formal value-engineering review was the identification of specific extreme technical specifications that were complicating the design of the process water system. The CFS group worked with the various accelerator systems groups to relax some of the more stringent criteria of the identified specifications. In doing so a more uniform and simplified design was achieved and costs were reduced.

During the course of the TDP, the CFS group used a similar process to review all of the mechanical and electrical criteria provided by the various accelerator systems design teams. Improvements were identified in all areas and included into the formal mechanical- and electrical-design process in both the Americas and Asian Regions. This effort produced simplified designs and substantial cost savings. In each case the designs reflect the improved criteria and local conditions and were supported by independent consultant studies.

## 5.2 Descriptions of studies

The conventional facilities design for the TDR represents a much more complete effort in the Americas and Asian Regions. In all cases, the work represents significant improvement over the RDR design. The conventional facilities work to date provides a global approach to understanding the implications of site-specific design solutions and the relative cost for each of those solutions. The following is a listing of various consultant studies that were conducted in each of the three ILC Project Regions. These studies were conducted in the support of the work provided by the in-house Conventional facilities and Siting Group. They provide added depth and independent expertise for the overall Convention Facilities and Siting design effort.

### 5.2.1 Americas Region

#### 5.2.1.1 ILC Tunnel Configuration Study

The purpose of this study [308] was to analyse various options for main-linac tunnel configuration for the Americas Region sample site. The original RDR deep twin-tunnel configuration was used as the baseline, alternative configurations were considered and cost estimates were provided. The alternatives were based on a single tunnel using the KCS system and investigated both deep-tunnel and near-surface solutions.

#### 5.2.1.2 Tunnel Cross-Section Configuration Study

The purpose of the *Tunnel Cross-Section Configuration Study Technical Design Report* [309] was to evaluate tunnel-lining/ground-support systems; and systems to support equipment and utilities. Five tunnel-lining alternatives were compared to determine their relative cost and schedule. Supports for equipment and utilities were compared on the basis of cost. This study also compared a pre-cast concrete flooring system to a cast-in-place concrete floor.

#### 5.2.1.3 Black and Veach: ILC Constructibility Study

The Constructability Study [310] reviewed the current method of construction considered for the Americas Region sample site including both tunnel boring and drill and blast methods for various parts of the underground enclosures, shaft construction and location, muck removal, disposal and alternatives for the finished tunnel lining. Other aspects of the construction process including community considerations, impact of construction noise and traffic were also considered.

#### 5.2.1.4 Holabird and Root: Fermilab ILC Programming Study

The Surface Building study [311] developed a comprehensive plan for the size and arrangement for the surface buildings required for the Americas Region sample site using the KCS system. Plans were developed for the surface buildings required at both major and minor access shafts with respect to Klystron Service and Cryogenic Buildings.

#### 5.2.1.5 Americas Region: Life-Safety/Fire-Protection Analysis for the ILC

The *Life-Safety/Fire-Protection Analysis* [312] determines life-safety requirements for the ILC single-tunnel designs for both the 30 m- and 100 m-depth designs. The analysis was prepared in accordance with the National Fire Protection Association (NFPA) 520-2005 Standard for Subterranean Spaces [313]. NFPA 205 categorizes underground facilities using two designations: "building" and "common space". Building portions of subterranean spaces are areas that are occupied. Common space portions are all other areas.

#### 5.2.1.6 Americas Region: "Fire Egress Analysis for the ILC"

The *Fire and Egress Analysis for the International Linear Collider (ILC)* [314] provides a performance based analysis of the feasibility of the single-tunnel designs for the ILC from a life-safety standpoint. The basis for the analysis was the *Life-Safety/Fire-Protection Analysis* prepared by Hughes Associates, Inc., May 21, 2010 [312].

#### 5.2.1.7 ILC Process Water and Air Treatment VE Cost Evaluation

The Value Engineering (VE) effort [315] was implemented with the intent of understanding the process water cost-driver elements and providing for a lower first-cost alternative solution for the process water and its impact on other systems. The effort was started with the VE session facilitated by the United States Army Corps of Engineers, which produced a number of potential VE lists for evaluation. The VE items selected were evaluated and the resulting report summarised the results of the assessments in terms of a matrix of cost savings versus impacts. Applicable elements from the study were incorporated into the current baseline design.

#### 5.2.1.8 Mechanical Design and Cost Report

The Mechanical Design Study [316] provided the first comprehensive effort to develop a full design for the air-treatment equipment, piped utilities and process cooling water for the Americas Region sample site using the KCS scheme. Alternatives were evaluated and a cost-effective solution was developed.

#### 5.2.1.9 Electrical Design and Cost Report

The Electrical Design Study [317] provided the first comprehensive effort to develop a full design for the electrical power and distribution for the Americas Region sample site using KCS. Alternatives were evaluated and a cost effective solution was developed.

### 5.2.2 Asian Region

#### 5.2.2.1 AAA Report, "Investigating the Single Tunnel Proposal in a Japanese Mountainous Site"

The Advanced Accelerator Association Report [318] for tunnelling methods used for a mountain site was the first study completed for sites in the Asian Region. The study provided guidance for tunnel methods previously used for transportation tunnels in Japan's mountainous areas. It described primarily tunnel construction experience using tunnel-boring machines and methods used for dewatering tunnels in mountainous regions.

#### 5.2.2.2 J-Power Asian Tunnel Configuration Study (FY2010)

Construction costs and schedules are studied in eight cases for different combinations of high-level RF systems, tunnel configurations and excavation methods including TBM and NATM [319].

#### 5.2.2.3 J-Power Asian Tunnel Configuration Study (FY2011)

Case 8 in the FY2010 study was developed and revised [320]. A "Kamaboko" shape was selected for the Main Linac single-tunnel cross section and other tunnel and cavern designs were also updated.

#### 5.2.2.4 J-Power Asian Detector-Hall Study

A structural design, construction costs and schedule were described [321]. An analysis for structural behavior for both detectors, ILD (14.700 t) and SiD (8.600 t) was shown.

#### 5.2.2.5 Nikken-Sekkei Electrical and Mechanical Design Study

Electrical and Mechanical (process cooling-water system, piping system and air-ventilation system) designs were developed including an equipment layout in the underground caverns [322]. Construction cost estimates were also developed as part of this study.

#### 5.2.2.6 Asian Region Fire Safety for ILC Single Tunnel

Fire safety was reviewed for various tunnel configurations including double- and single-tunnel configurations in Americas, Asian and European sample sites [323].

### 5.2.3 European Region

#### 5.2.3.1 Site-Selection Criteria for European ILC Sites

The "Site Selection Criteria for European ILC Sites" [324] provided an overview of the site investigation work conducted in the European Region. Three sites, located at high energy physics laboratories, were considered as prospects for a European Region sample site and are described in this study. Sites included in the study were located at the DESY Laboratory in Hamburg, Germany, the JINR Laboratory in Dubna, Russia and the CERN Laboratory located near Geneva, Switzerland.

#### 5.2.3.2 Siting Study for European ILC Sites

The "Siting Study for European ILC Sites" [325] described a set of selection criteria for possible sites in the European Region. A general description of the overall parameters and site requirements for the construction of the ILC project was provided. In addition, geological properties, environmental impact issues, and electrical and cooling requirements were also identified.

#### 5.2.3.3 Dubna Site Investigation

This report [326] was developed using the information of the site specific geologic study that was conducted at the proposed ILC sample site near the JINR Laboratory in Dubna, Russia. Based on geologic information obtained, alternatives for the construction of the ILC tunnels and enclosures were considered.

#### 5.2.3.4 Report on the Results of the Preliminary Geological Engineering Surveys Along the Proposed Route of the International Linear Collider (ILC) in the Taldom Area of the Moscow Region

In conjunction with the Joint Institute for Nuclear Research (JINR), the Russian Governmental Unitary Enterprise and the State Specialised Projecting Institute conducted a geological study of the proposed European Sample site near Dubna, Russia [327]. Overall geologic conditions were identified and along the proposed ILC alignment, vertical electric soundings were conducted that identified optimum locations for three soil-boring samples. This study was a preliminary investigation for a possible European sample site.

#### 5.2.3.5 Linear Collider Interaction Region Design Studies. Review of Interaction Region Cavern Layout Design

This report [328] was completed in conjunction with the Americas Region. The purpose of the report was twofold. First, a computerised geotechnical model was developed using the ILC Interaction Region configuration and the criteria of the geology of the European sample site. This model provided an analysis of the anticipated movement to the interaction region due to the short-term and long-term effects of the excavation of the interaction region. Second, the study developed a solution for the platform and movement system to allow the two detectors planned for both the ILC and CLIC projects to move in and out of the interaction point in an alternating data-taking mode. Designs were provided for platforms for both detectors that met the criteria for both deflection and repositioning and alignment requirements during the "push-pull" operation of the two detectors.

#### 5.2.3.6 Guidelines and Criteria for an Environmental Impact Assessment for the Linear Collider Project at CERN

The report [329] is a preliminary manual for an environmental-impact statement (EIA) which will have to be conducted for the linear collider project. It describes the process for conducting an EIA and the environmental impact criteria that will have to be evaluated for an environmental impact study.

# Chapter 6
# Post-TDR R&D

## 6.1 The R&D Program

The realisation of the ILC requires the global high-energy physics community to agree on the high priority of the project. During the time required to prepare recommendations and await government actions, a global ILC core accelerator and technology team must remain in being and continue development of the project in collaboration with the physics community. As it has in the past, ILC R&D will continue to benefit other projects having significant technical or design overlaps. The ILC project team will continue its partnership with those working on such projects, ensuring that new developments can modify the ILC design as appropriate.

The post-TDR program will have a different flavour than the GDE R&D efforts, which were primarily aimed at demonstrating project viability. The next phase will be more closely coupled to preparing for a possible construction project and thus will be more D than R. This work will cover any remaining technical elements needing further work, such as the positron target. Some effort will be made to simplify component designs to streamline industrial production. It is not expected that the baseline design presented in this document will change significantly, although a site specific design will require customisation of the baseline concept in certain systems.

A small cavity R&D program will be maintained with the goal of increasing the effective accelerating gradient aimed at 1 TeV operation (see Section 2.3.4). Cavity improvements are essentially independent of the cryostat design and as such can be adopted at almost any point in the program. Cavity gradient R&D is viewed as an independent effort from preparation for a construction project.

The main missions of the post-TDR R&D effort will be:

1. Accelerator Design and Integration (AD&I) (with physics and detector groups);

   The AD&I team will further develop the machine design, including:
   a) development of (potentially phased) options for running at other energies;
   b) incorporating new R&D results, and programmatic synergies;
   c) analysing specific siting choices.

2. coordination of R&D on improving the performance and reducing the cost of the superconducting main linac;

3. continuation of the industrialisation programme for mass production of key technical components, especially cavities and cryomodules.

As during the Technical Design Phase, the R&D Program for the ILC is a central focus for the community and will also produce any technical information requested by the contracting governments in order to proceed to approval of the project.

The post-TDR program will feature closer collaboration with the CLIC project under the new Linear Collider Collaboration (LCC). Many elements of the CLIC activities such as the beam-delivery system pose similar challenges.

## 6.2 Accelerator Design and Integration

### 6.2.1 Physics Requirements

The ILC baseline Technical Design satisfies criteria published by the ILC Steering Committee in the 'Parameters for the Linear Collider' document in 2003 (updated in 2006) [330, 331]. In the next few years the LHC will provide insight into some of the processes the ILC is intended to study, which may alter the physics requirements. The ILC design, because of the maturity and high performance of its underlying technology, will be able to satisfy the needs of a broad range of collider physics, in terms of energy reach, luminosity, polarisation and energy-scan flexibility.

The recent discovery at the LHC of a Higgs particle at 125 GeV is highly significant for the ILC. It provides a guaranteed physics program, ranging from precision measurements of Higgs branching ratios via $ZH$ associated production, requiring a 250 GeV centre-of-mass energy, to measurements of $ZHH$ (the Higgs self-coupling), which requires at least 500 GeV. This scenario suggests that a phased energy approach to a linear collider could be attractive. This dove-tails well with the possibility of a 1 TeV energy upgrade. This was specified in the ILCSC parameters document as a secondary goal of the TDP. The machine implications of energy scaling in a phased fashion both up and down from the 500 GeV baseline design will be the highest priority for the AD&I program after the TDP.

The scientific requirements dictated in the ILCSC parameters document include references to options beyond the baseline linear collider. These include operation at the $Z$-boson resonance, an electron-electron collider, operation at the $WW$ threshold, and a polarised gamma-gamma collider. It appears unlikely that sufficient resources will be available to do significant work on any of these topics in the near future but should the physics program dictate, then AD&I priorities would change accordingly.

### 6.2.2 Programmatic Synergies

Few aspects of accelerator development have remained untouched by work done in the ILC programme in the last two decades. It is not surprising that this impact has been greatest on linac technology, as this is the primary cost-driver of the linear collider. However, further technological developments are to be expected from ongoing linear-accelerator projects, for example to provide high-intensity photon beams (XFEL) [6, 332, 333] or high-intensity proton beams (ESS [334, 335] and Project X [336]). Certainly, experience with large-scale production of SRF technology for the above projects will yield valuable lessons on how best to manufacture it for the ILC.

The impact of ILC R&D has also been large outside linac technology. Examples include control of collective effects (electron cloud), which has directly affected the B-factories Super KEK-B and Super B [337, 338] and implementation of precision optics, which has improved the performance of synchrotron-radiation sources. Both B factories have adopted electron-cloud mitigation schemes in their designs, and will provide valuable lessons for the ILC from the production, deployment and operation of this mitigation hardware. Precision optics, supported by multi-step correction sequences and ultra-high-resolution beam monitors, have been deployed at synchrotron-radiation sources, inspired by work on the linear-collider beam delivery. These developments have allowed these accelerators to produce beams with quantum-limited emittance [117]. The operation of these machines and the completion of new ones will deliver valuable experience with ultra-low-emittance tuning.

Other linear-collider beam-development work outside the ILC community is likely to yield information and experience valuable and relevant enough to be applied to the ILC design. Such work includes targetry, sources, feedback and low-emittance transport.

The knowledge gained from these developments must be understood and assimilated by the full AD&I team in order to translate it into improvements to the overall ILC design.

### 6.2.3 Site Studies

Studies of two specific ILC sites in Japan have started and will continue beyond the publication of the TDR. These possible sites have quite different characteristics from the sample sites considered during the preparation of the Reference Design (2007). They are in relatively remote hilly (or mountainous) locations; detailed ongoing dialog between the AD&I team and the site-study teams will be required to understand the implications. Planned site-development work includes geotechnical exploration and alignment optimisation. Both are intended to allow the tunnel and surface-building complex to be configured in the optimal, cost-effective manner in the rugged Japanese topography. While the deployment of ILC technology is largely site-independent, a few key systems, namely linac high-power RF and cryogenics, are not and these must be optimised together with their housing and support utilities. These SCRF linac subsystems are expensive and will require ongoing evaluation as site decisions proceed.

The topography of the sites suggests that the optimum location of utility equipment may be underground, since it may be cumbersome and intrusive to provide adequate surface facilities with appropriate easements. Underground location of high-power utilities, (power transformers, pumps, cryogen compressors etc.) has associated drawbacks that will require study by the AD&I team. Other sites that might be proposed will have specific characteristics that will require investigation.

## 6.3 Main-Linac Technical Components

The ILC Technical Design Phase has seen an unprecedented global technology transfer by the R&D teams who did pioneering work in the decade or so leading up to the ITRP choice of SCRF technology in 2004. In that decade, the TESLA Collaboration R&D on 1.3 GHz technology succeeded in reducing the cost per MeV by a large factor over the early 1990's state-of-the-art SCRF and demonstrated operation of a high-current pulsed SCRF linac. During the TDP, the ILC GDE established and deployed a reliable, industrial process for consistent production of 35 MV/m cavities worldwide. This critical step, implemented through the global GDE SCRF team, was the single most important action taken, enabling the community to move beyond an unreliable R&D fabrication process toward a mature technology suitable for a practical project plan. With mature infrastructure and industry and projects underway in each region, progress is expected to move at an increasing pace in the next few years. The role of the ILC team will be to coordinate and provide paths for communication between teams working on key aspects (gradient, Q0 and cost) of the technology.

### 6.3.1 Test Facilities – Superconducting Linac Technology

The most substantial and promising infrastructures built or under construction during the Technical Design Phase are the beam facilities: VUVFEL FLASH and XFEL (DESY), NML/ASTA (Fermilab) and STF (KEK).

Beam tests at FLASH were used to demonstrate main-linac parameters and show the effectiveness of the control schemes planned. The next step is to evaluate cost-performance trade-offs through the process of characterisation and analysis. There has not been enough time or resources to complete a main-linac technology value-engineering cycle using the new facilities. Given the R&D progress of the last few years, this effort is expected to be an important component of post–TDR work.

In Europe, the 20 GeV XFEL project is midway through construction and is expected to accelerate first beams in 2015 with almost 1000 TESLA-like cavities in operation. The scale of this production is a natural intermediate step between the recently completed technology transfer and the order-of-magnitude larger-scale production needed for ILC. Although the details are different, the XFEL beam-performance parameters, cavity and cryomodule-fabrication process overlap that foreseen for the ILC to an extent sufficient to allow experience to directly feed into the project.

In Asia, the STF and the Cavity Fabrication Facility at KEK uniquely provide in one location all the tools needed for significant advances in the standard production process, allowing studies and associated improvements of a sort not practical to explore in a construction project like the XFEL.

### 6.3.2 Cavity gradient

During the TDP, some 90 cavities were built for the purpose of demonstrating the ILC gradient. Of these, more than half were subjected to the defined standard process as described in Section 2.3.2, while the others were used for developing the process. Under the prioritised R&D guidance set by the ILC-GDE, the baseline cavity-processing recipe was fixed and improved in the so-called S0 study. The success of the study, (including the successful qualification of production and test infrastructures), has provided fresh insight into basic SCRF processes. With the ability to build high-performance cavities reliably, an R&D program aimed at higher gradients than in the baseline parameter set, based on new fabrication procedures, can be crafted. The techniques used reflect a sound understanding of the basic physics processes. These techniques will include studies of new welding technology, coating technology, new surface-preparation techniques and basic material parameters. Alternate cavity shapes will also be investigated.

### 6.3.3 Cryomodule

Around 10 cryomodules, three of which were assembled by teams outside Europe, were produced during the TDP. Many more will be made in the next three years as the roughly 100 cryomodules for the XFEL (DESY) are built and tested. In contrast to earlier cryomodule experience and to cavity-performance improvements, cryomodule performance suffers significantly from gradient degradation. It is presumed that this is associated with field emission due to contamination that happens during string assembly. Attention will be focused on this problem post-TDR, in each region.

### 6.3.4 Industrialisation

The pace of progress is evident in the fostering of cavity fabrication companies in each region. Four institutional cavity process-and-test facilities (one in Europe, two in the Americas and one in Asia), are actively providing ILC cavities fabricated by four companies (two in Europe, one in the Americas and one in Asia). This number will be roughly doubled soon after the TDP. Fully functional high-technology cavity-production capability in each region is mandatory for providing the ILC project with a strong global technology basis.

Up to now the industrialisation process has focused on cavity production. Post-TDR, the practicality of having cavities assemblies produced in industry will be examined. These assemblies, called dressed cavities, include a cavity together with its helium tank, 2-phase helium-supply line, the cold part of the high-power coupler, and the mount for the tuner. This step requires that dressed cavities can undergo the standard process and test cycle in a similar fashion to the bare cavities. Maximising the industrial content in a cryomodule will lead to cheaper and more reliable assembles.

## 6.4 Beam-Test Facilities

Beam tests at FLASH were used to demonstrate main-linac parameters and show the effectiveness of the control schemes planned. The next step is to evaluate cost-performance trade-offs through the process of characterisation and analysis. There has not been enough time or resources to complete a main-linac technology value-engineering cycle using the new facilities. Given the R&D progress of the last few years, this effort is expected to be an important component of post–TDR work.

Two large-scale beam-test facilities intended for the study of beam dynamics were built and commissioned during the TDP: the CESR Test Accelerator at Cornell [339] and the Accelerator Test Facility – BDS test 'ATF2' at KEK [204]. Both of these will remain active for ILC-related studies following the TDR. The CesrTA program in support of ILC is largely completed [8] and a smaller program will be targeted on small-emittance-beam techniques, but the work at ATF2 has been delayed by roughly a year due to damage caused by the Great East Japan Earthquake of 2011. The aim is to complete the ATF2 program.

The most noteworthy partnership in beam dynamics is with the CLIC Study. Although the two-beam CLIC linac design is quite different from the ILC superconducting linac, other aspects of the collider complex are very similar – in some cases almost identical. This includes for example, beam emittance, precision optics and collective effects.

### 6.4.1 Main Linac Technology - FLASH, XFEL, STF, NML

It is especially important that each region deploy a full superconducting-linac system, including cryomodules, beam generation and handling, and RF power source and distribution systems, to integrate the accelerator technology and gain sufficient experience in that region. NML at Fermilab and STF at KEK were conceived as ILC beam-test facilities for the Americas and Asian regions, respectively. During the Technical Design phase, both operated as single-cryomodule test facilities supported the S1 programme. Post TDR, both NML and STF should be upgraded with additional ILC-type cryomodules capable of ILC average gradients and subsequently be capable of accelerating ILC-like beams.

Continuing the programme of joint studies at FLASH will allow characterisation of operation at the limits of RF-power overhead and gradient margins to evaluate cost-performance and operability trade-offs. Experience at FLASH will be transferred to the European XFEL, which in turn will provide a wealth of experience for ILC.

### 6.4.2 Electron Cloud – Cesr Test Accelerator (CesrTA)

The next phase of the CesrTA program will involve an investigation of low-emittance beams. As is typical in precision storage rings, the low-emittance tuning procedure is limited by systematics associated with measurements of vertical dispersion. The CesrTA group is developing a beam-based technique for compensating for systematic errors in beam-position monitors. This, together with tools for measuring horizontal beam size, beam energy spread and bunch length, will allow the simultaneous measurement of vertical, horizontal and longitudinal phase space. This is essential to the study of emittance dilution.

Measurement and analysis of intra-beam scattering is an important component of the next phase of the CesrTA program. The instrumentation described above will provide a complete set of measurements of the equilibrium charge density. As that equilibrium has a strong beam-energy dependence, measurements over a range of energies will help to distinguish IBS from other emittance-diluting effects. Measurements with electron as well as positron beams will isolate contributions from ions and electron cloud.

CesrTA is also an excellent laboratory for investigating ion effects in electron beams, and in particular the fast ion instability in bunch trains in the ultra-low vertical-emittance regime. The turn-by-turn spectra gathered with the high-bandwidth beam-position and vertical-beam-size monitors for each bunch in a train can provide signatures of ion-beam coupling and emittance dilution. Measurement of instability thresholds as a function of vacuum pressure, both for positrons as well as electrons, will be used to isolate the effects of ions from other collective phenomena.

### 6.4.3 Beam Delivery – Accelerator Test Facility (ATF / ATF2)

The challenge of colliding nanometre-sized beams at the interaction point involves three distinct issues:

- creating small emittance beams in the damping ring;
- preserving the emittance during acceleration and transport;
- focusing and stabilising the beams to nanometres before colliding them.

These three issues are being addressed at the Accelerator Test Facility (ATF/ATF2) at KEK. The prototype damping ring (ATF) is used as a beam injector for the final-focus test beam line, ATF2, built and commissioned during the TDP, but with different beam-line optics based on a scheme of local chromaticity correction [198]. The purpose of ATF2 is to demonstrate and characterise the performance of this scheme.

Regarding the first of the three issues highlighted above, R&D work needed to support the damping-ring design includes the development of the fast injection/extraction kicker system and the beam-optics tuning to achieve 2 pm-rad extracted-beam vertical emittance.

A vertical emittance of 2 pm-rad has been achieved in several light-source storage rings around the world. However, typical beam sizes in these rings are small enough to make verifying low-emittance performance difficult and these measurements were made using indirect methods [117]. Therefore, a key goal to measure directly an extracted-beam emittance of 2 pm-rad or less remains. This can only be done at the ATF, which is the only low-emittance storage ring with a properly instrumented extraction line. An important challenge is to transport the beam through the extraction kickers and septum magnets without causing emittance growth due to $x$-$y$ coupling or wakefields.

The remaining two main goals of ATF2 are:

1. achieving the 37 nm vertical beam size of the ILC design;
2. stabilizing the beam at the nanometre level.

These goals were not achieved during the Technical Design Phase and work toward them is ongoing. Achieving the first goal requires the development and implementation of a variety of methods to validate the design optics in the presence of imperfections, in particular beam measurement and tuning techniques to cancel distortions of the beam phase space. For the second goal, aimed at characterising and improving beam-line stability, the collaboration pursues hardware developments of particular relevance to future linear colliders. These include long baseline ultra-high-precision laser-alignment schemes, low-latency high-precision beam feedback and iterative algorithms for beam-optics aberration tuning.

## 6.5 Remaining R&D at other facilities

### 6.5.1 Positron source

The ILC positron source must create positrons efficiently and reliably at a rate close to the existing state of the art. The R&D program during the TDP had a two-pronged aspect that included R&D on critical components for the baseline undulator-driven source and R&D on alternative source technologies that might be applied to an auxiliary source or, possibly, in different collider configurations.

Following the completion of the TDR, work will continue on target and pulsed-magnet flux-concentrator component testing for several years. Studies aimed toward superconducting undulators with shorter periods, perhaps using Nb3Sn conductor, are also underway and are expected to continue for several years.

The high-speed rotating-target mechanism and the tapered-solenoid flux-concentrator pulsed-magnet R&D studies include the construction of a fully functional test device capable of operating at nominal ILC parameters. Specifically, these tests will prove the performance of a high-vacuum rotary seal and a full-power flux concentrator.

### 6.5.2 Beam-Delivery System (BDS) and Machine-Detector Interface (MDI)

The primary BDS R&D goals, preserving beam emittance and focusing the beams to nanometres, are closely linked to the ATF2 Beam Test Facility program. In addition, specific studies for BDS technical components include the design and testing of the superconducting final-doublet focusing magnet and development of a design for the high-power beam dump [214]. The design and prototyping of a superconducting final doublet that includes the required specialised correction coils has been finished but laboratory testing remains to be done. The integrated design of the final doublet and detectors must also be completed.

# List of Signatories

The following list of signatories represents a comprehensive list of those people who have contributed to the R&D and design work, for both the accelerator and the detectors, which is summarised in this report. The list also includes those people who wish to indicate their support for the next phases of the worldwide ILC effort.

It should be noted that inclusion in this list does not indicate any formal commitment by the signatories. It does not indicate commitment to the specific detector designs presented, nor exclusive support for ILC over other collider programs.

A. Abada[171], T. Abe[24], T. Abe[236], J. M. Abernathy[379], H. Abramowicz[265], A. Abusleme[231], S. Aderhold[47], O. Adeyemi[333], E. Adli[357,251], C. Adloff[164], C. Adolphsen[251], K. Afanaciev[209], M. Aguilar[31], S. Ahmad[93], A. Ahmed[382], H. Aihara[375], R. Ainsworth[237,139], S. Airi[154], M. Aizatskyi[208], T. Akagi[73], M. Akemoto[71], A. Akeroyd[367], J. Alabau-Gonzalvo[108], C. Albajar[46], J. E. Albert[379], C. Albertus[281], J. Alcaraz Maestre[31], D. Alesini[174], B. Alessandro[128], G. Alexander[265], J. P. Alexander[43], A. Alhaidari[243], N. Alipour Tehrani[33], B. C. Allanach[323], O. Alonso[311], J. Alwall[210], J. W. Amann[251], Y. Amhis[167], M. S. Amjad[167], B. Ananthanarayan[83], A. Andreazza[386,122], N. Andreev[58], L. Andricek[186], M. Anduze[172], D. Angal-Kalinin[258], N. Anh Ky[106,394], K. A. Aniol[18], K. I. Aoki[148], M. Aoki[148], H. Aoyagi[137], S. Aoyama[250], S. J. Aplin[47], R. B. Appleby[343,40], J. Arafune[96], Y. Arai[71], S. Araki[71], L. Arazi[404], A. Arbey[50], D. Ariza[47], T. Arkan[58], N. D. Arnold[7], D. Arogancia[194], F. Arteche[113], A. Aryshev[71], S. Asai[375], T. Asaka[137], T. Asaka[212], E. Asakawa[220], M. Asano[333], F. B. Asiri[58], D. Asner[225], M. Asorey[286], D. Attié[21], J. E. Augustin[169], D. B. Augustine[58], C. S. Aulakh[226], E. Avetisyan[47], V. Ayvazyan[47], N. Azaryan[142], F. Azfar[358], T. Azuma[246], O. Bachynska[47], H. Baer[355], J. Bagger[141], A. Baghdasaryan[407], S. Bai[102], Y. Bai[384], I. Bailey[40,176], V. Balagura[172,107], R. D. Ball[329], C. Baltay[405], K. Bamba[198], P. S. Bambade[167], Y. Ban[9], E. Banas[268], H. Band[384], K. Bane[251], M. Barbi[362], V. Barger[384], B. Barish[17,65], T. Barklow[251], R. J. Barlow[392], M. Barone[58,65], I. Bars[368], S. Barsuk[167], P. Bartalini[210], R. Bartoldus[251], R. Bates[332], M. Battaglia[322,33], J. Baudot[94], M. Baylac[170], P. Bechtle[295], U. Becker[185,33], M. Beckmann[47], F. Bedeschi[126], C. F. Bedoya[31], S. Behari[58], O. Behnke[47], T. Behnke[47], G. Belanger[165], S. Belforte[129], I. Belikov[94], K. Belkadhi[172], A. Bellerive[19], C. Belver Aguilar[108], A. Belyaev[367,259], D. Benchekroun[184], M. Beneke[57], M. Benoit[33], A. Benot-Morell[33,108], S. Bentvelsen[213], L. Benucci[331], J. Berenguer[31], T. Bergauer[224], S. Berge[138], E. Berger[7], J. Berger[251], C. M. U. Berggren[47], Z. Bern[318], J. Bernabeu[108], N. Bernal[295], G. Bernardi[169], W. Bernreuther[235], M. Bertucci[119], M. Besancon[21], M. Bessner[47], A. Besson[94,306], D. R. Bett[358,140], A. J. Bevan[234], A. Bhardwaj[326], A. Bharucha[333], G. Bhattacharyya[241], B. Bhattacherjee[150], B. Bhuyan[86], M. E. Biagini[174], L. Bian[102], F. Bianchi[128], O. Biebel[182], T. R. Bieler[190], C. Biino[128], B. Bilki[7,337], S. S. Biswal[221], V. Blackmore[358,140], J. J. Blaising[164], N. Blaskovic Kraljevic[358,140], G. Blazey[217], I. Bloch[48], J. Bluemlein[48], B. Bobchenko[107], T. Boccali[126], J. R. Bogart[251], V. Boisvert[237], M. Bonesini[121], R. Boni[174], J. Bonnard[168], G. Bonneaud[169], S. T. Boogert[237,139], L. Boon[233,7],

G. Boorman[237,139], E. Boos[180], M. Boronat[108], K. Borras[47], L. Bortko[48], F. Borzumati[272], M. Bosman[294], A. Bosotti[119], F. J. Botella[108], S. Bou Habib[401], P. Boucaud[171], J. Boudagov[142], G. Boudoul[89], V. Boudry[172], D. Boumediene[168], C. Bourgeois[167], A. Boveia[52], A. Brachmann[251], J. Bracinik[315], J. Branlard[47,58], B. Brau[346], J. E. Brau[356], R. Breedon[320], M. Breidenbach[251], A. Breskin[404], S. Bressler[404], V. Breton[168], H. Breuker[33], C. Brezina[295], C. Briegel[58], J. C. Brient[172], T. M. Bristow[329], D. Britton[332], I. C. Brock[295], S. J. Brodsky[251], F. Broggi[119], G. Brooijmans[42], J. Brooke[316], E. Brost[356], T. E. Browder[334], E. Brücken[69], G. Buchalla[182], P. Buchholz[301], W. Buchmuller[47], P. Bueno[111], V. Buescher[138], K. Buesser[47], E. Bulyak[208], D. L. Burke[251], C. Burkhart[251], P. N. Burrows[358,140], G. Burt[40], E. Busato[168], L. Butkowski[47], S. Cabrera[108], E. Cabruja[32], M. Caccia[389,122], Y. Cai[251], S. S. Caiazza[47,333], O. Cakir[6], P. Calabria[89], C. Calancha[71], G. Calderini[169], A. Calderon Tazon[110], S. Callier[93], L. Calligaris[47], D. Calvet[168], E. Calvo Alamillo[31], A. Campbell[47], G. I. E. Cancelo[58], J. Cao[102], L. Caponetto[89], R. Carcagno[58], M. Cardaci[201], C. Carloganu[168], S. Caron[97,213], C. A. Carrillo Montoya[123], K. Carvalho Akiba[291], J. Carwardine[7], R. Casanova Mohr[311], M. V. Castillo Gimenez[108], N. Castro[175], A. Cattai[33], M. Cavalli-Sforza[294], D. G. Cerdeno[111], L. Cerrito[234], G. Chachamis[108], M. Chadeeva[107], J. S. Chai[260], D. Chakraborty[217], M. Champion[254], C. P. Chang[201], A. Chao[251], Y. Chao[210], J. Charles[28], M. Charles[358], B. E. Chase[58], U. Chattopadhyay[81], J. Chauveau[169], M. Chefdeville[164], R. Chehab[89], A. Chen[201], C. H. Chen[203], J. Chen[102], J. W. Chen[210], K. F. Chen[210], M. Chen[330,102], S. Chen[199], Y. Chen[1], Y. Chen[102], J. Cheng[102], T. P. Cheng[351], B. Cheon[66], M. Chera[47], Y. Chi[102], P. Chiappetta[28], M. Chiba[276], T. Chikamatsu[193], I. H. Chiu[210,210], G. C. Cho[220], V. Chobanova[186], J. B. Choi[36,36], K. Choi[157], S. Y. Choi[37], W. Choi[375,248], Y. I. Choi[260], S. Choroba[47], D. Choudhury[326], D. Chowdhury[83], G. Christian[358,140], M. Church[58], J. Chyla[105], W. Cichalewski[263,47], R. Cimino[174], D. Cinca[337], J. Clark[58], J. Clarke[258,40], G. Claus[94], E. Clement[316,259], C. Clerc[172], J. Cline[187], C. Coca[206], T. Cohen[251], P. Colas[21], A. Colijn[213], N. Colino[31], C. Collard[94], C. Colledani[94], N. Collomb[258], J. Collot[170], C. Combaret[89], B. Constance[33], C. A. Cooper[58], W. E. Cooper[58], G. Corcella[174], E. Cormier[29], R. Cornat[172], P. Cornebise[167], F. Cornet[281], G. Corrado[123], F. Corriveau[187], J. Cortes[286], E. Cortina Gil[303], S. Costa[308], F. Couchot[167], F. Couderc[21], L. Cousin[94], R. Cowan[185], W. Craddock[251], A. C. Crawford[58], J. A. Crittenden[43], J. Cuevas[283], D. Cuisy[167], F. Cullinan[237], B. Cure[33], E. Currás Rivera[110], D. Cussans[316], J. Cvach[105], M. Czakon[235], K. Czuba[401], H. Czyz[365], J. D'Hondt[399], W. Da Silva[169], O. Dadoun[167], M. Dahiya[327], J. Dai[102], C. Dallapiccola[346], C. Damerell[259], M. Danilov[107], D. Dannheim[33], N. Dascenzo[47,238], S. Dasu[384], A. K. Datta[67], T. S. Datta[115], P. Dauncey[80], T. Davenne[259], J. David[169], M. Davier[167], W. De Boer[90], S. De Cecco[169], S. De Curtis[120], N. De Groot[97,213], P. De Jong[213], S. De Jong[97,213], C. De La Taille[93], G. De Lentdecker[307], S. De Santis[177], J. B. De Vivie De Regie[167], A. Deandrea[89], P. P. Dechant[49], D. Decotigny[172], K. Dehmelt[257], J. P. Delahaye[251,33], N. Delerue[167], O. Delferriere[21], F. Deliot[21], G. Della Ricca[388], P. A. Delsart[170], M. Demarteau[7], D. Demin[142], R. Dermisek[87], F. Derue[169], K. Desch[295], S. Descotes-Genon[171], A. Deshpande[252], A. Dexter[40], A. Dey[81], S. Dhawan[405], N. Dhingra[226], V. Di Benedetto[58,123], B. Di Girolamo[33], M. A. Diaz[231], A. Dieguez[311], M. Diehl[47], R. Diener[47], S. Dildick[331], M. O. Dima[206], P. Dinaucourt[167,93], M. S. Dixit[19], T. Dixit[252], L. Dixon[251], A. Djouadi[171], S. Doebert[33], M. Dohlus[47], Z. Dolezal[34], H. Dong[102], L. Dong[102], A. Dorokhov[94], A. Dosil[112], A. Dovbnya[208], T. Doyle[332], G. Doziere[94], M. Dragicevic[224], A. Drago[174], A. J. Dragt[345], Z. Drasal[34], I. Dremin[179], V. Drugakov[209], J. Duarte Campderros[110], F. Duarte Ramos[33], A. Dubey[272], A. Dudarev[142], E. Dudas[171,171], L. Dudko[180], C. Duerig[47], G. Dugan[43], W. Dulinski[94], F. Dulucq[93], L. Dumitru[206], P. J. Dunne[80], A. Duperrin[27], M. Düren[147], D. Dzahini[170], H. Eberl[224], G. Eckerlin[47], P. Eckert[297], N. R. Eddy[58],

W. Ehrenfeld[295], G. Eigen[314], S. Eisenhardt[329], L. Eklund[332], L. Elementi[58], U. Ellwanger[171], E. Elsen[47], I. Emeliantchik[209], L. Emery[7], K. Enami[71], K. Endo[71], M. Endo[375], J. Engels[47], C. Englert[49], S. Eno[345], A. Enomoto[71], S. Enomoto[197], F. Eozenou[21], R. Erbacher[320], G. Eremeev[269], J. Erler[287], R. Escribano[294], D. Esperante Pereira[108], D. Espriu[311], E. Etzion[265,33], S. Eucker[47], A. Evdokimov[336,107], E. Ezura[71], B. Faatz[47], G. Faisel[201], L. Fano[125], A. Faraggi[340], A. Fasso[251], A. Faus-Golfe[108], L. Favart[307], N. Feege[257], J. L. Feng[321], T. Ferber[47], J. Ferguson[33], J. Fernández[283], P. Fernández Martínez[108], E. Fernandez[294,293], M. Fernandez Garcia[110], J. L. Fernandez-Hernando[130], P. Fernandez-Martinez[32], J. Fernandez-Melgarejo[33], A. Ferrer[108], F. Ferri[21], S. Fichet[117], T. Fifield[347], L. Filkov[179], F. Filthaut[97,213], A. Finch[176], H. E. Fisk[58], T. Fiutowski[2], H. Flaecher[316], J. W. Flanagan[71], I. Fleck[301], M. Fleischer[47], C. Fleta[32], J. Fleury[93], D. Flores[32], M. Foley[58], M. Fontannaz[171], K. Foraz[33], N. Fornengo[128], F. Forti[126,391], B. Foster[47,140], M. C. Fouz[31], P. H. Frampton[353], K. Francis[7], S. Frank[224], A. Freitas[361], A. Frey[64], R. Frey[356], M. Friedl[224], C. Friedrich[48], M. Frigerio[163], T. Frisson[167], M. Frotin[172], R. Frühwirth[224], R. Fuchi[378], E. Fuchs[47], K. Fujii[71], J. Fujimoto[71], H. Fuke[134], B. Fuks[94,33], M. Fukuda[71], S. Fukuda[71], K. Furukawa[71], T. Furuya[71], T. Fusayasu[195], J. Fuster[108], N. Fuster[108], Y. Fuwa[95,159], A. Gaddi[33], K. Gadow[47], F. Gaede[47], R. Gaglione[164], S. Galeotti[126], C. Gallagher[356], A. A. Gallas Torreira[112], L. Gallin-Martel[170], A. Gallo[174], D. Gamba[140,33], D. Gamba[128], J. Gao[102], Y. Gao[24], P. H. Garbincius[58], F. Garcia[69], C. Garcia Canal[288], J. E. Garcia Navarro[108], P. Garcia-Abia[31], J. J. Garcia-Garrigos[108], L. Garcia-Tabares[31], C. García[108], J. V. García Esteve[286], I. García García[108], S. K. Garg[408], L. Garrido[311], E. Garutti[333], T. Garvey[167,261], M. Gastal[33], F. Gastaldi[172], C. Gatto[58,123], N. Gaur[326], D. Gavela Pérez[31], P. Gay[168], M. B. Gay Ducati[109], L. Ge[251], R. Ge[102], A. Geiser[47], A. Gektin[100], A. Gellrich[47], M. H. Genest[170], R. L. Geng[269], S. Gentile[387,127], A. Gerbershagen[33,358], R. Gerig[7], S. German[111], H. Gerwig[33], S. Ghazaryan[47], P. Ghislain[169], D. K. Ghosh[81], S. Ghosh[115], S. Giagu[387,127], L. Gibbons[43], S. Gibson[139,33], V. Gilewsky[143], A. Gillespie[328], F. Gilman[20], B. Gimeno Martínez[108], D. M. Gingrich[312,279], C. M. Ginsburg[58], D. Girard[164], J. Giraud[170], G. F. Giudice[33], L. Gladilin[180], P. Gladkikh[208], C. J. Glasman[343,40], R. Glattauer[224], N. Glover[49], J. Gluza[365], K. Gnidzinska[263], R. Godbole[83], S. Godfrey[19], F. Goertz[54], M. Goffe[94], N. Gogitidze[179,47], J. Goldstein[316], B. Golob[144,341], G. Gomez[110], V. Goncalves[290], R. J. Gonsalves[256], I. González[283], S. González De La Hoz[108], F. J. González Sánchez[110], G. Gonzalez Parra[294], S. Gopalakrishna[103], I. Gorelov[352], D. Goswami[86], S. Goswami[229], T. Goto[71], K. Gotow[398], P. Göttlicher[47], M. Götze[385], A. Goudelis[165], P. Goudket[258], S. Gowdy[33], O. A. Grachov[309], N. A. Graf[251], M. Graham[251], A. Gramolin[15], R. Granier De Cassagnac[172], P. Grannis[257], P. Gras[21], M. Grecki[47], T. Greenshaw[339], D. Greenwood[181], C. Grefe[33], M. Grefe[111], I. M. Gregor[47], D. Grellscheid[49], G. Grenier[89], M. Grimes[316], C. Grimm[58], O. Grimm[53], B. Grinyov[100], B. Gripaios[323], K. Grizzard[141], A. Grohsjean[47], C. Grojean[294,33], J. Gronberg[178], D. Grondin[170], S. Groote[370], P. Gros[240], M. Grunewald[310], B. Grzadkowski[382], J. Gu[102], M. Guchait[262], S. Guiducci[174], E. Guliyev[172], J. Gunion[320], C. Günter[47], C. Gwilliam[339], N. Haba[75], H. Haber[322], M. Hachimine[197], Y. Haddad[172], L. Hagge[47], M. Hagihara[378], K. Hagiwara[71,158], J. Haley[216], G. Haller[251], J. Haller[333], K. Hamaguchi[375], R. Hamatsu[276], G. Hamel De Monchenault[21], L. L. Hammond[58], P. Hamnett[47], L. Han[364], T. Han[361], K. Hanagaki[223], J. D. Hansen[211], K. Hansen[47], P. H. Hansen[211], X. Q. Hao[70], K. Hara[71], K. Hara[378], T. Hara[71], D. Harada[83], K. Harada[161], K. Harder[259], T. Harion[297], R. V. Harlander[385], E. Harms[58], M. Harrison[13], O. Hartbrich[47,385], A. Hartin[47], T. Hartmann[300], J. Harz[47], S. Hasegawa[197], T. Hasegawa[71], Y. Hasegawa[248], M. Hashimoto[38], T. Hashimoto[62], C. Hast[251], S. Hatakeyama[135], J. M. Hauptman[118], M. Hauschild[33], M. Havranek[105], C. Hawkes[315], T. Hayakawa[197], H. Hayano[71], K. Hayasaka[198], M. Hazumi[71,253], H. J. He[24], C. Hearty[317,104], H. F. Heath[316],

T. Hebbeker[235], M. Heck[90], V. Hedberg[183], D. Hedin[217], S. M. Heindl[90], S. Heinemeyer[110], I. Heinze[47], A. Hektor[205], S. Henrot-Versille[167], O. Hensler[47], A. Heo[23], J. Herbert[258], G. Herdoiza[138], B. Hermberg[47], J. J. Hernández-Rey[108], M. J. Herrero[111], B. Herrmann[165], A. Hervé[384], J. Hewett[251], S. Hidalgo[32], B. Hidding[333,318], N. Higashi[375], N. Higashi[71], T. Higo[71], E. Higón Rodríguez[108], T. Higuchi[150], M. Hildreth[354], C. T. Hill[58], S. Hillert[295], S. Hillier[315], T. Himel[251], A. Himmi[94], S. Himori[272], Z. Hioki[374], B. Hippolyte[94], T. Hiraki[159], K. Hirano[136], S. Hirano[197], K. Hirata[71], T. Hirose[276], M. Hirsch[108], J. Hisano[197], P. M. Ho[210], A. Hoang[302], A. Hocker[58], A. Hoecker[33], M. Hoeferkamp[352], M. Hoffmann[47], W. Hollik[186], K. Homma[72], Y. Homma[154], S. Honda[378], T. Honda[71], Y. Honda[71], N. T. Hong Van[106], K. Honkavaara[47], T. Honma[71], T. Hori[236], T. Horiguchi[272], Y. Horii[197], A. Horio[196], R. Hosaka[377], Y. Hoshi[271], H. Hoshino[197], K. Hosoyama[71], J. Y. Hostachy[170], G. W. Hou[210], M. Hou[102], A. Hoummada[184], M. S. Hronek[58], T. Hu[102], C. Hu-Guo[94], M. Huang[24], T. Huang[102], E. Huedem[58], F. Hügging[295], J. L. Hugon[330], C. Hugonie[173], K. Huitu[335], P. Q. Hung[380], C. Hunt[80], U. Husemann[90], G. Hussain[24], D. Hutchcroft[339], Y. Hyakutake[79], J. C. Ianigro[89], L. E. Ibanez[111], M. Ibe[96], M. Idzik[2], H. Igarashi[78], Y. Igarashi[71], K. Igi[236], A. Ignatenko[209], O. Igonkina[213], T. Iijima[198,197], M. Iinuma[73], Y. Iiyama[20], H. Ikeda[134], K. Ikeda[71], K. Ikematsu[301], J. I. Illana[281], V. A. Ilyin[207,180], A. Imhof[333], T. Inagaki[236], T. Inagaki[197], K. Inami[197], S. Inayoshi[248], K. Inoue[161], A. Irles[108], S. Isagawa[71], N. Ishibashi[378], A. Ishida[375], K. Ishida[212], N. Ishihara[71], S. Ishihara[77], K. Ishii[71], A. Ishikawa[272], K. Ishikawa[375], K. I. Ishikawa[72], K. Ishikawa[75], T. Ishikawa[71], M. Ishitsuka[275], K. Ishiwata[17], G. Isidori[174], A. Ismail[251], S. Iso[71], T. Isogai[197], C. Issever[358], K. Itagaki[272], T. Itahashi[223], A. Ito[275], F. Ito[378], S. Ito[272], R. Itoh[71], E. Itou[71], M. I. Ivanyan[26], G. Iwai[71], S. Iwamoto[375], T. Iwamoto[116], H. Iwasaki[71], M. Iwasaki[71], Y. Iwashita[95], S. Iwata[71], S. Iwata[276], T. Izubuchi[13,236], Y. Izumiya[272], S. Jablonski[401], F. Jackson[258], J. A. Jacob[316], M. Jacquet[167], P. Jain[40], P. Jaiswal[59], W. Jalmuzna[263], E. Janas[401], R. Jaramillo Echeverría[110], J. Jaros[251], D. Jatkar[67], D. Jeans[375], R. Jedziniak[58], M. J. Jenkins[176,40], K. Jensch[47], C. P. Jessop[354], T. Jezynski[47], M. Jimbo[35], S. Jin[102], O. Jinnouchi[275], M. D. Joergensen[211], A. S. Johnson[251], S. Jolly[309], D. T. Jones[340], J. Jones[258,40], R. M. Jones[343,40], L. Jönsson[183], N. Joshi[237], C. K. Jung[257,150], N. Juntong[343,40], A. Juste[88,294], W. Kaabi[167], M. Kadastik[205], M. Kado[167,33], K. Kadota[197], E. Kajfasz[27], R. Kajikawa[197], Y. Kajiura[197], M. Kakizaki[377], E. Kako[71], H. Kakuhata[377], H. Kakuno[276], A. Kalinin[258], J. Kalinowski[382], G. E. Kalmus[259], K. Kamada[47], J. Kaminski[295], T. Kamitani[71], Y. Kamiya[116], Y. Kamiya[71], R. Kammering[47], T. Kamon[266], J. I. Kamoshita[60], T. Kanai[275], S. Kananov[265], K. Kanaya[378], M. Kaneda[33], T. Kaneko[71,253], S. Kanemura[377], K. Kaneta[75], W. Kang[102], D. Kanjilal[115], K. Kannike[205], F. Kapusta[169], D. Kar[332], P. Karataev[237,139], P. E. Karchin[403], D. Karlen[379,279], S. Karstensen[47], Y. Karyotakis[164], M. Kasemann[47], V. S. Kashikhin[58], S. Kashiwagi[273], A. Kataev[98], V. Katalev[48], Y. Kataoka[116], N. Katayama[150], R. Katayama[375], E. Kato[272], K. Kato[155], S. Kato[71], Y. Kato[153], T. Katoh[71], A. Kaukher[47], S. Kawabata[71], S. I. Kawada[73], K. Kawagoe[161], M. Kawai[71], T. Kawamoto[116], H. Kawamura[71], M. Kawamura[71], Y. Kawamura[248], S. Kawasaki[71], T. Kawasaki[212], H. Kay[47], S. Kazama[375], L. Keegan[111], J. Kehayias[150], L. Keller[251], M. A. Kemp[251], J. J. Kempster[237], C. Kenney[251], I. Kenyon[315], R. Kephart[58], J. Kerby[7], K. Kershaw[33], J. Kersten[333], K. Kezzar[151], V. G. Khachatryan[26], M. A. Khan[23], S. A. Khan[242], Y. Khoulaki[184], V. Khoze[49], H. Kichimi[71], R. Kieffer[33], C. Kiesling[186], M. Kikuchi[377], Y. Kikuta[71], M. Killenberg[47], C. S. Kim[408], D. W. Kim[63], D. Kim[23], E. J. Kim[36,36], E. S. Kim[23], G. Kim[23], H. S. Kim[23], H. D. Kim[245], J. Kim[63], S. H. Kim[378], S. K. Kim[245], S. G. Kim[87], Y. I. Kim[358,140], Y. Kimura[71], R. E. Kirby[251], F. Kircher[21], Y. Kishimoto[96], L. Kisslinger[20], T. Kitahara[375], R. Kitano[272], Y. Kiyo[146,71], C. Kleinwort[47], W. Klempt[33], P. M. Kluit[213], V. Klyukhin[180,33], M. Knecht[28], J. L. Kneur[163], B. A. Kniehl[333], K. Ko[251], P. Ko[158], D. Kobayashi[275], M. Kobayashi[71], N. Kobayashi[71],

T. Kobayashi[116], M. Koch[295], P. Kodys[34], U. Koetz[47], E. N. Koffeman[213], M. Kohda[210], S. Koike[71], Y. Kojima[71], K. Kolodziej[365], Y. Kolomensky[319,177], S. Komamiya[375], T. Kon[244], P. Konar[229], Y. Kondou[71], D. Kong[23], K. Kong[338], O. C. Kong[201], T. Konno[275,275], V. Korbel[47], J. G. Körner[138], S. Korpar[344,144], S. R. Koscielniak[279], D. Kostin[47], K. Kotera[248], W. Kotlarski[382], J. Kotula[268], E. Kou[167], V. Kovalenko[333], S. V. H. Kox[170], K. Koyama[75], M. Krämer[235], S. Kraml[170], M. Krammer[224], M. W. Krasny[169], F. Krauss[49], T. Krautscheid[295], M. Krawczyk[382], K. Krempetz[58], P. Križan[341,144], B. E. Krikler[80], A. Kronfeld[58], K. Kruchinin[237,139], D. Krücker[47], K. Krüger[47], B. Krupa[268], Y. P. Kuang[24], K. Kubo[71], T. Kubo[71], T. Kubota[347], T. Kubota[275], Y. Kubyshin[298,180], V. Kuchler[58], I. M. Kudla[202], D. Kuehn[47], J. H. Kuehn[92], C. Kuhn[94], S. Kulis[2], S. Kulkarni[170], A. Kumar[10], S. Kumar[86], T. Kumita[276], A. Kundu[16], Y. Kuno[223], C. M. Kuo[201], M. Kurachi[198], A. Kuramoto[253], M. Kurata[375], Y. Kurihara[71], M. Kuriki[73,71], T. Kurimoto[377], S. Kuroda[71], K. Kurokawa[71], S. I. Kurokawa[71], H. Kuwabara[276], M. Kuze[275], J. Kvasnicka[105], P. Kvasnicka[34], Y. Kwon[408], L. Labun[210], C. Lacasta[108], T. Lackowski[58], D. Lacour[169], V. Lacuesta[108], R. Lafaye[164], G. Lafferty[343], B. Laforge[169], I. Laktineh[89], R. L. Lander[320], K. Landsteiner[111], S. Laplace[169], K. J. Larsen[213], R. S. Larsen[251], T. Lastovicka[105], J. I. Latorre[311], S. Laurien[333], L. Lavergne[169], S. Lavignac[22], R. E. Laxdal[279], A. C. Le Bihan[94], F. R. Le Diberder[167], A. Le-Yaouanc[171], A. Lebedev[13], P. Lebrun[33], T. Lecompte[7], T. Leddig[300], F. Ledroit[170], B. Lee[25], K. Lee[158], M. Lee[177], S. H. Lee[260], S. W. Lee[267], Y. H. Lee[210], J. Leibfritz[58], K. Lekomtsev[71], L. Lellouch[28], M. Lemke[47], F. R. Lenkszus[7], A. Lenz[49,33], O. Leroy[27], C. Lester[323], L. Levchuk[208], J. Leveque[164], E. Levichev[15], A. Levy[265], I. Levy[265], J. R. Lewandowski[251], B. Li[24], C. Li[364], C. Li[102], D. Li[102], H. Li[380], L. Li[195], L. Li[247], L. Li[364], S. Li[102], W. Li[102], X. Li[102], Y. Li[24], Y. Li[24], Y. Li[24], Z. Li[251], Z. Li[102], J. J. Liau[210], V. Libov[47], L. Lilje[47], J. G. Lima[217], C. J. D. Lin[204], C. M. Lin[154], C. Y. Lin[201], H. Lin[102], H. H. Lin[210], F. L. Linde[213], R. A. Lineros[108], L. Linssen[33], R. Lipton[58], M. Lisovyi[47], B. List[47], J. List[47], B. Liu[24], J. Liu[364], R. Liu[102], S. Liu[167], S. Liu[247], W. Liu[7], Y. Liu[102], Y. Liu[337,58], Z. Liu[361], Z. Liu[102], Z. Liu[102], A. Lleres[170], N. S. Lockyer[279,317], W. Lohmann[48,12], E. Lohrmann[333], T. Lohse[76], F. Long[102], D. Lontkovskyi[47], M. A. Lopez Virto[110], X. Lou[102,372], A. Lounis[167], M. Lozano Fantoba[32], J. Lozano-Bahilo[281], C. Lu[232], R. S. Lu[210], S. Lu[47], A. Lucotte[170], F. Ludwig[47], S. Lukic[396], O. Lukina[180], N. Lumb[89], B. Lundberg[183], A. Lunin[58], M. Lupberger[295], B. Lutz[47], P. Lutz[21], T. Lux[294], K. Lv[102], M. Lyablin[142], A. Lyapin[237,139], J. Lykken[58], A. T. Lytle[262], L. Ma[258], Q. Ma[102], R. Ma[312], X. Ma[102], F. Machefert[167], N. Machida[377], J. Maeda[276], Y. Maeda[159], K. Maeshima[58], F. Magniette[172], N. Mahajan[229], F. Mahmoudi[168,33], S. H. Mai[201], C. Maiano[119], H. Mainaud Durand[33], S. Majewski[356], S. K. Majhi[81], N. Majumdar[241], G. Majumder[262], I. Makarenko[47], V. Makarenko[209], A. Maki[71], Y. Makida[71], D. Makowski[263], B. Malaescu[169], J. Malcles[21], U. Mallik[337], S. Malvezzi[121], O. B. Malyshev[258,40], Y. Mambrini[171], A. Manabe[71], G. Mancinelli[27], S. K. Mandal[150], S. Mandry[309,186], S. Manen[168], R. Mankel[47], S. Manly[363], S. Mannai[303], Y. Maravin[149], G. Marchiori[169], M. Marcisovsky[105,45], J. Marco[110], D. Marfatia[338], J. Marin[31], E. Marin Lacoma[251], C. Marinas[295], T. W. Markiewicz[251], O. Markin[107], J. Marshall[323], S. Martí-García[108], A. D. Martin[49], V. J. Martin[329], G. Martin-Chassard[93], T. Martinez De Alvaro[31], C. Martinez Rivero[110], F. Martinez-Vidal[108], H. U. Martyn[235,47], T. Maruyama[251], A. Masaike[159], T. Mashimo[116], T. Masubuchi[116], T. Masuda[159], M. Masuzawa[71], Z. Mateusz[401], A. Matheisen[47], H. Mathez[89], J. Matias[293], H. Matis[177], T. Matsubara[276], T. Matsuda[71], T. Matsui[377], S. Matsumoto[161], S. Matsumoto[150], Y. Matsumoto[220], H. Matsunaga[71], T. Matsushita[154], T. S. Mattison[317], V. A. Matveev[142], U. Mavric[47], G. Mavromanolakis[33], K. Mawatari[399], S. J. Maxfield[339], K. Mazumdar[262], A. Mazzacane[58,123], R. L. Mccarthy[257], D. J. Mccormick[251], J. Mccormick[251], K. T. Mcdonald[232], R. Mcduffee[324], P. Mcintosh[258], B. Mckee[251], M. Medinnis[47], S. Mehlhase[211], T. Mehrling[47,333],

A. Mehta[339], B. Mele[127], R. E. Meller[43], I. A. Melzer-Pellmann[47], L. Men[102], G. Mendiratta[83], Z. Meng[316], M. H. Merk[213,400], M. Merkin[180], A. Merlos[32], L. Merminga[279], A. B. Meyer[47], A. Meyer[235], N. Meyners[47], Z. Mi[102], P. Michelato[119], S. Michizono[71], S. Mihara[71], A. Mikhailichenko[43], D. J. Miller[309], C. Milstene[403], Y. Mimura[210], D. Minic[398], L. Mirabito[89], S. Mishima[387], T. Misumi[13], W. A. Mitaroff[224], T. Mitsuhashi[71], S. Mitsuru[71], K. Miuchi[154], K. Miyabayashi[200], A. Miyamoto[71], H. Miyata[212], Y. Miyazaki[161], T. Miyoshi[71], R. Mizuk[107], K. Mizuno[3], U. Mjörnmark[183], J. Mnich[47], G. Moeller[47], W. D. Moeller[47], K. Moenig[48], K. C. Moffeit[251], P. Mohanmurthy[269], G. Mohanty[262], L. Monaco[119], S. Mondal[81], C. Monini[170], H. Monjushiro[71], G. Montagna[359,124], S. Monteil[168], G. Montoro[298], I. Montvay[47], F. Moortgat[53], G. Moortgat-Pick[333,47], P. Mora De Freitas[172], C. Mora Herrera[30], G. Moreau[171], F. Morel[94], A. Morelos-Pineda[280], M. Moreno Llacer[108], S. Moretti[367,259], V. Morgunov[47,107], T. Mori[71], T. Mori[272], T. Mori[116], Y. Morita[71], S. Moriyama[96,150], L. Moroni[121], Y. Morozumi[71], H. G. Moser[186], A. Moszczynski[268], K. Motohashi[275], T. Moulik[249], G. Moultaka[163], D. Moya Martin[110], S. K. Mtingwa[215], G. S. Muanza[27], M. Mühlleitner[91], A. Mukherjee[85], S. Mukhopadhyay[241], M. Mulders[33], D. Muller[251], F. Müller[47], T. Müller[90], V. S. Mummidi[83], A. Münnich[47], C. Munoz[111], F. J. Muñoz Sánchez[110], H. Murayama[375,319], R. Murphy[7], G. Musat[172], A. Mussgiller[47], R. Muto[71], T. Nabeshima[377], K. Nagai[71], K. Nagai[378], S. Nagaitsev[58], T. Nagamine[272], K. Nagano[71], K. Nagao[71], Y. Nagashima[223], S. C. Nahn[185], P. Naik[316], D. Naito[159], T. Naito[71], H. Nakai[71], K. Nakai[71,223], Y. Nakai[161], Y. Nakajima[177], E. Nakamura[71], H. Nakamura[71], I. Nakamura[71], S. Nakamura[159], S. Nakamura[71], T. Nakamura[116], K. Nakanishi[71], E. Nakano[222], H. Nakano[212], H. Nakano[272], Y. Namito[71], W. Namkung[230], H. Nanjo[159], C. D. Nantista[251], O. Napoly[21], Y. Nara[4], T. Narazaki[272], S. Narita[131], U. Nauenberg[324], T. Naumann[48], S. Naumann-Emme[47], J. Navarro[108], A. Navitski[47], H. Neal[251], K. Negishi[272], K. Neichi[270], C. A. Nelson[255], T. K. Nelson[251], S. Nemecek[105], M. Neubert[138], R. Neuhaus[218], L. J. Nevay[237], D. M. Newbold[316,259], O. Nezhevenko[58], F. Nguyen[175], M. Nguyen[172], M. N. Nguyen[251], T. T. Nguyen[106], R. B. Nickerson[358], O. Nicrosini[124], C. Niebuhr[47], J. Niehoff[58], M. Niemack[43], U. Nierste[92], H. Niinomi[159], I. Nikolic[169], H. P. Nilles[295], S. Nishida[71], H. Nishiguchi[71], K. Nishiwaki[67], O. Nitoh[277], L. Niu[24], R. Noble[251], M. Noji[71], M. Nojiri[71,150], S. Nojiri[197,198], D. Nölle[47], A. Nomerotski[358], M. Nomura[135], T. Nomura[201], Y. Nomura[319,177], C. Nonaka[198], J. Noonan[7], E. Norbeck[337], Y. Nosochkov[251], D. Notz[47], O. Novgorodova[48,12], A. Novokhatski[251], J. A. Nowak[349], M. Nozaki[71], K. Ocalan[192], J. Ocariz[169], S. Oda[161], A. Ogata[71], T. Ogawa[248], T. Ogura[248], A. Oh[343], S. K. Oh[156], Y. Oh[23], K. Ohkuma[61], T. Ohl[145], Y. Ohnishi[71], K. Ohta[189], M. Ohta[71,253], S. Ohta[71,253], N. Ohuchi[71], K. Oishi[161], R. Okada[71], Y. Okada[71,253], T. Okamura[71], H. Okawa[13], T. Okugi[71], T. Okui[59], K. I. Okumura[161], Y. Okumura[52,58], L. Okun[107], H. Okuno[236], C. Oleari[390], C. Oliver[31], B. Olivier[186], S. L. Olsen[245], M. Omet[253,71], T. Omori[71], Y. Onel[337], H. Ono[214], Y. Ono[272], D. Onoprienko[251], Y. Onuki[116,150], P. Onyisi[371], T. Oogoe[71], Y. Ookouchi[159], W. Ootani[116], M. Oreglia[52], M. Oriunno[251], M. C. Orlandea[206], J. Orloff[168], M. Oroku[375,71], R. S. Orr[376], J. Osborne[33], A. Oskarsson[183], P. Osland[314], A. Osorio Oliveros[282], L. Österman[183], H. Otono[223], M. Owen[343], Y. Oyama[71], A. Oyanguren[108], K. Ozawa[71,375], J. P. Ozelis[58,191], D. Ozerov[47], G. Pásztor[304,99], H. Padamsee[43], C. Padilla[294], C. Pagani[119,386], R. Page[316], R. Pain[169], S. Paktinat Mehdiabadi[101], A. D. Palczewski[269], S. Palestini[33], F. Palla[126], M. Palmer[58], F. Palomo Pinto[285], W. Pan[102], G. Pancheri[174], M. Pandurovic[396], O. Panella[125], A. Pankov[227], Y. Papaphilippou[33], R. Paparella[119], A. Paramonov[7], E. K. Park[75], H. Park[23], S. I. Park[23], S. Park[260], S. Park[373], W. Park[23], A. Parker[323], B. Parker[13], C. Parkes[343], V. Parma[33], Z. Parsa[13], R. Partridge[251], S. Pastor[108], E. Paterson[251], M. Patra[83], J. R. Patterson[43], M. Paulini[20], N. Paver[129], S. Pavy-Bernard[167],

B. Pawlik[268], A. Pérez Vega-Leal[285], B. Pearson[355], J. S. Pedersen[211], A. Pedicini[397], S. Pedraza López[108], G. Pei[102], S. Pei[102], G. Pellegrini[32], A. Pellegrino[213], S. Penaranda[286], H. Peng[364], X. Peng[102], M. Perelstein[43], E. Perez[112], M. A. Perez-Garcia[284,114], M. Perez-Victoria[281], S. Peris[293], D. Perret-Gallix[164], H. Perrey[47], T. M. Perry[384], M. E. Peskin[251], P. Petagna[33], R. Y. Peters[64,47], T. C. Petersen[211], D. P. Peterson[43], T. Peterson[58], E. Petrakou[210], A. A. Petrov[403], A. Petrukhin[89,107], S. Pfeiffer[47], H. Pham[94], K. H. Phan[395,71], N. Phinney[251], F. Piccinini[124], A. Pich[108], R. Pichai[85], J. Piedra[283], J. Piekarski[401], A. Pierce[348], P. Pierini[119], N. Pietsch[333,47], A. Pineda[293], J. Pinfold[312,152], A. Piotrowski[263], Y. Pischalnikov[58], R. Pittau[281], M. Pivi[251], W. Placzek[132], T. Plehn[296], M. A. Pleier[13], M. Poelker[269], L. Poggioli[167], I. Pogorelsky[13], V. Poireau[164], M. E. Pol[30], I. Polak[105], F. Polci[169], M. Polikarpov[107], T. Poll[316,259], M. W. Poole[258,40], W. Porod[145], F. C. Porter[17], S. Porto[333], J. Portolés[108], R. Pöschl[167], S. Poss[33], C. T. Potter[356], P. Poulose[86], K. T. Pozniak[401], V. Prahl[47], R. Prepost[384], C. Prescott[251], D. Price[87], T. Price[315], P. S. Prieto[58], D. Protopopescu[332], D. Przyborowski[2], K. Przygoda[263], H. Przysiezniak[164], F. Ptochos[325], J. Puerta-Pelayo[31], C. Pulvermacher[90], M. Purohit[366], Q. Qin[102], F. Qiu[102], H. Qu[102], A. Quadt[64], G. Quast[90], D. Quirion[32], M. Quiros[88], J. Rademacker[316], R. Rahmat[350], S. Rai[67], M. Raidal[205], S. Rakshit[84], M. Ramilli[333], F. Rarbi[170], P. Ratoff[176], T. Raubenheimer[251], M. Rauch[91], L. Raux[93], G. Raven[400,213], P. Razis[325], V. Re[124], S. Redford[33], C. E. Reece[269], I. Reichel[177], A. Reichold[358,140], P. Reimer[105], M. Reinecke[47], A. Rekalo[100], J. Repond[7], J. Resta-Lopez[108], J. Reuter[47], J. T. Rhee[156], P. M. Ribeiro Cipriano[47], A. Ribon[33], G. Ricciardi[292,123], F. Richard[167], E. Richter-Was[132], G. Riddone[33], S. Riemann[48], T. Riemann[48], M. Rijssenbeek[257], K. Riles[348], C. Rimbault[167], R. Rimmer[269], S. D. Rindani[229], A. Ringwald[47], L. Rinolfi[33], I. Ripp-Baudot[94], I. Riu[294], T. G. Rizzo[251], P. Robbe[167], J. Roberts[140,33], A. Robson[332], G. Rodrigo[108], P. Rodriguez[251], P. Rodriguez Perez[112], K. Rolbiecki[111], P. Roloff[33], R. S. Romaniuk[401], E. Romero Adam[108], A. Ronzhin[58], L. Roos[169], E. Ros[108], A. Rosca[47], C. Rosemann[47], J. Rosiek[382], M. C. Ross[251], R. Rossmanith[90], S. Roth[235], J. Rouëné[167], A. Rowe[58], P. Rowson[251], A. Roy[115], L. Royer[168], P. Royole-Degieux[93], C. Royon[21], A. Rozanov[27], M. Ruan[172], D. L. Rubin[43], I. Rubinskiy[47], R. Rückl[145], R. Ruiz[361], R. Ruiz De Austri[108], P. Ruiz Valls[108], P. Ruiz-Femenía[108], A. Ruiz-Jimeno[110], R. Ruland[251], V. Rusinov[107], J. J. Russell[251], I. Rutkowski[401], V. Rybnikov[47], A. Ryd[43], A. Sabio Vera[111], B. Sabirov[142], J. J. Saborido Silva[112], H. F. W. Sadrozinski[322], T. Saeki[71], B. Safarzadeh[101], P. Saha[305], H. Sahoo[7], A. Sailer[33], N. Saito[71], T. Saito[272], T. Sakaguchi[13], H. Sakai[71], K. Sakai[71], K. Sakaue[402], K. Sakurai[47], R. Salerno[172], J. Salfeld-Nebgen[47], J. Salt[108], L. Sanchez[31], M. A. Sanchis Lozano[108], J. Sandweiss[405], A. Santa[377], A. Santagata[286], A. Santamaria[108], P. Santorelli[292], T. Sanuki[272], A. A. Sapronov[142], M. Sasaki[96], H. Sato[248], N. Sato[71], Y. Sato[272], M. Satoh[71], E. Sauvan[164], V. Saveliev[238,47], A. Savoy-Navarro[166,126], M. Sawabe[71], R. Sawada[116], H. Sawamura[402], L. Sawyer[181], O. Schäfer[300,47], R. Schäfer[47], J. Schaffran[47], T. Schalk[322,251], R. D. Schamberger[257], J. Scheirich[34], G. Schierholz[47], F. P. Schilling[90], F. Schirra[89], F. Schlander[47], H. Schlarb[47], D. Schlatter[33], P. Schleper[333], J. L. Schlereth[7], R. D. Schlueter[177], C. Schmidt[47], U. Schneekloth[47], S. Schnetzer[239], T. Schoerner-Sadenius[47], M. Schram[225], H. J. Schreiber[48], S. Schreiber[47], K. P. Schüler[47], D. Schulte[33], H. C. Schultz-Coulon[297], M. Schumacher[5], S. Schumann[64], B. A. Schumm[322], M. H. Schune[167], S. Schuwalow[333,48], C. Schwanda[224], C. Schwanenberger[343], F. Schwartzkopff[295], D. J. Scott[258,58], F. Sefkow[47,33], A. Segui[286], N. Seguin-Moreau[93], S. Seidel[352], Y. Seiya[222], J. Sekaric[338], K. Seki[197], S. Sekmen[33], S. Seletskiy[13], S. Sen[337], E. Senaha[158], K. Senyo[406], S. Senyukov[94], I. Serenkova[227], D. A. Sergatskov[58], H. Sert[47,333], D. Sertore[119], A. Seryi[358,140], O. Seto[74], R. Settles[186], P. Sha[102], S. Shahid[301], A. Sharma[33], G. Shelkov[142], W. Shen[297], J. C. Sheppard[251], M. Sher[41], C. Shi[102], H. Shi[102], T. Shidara[71], W. Shields[237,139], M. Shimada[71], H. Shimizu[71], Y. Shimizu[272],

A. Vicente[171], J. Vidal-Perona[108], H. L. R. Videau[172], I. Vila[110], X. Vilasis-Cardona[299], E. Vilella[311], A. Villamor[32], E. G. Villani[259], J. A. Villar[286], M. A. Villarejo Bermúdez[108], D. Vincent[169], P. Vincent[169], J. M. Virey[28], A. Vivoli[58], V. Vogel[47], R. Volkenborn[47], O. Volynets[47], F. Von Der Pahlen[110], E. Von Toerne[295], B. Vormwald[47], A. Voronin[180], M. Vos[108], J. H. Vossebeld[339], G. Vouters[164], Y. Voutsinas[94,47], V. Vrba[105,45], M. Vysotsky[107], D. Wackeroth[256], A. Wagner[47], C. E. Wagner[7,52], R. Wagner[7], S. R. Wagner[324], W. Wagner[385], J. Wagner-Kuhr[90], A. P. Waite[251], M. Wakayama[197], Y. Wakimoto[276], R. Walczak[358,140], R. Waldi[300], D. G. E. Walker[251], N. J. Walker[47], M. Walla[47], C. J. Wallace[49], S. Wallon[171,393], D. Walsh[328], S. Walston[178], W. A. T. Wan Abdullah[342], D. Wang[102], G. Wang[102], J. Wang[251], L. Wang[251], L. Wang[52], M. H. Wang[251], M. Z. Wang[210], Q. Wang[102], Y. Wang[102], Z. Wang[24], R. Wanke[138], C. Wanotayaroj[356], B. Ward[8], D. Ward[323], B. Warmbein[47], M. Washio[402], K. Watanabe[71], M. Watanabe[212], N. Watanabe[71], T. Watanabe[155], Y. Watanabe[71], S. Watanuki[272], Y. Watase[71], N. K. Watson[315], G. Watts[383], M. M. Weber[90], H. C. Weddig[47], H. Weerts[7], A. W. Weidemann[251], G. Weiglein[47], A. Weiler[47], S. Weinzierl[138], H. Weise[47], A. Welker[138], N. Welle[47], J. D. Wells[33,348], M. Wendt[58,33], M. Wenskat[47], H. Wenzel[58], N. Wermes[295], U. Werthenbach[301], W. Wester[58], L. Weuste[186,57], A. White[373], G. White[251], K. H. Wichmann[47], M. Wielers[183], R. Wielgos[58], W. Wierba[202], T. Wilksen[47], S. Willocq[346], F. F. Wilson[259], G. W. Wilson[338], P. B. Wilson[251], M. Wing[309], M. Winter[94], K. Wittenburg[47], P. Wittich[43], M. Wobisch[181], A. Wolski[339,40], M. D. Woodley[251], M. B. Woods[251], M. Worek[385], S. Worm[33,259], G. Wormser[167], D. Wright[178], Z. Wu[251], C. E. Wulz[224], S. Xella[211], G. Xia[40,343], L. Xia[7], A. Xiao[7], L. Xiao[251], M. Xiao[102], Q. Xiao[102], J. Xie[7], C. Xu[102], F. Xu[210], G. Xu[102], K. Yagyu[201], U. A. Yajnik[85], V. Yakimenko[251], S. Yamada[71,116], S. Yamada[71], Y. Yamada[272], Y. Yamada[402], A. Yamaguchi[274], D. Yamaguchi[275], M. Yamaguchi[272], S. Yamaguchi[272], Y. Yamaguchi[375], Y. Yamaguchi[75], A. Yamamoto[71,375], H. Yamamoto[272], K. Yamamoto[222], K. Yamamoto[118], M. Yamamoto[71], N. Yamamoto[197], N. Yamamoto[71], Y. Yamamoto[71], Y. Yamamoto[375], T. Yamamura[375], T. Yamanaka[116], S. Yamashita[116], T. Yamashita[3], Y. Yamashita[214], K. Yamauchi[197], M. Yamauchi[71], T. Yamazaki[375], Y. Yamazaki[154], J. Yan[375,71], W. Yan[364], C. Yanagisawa[257,11], H. Yang[247], J. Yang[56], U. K. Yang[245,343], Z. Yang[24], W. Yao[177], S. Yashiro[71], F. Yasuda[375], O. Yasuda[276], I. Yavin[188,228], E. Yazgan[331], H. Yokoya[377], K. Yokoya[71], H. Yokoyama[375], S. Yokoyama[275], R. Yonamine[71], H. Yoneyama[240], M. Yoshida[71], T. Yoshida[62], K. Yoshihara[116,33], S. Yoshihara[116,33], M. Yoshioka[71,272], T. Yoshioka[161], H. Yoshitama[73], C. C. Young[251], H. B. Yu[348], J. Yu[373], C. Z. Yuan[102], F. Yuasa[71], J. Yue[102], A. Zabi[172], W. Zabolotny[401], J. Zacek[34], I. Zagorodnov[47], J. Zalesak[105,58], A. F. Zarnecki[381], L. Zawiejski[268], M. Zeinali[101], C. Zeitnitz[385], L. Zembala[401], K. Zenker[47], D. Zeppenfeld[91], D. Zerwas[167], P. Zerwas[47], M. Zeyrek[192], A. Zghiche[164], J. Zhai[102], C. Zhang[102], J. Zhang[102], J. Zhang[7], Y. Zhang[24,33], Z. Zhang[167], F. Zhao[102], F. Zhao[102], T. Zhao[102], Y. Zhao[251], H. Zheng[102], Z. Zhengguo[364], L. Zhong[24], F. Zhou[251], X. Zhou[364,102], Z. Zhou[102], R. Y. Zhu[17], X. Zhu[24], X. Zhu[102], M. Zimmer[47], F. Zomer[167], T. Zoufal[47], R. Zwicky[329] 1 Academia Sinica - 128 Sec. 2, Institute of Physics, Academia Rd., Nankang, Taipei 11529, Taiwan, R.O.C.

2 AGH University of Science and Technology, Akademia Gorniczo-Hutnicza im. Stanislawa Staszica w Krakowie, Al. Mickiewicza 30 PL-30-059 Cracow, Poland

3 Aichi Medical University, Nagakute, Aichi, 480-1195, Japan

4 Akita International University, Yuwa, Akita City, 010-1292, Japan

5 Albert-Ludwigs Universität Freiburg, Physikalisches Institut, Hermann-Herder Str. 3, D-79104 Freiburg, Germany

6 Ankara Üniversitesi Fen Fakültesi, Fizik Bölümü, Dögol Caddesi, 06100 Tandoĝan Ankara, Turkey

7 Argonne National Laboratory (ANL), 9700 S. Cass Avenue, Argonne, IL 60439, USA

8 Baylor University, Department of Physics, 101 Bagby Avenue, Waco, TX 76706, USA

9 Beijing University, Department of Physics, Beijing, China 100871

10 Benares Hindu University, Benares, Varanasi 221005, India

11 Borough of Manhattan Community College, The City University of New York, Department of Science, 199 Chambers Street, New York, NY 10007, USA

12 Brandenburg University of Technology, Postfach 101344, D-03013 Cottbus, Germany

13 Brookhaven National Laboratory (BNL), P.O.Box 5000, Upton, NY 11973-5000, USA

14 Brown University, Department of Physics, Box 1843, Providence, RI 02912, USA

15 Budker Institute for Nuclear Physics (BINP), 630090 Novosibirsk, Russia

16 Calcutta University, Department of Physics, 92 A.P.C. Road, Kolkata 700009, India

17 California Institute of Technology, Physics, Mathematics and Astronomy (PMA), 1200 East California Blvd, Pasadena, CA 91125, USA

18 California State University, Los Angeles, Dept. of Physics and Astronomy, 5151 State University Dr., Los Angeles, CA 90032, USA

19 Carleton University, Department of Physics, 1125 Colonel By Drive, Ottawa, Ontario, Canada K1S 5B6

20 Carnegie Mellon University, Department of Physics, Wean Hall 7235, Pittsburgh, PA 15213, USA

21 CEA Saclay, IRFU, F-91191 Gif-sur-Yvette, France

22 CEA Saclay, Service de Physique Théorique, CEA/DSM/SPhT, F-91191 Gif-sur-Yvette Cedex, France

23 Center for High Energy Physics (CHEP) / Kyungpook National University, 1370 Sankyuk-dong, Buk-gu, Daegu 702-701, Republic of Korea

24 Center for High Energy Physics (TUHEP), Tsinghua University, Beijing, China 100084

25 Center For Quantum Spacetime (CQUeST), Sogang University, 35 Baekbeom-ro, Mapo-gu, Seoul 121-742, Republic of Korea

26 Center for the Advancement of Natural Discoveries using Light Emission (CANDLE), Acharyan 31, 0040, Yerevan, Armenia

27 Centre de Physique des Particules de Marseille (CPPM), Aix-Marseille Université, CNRS/IN2P3, 163, Avenue de Luminy, Case 902, 13288 Marseille Cedex 09, France

28 Centre de Physique Theorique, CNRS - Luminy, Universiti d"Aix - Marseille II, Campus of Luminy, Case 907, 13288 Marseille Cedex 9, France

29 Centre Lasers Intenses et Applications (CELIA), Université Bordeaux 1 - CNRS - CEA, 351 Cours de la Libération, 33405 Talence Cedex, France

30 Centro Brasileiro de Pesquisas Físicas (CBPF), Rua Dr. Xavier Sigaud, n.150 22290-180, Urca - Rio de Janeiro, RJ, Brazil

31 Centro de Investigaciones Energéticas, Medioambientales y Tecnológicas, CIEMAT, Avenida Complutense 22, E-28040 Madrid, Spain

32 Centro Nacional de Microelectrónica (CNM), Instituto de Microelectrónica de Barcelona (IMB), Campus UAB, 08193 Cerdanyola del Vallès (Bellaterra), Barcelona, Spain

33 CERN, CH-1211 Genève 23, Switzerland

34 Charles University, Institute of Particle & Nuclear Physics, Faculty of Mathematics and Physics, V Holesovickach 2, CZ-18000 Prague 8, Czech Republic

35 Chiba University of Commerce, 1-3-1 Konodai, Ichikawa-shi, Chiba, 272-8512, Japan

36 Chonbuk National University, Division of Science Education, Jeonju 561-756, Republic of Korea

37 Chonbuk National University, Physics Department, Jeonju 561-756, Republic of Korea

38 Chubu University, 1200 Matsumoto-cho, Kasugai-shi, Aichi, 487-8501, Japan

39 Chung Yuan Christian University, Department of Physics, 200 Chung Pei Rd., Chung Li 32023 Taiwan, R.O.C

40 Cockcroft Institute, Daresbury, Warrington WA4 4AD, UK

41 College of William and Mary, Department of Physics, Williamsburg, VA, 23187, USA

42 Columbia University, Department of Physics, New York, NY 10027-6902, USA

43 Cornell University, Laboratory for Elementary-Particle Physics (LEPP), Ithaca, NY 14853, USA

44 Czech Technical University in Prague, Institute of Experimental and Applied Physics (IEAP), Horska 3a/22, 12800 Prague 2, Czech Republic

45 Czech Technical University, Faculty of Nuclear Science and Physical Engineering, Brehova 7, CZ-11519 Prague 1, Czech Republic

46 Departamento de Física Teórica, Facultad de Ciencias, Módulo 15 (antiguo C-XI) y Módulo 8, Universidad Autónoma de Madrid, Campus de Cantoblanco, 28049 Madrid, Spain

47 Deutsches Elektronen-Synchrotron DESY, A Research Centre of the Helmholtz Association, Notkestrasse 85, 22607 Hamburg, Germany (Hamburg site)

48 Deutsches Elektronen-Synchrotron DESY, A Research Centre of the Helmholtz Association, Platanenallee 6, 15738 Zeuthen, Germany (Zeuthen site)

49 Durham University, Department of Physics, Ogen Center for Fundamental Physics, South Rd., Durham DH1 3LE, UK

50 École Normale Supérieure de Lyon, 46 allée d'Italie, 69364 Lyon Cedex 07, France

51 Ege University, Department of Physics, Faculty of Science, 35100 Izmir, Turkey

52 Enrico Fermi Institute, University of Chicago, 5640 S. Ellis Avenue, RI-183, Chicago, IL 60637, USA
53 ETH Zürich, Institute for Particle Physics (IPP), Schafmattstrasse 20, CH-8093 Zürich, Switzerland
54 ETH Zürich, Institute for Theoretical Physics (ITP), Wolfgang-Pauli-Str. 27, Zürich, Switzerland
55 European Spallation Source ESS AB, Box 176, 221 00 Lund, Sweden
56 Ewha Womans University, 11-1 Daehyun-Dong, Seodaemun-Gu, Seoul, 120-750, Republic of Korea
57 Excellence Cluster Universe, Technische Universität München, Boltzmannstr. 2, 85748 Garching, Germany
58 Fermi National Accelerator Laboratory (FNAL), P.O.Box 500, Batavia, IL 60510-0500, USA
59 Florida State University, Department of Physics, 77 Chieftan Way, Tallahassee, FL 32306-4350, USA
60 Fujita Gakuen Health University, Department of Physics, Toyoake, Aichi 470-1192, Japan
61 Fukui University of Technology, 3-6-1 Gakuen, Fukui-shi, Fukui 910-8505, Japan
62 Fukui University, Department of Physics, 3-9-1 Bunkyo, Fukui-shi, Fukui 910-8507, Japan
63 Gangneung-Wonju National University, 210-702 Gangneung Daehangno, Gangneung City, Gangwon Province, Republic of Korea
64 Georg-August-Universität Göttingen, II. Physikalisches Institut, Friedrich-Hund-Platz 1, 37077 Göttingen, Germany
65 Global Design Effort
66 Hanyang University, Department of Physics, Seoul 133-791, Republic of Korea
67 Harish-Chandra Research Institute, Chhatnag Road, Jhusi, Allahabad 211019, India
68 Helmholtz-Zentrum Berlin für Materialien und Energie (HZB), Wilhelm-Conrad-Röntgen Campus, BESSY II, Albert-Einstein-Str. 15, 12489 Berlin, Germany
69 Helsinki Institute of Physics (HIP), P.O. Box 64, FIN-00014 University of Helsinki, Finland
70 Henan Normal University, College of Physics and Information Engineering, Xinxiang, China 453007
71 High Energy Accelerator Research Organization, KEK, 1-1 Oho, Tsukuba, Ibaraki 305-0801, Japan
72 Hiroshima University, Department of Physics, 1-3-1 Kagamiyama, Higashi-Hiroshima, Hiroshima 739-8526, Japan
73 Hiroshima University, Graduate School of Advanced Sciences of Matter, 1-3-1 Kagamiyama, Higashi-Hiroshima, Hiroshima 739-8530, Japan
74 Hokkai-Gakuen University, 4-1-40 Asahimachi, Toyohira-ku, Sapporo 062-8605, Japan
75 Hokkaido University, Department of Physics, Faculty of Science, Kita, Kita-ku, Sapporo-shi, Hokkaido 060-0810, Japan
76 Humboldt Universität zu Berlin, Fachbereich Physik, Institut für Elementarteilchenphysik, Newtonstr. 15, D-12489 Berlin, Germany
77 Hyogo University of Teacher Education, 942-1 Shimokume, Kato-city, Hyogo 673-1494, Japan
78 Ibaraki National College of Technology, 866 Nakane, Hitachinaka, Ibaraki 312-8508, Japan
79 Ibaraki University, College of Technology, Department of Physics, Nakanarusawa 4-12-1, Hitachi, Ibaraki 316-8511, Japan
80 Imperial College, Blackett Laboratory, Department of Physics, Prince Consort Road, London, SW7 2BW, UK
81 Indian Association for the Cultivation of Science, Department of Theoretical Physics and Centre for Theoretical Sciences, Kolkata 700032, India
82 Indian Institute of Science Education and Research (IISER) Kolkata, Department of Physical Sciences, Mohanpur Campus, PO Krishi Viswavidyalaya, Mohanpur 741252, Nadia, West Bengal, India
83 Indian Institute of Science, Centre for High Energy Physics, Bangalore 560012, Karnataka, India
84 Indian Institute of Technology Indore, IET Campus, M-Block, Institute of Engineering and Technology (IET), Devi Ahilya Vishwavidyalaya Campus, Khandwa Road, Indore - 452017, Madhya Pradesh, India
85 Indian Institute of Technology, Bombay, Powai, Mumbai 400076, India
86 Indian Institute of Technology, Guwahati, Guwahati, Assam 781039, India
87 Indiana University, Department of Physics, Swain Hall West 117, 727 E. 3rd St., Bloomington, IN 47405-7105, USA
88 Institucio Catalana de Recerca i Estudis, ICREA, Passeig Lluis Companys, 23, Barcelona 08010, Spain
89 Institut de Physique Nucléaire de Lyon (IPNL), Domaine scientifique de la Doua, Bâtiment Paul Dirac 4, rue Enrico Fermi, 69622 Villeurbanne, Cedex, France
90 Institut für Experimentelle Kernphysik, KIT,Universität Karlsruhe (TH), Wolfgang-Gaede-Str. 1, Postfach 6980, 76128 Karlsruhe, Germany
91 Institut für Theoretische Physik (ITP), Karlsruher Institut für Technologie (KIT), Fakultät für Physik, Postfach 6980, 76049 Karlsruhe, Germany
92 Institut für Theoretische Teilchenphysik, Campus Süd, Karlsruher Institut für Technologie (KIT), 76128 Karlsruhe, Germany
93 Institut National de Physique Nucleaire et de Physique des Particules, 3, Rue Michel- Ange, 75794 Paris Cedex 16, France
94 Institut Pluridisciplinaire Hubert Curien, 23 Rue du Loess - BP28, 67037 Strasbourg Cedex 2, France
95 Institute for Chemical Research, Kyoto University, Gokasho, Uji, Kyoto 611-0011, Japan
96 Institute for Cosmic Ray Research, University of Tokyo, 5-1-5 Kashiwa-no-Ha, Kashiwa, Chiba 277-8582, Japan
97 Institute for Mathematics, Astrophysics and Particle Physics (IMAPP), P.O. Box 9010, 6500 GL Nijmegen, Netherlands
98 Institute for Nuclear Research, Russian Academy of Sciences (INR RAS), 60-th October Anniversary Prospect 7a, 117312, Moscow, Russia
99 Institute for Particle and Nuclear Physics, Wigner Research Centre for Physics, Hungarian Academy of Sciences, P.O. Box 49, 1525 Budapest, Hungary
100 Institute for Scintillation Materials (ISMA), 60 Lenina Ave, 61001, Kharkiv, Ukraine
101 Institute for studies in fundamental sciences (IPM), Niavaran Square, P.O. Box 19395-5746, Tehran, Iran
102 Institute of High Energy Physics - IHEP, Chinese Academy of Sciences, P.O. Box 918, Beijing, China 100049
103 Institute of Mathematical Sciences, Taramani, C.I.T. Campus, Chennai 600113, India
104 Institute of Particle Physics, Canada

105 Institute of Physics, ASCR, Academy of Science of the Czech Republic, Division of Elementary Particle Physics, Na Slovance 2, CZ-18221 Prague 8, Czech Republic

106 Institute of Physics, Vietnam Academy of Science and Technology (VAST), 10 Dao-Tan, Ba-Dinh, Hanoi 10000, Vietnam

107 Institute of Theoretical and Experimetal Physics, B. Cheremushkinskawa, 25, RU-117259, Moscow, Russia

108 Instituto de Fisica Corpuscular (IFIC), Centro Mixto CSIC-UVEG, Edificio Investigacion Paterna, Apartado 22085, 46071 Valencia, Spain

109 Instituto de Física da Universidade Federal do Rio Grande do Sul (UFRGS), Av. Bento Gonçalves 9500, Caixa Postal 15051, CEP 91501-970, Porto Alegre, RS, Brazil

110 Instituto de Fisica de Cantabria, (IFCA, CSIC-UC), Facultad de Ciencias, Avda. Los Castros s/n, 39005 Santander, Spain

111 Instituto de Física Teórica UAM/CSIC, C/ Nicolás Cabrera 13-15, Universidad Autónoma de Madrid, Cantoblanco, 28049 Madrid, Spain

112 Instituto Galego de Fisica de Altas Enerxias (IGFAE,USC) Facultad de Fisica, Campus Sur E-15782 Santiago de Compostela, Spain

113 Instituto Tecnológico de Aragón (ITA), C/ María de Luna 7-8, 50018 Zaragoza, Spain

114 Instituto Universitario de Física Fundamental y Matemáticas de la Universidad de Salamanca (IUFFyM), Casas del Parque, 37008 Salamanca, Spain

115 Inter-University Accelerator Centre, Aruna Asaf Ali Marg, Post Box 10502, New Delhi 110067, India

116 International Center for Elementary Particle Physics, University of Tokyo, Hongo 7-3-1, Bunkyo District, Tokyo 113-0033, Japan

117 International Institute of Physics, Federal University of Rio Grande do Norte, Av. Odilon Gomes de Lima, 1722 - Capim Macio - 59078-400 - Natal-RN, Brazil

118 Iowa State University, Department of Physics, High Energy Physics Group, Ames, IA 50011, USA

119 Istituto Nazionale di Fisica Nucleare (INFN), Laboratorio LASA, Via Fratelli Cervi 201, 20090 Segrate, Italy

120 Istituto Nazionale di Fisica Nucleare (INFN), Sezione di Firenze, Via G. Sansone 1, I-50019 Sesto Fiorentino (Firenze), Italy

121 Istituto Nazionale di Fisica Nucleare (INFN), Sezione di Milano Bicocca, Piazza della Scienza 3, I-20126 Milano, Italy

122 Istituto Nazionale di Fisica Nucleare (INFN), Sezione di Milano, Via Celoria 16, I-20133 Milano, Italy

123 Istituto Nazionale di Fisica Nucleare (INFN), Sezione di Napoli, Complesso Universitá di Monte Sant'Angelo,via, I-80126 Naples, Italy

124 Istituto Nazionale di Fisica Nucleare (INFN), Sezione di Pavia, Via Bassi 6, I-27100 Pavia, Italy

125 Istituto Nazionale di Fisica Nucleare (INFN), Sezione di Perugia, Via A. Pascoli, 06123 Perugia, Italy

126 Istituto Nazionale di Fisica Nucleare (INFN), Sezione di Pisa, Edificio C - Polo Fibonacci Largo B. Pontecorvo, 3, I-56127 Pisa, Italy

127 Istituto Nazionale di Fisica Nucleare (INFN), Sezione di Roma, c/o Dipartimento di Fisica - Università degli Studi di Roma "La Sapienza", P.le Aldo Moro 2, I-00185 Roma, Italy

128 Istituto Nazionale di Fisica Nucleare (INFN), Sezione di Torino, c/o Universitá di Torino, facoltá di Fisica, via P Giuria 1, 10125 Torino, Italy

129 Istituto Nazionale di Fisica Nucleare (INFN), Sezione di Trieste, Padriciano 99, I-34012 Trieste (Padriciano), Italy

130 ITER Organization, Route de Vinon-sur-Verdon, 13115 St. Paul-lez-Durance, France

131 Iwate University, 4-3-5 Ueda, Morioka, Iwate, 020-8551, Japan

132 Jagiellonian University, Institute of Physics, Ul. Reymonta 4, PL-30-059 Cracow, Poland

133 Jamia Millia Islamia, Department of Physics, Jamia Nagar, New Delhi 110025, India

134 Japan Aerospace Exploration Agency, Sagamihara Campus, 3-1-1 Yoshinodai, Sagamihara, Kanagawa 220-8510 , Japan

135 Japan Atomic Energy Agency, 4-49 Muramatsu, Tokai-mura, Naka-gun, Ibaraki 319-1184, Japan

136 Japan Atomic Energy Agency, Tokai Research and Development Center, 2-4 Shirane Shirakata, Tokai-mura, Naka-gun, Ibaraki 319-1195, Japan

137 Japan Synchrotron Radiation Research Institute (JASRI), 1-1-1, Kouto, Sayo-cho, Sayo-gun, Hyogo 679-5198, Japan

138 Johannes Gutenberg Universität Mainz, Institut für Physik, 55099 Mainz, Germany

139 John Adams Institute for Accelerator Science at Royal Holloway University of London, Egham Hill, Egham, Surrey TW20 0EX, UK

140 John Adams Institute for Accelerator Science at University of Oxford, Denys Wilkinson Building, Keble Road, Oxford OX1 3RH, UK

141 Johns Hopkins University - Henry A. Rowland Department of Physics & Astronomy 3701 San Martin Drive, Baltimore, Maryland (MD) 21218, USA

142 Joint Institute for Nuclear Research (JINR), Joliot-Curie 6, 141980, Dubna, Moscow Region, Russia

143 Joint Institute for Power and Nuclear Research "Sosny" at National Academy of Sciences of Belarus, 99 Academician A.K.Krasin Str., Minsk BY-220109, Belarus

144 Jozef Stefan Institute, Jamova cesta 39, 1000 Ljubljana, Slovenia

145 Julius-Maximilians-Universität Würzburg, Fakultät für Physik und Astronomie, Am Hubland, 97074 Würzburg, Germany

146 Juntendo University, School of Medicine, Dept. of Physics, Hiraga-gakuendai 1-1, Inzai-shi, Chiba 270-1695, Japan

147 Justus-Liebig-Universität Gießen, II. Physikalisches Institut, Heinrich-Buff-Ring 16, 35392 Gießen, Germany

148 Kanazawa University, Institute for Theoretical Physics (KITP), School of Mathematics and Physics, College of Science and Engineering, Kakuma-machi, Kanazawa city, Ishikawa 920-1192, Japan

149 Kansas State University, Department of Physics, 116 Cardwell Hall, Manhattan, KS 66506, USA

150 Kavli Institute for the Physics and Mathematics of the Universe (Kavli IPMU), University of Tokyo, 5-1-5 Kashiwanoha, Kashiwa, 277-8583, Japan

151 King Saud University (KSU), Dept. of Physics, P.O. Box 2454, Riyadh 11451, Saudi Arabia

152 King's College London - Department of physics, Strand, London WC2R 2LS, London, UK

153 Kinki University, Department of Physics, 3-4-1 Kowakae, Higashi-Osaka, Osaka 577-8502, Japan

154 Kobe University, Department of Physics, 1-1 Rokkodai-cho, Nada-ku, Kobe, Hyogo 657-8501, Japan

155 Kogakuin University, Department of Physics, Shinjuku Campus, 1-24-2 Nishi-Shinjuku, Shinjuku-ku, Tokyo 163-8677, Japan
156 Konkuk University, 93-1 Mojin-dong, Kwanglin-gu, Seoul 143-701, Republic of Korea
157 Korea Advanced Institute of Science & Technology, Department of Physics, 373-1 Kusong-dong, Yusong-gu, Taejon 305-701, Republic of Korea
158 Korea Institute for Advanced Study (KIAS), School of Physics, 207-43 Cheongryangri-dong, Dongdaemun-gu, Seoul 130-012, Republic of Korea
159 Kyoto University, Department of Physics, Kitashirakawa-Oiwakecho, Sakyo-ku, Kyoto 606-8502, Japan
160 Kyoto University, Yukawa Institute for Theoretical Physics, Kitashirakawa-Oiwakecho, Sakyo-Ku, Kyoto 606-8502, Japan
161 Kyushu University, Department of Physics, 6-10-1 Hakozaki, Higashi-ku, Fukuoka 812-8581, Japan
162 L.P.T.A., UMR 5207 CNRS-UM2, Université Montpellier II, Case Courrier 070, Bât. 13, place Eugène Bataillon, 34095 Montpellier Cedex 5, France
163 Laboratoire Charles Coulomb UMR 5221 CNRS-UM2, Université Montpellier 2, Place Eugène Bataillon - CC069, 34095 Montpellier Cedex 5, France
164 Laboratoire d'Annecy-le-Vieux de Physique des Particules (LAPP) , Université de Savoie, CNRS/IN2P3, 9 Chemin de Bellevue, BP 110, F-74941 Annecy-Le-Vieux Cedex, France
165 Laboratoire d'Annecy-le-Vieux de Physique Theorique (LAPTH), Chemin de Bellevue, BP 110, F-74941 Annecy-le-Vieux Cedex, France
166 Laboratoire d'AstroParticules et Cosmologie (APC), Université Paris Diderot-Paris 7 - CNRS/IN2P3, Bâtiment Condorcet, Case 7020, 75205 Paris Cedex 13, France
167 Laboratoire de l'Accélérateur Linéaire (LAL), Université Paris-Sud 11, Bâtiment 200, 91898 Orsay, France
168 Laboratoire de Physique Corpusculaire de Clermont-Ferrand (LPC), Université Blaise Pascal, I.N.2.P.3./C.N.R.S., 24 avenue des Landais, 63177 Aubière Cedex, France
169 Laboratoire de Physique Nucléaire et des Hautes Energies (LPNHE), UPMC, UPD, IN2P3/CNRS, 4 Place Jussieu, 75005, Paris Cedex 05, France
170 Laboratoire de Physique Subatomique et de Cosmologie (LPSC), Université Joseph Fourier (Grenoble 1), CNRS/IN2P3, Institut Polytechnique de Grenoble, 53 rue des Martyrs, F-38026 Grenoble Cedex, France
171 Laboratoire de Physique Theorique, Université de Paris-Sud XI, Batiment 210, F-91405 Orsay Cedex, France
172 Laboratoire Leprince-Ringuet (LLR), École polytechnique – CNRS/IN2P3, Route de Saclay, F-91128 Palaiseau Cedex, France
173 Laboratoire Univers et Particules de Montpellier (LUPM) - UMR5299, Université de Montpellier II, Place Eugène Bataillon - Case courrier 72, 34095 Montpellier Cedex 05, France
174 Laboratori Nazionali di Frascati, via E. Fermi, 40, C.P. 13, I-00044 Frascati, Italy
175 Laboratório de Instrumentação e Física Experimental de Partículas (LIP LISBOA), Av. Elias Garcia 14 - 1°, 1000-149 Lisbon, Portugal
176 Lancaster University, Physics Department, Lancaster LA1 4YB, UK
177 Lawrence Berkeley National Laboratory (LBNL), 1 Cyclotron Rd, Berkeley, CA 94720, USA
178 Lawrence Livermore National Laboratory (LLNL), Livermore, CA 94551, USA
179 Lebedev Physical Institute, Leninsky Prospect 53, RU-117924 Moscow, Russia
180 Lomonosov Moscow State University, Skobeltsyn Institute of Nuclear Physics (MSU SINP), 1(2), Leninskie gory, GSP-1, Moscow 119991, Russia
181 Louisiana Tech University, Department of Physics, Ruston, LA 71272, USA
182 Ludwig-Maximilians-Universität München, Fakultät für Physik, Am Coulombwall 1, D - 85748 Garching, Germany
183 Lunds Universitet, Fysiska Institutionen, Avdelningen för Experimentell Högenergifysik, Box 118, 221 00 Lund, Sweden
184 L'Université Hassan II, Aïn Chock, "Réseau Universitaire de Physique des Hautes Energies" (RUPHE), Département de Physique, Faculté des Sciences Aïn Chock, B.P 5366 Maarif, Casablanca 20100, Morocco
185 Massachusetts Institute of Technology (MIT), Laboratory for Nuclear Science, 77 Massachusetts Avenue, Cambridge, MA 02139, USA
186 Max-Planck-Institut für Physik (Werner-Heisenberg-Institut), Föhringer Ring 6, 80805 München, Germany
187 McGill University, Department of Physics, Ernest Rutherford Physics Bldg., 3600 University Street, Montreal, Quebec, H3A 2T8 Canada
188 McMaster University, Department of Physics & Astronomy, 1280 Main Street West, Hamilton, ON, L8S 4M1, Canada
189 Meiji Gakuin University, Department of Physics, 2-37 Shirokanedai 1-chome, Minato-ku, Tokyo 244-8539, Japan
190 Michigan State University, Department of Chemical Engineering & Materials Science, 2527 Engineering Building East Lansing, MI 48824-1226, USA
191 Michigan State University, Department of Physics and Astronomy, East Lansing, MI 48824, USA
192 Middle East Technical University, Department of Physics, TR-06531 Ankara, Turkey
193 Miyagi Gakuin Women's University, Faculty of Liberal Arts, 9-1-1 Sakuragaoka, Aoba District, Sendai, Miyagi 981-8557, Japan
194 MSU-Iligan Institute of Technology, Department of Physics, Andres Bonifacio Avenue, 9200 Iligan City, Phillipines
195 Nagasaki Institute of Applied Science, 536 Abamachi, Nagasaki-Shi, Nagasaki 851-0193, Japan
196 Nagoya University, Department of Materials Science and Engineering, Furo-cho, Chikusa-ku, Nagoya, 464-8603, Japan
197 Nagoya University, Department of Physics, School of Science, Furo-cho, Chikusa-ku, Nagoya, Aichi 464-8602, Japan
198 Nagoya University, Kobayashi-Maskawa Institute for the Origin of Particles and the Universe (KMI), Furo-cho, Chikusa-ku, Nagoya Aichi 464-8602, Japan
199 Nanjing University, Department of Physics, Nanjing, China 210093
200 Nara Women's University, High Energy Physics Group, Kitauoya-Nishimachi, Nara 630-8506, Japan
201 National Central University, High Energy Group, Department of Physics, Chung-li, Taiwan 32001, R.O.C
202 National Centre of Nuclear Research (NCBJ), ul. Andrzeja Soltana 7, 05-400 Otwock-Swierk, Poland
203 National Cheng Kung University, Physics Department, 1 Ta-Hsueh Road, Tainan, Taiwan 70101, R.O.C

204 National Chiao-Tung University, Institute of Physics, 1001 Ta Hsueh Rd, Hsinchu, Taiwan 300, R.O.C.

205 National Institute of Chemical Physics and Biophysics (NICPB), Ravala pst 10, 10143 Tallinn, Estonia

206 National Institute of Physics and Nuclear Engineering “Horia Hulubei” (IFIN-HH), Str. Reactorului no.30, P.O. Box MG-6, R-76900 Bucharest - Magurele, Romania

207 National Research Centre “Kurchatov Institute”, 1 Akademika Kurchatova pl., Moscow, 123182, Russia

208 National Science Center - Kharkov Institute of Physics and Technology (NSC KIPT), Akademicheskaya St. 1, Kharkov, 61108, Ukraine

209 National Scientific & Educational Centre of Particle & High Energy Physics (NCPHEP), Belarusian State University, M.Bogdanovich street 153, 220040 Minsk, Belarus

210 National Taiwan University, Physics Department, Taipei, Taiwan 106, R.O.C

211 Niels Bohr Institute (NBI), University of Copenhagen, Blegdamsvej 17, DK-2100 Copenhagen, Denmark

212 Niigata University, Department of Physics, Ikarashi, Niigata 950-218, Japan

213 Nikhef, National Institute for Subatomic Physics, P.O. Box 41882, 1009 DB Amsterdam, Netherlands

214 Nippon Dental University School of Life Dentistry at Niigata, 1-8 Hamaura-cho, Chuo-ku, Niigata 951-1500, Japan

215 North Carolina A&T State University, 1601 E. Market Street, Greensboro, NC 27411, USA

216 Northeastern University, Physics Department, 360 Huntington Ave, 111 Dana Research Center, Boston, MA 02115, USA

217 Northern Illinois University, Department of Physics, DeKalb, Illinois 60115-2825, USA

218 Northwestern University, Department of Physics and Astronomy, 2145 Sheridan Road., Evanston, IL 60208, USA

219 Novosibirsk State University (NGU), Department of Physics, Pirogov st. 2, 630090 Novosibirsk, Russia

220 Ochanomizu University, Department of Physics, Faculty of Science, 1-1 Otsuka 2, Bunkyo-ku, Tokyo 112-8610, Japan

221 Orissa University of Agriculture & Technology, Bhubaneswar 751003, Orissa, India

222 Osaka City University, Department of Physics, Faculty of Science, 3-3-138 Sugimoto, Sumiyoshi-ku, Osaka 558-8585, Japan

223 Osaka University, Department of Physics, 1-1 Machikaneyama, Toyonaka, Osaka 560-0043, Japan

224 Österreichische Akademie der Wissenschaften, Institut für Hochenergiephysik, Nikolsdorfergasse 18, A-1050 Vienna, Austria

225 Pacific Northwest National Laboratory, (PNNL), PO Box 999, Richland, WA 99352, USA

226 Panjab University, Chandigarh 160014, India

227 Pavel Sukhoi Gomel State Technical University, ICTP Affiliated Centre & Laboratory for Physical Studies, October Avenue, 48, 246746, Gomel, Belarus

228 Perimeter Institute for Theoretical Physics, 31 Caroline Street North, Waterloo, Ontario N2L 2Y5, Canada

229 Physical Research Laboratory, Navrangpura, Ahmedabad 380 009, Gujarat, India

230 Pohang Accelerator Laboratory (PAL), San-31 Hyoja-dong, Nam-gu, Pohang, Gyeongbuk 790-784, Republic of Korea

231 Pontificia Universidad Católica de Chile, Avda. Libertador Bernardo OHiggins 340, Santiago, Chile

232 Princeton University, Department of Physics, P.O. Box 708, Princeton, NJ 08542-0708, USA

233 Purdue University, Department of Physics, West Lafayette, IN 47907, USA

234 Queen Mary, University of London, Mile End Road, London, E1 4NS, United Kingdom

235 Rheinisch-Westfälische Technische Hochschule (RWTH), Physikalisches Institut, Physikzentrum, Otto-Blumenthal-Straße, 52056 Aachen

236 RIKEN, 2-1 Hirosawa, Wako, Saitama 351-0198, Japan

237 Royal Holloway, University of London (RHUL), Department of Physics, Egham, Surrey TW20 0EX, UK

238 Russian Academy of Science, Keldysh Institute of Applied Mathematics, Muiskaya pl. 4, 125047 Moscow, Russia

239 Rutgers, The State University of New Jersey, Department of Physics & Astronomy, 136 Frelinghuysen Rd, Piscataway, NJ 08854, USA

240 Saga University, Department of Physics, 1 Honjo-machi, Saga-shi, Saga 840-8502, Japan

241 Saha Institute of Nuclear Physics, 1/AF Bidhan Nagar, Kolkata 700064, India

242 Salalah College of Technology (SCOT), Engineering Department, Post Box No. 608, Postal Code 211, Salalah, Sultanate of Oman

243 Saudi Center for Theoretical Physics, King Fahd University of Petroleum and Minerals (KFUPM), Dhahran 31261, Saudi Arabia

244 Seikei University, Faculty of Science and Technology, 3-3-1 Kichijoji-Kitamachi, Musashino-shi, Tokyo 180-8633, Japan

245 Seoul National University, San 56-1, Shinrim-dong, Kwanak-gu, Seoul 151-742, Republic of Korea

246 Setsunan University, Institute for Fundamental Sciences, 17-8 Ikeda Nakamachi, Neyagawa, Osaka, 572-8508, Japan

247 Shanghai Jiao Tong University, Department of Physics, 800 Dongchuan Road, Shanghai, China 200240

248 Shinshu University, 3-1-1, Asahi, Matsumoto, Nagano 390-8621, Japan

249 Shiv Nadar University, Village Chithera, Tehsil Dadri, District Gautam Budh Nagar, 203207 Uttar Pradesh, India

250 Shizuoka University, Department of Physics, 836 Ohya, Suruga-ku, Shizuoka 422-8529, Japan

251 SLAC National Accelerator Laboratory, 2575 Sand Hill Road, Menlo Park, CA 94025, USA

252 Society for Applied Microwave Electronics Engineering and Research (SAMEER), I.I.T. Campus, Powai, Post Box 8448, Mumbai 400076, India

253 Sokendai, The Graduate University for Advanced Studies, Shonan Village, Hayama, Kanagawa 240-0193, Japan

254 Spallation Neutron Source (SNS), Oak Ridge National Laboratory (ORNL), P.O. Box 2008 MS-6477, Oak Ridge, TN 37831-6477, USA

255 State University of New York at Binghamton, Department of Physics, PO Box 6016, Binghamton, NY 13902, USA

256 State University of New York at Buffalo, Department of Physics & Astronomy, 239 Franczak Hall, Buffalo, NY 14260, USA

257 State University of New York at Stony Brook, Department of Physics and Astronomy, Stony Brook, NY 11794-3800, USA

258 STFC Daresbury Laboratory, Daresbury, Warrington, Cheshire WA4 4AD, UK

259 STFC Rutherford Appleton Laboratory, Chilton, Didcot, Oxon OX11 0QX, UK

260 Sungkyunkwan University (SKKU), Natural Science Campus 300, Physics Research Division, Chunchun-dong, Jangan-gu, Suwon, Kyunggi-do 440-746, Republic of Korea

261 Swiss Light Source (SLS), Paul Scherrer Institut (PSI), PSI West, CH-5232 Villigen PSI, Switzerland

262 Tata Institute of Fundamental Research, School of Natural Sciences, Homi Bhabha Rd., Mumbai 400005, India

263 Technical University of Lodz, Department of Microelectronics and Computer Science, al. Politechniki 11, 90-924 Lodz, Poland

264 Technische Universität Dresden, Institut für Kern- und Teilchenphysik, D-01069 Dresden, Germany

265 Tel-Aviv University, School of Physics and Astronomy, Ramat Aviv, Tel Aviv 69978, Israel

266 Texas A&M University, Physics Department, College Station, 77843-4242 TX, USA

267 Texas Tech University, Department of Physics, Campus Box 41051, Lubbock, TX 79409-1051, USA

268 The Henryk Niewodniczanski Institute of Nuclear Physics, Polish Academy of Sciences (IFJ PAN), ul. Radzikowskiego 152, PL-31342 Cracow, Poland

269 Thomas Jefferson National Accelerator Facility (TJNAF), 12000 Jefferson Avenue, Newport News, VA 23606, USA

270 Tohoku Gakuin University, Department of Business Administration, 1-3-1 Tsuchitoi, Aoba-ku Sendai, Miyagi 980-8511, Japan

271 Tohoku Gakuin University, Faculty of Technology, 1-13-1 Chuo, Tagajo, Miyagi 985-8537, Japan

272 Tohoku University, Department of Physics, Aoba District, Sendai, Miyagi 980-8578, Japan

273 Tohoku University, Research Center for Electron Photon Science, Taihaku District, Sendai, Miyagi 982-0826, Japan

274 Tohoku University, Research Center for Neutrino Science, Aoba District, Sendai, Miyagi 980-8578, Japan

275 Tokyo Institute of Technology, Department of Physics, 2-12-1 O-Okayama, Meguro, Tokyo 152-8551, Japan

276 Tokyo Metropolitan University, Faculty of Science and Engineering, Department of Physics, 1-1 Minami-Osawa, Hachioji-shi, Tokyo 192-0397, Japan

277 Tokyo University of Agriculture Technology, Department of Applied Physics, Naka-machi, Koganei, Tokyo 183-8488, Japan

278 Toyama Prefectural University, Department of Mathematical Physics, 5180 Kurokawa Imizu-shi, Toyama, 939-0398, Japan

279 TRIUMF, 4004 Wesbrook Mall, Vancouver, BC V6T 2A3, Canada

280 Universidad Autónoma de San Luis Potosí, Alvaro Obregon 64, Col. Centro, San Luis Potosí, S.L.P. 78000, México

281 Universidad de Granada, Departamento de Física Teórica y del Cosmos, Campus de Fuentenueva, E-18071 Granada, Spain

282 Universidad de los Andes, Faculty of Science, Department of Physics, Carrera 1 18A-10, Bloque Ip. Bogotá, Colombia

283 Universidad de Oviedo, Departamento de Física, Campus de Llamaquique. C/ Calvo Sotelo, s/n 33005 Oviedo, Spain

284 Universidad de Salamanca, Departamento de Física Fundamental, Plaza de la Merced, s/n., 37008 Salamanca, Spain

285 Universidad de Sevilla, Escuela Técnica Superior de Ingeniería, Departamento Ingeniería Electrónica, Camino de los Descubrimientos s/n, 41092 Sevilla, Spain

286 Universidad de Zaragoza - Departamento de Física Teórica, Pedro Cerbuna 12, E-50009 Zaragoza, Spain

287 Universidad Nacional Autónoma de México, Instituto de Física, Circuito de la Investigación Cientifica s/n, Ciudad Universitaria, CP 04510 México D.F., Mexico

288 Universidad Nacional de La Plata, Departamento de Física, Facultad de Ciencias Exactas, C.C. N 67, 1900 La Plata, Argentina

289 Universidade do Estado do Rio de Janeiro (UERJ), Rio de Janeiro, RJ - Brasil 20550-900, Brazil

290 Universidade Federal de Pelotas, Instituto de Física e Matemática, Campus Universitário, Caixa Postal 354, 96010-900 Pelotas, RS, Brazil

291 Universidade Federal do Rio de Janeiro (UFRJ), Instituto de Física, Av. Athos da Silveira Ramos 149, Centro de Tecnologia - Bloco A, Cidade Universitária, Ilha do Fundão, Rio de Janeiro, RJ, Brazil

292 Universitá degli Studi di Napoli "Federico II", Dipartimento di Fisica, Via Cintia, 80126 Napoli, Italy

293 Universitat Autònoma de Barcelona, Departament de Física, Edifici C, 08193 Bellaterra, Barcelona, Spain

294 Universitat Autònoma de Barcelona, Institut de Fisica d'Altes Energies (IFAE), Campus UAB, Edifici Cn, E-08193 Bellaterra, Barcelona, Spain

295 Universität Bonn, Physikalisches Institut, Nußallee 12, 53115 Bonn, Germany

296 Universität Heidelberg, Institut für Theoretische Physik, Philosophenweg 16, 69120 Heidelberg, Germany

297 Universität Heidelberg, Kirchhoff-Institut für Physik, Im Neuenheimer Feld 227, 69120 Heidelberg, Germany

298 Universitat Politècnica de Catalunya, Institut de Tècniques Energètiques, Campus Diagonal Sud, Edifici PC (Pavelló C). Av. Diagonal, 647 08028 Barcelona, Spain

299 Universitat Ramon Llull, La Salle, C/ Quatre Camins 2, 08022 Barcelona, Spain

300 Universität Rostock, 18051 Rostock, Germany

301 Universität Siegen, Naturwissenschaftlich-Technische Fakultät, Department Physik, Emmy Noether Campus, Walter-Flex-Str.3, 57068 Siegen, Germany

302 Universität Wien - Theoretische Physik Boltzmanngasse 5, A-1090 Vienna, Austria

303 Université catholique de Louvain, Centre for Cosmology, Particle Physics and Phenomenology (CP3), Institute of Mathematics and Physics, 2 Chemin du Cyclotron, 1348 Louvain-la-Neuve, Belgium

304 Université de Genève, Section de Physique, 24, quai E. Ansermet, 1211 Genève 4, Switzerland

305 Université de Montréal, Département de Physique, Groupe de Physique des Particules, C.P. 6128, Succ. Centre-ville, Montréal, Qc H3C 3J7, Canada

306 Université de Strasbourg, UFR de Sciences Physiques, 3-5 Rue de l'Université, F-67084 Strasbourg Cedex, France

307 Université Libre de Bruxelles, Boulevard du Triomphe, 1050 Bruxelles, Belgium

308 Universittà di Catania, Dipartimento di Fisica e Astronomia, Via Santa Sofia 64, 95123 Catania, Italy

309 University College London (UCL), High Energy Physics Group, Physics and Astronomy Department, Gower Street, London WC1E 6BT, UK

310 University College, National University of Ireland (Dublin), Department of Experimental Physics, Science Buildings, Belfield, Dublin 4, Ireland
311 University de Barcelona, Facultat de Física, Av. Diagonal, 647, Barcelona 08028, Spain
312 University of Alberta - Faculty of Science, Department of Physics, 4-181 CCIS, Edmonton AB T6G 2E1, Canada
313 University of Arizona, Department of Physics, 1118 E. Fourth Street, PO Box 210081, Tucson, AZ 85721, USA
314 University of Bergen, Institute of Physics, Allegaten 55, N-5007 Bergen, Norway
315 University of Birmingham, School of Physics and Astronomy, Particle Physics Group, Edgbaston, Birmingham B15 2TT, UK
316 University of Bristol, H. H. Wills Physics Lab, Tyndall Ave., Bristol BS8 1TL, UK
317 University of British Columbia, Department of Physics and Astronomy, 6224 Agricultural Rd., Vancouver, BC V6T 1Z1, Canada
318 University of California (UCLA), Los Angleles, CA 90095, US
319 University of California Berkeley, Department of Physics, 366 Le Conte Hall, #7300, Berkeley, CA 94720, USA
320 University of California Davis, Department of Physics, One Shields Avenue, Davis, CA 95616-8677, USA
321 University of California Irvine, Department of Physics and Astronomy, High Energy Group, 4129 Frederick Reines Hall, Irvine, CA 92697-4575 USA
322 University of California Santa Cruz, Institute for Particle Physics, 1156 High Street, Santa Cruz, CA 95064, USA
323 University of Cambridge, Cavendish Laboratory, J J Thomson Avenue, Cambridge CB3 0HE, UK
324 University of Colorado at Boulder, Department of Physics, 390 UCB, University of Colorado, Boulder, CO 80309-0390, USA
325 University of Cyprus, Department of Physics, P.O.Box 20537, 1678 Nicosia, Cyprus
326 University of Delhi, Department of Physics and Astrophysics, Delhi 110007, India
327 University of Delhi, S.G.T.B. Khalsa College, Delhi 110007, India
328 University of Dundee, Department of Physics, Nethergate, Dundee, DD1 4HN, Scotland, UK
329 University of Edinburgh, School of Physics, James Clerk Maxwell Building, The King's Buildings, Mayfield Road, Edinburgh EH9 3JZ, UK
330 University of Florida, Department of Physics, Gainesville, FL 32611, USA
331 University of Ghent, Department of Subatomic and Radiation Physics, Proeftuinstraat 86, 9000 Gent, Belgium
332 University of Glasgow, SUPA, School of Physics & Astronomy, University Avenue, Glasgow G12 8QQ, Scotland, UK
333 University of Hamburg, Physics Department, Luruper Chaussee 149, 22761 Hamburg, Germany
334 University of Hawaii, Department of Physics and Astronomy, HEP, 2505 Correa Rd., WAT 232, Honolulu, HI 96822-2219, USA
335 University of Helsinki, Department of Physical Sciences, P.O. Box 64 (Vaino Auerin katu 11), FIN-00014, Helsinki, Finland
336 University of Illinois at Chicago, Department Of Physics, 845 W Taylor St., Chicago IL 60607, USA
337 University of Iowa, Department of Physics and Astronomy, 203 Van Allen Hall, Iowa City, IA 52242-1479, USA
338 University of Kansas, Department of Physics and Astronomy, Malott Hall, 1251 Wescoe Hall Drive, Room 1082, Lawrence, KS 66045-7582, USA
339 University of Liverpool, Department of Physics, Oliver Lodge Lab, Oxford St., Liverpool L69 7ZE, UK
340 University of Liverpool, Division of Theoretical Physics, Department of Mathematical Sciences, Chadwick Building, Liverpool L69 3BX, UK
341 University of Ljubljana, Faculty of Mathematics and Physics, Jadranska ulica 19, 1000 Ljubljana, Slovenia
342 University of Malaya, Faculty of Science, Department of Physics, 50603 Kuala Lumpur, Malaysia
343 University of Manchester, School of Physics and Astronomy, Schuster Lab, Manchester M13 9PL, UK
344 University of Maribor, Faculty of Chemistry and Chemical Engineering (FKKT), Smetanova ulica 17, 2000 Maribor, Slovenia
345 University of Maryland, Department of Physics and Astronomy, Physics Building (Bldg. 082), College Park, MD 20742, USA
346 University of Massachusetts - Amherst, Department of Physics, 1126 Lederle Graduate Research Tower (LGRT), Amherst, MA 01003-9337, USA
347 University of Melbourne, School of Physics, Victoria 3010, Australia
348 University of Michigan, Department of Physics, 500 E. University Ave., Ann Arbor, MI 48109-1120, USA
349 University of Minnesota, 148 Tate Laboratory Of Physics, 116 Church St. S.E., Minneapolis, MN 55455, USA
350 University of Mississippi, Department of Physics and Astronomy, 108 Lewis Hall, PO Box 1848, Oxford, Mississippi 38677-1848, USA
351 University of Missouri – St. Louis, Department of Physics and Astronomy, 503 Benton Hall One University Blvd., St. Louis Mo 63121, USA
352 University of New Mexico, New Mexico Center for Particle Physics, Department of Physics and Astronomy, 800 Yale Boulevard N.E., Albuquerque, NM 87131, USA
353 University of North Carolina at Chapel Hill, Department of Physics and Astronomy, Phillips Hall, CB #3255, 120 E. Cameron Ave., Chapel Hill, NC 27599-3255, USA
354 University of Notre Dame, Department of Physics, 225 Nieuwland Science Hall, Notre Dame, IN 46556, USA
355 University of Oklahoma, Department of Physics and Astronomy, Norman, OK 73071, USA
356 University of Oregon, Department of Physics, 1371 E. 13th Ave., Eugene, OR 97403, USA
357 University of Oslo, Department of Physics, P.O box 1048, Blindern, 0316 Oslo, Norway
358 University of Oxford, Particle Physics Department, Denys Wilkinson Bldg., Keble Road, Oxford OX1 3RH England, UK
359 University of Pavia, Department of Physics, via Bassi 6, I-27100 Pavia, Italy
360 University of Pennsylvania, Department of Physics and Astronomy, 209 South 33rd Street, Philadelphia, PA 19104-6396, USA
361 University of Pittsburgh, Department of Physics and Astronomy, 100 Allen Hall, 3941 O'Hara St, Pittsburgh PA 15260, USA
362 University of Regina, Department of Physics, Regina, Saskatchewan, S4S 0A2 Canada

363 University of Rochester, Department of Physics and Astronomy, Bausch & Lomb Hall, P.O. Box 270171, 600 Wilson Boulevard, Rochester, NY 14627-0171 USA
364 University of Science and Technology of China, Department of Modern Physics (DMP), Jin Zhai Road 96, Hefei, China 230026
365 University of Silesia, Institute of Physics, Ul. Uniwersytecka 4, PL-40007 Katowice, Poland
366 University of South Carolina, Department of Physics and Astronomy, 712 Main Street, Columbia, SC 29208, USA
367 University of Southampton, School of Physics and Astronomy, Highfield, Southampton S017 1BJ, England, UK
368 University of Southern California, Department of Physics & Astronomy, 3620 McClintock Ave., SGM 408, Los Angeles, CA 90089-0484, USA
369 University of Sydney, Falkiner High Energy Physics Group, School of Physics, A28, Sydney, NSW 2006, Australia
370 University of Tartu, Institute of Physics, Riia 142, 51014 Tartu, Estonia
371 University of Texas at Austin, Department of Physics, 1 University Station C1600, Austin, Texas 78712, USA
372 University of Texas at Dallas, Department of Physics, 800 West Campbell Road, Richardson, Texas 75080, USA
373 University of Texas, Center for Accelerator Science and Technology, Arlington, TX 76019, USA
374 University of Tokushima, Institute of Theoretical Physics, Tokushima-shi 770-8502, Japan
375 University of Tokyo, Department of Physics, 7-3-1 Hongo, Bunkyo District, Tokyo 113-0033, Japan
376 University of Toronto, Department of Physics, 60 St. George St., Toronto M5S 1A7, Ontario, Canada
377 University of Toyama, Department of Physics, 3190 Gofuku, Toyama 930-8555, Japan
378 University of Tsukuba, Faculty of Pure and Applied Sciences, 1-1-1 Ten'nodai, Tsukuba, Ibaraki 305-8571, Japan
379 University of Victoria, Department of Physics and Astronomy, P.O.Box 3055 Stn Csc, Victoria, BC V8W 3P6, Canada
380 University of Virginia, Department of Physics, 382 McCormick Rd., PO Box 400714, Charlottesville, VA
381 University of Warsaw, Institute of Experimental Physics, Ul. Hoza 69, PL-00 681 Warsaw, Poland
382 University of Warsaw, Institute of Theoretical Physics, Ul. Hoza 69, PL-00 681 Warsaw, Poland
383 University of Washington, Department of Physics, PO Box 351560, Seattle, WA 98195-1560, USA
384 University of Wisconsin, Physics Department, Madison, WI 53706-1390, USA
385 University of Wuppertal, Gaußstraße 20, D-42119 Wuppertal, Germany
386 Università degli Studi di Milano, Dipartimento di Fisica, Via Celoria 16, 20133 Milano, Italy
387 Università degli Studi di Roma "La Sapienza", Dipartimento di Fisica, Piazzale Aldo Moro 5, 00185 Roma, Italy
388 Università degli Studi di Trieste, Dipartimento di Fisica, via A. Valerio 2, I-34127 Trieste, Italy
389 Università dell'Insubria in Como, Dipartimento di Scienze CC.FF.MM., via Vallegio 11, I-22100 Como, Italy
390 Università di Milano-Bicocca, Dipartimento di Fisica"G. Occhialin", Piazza della Scienza 3, 20126 Milano, Italy
391 Università di Pisa, Departimento di Fisica "Enrico Fermi", Largo Bruno Pontecorvo 3, I-56127 Pisa, Italy
392 Universiy of Huddersfield, International Institute for Accelerator Applications, Queensgate Campus, Huddersfield HD1 3DH, UK
393 UPMC Univ. Paris 06, Faculté de Physique (UFR 925), 4 Place Jussieu, 75252 Paris Cedex 05, France
394 Vietnam National University, Laboratory of High Energy Physics and Cosmology, Faculty of Physics, College of Science, 334 Nguyen Trai, Hanoi, Vietnam
395 Vietnam National University, University of Natural Sciences, 227 Nguyen Van Cu street, District 5, Ho Chi Minh City, Vietnam
396 VINCA Institute of Nuclear Sciences, Laboratory of Physics, PO Box 522, YU-11001 Belgrade, Serbia
397 Virginia Commonwealth University, Department of Physics, P.O. Box 842000, 701 W. Grace St.,Richmond, VA. 23284-2000, USA
398 Virginia Polytechnic Institute and State University, Physics Department, Blacksburg, VA 2406, USA
399 Vrije Universiteit Brussel, Pleinlaan 2, 1050 Brussels, Belgium
400 Vrije Universiteit, Department of Physics, Faculty of Sciences, De Boelelaan 1081, 1081 HV Amsterdam, Netherlands
401 Warsaw University of Technology, The Faculty of Electronics and Information Technology, ul. Nowowiejska 15-19, 00-665 Warsaw, Poland
402 Waseda University, Advanced Research Institute for Science and Engineering, Shinjuku, Tokyo 169-8555, Japan
403 Wayne State University, Department of Physics, Detroit, MI 48202, USA
404 Weizmann Institute of Science, Department of Particle Physics, P.O. Box 26, Rehovot 76100, Israel
405 Yale University, Department of Physics, New Haven, CT 06520, USA
406 Yamagata University, 1-4-12 Kojirakawa-cho, Yamagata-shi, Yamagata, 990-8560, Japan
407 Yerevan Physics Institute, 2 Alikhanyan Brothers St., Yerevan 375036, Armenia
408 Yonsei University, Department of Physics, 134 Sinchon-dong, Sudaemoon-gu, Seoul 120-749, Republic of Korea

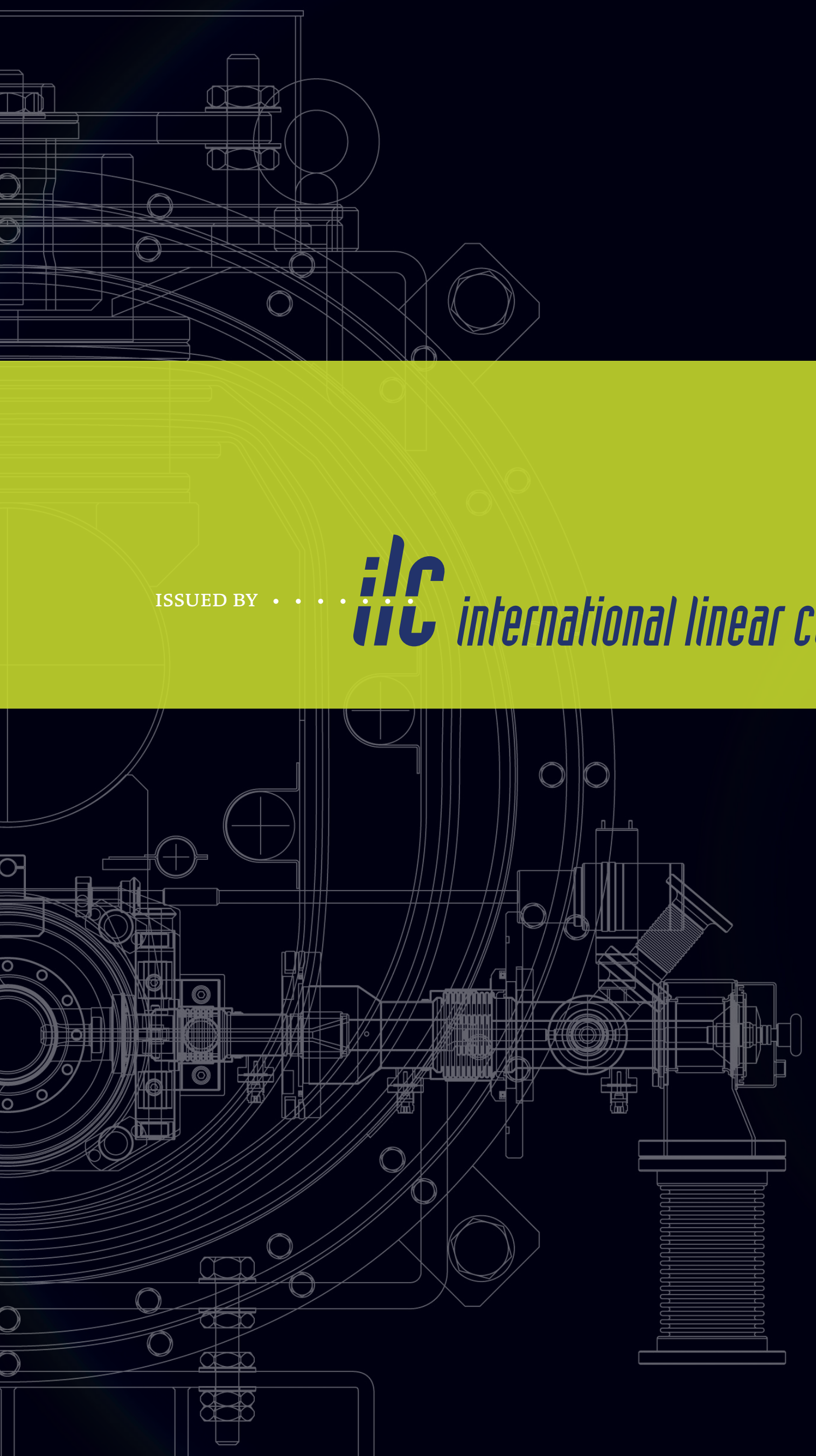

ISSUED BY · · · · · · · ilc international linear collider

WWW.LINEARCOLLIDER.ORG